\documentclass[a4paper, 12pt, twoside]{Thesis}
\usepackage{graphicx}
\usepackage[dvipsnames]{xcolor}
\usepackage{threeparttable}
\usepackage[justification=justified,singlelinecheck=false]{caption}
\usepackage{pdflscape}
\usepackage{enumerate}
\usepackage{rotating}
\usepackage{longtable}
\usepackage{mathrsfs}
\usepackage{addfont}
\usepackage{wasysym}
\usepackage{pifont}
\usepackage{afterpage} 
\usepackage{addfont}
\addfont[1.25]{OT1}{hge}{\hge}
\graphicspath{{./Figures/}}
\usepackage{natbib}
\bibpunct{(}{)}{;}{a}{}{,}
\usepackage{bigints}
\usepackage{verbatim} 
\usepackage{hyperref}
\hypersetup{colorlinks, citecolor=blue, filecolor=black, linkcolor=blue, urlcolor=blue}
\usepackage{ftnxtra}
\usepackage{fnpos}
\usepackage{sidecap}
\usepackage{epigraph}
\usepackage{adjustbox}
\usepackage{comment}
\usepackage{bm}
\usepackage{bibentry}
\usepackage{color, soul}
\usepackage{caption}
\usepackage{mwe}
\usepackage{styles/srk_thesis_ref}
\usepackage{multirow}
\usepackage{multicol}
\usepackage{tabularx}
\usepackage{adjustbox}
\usepackage{threeparttable}
\usepackage{enumitem}

\DeclareUnicodeCharacter{2212}{-}
\def\myfnt{\ifx\protect\@typeset@protect\expandafter\footnote\else\expandafter\@gobble\fi}

\newcommand\ion[2]{#1$\;${%
\ifx\@currsize\normalsize\small \else
\ifx\@currsize\small\footnotesize \else
\ifx\@currsize\footnotesize\scriptsize \else
\ifx\@currsize\scriptsize\tiny \else
\ifx\@currsize\large\normalsize \else
\ifx\@currsize\Large\large
\fi\fi\fi\fi\fi\fi
\uppercase\expandafter{\romannumeral #2\relax}}}%

\usepackage{charter} 
\usepackage{newtxmath} 
\usepackage{booktabs}
\usepackage{makecell}
\usepackage{lettrine}

\usepackage{array}    
\newcolumntype{R}[1]{>{\raggedright\arraybackslash}p{#1}}

\newcommand{\smallheader}[1]{%
  \vspace{0.5cm}%
  {\large \textbf{#1} \vspace{0.2cm} \\ }%
}

\usepackage{amssymb}

\begin{document}

\frontmatter         
\pagenumbering{alph}
		 




\thispagestyle{empty}

\begin{center}
    \noindent\rule{\linewidth}{2pt}

    {\huge \bfseries \setstretch{1.2}
    Kinematics and Thermodynamics of \\[0.2cm]
    Coronal Mass Ejections}

    \vspace{0.3cm}
    \noindent\rule{\linewidth}{2pt}

    \vspace{0.5cm}
    {\Large A thesis \\[0.2cm]
    submitted for the degree of} \\[0.2cm]
    {\normalfont\hge {\Large{{Doctor of Philosophy}}}}\\ 

    \vspace{0.2cm}
    {\Large in \\
    Department of Physics \\ [0.2cm]
    Pondicherry University, Puducherry 605014, India}

    \vspace{0.4cm}
    \includegraphics[width=0.18\textwidth]{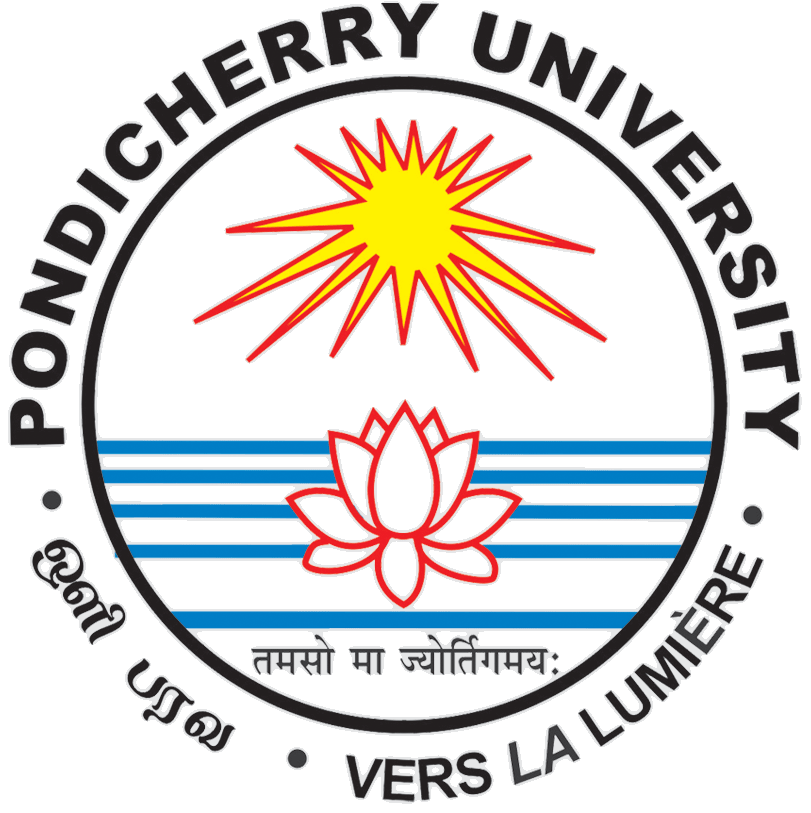}

    \vspace{0.4cm}
    {\Large  by}

    \vspace{0.2cm}
    {\huge \bfseries Soumyaranjan Khuntia}

    \vspace{0.2cm}
    {\Large Indian Institute of Astrophysics \\[0.2cm]
    Bengaluru 560034, India}

    \vspace{0.5cm}
    \includegraphics[width=0.17\textwidth]{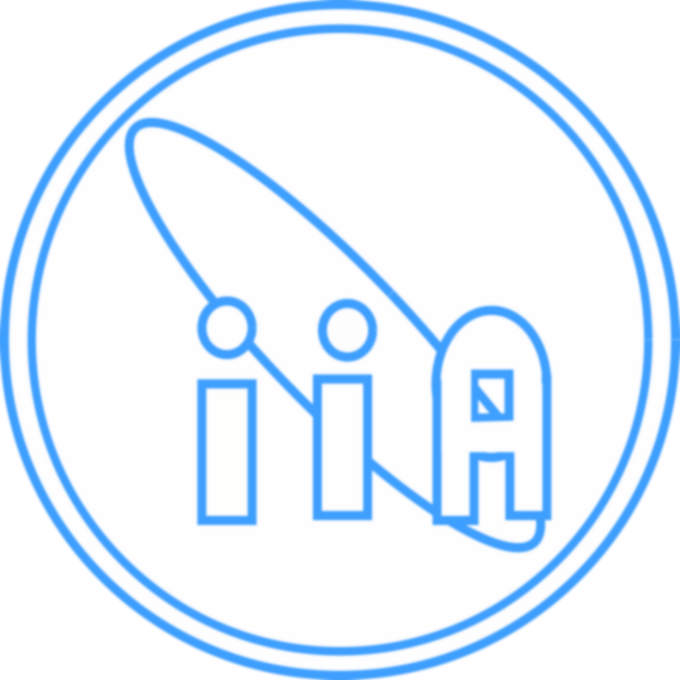}

    \vfill
    {\Large September 2025}
\end{center}
\pagestyle{empty}
\clearpage\null\newpage
\setboolean{@twoside}{true}




\thispagestyle{empty}

\begin{center}
\noindent

\vspace{0.6 cm}
\setstretch{1.5}
\textbf {\Huge {Kinematics and Thermodynamics of Coronal Mass Ejections}}
\\
\vspace{4.0 cm}
{\huge {Soumyaranjan Khuntia}}\\ 
\textit{ \large {Indian Institute of Astrophysics }} \\
\vspace{11.0cm}
        \hspace{0.02cm}   \includegraphics[width=0.170\textwidth]{logo_IIA} 

\setstretch{1.2}
\vspace{-0.5cm}
\begin{center}\large Indian Institute of Astrophysics \\  Bengaluru 560034, India\\ \end{center}
\large
\end{center}

\pagestyle{empty}             
\clearpage\null\newpage
\thispagestyle{plain}
\onehalfspacing

\thispagestyle{plain}
\onehalfspacing

   
 



\vspace*{3cm}
\noindent\rule{\linewidth}{1.5pt}
\vspace{0.2cm}\\
\renewcommand{\arraystretch}{1.2} 
\begin{tabular}{@{} p{3.cm} p{0.2cm} p{8.5cm} @{}}
\textbf{Thesis Title} & : & \textbf{\large Kinematics and Thermodynamics of Coronal Mass Ejections} \\ \\

\textbf{Author} & : & \textbf{\large Soumyaranjan Khuntia} \\

\textbf{Address} & : & Indian Institute of Astrophysics \\
                 &   & II Block, Koramangala \\
                 &   & Bengaluru 560034, India \\

\textbf{Email} & : & \href{mailto:soumyaranjan.khuntia@iiap.res.in}{soumyaranjan.khuntia@iiap.res.in} \\
               &   & \href{mailto:khuntias133@gmail.com}{khuntias133@gmail.com} \\ \\

\textbf{Supervisor} & : & \textbf{\large Dr. Wageesh Mishra} \\ 

\textbf{Address} & : & Indian Institute of Astrophysics \\
                 &   & II Block, Koramangala \\
                 &   & Bengaluru 560034, India \\

\textbf{Email} & : & \href{mailto:wageesh.mishra@iiap.res.in}{wageesh.mishra@iiap.res.in} \\
              &   & \href{mailto:m.wageesh30@gmail.com}{m.wageesh30@gmail.com} \\

\end{tabular}

\vspace{0.3cm}
\noindent\rule{\linewidth}{1.5pt}
\thispagestyle{empty}

\clearpage\null

\phantomsection
\normalfont
\setstretch{1.3} 

\fancyhead{}  
\rhead{}
\lhead{}
\pagestyle{empty}

\noindent
\vspace{2.5cm}
\newpage
\vspace*{2.5cm}
\begin{center}{\huge\bf Certificate}\end{center}
\thispagestyle{empty}
This is to certify that the thesis titled \textbf{``Kinematics and Thermodynamics of Coronal Mass Ejections''}, submitted to Pondicherry University by Mr. Soumyaranjan Khuntia for the award of the degree of Doctor of Philosophy, is based on the results of investigations carried out by him under my supervision and guidance at the Indian Institute of Astrophysics. To the best of our knowledge, the thesis is an original work and has not been submitted, either in this university or in any other university or institution in India or abroad, for the award of any degree, diploma, fellowship, or equivalent qualification. The thesis is worthy of consideration for the award of the degree of Doctor of Philosophy (Ph.D.) in Physics.

\vspace{1.5cm}

\textbf{Dr. Wageesh Mishra}\\
Thesis Supervisor,\\
Indian Institute of Astrophysics,\\
Bengaluru 560034, India

\vspace{2.cm}

 

\noindent{Signature:}  \\[1.2cm]
\noindent{Date:}

\clearpage

\clearpage\null\newpage

\noindent
\vspace*{2.5cm}
\begin{center}{\huge\bf Declaration of Authorship}\end{center}
\thispagestyle{empty}
I hereby declare that the research contained in this thesis is the result of the investigations carried out by me at the Indian Institute of Astrophysics, Bengaluru, under the supervision of Dr. Wageesh Mishra. This work has not been submitted for the award of any degree or equivalent qualification at this university or at any other university or institution in India or abroad.

\vspace{1.5cm}
\textbf{Soumyaranjan Khuntia}\\
Ph.D. Student, \\
Indian Institute of Astrophysics,\\
Bengaluru 560034, India

\vspace{2.cm}



\noindent{Signature:}  \\[1.2cm]
\noindent{Date:}

\clearpage  
\thispagestyle{empty}

\clearpage\null\newpage

\pagestyle{empty}
\setstretch{1.3} 
\pagestyle{empty}
\addtocontents{toc}{\protect\setcounter{tocdepth}{-1}}

\acknowledgements{
\thispagestyle{empty}
\setstretch{1.3} 

It is a privilege to express my heartfelt gratitude to all those who have supported and guided me during my doctoral journey.

First and foremost, I am deeply indebted to my PhD supervisor, Dr. Wageesh Mishra, whose unwavering guidance and support steered me through the complexities of research, laying a strong foundation for my future as well. His clarity of thought and patience helped me refine even the most half-formed ideas into something substantial. His ability to simplify complex ideas, his patience in addressing my doubts, and his unwavering belief in my potential kept me moving forward even in challenging times.

I am also grateful to my collaborators, whose diverse strengths enriched this research. Dr. Sudheer K. Mishra patiently taught me the basics at the outset; Prof. Yuming Wang, with his brilliance and prompt suggestions, kept our project on track and taught me a great deal about dedication. Prof. Jie Zhang, with his humility and insightful feedback, helped me look deeper into the problems at hand. Prof. Teresa Nieves-Chinchilla provided precise and valuable feedback that always struck at the core of an issue. I also thank Shaoyu Lyu and Anjali Agarwal for their collaboration and support, which made our joint work both productive and enjoyable.

I sincerely thank my doctoral committee, Prof. S. P. Rajaguru and Prof. K. V. P. Latha, for their insightful feedback and constructive suggestions throughout my PhD. I am also grateful to my comprehensive examination external examiner, Prof. Abhishekh K. Srivastava, for his valuable comments and guidance.  

It has been a privilege to be part of this institute. I extend my thanks to the Director, the Dean, the BGS Chair, and the BGS office for their constant support and encouragement. I also deeply appreciate the time and effort they invested in reviewing my work and in supporting my academic progress.  I am equally thankful to all the Bhaskara staff for their care, for providing healthy food, and for their thoughtful understanding of students’ needs, which made day-to-day life smoother.

I am fortunate to have had seniors such as Satabdwa, Ankit, Raghubar, Swastik, Suman, Sonith, Priya, Partha, Vishnu, Jyoti, Rishabh, Ravi, Anohita, and Pallavi, who were always generous with their advice, answered my silly questions, and made life at the institute and Bhaskara more welcoming and enjoyable.

My batchmates have been an integral part of this journey. I thank Neeraj, my roommate, for maintaining a peaceful room environment; Sipra, with whom I have shared not just academic struggles but daily meals over the past four years; Amrutha, Judhajeet, and Parvathy, for sharing countless memorable times; Anisha, Payel, Amlan, and Saraswati, for their companionship, which made these years memorable. I appreciate the sense of family we built together.

I am also grateful to my friends from other institutes, Debesh, Sandeep, Ayushi, Dipali, Dibyakriti, Upasna, Jyoti, and Nitin, for the warmth of their friendship. Though we were often miles apart, our discussions and small moments during some conferences always brought happiness.

The energy and enthusiasm of my juniors, Preity, Saili, Renu, Shivani, Rupesh, Swagata, Sunit, Saurav, Pravash, Nitish, Puja, Chandan, Annu, Shatakshi, and Lupa, added much to my PhD life. Interacting with them, whether in research discussions or on the badminton court or on outreach activities, has been a source of both learning and joy. I have always tried to share with them not just academic guidance but also small lessons about life, hoping they will carry them forward in their own journeys.  

I owe a special debt of gratitude to my college teachers. Ms. Gita and Mr. Prafula recognized my potential and encouraged me to pursue a career in research. Their early belief in me lit the spark that eventually led me to where I am today.

I also cherish my time with the DOOT magazine team. Working on outreach projects, taking on leadership roles, conducting interviews, and editing articles taught me skills beyond academic research. The creativity and teamwork of this group provided a refreshing break from equations and made my PhD journey more well-rounded.

Finally, none of this would have been possible without the unconditional love and support of my family. My parents have been my strongest pillars, giving me patience, encouragement, and faith at every step. My sister and brother-in-law, along with my brothers and sisters, have been constant sources of strength. I am especially grateful to my paternal uncle and aunt, whom I regard as my second parents, for their endless care and guidance. To all of them, I owe everything. 

\flushright -- Soumyaranjan Khuntia

\pagestyle{empty}
\setstretch{1.5} 
}

\dedicatory{\thispagestyle{empty}
\vspace*{2cm}

\begin{center}
\textit{To the roots that held me firm and the wings that let me fly...}
\vspace{2em}

{\LARGE \textit{\textbf{To my father and mother,}}}\\
\textit{You taught me the meaning of hard work, humility, empathy,\\
and how to dream with my feet on the ground.}

\vspace{1.5em}

{\LARGE \textbf{\textit{To my sister,}}}\\
\textit{The pole star in my sky. Your cheer, patience,\\
and silent strength have been my constant companions.}

\vspace{1.5em}

\textit{This journey is mine, but its foundation is yours.}

\vspace{2em}

\scalebox{2}{\color{red}$\heartsuit$}

\end{center}
}
\clearpage

\addtocontents{toc}{\protect\setcounter{tocdepth}{2}}
\clearpage

\pagenumbering{roman}
\frontmatter

\Abstract{
\addtocontents{toc}{}

Coronal Mass Ejections (CMEs) are powerful solar eruptions that expel magnetized plasma into the heliosphere, often triggering severe space weather disturbances at Earth. While their large-scale kinematics and magnetic structures are well studied, the internal thermodynamic evolution of CMEs, from the low corona to 1~au, remains inadequately understood. This thesis aims to bridge this gap through a comprehensive investigation that integrates analytical modeling, remote observations, in situ measurements, and statistical analysis.

One of the core focuses of the thesis is the development and refinement of the Flux Rope Internal State (FRIS) model, an analytical framework that derives internal thermodynamic parameters, such as polytropic index, heating rate, and internal forces, using the remotely measurable kinematic quantities from coronagraphs (SOHO/LASCO and STEREO/COR). The original formulation of FRIS is critically re-evaluated, and key mathematical inconsistencies are corrected to ensure physical self-consistency in the derived quantities. This improved model is then applied to selected CMEs with contrasting kinematics, revealing detailed insights into their thermal state evolution and internal force balance during the early propagation phase. These results are further compared with in situ analyses of two selected ICMEs, where sustained heating is indicated by near-isothermal effective polytropic indices ($\Gamma\approx $ 0.88 \& 0.76) and supported by turbulence diagnostics, such as inertial and dissipation-scale spectral slopes, low magnetic compressibility, and enhanced intermittency in the sheath and post-ICME regions, reinforcing the continuity of heating from the low corona to 1 au.

Beyond case-specific implementation, a focused investigation of nine Earth-direc\\ted, fast CMEs using the improved FRIS model and 3D kinematics derived from the Graduated Cylindrical Shell (GCS) reconstruction reveals a two-phase thermal evolution: an initial heat-release phase ($\Gamma > 5/3$) followed by a transition into a nearly isothermal heating state ($\Gamma \approx 0.8$--$1.2$) at heliocentric distances between 3--7\,$R_\odot$. This evolution is closely coupled with the CMEs' expansion behavior, where CMEs with slower expansion acceleration show less pronounced temperature drops before reaching isothermal conditions. A detailed internal force balance analysis indicates that Lorentz forces act to suppress expansion during the early propagation phase, whereas thermal pressure and centrifugal forces drive the expansion, with thermal pressure becoming the dominant contributor at larger heliocentric distances. Differential emission measure (DEM) analysis of the CME source regions using SDO/AIA EUV data corroborates the presence of intrinsically hot plasma (2.8--7.2\,MK), in agreement with the FRIS-derived heating trends. These findings challenge the often assumed constancy of $\Gamma$ in CME modeling and highlight the necessity of incorporating its dynamic evolution into global MHD simulations and predictive tools.

The statistical component of the thesis analyzes CMEs across Solar Cycles  23, 24, and the ascending phase of Cycle 25 using in situ observations at 1 au from missions such as Wind and ACE. Within a polytropic framework, the thermal state of magnetic ejecta (MEs) is characterized using event-wise median proton polytropic indices ($\Gamma_p$), allowing classification into heating and cooling categories. MEs predominantly exhibit non-equilibrium thermal behavior, with a substantial fraction ($\sim$45\%) undergoing net heating during interplanetary propagation, especially near solar maxima. In contrast, cooling MEs maintain nearly constant and elevated $\Gamma_p$ values ($\sim$2), indicating enhanced energy loss beyond adiabatic expectations or possible thermal retention from their eruption. A distinct solar-cycle modulation is observed in $\Gamma_p$, proton temperature, and expansion speed, with the median $\Gamma_p$ increasing from 1.46 in Solar Cycle 23 to 1.87 in Cycle 24, suggesting a shift toward more cooling-dominated states over time. Despite thermodynamic similarities between magnetic clouds (MCs) and non-cloud ejecta, MCs consistently emerge as the most geoeffective structures, characterized by enhanced magnetic fields, low plasma beta, and suppressed $\Gamma_p$ values. Further, superposed epoch analyses (SEA) are performed on categorized ICME events, which reveal systematic variations in thermal state, plasma temperature, and magnetic field strength across distinct ICME substructures, including the pre-ICME, sheath, magnetic ejecta, and post-ICME regions. Notably, High-impact ICMEs are predominantly found to be thermally heated, with lower $\Gamma_p$ values, stronger sheath compression, elevated bulk speeds, and trailing high-speed solar wind, all contributing to enhanced geomagnetic storm potential.

Finally, a detailed study of CME--CME interactions associated with the 10 May 2024 great storm shows that such interactions significantly modify both kinematics and thermodynamics of the involved structures. The resulting complex ejecta at 1~au exhibit enhanced magnetic fields, heat-release-dominated electrons, and bimodal ion thermal states. FRIS model analysis reveals diverse thermal behaviors among the individual CMEs, with most exhibiting a transition to an isothermal state by 6--9\,$R_\odot$, except in cases of suppressed expansion. The findings underscore that CME--CME interactions can obscure one-to-one comparisons between solar-origin properties and in situ measurements, while also enhancing internal heating, structural complexity, and geoeffectiveness.

This thesis concludes by synthesizing results from analytical modeling, selected case studies, and long-term statistical analyses to construct a unified understanding of CME thermodynamics in the low corona and at 1~au. The findings reveal that CME thermal evolution is highly variable and depends on factors such as expansion dynamics, Solar Cycle phase, and interplanetary interactions. By integrating remote-sensing and in situ observations within a physically consistent thermodynamic framework, this work successfully bridges the gap between observational regimes, revealing the dynamic interplay between CME kinematics and internal thermal evolution. These insights not only advance the fundamental understanding of CME physics but also offer new pathways for improving space weather forecasting through diagnostics based on CME thermal state.

}

\clearpage
\pagestyle{fancy}
\lhead{\emph{Contents}}
\tableofcontents

\lhead{\emph{List of Figures}}
\listoffigures 

\lhead{\emph{List of Tables}}
\listoftables 

\clearpage

\fancyhead{}  
\rhead{}
\lhead{}

\thispagestyle{empty}
\noindent
\pagestyle{plain}
\addtotoc{List of Publications}
{\addtocontents{toc}{}

\vspace*{0.1cm}
\begin{center}
{\huge\bf List of Publications}
\end{center}
\begin{enumerate}[leftmargin=0pt]

    \item Wageesh Mishra, Yuming Wang, Shaoyu Lyu, and {\bf {\color{purple} Soumyaranjan Khuntia}}, 2023, Erratum: Modeling the Thermodynamic Evolution of Coronal Mass Ejections Using Their Kinematics (2018, ApJ, 865, 50), \textit{\textbf{The Astrophysical Journal}, Volume 952, Number 2, Page 173,} \href{http://doi.org/10.3847/1538-4357/ace691}{DOI:10.3847/1538-4357/ace691}

    \item {\bf {\color{purple} Soumyaranjan Khuntia}}, Wageesh Mishra, Sudheer K. Mishra, Yuming Wang, Jie Zhang, and Shaoyu Lyu, 2023, Unraveling the Thermodynamic Enigma between Fast and Slow Coronal Mass Ejections, \textit{\textbf{The Astrophysical Journal}, Volume 958, Number 1, Page 92,} \href{http://doi.org/10.3847/1538-4357/ad00ba}{DOI:10.3847/1538-4357/ad00ba}

    \item Wageesh Mishra, Preity Sukla Sahani, {\bf {\color{purple} Soumyaranjan Khuntia}}, and Dibyendu Chakrabarty, 2024, Distribution and Recovery Phase of Geomagnetic Storms During Solar Cycles 23 and 24, \textit{\textbf{Monthly Notices of the Royal Astronomical Society}, Volume 530, Issue 3, Page 3171,} \href{https://doi.org/10.1093/mnras/stae1045}{DOI: 10.1093/mnras/stae1045}

    \item {\bf {\color{purple} Soumyaranjan Khuntia}}, Wageesh Mishra, Yuming Wang, Sudheer K. Mishra, Teresa Nieves-Chinchilla and Shaoyu Lyu, 2024, Deciphering the Evolution of Thermodynamic Properties and their Connection to the Global Kinematics of High-speed Coronal Mass Ejections using FRIS Model, \textit{\textbf{Monthly Notices of the Royal Astronomical Society}, Volume 535, Issue 3, Page 2585,} \href{https://doi.org/10.1093/mnras/stae2523}{DOI: 10.1093/mnras/stae2523}

    \item {\bf {\color{purple} Soumyaranjan Khuntia}}, Wageesh Mishra, and Anjali Agarwal, 2025a, Evolution of the Interacting Coronal Mass Ejections that Drove the Great Geomagnetic Storm of 10 May 2024,  \textit{\textbf{Astronomy and Astrophysics}, Volume 698, Article Number A79, Pages 14,} \href{https://doi.org/10.1051/0004-6361/202452866}{DOI: 10.1051/0004-6361/202452866}


     \item {\bf {\color{purple} Soumyaranjan Khuntia}} and Wageesh Mishra, 2025b, Thermal and Turbulence Characteristics of Fast and Slow Coronal Mass Ejections at 1 AU, \textit{\textbf{Journal of Astrophysics and Astronomy}, Volume 46, Issue 2, Page 70}, \href{https://doi.org/10.1007/s12036-025-10085-5}{DOI: 10.1007/s12036-025-10085-5}

    \item {\bf {\color{purple} Soumyaranjan Khuntia}} and Wageesh Mishra, 2025c, Thermal properties of interplanetary coronal mass ejections at 1 AU and their connection to geoeffectiveness across solar cycles 23-25, \textit{\textbf{Monthly Notices of the Royal Astronomical Society}, Volume 545, Issue 4, } \href{https://doi.org/10.1093/mnras/staf2242}{DOI: 10.1093/mnras/staf2242}

\end{enumerate}
}

\clearpage\null
\pagestyle{empty}
\newpage

\thispagestyle{empty}
\pagestyle{plain}
\addtotoc{List of Conference \& Workshop}
\vspace*{0.7cm}
\begin{center}
{\huge\bf Conferences \& Workshops}
\end{center}
\vspace{0.7cm}
{\large \bf \it A) International}
\begin{enumerate}[leftmargin=0pt]

\item \textbf{\color{purple}Poster} presented on ``\textbf{Thermal Evolution of Earth-Directed CMEs Driving the Extreme Geomagnetic Storm on May 10, 2024}," in the \textit{International Conference on Sun, Space Weather, and Solar-Stellar Connections}, organized by \textit{IIA} in January 2025.

\item \textbf{\color{purple}Online participant} in the \textbf{May 2024 Solar and Geospace Superstorm Workshop}, organized by \textit{Johns Hopkins Applied Physics Laboratory (APL), US} in October 2024.

\item \textbf{\color{purple}Contributed talk} presented on ``\textbf{Integrating Kinematics and Thermodynamics of Coronal Mass Ejections Through Observations and Analytical Modeling}," in the \textit{IAU Symposium 388: Solar and Stellar Coronal Mass Ejections}, organized at \textit{Astronomical Observatory of the Jagiellonian University, Krakow, Poland} in May 2024.

\item \textbf{\color{purple}Online poster} presented on ``\textbf{Understanding the Thermal Properties of Fast CMEs by Integrating White-light Observations and Analytical Modeling}," in the \textit{European Geosciences Union (EGU) General Assembly}, organized at the \textit{Austria Center Vienna (ACV), Vienna, Austria} in April 2024.

\item \textbf{\color{purple}Contributed talk} presented online on ``\textbf{Analytical Modeling Approach to Understand the Thermodynamic Evolution of Coronal Mass Ejections}," in the \textit{30th Young Scientists' Conference on Astronomy and Space Physics} at Kyiv, Ukraine in April 2024.

\item \textbf{\color{purple}Contributed talk} presented online on ``\textbf{Exploring the Thermodynamic Evolution of  Coronal Mass Ejections: An Analytical Modeling Approach Using Global Kinematics}," in the \textit{European Solar Physics Online Seminars (ESPOS)} in February 2024.

\end{enumerate}

\vfill
\hfill \textit{Continued on next page}

\clearpage\null


{\large \bf \it B) National}
\begin{enumerate}[leftmargin=0pt]
   \item \textbf{\color{purple}Lecture} on ``\textbf{The Sun}," in the \textit{Advanced Summer School} at MP Birla Institute of Fundamental Research, India, in May 2025.

   \item \textbf{\color{purple}Conducted workshop} on ``\textbf{Remote and in situ observation of coronal mass ejections (CMEs)}," in the \textit{Kodaikanal Winter School} at Kodaikanal Solar Observatory (KSO), India in March 2025.
   
   \item \textbf{\color{purple}Poster} presented on ``\textbf{Investigating the Thermodynamics of CMEs and Insights From the Successive CMEs Driving the 11 May 2024 Great Geomagnetic Storm}," in the \textit{Astronomical Society of India (ASI) meeting} at NIT Rourkela, India, in March 2025.
   
   \item \textbf{\color{purple}Poster} presented on ``\textbf{Modeling the Thermodynamic State of Fast Coronal Mass Ejections During Their Propagation}," in the \textit{Young Astronomers’ Meet (YAM)} at CHRIST (Deemed to be University), India in February 2024.
   
   \item \textbf{\color{purple}Contributed talk} presented on ``\textbf{Modeling the Thermodynamic Enigma of Coronal Mass Ejections During Their Propagation towards Earth}," in the \textit{National Space Science Symposium (NSSS)} at Goa University, India, in February 2024.

   \item \textbf{\color{purple}Contributed workshop talk} presented on ``\textbf{Understanding the Thermodynamic Evolution of CMEs through Analytical Modeling}," in the \textit{Astronomical Society of India (ASI) meeting} at IISc Bangalore, India in February 2024.

   \item \textbf{\color{purple}Conducted workshop} on ``\textbf{Kinematics of CMEs from remote and in situ observation}," in the \textit{Kodaikanal Winter School} at Kodaikanal Solar Observatory (KSO), India, in January 2024.

   \item \textbf{\color{purple}Poster} presented on ``\textbf{Unravelling the internal thermal properties of coronal mass ejections (CMEs): implications for CME kinematics and space weather prediction}," in the \textit{USO-PRL Solar Physics Workshop} at Udaipur Solar Observatory, India, in April 2023.
       
\end{enumerate}

\vfill
\hfill \textit{Continued on next page}

\clearpage\null


\begin{enumerate}[leftmargin=0pt]
\setcounter{enumi}{8}
   \item \textbf{\color{purple}Participant} in the \textbf{second Aditya-L1 Support Cell (AL1SC) workshop} at Manipal Academy of Higher Education (MAHE), India in November 2022.

   \item \textbf{\color{purple}Contributed talk} presented on ``\textbf{ Evolution of the internal thermodynamic state of Coronal Mass Ejections}," in the \textit{Young Astronomers’ Meet (YAM)} at ARIES, Nainital, India in November 2022.

   \item \textbf{\color{purple}Participant} in the \textbf{Aditya-L1/VELC workshop} at Kodaikanal Solar Observatory (KSO), India in 2022.

   \item \textbf{\color{purple}Poster} presented online on ``\textbf{ Probing the thermodynamical state of a CME during its propagation in the heliosphere}," in the \textit{Astronomical Society of India (ASI) meeting} at IIT Roorkee, India in March 2022.
       
\end{enumerate}


\clearpage
\pagestyle{empty}
\newpage


\clearpage
\thispagestyle{empty}
\newpage

\addtocontents{toc}{\vspace{2em}}
\setboolean{@twoside}{true}  
\mainmatter
\pagestyle{fancy}

\lhead{\emph{Chapter 1}}

\quotechapter{1}{`` The true sign of intelligence is not knowledge\\but imagination.''\\ \smallskip — Albert Einstein}{Introduction}

\lhead{\emph{\textcolor{darkgray} Chapter 1: Introduction}}
\vspace{-1.cm}

\lettrine[lines=2, loversize=0.1]{S}{}ince the dawn of human consciousness, our Sun has commanded reverence and curiosity across all civilizations. Ancient Egyptians worshipped Ra as the supreme solar deity, while the Greeks personified the Sun as Helios, driving his chariot across the heavens. In Indian tradition, the Sun (Surya) has been worshiped since Vedic times as a symbol of life, energy, and spiritual awakening. Ancient texts, rituals like Surya Namaskar, and Sun temples such as Konark reflect this enduring cultural reverence. These spiritual and mythological frameworks, while scientifically primitive, reflected humanity's intuitive understanding that the Sun was not merely a celestial body but the fundamental source of all earthly energy and life \citep[chapter~1]{Bhatnagar2005}.

The transition from mythology to scientific understanding began with many philosophers all around the world. Nicolaus Copernicus proposed the heliocentric model in 1543, which fundamentally shifted our perspective, placing the Sun at the center of our cosmic neighborhood. The advent of telescopic astronomy in the early 17th century and Galileo Galilei's sunspot observations marked the beginning of systematic solar physics, a field that would eventually reveal the Sun not as a static orb, but as a complex system of interacting magnetic fields, plasma flows, and explosive phenomena. The discovery of the 11-year Solar Cycle by Heinrich Schwabe in 1843 and George Ellery Hale's identification of sunspot magnetic fields in 1908 opened the door to understanding the magnetic nature of solar activity. The 19th century brought spectroscopy, allowing scientists like Joseph von Fraunhofer to decode the Sun's chemical composition through its absorption lines. The discovery of helium in the solar spectrum before its terrestrial detection exemplified how solar studies could reveal fundamental properties of matter itself.

The 20th century witnessed an explosion of solar understanding. Hans Bethe's detailed fusion calculations, which earned him the Nobel Prize in 1967, solved the long-standing puzzle of the solar energy source \citep{Wijers2007}. The launch of Skylab in 1973 provided the first comprehensive view of the Sun in X-rays, unveiling a dynamic and structured solar atmosphere. The first white-light detection of a solar flare in 1859 and coronal mass ejection (CME) in 1971 \citep{Howard2023} fundamentally changed our understanding of solar-terrestrial relationships, showing that the Sun could have potentially catastrophic consequences at Earth. Today, we understand the Sun as a complex magnetohydrodynamic laboratory, and its influence extends far beyond providing light and warmth; it shapes the entire heliosphere. Particularly, the study of CMEs has evolved from serendipitous white-light observations to a sophisticated interdisciplinary field combining observations and advanced theoretical modeling \citep{Gopalswamy2016hist}. As our society becomes increasingly dependent on space-based technologies, understanding the complete dynamics of CMEs, from their magnetic origins in the Sun to their eventual interaction with planetary magnetospheres, has transformed from academic curiosity to a technological imperative.

\section{The Sun and its Atmosphere}

The Sun stands as a typical G2V-type spectral class main-sequence star, representing a middle-aged stellar object approximately 4.6 billion years into its 10-billion-year hydrogen burning lifetime.  With a mass of $1.99 \times 10^{30}$ kg and a radius of $6.96 \times 10^8$ m \citep[page~24]{Lang2006}, the Sun contains 99.86\% of the solar system's total mass, creating the gravitational architecture that governs planetary motion and maintains the orbital stability necessary for life on Earth.

\subsection{Solar Interior}
The Sun's internal structure consists of distinct regions, each characterized by unique physical processes and conditions that collectively drive the complex phenomena observed at the surface. Our current understanding of the solar interior has been revolutionized by three major scientific advances: the development of the standard solar model (SSM; \citealt{Bahcall1982}), the emergence of helioseismology as a precision diagnostic tool \citep{Leibacher1985, Basu2016}, and the detection and interpretation of solar neutrinos \citep{Bahcall2001}. The SSM provides theoretical predictions of the Sun's internal structure based on fundamental physics, stellar evolution theory, and observational constraints. Helioseismology, utilizing acoustic oscillations to probe the solar interior, has enabled precise measurements of the internal rotation profile and the location of the convection zone boundary with unprecedented accuracy. Solar neutrino observations have validated our understanding of nuclear fusion processes in the core and provided direct constraints on the central temperature and density conditions.

\begin{figure}[hbt]
    \centering
    \includegraphics[width=0.8\linewidth]{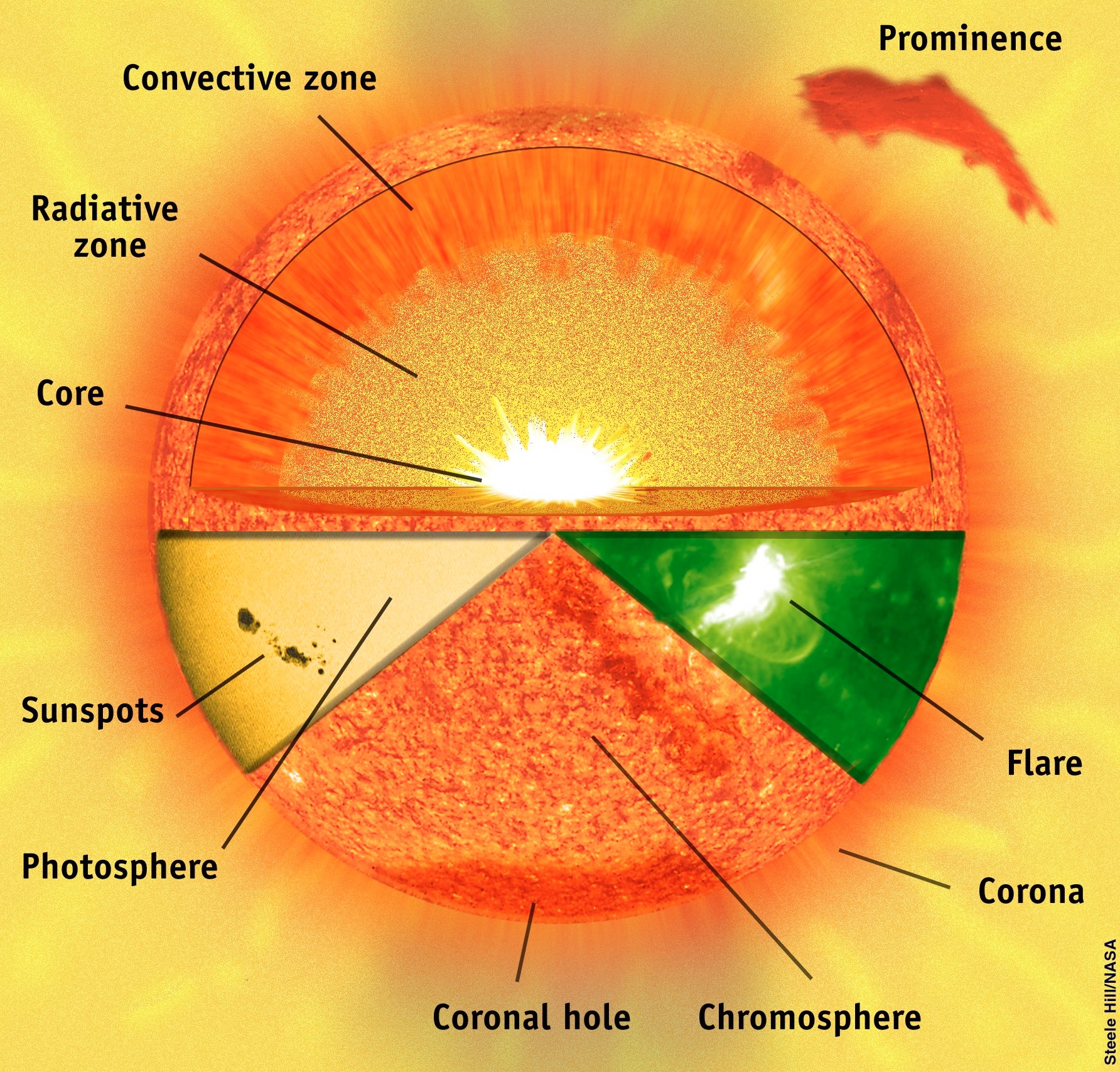}
    \caption[Schematic illustration of the Sun's internal and external]{Schematic illustration of the Sun's internal and external layers along with some notable features. [\textit{Image credit: \href{https://www.thesuntoday.org/sun/solar-structure/}{\textcolor{blue}{NASA}}}]}
    \label{chap1_fig:sun_structure}
\end{figure}

The solar interior comprises three primary zones: the core, radiative zone, and convective zone. The Sun's energy generation occurs through nuclear fusion in its core, which extends from the center to $ 0.25 $ $R_\odot$. The core maintains extreme physical conditions with temperatures reach approximately $1.56\times 10^{7}$  K and mass density exceed $1.5\times 10^{5}$ kg~m$^{-3}$ \citep[page~24]{Lang2006}. Under these conditions, the primary fusion mechanism follows the proton-proton chain reaction, converting hydrogen nuclei into helium through a series of nuclear reactions, releasing $3.8\times 10^{26}$ watts per second in all directions continuously \citep{Bhatnagar2005}. The energy release occurs primarily through gamma-ray photons, positrons, and neutrinos, with the neutrinos escaping directly, while the electromagnetic energy must diffuse through the overlying layers.

The radiative zone extends from 0.25 to approximately 0.7 $R_\odot$, where photons undergo countless absorption and re-emission events. This results in photon transport timescales of $10^{4}$ to $10^{5}$ years before energy reaches the convective zone \citep{Mitalas1992, Stix2003}. The radiative zone maintains a stable stratification with the Schwarzschild criterion for convective stability being satisfied ($[dT/dr]_{radiative} < [dT/dr]_{adiabatic}$), where the actual radiative temperature gradient is less than the adiabatic gradient. This stability is crucial for maintaining the sharp transition to the convective zone and the formation of the tachocline.

The solar convection zone extends from approximately $0.7\,R_\odot$ to the photosphere, with its depth precisely constrained by helioseismology measurements \citep{Thompson2004}. In this region, the increasing opacity inhibits radiative energy transport, thereby triggering convective instability. The convection zone is dominated by turbulent motions spanning a wide range of spatial scales, from granulation ($\sim$1~Mm) to supergranulation ($\sim$30~Mm), which permeate the entire layer. These convective flows not only transport thermal energy from the radiative interior to the solar surface on timescales of weeks to months but also facilitate the upward advection of magnetic flux tubes, leading to the formation of active regions on the photosphere.

The interface between the radiative and convective zones, named the tachocline, represents a thin shear layer ($\approx$0.04 $R_\odot$) where the transition from differential rotation in the convection zone to nearly uniform rotation in the radiative interior occurs. This region is fundamental to the solar dynamo mechanism, as it provides the strong radial shear necessary for generating toroidal magnetic fields from poloidal components.

\begin{figure}
    \centering
    \includegraphics[width=1.1\linewidth]{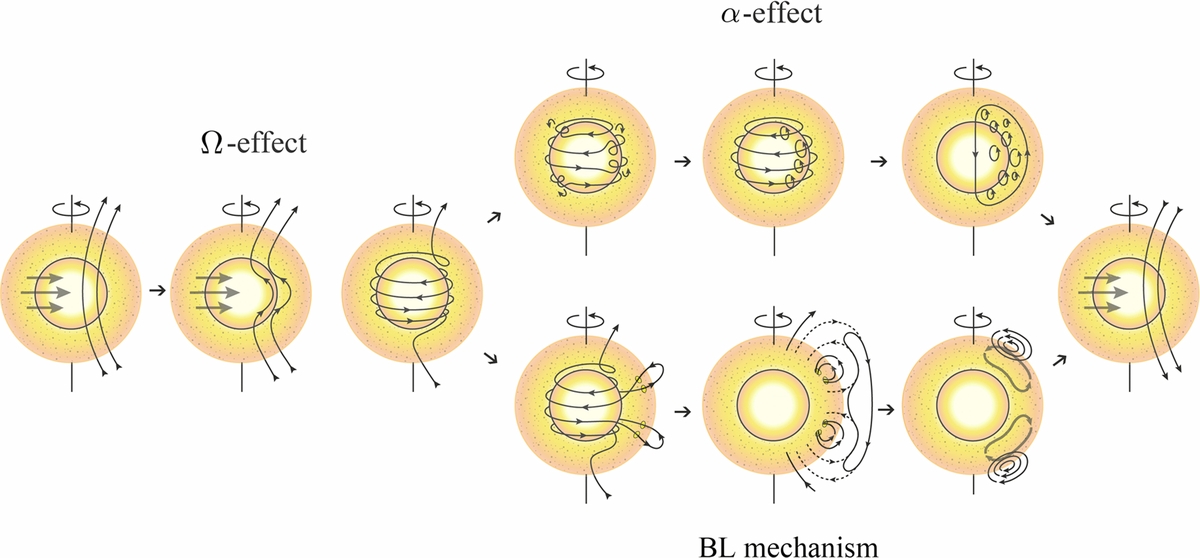}
    \caption[Schematic illustration of the solar dynamo process]{Schematic illustration of the solar dynamo process showing the $\Omega$-effect (left) generating toroidal field from poloidal field, and poloidal field regeneration via the $\alpha$-effect (top) and Babcock--Leighton mechanism (bottom). [\textit{Image Credit: \citealt{Sanchez2014}}]}
    \label{chap1_fig:dynamo}
\end{figure}

The convective motion plays a crucial role in generating and maintaining the Sun's magnetic field through the solar dynamo mechanism \citep{Charbonneau2020}, a magnetohydrodynamic process involving the complex interaction between differential rotation, meridional circulation, and convective turbulence. Figure~\ref{chap1_fig:dynamo} illustrates the commonly described solar dynamo framework of $\alpha$– $\Omega$ dynamo theory \citep{Charbonneau2020, Cameron2017, Dikpati2005}. The $\Omega$-effect (left) illustrates the transformation of a primary poloidal magnetic field into a toroidal field because of solar differential rotation. The regeneration of the poloidal field is subsequently achieved either via the $\alpha$-effect or the Babcock--Leighton (BL) mechanism. In the case of the $\alpha$-effect, the toroidal field located near the base of the convection zone is subjected to cyclonic convective turbulence, which generates small-scale poloidal loops. When averaged over large scales, these loops contribute to the reformation of a coherent poloidal field. Alternatively, the BL mechanism operates at the solar surface, where buoyant toroidal flux tubes rise through the convection zone and emerge as sunspot pairs. As these active regions evolve, the magnetic flux from sunspots near the equator tends to cancel across hemispheres, while the flux from those at higher latitudes migrates poleward. This high-latitude flux, possessing polarity opposite to the existing polar field, leads to polarity reversal. The newly established polar field is then advected by meridional circulation toward the base of the convection zone, completing the regeneration of the large-scale poloidal field. This cyclical process produces the well-known 22-year magnetic cycle. The dynamo mechanism explains not only the periodic nature of solar activity but also governs the emergence of active regions, the formation of sunspots, and the generation of complex magnetic field configurations. This intricate coupling between the solar interior dynamics and magnetic field generation establishes the foundation for understanding how subsurface processes influence the solar atmosphere and the broader heliospheric environment \citep{Low2001, Pinto2011, Petrie2013}.

\subsection{Solar Atmosphere}

\begin{figure}[hbt]
    \centering
    \includegraphics[width=\linewidth]{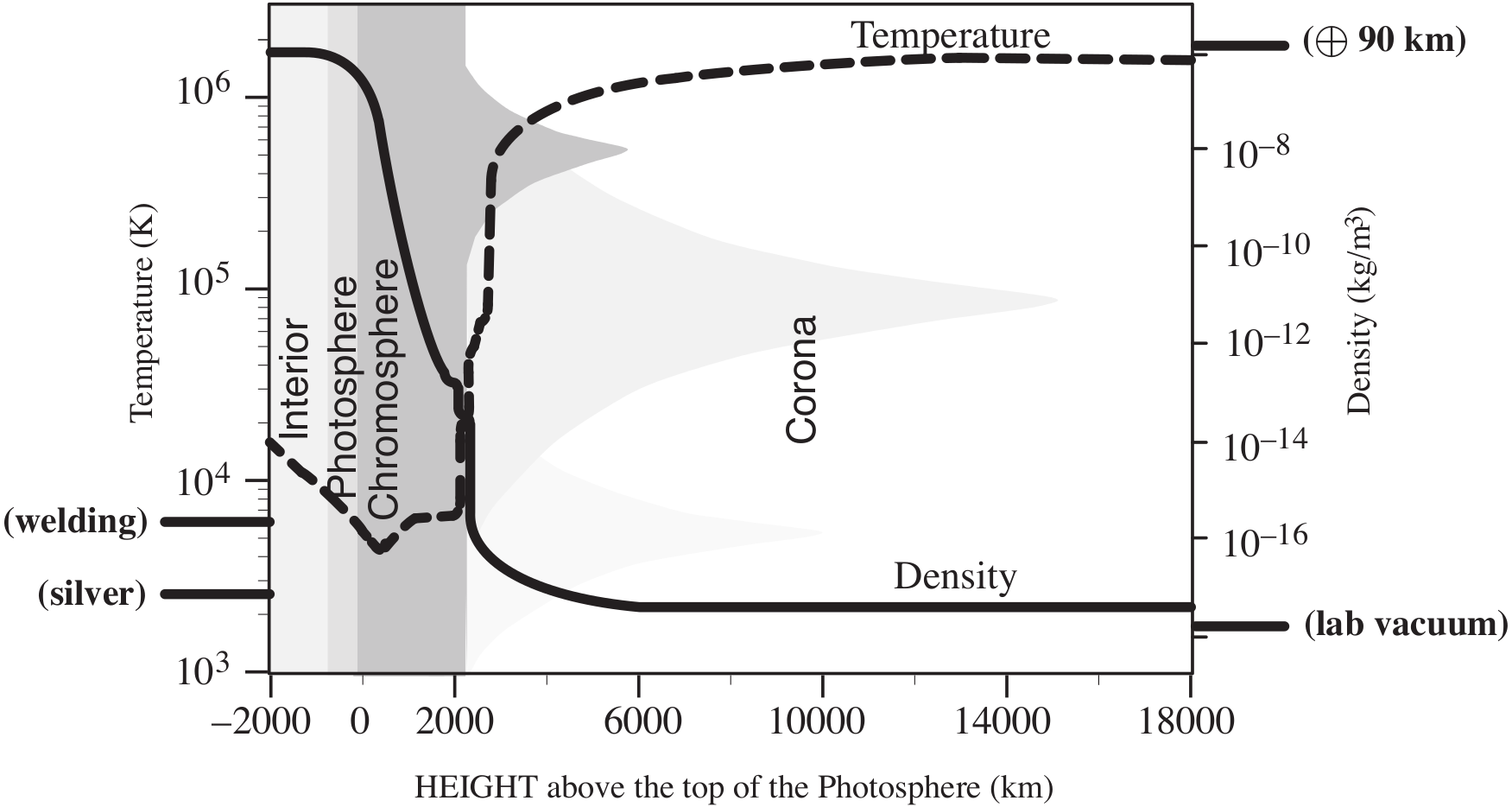}
    \captionsetup{width=\textwidth}
    \caption[Temperature (dashed line) and mass density (solid line)]{Temperature (dashed line) and mass density (solid line) profiles across the different layers of the solar atmosphere. [\textit{Image Credit:} \href{https://www.swpc.noaa.gov/content/solar-physics-and-terrestrial-effects-curriculum-guide-teachers-grades-7-12}{\textcolor{blue}{SWPC/NOAA}}]}
    \label{chap1_fig:sun_tempe_dens}
\end{figure}

The solar atmosphere extends from the visible surface outward into interplanetary space, exhibiting a complex stratified structure with distinct physical and thermal properties. This atmospheric hierarchy comprises the photosphere, chromosphere, transition region, and corona, each characterized by unique dominant physical processes that collectively govern solar radiation. The solar atmosphere exhibits a remarkable temperature structure that defies simple expectations, with temperatures initially decreasing outward from the interior before undergoing a dramatic reversal in the upper atmospheric layers (Figure~\ref{chap1_fig:sun_tempe_dens}). This temperature inversion represents one of the most fundamental unsolved problems in solar physics \citep{Moortel2015}.

\smallheader{The Photosphere}
The photosphere, literally meaning ``sphere of light," constitutes the Sun's visible surface and represents the layer from which most solar radiation escapes to space. It exhibits a temperature that decreases from about 6500 K at its base to approximately 4000 K at its upper boundary \citep{Foukal2004}. The photospheric magnetic field exhibits a hierarchical structure ranging from small-scale magnetic elements in the quiet Sun to large-scale active regions. Quiet Sun magnetic fields, with typical strengths of 10--100 G \citep{Rubio2019}, are organized into a network structure that outlines supergranular convection cells. Active regions, characterized by magnetic field strengths exceeding 1000 G, emerge through the buoyant rise of magnetic flux tubes from the base of the convection zone \citep{Abramenko2005}.

More often, the active regions manifested as Sunspots, a key photospheric phenomenon, representing regions of intense magnetic field concentration where the magnetic pressure suppresses convective energy transport, resulting in cooler temperatures than the surrounding quiet photosphere, and that's why they look darker. The magnetic field structure of sunspots includes a central umbra with nearly vertical magnetic fields exceeding 3000 G and a surrounding penumbra characterized by more inclined fields with complex fine structure \citep{Borrero2011, Rezaei2012}. The complexity of active region magnetic fields, including the presence of magnetic shear, twist, and current sheets, creates the conditions necessary for explosive magnetic reconnection events that drive solar flares and coronal mass ejections.


\smallheader{The Chromosphere}
Above the photosphere lies the chromosphere, a dynamic layer extending from approximately 400 to 2100 km above the visible surface, where the temperature rises from 4000 K at its base to 8000 K at its top \citep{Foukal2004}. This temperature increase, contrary to thermodynamic expectations, indicates the presence of non-radiative heating mechanisms that become increasingly important with height in the solar atmosphere.

The chromosphere exhibits a rich variety of dynamic phenomena, including spicules, mottles, fibrils, and prominences, all of which reflect the complex interaction between magnetic fields and plasma flows. Prominences represent particularly dramatic chromospheric phenomena, consisting of cool, dense plasma suspended in the hot corona by magnetic field configurations. These structures, which can extend tens of thousands of kilometers above the solar surface and persist for weeks or months, demonstrate the ability of magnetic fields to support material against gravity and thermal pressure gradients. Moreover, the chromosphere serves as a critical interface where photospheric magnetic fields undergo significant reconfiguration before extending into the corona, influencing the formation and evolution of coronal magnetic structures.


\smallheader{{The Transition Region}}
Perhaps the most dramatic feature of the solar atmosphere is the transition region, a thin layer of only 100-200 km in thickness where the temperature increases abruptly and reaches $10^4-10^6$ K \citep[page 300]{Foukal2004}. The transition region's structure is highly inhomogeneous, consisting of magnetic flux tubes embedded in a more tenuous medium. These flux tubes channel energy from the chromosphere to the corona and serve as conduits for mass exchange between atmospheric layers. The transition region frequently exhibits red-shifted spectral lines indicating downflows of material, suggesting that much of the transition region plasma is actually cooling coronal material rather than heated chromospheric material \citep{Brynildsen2001, Feldman2011}.


\smallheader{The Corona}
The solar corona represents the Sun's outermost atmospheric layer, extending millions of kilometers into space with temperatures exceeding one million Kelvin \citep[page 310]{Foukal2004}. This extreme temperature constitutes one of the most significant unsolved problems in solar physics and has driven decades of theoretical and observational investigations into coronal heating mechanisms \citep{Klimchuk2006, Moortel2015}. The coronal magnetic field structure also exhibits significant complexity with magnetic field lines extending from active regions to form loops, arcades, coronal holes, streamers, and large-scale structures that define the coronal architecture. These magnetic structures channel plasma flows, guide energy transport, and provide the framework for solar wind acceleration and coronal mass ejection formation. The magnetic field topology includes both closed field regions, where magnetic field lines connect different areas of the photosphere, and open field regions, where field lines extend indefinitely into the heliosphere, forming the source regions of the fast solar wind. Moreover, the corona exhibits significant temporal variability, ranging from rapid changes associated with solar flares and coronal mass ejections to long-term variations linked to the Solar Cycle.


\begin{figure}[ht]
    \centering
    \includegraphics[width=1.1\linewidth]{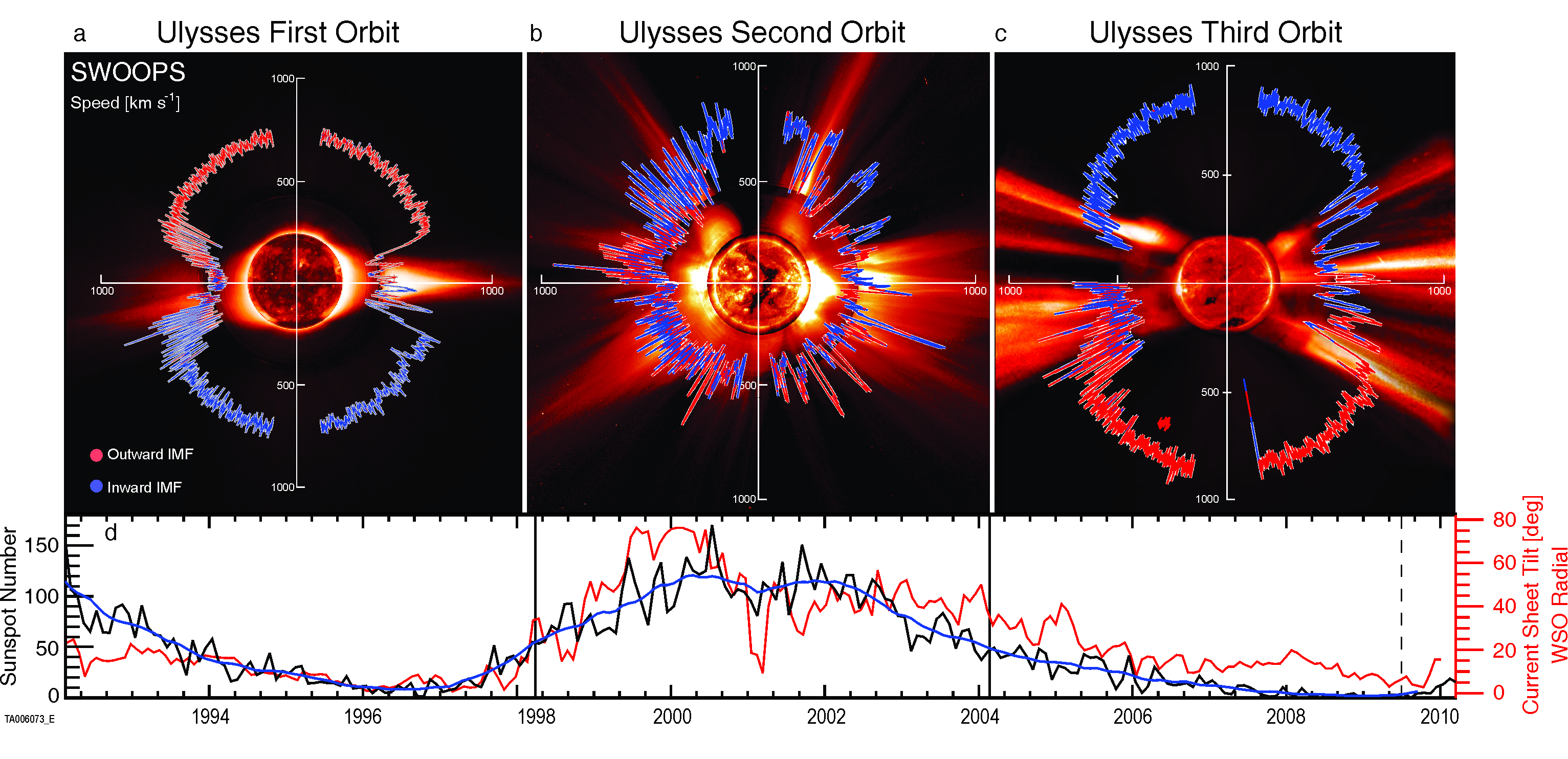}
    \caption[The Ulysses spacecraft measured solar wind speed]{The Ulysses spacecraft measured solar wind speed (indicated by red and blue lines in the radial plots), overlaid on representative coronal images corresponding to different epochs. These measurements are compared with sunspot numbers shown at the bottom of the panel. [\textit{Image credit: \citealt{McComas2013}}]}
    \label{chapt1_fig:ulysses_SW}
\end{figure}

\smallheader{Solar Wind and Heliosphere}
The solar wind is a continuous outflow of magnetized plasma from the solar corona that expands supersonically into interplanetary space \citep{Parker1958}. It comprises two primary components: the fast solar wind ($\sim$750~km~s$^{-1}$), originating from open magnetic field regions such as coronal holes, and the slow solar wind ($\sim$400~km~s$^{-1}$), typically associated with closed magnetic field regions near the streamer belt \citep{Habbal1997, Woo2000, Feldman2005}. The Ulysses spacecraft, through its unique polar orbit around the Sun, delivered the first comprehensive measurements of solar wind speed at high heliographic latitudes (Figure~\ref{chapt1_fig:ulysses_SW}). It observed persistent fast wind streams from polar coronal holes during solar minimum, and slower, more variable flows near the equatorial regions. During solar maximum, as polar coronal holes diminish, the solar wind becomes more isotropic, with slow wind dominating across all latitudes. These observations firmly established the Solar Cycle’s role in modulating the global structure of the solar wind. An important heliospheric phenomenon, known as a stream interaction region (SIR), arises when fast solar wind overtakes slower wind ahead of it, producing compression regions characterized by enhanced magnetic fields and energetic particle acceleration. When such structures persist over multiple solar rotations due to the Sun’s rotation, they are referred to as corotating interaction regions (CIRs) \citep{Gosling1999c, Richardson2018, Allen2021}.

The solar wind shapes the heliosphere, a vast, magnetized bubble of solar influence that extends well beyond the orbit of the outer planets and serves as a dynamic interface between the solar wind and the local interstellar medium (LISM). The outer boundary of the heliosphere, known as the heliopause (located at $\sim$110--150~au), marks the point where the outward pressure of the solar wind is balanced by the inward pressure of the LISM \citep{Suess1990}. Just inside this boundary lies the termination shock (at $\sim$ 84--94~au), where the supersonic solar wind decelerates abruptly to subsonic speeds due to interaction with the interstellar plasma \citep{Stone2005, Stone2008}. Between the termination shock and the heliopause lies the heliosheath, a turbulent region where the subsonic solar wind is compressed, heated, and mixed with interstellar particles. The global structure of the heliosphere is also shaped by the Sun’s motion through the LISM and by variations in solar magnetic activity. Furthermore, the heliosphere acts as a modulator of galactic cosmic rays (GCRs), with the solar wind and its embedded magnetic field scattering and decelerating incoming high-energy particles. This modulation leads to the observed 11-year cosmic ray cycle, which is anti-correlated with solar activity \citep{Balasubrahmanyan1969}.

\section{CME: Observations}

Coronal Mass Ejections (CMEs) represent energetic and consequential eruptive phenomena in our solar system, constituting massive expulsions of magnetized plasma from the solar corona that can fundamentally alter the space environment throughout the heliosphere. These dramatic events typically involve the sudden release of $10^{30}$ ergs of energy, equivalent to billions of hydrogen bombs, and the ejection of $10^{15}$ grams of coronal plasma into interplanetary space at speeds ranging from 100--3000 km~s$^{-1}$ \citep{Howard2011, Webb2012, Chen2011}.

The discovery of CMEs emerged from early coronagraph observations in the 1970s, when the white-light coronagraph aboard the seventh Orbiting Solar Observatory (OSO-7; \citealt{Koomen1975}) first revealed these spectacular eruptions as expanding brightness enhancements propagating outward through the corona \citep{Tousey1973}. The systematic cataloging of CMEs began with the Solar Maximum Mission (SMM; \citealt{Chipman1981}) and continued with the SOHO spacecraft, which has observed over 30,000 CMEs since 1996, providing unprecedented statistical insights into their occurrence rates, morphologies, and properties \citep{Hundhausen1984, Hundhausen1999, Gopalswamy2009, Gopalswamy2009cdaw, Webb2012, Gopalswamy2024cata}.

\subsection{Remote Observations of CME Plasma}

Our understanding of CME plasma properties derives primarily from remote sensing observations across the electromagnetic spectrum. Each wavelength regime provides unique insights into different aspects of CME structure, initiation, composition, and evolution, creating a comprehensive picture of these complex plasma systems.

\smallheader{White-Light Coronagraph}
White-light coronagraphs remain the primary tool for CME detection and characterization, utilizing Thomson scattering of photospheric light by free electrons to image plasma density structures in the corona \citep{Minnaert1930}. For the near-Sun coronagraph field of view, the scattered intensity is maximized for electrons located near the plane of the sky due to the angular dependence of the Thomson scattering cross-section \citep{Hundhausen1993, Vourlidas2013}. Although CMEs and background coronal densities are often comparable, CMEs become distinguishable through background subtraction techniques that isolate transient structures such as CMEs from the relatively static corona.



\begin{figure}
    \centering
    \includegraphics[width=0.49\linewidth]{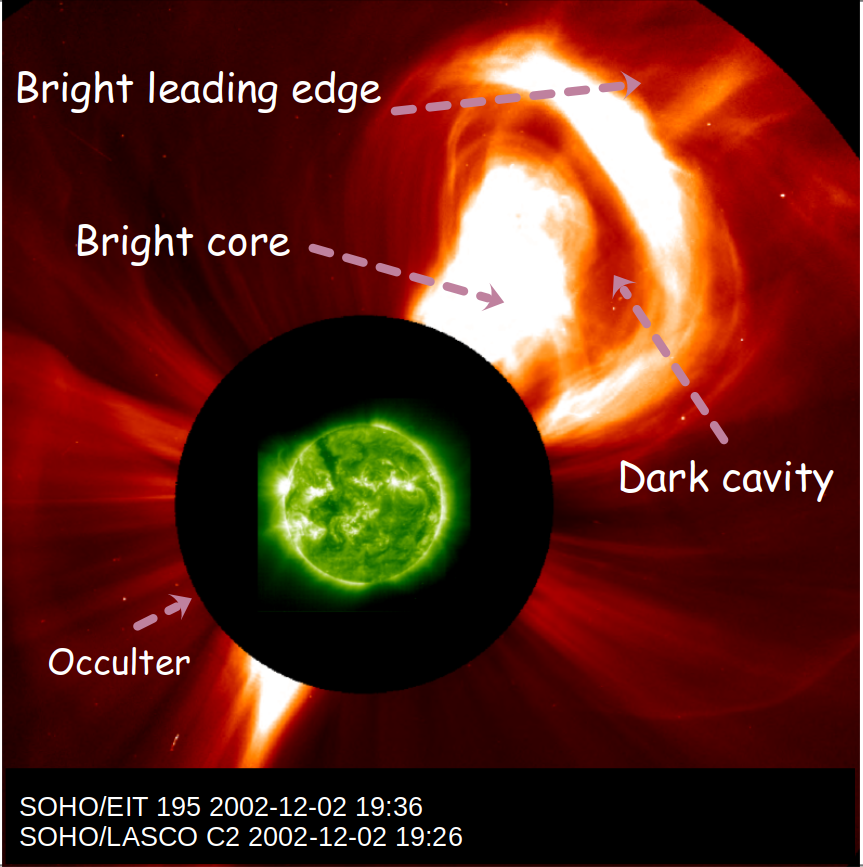}
    \includegraphics[width=0.488\linewidth]{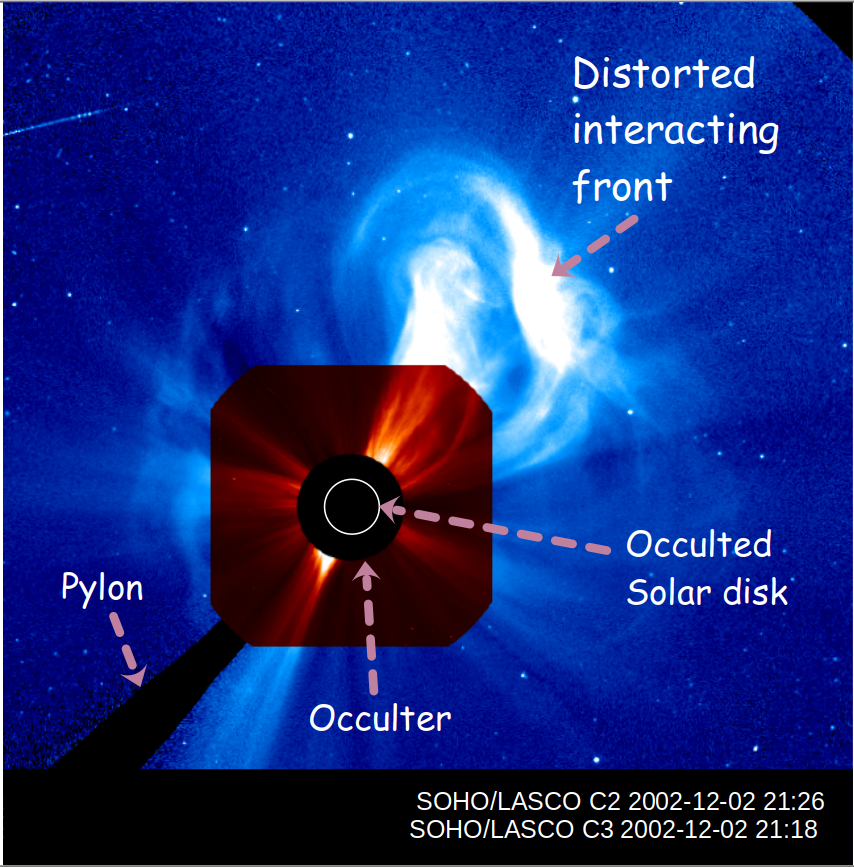}
    \caption[A classical three-part CME observed on 2 December 2002]{A classical three-part CME observed on 2 December 2002, captured in the LASCO C2 (overlaid with SOHO/EIT 195 Å) and C3 (overlaid with C2) fields of view. The structure clearly shows a bright leading edge, a dark central cavity, and a bright core. [\textit{Image generated using Helioviewer:} \url{https://gs671-suske.ndc.nasa.gov/}]}
    \label{chap1-fig:CME_lasco}
\end{figure}

The Large Angle and Spectrometric Coronagraph (LASCO; \citealt{Brueckner1995}) aboard SOHO has provided the most comprehensive CME database. Several CME catalogs have been developed using LASCO observations, each employing different detection methods. The CDAW catalog, the only manually compiled list, covers events from 1996 onward \citep{Gopalswamy2009cat}. In contrast, automated catalogs like CACTUS \citep{Robbrecht2009}, SEEDS \citep{Olmedo2008}, and CORIMP \citep{Byrne2015} use image processing techniques for faster and consistent detection. These observations reveal the characteristic three-part structure of many CMEs (Figure~\ref{chap1-fig:CME_lasco}): a bright leading front representing compressed plasma and often associated with a shock wave, a dark cavity corresponding to the low-density interior of a magnetic flux-rope, and a bright core typically consisting of erupting prominence or filament material \citep{Illing1985, Cremades2004, Vourlidas2013, Song2022}. Furthermore, the STEREO mission, with its twin spacecraft, enabled stereoscopic views of CMEs, greatly improving our understanding of their 3D structure and propagation. It helped eliminate projection effects, revealing wider CME widths and non-radial trajectories, with significant implications for space weather forecasting \citep{Mierla2010, NSrivastava2010, Wood2017, Mishra2021}.

\smallheader{Extreme Ultraviolet Observations}
Extreme ultraviolet (EUV) observations provide crucial information about CME initiation and early evolution in the low corona, where white-light coronagraphs cannot observe due to scattered light limitations. The Extreme Ultraviolet Imaging Telescope (EIT; \citealt{Delaboudinire1995}) onboard SOHO, Extreme UltraViolet Imager (EUVI; \citealt{Wuelser2004}) onboard STEREO, and Atmospheric Imaging Assembly (AIA; \citealt{Lemen2012}) onboard Solar Dynamics Observatory (SDO; \citealt{Pesnell2012}) images the corona in multiple EUV wavelengths corresponding to different temperature ranges, revealing the complex thermal structure of erupting plasma systems.

\begin{figure}
    \centering
    \includegraphics[width=1\linewidth]{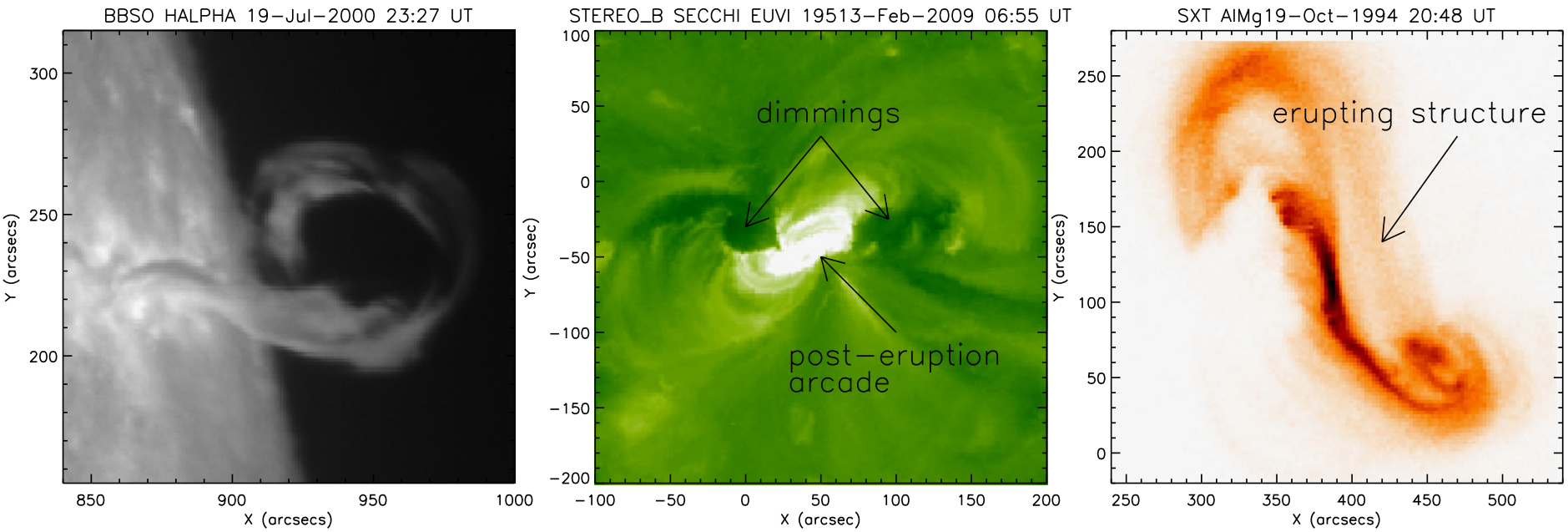}
    \caption[Lower coronal signatures of CME eruptions, including]{Lower coronal signatures of CME eruptions, including filament eruption (left), post-eruption arcades and coronal dimming (middle), and rising EUV/soft X-ray structures (right). [\textit{Image credit: \citealt{Green2018}}]}
    \label{chap1_fig:euv_cme}
\end{figure}

EUV observations frequently reveal multiple coronal signatures associated with CME eruptions, offering valuable diagnostics of the eruption process (Figure~\ref{chap1_fig:euv_cme}). One prominent feature is coronal dimming, regions of reduced EUV emission caused by plasma evacuation during the rapid CME eruption \citep{Hudson2001, Cheng2016a}. These dimming regions often correspond to the footprints of the erupting magnetic flux-rope, thus mapping the source region’s magnetic topology. Another key signature is filament or prominence eruption, visible as the sudden rise and disappearance of cool, dense plasma structures in EUV and H$\alpha$ wavelengths. These filaments typically reside within sheared or twisted magnetic field lines that form the core of the erupting flux-rope. Post-eruption arcades, consisting of bright, loop-like structures that form below the erupting CME, are interpreted as the result of magnetic reconnection in the wake of the eruption. Together, these EUV signatures serve as crucial observational markers for identifying CME onset and constraining eruption dynamics and energetics.

\smallheader{Radio Observations}
CMEs generate radio emissions as a result of shock formation, magnetic reconnection, and plasma compression, which accelerate electrons to non-thermal energies. These electrons excite plasma waves that convert into radio signals near the local plasma frequency or its harmonics. Radio observations thus provide unique diagnostics of CME-driven shocks, particle acceleration, and evolving magnetic structures in the corona and interplanetary space \citep{Pick2008, Gopalswamy2006ra, Morosan2021}. 

Several distinct types of radio bursts are associated with CMEs, particularly type II and type IV emissions. Type II radio bursts, characterized by narrow-band, slowly drifting features in dynamic spectra, are produced by shock-accelerated electrons and trace the shock’s propagation through the corona \citep{Roberts1959, Gopalswamy2006ra, Ramesh2010, Kumari2023}. Type IV radio bursts, in contrast, appear as broadband continuum emissions at decimetric to metric wavelengths. These can originate from stationary sources near post-eruption arcades or from moving sources associated with the radial propagation of CME flux ropes or lateral expansion of CME flanks \citep{Bastian2001, Morosan2020, Kumari2021}. Together, multi-type radio observations offer a time-resolved, frequency-dependent view of CME evolution and associated energy release processes, complementing white-light and EUV observations.

\subsection{In Situ Measurements}

When CMEs propagate through the interplanetary medium and reach in situ spacecraft locations, they are detected as interplanetary coronal mass ejections (ICMEs). In situ measurements provide crucial insights into the internal structure, composition, and thermodynamic properties of CMEs that cannot be obtained with certainty through remote sensing observations alone. A typical ICME structure consists of three distinct regions (Figure~\ref{chapt1_fig:cme_insitu}): a shock front followed by a turbulent sheath region, and the magnetic ejecta (ME) or flux-rope core \citep{Richardson2010, Zhang2007}. Each region within ICME is characterized by a distinctive set of plasma and magnetic field signatures that distinguish it from the ambient solar wind \citep{Richardson2010, Kilpua2017}. Here, we briefly outline some of the key criteria used for identification:

\begin{figure}
    \centering
    \includegraphics[width=\linewidth]{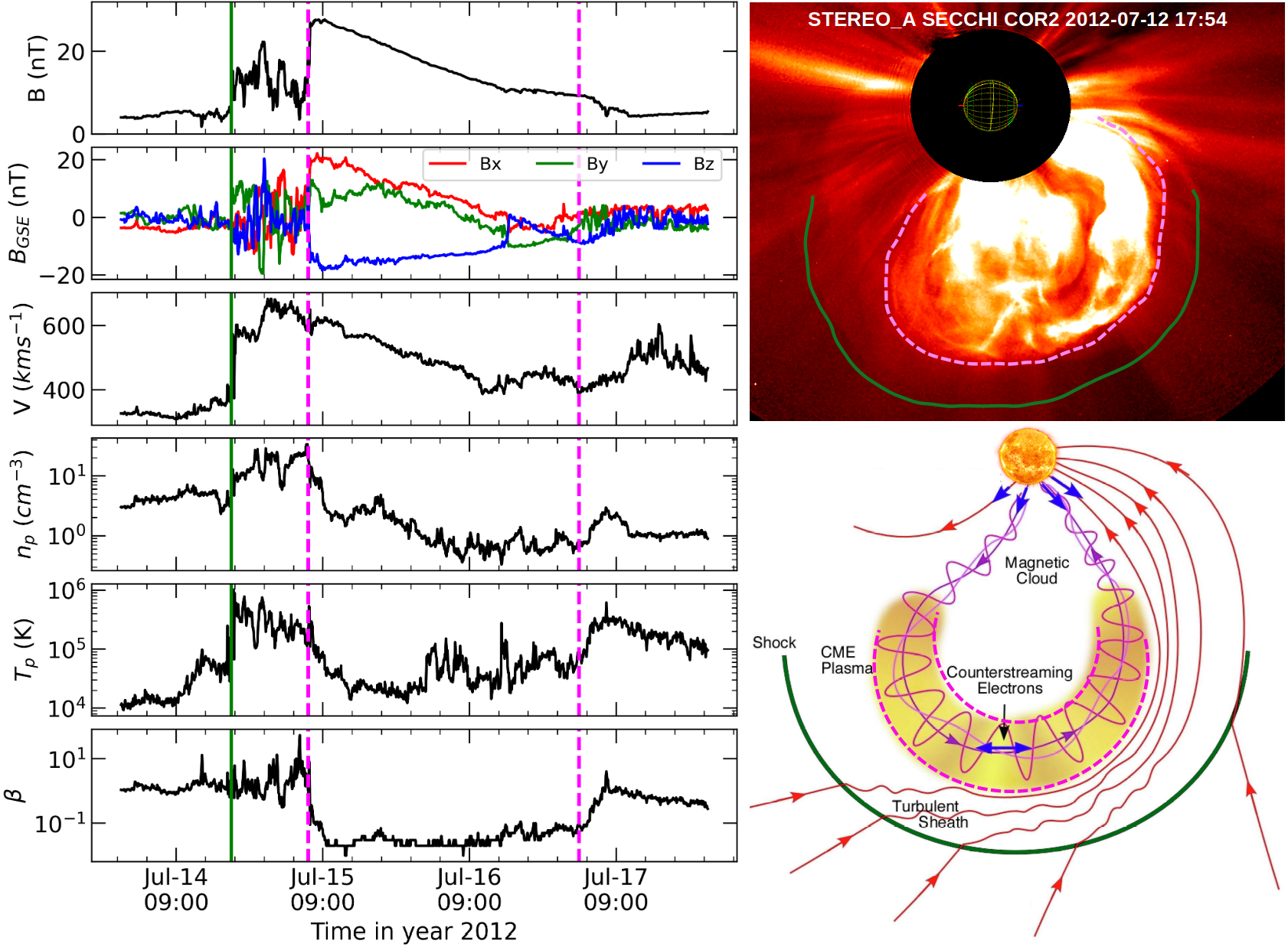}
    
    \caption[This figure illustrates the connection between]{This figure illustrates the connection between remote-sensing and in situ observations, highlighting the temporal and structural correspondence between CME signatures near the Sun and at 1 au. Left: In situ measurements of an ICME near Earth observed by the Wind spacecraft, showing the magnetic field strength, speed, proton number density, proton temperature, and plasma beta. The vertical green line marks the arrival of the shock, while the two vertical magenta lines denote the start and end of the CME ejecta. Top-right: Corresponding CME observed in the STEREO-A/COR2 coronagraph, with the shock front and leading edge of ejecta outlined by green and magenta curves, respectively. Note that it is often difficult to unambiguously identify the trailing edge of a CME/ME in remote observations, and therefore, it is not marked in the top-right image. Bottom-right: a schematic of global ICME structures, including shock (green line), sheath, and ME with its leading and trailing edges (magenta dashed curves). [\textit{Image credit: \citealt{Zurbuchen2006}}]}
    \label{chapt1_fig:cme_insitu}
\end{figure}

\textbf{Magnetic Field Signatures:} The most defining characteristic of ICMEs is the presence of large-scale, coherent magnetic field rotations, often exhibiting flux-rope topology. These rotations typically occur over timescales of several hours to days, with enhanced magnetic field strength (typically 10-30 nT) compared to the ambient solar wind ($\sim$5 nT) \citep{Lepping1990, Lynch2003}.

\textbf{Plasma Parameters:} ICMEs exhibit distinctive plasma characteristics including reduced proton temperature relative to solar wind, low plasma beta ($\beta < 0.1$), enhanced alpha-to-proton ratios, and often bidirectional suprathermal electron flows indicating closed magnetic field lines \citep{Richardson1995, Gosling1987}. The proton temperature depression, quantified by the temperature ratio $T_p/T_{exp} < 0.5$, where $T_{exp}$ is the expected temperature based on solar wind speed, serves as a reliable ICME indicator \citep{Richardson1996}.

\textbf{Compositional Anomalies:} ICMEs frequently display enhanced heavy ion abundances and elevated charge states (e.g., O$^{7+}$/O$^{6+}$ ratios $> 0.1$) that reflect their coronal origin and the high-temperature environment during eruption \citep{Henke1998,Gopalswamy2013, Zurbuchen2006}. Iron charge states, particularly Fe$^{16+}$, provide diagnostics of the coronal source temperature at the time of CME initiation.

\textbf{Expansion Signatures:} The magnetic ejecta within ICMEs exhibit characteristic kinematic signatures. The proton speed typically shows an elevated profile at the ME's leading edge boundaries and a systematic decreasing trend toward the trailing edge. This velocity pattern reflects the radial expansion of the flux-rope structure and serves as a key identification criterion for magnetic ejecta \citep{Lepping1990, Lepri2001, Richardson2010}.

\citet{Burlaga1981} first identified magnetic flux ropes in ICMEs using Voyager data, establishing the fundamental connection between CME flux-rope structure and in situ observations. Key ICME discoveries have been enabled by successive space missions: IMP-8 provided the first magnetic cloud catalog \citep{Lepping1990}, Wind and ACE established fundamental plasma and compositional characteristics \citep{Richardson2010, Lepri2001}, STEREO revealed multi-point structural variations \citep{Kilpua2011}, and recent missions such as Parker Solar Probe \citep{fox2016} and Solar Orbiter \citep{muller2020} provide comprehensive multi-point radial evolution of flux-rope during propagation \citep{Good2019, Davies2021, Davies2024, Liu2024b}.

In situ measurements thus provide the essential ground truth for understanding CME internal structure and evolution, complementing remote sensing observations and serving as crucial validation datasets for theoretical models and simulations of CME thermodynamics and kinematics.

\section{CME Physics: Initiation and Evolution}
 
The initiation and evolution of CMEs are governed by fundamental magnetic and plasma processes in the solar atmosphere and heliosphere. This section focuses briefly on the physical mechanisms behind their initiation and the characteristics of their evolution in interplanetary space, influencing space weather conditions.

\vspace{-0.5cm}
\subsection{CME Initiation}

Understanding CME initiation remains a central challenge in solar physics, owing to the complexity of coronal magnetic structures and the limited direct observational access to the pre-eruptive phase. Despite this, theoretical models, constrained by coronal and heliospheric observations, have provided a rich framework to describe CME onset and initial kinematics.

\vspace{0.5cm}
{\large \textbf{Pre-conditions for Eruption}}

The initiation of a CME requires access to a large reservoir of free magnetic energy, which is primarily associated with non-potential coronal magnetic fields. In the low plasma-$\beta$ coronal environment, magnetic forces dominate over thermal and inertial forces, making magnetic stress and topology the key drivers of eruptive behavior. Photospheric motions, such as shear flows, rotation, and flux emergence, inject helicity and stress the magnetic field, forming sheared arcades or flux ropes. Over time, the system evolves toward a critical state where equilibrium between the upward magnetic pressure and the downward magnetic tension of the overlying field breaks down, the system becomes unstable, and erupts \citep{Forbes2006}. The pre-eruption configuration is typically modeled as a two-component magnetic structure: a low-lying core field (often a flux-rope) and an overlying arcade that restrains its expansion \citep{Rachmeler2009, Forbes2006}. The precondition for CME eruption is not merely energy accumulation, but also the evolution toward a magnetic topology susceptible to instability or reconnection, either through a slow loss of equilibrium or a rapid transition triggered by a disturbance such as flux emergence, cancellation, or internal reconnection.

\smallheader{Eruption Mechanisms}
Observationally, CMEs are generally associated with either (1) streamer blowouts and eruptive prominences or (2) flares originating in active regions \citep{Webb2012}. The former tend to evolve gradually into the solar wind, while the latter exhibit more explosive dynamics, reflecting fundamental differences in their initiation mechanisms. Regardless of origin, CME onset involves a disruption of magnetic equilibrium, driven by either ideal MHD instabilities or magnetic reconnection. Notably, the locations of onset and energy release may not be co-spatial or directly linked, but both processes collectively govern the eruption and early evolution of the CME

\smallheader{Models Not Requiring Magnetic Reconnection:}
Here, we briefly mention several CME initiation models that rely solely on ideal MHD processes, not invoking magnetic reconnection explicitly.
\begin{itemize}
     \item \textbf{Magnetic Buoyancy:} Eruption results from coronal expansion driven by thermal and magnetic pressure overcoming gravity, often triggered by new flux emergence \citep{Low1994, Wolfson1997}.
    
    \item \textbf{Flux Injection (Toroidal Instability):} Additional poloidal flux enhances toroidal current in a flux-rope, producing an upward Lorentz force \citep{Chen1989, Krall2001}. However, criticisms include the absence of observational support for the required energy input from the photosphere \citep{Schuck2010}.

    \item \textbf{Kink Instability:} A highly twisted flux-rope becomes unstable, leading to a helical deformation, the formation of current sheets, and potentially to eruption \citep{Torok2003, Fan2004, Rachmeler2009}. This model is consistent with features like X-ray sigmoids and some confined eruptions \citep{Kliem2006, Gerrard2003}.
    
    \item \textbf{Mass Loading/Unloading:} Depending on magnetic field configuration, Loss or redistribution of mass in a prominence can upset force balance, leading to eruption \citep{Zhang2004, Fong2002}.

\end{itemize}

\smallheader{Models Requiring Magnetic Reconnection:}
Here, we briefly mention other CME initiation models that posit magnetic reconnection as essential for both triggering the eruption and releasing energy.
\begin{itemize}
    \item \textbf{Tether-Cutting:} Internal reconnection within a sheared core field weakens the overlying strapping field, enabling eruption and sustained acceleration \citep{Sturrock1989, Moore2001}. Effective for flare-associated events, though its standalone sufficiency is debated \citep{Antiochos1999}.

    \item \textbf{Magnetic Breakout:} In a quadrupolar configuration, reconnection above the core removes overlying flux, triggering eruption. A secondary flare-related reconnection follows beneath the core, flare-CME timings \citep{Antiochos1999, Lynch2008, Aulanier2000, Lynch2004}.
    
    \item \textbf{Flux Cancellation (Catastrophe Model):} Cancellation of opposite polarity magnetic fields at the photosphere leads to flux-rope formation and eventual loss of equilibrium \citep{Martin1985, Forbes1991, Lin1998}. This model is consistent with disappearing filament observations \citep{Linker2003}

\end{itemize}

No single model can fully account for all CME-associated phenomena. For instance, some CMEs may result from magnetic buoyancy and flux emergence, whereas others, flare-associated CMEs, likely require reconnection-based models. For a comprehensive overview of CME initiation mechanisms, the review articles by \citet{Low1996} and \citet{Forbes2006} offer foundational insights into the magnetic and dynamic aspects of eruptions, while a more recent synthesis of theoretical models and observational constraints is provided by \citet{Lin2010}. Comparative modeling and high-resolution coronal observations, such as from SDO and SolO, will be essential to disentangle these mechanisms. Ultimately, the Sun may exploit any available instability to trigger a CME, making the study of their initiation both a challenging and essential pursuit in heliophysics.

\vspace{-0.5cm}
\subsection{CME Evolution}

The evolution of CMEs through the interplanetary space represents a complex interplay of forces and processes that fundamentally alter the properties of these structures. Understanding this evolution is crucial for interpreting the relationship between remote observations near the Sun and in situ measurements at 1 au and beyond, and for predicting the space weather effects of Earth-directed CMEs.

\smallheader{Initial Acceleration}
CMEs are typically observed with a broad range of initial speeds (100--3000 km~s$^{-1}$), acquired within the first few solar radii ($\sim$20~$R_\odot$) \citep{Yashiro2004, Gopalswamy2009}. The acceleration profile is closely linked with flare-associated reconnection processes. Peak acceleration often coincides with the impulsive phase of solar flares \citep{Zhang2001fla, Zhang2006fla, Temmer2008}. Observations indicate that about 74\% of CMEs achieve peak acceleration below 1.5 $R_\odot$ \citep{Bein2011}, often before the flare peak in soft X-rays \citep{Marivic2007}. During the initial rapid acceleration phase, CMEs often attain their peak speeds before transitioning to a phase of more gradual evolution \citep{Zhang2006fla}. The acceleration is primarily driven by the Lorentz force ($\vec{J} \times \vec{B}$), arising from the CME's internal magnetic field, as well as contributions from thermal pressure gradients, gravity, and aerodynamic drag \citep{Vrsnak2004}. As CMEs ascend beyond $\sim$20~$R_\odot$, the decreasing magnetic dominance ($\beta > 1$ in ambient plasma) allows solar wind drag to govern CME kinematics \citep{Howard2011}.

\smallheader{Drag Force}
As CMEs propagate through interplanetary space, their interaction with the ambient solar wind becomes the dominant influence on its dynamics, beyond the low corona ($>$15~$R_\odot$) \citep{Vrsnak2004}. The heliocentric distance where drag dominates varies between events and depends on CME and solar wind properties. Typically, the drag effect becomes significant beyond 15--30 $R_\odot$, and dominates CME kinematics out to 1 au \citep{Temmer2011}. Some extreme CMEs (e.g., 2012 July 23) show negligible deceleration, suggesting exceptionally low ambient density and weak drag \citep{Temmer2015}. 

The most general form of the drag force is expressed as:
\begin{equation} 
F_D = -\rho_{sw} A C_D (v - v_{sw})|v - v_{sw}|,
\end{equation}
where $A$ is the CME cross-sectional area, $v$ and $v_{sw}$ are the CME and solar wind velocities respectively, $\rho_{sw}$ is the ambient solar wind density, and $C_D$ is a dimensionless drag coefficient \citep{Cargill2004}.

The efficiency of drag also depends on CME mass: more massive CMEs experience less acceleration/deceleration due to their larger inertia \citep{Vrsnak2010}. More refined models incorporate solar wind pile-up effects, such as the snow-plough model \citep{Tappin2006}, or provide analytical solutions for real-time forecasts \citep{Vrsnak2013}. This force acts to equalize the CME speed with that of the surrounding solar wind. Fast CMEs ($v > v_{\mathrm{sw}}$) are decelerated, while slower ones are accelerated, causing a speed convergence toward typical solar wind velocities \citep{Vrsnak2004, Gopalswamy2000}. Overall, drag-based models remain central to understanding and forecasting CME propagation in interplanetary space. However, accurate application requires knowledge of CME mass, size, and ambient solar wind parameters, which are often difficult to constrain precisely.

\smallheader{Expansion and Geometric Evolution}
As CMEs propagate through the heliosphere, they undergo continuous expansion driven by internal overpressure. This expansion is generally non-self-similar and is influenced by external solar wind pressure, internal magnetic tension, and aerodynamic drag \citep{Liu2005, Manchester2004b}. It occurs both radially (along the Sun–CME line) and laterally (perpendicular to the propagation direction), with typical expansion speeds ranging from 50–200km~s$^{-1}$, decreasing with heliocentric distance \citep{Farrugia1993a,Liu2005, Savani2011, Good2016ex, Mishra2021a}.

Simultaneously, the 3D geometry and orientation of CMEs evolve due to interactions with the structured solar wind, magnetic field draping, and internal instabilities \citep{Odstrcil1999, Manchester2004, Braga2022}. Stereoscopic observations and 3D reconstructions (e.g., from STEREO) reveal that CMEs often undergo deflection, rotation, and non-radial propagation, with common shifts toward the heliospheric equator and western longitudes due to large-scale magnetic gradients \citep{Cremades2004, Liu2010, Gui2011, Vourlidas2013, Kay2017p, Kumar2023}. These geometric changes affect the magnetic field profiles measured in situ, thereby influencing CME's impact.

\smallheader{Shock Formation and Sheath Dynamics}
Fast CMEs, typically those exceeding $500$~km~s$^{-1}$, can generate shock waves when their speed surpasses the local fast-mode magnetosonic speed in the ambient solar wind \citep{Sheeley1985}. These CME-driven shocks are collisionless in nature and serve as efficient agents for transferring energy and momentum between the CME and the surrounding solar wind. Observational evidence from metric and decametric type II radio bursts suggests that shock formation generally occurs within a few solar radii from the Sun \citep{Cane1987, Gopalswamy2001}. As these shocks propagate outward through the heliosphere, they gradually weaken due to expansion and energy dissipation. While the shock compression ratio and associated magnetic field amplification tend to decrease with heliocentric distance, strong shocks can remain well-defined even at 1 au and beyond \citep{Tsurutani2003}.

Trailing the shock is a sheath region composed of compressed, heated, and turbulent plasma. The sheath thickness is determined by the speed and physical characteristics of the driving ICME, such as its geometry and radius of curvature, as well as the shock's compression ratio \citep{Russell2002}. This region significantly alters the upstream solar wind and can enhance geoeffectiveness due to its elevated magnetic field strength, density, and dynamic pressure \citep{Lugaz2012, Kilpua2017}. These sheath regions can often be more geoeffective than the CME ejecta itself, especially when they contain strong southward magnetic fields, making them a critical component of CME-driven space weather impacts.

\smallheader{CME-CME and CME-Solar Wind Interactions}
CMEs do not propagate in isolation; rather, they continuously interact with the ambient solar wind, SIRs/CIRs, and other CMEs launched in close succession. CME-CME interactions occur when multiple CMEs are launched from the Sun within time intervals shorter than their transit time to 1 au. This collision or merging, sometimes referred to as ``CME cannibalism," can produce a single, complex structure characterized by enhanced magnetic field strength, increased plasma temperature, and irregular field rotations \citep{Gopalswamy2001, Lugaz2013, Mishra2015, Wang2003}. Such events are often difficult to interpret using conventional single-flux-rope models and have been linked to prolonged southward magnetic fields and intensified geomagnetic storms \citep{Gopalswamy2001, Lugaz2017, Scolini2020}. Multi-spacecraft observations and MHD simulations have shown that these merged structures may retain identifiable substructures or form fully blended ejecta depending on the interaction geometry and timing \citep{Lugaz2014}.

Similarly, interactions between CMEs and solar wind structures like SIRs/CIRs can also result in significant dynamical changes. Depending on the relative speeds and orientations, these interactions can lead to CME acceleration or deceleration, increased compression, and additional heating within both the CME and the interacting stream \citep{Prise2015, Matamba2018}. These processes can modify the magnetic topology and plasma parameters of the CME, thereby influencing its ability to couple with the Earth's magnetosphere. Overall, such interactions contribute to the complexity and variability of CME structures observed at 1 au and represent a major challenge for both in situ interpretation and space weather prediction \citep{Lugaz2017}.

\subsection{CME Space Weather Impacts}

CMEs are among the most impactful drivers of space weather disturbances at Earth. When Earth-directed, CMEs can trigger intense geomagnetic storms primarily through magnetic reconnection facilitated by a strong and sustained southward magnetic field component \citep{Zhang2007, Wu2011, Richardson2012, Gopalswamy2015}. This reconnection injects energy into the magnetosphere, enhancing convection, heating plasma, and driving phenomena such as ring current intensification, radiation belt enhancements, and auroral activity \citep{Gonzalez1994}. CME-driven shocks can also cause sudden storm commencements (SSCs) and compress the magnetosphere, further intensifying geomagnetic effects \citep{Curto2007}.

Fast CMEs are also closely associated with SEP events, which originate near CME-driven shocks or flare sites \citep{Gopalswamy2005, Yashiro2014, Kihara2020}. SEPs pose serious radiation hazards to spacecraft, astronauts, and high-altitude aviation by damaging electronics, degrading solar panels, and disrupting HF communications during polar cap absorption (PCA) events \citep{Fiori2022, Waterfall2023}. The onset and severity of CME-driven SEP events depend on CME speed, shock geometry, and IMF conditions, making accurate forecasting challenging.

The societal and technological consequences of CME-driven storms are substantial. Geomagnetically induced currents (GICs) can disrupt power grids, as demonstrated by the 1989 Hydro-Quebec and 2003 Swedish blackouts \citep{Boteler2019, Pulkkinen2005, Liu2024g}. Satellites face increased surface charging, internal discharges, and atmospheric drag, shortening operational lifespans \citep{Parker2024}. Radio communications, GNSS reliability, and aviation safety are also affected. Despite advances in heliospheric imaging and in situ monitoring, predicting CME arrival times and internal magnetic structure remains a key challenge for space weather forecasting. Addressing this challenge requires a deeper understanding of CME propagation through the heliosphere, motivating the development of a variety of models that describe their large-scale evolution from the Sun to Earth.

\subsection{Models for CME evolution}
Space‐weather models of CME evolution fall into two broad classes: full 3D MHD simulations (physics‐based models) and more empirical or geometric approaches. Physics‐based models solve the MHD equations for the corona/heliosphere (often including CME injection), while empirical or geometric models use simplified kinematics or fits to observations. Below, we have briefly mentioned some widely used CME models.

\vspace{0.5cm}
{\large \textbf{Global MHD models}}
\begin{itemize}[leftmargin=0pt]
\item \textbf{CORHEL/MAS:} The MAS (Magnetohydrodynamic Algorithm outside a Sphere) code is a 3D MHD model forming the core of the CORHEL framework, which couples the solar corona and heliosphere \citep{Riley2012}. It solves ideal MHD equations on an adaptive grid with thermodynamics to generate steady solar wind solutions. CMEs are initiated by inserting analytic flux ropes (e.g., Titov–Démoulin or Gibson–Low) \citep{Titov1999} and simulating their eruption self-consistently \citep{Lionello2013, Torok2018}. CORHEL can model CME evolution from the Sun to 1~au \citep{Linker2003, Linker2024}, though its high computational cost limits real-time applications.

\item \textbf{SWMF (BATS-R-US/AWSoM):} The Space Weather Modeling Framework (SWMF) \citep{Toth2005, Toth2012} is a modular MHD framework that couples solar, heliospheric, and magnetospheric models. Its solar component, BATS-R-US, is a 3D MHD solver \citep{Gombosi2021}, while AWSoM adds Alfvén-wave–driven coronal heating \citep{Van2014, Meng2015}. Driven by synoptic magnetograms, SWMF/AWSoM simulates the background wind from the chromosphere to the inner heliosphere. CMEs are inserted via analytic flux ropes (e.g., Gibson–Low) using tools like Eruptive Event Generator–Gibson–Low (EEGGL) \citep{Borovikov2017}. The framework supports end-to-end Sun-to-Earth simulations.

\item \textbf{WSA–ENLIL–Cone:} This operational model combines the empirical Wang Sheeley Arge (WSA) model for background solar wind with the 3D MHD ENLIL model to simulate CME propagation \citep{Arge2000, Odstrcil2003, Odstrcil2023}. Driven by synoptic magnetograms, it inserts CMEs as unmagnetized hydrodynamic cones based on coronagraph-derived parameters (speed, direction, width). While it cannot model internal magnetic fields, it enables fast event-specific forecasts and is widely used by NOAA/SWPC, NASA/CCMC, and other agencies. Arrival-time predictions are typically accurate within 12–14 hours \citep{Pizzo2011, Mays2020}.

\item \textbf{EUHFORIA:} The European Heliospheric Forecasting Information Asset is a physics-based model for space weather forecasting \citep{Pomoell2018, Poedts2020}. It couples a data-driven coronal model (PFSS + empirical) with a time-dependent 3D MHD heliospheric model extending to ~2 au. CMEs are inserted as either unmagnetized cones or magnetized flux ropes (e.g., spheromak or FRi3D), enabling more realistic simulation of CME propagation and shock formation compared to kinematic models \citep{Maharana2022}.

\item \textbf{SWASTi-CME:} A newly developed MHD-based CME module within the SWASTi (Space Weather Adaptive Simulation) framework \citep{Mayank2023}. SWASTi-CME offers both an unmagnetized elliptic cone and a magnetized flux‐rope CME model. In recent studies, it has been applied to multiple-CME and single-CME cases \citep{Mayank2023, Mayank2024}. Results highlight that ambient solar wind heterogeneity can cause asymmetric CME deformation and variable drag forces. As a new model, it is still being validated; it demonstrates the trend towards detailed physics-based modeling of CME–solar wind interaction.

\end{itemize}

{\large \textbf{Geometric and semi-empirical CME models}}
\begin{itemize}[leftmargin=0pt]
\item \textbf{Graduated Cylindrical Shell (GCS):} A geometric forward-fitting model that reconstructs CME 3D morphology using multi-view coronagraph images \citep{Thernisien2006, Thernisien2011}. It models CMEs as a flux-rope-shaped structure manually adjusted through parameters like latitude, longitude, tilt, height, half-angle, and aspect ratio. GCS is purely kinematic, providing initial geometry without modeling CME dynamics or internal physics, and is useful for initializing MHD or drag-based models. In this thesis, we use GCS to derive CME 3D kinematics.In this thesis, we employ the GCS model to derive 3D kinematics of CMEs, which are further discussed in subsequent sections.

\item \textbf{Drag-Based Ensemble Model (DBEM):} An extension of the Drag-Based Model \citep{Vrsnak2013} that provides probabilistic CME forecasts by running multiple realizations with varied inputs \citep{Dumbovic2018}. It outputs arrival time distributions with confidence intervals. DBEM is fast and captures uncertainty, but neglects magnetic forces and CME interactions. Its performance is comparable to ensemble WSA–ENLIL runs, with mean errors around 14 hours.

\item \textbf{FRi3D (Flux Rope in 3D):} An advanced 3D flux-rope model \citep{Isavnin2016} that includes writhed legs and distorted cross-sections. Integrated into EUHFORIA \citep{Maharana2022}, it improved magnetic field predictions by 30–70\% for the 12 July 2012 event. While offering realistic geometry and field structure, FRi3D involves many free parameters and a higher computational cost than a cone model.

\item \textbf{3DCORE (3D Coronal Rope Ejection):} A semi-empirical model that uses a prescribed toroidal flux-rope geometry with self-similar expansion to simulate CME propagation and magnetic profiles \citep{Mostl2018}. It generates synthetic in situ time series along spacecraft tracks and can incorporate GCS-derived parameters. While computationally efficient and able to estimate geomagnetic indices, its accuracy is highly sensitive to initial orientation and relies on fixed expansion laws.

\end{itemize} 

Overall, the physics-based models (e.g., CORHEL, SWMF, ENLIL, EUHFORIA, SWASTi) offer realistic CME simulations but are computationally intensive and sensitive to input quality. Empirical and geometric models (e.g., WSA, GCS, drag-based) are faster and operationally used but rely on simplified assumptions. Recent advances include hybrid models, ensemble forecasting, and improved flux-rope representations (e.g., FRi3D, 3DCORE). Ongoing efforts focus on better capturing CME magnetic structure and reducing uncertainties through validation with in situ and remote-sensing data (often via NASA CCMC/DONKI campaigns).

Having discussed the fundamental physics of CMEs, their observational signatures, heliospheric evolution, and space weather impacts, along with the current state of modeling efforts, it is evident that significant advancements have been made in understanding these complex solar phenomena. However, despite these advances, key challenges remain, particularly in accurately predicting CME arrival times, magnetic field structures at Earth, and their geoeffectiveness. The limitations of existing observational capabilities, model assumptions, and data coverage underscore the need for further investigation. In the following section, we identify the major research gaps that motivate the present study and outline the specific questions this thesis seeks to address.

\section{Identification of Research Gaps}\label{sec:chapt1-gaps}

Research on CMEs has extensively explored their source-region dynamics, propagation characteristics, in situ signatures, and geo-effectiveness \citep{Harrison1995, Webb2000, Gopalswamy2009, Mishra2013, Lugaz2017}. Moreover, the kinematics of CMEs, including their direction, speed, acceleration, propagation, and arrival time prediction, have been the focus of sustained observational, modeling, and theoretical efforts since their discovery \citep{Zhang2004, Temmer2012, Byrne2013, Riley2018}. Over time, the availability of coronagraphic observations (e.g., from SOHO/LASCO, STEREO/COR), heliospheric imagers (e.g., STEREO/HI), and in situ measurements (e.g., Wind, ACE, Parker Solar Probe, SolO) has significantly advanced our ability to track CME motion and estimate their arrival at Earth. However, a key aspect of CME evolution, their thermodynamic behavior, remains less understood. Important questions persist regarding how CMEs gain or lose thermal energy during propagation, how internal plasma properties evolve with heliocentric distance, and how these thermodynamic changes influence their interaction with the heliosphere and Earth's magnetosphere. This is a critical limitation because CMEs are magnetized plasma structures whose dynamics and thermodynamics are intrinsically coupled, internal pressure drives expansion, and evolving kinematic conditions in turn influence the plasma energy budget \citep{Vrsnak2004, Liu2006, Manchester2017}. Therefore, a comprehensive understanding of CME propagation requires not only kinematic modeling but also a detailed examination of their thermodynamic evolution.

This knowledge gap is largely driven by observational limitations. EUV imaging and spectroscopic diagnostics, which can probe thermal properties in the low corona, are constrained to heights below $\approx$ 2 $R_\odot$ and are subject to line-of-sight ambiguities \citep{Ciaravella2000, Akmal2001, Kohl2006, Landi2010}. As CMEs propagate beyond a few solar radii, direct thermal measurements become sparse. White-light coronagraphs, while excellent at tracking CME morphology for a wide field of view, offer no direct plasma thermal information. Several studies have used the polytropic index ($\Gamma$) as a proxy to infer the net heating or cooling experienced by CMEs during propagation from in situ measurements \citep{Belmont1992, Osherovich1993, Gosling1999, Liu2005}. This approach, while useful, has its limitations. The physical interpretation of the heating mechanism through $\Gamma$ remains ambiguous. Moreover, in situ spacecraft like Wind, ACE, and PSP sample only localized slices of CME structures at discrete radial distances, making it difficult to reconstruct continuous thermodynamic evolution from the Sun to Earth. Bridging the gap between remote sensing and in situ observations remains a fundamental challenge, as this disconnect has long been recognized as a major limitation in understanding CME evolution \citep{Liu2010, Rouillard2010, Mishra2021}.

To bridge the remote-in situ divide, modeling has become an essential tool. Global MHD models, such as ENLIL, EUHFORIA, SWMF, and CORHEL, simulate CME propagation through the inner heliosphere \citep{Odstrcil2003, Riley2012, Toth2012, Pomoell2018}. However, these models typically adopt fixed thermodynamic assumptions, such as adiabatic or constant $\Gamma$ value, thereby neglecting spatial and temporal variations in the internal thermal state of CMEs.  For example, many MHD simulations assume $\Gamma < 5/3$ to retain some level of internal heating \citep{Odstrcil2002, Riley2003, Manchester2004, Desai2020, Liu2022polymhd, Cai2025}, while others simplify the treatment even further with an isothermal assumption ($\Gamma = 1$) \citep{Lynch2016}. Furthermore, studies have demonstrated that variations in $\Gamma$ can significantly influence the simulated evolution of both the solar wind and CMEs \citep{Manchester2004, Wu2004, Liou2014, Mayank2022, Liu2022polymhd}. More physically motivated models, such as ANTEATR-PARADE, attempt to account for CME expansion and thermal evolution more realistically. Using this framework, \citet{Kay2021b} demonstrated that $\Gamma$ is strongly correlated with key CME parameters such as expansion speed, in situ duration, temperature, and magnetic field strength. Therefore, constraining CME's polytropic index near and far from the Sun from remote and in situ observations, especially across different CME types and solar cycles, offers a promising path toward understanding the coupling between CME kinematics and thermodynamics, and its implications for space weather impact.

Furthermore, many observational studies suggest that CMEs do not evolve adiabatically. Near-Sun spectroscopic observations using SOHO/UVCS have revealed that CMEs experience significant plasma heating during their early evolution phases \citep{Ciaravella2000, Akmal2001, Bemporad2010, Reva2023}. EUV imaging studies have similarly reported elevated temperatures in CME cores and surrounding structures \citep{Filippov2002, Lee2017}. At larger heliocentric distances, in situ measurements at 1 au show that CMEs, especially magnetic clouds, often retain higher temperatures than expected from adiabatic cooling, implying ongoing heating or inefficient cooling mechanisms during propagation \citep{Rakowski2007, Lepri2012}. Importantly, this prolonged heating can influence CME expansion, magnetic field retention, and hence geoeffectiveness. For instance, \citet{Manchester2004} showed that internal thermal pressure can maintain CME expansion and magnetic flux-rope integrity, affecting arrival times and impact strength at Earth. Additionally, heating influences charge-state composition, which in turn serves as a diagnostic for reconstructing CME initiation and early acceleration histories \citep{Gruesbeck2011, Reinard2012}. Despite this, systematic statistical investigations into CME thermal evolution across heliocentric distances remain scarce.

In summary, despite significant progress in understanding the dynamics of CMEs, such as their kinematics, propagation, and arrival-time forecasting, their internal thermodynamic evolution remains one of the least constrained aspects of heliospheric science. Major knowledge gaps persist due to (i) the absence of continuous and reliable thermal diagnostics beyond the low corona, (ii) the simplified treatment of thermodynamics in models, including the use of constant or ad hoc polytropic index, (iii) the limited incorporation of realistic energy transport processes (e.g., heating, conduction, radiation) in global MHD simulations, (iv) a lack of comprehensive statistical studies quantifying CME thermal properties across different heliocentric distances and solar cycles, and (v) inadequate coupling between remote-sensing and in situ observations to reconstruct the full thermal trajectory of CMEs. These limitations hinder our ability to link CME thermal state to expansion behavior, magnetic field evolution, and ultimately, their geoeffectiveness at Earth. Bridging these gaps is essential, not only to advance our fundamental understanding of CME plasma physics but also to improve the fidelity of space weather forecasting.

\section{Thesis Objectives and Organization}

Building on the discussed research gaps in understanding CME kinematics and thermodynamics, this section outlines the core objectives of the thesis and how it is structured to address them. The work combines analytical modeling, case studies, and statistical analyses to investigate CME internal evolution using both remote and in situ observations.

\subsection{Thesis Objectives}

The primary objectives of this thesis are:
\begin{itemize}[leftmargin=0.5pt]
    
\item To develop and revise an analytical framework, the Flux Rope Internal State (FRIS) model, that enables a self-consistent description of the internal thermodynamic evolution of CMEs, including the estimation of the polytropic index, heating rates, and plasma parameters from remotely observed CME kinematics.

\item To implement the FRIS model to individual CME events with varying kinematic profiles, in order to derive their internal thermal state and force balance during propagation, and to establish meaningful connections between model-derived parameters and corresponding in situ measurements at 1 au.

\item To investigate the relationship between CME internal thermodynamic properties (e.g., polytropic index, temperature) and global kinematic behavior (e.g., speed, expansion) using both modeled outputs and observational data.

\item To statistically examine the thermal and plasma characteristics of ICMEs at 1 au, with emphasis on differences between magnetic clouds (MCs) and non-magnetic clouds (Non-MCs), and to classify ICMEs into heating or cooling categories based on the event-wise polytropic index as a diagnostic tool.

\item To explore the variability of thermodynamic properties across different ICME substructures, pre-ICME solar wind, sheath region, magnetic ejecta (ME), and post-ICME solar wind, and to analyze how these variations relate to their geoeffectiveness.

\item To characterize the recovery phase of geomagnetic storms across Solar Cycles 23 and 24, evaluate its dependence on storm intensity, morphology, and interplanetary drivers.

\item To investigate CME–CME interaction events and assess how such interactions alter the kinematic and thermodynamic evolution of the CMEs involved, particularly in terms of plasma heating, magnetic field compression, and enhanced geomagnetic impact potential.

\item To integrate findings from individual case studies and large-scale statistical analyses, combining remote-sensing and in situ observations, to construct a unified understanding of CME thermodynamic behavior near the Sun and at 1 au.

\end{itemize}

\subsection{Thesis Organization}

To address these objectives, the thesis is structured into six chapters, each contributing to a comprehensive investigation of CME kinematics and thermodynamics. A brief outline of each chapter is provided below.

\textbf{Chapter 1: Introduction}\\
Provides an overview of the Sun and its atmosphere, the observational signatures of CMEs, and their initiation and evolution. It reviews existing modeling approaches, identifies key research gaps, and outlines the objectives of this thesis, thereby setting the foundation for the subsequent chapters.

\textbf{Chapter 2: Model Development, Methodology, and Observational Data}\\ 
Presents the correction and advancement of the Flux Rope Internal State (FRIS) model for analyzing CME thermodynamics. It also describes the observational datasets used (SDO, SOHO, STEREO, Wind, and ACE) and details the implementation methods. The model development part in this chapter is based on the work published in \citet{Mishra_E2023} and \citet{Khuntia2023}.

\textbf{Chapter 3: Thermal Evolution -- Case Studies}\\ 
Applies the revised FRIS model to selected CME events with contrasting kinematic properties. It derives their internal thermal states and force profiles near the Sun, and compares these with in situ plasma, magnetic, and turbulence signatures at 1 au. This chapter is based on the work published in \citet{Khuntia2023} and \citet{Khuntia2025b}.

\textbf{Chapter 4: Thermal Evolution -- Statistics}\\ 
Presents a three-part statistical investigation. The first part applies the revised FRIS model with 3D kinematics to explore the thermodynamic evolution of fast CMEs near the Sun. The second part provides a statistical analysis of ICME plasma and thermal properties over Solar Cycles 23–25, examining solar-cycle trends of the CME thermal state and their relation to geoeffectiveness. The third part analyzes the recovery phase of intense and stronger-than-intense geomagnetic storms during Solar Cycles 23 and 24, highlighting the magnetospheric response and the role of different interplanetary drivers. This chapter is based on the work published in \citet{Khuntia2024}, \citet{Mishra2024}, and Khuntia et al. (2025c) (under review).

\textbf{Chapter 5: Thermo-Kinematic Evolution of Interacting CMEs}\\  
Investigates the thermodynamic and magnetic consequences of CME–CME interactions, focusing on the series of eruptions that preceded the 10 May 2024 geomagnetic storm. It identifies complex ejecta structures and compares the pre- and post-interaction properties. This chapter is based on the work published in \citet{Khuntia2025a}.

\textbf{Chapter 6: Conclusion and Future Work}\\ 
Synthesizes the insights gained from the analytical modeling, case studies, and statistical analyses. It revisits the research objectives, summarizes the key contributions of the thesis, highlights its limitations, and outlines promising directions for future work.

Through the combination of theoretical modeling, detailed case analyses, and long-term statistical studies, this thesis aims to advance the physical understanding of CME thermodynamics, an often overlooked but crucial component in heliophysics. By building a consistent framework utilizing remote and in situ observations, the research can contribute not only to fundamental solar astrophysics but also to the operational goal of improving space weather forecasting.

\lhead{\emph{Chapter 2}}  

\quotechapter{2}{`` All models are wrong, but some are useful.''\\ \smallskip — George E. P. Box}{Model Development, Methodology and Observational Data}


\lhead{{ \emph{Chapter 2: Model Development, Methodology and Observational Data}}}

\section{Introduction} \label{Ch2-sec:intro}

As discussed earlier, the core objective of this thesis is to understand the evolution of the thermal properties of CMEs. While several existing models and methods focus on understanding the large-scale kinematics and geometric structure of CMEs, only a few attempt to quantify their internal plasma conditions, such as pressure, temperature, thermal state, and heating rate. The Flux Rope Internal State (FRIS) model provides a data-driven analytical framework that connects remotely observable kinematic parameters to internal thermodynamic properties by treating the CME as a self-similarly expanding flux-rope \citep{Wang2009, Mishra2018, Mishra2020}. In this chapter, we revisit the original FRIS formulation, address mathematical inconsistencies, and extend the model to enable a physically consistent treatment of thermal state, heating rate, and internal force balance. This refined framework forms the basis for both event-specific and statistical applications presented in the subsequent chapters.

\section{Flux-Rope Internal State (FRIS) Model}\label{metho:fris_derivation}

On global scales, CMEs often exhibit a large loop-like structure with both ends anchored to the solar surface. However, on local scales, they can be approximated as axisymmetric cylindrical flux ropes. In cylindrical coordinates \((r, \phi, z)\), with the origin located along the axis of the flux-rope, we assume symmetry such that $\frac{\partial}{\partial \phi} = \frac{\partial}{\partial z} = 0$. The cross-section of the flux-rope is considered circular, with a time-dependent radius \(R(t)\). The expansion speed at the boundary is given by
$v_e(t) = \frac{dR(t)}{dt}$, and the corresponding expansion acceleration is $a_e(t) = \frac{dv_e(t)}{dt}$.

\begin{figure}
    \centering
    \includegraphics[width=0.9\linewidth]{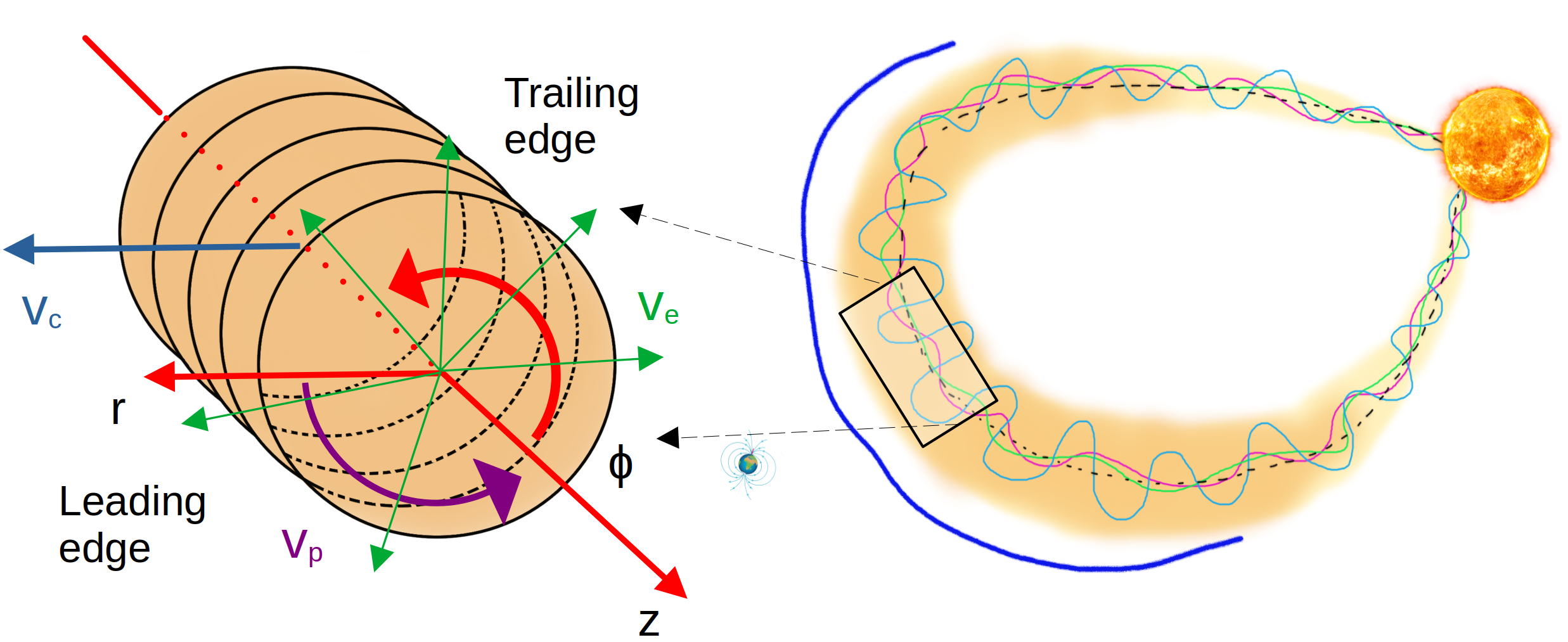}
    \caption[Schematic of a flux-rope CME at the local scale in a]{Schematic of a flux-rope CME at the local scale in a cylindrical coordinate system (i.e., \(r\), \(\phi\), and \(z\), indicated by red arrows), illustrating the propagation speed (\(v_c\)) of the flux-rope's axis, the expansion speed (\(v_e\)), and the poloidal speed (\(v_p\)), represented by blue, green, and violet arrows, respectively.}
    \label{chatp2_fig:fris_image}
\end{figure}

Considering a self-similar expansion of the flux-rope, we define a normalized radial coordinate \(x\), which is independent of time and given by $x = \frac{r}{R(t)}$, where \(x = 1\) corresponds to the boundary of the flux-rope. Under this assumption, the radial and azimuthal components of the plasma velocity can be expressed as functions of \(x\) as,
\begin{equation}
 v_r( t,x ) = \frac{dr}  {dt} =x \frac{dR(t)}  {dt}=xv_e( t )\label{eqn:Vr}
\end{equation}
\begin{equation}
  v_\phi ( t,x )=r \frac{d\phi(t)}  {dt}=f_{v\phi }( x ) v_p( t )\label{eqn:VrVphi}    
\end{equation}
Where $v_p$ is the poloidal speed at the boundary of the flux-rope, and $f_{v\phi }( x )$ is an unknown distribution function for the azimuthal component of the propagation velocity.

\subsection{Conservation of Total Mass and Angular Momentum}
Considering that the magnetic field lines are frozen into the plasma and the flux-rope expands in a self-similar manner, the radial distribution of density remains fixed and can be represented by a function \(f_\rho(x)\). Thus, the plasma density inside the flux-rope can be expressed as $\rho(t, x) = f_\rho(x)\, \rho_0(t)$, where \(\rho_0(t) = \bar{\rho} = \frac{M}{\pi l R^2(t)}\) is the average mass density of the flux-rope, \(M\) is the total mass, \(l\) is the axial length, and \(R(t)\) is the time-dependent radius of the flux-rope.

The total mass of a CME can be written as 
\begin{equation}
\begin{array}{rl}
   &M = \int_V \rho \,dV=\int_V \rho \,rdrd\phi dz {\nonumber}\\
   &\Rightarrow M = 2\pi R^2 l \rho_{0}(t) \int_{0}^{1}   f_\rho( x ) \,xdx
   \end{array}
\end{equation}

The total angular momentum of the flux-rope is given by
\begin{equation*}
   \begin{array}{rl}
  & L_A = \int_V \rho rv_\phi \,rdr d\phi dz\\ 
  & \Rightarrow L_A = 2MR v_p(t) \int_{0}^{1} {f_\rho(x) f_{v\phi}(x) x^2} \,dx
   \end{array} 
\end{equation*}

\begin{equation}
  \Rightarrow v_p ( t )= \frac{L_A}  {2MR \int_{0}^{1} {f_\rho(x) f_{v\phi}(x) x^2} \,dx } \label{eqn:vp}
\end{equation}

Putting the Equation~(\ref{eqn:vp}) in Equation~(\ref{eqn:VrVphi}), we will have
\begin{align}
\label{eqn:vphi}
v_\phi (t,x )= f_{v\phi} ( x ) v_p ( t )\implies v_\phi ( t,x) = \frac{L_A k_1 } { MR } f_{v\phi} ( x )
\end{align}  
where, $k_1 = \frac{1}  {2\int_{0}^{1} f_\rho(x) f_{v\phi}(x) x^2 \,dx } $ is an unknown integral constants.

\subsection{Equations of Thermodynamics}
For an ideal gas undergoing reversible processes, the first and second laws of thermodynamics are given, respectively, by:
\begin{align}
\label{eqn:thermo}
    du = dQ - pd\left(\frac{1}{\rho}\right)\text{ and}\hspace{0.1cm}ds = \frac{dQ}{T}
\end{align}
where $u = \frac{p}{(\gamma - 1)\rho}, \quad p = nk(T_p + T_e) = 2nkT, \quad n = \frac{\rho}{m_p}$ are the internal energy per unit mass, thermal pressure, and proton number density, respectively. Here, \(m_p\) is the proton mass, \(\rho\) is the mass density of protons, and \(\gamma\) is the adiabatic index, which is \(5/3\) for a monoatomic ideal gas. \(T_p\) and \(T_e\) are the temperatures of the protons and electrons, respectively, and it is assumed that \(T_p \approx T_e = T\).

Solving Equation~(\ref{eqn:thermo}), we will get
\begin{align*}
    & ln \left( \frac{p}{{\rho}^{\gamma}} \right) =\frac {  (\gamma -1)T \rho s}{ p }= \frac{  (\gamma -1)m_p s}{ 2k } \\
    &\Rightarrow  p = e^{\left[\frac {(\gamma -1)m_p s}{2k} \right]}\rho ^{\gamma}
\end{align*}
putting \hspace{0.2cm}  $\frac{(\gamma -1)m_p }{2k} = \sigma =\text{constant}$ \hspace{0.1cm} in the above equation, we will get \hspace{0.5cm} 
\begin{equation}\label{eqn:pres}
    p(t,r)=e^{ \sigma s } \rho^{\gamma}
\end{equation}

\subsection{Internal Forces}

The Magnetohydrodynamic equation of motion for a magnetised fluid element is given by
\begin{equation}
\label{eqn:MHD}
    \rho \left( \frac{\partial \vec{V}}{\partial t} \right) + \rho (\vec{V} \cdot \nabla) \vec{V} = - \nabla p + \vec{j} \times \vec{B}
\end{equation}
Here, \(p\) denotes the plasma thermal pressure, \(\vec{B} = (0, B_\phi, B_z)\) is the magnetic field, and the current density is given by $\vec{j} = \frac{\nabla \times \vec{B}}{\mu_0}$. Equation~(\ref{eqn:MHD}) represents the equation of motion for the plasma in the inertial frame of the flux-rope. Consequently, it characterizes the expansion dynamics of the flux-rope.

The radial component of the MHD Equation~(\ref{eqn:MHD}) is given by  
\begin{equation*}
   \rho \frac { \partial v_r}{ \partial t}+\rho \left({ v_r \frac{\partial v_r}{\partial r}}-\frac{{v_\phi}^2}{r}\right)=- \frac{ \partial p}{\partial r} + (\vec {j} \times   \vec {B} )_r
\end{equation*}

By using Equations~(\ref{eqn:Vr}) and (\ref{eqn:vphi}) in the above equation, we get
\begin{equation}
\label{eqn:fem_r}
    ( \vec{j} \times \vec {B} )_r =\rho \left({a_e x - \frac{{k_1}^2 {L_A}^2 {f_\phi}^2}{M^2 x R^3} }\right) +\frac{ \partial p}{\partial r }
\end{equation}
By integrating Equation~(\ref{eqn:fem_r}) over the cross-section of the flux-rope, we can calculate the average Lorentz force and is given by
\begin{align}
   &{{\bar f}_{em}} =\frac{1}{ \pi R^2} \int_{0}^{R} \int_{0}^{2\pi} ( \vec{j} \times \vec {B} )_r \,rdr \,d\phi{\nonumber}\\
    &={2} \int_{0}^{1} {\rho_0 f_\rho(x)(a_e x - \frac{k_1 ^2 L_A ^2 f_\phi ^2}{M^2 x R^3})} \,xdx + \frac{ 2}{R } \int_{0}^{1} {x \frac{\partial p}{\partial x}}\,dx
\end{align}
Using Equation~(\ref{eqn:pres}) and $\rho_0 = \frac{M}{ \pi l R^2} $ in the above equation and by solving, we will get
\begin{align}
\label{eqn:fem}
   {{\bar f}_{em}} = \frac{ k_2 M a_e}{lR^2 }-\frac{ k_1^2 k_3 L_A^2}{MlR^5 }-{\frac{M^\gamma e^{\sigma s}}{l^\gamma R^{2\gamma +1}} } k_4
\end{align}
Where, $k_2 = \frac{ 2}{\pi } \int_{0}^{1} {f_\rho( x ) x^2} \,dx>0, k_3 =\frac{ 2}{\pi } \int_{0}^{1} {f_\rho( x )f_{v\phi}^2( x )} \,dx\ge 0 $ and \\$k_4 =\frac{2}{\pi^\gamma}\left[  \int_{0}^{1} {f_\rho^\gamma(x)}\,dx - f_\rho^\gamma(1)\right]$ are unknown integral constants. The first, second, and third terms on the right-hand side of Equation~(\ref{eqn:fem}) represent the average total force associated with the expansion, the average force arising from the poloidal motion, and the average force due to the thermal pressure, respectively.

The magnetic field for an axisymmetric cylindrical flux-rope can be written as
\begin{equation*}
\vec{B} = B_\phi \hat{\Phi} + B_z  \hat {z} = \nabla \times \vec {A} \hspace{0.3cm}  \text{with} 
\end{equation*}
\begin{equation*}
B_\phi = - \frac{ \partial A_z}{\partial r }\text{and} \hspace{0.1cm} B_z =\frac{ 1}{r }\frac{\partial(rA_\phi) }{ \partial r}
\end{equation*}

Under the assumption of self-similar expansion, the magnetic fluxes are conserved in the \(\phi\) and \(z\) directions, and can be expressed as,
\begin{equation}
\label{eqn:magFlux}
\left.
  \begin{array}{rl}
&F_\phi = -l \int_{0}^{R} {\frac {\partial A_z}{\partial r}}\, dr =  l\left([A_z(0)-A_z(R)\right]\\
&F_z = 2\pi \int_{0}^{R} { \frac{\partial( r A_\phi)}{\partial r} }\,dr =  2\pi RA_\phi(R)
\end{array}
\right\}  
\end{equation}

The magnetic vector potentials \(A_\phi\) and \(A_z\) also preserve their spatial distributions under self-similar expansion, and can be expressed as
\begin{equation}
\label{eqn:magVec}
\left.
  \begin{array}{rl}
&A_\phi = f_{A\phi}( x )A_{ \phi o }( t )\\
&A_z = f_{Az}( x )A_{ z o }( t )
\end{array}
\right\}  
\end{equation}

Solving Equations~(\ref{eqn:magFlux}) and (\ref{eqn:magVec}), we will get
\begin{equation}
\label{eqn:magVec2}
 A_\phi(t,x) = \frac{ f_{A\phi}(x)}{R }\hspace{0.1cm}\text{and}\hspace{0.1cm}A_z (t,x) = \frac{f_{Az}( x )}{l}  
\end{equation}

Using Equations~(\ref{eqn:magVec}) and (\ref{eqn:magVec2}), we can find out the average Lorentz force over the cross-section of the flux-rope as 
\begin{align}
\label{eqn:fem2}
 {{\bar f}_{em}} &=\frac{1}{ \pi R^2} \int_{0}^{R} \int_{0}^{2\pi} ( \vec{j} \times \vec {B} )_r \,rdr \,d\phi=\frac{2}{ R^2} \int_{0}^{R} {( \vec{j} \times \vec {B} )_r} \,rdr {\nonumber}\\
&=\frac{2}{ R^2}\int_{0}^{R}  {\biggl[ \frac{ 1} {\mu _0 }(-\frac{\partial B_z}{\partial r} \hat{\phi} +\frac{\partial rB_\phi}{\partial r} \hat {z})\times (B_\phi \hat {\phi} + B_z \hat {z}) \biggr]_r} \,rdr{\nonumber}\\
 \Rightarrow {{\bar f}_{em}}&={-\frac{k_5 } {\mu_0 R^5 } }-{ \frac{k_6}{\mu_0 l^2 R^3} }
\end{align}
Where, $k_5 =2 {\int_{0}^{1} {[\frac{\partial( x f_{A \phi})}{ \partial x}][\partial \frac{(\frac{\partial(x f_{A \phi})}{x \partial x})}{\partial x}]}\,dx}$ and $k_6 ={2 \int_{0}^{1}{[\frac{\partial f_{Az}}{ \partial x}][\frac{\partial (x\frac{\partial f_{Az}}{ \partial x})}{\partial x}]}\,dx}$ are unknown integral constants.

With current observational capabilities, it is challenging to accurately determine the axial length \(l\) of the flux-rope. However, under the assumption that \(l\) is proportional to the distance \(L\) between the flux-rope axis and the solar surface, where \(L\) can be estimated from imaging observations of CMEs, we can write
\begin{equation}
    \label{eqn:L}
    \text{ $l=k_7 L$\hspace{0.3cm}, where $k_7$ is a positive constant.}
\end{equation}
Using Equation~(\ref{eqn:L}) and equating Equations~(\ref{eqn:fem}) and (\ref{eqn:fem2}), we will get
\begin{equation}
\label{eqn:motion1}
    a_e - c_1 R^{-3} = c_2 LR^{-3} + c_3 L^{-1} R^{-1} + \lambda ( t ) L^{ 1-\gamma } R^{ 1-2\gamma }
\end{equation}
This is the equation for the radial expansion of the flux-rope, where $c_1 = \frac{ k_1^2 k_3 L_A^2}{k_2 M^2 }\geq 0 $, $c_2 = \frac{ -k_5 k_7}{\mu_0 k_2 M }$, and $ c_3=\frac{  -k_6}{\mu_0 k_2 k_7 M }\leq 0$ are constants. $\lambda(t)= c_0 e^{ \sigma s(t) }$ is a time-depedent parameter with $c_0 = \frac{ k_4 M^{\gamma -1}}{k_2 k_7^{\gamma -1} }$ is a constant. The first and second terms on the left-hand side of Equation~(\ref{eqn:motion1}) represent the accelerations associated with the radial and poloidal motions, respectively. On the right-hand side, the first two terms correspond to the components of the Lorentz force, while the third term accounts for the thermal pressure force.

The average Lorentz force, the average thermal pressure force, and the average centrifugal force can be expressed in terms of the measurable parameters \(L(t)\) and \(R(t)\), the unknown dimensionless constants \(c_1\)–\(c_3\), and the factor \(\lambda(t)\) as
\begin{align}
    &{{\bar f}_{em}} ={-\frac{k_5 } {\mu_0 R^5 } }-{ \frac{k_6}{\mu_0 l^2 R^3} }= \frac{k_2 M}{ k_7 }{[ {c_2 R^{-5}}+{c_3 L^{-2} R^{-3}} ]}\label{eqn:fem_final}\\ 
    &{{\bar f}_{th} }= \frac{k_4 M^\gamma e^{ \sigma s  }} { l^\gamma R^{2\gamma +1}} = {\frac{ k_2 M}{k_7 }}{ [\lambda(t) L^{-\gamma} R^{-2\gamma -1}] }\label{eqn:fth_final}\\ 
    &{{\bar f}_p} = \frac{k_1^2 k_3 L_A^2 } {MlR^5} = {\frac{k_2 M}{ k_7}}{ [c_1 R^{-5} L^{-1}] }\label{eqn:fp_final}
\end{align}

\subsection{Density, Temperature and Thermal pressure}

The average proton mass density, average proton number density, average thermal pressure, and average temperature can be expressed, respectively, as
\begin{align}
    &\bar {\rho} = \frac{M  }  { \pi l R^2 } =\frac{M  }  { \pi k_7  } ( L R^2 )^{-1} \\
    &{{\bar n}_p} =\frac{\bar \rho} {m_p}  = \frac{M  }  { \pi k_7 m_p  } ( L R^2 )^{-1}\label{eqn:np}\\
    &{\bar p}( t ) = \frac{ 2}{ R^2} \int_{0}^{R} p \,rdr = \frac{ k_2 k_8 M  } {k_4 k_7  }{ \lambda (LR^2)^{-\gamma} }\hspace{0.3cm}\\
    &\text{where, } k_8 = {\frac{2} {\pi ^\gamma }\int_{0}^{1}  {f_\rho ^\gamma(x)}   \,xdx}{\nonumber}\\
    &\bar {T }=\frac{\bar {p} } { 2 \bar {n_p} k}=\frac{\pi \sigma}   {(\gamma -1)} \frac{ k_2 k_8}{k_4 }{\lambda (LR^2)^{1-\gamma}  }\label{eqn:T}
\end{align}

\subsection{Polytropic Index }
During its propagation, the CME may undergo a polytropic process involving multiple phases of expansion and compression, which may include heat exchange with the surroundings. Polytropic processes offer a useful framework to study plasma thermodynamics by relating macroscopic parameters such as pressure and density. A polytropic process follows the relation $p = b(t)\, \rho^{\Gamma}$, where \(p\) is the thermal pressure, \(\rho\) is the plasma density, \(\Gamma\) is the polytropic index, and \(b(t)\) is a time-dependent coefficient. 

Substituting this expression into Equation~(\ref{eqn:pres}), we obtain:
\begin{equation}
\label{eqn:sigma}
     e^{ \sigma s }=b{\rho} ^{\Gamma-\gamma} 
\end{equation} 

Using Equation~(\ref{eqn:sigma}), the expression for $\lambda(t)$ can be written as
\begin{align}
  \label{eqn:lamb1}
    &\lambda( t ) = c_o e^{ \sigma s } =c_o b( t ) \rho^{ \Gamma -\gamma }{\nonumber}\\
    &\Rightarrow  \Gamma = \gamma + \frac{ln[\lambda(t)] -ln[c_o b(t)]}  {ln[\rho(t)]} 
\end{align}

Assuming the thermodynamic process to be quasi-static between two measurement points, i.e., from time \(t\) to \(t + \Delta t\), where \(\Delta t\) is small compared to the overall timescale of the CME propagation, Equation~(\ref{eqn:lamb1}) can be written at time \(t + \Delta t\) as
\begin{equation}
\label{eqn:lamb2}
    \Gamma \approx \gamma + \frac{ln[\lambda(t+\Delta t)] -ln[c_o b(t\Delta t)]}  {ln(\rho(t))} 
\end{equation}

Where the values of \(\Gamma\) and \(b\) remain approximately constant over the short interval \(\Delta t\). Solving Equations~(\ref{eqn:lamb1}) and~(\ref{eqn:lamb2}) yields the final expression for the polytropic index:
\begin{equation}
\label{eqn:lamb}
    {\Gamma} \approx \gamma +  { \frac{ln\left[{\frac{\lambda (t)}  {\lambda (t+\Delta t)}}\right]} {ln\left[\frac{L(t+\Delta t)}{L(t)}\left(\frac{R(t+\Delta t)}  {R(t)}\right)^2\right]} }
\end{equation}

\subsection{Rate of Change of Entropy and Heating Rate}
The rate of change of entropy per unit mass of the CME can be derived from the expression of $\lambda(t)$ as follows,
\begin{align}
    \label{eqn:entropy}
    &\lambda( t ) = c_o e^{ \sigma s } {\nonumber}\\
    &\Rightarrow \frac{d\lambda }{dt } = {\sigma \lambda}\frac{ds }{dt }{\nonumber}\\
    &\Rightarrow \frac{ds }{dt } = \frac{1}{{\sigma \lambda}}\frac{d\lambda }{dt } 
\end{align}

The average heating rate per unit mass of the CME can be written as
\begin{align}
    \label{eqn:heating}
   &\bar {\kappa} ( t ) = \frac{dQ(t) }{dt }  = \bar{T} \frac{ ds} {dt } {\nonumber}\\
   &\Rightarrow \bar {\kappa} ( t )) =\frac{\pi } {(\gamma -1)} \frac{ k_2 k_8}{k_4 }{ (LR^2)^{1-\gamma}  }\frac{d\lambda }{dt }
\end{align}

{ 
\renewcommand{\arraystretch}{1.6} 

\begin{table*}[ht]
\caption[List of derived internal plasma parameters from the FRIS Model]{\label{chap2_tab:FRIS_parameters} List of derived internal plasma parameters from the FRIS Model.}
\begin{adjustbox}{width=\textwidth}
\begin{tabular}{lccc}
\hline
\textbf{Quantities} & \textbf{Factors} & \textbf{Values} & \textbf{SI Units} \\[3pt]
\hline
Lorentz Force (${{\bar f}_{em}}$) & \(\displaystyle \frac{ k_2 M}{k_7 }\) & \(\displaystyle { {c_2 R^{-5}}+{c_3 L^{-2} R^{-3}} }\) &$Pa.m^{-1}$  \\[3pt]

Thermal pressure Force (${{\bar f}_{th}}$) & \(\displaystyle \frac{ k_2 M}{k_7 }\) & \(\displaystyle { \lambda(t) L^{-\gamma} R^{-2\gamma -1} }\) &$Pa.m^{-1}$  \\[3pt]

Centrifugal Force (${{\bar f}_{p}}$) & \(\displaystyle \frac{ k_2 M}{k_7 }\) &\(\displaystyle { c_1 R^{-5} L^{-1} }\)& $Pa.m^{-1}$  \\[3pt]

Proton number density (${{\bar n}_{p}}$) & \(\displaystyle \frac{ M}{k_7 }\) & \(\displaystyle \frac{1}  {\pi  m_p} ( L R^2 )^{-1}\)& $m^{-3}$ \\[3pt]

Thermal pressure ($\bar {p}$) & \(\displaystyle \frac{ k_2 k_8 M}{k_4 k_7 }\) &\(\displaystyle { \lambda (LR^2)^{-\gamma} }\) & $Pa$\\[3pt]

Temperature ($\bar {T}$) & \(\displaystyle \frac{ k_2 k_8 }{k_4 }\) & \(\displaystyle \frac{\pi \sigma}   {(\gamma -1)} {\lambda (LR^2)^{1{-\gamma}}  }\) &K\\[3pt]

Rate of change of entropy ($\frac{ds}{dt}$) & & \(\displaystyle \frac{1}{{\sigma \lambda}}\frac{d\lambda }{dt }\) & $J.K^{-1}.kg^{-1}.s^{-1}$ \\[3pt]

Heating rate ($\bar{\kappa}$) & \(\displaystyle \frac{ k_2 k_8 }{k_4 }\) & \(\displaystyle \frac{\pi }   {(\gamma -1)} { (LR^2)^{1{-\gamma}}  }\frac{d\lambda }{dt }\) & $J.kg^{-1}.s^{-1}$ \\[3pt]

Polytropic Index ($\Gamma$) & & \(\displaystyle \gamma +  \frac{\ln{\frac{\lambda (t)}  {\lambda (t+\Delta t)}}}  {\ln\left[\left(\frac{L(t+\Delta t)}  {L(t)}\right)\left(\frac{R(t+\Delta t)}  {R(t)}\right)^2\right]} \) &  \\ \vspace{-1cm} \\
\hline
\end{tabular}
\end{adjustbox}
\begin{tablenotes}
\it
\small
\item Note -- The factors are unknown constants that can not be estimated from the FRIS model.
\end{tablenotes}

\end{table*}
}

\subsection{Equation of Radial Expansion }
The equation of motion (\ref{eqn:motion1}) can be written as
\begin{equation}
\label{eqn:motion2}
    \lambda ( t )= L^{ \gamma -1 } R^{ 2\gamma -1 }( a_e - c_1 R^{-3} - c_2 LR^{-3} - c_3 L^{-1} R^{-1} )
\end{equation}

Using only the measurable parameters \(L\), \(R\), and their time derivatives, Equation~(\ref{eqn:motion2}) alone is insufficient to determine the unknown parameters \(c_1\)–\(c_3\) and \(\lambda(t)\). Therefore, an additional constraint on \(\lambda(t)\) is required. 

To provide this constraint, we assume that, regardless of the specific heating mechanisms within CMEs, the heating rate per unit mass can be equivalently represented as the result of heat flow. So
\begin{align*}
 &{\bar {\kappa}}  \propto {\frac{Area \cdot \Delta T}{length} }\Rightarrow {\bar {\kappa}} = k_{11}{\frac{T_a - {\bar{T}}}{{\bar{\rho}}L^2 }} \\
 &\Rightarrow \frac{ \pi k_2 k_8}{(\gamma -1)k_4 }(LR^2)^{1-\gamma}\frac{d\lambda }{dt}= k_{11} \left[ \frac{{T_a}-{\frac{\pi \sigma k_2 k_8}{(\gamma -1)k_4}\lambda(LR^2)^{1-\gamma}}} {\frac{M}{\pi k_7} (LR^2)^{-1} L^2} \right]
\end{align*}
\begin{equation}
\label{eqn:motion3}
 \Rightarrow(LR^2 ) ^{ \gamma -1 } = c_4 LR^{-2} \frac{d\lambda }{dt } + c_5 \lambda 
\end{equation}
 Where $k_{11}$ is an unknown constant, $T_a$ is the average temperature of the ambient solar wind, $c_4 = \frac{ k_2 k_8 M}{(\gamma -1) k_4 k_7 k_{11} T_a}$ and $c_5 = \frac{\pi \sigma k_2 k_8 }{ (\gamma -1) k_4 T_a}$ are unknown constant integrals.

On integrating Equation~(\ref{eqn:motion2}), we will get 
\begin{align}
\label{eqn:dlambda}
 \frac{ d\lambda}{dt }=L^{ \gamma -1 }R^{2 \gamma -1 }& \biggl[\frac{da_e}{dt}+{(2\gamma -1){a_e v_e R^{-1}}+(\gamma -1) a_e v_c L^{-1}}\biggr.{\nonumber}\\
 &\left.+ c_1 ({(4-2\gamma)v_e R^{-4}}+{(1-\gamma)v_c L^{-1} R^{-3}})\right.{\nonumber}\\
 &\left.+ c_2 ({(4-2\gamma )v_e L R^{-4} - {\gamma v_c R^{-3}}})\right.{\nonumber}\\
 &\biggl.+ c_3 ({( 2-2\gamma ) v_e L^{-1} R^{-2}} +(2-\gamma) v_c L^{-2} R^{-1})\biggr] 
\end{align}

Solving Equation~(\ref{eqn:motion2}), (\ref{eqn:dlambda}) and (\ref{eqn:motion3}), and solving, we will get
\begin{align}
&(LR^2)^{\gamma -1} = L^{\gamma -1} R^{2\gamma -1}\biggl[ ({c_5 a_e}-{c_3 c_5 L^{-1} R^{-1}}+({{c_4 L \frac {da_e}{dt}}+{(\gamma -1)c_4 a_e v_c}}) R^{-2} )  \biggr.{\nonumber}\\
&\left.+  ({(2-\gamma)c_3 c_4 v_c L^{-1}}+{(2\gamma -1)c_4 a_e v_e L}-{c_2 c_5 L}-{c_1 c_5}  )R^{-3} + ((2-2\gamma)c_3 c_4 v_e  )R^{-4} \right.{\nonumber}\\
&\biggl.+ ({(1-\gamma)c_1 c_4 v_c}-{\gamma c_2 c_4 v_c L} )R^{-5} + ({(4-2\gamma)c_1 c_4 v_e L}+{(4-2\gamma)c_2 c_4 v_e L^2})R^{-6}\biggr]
\end{align}

Rearranging the above equation, we can write it further as follows. 

\begin{align}
\frac {R}{L} =  &{c_5 \biggl[\frac {a_e R^2}{L}\biggr]}-{c_3 c_5 \biggl[\frac {R}{L^2}\biggr]}-{c_2 c_5\biggl[\frac {1}{R}\biggr]}-{c_1 c_5\biggl[\frac {1}{LR}\biggr]} {\nonumber}\\
&+c_4 \biggl[\frac {da_e}{dt}+{\frac {(\gamma -1)a_e v_c}{L}}+\frac{(2\gamma -1)a_e v_e}{R}\biggr] + c_3 c_4\biggl[ \frac{(2-\gamma) v_c }{L^2 R}+{\frac{(2-2\gamma) v_e }{LR^2}}\biggr] {\nonumber}\\ 
&+c_2 c_4\biggl[{\frac{(4-2\gamma) v_e L}{R^4}}-{\frac{\gamma v_c}{R^3}}\biggr] + c_1 c_4\biggl[{\frac{(4-2\gamma) v_e}{R^4}}+{\frac{(1-\gamma)v_c}{LR^3}}\biggr]
\label{fitting}
\end{align}

The above Equation~(\ref{fitting}) represents the final equation of motion describing the radial expansion of the CME. The unknown constants \(c_1\)–\(c_5\) and the function \(\lambda(t)\) can be determined by fitting Equation~(\ref{fitting}) to the observational measurements of \(L(t)\), \(R(t)\), and their time derivatives.

\section{Methodology}

As the application of the FRIS model relies on accurate measurements of CME kinematics, sky-projected values obtained directly from coronagraph images are insufficient for this purpose due to line-of-sight projection effects. To address this limitation and to provide reliable model inputs, additional methodologies are employed. The Graduated Cylindrical Shell model is used to reconstruct the true three-dimensional geometry and kinematic properties of CMEs by correcting for projection effects in multi-viewpoint coronagraph data.

To complement the FRIS results and to capture the early thermal evolution of the CME flux-rope, Differential Emission Measure analysis is applied to EUV observations, enabling estimates of temperature and emission measure shortly after eruption. Additionally, to characterize the thermal state during the interplanetary phase, in situ measurements of plasma and magnetic field parameters near 1 au are utilized. Together, these complementary approaches provide a multi-scale, multi-instrument basis for a consistent and physically grounded reconstruction of CME thermodynamics from eruption through heliospheric propagation.

\subsection{Graduated Cylindrical Shell Model}\label{sec:chapt2-metho-gcs}

To derive the true kinematic and geometric parameters of CMEs, this thesis employs the Graduated Cylindrical Shell (GCS) model \citep{Thernisien2006, Thernisien2009}. The GCS model is a forward-fitting, geometrical model designed to reproduce the large-scale morphology of CMEs observed in coronagraph images from multiple viewpoints. It represents a flux-rope-like CME as a combination of conical legs and a curved front, forming a hollow, croissant-shaped shell that closely resembles CME structures seen in white-light observations. By adjusting six free parameters such as latitude, longitude, tilt angle, height, aspect ratio, and half-angle, the model is fit to simultaneous images from coronagraphs on SOHO/LASCO and STEREO/SECCHI/COR, enabling the extraction of the CME’s true propagation direction, angular width, and deprojected height–time profile.

The GCS model has been widely used in CME studies to overcome the limitations of single-viewpoint projection \citep{Kilpua2012, Vourlidas2013, Wang2014, Mishra2015}.  These studies highlight the model's utility in improving our understanding of CME trajectories, and arrival time predictions at 1 au. In this thesis, the GCS-derived kinematics provide key inputs for the FRIS model, particularly for accurate measurements of CME radial distance, expansion rate, and aspect ratio.

\begin{figure}
    \centering
    \includegraphics[width=1.1\linewidth]{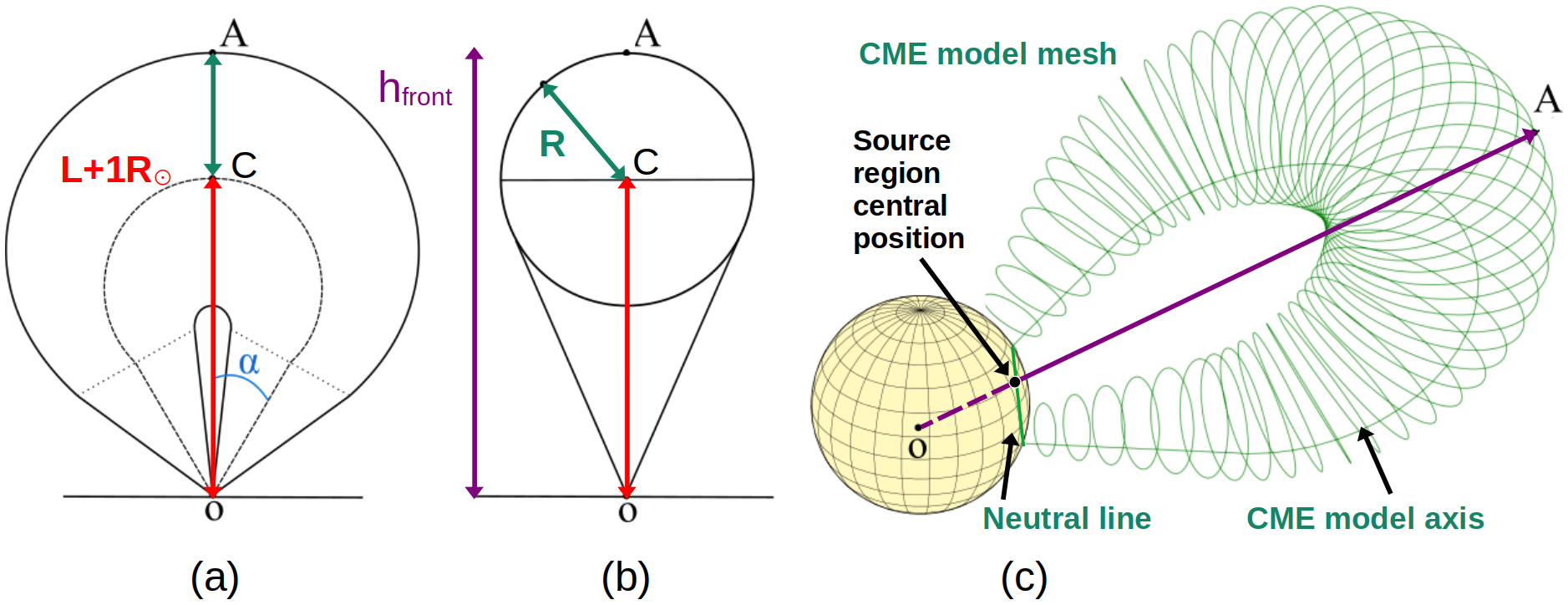}
    \caption[Schematic illustration of the GCS model showing]{Schematic illustration of the GCS model showing (a) a face-on view and (b) an edge-on view of the CME structure, where `O' marks the solar center and `C' denotes the apex center. The primary axis of the model extends from the Sun's center to the CME apex. Panel (c) displays the 3D mesh of the CME along with the neutral line connecting the two footpoints on the solar surface, including the central location of the source region. [\textit{Image Credit: reproduce from \citealt{Thernisien2006} and \citealt{ Kouloumvakos2022}}]}
    \label{chap2_fig:GCS}
\end{figure}

Despite several advantages of the GCS model, it has certain limitations. It assumes a symmetric and idealized flux-rope geometry, which may not capture the true complexity of highly distorted or asymmetric CMEs. The fitting process is also subjective and depends on the visual judgment of the observer, although experience and multi-viewpoint constraints help mitigate this. Furthermore, it does not capture internal CME structure (e.g., current sheets, filament material), nor does it provide information on plasma thermodynamics. Nevertheless, the GCS model remains one of the most effective tools for obtaining three-dimensional CME parameters when multi-spacecraft coronagraph observations are available, and its integration with the FRIS framework serves our thesis purpose.


\subsection{Differential Emission Measure}\label{chapt2-metho-dem}

To understand the early heating characteristics of CMEs, it is essential to investigate their thermal state in the lower corona, particularly below the heights at which they are first detected in white-light coronagraphic observations. At these early stages, the CME core typically corresponds to a magnetic flux-rope (MFR), a coherent bundle of helical magnetic field lines wrapped around a common axis \citep{Chen2017, Wang2017, Wang2018, Green2018}. In many ideal magnetohydrodynamic (MHD) models, including those based on torus and kink instabilities \citep{Kliem2006, Torok2010, Amari2014}, MFRs naturally emerge due to the quasi-force-free nature of the coronal magnetic field, wherein the electric currents are largely aligned with magnetic field lines. These aligned currents induce poloidal flux, which introduces twist in the surrounding field lines and results in the formation of a stable helical structure. The presence of MFRs is strongly supported by coronal magnetic field reconstructions and multi-wavelength observational studies \citep{Chen2017, Wang2017, Gou2019, Duan2019}. Given this fundamental role of MFRs in CME initiation and evolution, a detailed thermodynamic diagnosis of the flux-rope structure at the source region is crucial for capturing the early energy release and plasma heating.

\begin{figure}[hbt]
    \centering
    \includegraphics[width=0.8\linewidth]{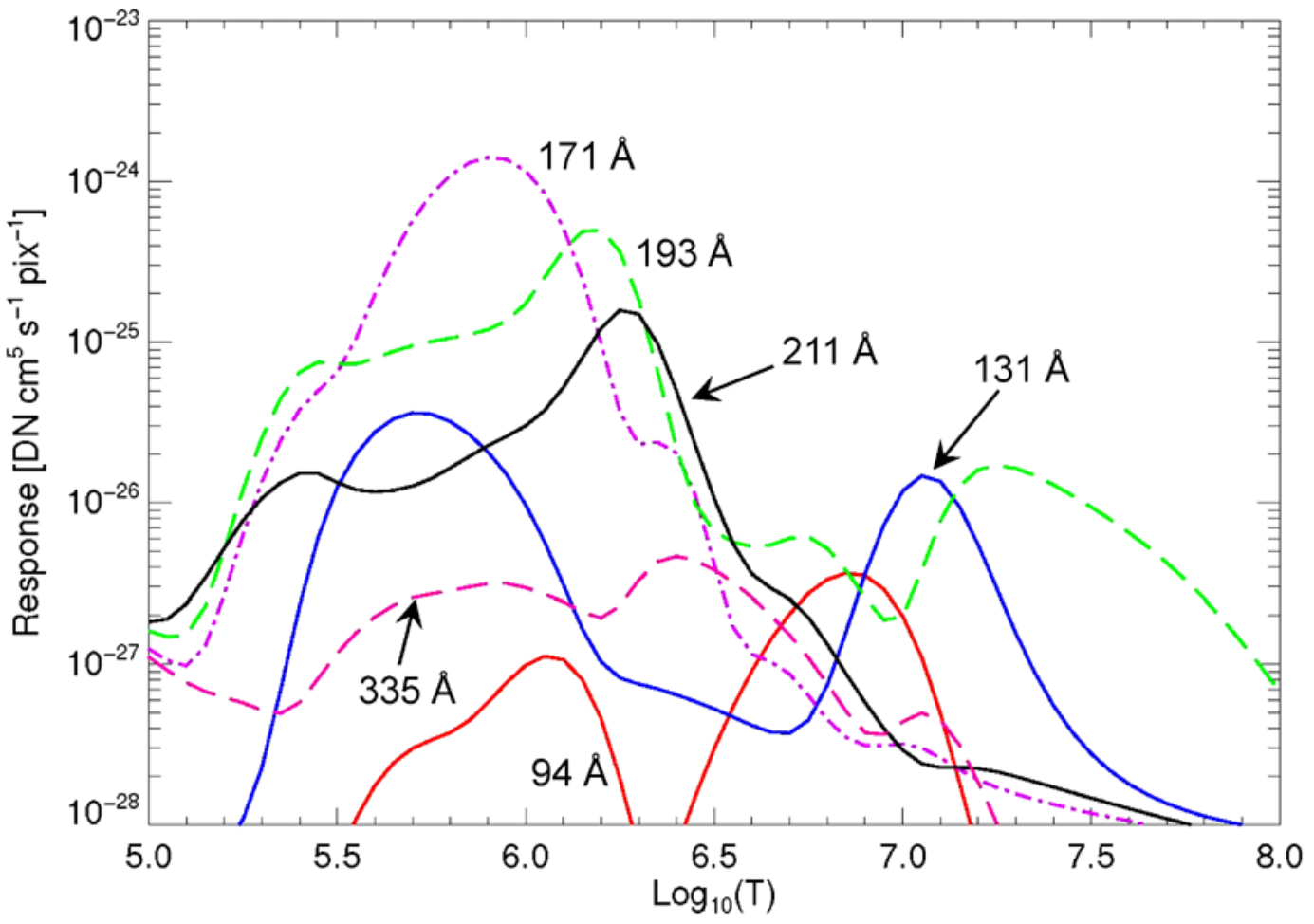}
    \caption[Temperature response functions of the six EUV channels]{Temperature response functions of the six EUV channels observed by SDO/AIA, illustrating their sensitivity to different coronal temperature ranges. [\textit{Image Credit: \citealt{Lemen2012}}]}
    \label{chap2_fig:aia_respons}
\end{figure}

To investigate this, we employed multi-wavelength EUV observations from the Atmospheric Imaging Assembly (AIA; \citealt{Lemen2012}) onboard the Solar Dynamics Observatory (SDO; \citealt{Pesnell2012}). AIA provides full-disk imaging of the solar corona with high temporal and spatial resolution in six EUV channels (94, 131, 171, 193, 211, and 335 \AA), each sensitive to different temperature regimes. The temperature response functions of these EUV channels are shown in Figure~\ref{chap2_fig:aia_respons}, indicating their coverage across a broad temperature range. This multi-channel capability enables the application of differential emission measure (DEM) analysis, which reconstructs the distribution of plasma as a function of temperature along the line of sight. Mathematically, the observed intensity in a given EUV channel can be written as:

$$
I_i = \int G_i(T)\, DEM(T)\, dT,
$$

where $I_i$ is the observed intensity in the $i$-th channel, $G_i(T)$ is the instrument’s temperature response function, and $DEM(T)$ is the differential emission measure, the key physical quantity that represents the amount of plasma per unit temperature interval along the line of sight. Recovering $DEM(T)$ from this integral inversion is an ill-posed problem. To address this, we used the sparse inversion code developed by \citet{Cheung2015}, which applies a regularized optimization technique to recover stable and physically meaningful DEM solutions. This method minimizes the number of non-zero elements in the DEM solution (sparsity), promoting simpler distributions that are consistent with observed data.

In our study, DEM solutions were calculated over a temperature range spanning log $T$ = 5.5 to 7.5, with bin widths of log $T \leq 0.1$. For each CME event, we identified five regions of interest (ROIs) with dimensions of $10'' \times 10''$ within the observed EUV flux-rope structures. The selected ROIs were carefully placed to minimize contamination from nearby active regions or post-eruptive loops and to ensure that the derived properties reflected the thermal state of the flux-rope itself. Although these ROIs cover a limited portion of the flux-rope, we consider them representative due to careful manual selection based on morphology and temporal evolution. To associate the EUV signatures with the corresponding CME source region, we employed well-established indicators of CME activity such as flare ribbons, post-eruption arcades, and rising loop systems \citep{Liu2010a, Cheng2011}.

From the derived DEM profiles, we calculated the total emission measure (EM) by integrating over the temperature range and determined the DEM-weighted average electron temperature using the expression:

$$
\bar{T} = \frac{\sum DEM(T)\, T\, \Delta T}{\sum DEM(T)\, \Delta T}.
$$

This average temperature provides a characteristic thermal state of the flux-rope plasma shortly after eruption. These DEM-derived thermal parameters are used to constrain the early-stage energy content of CMEs and serve as a key observational anchor for validating the thermodynamic predictions of the FRIS model.

\subsection{Measurement of Thermal State from In Situ}\label{sec:chapt2-metho-poly}

The polytropic index ($\Gamma$) serves as a key diagnostic parameter to characterize the thermodynamic state of CME plasma as it propagates through interplanetary space. By considering that the plasma evolves through a general polytropic process, a relation between temperature ($T$) and number density ($n$) can be expressed as $T \, n^{1-\Gamma} = \text{constant}$. This implies that a linear fit between the logarithmic values of temperature and density can be used to estimate $\Gamma$. Specifically, taking the logarithm of both sides yields a linear form: $\log T = (\Gamma - 1) \log n + \text{constant}$, enabling the direct extraction of $\Gamma$ from the slope of the fit. 

Given the potential presence of fine-scale thermal inhomogeneities within CMEs, especially due to processes such as magnetic reconnection, turbulence, or solar wind interactions, it is likely that different regions of a single ICME or complex ejecta may not share the same thermal evolution. Consequently, fitting a single linear relation over the full duration of an ICME may obscure these variations and result in poor correlation. To address this, we implemented a moving window approach where the $\log T$ vs. $\log n$ fit was performed over short, overlapping sub-intervals, allowing for localized estimation of the polytropic index. This method has also been employed in prior solar wind and ICME studies \citep{Nicolaou2014, Nicolaou2020, Dayeh2022}, revealing regional variability in the thermal behavior, such as adiabatic conditions in the sheath and solar wind, and near-adiabatic heating in magnetic ejecta (ME) regions.

\begin{figure}
    \centering
    \includegraphics[width=0.65\linewidth,trim={5pt 5pt 0pt 10pt}, clip]{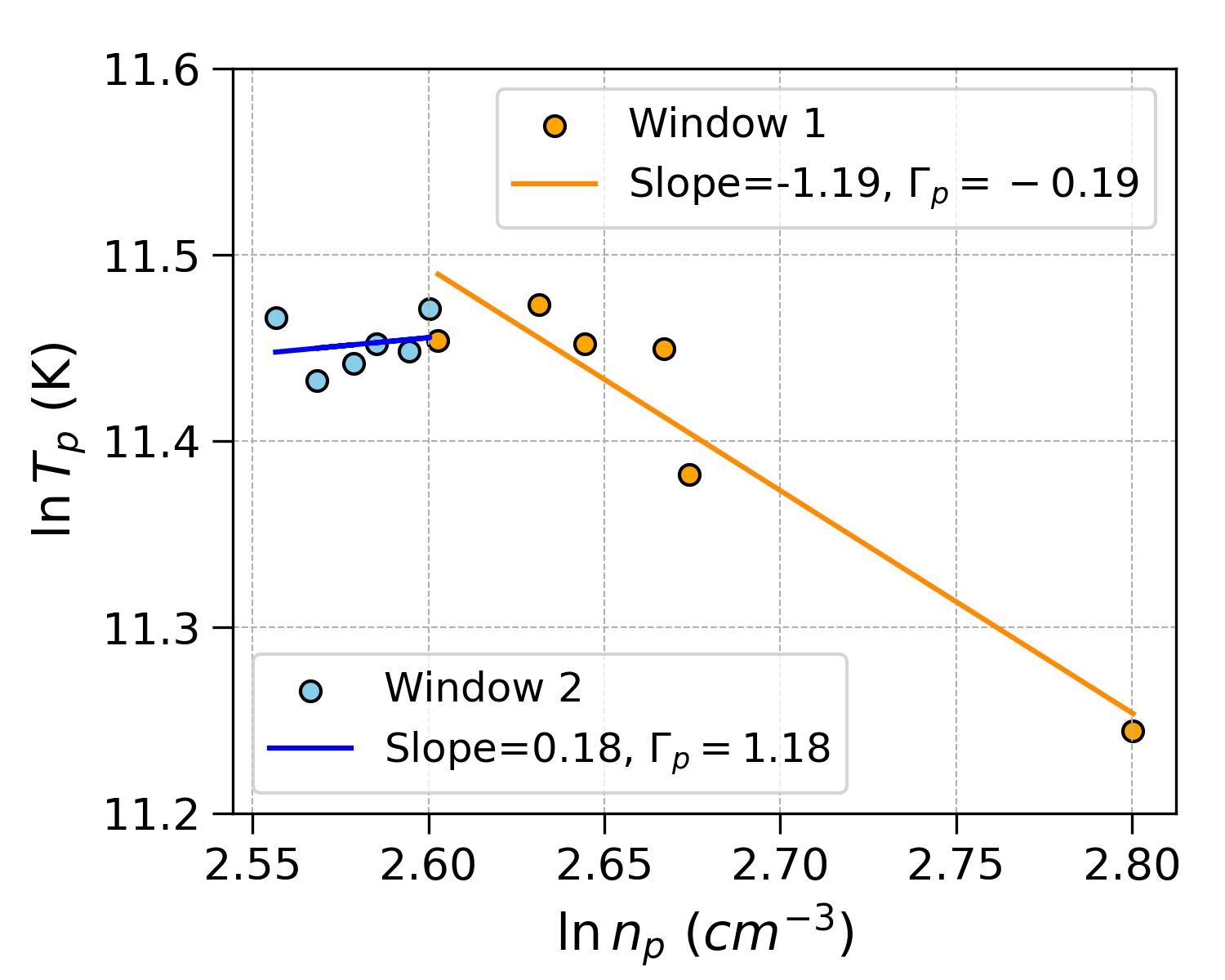}
    \caption[Example of proton polytropic index ($\Gamma_p$) estimation]{Example of proton polytropic index ($\Gamma_p$) estimation for two random subintervals, each consisting of 6 data points. Linear fits between $\log T_p$ and $\log n_p$ are shown to illustrate the method used for deriving $\Gamma_p$.}
    \label{chapt2_fig:demo_gamma}
\end{figure}

In our analysis, we used in situ data from multiple sources and at varying temporal resolutions. For protons, 92-second resolution data for proton number density ($n_p$) and temperature ($T_p$) were obtained from the Solar Wind Experiment (SWE; \citealt{Ogilvie1995}) onboard the Wind spacecraft. For electrons, 9-second resolution measurements of electron density ($n_e$) and temperature ($T_e$) from Wind/SWE were used. Additionally, for longer statistical surveys and large-scale event sampling, we utilized 1-minute averaged proton density and temperature data from the OMNI database (\url{https://cdaweb.gsfc.nasa.gov/}). These datasets collectively enabled us to analyze both fine-scale and bulk thermal behavior of CMEs.

To determine the optimal sub-interval length for the fitting, we tested durations ranging from 3 to 10 data points and found that a 6-point interval consistently yielded the highest number of statistically robust estimates, defined by Pearson correlation coefficient (CC) $\geq$ 0.8 and p-value $\leq$ 0.05 (Figure~\ref{chapt2_fig:demo_gamma}). This window corresponds to approximately 9.2 minutes for protons (6 × 92 s), 54 seconds for electrons (6 × 9 s), and 6 minutes for the OMNI data (6 × 1 min). Sub-intervals that did not meet the correlation criteria were excluded from further analysis. This filtering ensured that only those segments where temperature and density exhibit a strong and statistically significant linear relationship were used in the final estimation of $\Gamma$.

\begin{figure*}
    \centering
    \includegraphics[width=1\linewidth]{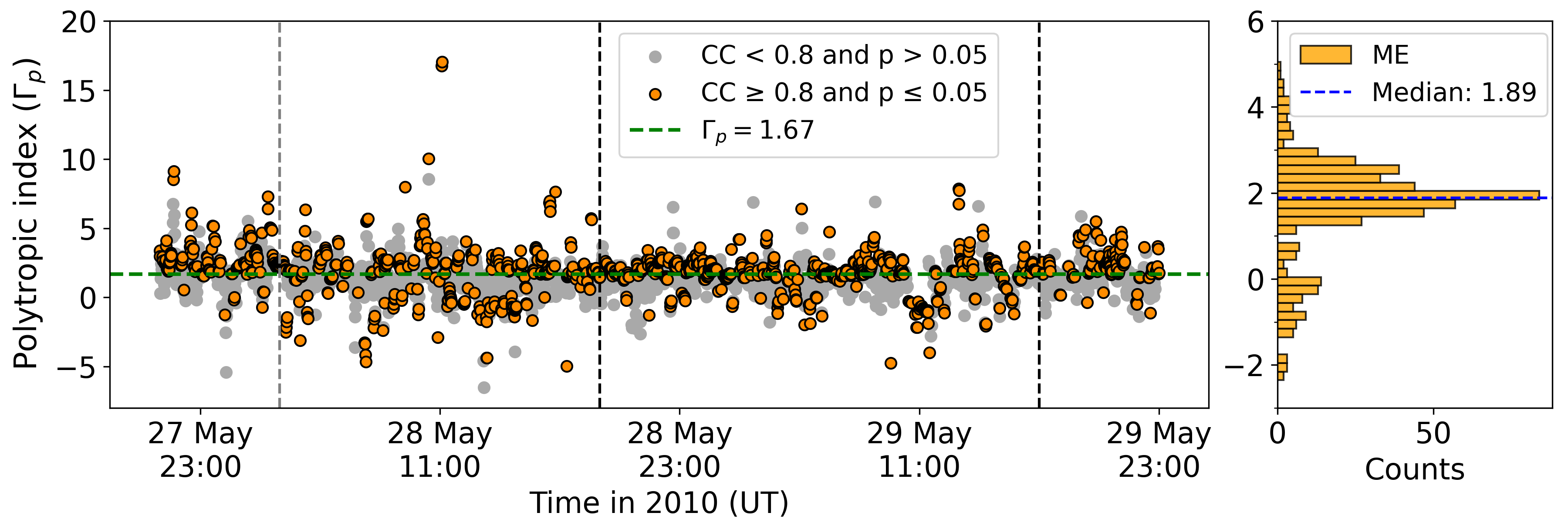}
    \caption[Derived $\Gamma_p$ values for the ICME event on 28 May 2010]{Derived $\Gamma_p$ values for the ICME event on 28 May 2010. The gray dashed vertical line indicates the onset of the sheath region, while the black dashed vertical lines mark the boundaries of the magnetic ejecta (ME). Orange dots correspond to statistically reliable $\Gamma_p$ estimates, and gray dots represent less reliable values excluded from analysis. The green dashed horizontal line denotes the adiabatic index ($\Gamma_p = 5/3$) for reference. The right-hand panel presents a histogram of the reliable $\Gamma_p$ values within the ME interval. }
    \label{chapt2_fig:one_gamma}
\end{figure*}

To maximize the coverage and spatial sampling, we applied the fitting procedure to moving sub-intervals using step sizes equal to the native resolution of the respective dataset, i.e., 9 s for electrons, 92 s for protons (Wind), and 1 min for OMNI. This approach produced a time series of localized $\Gamma$ values along the CME trajectory, enabling us to map variations in the thermal state both within and across different CME substructures (Figure~\ref{chapt2_fig:one_gamma}). We did not apply Bernoulli integral filtering, which is often used to ensure that the plasma parcel remains on a single streamline, as the high correlation thresholds imposed in our analysis already ensure that each sub-interval represents a coherent thermodynamic state. In addition to the magnetic ejecta (ME) region, we also analyzed the pre-ICME and post-ICME solar wind to compare the thermal behavior of CME plasma with the ambient solar wind environment.

This methodology provides a statistically robust and physically motivated framework for evaluating the local thermodynamic properties of CME plasma using in situ measurements. The localized approach adopted here not only improves the accuracy of $\Gamma$ estimates but also facilitates a more detailed interpretation of CME heating and expansion processes as they evolve through interplanetary space.

\subsection{Superposed Epoch Analysis}\label{sec:chapt2-metho-sea}

To characterize the average temporal evolution of plasma and magnetic field properties associated with ICMEs, particularly within their ME regions, we employed the Superposed Epoch Analysis (SEA) technique. SEA is a widely used statistical method in space physics that enables the extraction of typical patterns from a collection of events by aligning them to a common reference time and averaging over multiple realizations. This method has been extensively applied in the context of solar wind, CME, and geomagnetic studies \citep{Zhang1988sea, Winslow2015sea, MasiasMeza2016, Janvier2019, Regnault2020, Guo2021}. 

The core idea of SEA is to map events with varying durations and starting times onto a common, normalized timescale. This allows for statistical comparisons of event structures by reducing temporal variability. In our implementation, each ICME event was aligned such that its sheath start time (or disturbance time, as defined in the Richardson and Cane catalog) was mapped onto a common reference point. Subsequently, both the sheath and ME regions were rescaled proportionally to a standard time interval. Based on earlier findings that ME regions at 1 au are typically about three times longer than the sheath regions \citep{Zhang2008, MasiasMeza2016, Mishra2021a}, we assigned 25 time bins to the sheath and 75 time bins to the ME. Additionally, we included two auxiliary regions: a pre-ICME solar wind interval preceding the sheath and a post-ICME wake interval following the ME. These intervals were assigned durations equal to the sheath and ME, respectively, thereby preserving temporal symmetry around the main ICME structure.

After this normalization, each event was divided into equal-length time bins to enable consistent averaging across events. Since events have different natural durations and data cadences, binning ensures that each event contributes equally to the resulting average profiles. For each time bin across all events, we calculated the median values of various physical parameters, including magnetic field magnitude, plasma density and temperature, dynamic pressure, and the derived proton polytropic index $\Gamma_p$. The use of median values instead of means reduces the influence of outliers and enhances the robustness of the SEA-derived profiles.

To investigate differences in the thermal behavior, particularly in the distribution of $\Gamma_p$, across different ICME categories, we applied a suite of complementary statistical techniques. First, the \textit{Kolmogorov--Smirnov (KS) test} was used to compare the cumulative distributions of $\Gamma_p$ between groups. A KS statistic $D > 0.1$ combined with a $p$-value $< 0.05$ indicates a statistically significant difference in overall distribution shape. Second, the \textit{Mann--Whitney U test}, a non-parametric test, was applied to evaluate whether one group tends to produce higher or lower $\Gamma_p$ values than another. A $p$-value below 0.05 in this test implies a significant difference in the central tendency (typically the median). 

To quantify the magnitude and direction of these differences, we employed \textit{Cliff’s delta}, which measures the probability that a randomly chosen value from one distribution is greater than a randomly chosen value from another. Values of Cliff’s delta close to zero indicate substantial overlap, while values approaching $\pm1$ indicate strong effect sizes. Furthermore, we analyzed the \textit{percentage of SEA time bins} where the median $\Gamma_p$ value dropped below the adiabatic threshold ($\Gamma_p < 1.67$), offering insight into the duration and intensity of localized heating processes.

Lastly, to rigorously assess the statistical significance of observed differences in median $\Gamma_p$ values, we performed a \textit{permutation test}. This involved randomly reassigning category labels across the dataset to create a null distribution of median differences. If the observed difference in medians lies in the extreme tails of this null distribution (e.g., $p < 0.01$), we can conclude that the result is unlikely to have arisen by chance. Together, these statistical methods provide a comprehensive framework to determine not only whether thermal differences exist between ICME categories, but also the strength, directionality, and consistency of those differences throughout the ICME lifecycle.

\subsection{Turbulence and Intermittency Characteristics}\label{sec:chapt2-metho-turbu}

To investigate the small-scale dynamics and energy dissipation processes within ICME structures, we analyzed magnetic field fluctuations using high-resolution in situ data from the Magnetic Field Investigation (MFI; \citealt{Lepping1995}) instrument onboard the Wind spacecraft. Specifically, we used the 11 Hz (0.092 s) resolution magnetic field time series to resolve fine-scale turbulent structures and determine spectral properties with high accuracy. This high cadence is crucial for capturing fluctuations across the inertial and dissipation ranges of solar wind turbulence.

The time-series magnetic field fluctuations were computed as:
\[
\delta B(t) = B(t) - \langle B(t) \rangle, \quad \delta B_i(t) = B_i(t) - \langle B_i(t) \rangle, \quad i \in (x, y, z),
\]
where $\langle \cdot \rangle$ denotes the average over a selected interval (e.g., sheath, ME, or solar wind). Fast Fourier Transform (FFT) was then applied to convert the fluctuations $\delta B_i(t)$ into the frequency domain, $\delta B_i(f)$. The corresponding power spectral densities (PSDs) for each component were computed as:
\[
P_x = |\delta B_x(f)|^2, \quad P_y = |\delta B_y(f)|^2, \quad P_z = |\delta B_z(f)|^2,
\]
and the total and trace magnetic field PSDs were calculated as:
\[
P_t = |\delta B(f)|^2, \quad P_{\text{tr}} = P_x + P_y + P_z.
\]
The trace PSD $P_{\text{tr}}(f)$ quantifies how magnetic fluctuation energy is distributed across frequencies. The turbulence spectrum in magnetized plasma is typically characterized by a power-law dependence, $P_{\text{tr}}(f) \propto f^{\alpha_B}$, where $\alpha_B$ is the spectral slope. We applied a sliding three-point moving average to smooth the PSD curves and performed a linear fit in log-log space within the inertial and dissipation ranges to determine $\alpha_B$. The inertial range, typically spanning $-3 < \log_{10} f < -0.7$, reflects scale-invariant turbulent cascades with spectral slopes close to $-5/3$ or $-3/2$, while the dissipation range ($\log_{10} f > -0.3$) exhibits steeper slopes ($\sim -2$ or steeper), indicating energy dissipation via kinetic processes \citep{Bruno2013, Chen2020}.

To characterize the nature of the turbulent fluctuations, we also computed the \textit{magnetic compressibility}:
\[
C_B(f) = \frac{P_t(f)}{P_{\text{tr}}(f)},
\]
which reflects the degree to which compressible fluctuations contribute to the total turbulence energy. Low values of $C_B$ (typically $<0.2$) are indicative of Alfvénic, incompressible turbulence, whereas higher values suggest compressible modes, such as fast or slow magnetosonic waves \citep{Telloni2021}. In our analysis, we focus on the inertial range and compute the average $C_B$ over this frequency band, as marked in relevant figures.


In addition to turbulence spectra, we examined the intermittent nature of energy dissipation using the \textit{Partial Variance of Increments} (PVI) method. Intermittency, manifested as sporadic and localized intensifications in turbulence, plays a critical role in plasma heating and is linked to coherent magnetic structures and current sheets \citep{NOVIKOV1971, Greco2008}. The PVI method provides a robust measure for identifying such intermittent events in the time series of vector magnetic field data.

The magnetic field increment is defined as, 
$\Delta B(t, \tau) = B(t + \tau) - B(t)$, where $\tau$ is the time lag (scale). The PVI statistic is given by:
\[
\text{PVI}(t, \tau) = \frac{|\Delta B(t, \tau)|}{\sqrt{\langle |\Delta B(t, \tau)|^2 \rangle}},
\]
with the denominator representing the root-mean-square value of the magnetic field increments over a trailing window. This normalization enables detection of significant magnetic field gradients irrespective of their absolute magnitude. High PVI values identify locations of coherent structures or potential sites of enhanced dissipation, such as reconnection exhausts or shear layers. To assess intermittency across different spatial scales, we computed PVI for three characteristic time lags: 0.18 s (dissipation scale), 9.2 s (transition), and 92 s (inertial scale), following previous studies \citep{Kilpua2020, Emiliya2021}. The averaging in the denominator was performed over the full dataset, encompassing pre-ICME solar wind, sheath, magnetic ejecta, and post-ICME regions, to ensure statistical reliability while avoiding biases from localized structures.

This combined spectral and intermittency analysis enables a comprehensive investigation of energy cascade and dissipation processes within ICMEs, offering insights into their internal dynamics and potential plasma heating mechanisms.

\section{Observational Data}

In order to investigate the kinematic and thermodynamic evolution of CMEs through the heliosphere using the above discussed methodologies, this thesis employs a combination of remote-sensing and in situ observations of Earth-directed events. These complementary datasets enable a multi-perspective understanding of CME behavior across spatial and temporal scales.

Remote-sensing observations detect CMEs via Thomson-scattered white-light from CME plasma free electrons, providing information about their morphology, kinematics, direction of propagation, and mass near the Sun. In contrast, in situ measurements directly sample the plasma and magnetic field parameters as the CME passes over a spacecraft, offering insights into its local structure, composition, temperature, and magnetic topology at 1 au and beyond. Remote observations typically capture the early stages of CME evolution in the low and extended corona, whereas in situ observations reflect their evolved state after significant interaction with the ambient heliospheric environment.

The following sections provide an overview of the observational instruments and datasets used in this study.

\subsection{Remote-Sensing Observations}

To characterize three-dimensional geometry and true kinematics of CMEs, we utilized multi-point white-light coronagraph images from multiple spacecraft missions that continuously observe the solar corona. These observations allow us to track CME propagation in the heliosphere and reconstruct their 3D morphology.

\begin{itemize}
    \item \textbf{SOHO/LASCO C2 and C3:} SOHO, a joint project of ESA and NASA, has been in operation since its launch in December 1995. It is positioned at the Lagrangian point L1 and has provided uninterrupted solar observations since 1996. LASCO is one of the key instruments onboard SOHO, designed specifically for imaging the solar corona by blocking direct sunlight using an occulting disk. LASCO comprises three coronagraphs such as C1, C2, and C3, with a field of view of 1.1--3~$R_\odot$, 1.5--6~$R_\odot$, and 3.7--30~$R_\odot$respectively. Together, they provide detailed imaging of CMEs as they propagate through the outer corona.
    
    \item \textbf{STEREO-A/SECCHI COR1 and COR2:} STEREO mission, launched in 2006, consists of two nearly identical spacecraft (STEREO-Ahead and STEREO-Behind) placed in heliocentric orbits ahead and behind the Earth, respectively. This configuration allows stereoscopic imaging of the Sun and its eruptions. The Sun Earth Connection Coronal and Heliospheric Investigation (SECCHI; \citealt{Howard2008}) is a suite of instruments, an extreme ultraviolet imager (EUVI: 1--1.7~$R_\odot$), two coronagraphs (COR1: 1.5--4~$R_\odot$ and COR2: 2.5--15~$R_\odot$) and two heliospheric imagers (HI-1: 15--84~$R_\odot$ and HI-2: 66--318~$R_\odot$). These instruments provide an additional viewpoint, enabling the reconstruction of CME geometry, orientation, and propagation direction with reduced projection effects when used in combination with SOHO observations.
\end{itemize}

In this thesis, the GCS model was applied to these multi-viewpoint coronagraph images to derive the 3D structure, orientation, and propagation direction of selected Earth-directed CMEs. This forward-modeling approach is described in detail in Section~\ref{sec:chapt2-metho-gcs}.

To determine the thermal properties of CME flux ropes at their origin in the low corona, we employed EUV observations.

\begin{itemize}
    \item \textbf{SDO/AIA}: SDO mission under NASA’s \textit{Living With a Star (LWS)} program, launched on 11 February 2010. Its primary goal is to study the causes of solar variability and understand how solar activity influences the Earth and near-Earth environment. SDO is equipped with three key scientific instruments: the Atmospheric Imaging Assembly (AIA; \citealt{Lemen2012}), the Extreme Ultraviolet Variability Experiment (EVE; \citealt{Woods2012}), and the Helioseismic and Magnetic Imager (HMI; \citealt{Scherrer2012}).

In this thesis, data from the AIA instrument are primarily utilized. AIA consists of four co-aligned telescopes designed to capture high-resolution full-disk solar images across multiple ultraviolet (UV) and extreme ultraviolet (EUV) channels. Specifically, AIA observes in 10 different channels, seven in EUV (94, 131, 171, 193, 211, 335~\AA) and three in UV/visible light, each sensitive to plasma at specific characteristic temperatures, ranging from approximately $10^4$~K to over $10^7$~K. The instrument provides 4096~$\times$~4096 pixel images with a spatial resolution of 0.6 arcsec per pixel and a cadence of 12 seconds, enabling the continuous tracking of solar atmospheric phenomena with high temporal and spatial detail. This comprehensive wavelength coverage and high imaging cadence make AIA particularly suitable for observing dynamic coronal features such as flares, coronal loops, and CME eruptions.

\end{itemize}

To study the thermal structure of the CME flux-rope at its source region, this thesis applies Differential Emission Measure (DEM) analysis to SDO/AIA EUV observations, following the sparse inversion technique as described in Section~\ref{sec:chapt2-metho-gcs}. This approach provides estimates of emission measure and temperature critical for assessing the early thermal state of the eruption.

\subsection{In Situ Observations at 1 au}

In in situ spacecraft, plasma measurements are primarily carried out using electrostatic analyzers (ESAs), which separate charged particles based on their energy per charge (E/q) ratio. In an ESA, particles enter through a collimator and are deflected by an electric field between curved electrodes. Only particles with a specific E/q follow the correct trajectory to reach the detector, allowing for the construction of a differential energy spectrum. By sweeping across a range of voltages, the instrument captures distributions of electrons and ions, from which bulk parameters such as flow velocity, thermal speed (temperature), and number density can be derived (for more details, \citealt{Paschmann1998}).

Magnetic field measurements are typically performed using fluxgate magnetometers. These sensors consist of ferromagnetic cores wound with drive and sense coils. An alternating current in the drive coil periodically magnetizes the core, and in the presence of an ambient magnetic field, the symmetry of the induced signal is distorted. This distortion is detected by the sense coils and is proportional to the strength and orientation of the external magnetic field. The result is a continuous time series of vector magnetic field measurements with high temporal resolution, enabling detailed analyses of magnetic structures and turbulence (for more details, \citealt{Lepping1995}).

To investigate the thermodynamic evolution, turbulence, and heating characteristics of CMEs in interplanetary space near 1~au, we used high-resolution plasma and magnetic field data from the Wind spacecraft, along with solar wind data from the OMNI database. These in situ measurements are critical for diagnosing the local plasma state, magnetic configuration, and small-scale structures encountered as CMEs interact with the ambient solar wind and approach Earth.

\begin{itemize} [leftmargin=0pt]
    \item {\large \textbf{Wind:}} Launched in November 1994 as part of the NASA Global Geospace Science (GGS) initiative, the Wind spacecraft orbits the first Lagrangian point (L1) and has provided continuous high-quality solar wind monitoring. It carries multiple instruments that enable detailed studies of ICMEs, solar wind turbulence, and energetic particle events. In this thesis, data from the following instruments were used:
    
    \begin{itemize}
        \item  The \textbf{Solar Wind Experiment} (SWE; \citealt{Ogilvie1995}) instrument uses electrostatic analyzers and Faraday cups to measure the bulk properties of the solar wind plasma. Proton and electron measurements from the SWE instrument are available at multiple resolutions; however, for our selected ICME events, we primarily used proton data at 92-second cadence and electron data at 9-second cadence. This selection was based on the availability and quality of high-resolution data for the events under study. The finer cadence of the electron data allows for resolving small-scale thermal variations, while the 92-second proton data provide robust characterization of bulk plasma properties.

        \item The \textbf{Three-Dimensional Plasma Analyzer} (3DP;  \citealt{Lin1995}) instrument measures the full three-dimensional distribution functions of ions and electrons across a wide energy range, supporting advanced studies of particle kinetics and energy transport. In this thesis, we used 92-second resolution data from 3DP to derive the thermal state (i.e., polytropic index) of both electrons and protons for a complex ICME event characterized by interacting structures. 

        \item The \textbf{Magnetic Field Investigation} (MFI; \citealt{Lepping1995}) instrument provides high-resolution vector magnetic field measurements using dual triaxial fluxgate magnetometers. We have used data with a cadence of 11 Hz, making it ideal for spectral analysis of magnetic fluctuations, turbulence diagnostics, and intermittency studies within different ICME substructures.
    \end{itemize}

    \item {\large \textbf{OMNI database:}} The OMNI database is a composite, high-quality, continuous dataset of solar wind and interplanetary magnetic field (IMF) parameters at 1~au. It is compiled by NASA's Space Physics Data Facility and includes propagated measurements from multiple spacecraft, including Wind and ACE, mapped to the Earth's bow shock nose. The 1-minute resolution OMNI data were retrieved from the Coordinated Data Analysis Web (CDAWeb: \url{https://cdaweb.gsfc.nasa.gov/}). In this thesis, OMNI data were used primarily for statistical studies of CME thermal state (e.g., $\Gamma_p$) and for constructing SEA profiles across multiple ICME events.
\end{itemize}

The combination of high-resolution Wind data and well-calibrated OMNI data allows us to probe both fine-scale and large-scale characteristics of ICME plasma. Specifically, proton and electron temperature-density relationships from SWE and 3DP were used to derive polytropic indices, MFI data were used for turbulence and intermittency analyses, and OMNI data supported event selection and comparative statistical studies.

\vspace{1cm}
In summary, the synergistic application of our FRIS model with multi-instrument observations, white-light coronagraphs for reconstructing the 3D geometry and kinematics of CMEs, EUV imagers for probing their early thermal properties in the low corona, and in situ plasma and magnetic field measurements at 1 au, provides a robust and comprehensive observational dataset. This integrated approach enables a detailed investigation of the kinematic and thermodynamic evolution of CMEs from their initiation near the Sun to their propagation through interplanetary space. The results and interpretations derived from this framework will be discussed in detail in the following chapters.

\lhead{\emph{Chapter 3}}  

\quotechapter{3}{`` Science is the attempt to make the chaotic diversity of our sense-experience correspond to a logically uniform system of thought.''\\ \smallskip — Albert Einstein}{Thermal Evolution - Case Study}

\lhead{\emph{Chapter 3: Thermal Evolution - Case Study}}

\section{Introduction} \label{Ch3-sec:intro}

Despite decades of observational and theoretical efforts focused on CME dynamics and magnetic structure, the thermal evolution of CMEs remains poorly understood. While several studies have extensively investigated CME kinematics, arrival times, and in situ signatures, a continuous and detailed understanding of their internal thermodynamic evolution, from the low corona to interplanetary distances, is still lacking (as discussed in Section~\ref{sec:chapt1-gaps}).

Near the Sun, EUV spectroscopic and imaging diagnostics suggest that CME cavities can reach temperatures significantly higher than the ambient corona, implying strong localized heating \citep{Akmal2001, Filippov2002, Raymond2002, Kohl2006, Lee2009, Lee2017, Bemporad2022}. In situ observations at 1~au show that ICMEs generally exhibit lower proton temperatures, stronger magnetic fields, and higher ionization states compared to the background solar wind \citep{Zurbuchen2006, Rakowski2007}. These findings suggest that CME plasma often undergoes net heating, rather than the expected adiabatic cooling, during its interplanetary evolution. Such thermal states are frequently characterized using a polytropic framework, where the polytropic index $\Gamma$ serves as an effective parameter capturing the net energy exchange. Prior in situ analyses between 0.3--20 au have reported $\Gamma$ values typically in the range of 1.1--1.3, indicating quasi-isothermal behavior \citep{Osherovich1993, Liu2005, Liu2006}.

However, these investigations often suffer from a lack of continuity across heliocentric distances, limited spatial resolution, and assumptions of global uniformity. For instance, in many studies, a single thermal state is inferred for the entire CME plasma based on in situ measurements, without accounting for local spatial inhomogeneities across CME structure. As discussed in Section~\ref{sec:chapt1-gaps}, the inability of white-light coronagraphs to provide thermal diagnostics, combined with the disconnect between remote and in situ datasets, has hindered a unified physical understanding of CME thermodynamics.

Motivated by these research gaps, this chapter presents a two-stage comparative analysis of the thermodynamic evolution of two kinematically distinct CMEs: a fast event on 24 September 2011 (CME1) and a slow event on 20 August 2018 (CME2). These events were selected for their well-studied contrasting kinematic profiles and geoeffectiveness \citep{Wood2016, Palmerio2022, Gopalswamy2022}. The aim is to investigate and compare their thermodynamic evolution in the low corona and at 1~au. The first part of the chapter applies the revised FRIS model to derive near-Sun thermodynamic parameters up to $\sim 20\,R_\odot$ using coronagraph-based 3D kinematics. The second part investigates the corresponding ICMEs at 1~au using Wind spacecraft data, focusing on spatial variations in thermal and turbulence properties across different substructures. This includes estimating the proton and electron polytropic indices, as well as characterizing inertial and dissipation-scale magnetic field turbulence, compressibility, and intermittency.

By bridging remote and in situ diagnostics for fast and slow CMEs, this chapter aims to uncover how kinematic differences manifest as thermodynamic diversity, and how internal heating, expansion, and turbulent dissipation govern the thermal evolution of CMEs. The results offer physically consistent constraints on CME propagation models and enhance our understanding of magnetized ejecta in the heliosphere.

\section{Near-Sun Thermodynamic Analysis Using FRIS} \label{Ch3-sec:1}

The thermodynamic behavior of CMEs during their early propagation phase is governed by complex interactions between plasma heating, expansion, and internal forces. To understand how these internal properties evolve with height, we apply the revised FRIS model (Section~\ref{metho:fris_derivation}), which utilizes kinematic inputs to derive thermodynamic parameters such as the polytropic index, temperature, internal forces, and heating rate. This model enables us to trace the evolution of CME thermal state at heights typically inaccessible to direct thermal diagnostics. In this section, we perform a comparative thermodynamic analysis of two geoeffective CME events with distinct kinematic profiles. We begin by summarizing the source region characteristics and white-light evolution of each event before applying the FRIS model to quantify their internal thermodynamic evolution at higher heights.

\subsection{Methodology}\label{Ch3-sec:methods}

The two CME events selected for this study, 24 September 2011 (CME1) and 20 August 2018 (CME2), have been previously analyzed in detail for their kinematic and geoeffective properties \citep{Wood2016, Palmerio2022, Gopalswamy2022}. This prior characterization allows us to focus primarily on their internal thermodynamic evolution. These events were chosen for their contrasting propagation profiles: CME1 is a fast, impulsive event, while CME2 is a relatively slow, gradually accelerating event. By comparing a fast and a slow CME, we aim to investigate whether differences in global kinematics translate into contrasting thermal evolution patterns. Additionally, the FRIS model requires reliable 3D kinematics as input, which demands that the CME structure be clearly tracked through the coronagraph field of view without major observational uncertainties. Both selected events satisfy this criterion, ensuring accurate modeling of their internal thermodynamic parameters.

We utilized continuous white-light coronagraphic data to track the selected CMEs from their early stages. Since both CME1 and CME2 are Earth-directed, observations from SOHO/LASCO alone are subject to significant projection effects, typically underestimating speed and overestimating angular width. To overcome this, we employed multi-viewpoint observations, including STEREO/COR, and applied the GCS model (Section~\ref{sec:chapt2-metho-gcs}), which enables 3D reconstruction of CME geometry and kinematics with minimized projection uncertainties \citep{Thernisien2006, Mierla2010, Davies2013, Mishra2014}. Once the 3D kinematics are obtained, the final equation of motion (Equation~\ref{fitting}) in the FRIS model is fitted to derive the unknown coefficients \(c_1\)–\(c_5\). These coefficients are then used to compute the internal plasma parameters and force components acting on the CME, as summarized in Table~\ref{chap2_tab:FRIS_parameters}.

\subsection{Measured 3D Kinematics}
\label{Ch3-sec:results_kine}

CME1 was first observed by SOHO/LASCO-C2 on 24 September 2011 at 12:48~UT as a full halo Earth-directed CME, with a projected plane-of-sky (POS) linear speed of 1915~km~s$^{-1}$. Its interplanetary counterpart (ICME1) arrived at Earth on 26 September 2011 at 11:34~UT and triggered a geomagnetic storm with a minimum Dst of $-118$~nT \citep{Wood2016}. CME2, observed on 20 August 2018, was first detected by SOHO/LASCO-C2 at 21:24~UT with a much slower POS speed of 126~km~s$^{-1}$. It reached 1~au on 25 August 2018 at 02:00~UT, resulting in the third-largest geomagnetic storm of Solar Cycle 24, with a minimum Dst of $-174$~nT \citep{Chen2019, Gopalswamy2022}. Both events have been studied in previous works for their source region characteristics, propagation, and geoeffectiveness \citep{Wood2016, ChinChun2016, Mishra2019, Gopalswamy2022, Palmerio2022}.

\begin{figure*}[ht]
        \centering
        \includegraphics[width=0.42\linewidth,angle=-90]{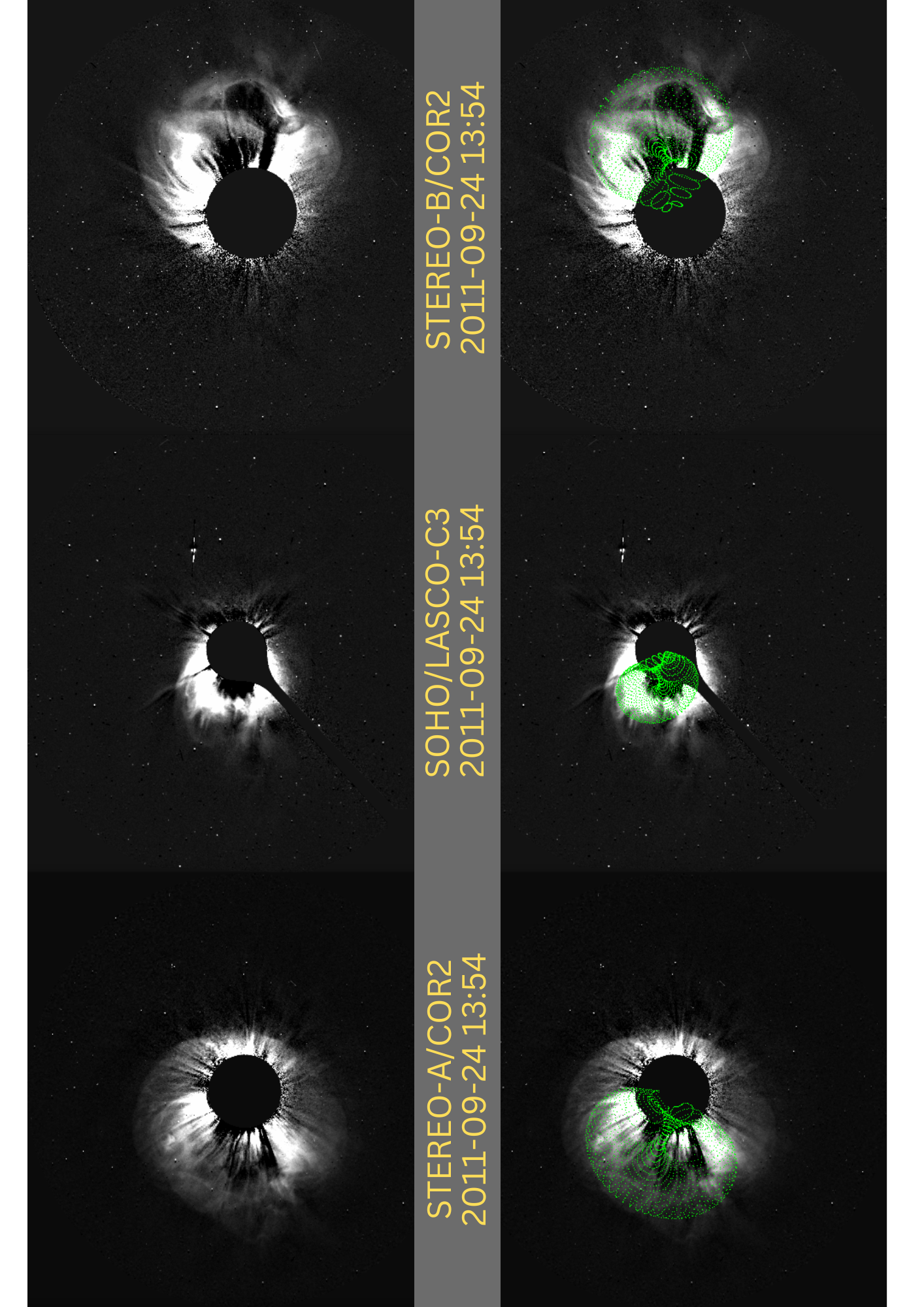}    
        \includegraphics[width=0.42\linewidth,angle=-90]{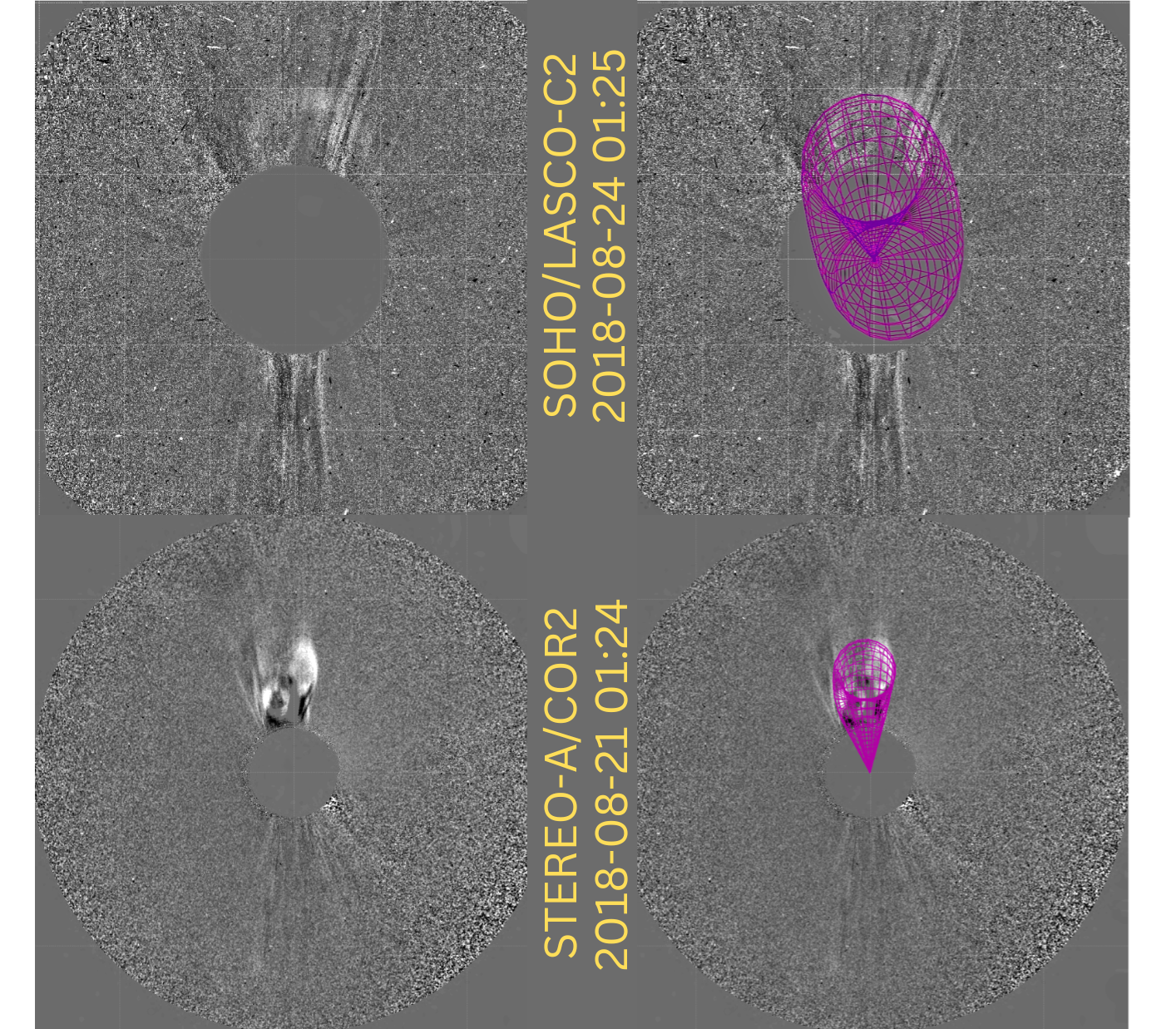}
        \caption[GCS model wireframe (green and pink) overlaid on ]{GCS model wireframe (green and pink) overlaid on simultaneous coronagraph images of CME1 (left panel: STEREO-A/COR2; center: SOHO/LASCO-C3; right panel: STEREO-B/COR2) and CME2 (right panel: left—STEREO-A/COR2; right—SOHO/LASCO-C2).}
        \label{fig:coronagraph_images}
\end{figure*}

To reconstruct the 3D structure and kinematics of both CMEs, we applied the GCS to simultaneous coronagraph images from SOHO/LASCO (C2 and C3) and STEREO/COR (COR1 and COR2) (Figure~\ref{fig:coronagraph_images}). Since STEREO-B has been unavailable since 2014, only STEREO-A was used for CME2. To reduce the degeneracy among GCS parameters, initial estimates of longitude and latitude were guided by the CME source region as seen in multi-wavelength EUV observations \citep{Mostl2014, Palmerio2018}. Although the tilt of a CME flux-rope can be estimated from the orientation of the polarity inversion line and the post-eruption arcade structure at the source region \citep{Marubashi2015, Palmerio2018}, CMEs often undergo significant deflection and rotation in the low corona. As a result, the initial values of longitude, latitude, and tilt derived from source region observations may not accurately represent the CME geometry at larger heights. Therefore, in this study, we did not use tilt values from the source region but instead relied on direct GCS model fitting to coronagraph images taken from multiple viewpoints.

The GCS model provides six parameters, longitude, latitude, height, tilt, aspect ratio, and half angular width, which we adjusted at each time step to best match the observed CME morphology. The fitted parameters at a representative time are listed in Table~\ref{chap3-tab:GCS}, while the full time series of fitted parameters is provided in Tables~\ref{chap3-tab:CME1_gcs_all} and \ref{chap3-tab:CME2_gcs_all}. For CME1, the best-fit longitude, latitude, aspect ratio, tilt, and half angle were found to be $-41^\circ$, $13^\circ$, 0.39, $-62^\circ$, and $26^\circ$, respectively. For CME2, the corresponding values were $10^\circ$, $5^\circ$, 0.27, $10^\circ$, and $16^\circ$. These parameters remained constant throughout the tracked evolution for both events, indicating no significant rotation or deflection and supporting the assumption of self-similar expansion, consistent with the requirements of the FRIS model.

\begin{table*}[ht]
\caption[Summary of GCS model-derived geometrical and]{\label{chap3-tab:GCS} Summary of GCS model-derived geometrical and positional parameters for CME1 (24 September 2011) and CME2 (20 August 2018).}

\begin{adjustbox}{width=\textwidth}
\begin{tabular} {lccccccc}
\hline
\textbf{Events} & \textbf{Time} & \textbf{Height}  & \textbf{Longitude} & \textbf{Latitude} & \textbf{Aspect}  & \textbf{Tilt}   & \textbf{Half} \\
 & \textbf{(UT)} &   \textbf{($R_\odot$)} & \textbf{(deg)} & \textbf{(deg)} & \textbf{Ratio} & \textbf{Angle} & \textbf{Angle}\\
  & & & & & &\textbf{(deg)} &\textbf{(deg)}\\
\hline
\hline
2011 September 24 & 13:54 & 12.9 & -41 & 13 & 0.39 & -62 &26\\
\hline
2018 August 20 & 1:24 & 8.5 & 10 & 5 & 0.27 & 10 & 16\\
\hline

\end{tabular}

\end{adjustbox}
\end{table*}

The GCS-derived longitudes suggest that CME1 propagated $\sim41^\circ$ east of the Sun–Earth line, while CME2 propagated $\sim10^\circ$ west. The aspect ratio of CME1 was approximately 44\% larger than CME2, and its half angular width was about 60\% greater. These geometric differences are consistent with earlier findings that link CME angular width to propagation speed \citep{Gopalswamy2009}. Using the fitted leading-edge height ($h$), the radius of the flux-rope ($R$) was computed as $R = \left(\frac{\kappa}{1 + \kappa}\right) h$, where $\kappa$ is the aspect ratio. The propagation speed ($v$) and expansion speed ($v_e$) were derived by taking the time derivatives of $h$ and $R$, respectively, using a three-point boxcar method for central points and a

\begin{table}[ht]
\parbox{.45\linewidth}{
\centering
\caption[GCS model-derived leading-edge height of CME1]{\label{chap3-tab:CME1_gcs_all}GCS model-derived leading-edge height of CME1 (24 September 2011) at each observational time step.}
\begin{tabular}{ccc}
\hline
\textbf{Date} &\textbf{Time} & \textbf{Height} \\
 & \textbf{(UT)} &   \textbf{($R_\odot$)} \\
\hline
\hline
2011 September 24 & 12:45 & 2.5 \\
\hline
2011 September 24 & 12:50 & 3.2\\
\hline
2011 September 24 & 12:55 & 4.0\\
\hline
2011 September 24 & 13:24 & 8.3\\
\hline
2011 September 24 & 13:30 & 9.2\\
\hline
2011 September 24 & 13:39 & 10.6\\
\hline
2011 September 24 & 13:54 & 12.9\\
\hline
2011 September 24 & 14:06 & 14.8\\
\hline
2011 September 24 & 14:18 & 16.7\\
\hline
2011 September 24 & 14:24 & 17.7\\
\hline
2011 September 24 & 14:30 & 18.6\\
\hline
2011 September 24 & 14:39 & 20.1\\
\hline
\end{tabular}
}
\hfill
\parbox{.45\linewidth}{
\centering
\caption[GCS model-derived leading-edge height of CME2]{\label{chap3-tab:CME2_gcs_all}GCS model-derived leading-edge height of CME2 (2018 August 20) at each observational time step.}
\begin{tabular}{ccc}
\hline
\textbf{Date} &\textbf{Time} & \textbf{Height} \\
 &\textbf{(UT)} &   \textbf{($R_\odot$)} \\
\hline
\hline
     2018 August 20 &    21:24 &     5.7 \\ 
      \hline
     2018 August 20 &   22:24  &    6.4 \\
      \hline
     2018 August 20 &   23:24 &    7.1 \\
      \hline
     2018 August 20 &   23:39  &    7.2 \\
      \hline
     2018 August 20 &   23:54  &    7.4 \\
      \hline
     2018 August 21 &   0:24  &    7.7 \\
      \hline
     2018 August 21 &   0:39  &    7.9 \\
      \hline
     2018 August 21 &   0:54  &    8.1 \\
      \hline
     2018 August 21 &   1:24  &    8.5 \\
      \hline
     2018 August 21 &   1:39  &    8.7 \\
      \hline
     2018 August 21 &   1:54  &    8.8 \\
      \hline
     2018 August 21 &   2:24  &    9.3 \\
      \hline
     2018 August 21 &   2:39  &    9.5 \\
      \hline
     2018 August 21 &  2:54  &    9.8 \\
      \hline
     2018 August 21 &   3:24  &    10.4 \\
      \hline
     2018 August 21 &   3:39  &    10.7 \\
      \hline
     2018 August 21 &   3:54  &    11.0 \\
      \hline
     2018 August 21 &   4:24  &    11.7 \\
      \hline
     2018 August 21 &  4:39  &    12.1 \\
      \hline
     2018 August 21 &  4:54   &   12.4 \\
      \hline
     2018 August 21 &   5:24   &   13.2 \\
      \hline
     2018 August 21 &  5:39  &    13.6 \\
      \hline
     2018 August 21 &  5:54  &    14.1 \\
      \hline
     2018 August 21 &   6:24  &    15.2 \\
\hline
\end{tabular}
}
\end{table}

\begin{figure*}[ht]
\centering
        \includegraphics[width=0.47\textwidth,trim={6cm 0cm 22cm 0cm}]{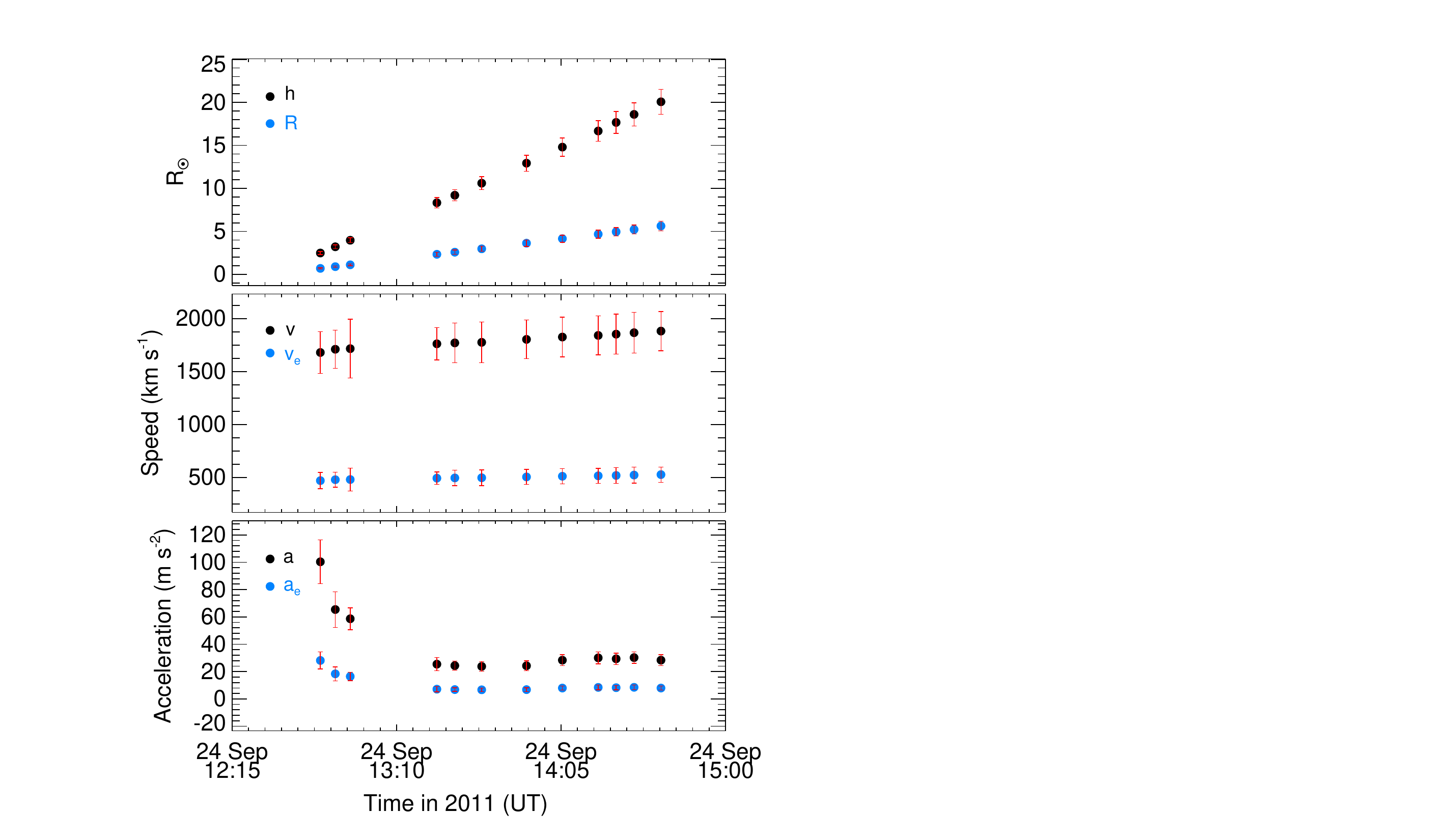}
        \includegraphics[width=0.47\textwidth,trim={4cm 0cm 24cm 0cm}]{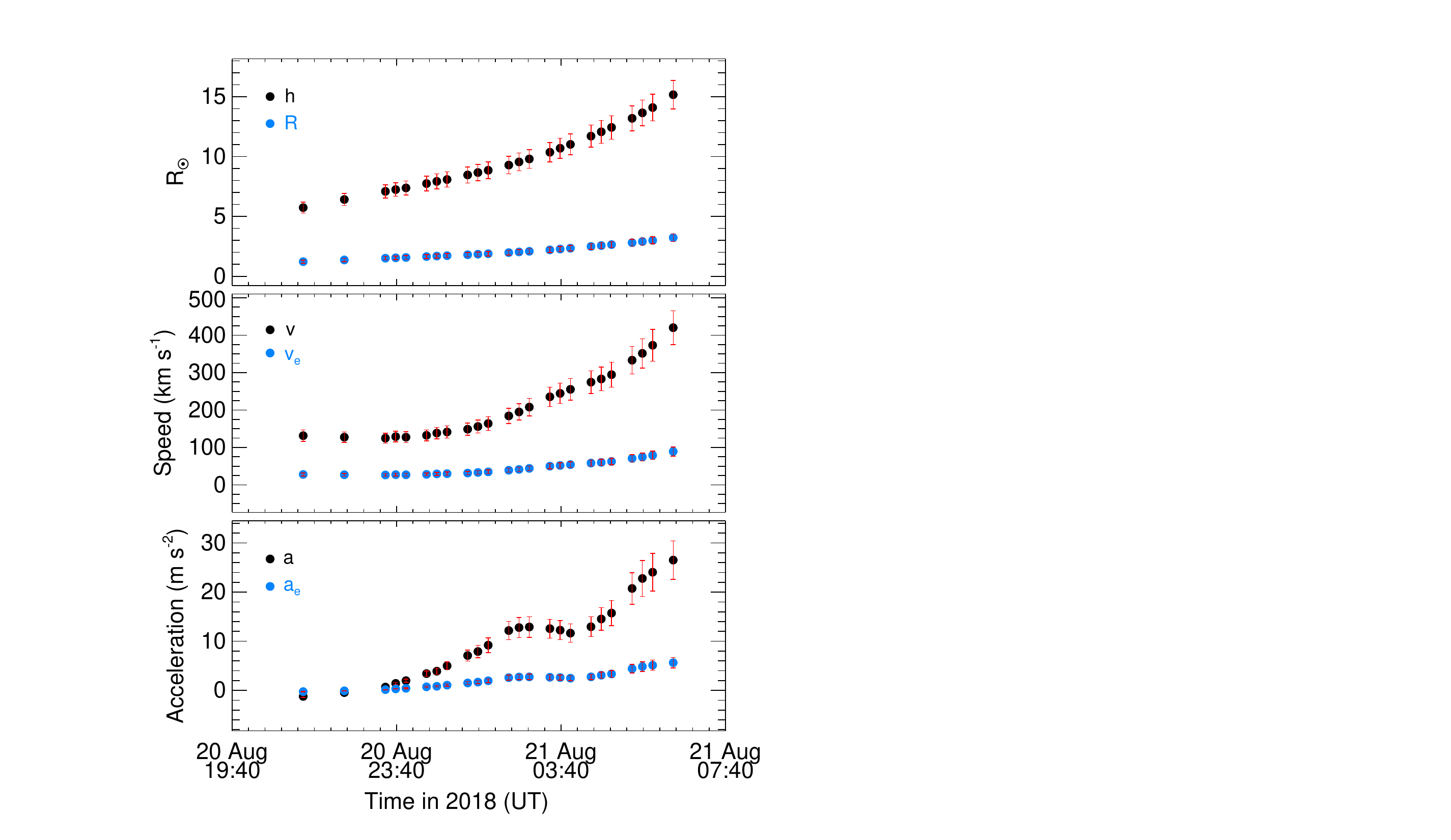}     
        \caption[Kinematic profiles of CME1 (left) and CME2 (right) derived]{Kinematic profiles of CME1 (left) and CME2 (right) derived from GCS model fits to coronagraph observations. Top panels: heliocentric distance ($h$) of the flux-rope leading edge and its radius ($R$). Middle panels: propagation speed ($v$) and expansion speed ($v_e$), obtained using three-point derivatives of $h$ and $R$, respectively. Bottom panels: propagation acceleration ($a$) and expansion acceleration ($a_e$), calculated as derivatives of $v$ and $v_e$. Red vertical bars represent 10\% uncertainties in the leading edge height ($h$) measurements.}
        \label{chap3-fig:kinematics}
\end{figure*}

\clearpage

\noindent two-point derivative at the boundaries. This technique preserves the resolution of the original dataset and provides realistic estimates of speed and acceleration (Figure~\ref{chap3-fig:kinematics}).

CME1 was tracked up to $\sim20\,R_\odot$, reaching a maximum $R \approx 5.6\,R_\odot$, with a peak $v$ of 1885~km~s$^{-1}$ and $v_e$ of 528~km~s$^{-1}$. The acceleration sharply decreased up to $\sim8.4\,R_\odot$ and then transitioned into a nearly constant-speed phase, in agreement with previous studies showing impulsive acceleration below $3\,R_\odot$ \citep{Zhang2004, Temmer2010}. In contrast, CME2 was tracked up to $\sim15\,R_\odot$, with a peak $R \approx 3.2\,R_\odot$, $v \approx 420$~km~s$^{-1}$, and $v_e \approx 90$~km~s$^{-1}$. It showed a gradual increase in both propagation and expansion acceleration. Thus, CME1 represents a fast, impulsively accelerating CME, while CME2 is a slow and gradually expanding one. We note that the initial stages of CME evolution, particularly below $\sim3\,R_\odot$, were not captured due to the occulting disk in coronagraphs and the Earth-directed geometry. However, the events became clearly visible once beyond this height, enabling accurate tracking. The derived kinematic profiles serve as critical input to the FRIS model, allowing us to quantify the associated thermodynamic evolution, which is discussed in the next section.

\subsection{FRIS-model-derived Parameters}
\label{Ch3-sec:results_fris}

\begin{figure*}[ht]
     \centering
        \includegraphics[width=0.47\textwidth,trim={10cm 1cm 12cm 2cm}]{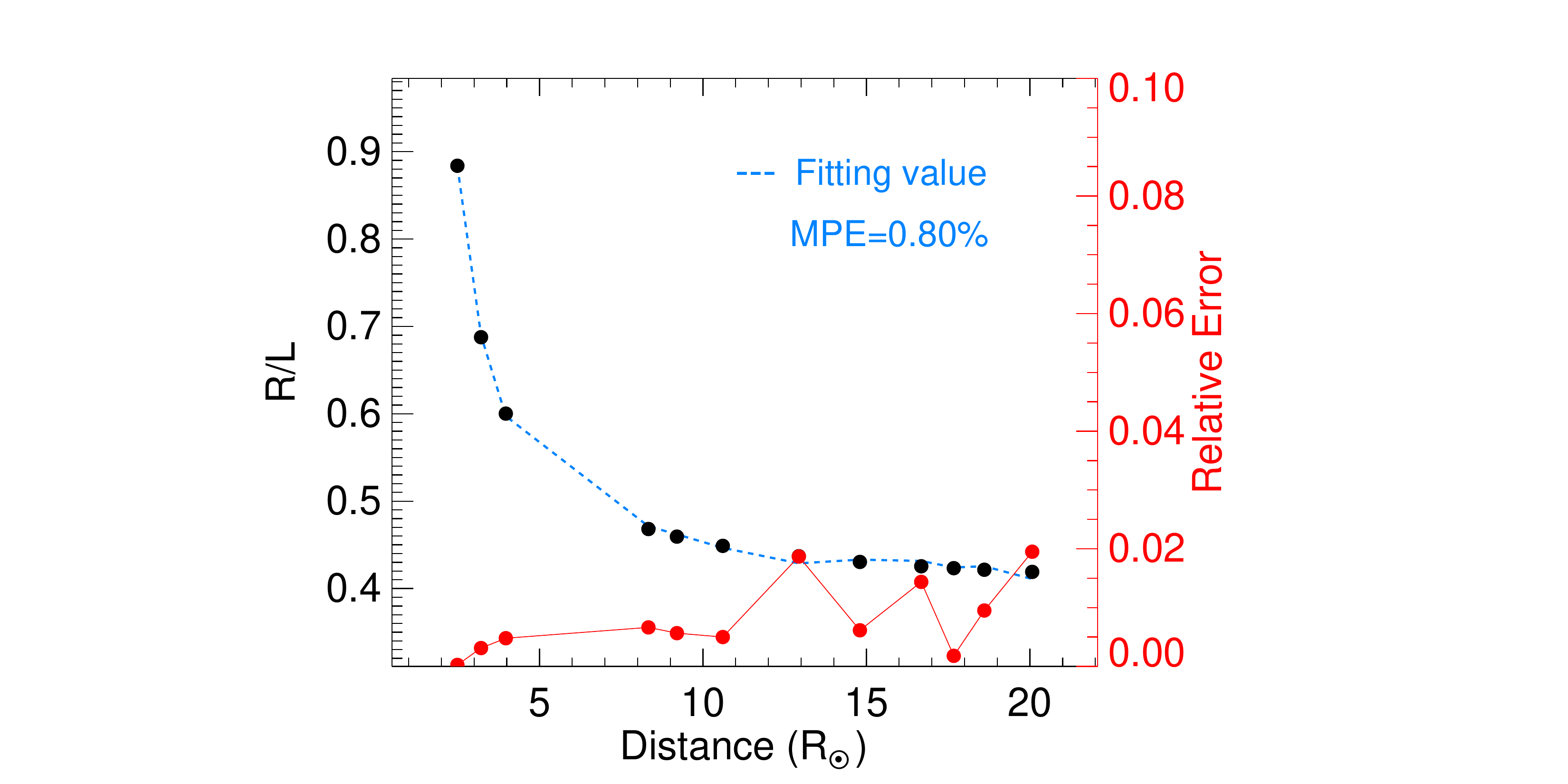}
        \includegraphics[width=0.47\textwidth,trim={8cm 1cm 14cm 2cm}]{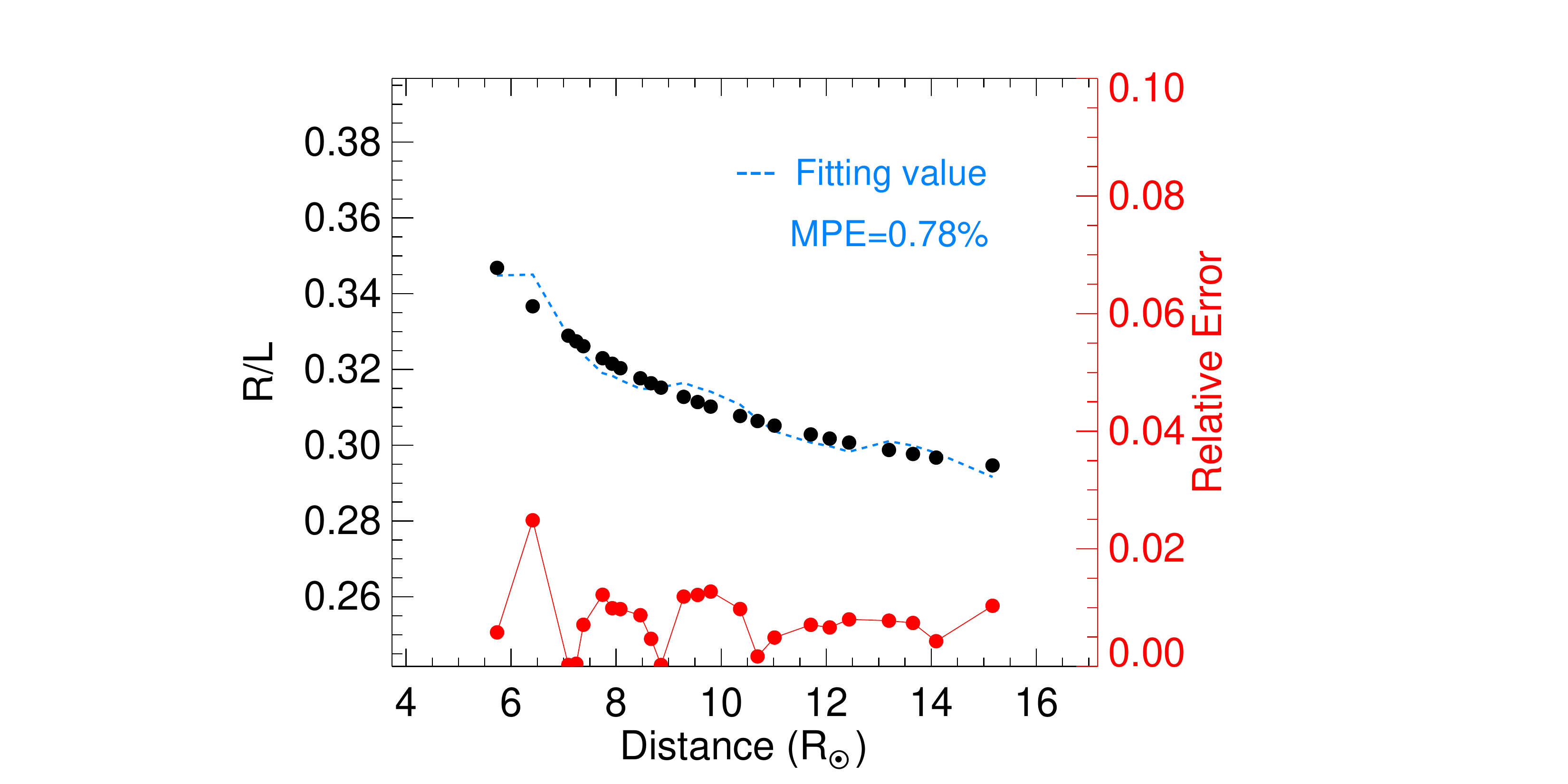}
        \caption[The profile of $R/L$ for CME1  and CME2 from the]{The profile of $R/L$ for CME1  and CME2 from the measurements (black; left-hand side of Equation \ref{fitting}), the model-fitted result for this parameter (blue; right-hand side of Equation \ref{fitting}), and the relative error (red) are shown. (MPE: Mean Percentage Error) }
        \label{fig:fitting}
\end{figure*}

As outlined in the FRIS model framework, the internal thermodynamic properties of CMEs are computed using their kinematic profiles. Specifically, the model requires the heliocentric distance of the flux-rope center ($L$), its propagation speed ($v_c = dL/dt$), the radius of the flux-rope ($R$), its expansion speed ($v_e$), and the time derivatives of these quantities. The center distance $L$ is calculated as $L = h - R - 1\,R_\odot$, and $R$ is derived from the GCS-fitted leading-edge height $h$ and the aspect ratio $\kappa$ using $R = (\kappa / (1 + \kappa)) h$ \citep{Thernisien2011}. To implement the FRIS model, we fit the equation of motion (Equation~\ref{fitting}) using the measured 3D kinematics as inputs, as shown in Figure~\ref{fig:fitting}. The fitting was performed using the \texttt{CURVE\_FIT} routine from \texttt{SciPy}, allowing us to extract the unknown coefficients $c_1$ to $c_5$. With these fitted coefficients and the observationally derived kinematic quantities, we estimate the internal forces and thermodynamic parameters of both CMEs, as summarized in Table~\ref{chap2_tab:FRIS_parameters}.

\subsubsection{Thermodynamic Parameters}\label{sec:results_thermo}

The FRIS model allows for the estimation of various thermodynamic and plasma parameters of CMEs(Table~\ref{chap2_tab:FRIS_parameters}). In this study, we focus on four key parameters that characterize the internal thermal state of the CME flux-rope: the polytropic index ($\Gamma$), average heating rate ($dQ/dt$), average temperature ($\bar{T}$), and average proton number density ($\bar{n}_p$). These quantities are derived using Equations~\ref{eqn:lamb}, \ref{eqn:heating}, \ref{eqn:T}, and \ref{eqn:np}, in conjunction with the 3D kinematic parameters obtained from GCS modeling.

\begin{figure*}[hbt]
         \includegraphics[width=0.48\textwidth,trim={8cm 0cm 16cm 2cm}]{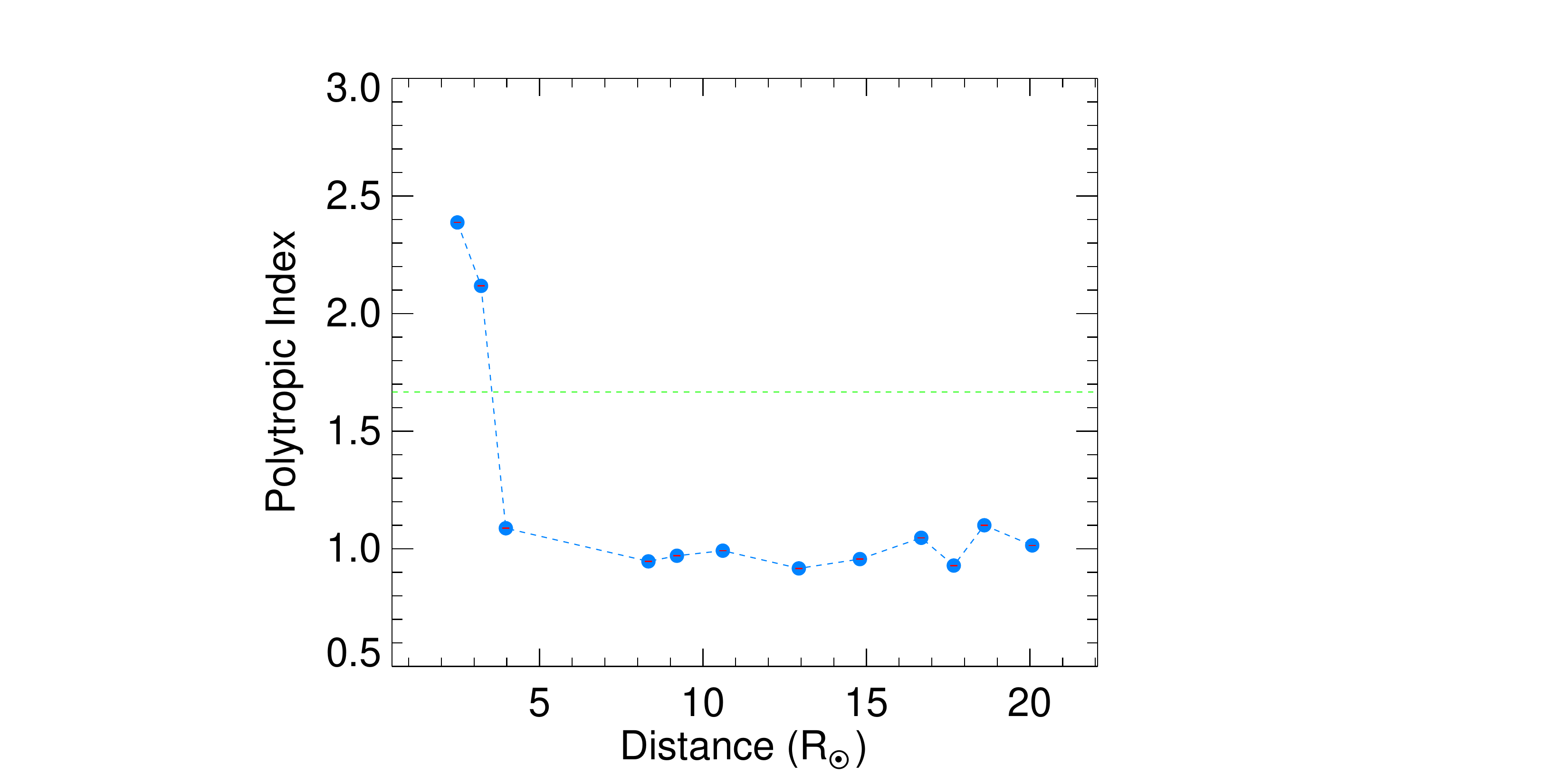}
         \includegraphics[width=0.48\textwidth,trim={6cm 0cm 18cm 2cm}]{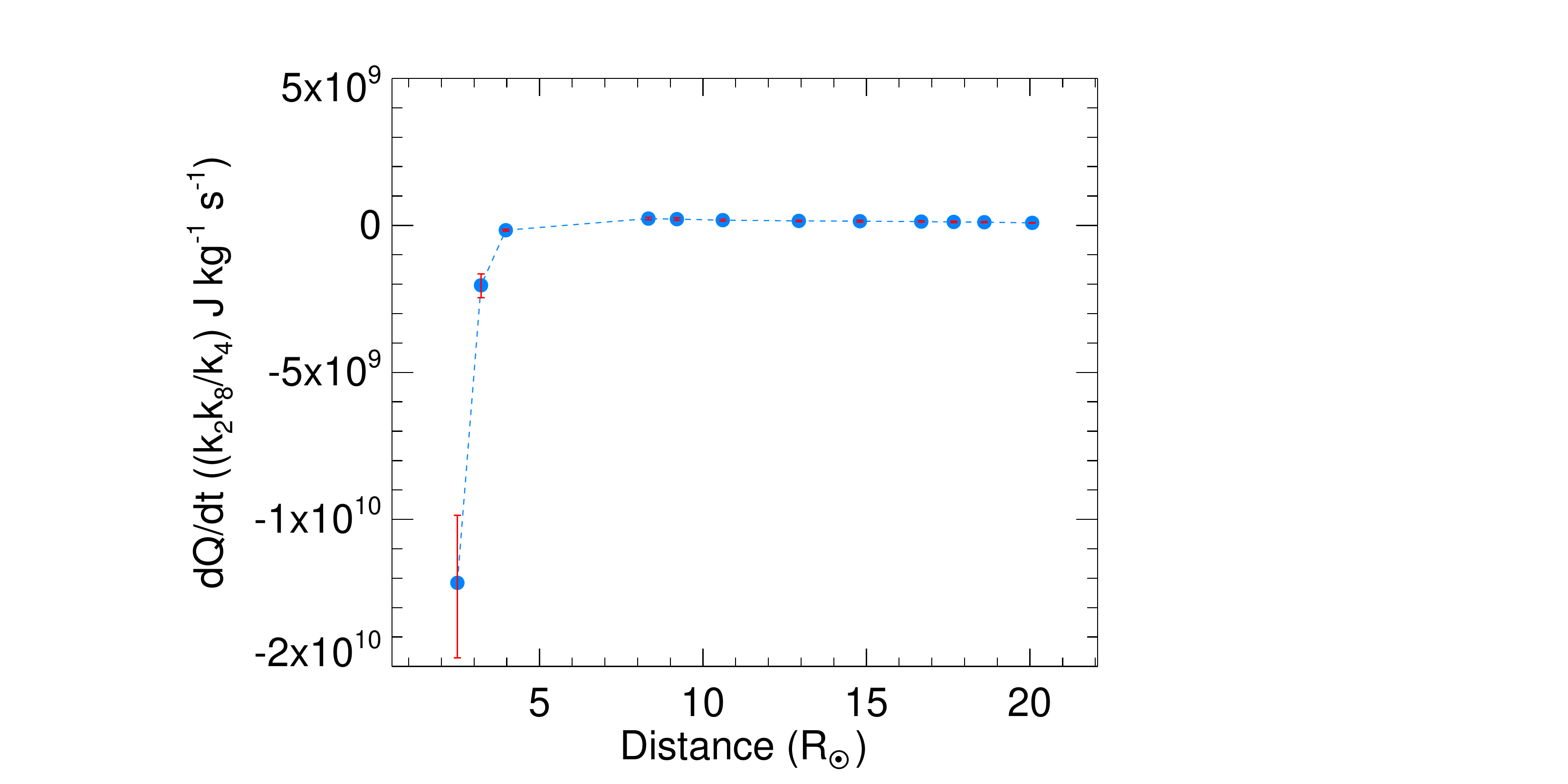}\\
         \includegraphics[width=0.48\textwidth,trim={8cm 1cm 16cm 2cm}]{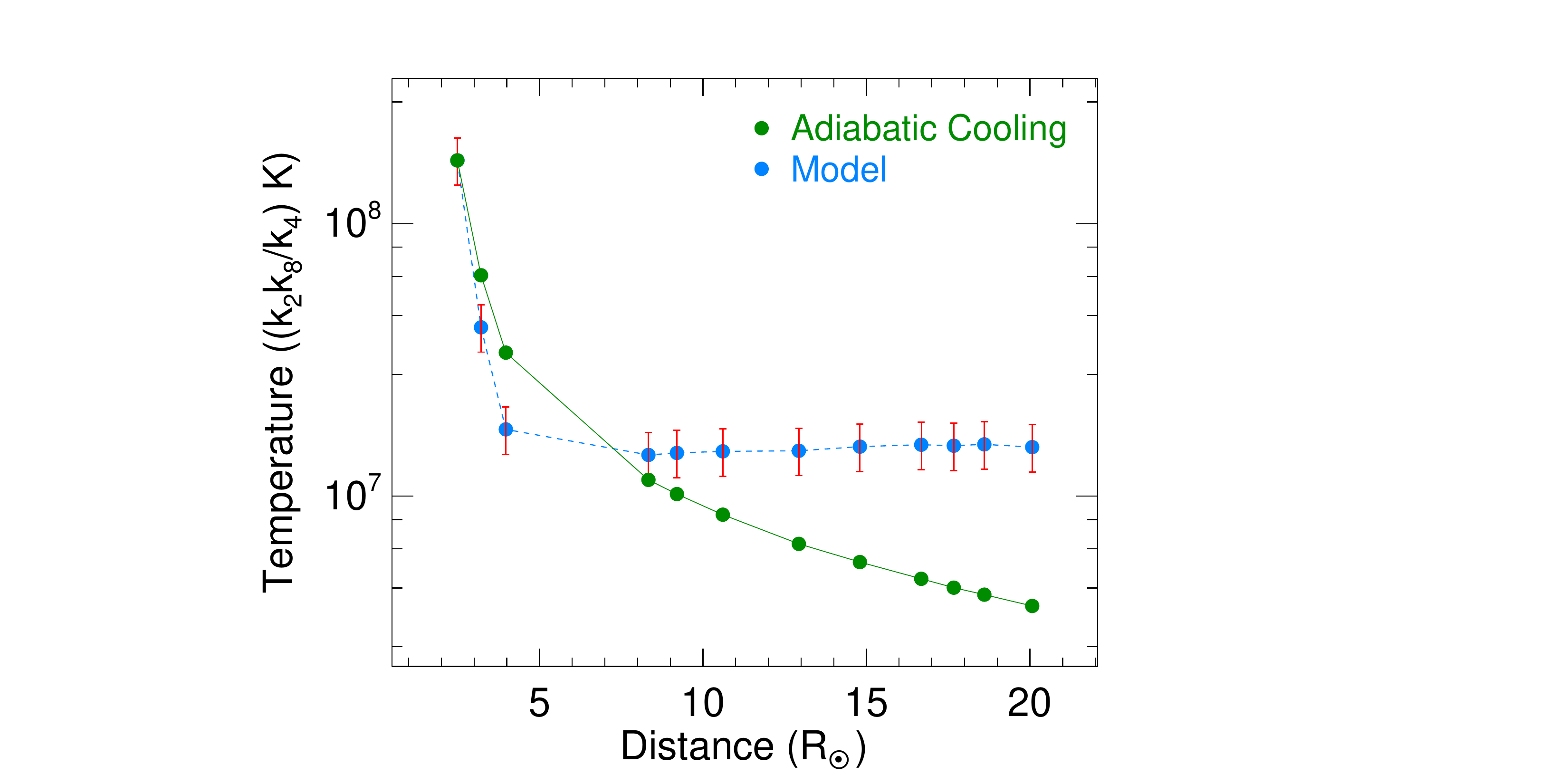}
         \includegraphics[width=0.48\textwidth,trim={6cm 1cm 18cm 2cm}]{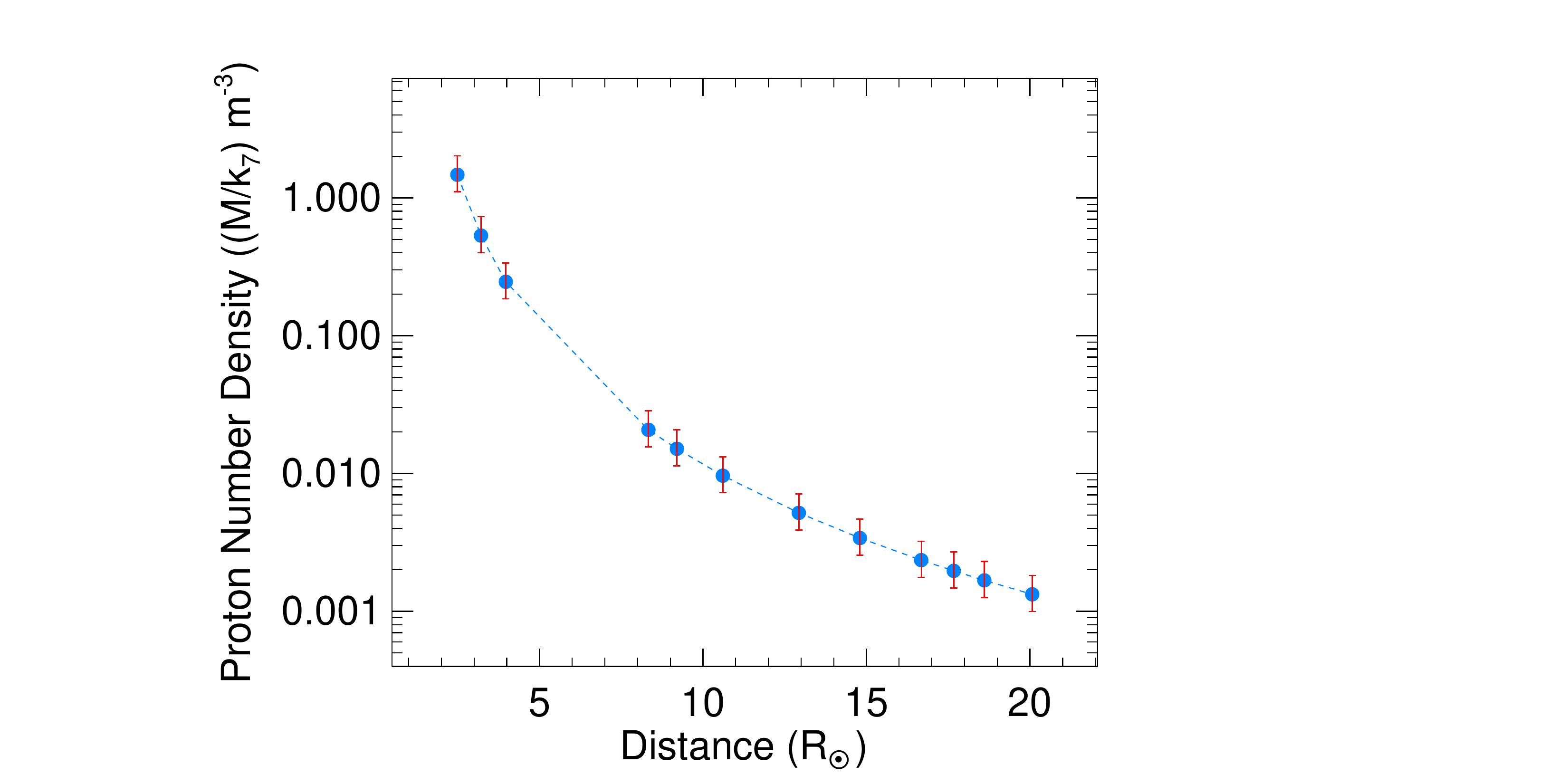}
         \caption[For the CME1: Variation of the polytropic index ($\Gamma$),]{For the CME1: Variation of the polytropic index ($\Gamma$), average heating rate (dQ/dt), average temperature $(\bar{T})$, and average proton number density (${\bar{n}}_{p}$) of the CME with the heliocentric distance of the CME’s leading edge (h) is shown in the top-left, top-right, bottom-left and bottom right, respectively. The red vertical lines at each data point show the error bars derived by considering an error of 10{\%} in measurements of the flux-rope's leading edge height ($h$).}
        \label{fig:Thermo_para_CME1}
\end{figure*}

Applying the FRIS model to CME1, we find that the polytropic index $\Gamma$ exceeds the adiabatic value ($\Gamma = 5/3$) at low heights between 2.5 and 3.2~$R_\odot$ (Figure~\ref{fig:Thermo_para_CME1}, top-left), indicating that the CME is initially releasing heat into the surrounding medium. As the CME evolves, $\Gamma$ decreases rapidly and reaches a value of $\sim$1.09 at $\sim$4~$R_\odot$, after which it remains nearly constant, hovering around unity. Beyond this point, with $\Gamma < 5/3$, the plasma behavior suggests net heat gain, implying the presence of in-transit heating processes. Notably, this thermal transition aligns with the CME's kinematic evolution: an early phase of rapid deceleration accompanied by heat release is followed by a gradual acceleration phase during which the CME approaches an isothermal state ($\Gamma \approx 1$). The sustained deviation from the adiabatic index at larger heights confirms that CME1 undergoes continuous thermal evolution during its propagation.

We also estimated the average heating rate of CME1, denoted by $\bar{\kappa} = dQ/dt$ (Figure~\ref{fig:Thermo_para_CME1}, top-right). A negative value of $dQ/dt$ implies heat release from the CME to the surrounding medium, while a positive value indicates heat injection into the system. The heating rate remains negative up to $\sim$3.2\,$R_\odot$, confirming the initial phase of heat release identified through the polytropic index analysis. Beyond 4\,$R_\odot$, $dQ/dt$ becomes positive and remains so up to 20\,$R_\odot$, although its magnitude gradually decreases beyond 8.4\,$R_\odot$, indicating a weakening heating rate with height. This behavior reinforces the finding that CME1 transitions into a heating phase during its outward expansion.

The temperature evolution of CME1, derived using the FRIS model (Figure~\ref{fig:Thermo_para_CME1}, bottom-left), shows a sharp decrease up to 4\,$R_\odot$, followed by a gradual increase at greater heights. This profile is consistent with an initial phase dominated by expansion-driven cooling, followed by a moderate rise due to sustained heating. The corresponding proton number density $\bar{n}_p$ (Figure~\ref{fig:multi-wavelength_CME1}, bottom-right) decreases rapidly up to 4\,$R_\odot$, aligning with the early fast expansion phase, and then falls off more slowly during the later gradual expansion. Notably, despite the onset of positive heating beyond 4\,$R_\odot$ and a suppressed $\Gamma$ ($<5/3$), the temperature continues to decline up to 20\,$R_\odot$. This suggests that the energy injected during 3.2–4\,$R_\odot$ may have only partially compensated for the strong expansion-driven cooling. However, beyond 8.4\,$R_\odot$, the heat input appears sufficient to counteract cooling, resulting in a slow temperature rise during the later phase of CME1’s propagation.

\begin{figure*}[hbt]    
        \includegraphics[width=0.48\textwidth,trim={8cm 0cm 16cm 2cm}]{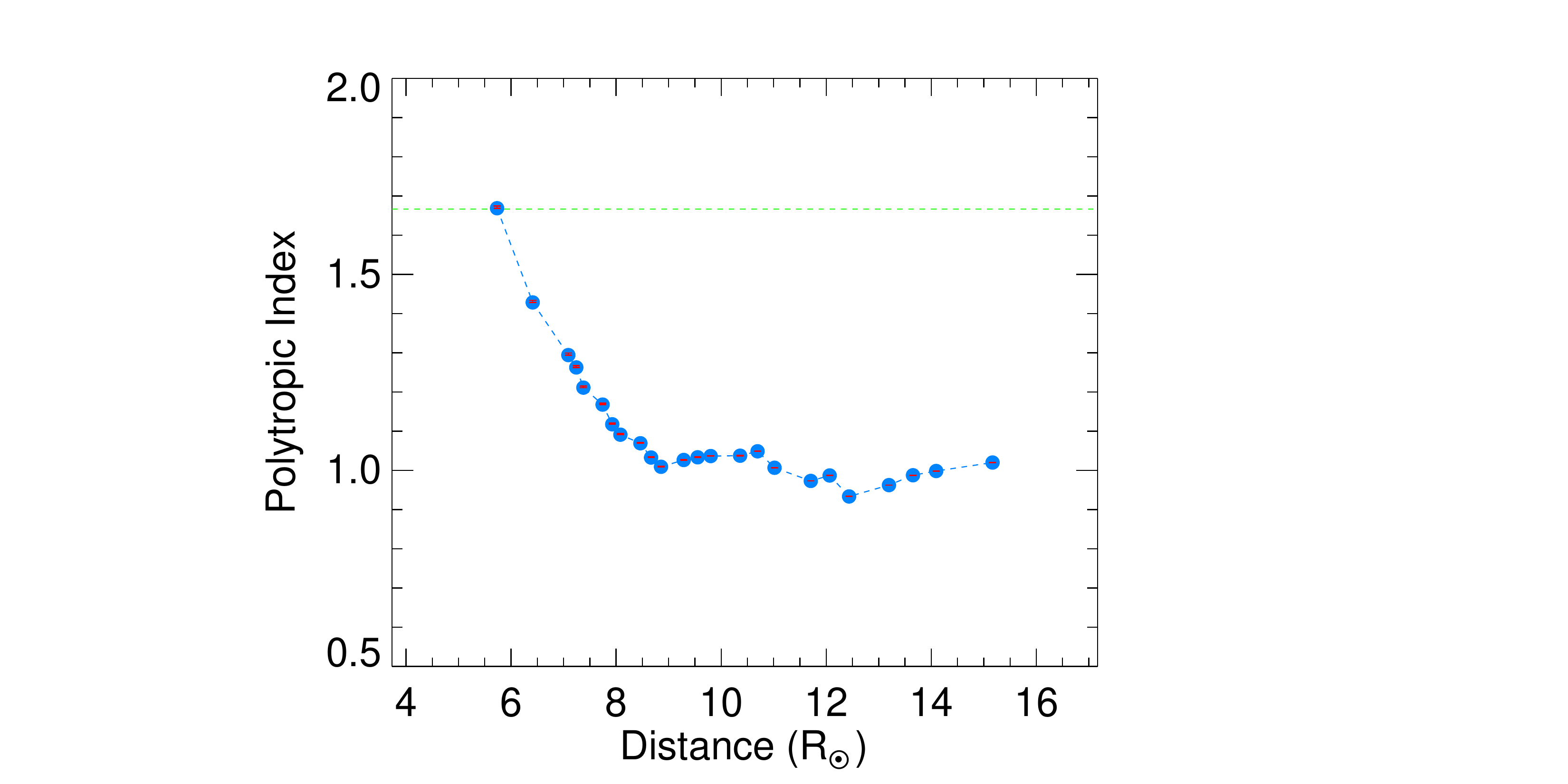}      
         \includegraphics[width=0.48\textwidth,trim={6cm 0cm 18cm 2cm}]{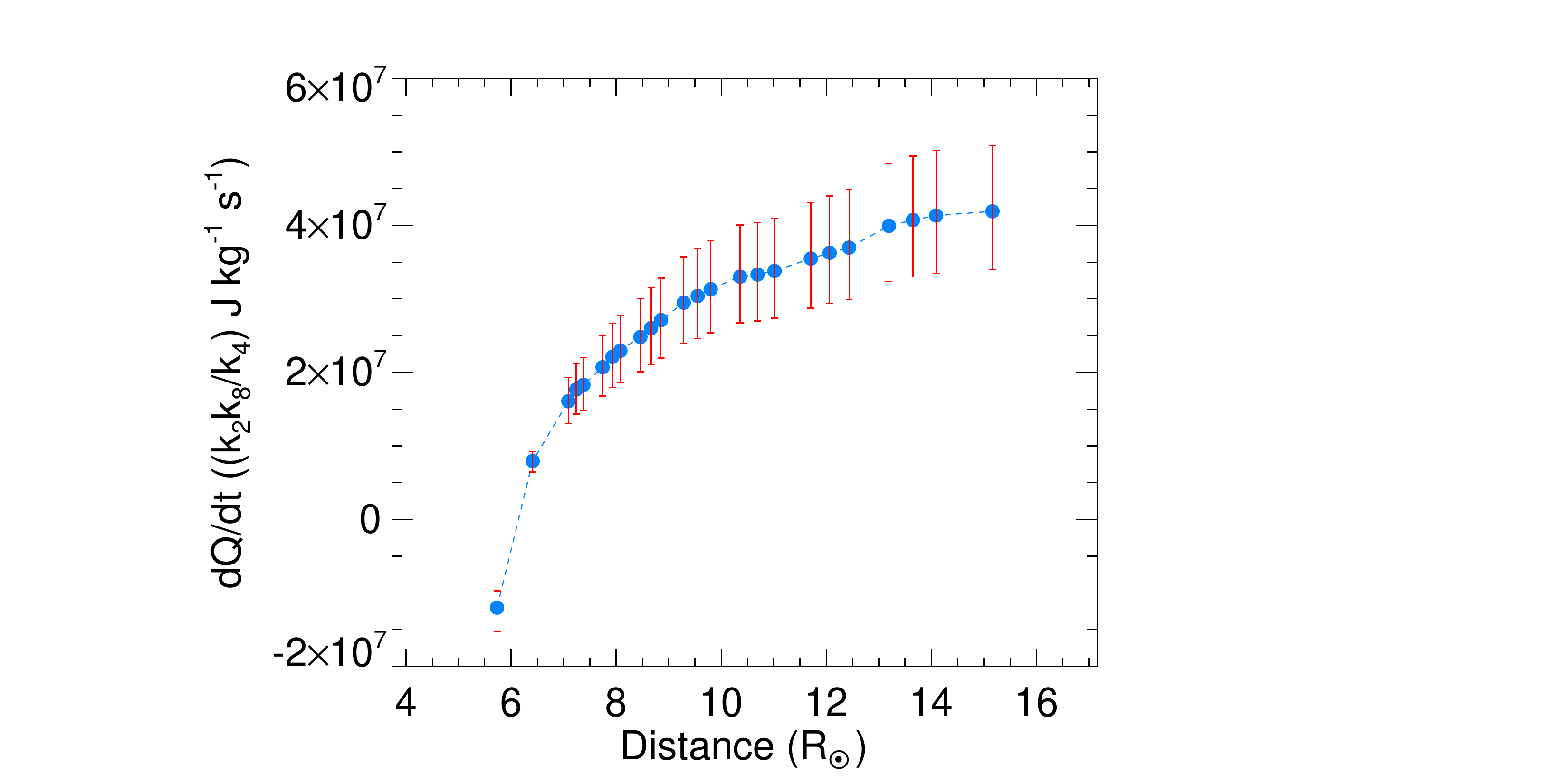}\\
         \includegraphics[width=0.48\textwidth,trim={8cm 1cm 16cm 2cm}]{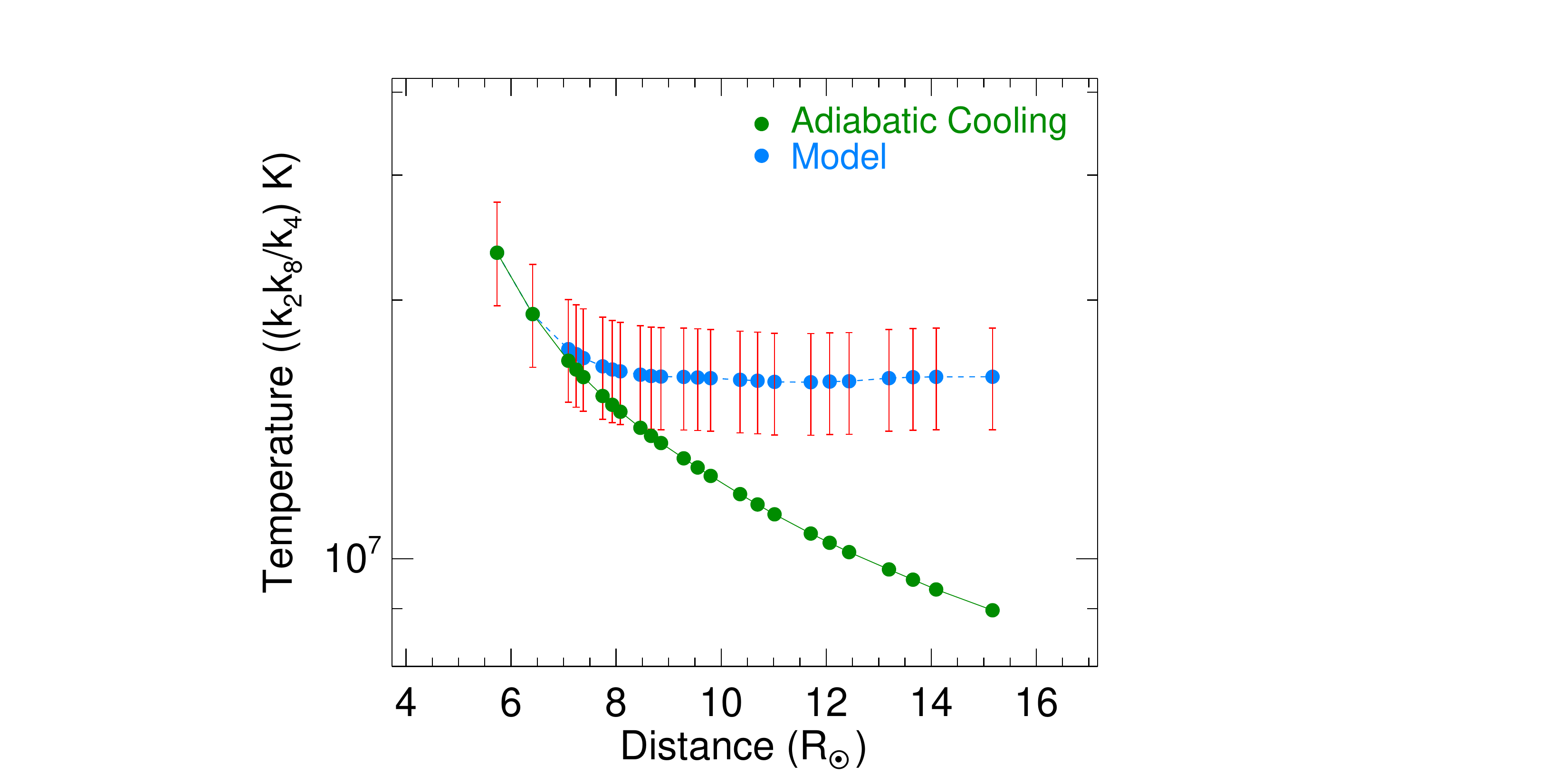}
         \includegraphics[width=0.48\textwidth,trim={6cm 1cm 18cm 2cm}]{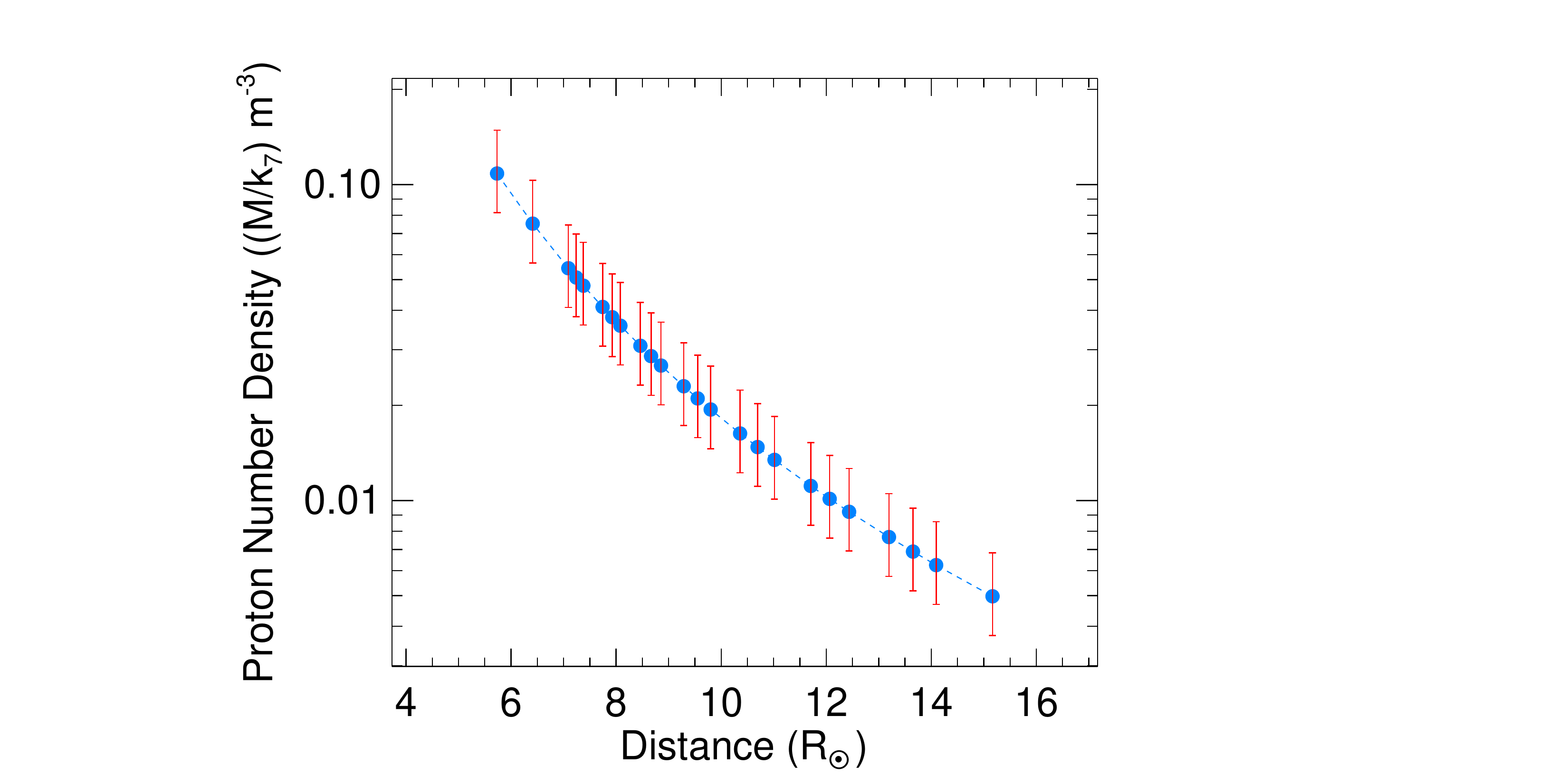}
         \caption[For the CME2: Variation of the Polytropic index ($\Gamma$), ]{For the CME2: Variation of the Polytropic index ($\Gamma$), average heating rate (dQ/dt), average temperature $(\bar{T})$, and average proton number density (${\bar{n}}_{p}$) of the CME with the heliocentric distance of the CME’s leading edge (h) is shown in the top-left, top-right, bottom-left and bottom right, respectively. The red vertical lines at each data point show the error bars derived by considering an error of 10{\%} in measurements of the flux-rope's leading edge height ($h$).}
        \label{fig:Thermo_para_CME2}
\end{figure*}

We further applied the FRIS model to CME2, and the resulting thermodynamic parameters are shown in Figure~\ref{fig:Thermo_para_CME2}. With the exception of the initial point at 5.6\,$R_\odot$, the polytropic index $\Gamma$ remains below the adiabatic value ($\Gamma = 5/3$) throughout the observed height range (top-left panel), indicating sustained heat injection into the CME plasma during its propagation. Beyond 8\,$R_\odot$, the $\Gamma$ value approaches unity, suggesting that the accumulated heating is sufficient to maintain a near-isothermal state despite the CME’s ongoing expansion. The corresponding heating rate $dQ/dt$ (Figure~\ref{fig:Thermo_para_CME2}, top-right) remains positive and increases with height, further supporting the presence of continuous energy input into the system. The derived temperature profile (bottom-left panel) shows a gradual decrease from the initial observation point up to 8\,$R_\odot$, after which it plateaus, indicating a thermally stabilized state. This behavior suggests that the increasing heat input compensates for the cooling expected from expansion beyond 8\,$R_\odot$. The average proton number density $\bar{n}_p$ (Figure~\ref{fig:Thermo_para_CME2}, bottom-right) exhibits a smooth decline with increasing heliocentric distance, consistent with the gradual expansion of the CME flux-rope.

In this study, we also compared the FRIS model-derived temperature profiles of both CMEs with their corresponding adiabatic temperature profiles, which assume no heat exchange with the surroundings. The adiabatic temperature evolution was computed using the relation $ \frac{T_{i+1}}{T_i} = \left( \frac{p_{i+1}}{p_i} \right)^{1 - \frac{1}{\gamma}}$, where $T_i$ and $T_{i+1}$ denote the temperatures, and $p_i$ and $p_{i+1}$ the pressures, at two successive time steps. Here, the pressure at each step is taken from FRIS model outputs, while the initial temperature $T_i$ is also taken from the FRIS estimate.

Under purely adiabatic conditions, the temperature should decrease in tandem with the plasma pressure as the CME expands. However, any deviation from this trend implies heating or cooling beyond adiabatic expectations. For CME1, the FRIS-derived temperature remains below the adiabatic profile up to $\sim$4\,$R_\odot$, consistent with an early phase of heat release. Beyond 8.4\,$R_\odot$, the FRIS-derived temperature exceeds the adiabatic temperature, indicating net heating during the later expansion phase (Figure~\ref{fig:Thermo_para_CME1}, bottom-left). Similarly, CME2 initially follows the adiabatic temperature profile, but as heating becomes more dominant, the FRIS-derived temperature exceeds the adiabatic expectation between 7 and 15\,$R_\odot$ (Figure~\ref{fig:Thermo_para_CME2}, bottom-left). These comparisons confirm that thermal evolution of both CMEs involves active heating and is not governed solely by adiabatic expansion, further reinforcing the need to account for non-adiabatic processes in CME thermodynamic modeling.

The physical mechanisms responsible for heating CMEs remain poorly understood, with several plausible explanations proposed. One possibility is heat transfer from the lower corona into the CME flux-rope, as its footpoints are believed to remain magnetically connected to the Sun \citep{Larson1997}. Alternatively, thermal energy could be continuously injected into the CME from the ambient solar wind in the outer corona. However, this process is constrained by the limited efficiency of cross-field diffusion of charged particles across the magnetic field \citep{Zhang2003}. Another potential mechanism involves the dissipation of the CME’s internal magnetic energy at varying rates with height, which could contribute to heating and affect its global kinematic behavior. For instance, untwisting of the flux-rope may provide energy for CME expansion and propagation \citep{Vourlidas2000}, while thermal energy can also be generated through partial conversion of internal magnetic energy \citep{Kumar1996}. This conversion may occur via Joule heating, described by the term $j^2/\sigma$, where $j$ is the current density and $\sigma$ is the electrical conductivity. However, the typically high value of $\sigma$ in the interplanetary medium renders Joule heating relatively ineffective. Additionally, magnetic reconnection, either internal within the CME structure \citep{Farrugia1993} or between the CME and the interplanetary magnetic field (IMF) \citep{Lugaz2013}, may play a critical role in transforming magnetic energy into thermal energy, thereby heating the plasma within the CME flux-rope. 

Our current study does not identify the exact physical mechanisms responsible for the inferred heating of the selected CMEs; rather, it focuses on their trend and dependencies on global kinematics. Therefore, future research should aim to quantify the efficiency of various heating processes and their contributions to the thermodynamic evolution of CMEs.

\subsubsection{Internal Forces}\label{sec:results_forces}

\begin{figure*}[ht]
\centering
        \includegraphics[width=0.6\textwidth]{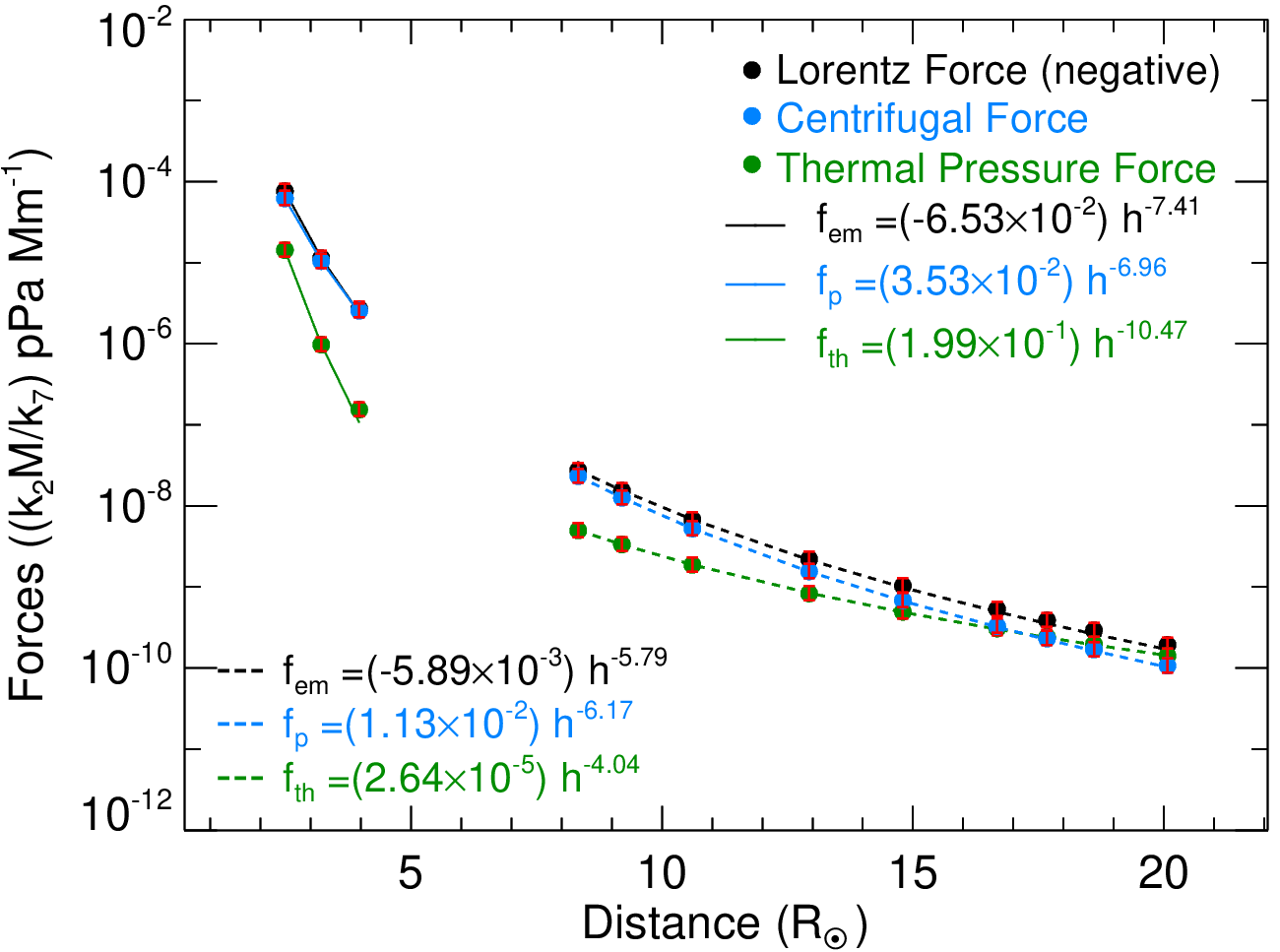}
         \includegraphics[width=0.6\textwidth]{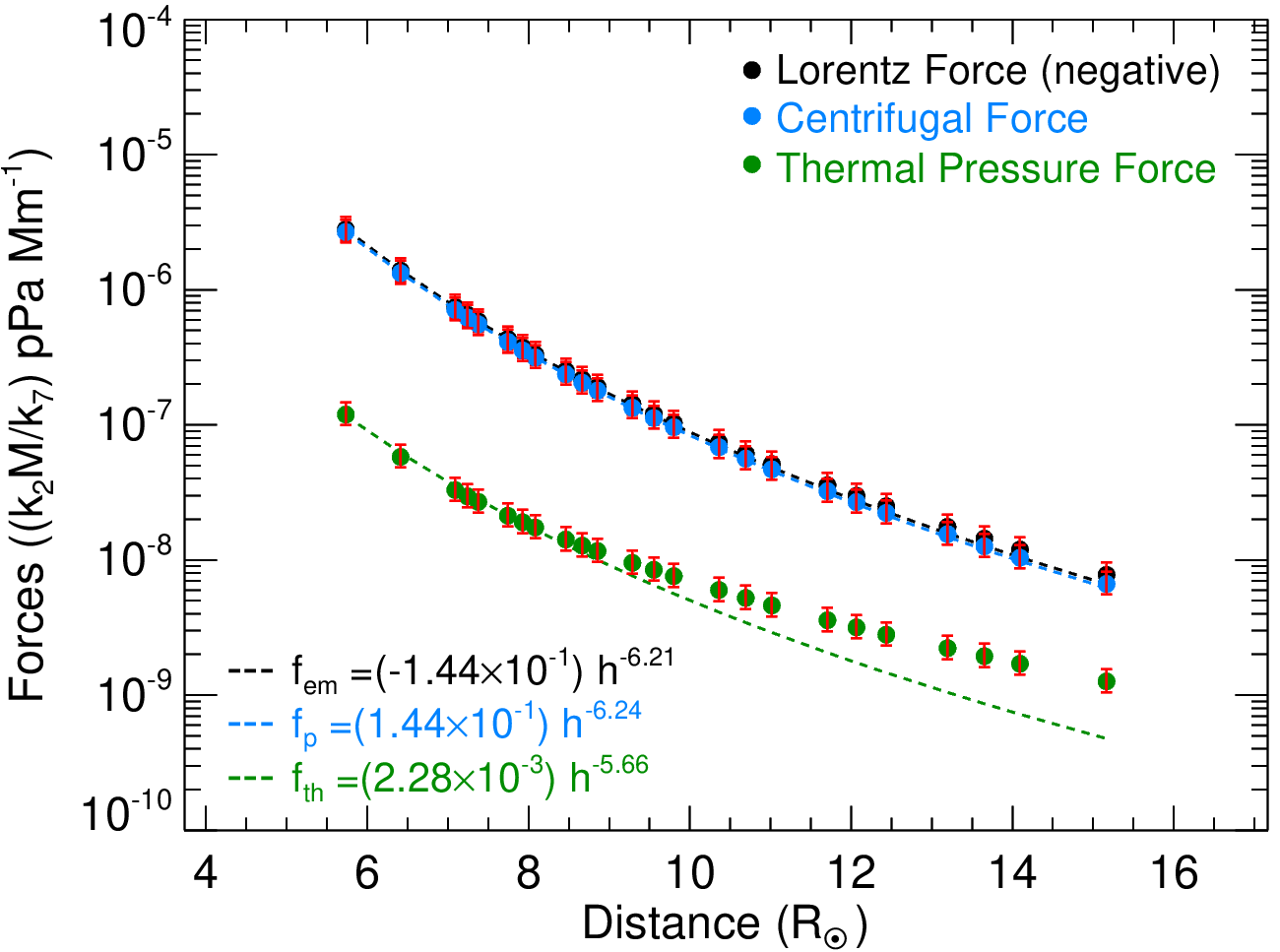}
        \caption[The FRIS model-derived average internal forces, such as ]{The FRIS model-derived average internal forces, such as Lorentz force, Thermal pressure force, and Centrifugal force, that are responsible for the radial expansion of the flux-rope of CME1 (left) and CME2 (right). The red vertical lines at each data point show the error bars derived by considering an error of 10{\%} in measurements of the leading edge height of the flux-rope ($h$). The solid and dash lines show the power-law fitted values for the model-derived internal forces.}
        \label{fig:internal_forces}
\end{figure*}

In addition to estimating thermodynamic parameters, the FRIS model also enables the derivation of key internal forces within the CME structure, namely, the average Lorentz force (${\bar f}_{em}$), the average thermal pressure force (${\bar f}_{th}$), and the average centrifugal force (${\bar f}_{p}$), calculated using Equations~(\ref{eqn:fem_final}), (\ref{eqn:fth_final}), and (\ref{eqn:fp_final}), respectively. For the selected CME1 and CME2 events, we analyzed the evolution of these three forces during their propagation away from the Sun, as illustrated in Figure~\ref{fig:internal_forces}. These forces govern the internal dynamical processes that influence the radial expansion of the CMEs. Our analysis shows that the Lorentz force consistently acts in the negative direction, whereas the thermal pressure and centrifugal forces act in the positive direction throughout the CME evolution. This implies that ${\bar f}_{em}$ acts inward, toward the flux-rope center, while ${\bar f}_{th}$ and ${\bar f}_{p}$ act outward, supporting expansion. This pattern holds true for both fast and slow CMEs, suggesting that ${\bar f}_{em}$ resists expansion while ${\bar f}_{th}$ and ${\bar f}_{p}$ promote it.

Our analysis shows that the Lorentz force consistently acts in the negative direction, whereas the thermal pressure and centrifugal forces act in the positive direction throughout the CME evolution. This implies that ${\bar f}_{em}$ acts inward, toward the flux-rope center, while ${\bar f}_{th}$ and ${\bar f}_{p}$ act outward, supporting expansion. This pattern holds true for both fast and slow CMEs, suggesting that ${\bar f}_{em}$ resists expansion while ${\bar f}_{th}$ and ${\bar f}_{p}$ promote it. We further observe that all three forces decrease rapidly in CME1 during the initial phase (up to 4~$R_\odot$), followed by a more gradual decline up to 20~$R_\odot$. In contrast, for CME2, the forces show a moderate and steady decline over the observed range (5–15~$R_\odot$). To quantify these trends, we fit a power-law of the form $y = m \cdot h^{-c}$ to the force profiles, where $m$ and $c$ are fitting coefficients and $h$ is the leading edge height of the CME. CME1 exhibits a two-phase kinematic behavior, an initial phase with rapid acceleration decrease, followed by a constant acceleration phase. This behavior is reflected in the evolution of the internal forces, necessitating two separate power-law fits. During the initial phase, ${\bar f}_{th}$ decreases most rapidly as ${\bar f}_{th} \propto h^{-10.47}$, while ${\bar f}_{em} \propto h^{-7.41}$ and ${\bar f}_{p} \propto h^{-6.96}$. In the second phase, the forces decrease more gradually, with ${\bar f}_{em} \propto h^{-5.79}$, ${\bar f}_{p} \propto h^{-6.17}$, and ${\bar f}_{th} \propto h^{-4.04}$ (left panel of Figure~\ref{fig:internal_forces}). For CME2, the internal forces are well represented by a single-phase power-law fit: ${\bar f}_{em} \propto h^{-6.21}$, ${\bar f}_{p} \propto h^{-6.24}$, and ${\bar f}_{th} \propto h^{-5.66}$ (right panel of Figure~\ref{fig:internal_forces}). Among these, ${\bar f}_{p}$ exhibits the steepest decline, while ${\bar f}_{th}$ decreases the slowest. Notably, the decrease in ${\bar f}_{th}$ becomes even more gradual beyond approximately 9~$R_\odot$.

These results highlight a key distinction: CMEs with gradual acceleration (like CME2) exhibit slowly decreasing internal forces, whereas those with rapid initial deceleration (like CME1) show a sharp drop in forces early on. Moreover, our findings demonstrate that the rate of force evolution can vary across different segments of a CME’s trajectory, yet these internal forces collectively shape the global kinematic behavior.

A comparison between CME1 and CME2 shows that ${\bar f}_{em}$ and ${\bar f}_{p}$ start with comparable magnitudes but ${\bar f}_{p}$ decreases faster as the CMEs expand outward. For CME1, ${\bar f}_{th}$ is initially lower in magnitude compared to the other two forces, yet it declines rapidly within the first 4~$R_\odot$. The relatively large inward ${\bar f}_{em}$ and its slower decline appear to drive the observed reduction in expansion acceleration during this early phase. In CME2, ${\bar f}_{th}$ remains much weaker than the other two forces, resulting in a net internal force ${\bar f}_{net} = {\bar f}_{em} + {\bar f}_{th} + {\bar f}_{p}$ that is negative up to about 6.6~$R_\odot$, causing a negative expansion acceleration. As the sum of the outward forces (${\bar f}_{th} + {\bar f}_{p}$) exceeds the inward ${\bar f}_{em}$ beyond this height, the net force becomes positive, allowing for expansion acceleration. In both CMEs, ${\bar f}_{p}$ dominates over ${\bar f}_{th}$ at lower heights, but this relationship may reverse farther out due to their differing decline rates. For CME1, ${\bar f}_{th}$ overtakes ${\bar f}_{p}$ beyond 18~$R_\odot$. From the observed trends, it is evident that at larger heliocentric distances (beyond the observed ranges), ${\bar f}_{th}$ will likely dominate the force balance. This suggests that thermal pressure may become the primary driver of expansion at greater distances from the Sun. These findings emphasize the necessity of tracking CMEs to larger heliocentric distances in order to fully understand the evolution of their internal forces, thermodynamics, and kinematics in the interplanetary medium.

\subsection{Multi-wavelength Imaging Observations}

Examining the thermodynamic state of a CME flux-rope at its source region is equally important, as it provides insights into the intrinsic thermal conditions of CMEs at the time of their formation. Understanding these initial conditions can help link the origin of CMEs to their evolving thermal characteristics at greater coronal heights. While we have already presented the FRIS model-derived estimates of thermodynamic parameters at heights where CME kinematics can be reliably determined from coronagraphic observations, we now turn our attention to analyzing the flux-rope’s thermal state at its origin in the lower corona. To achieve this, we utilize high-resolution multi-wavelength observations from the Atmospheric Imaging Assembly (AIA) and Helioseismic and Magnetic Imager (HMI) onboard the Solar Dynamics Observatory (SDO), along with data from STEREO, SOHO, and the Global Oscillation Network Group (GONG)/H$_{\alpha}$.

\vspace{0.5cm}
{\large \textbf{Source region of CME1}}
\vspace{0.2cm}

\begin{figure*}
  \includegraphics[trim = 0.0cm 0.0cm 1.0cm 0.0cm, scale=0.86]{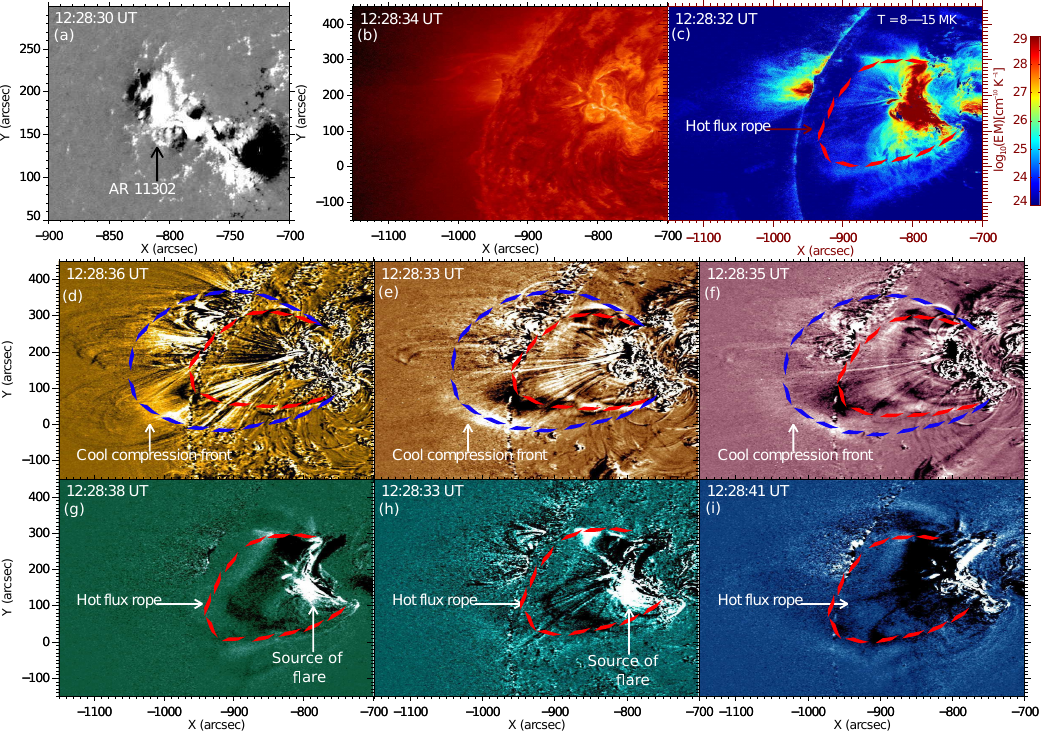}
\caption[Multi-wavelength view of magnetic flux-rope as seen]{Multi-wavelength view of magnetic flux-rope as seen in different wavebands of SDO/AIA. Top row (panels a--c): The hot flux-rope formed above an active region, AR 11302, as shown by SDO/HMI line-of-sight magnetograms (panel a). The SDO/AIA 304 {\AA} image depicts the absence of flux-rope/eruption (panel b). The panel (c) indicates the presence of a hot flux-rope (8--15 MK) in the DEM map.  Middle row (panels d--f): It shows the distinct feature of the magnetic flux-rope in its eruptive phase. The cyan dotted lines show the cool compression front, followed by the dark cavity region of the flux-rope, which best appears at coronal temperature (0.6--2 MK) in AIA 171 {\AA}, 193 {\AA}, and 211 {\AA} wavebands.  Bottom row (panels g--i): The hot channel of the flux-rope has been shown via dotted red lines in the hotter wavebands of AIA (2.5--10 MK, i.e., 94 {\AA}, 131 {\AA}, and 335 {\AA}). It should be noted that cool coronal images (middle panel) show the very faint signature of this hot channel of the flux-rope.}
 \label{fig:multi-wavelength_CME1}
\end{figure*}

To identify the source region of this eruption, we analyzed HMI magnetograms and found that CME1 originated from a $\beta\gamma$-type sunspot located in active region (AR) 11302 near the northeastern limb. This region was highly active, producing several M-class and one X-class flare within a 24-hour period (Figure~\ref{fig:multi-wavelength_CME1}, panel~a). The observed build-up of magnetic energy, along with the formation, evolution, and eventual eruption of the flux-rope, suggests that the CME1 structure was thermally hot \citep{Zhang2012}. This is supported by its association with intense flare activity, as evident in panels~g and h of Figure~\ref{fig:multi-wavelength_CME1}. On 2011 September 24, at approximately 12:33~UT, an M7.1-class flare was observed in AR 11302. This flare was accompanied by a plasma eruption that manifested as a fast halo CME, identified as CME1, in the LASCO field of view. Notably, the flare began around 12:33~UT. We marked the initiation region of this flare based on brightening enhancements in the hot AIA EUV channels, particularly 94~\AA{} and 131~\AA{} (see panels~g and h of Figure~\ref{fig:multi-wavelength_CME1}). Prior to the flare onset, the hot flux-rope became unstable and began to rise (panels~d--i of Figure~\ref{fig:multi-wavelength_CME1}).

Observations of CME1 clearly display the classic three-part structure: a bright leading front, a dark cavity, and a hot core channel corresponding to the flux-rope itself \citep{Zhang2012}. This eruptive flux-rope appears diffuse, so we employed base-difference images from SDO/AIA to enhance the contrast with the background corona. The hot channel of the flux-rope began to rise around 12:05~UT, roughly 20 minutes before the flare initiation. This hot component (manually outlined by red dotted lines) is observed only in AIA's high-temperature channels, namely 94~\AA{} (6.3~MK), 131~\AA{} (10~MK), and 335~\AA{} (2.5~MK), as shown in panels~g, h, and i of Figure~\ref{fig:multi-wavelength_CME1}. In contrast, the cool component of the flux-rope is not visible in lower-temperature passbands such as 171~\AA{}, 193~\AA{}, and 211~\AA{} (panels~d, e, and f of Figure~\ref{fig:multi-wavelength_CME1}).

Interestingly, the hot channel is seen pushing overlying magnetic field lines, and the resulting bright, cooler front, visible at coronal temperatures (171~\AA{} (0.6~MK), 193~\AA{} (1.2~MK), and 211~\AA{} (2~MK)), represents the compressed leading edge (LE) of the flux-rope. This LE is highlighted with blue dotted lines in panels~d, e, and f of Figure~\ref{fig:multi-wavelength_CME1}. Such observations are consistent with the onset of torus instability as the eruption trigger mechanism \citep{Kliem2006, Zhang2012}. A reconnecting current sheet may form between the overlying cooler front and the hot channel, potentially contributing to further heating of the flux-rope. To investigate the thermal structure of the eruptive flux-rope, we performed Differential Emission Measure (DEM) analysis \citep{Cheung2015} (Section~\ref{chapt2-metho-dem}). The analysis revealed that the hot channel spans a broad temperature range, reaching as high as $\approx$8--15~MK (panel~c of Figure~\ref{fig:multi-wavelength_CME1}; red dotted lines). In comparison, the temperature of the compressed cool front is estimated to be in the 1--3~MK range, based on its visibility in the 211~\AA{}, 193~\AA{}, and 171~\AA{} channels.

These results indicate that CME1’s flux-rope was already in a hot state at its source region. Such an inherently hot structure is expected to release thermal energy as it expands through the outer corona. This interpretation is consistent with our FRIS model-derived estimates of CME1’s thermodynamic parameters, which show a declining temperature profile up to 4~$R_\odot$, and a polytropic index that remains above the adiabatic value.

\begin{figure}[ht]
  \includegraphics[trim = 0.0cm 0.0cm 1.0cm 0.0cm, scale=0.88]{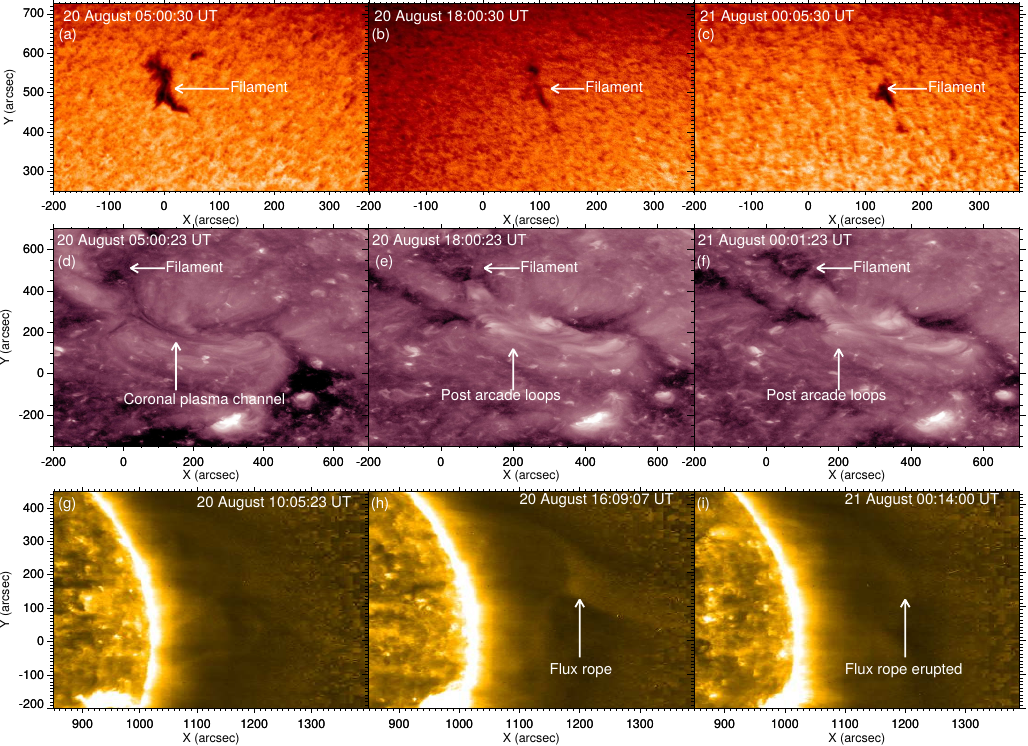}
  \caption[Top row (panels a--c): The sequence of images of GONG/H$_{\alpha}$]{Top row (panels a--c): The sequence of images of GONG/H$_{\alpha}$ shows the gradual eruption of a quiescent filament on 2018 August 20. It is associated with an elongated filament channel containing no cool component but a hot plasma channel. Middle row (panels d--f): Temporal evolution of coronal plasma channel (hot filament channel) lying between two coronal holes as observed by SDO/AIA 211 {\AA}.  Bottom row (panels g--i): The formation of flux-rope and its eruption lying over the coronal plasma channel is shown in STEREO-A/EUVI 171 {\AA} on 2018 August 20.}
  \label{fig:multi-wavelength_CME2}
  \end{figure}

\vspace{0.5cm}
{\large \textbf{Source region of CME2}}
\vspace{0.2cm}

To identify the source region of the selected CME2, which appeared as a slow and faint event in the LASCO field of view on 2018 August 20, we analyzed GONG/H$_{\alpha}$ observations along with high-resolution EUV imaging data. A detailed inspection suggests multiple components potentially contributing to the origin of CME2, including a draining filament, an associated coronal plasma channel, and an overlying flux-rope. The filament structure is evident in panels~a, b, and c of Figure~\ref{fig:multi-wavelength_CME2}, located between two trans-equatorial coronal holes (see panels~d, e, and f of Figure~\ref{fig:multi-wavelength_CME2}). An extended hot coronal plasma channel, also referred to as a filament channel, appears in the AIA coronal EUV filters corresponding to 2–4~MK (particularly in 193~\AA{} and 211~\AA{}). Around 08:00~UT, we observe the gradual drainage of the filament along with the eruption of the elongated plasma channel. While the filament erupts slowly, the associated plasma channel appears to accelerate more rapidly \citep{Chen2019, Gopalswamy2022, Mishra2019, Palmerio2022}. Notably, there are no significant signatures on the solar disk during either of these eruptions, highlighting the stealthy nature of this event \citep{Nitta2021, Mishra2019, Palmerio2022}.

However, several hours later, the formation of post-eruption arcades and coronal dimming becomes evident, implying a preceding eruption event, most likely involving an overlying diffuse flux-rope situated above the plasma channel. Upon close examination, this faint overlying flux-rope is faintly visible in the STEREO-A/EUVI 171~\AA{} passband (see panels~g, h, and i of Figure~\ref{fig:multi-wavelength_CME2}). A small portion of the filament located above the coronal hole is also seen erupting in a jet-like fashion around 19:00~UT \citep{Mishra2019, Palmerio2022}, along with the overlying flux-rope. Together, these features contributed to the formation of CME2.

CME2 exhibits characteristics typical of a stealth CME, showing no distinct on-disk signatures and lacking evidence of a hot flux-rope, unlike CME1. Additionally, the kinematic profile of this faint CME does not indicate rapid expansion in the higher corona. The absence of significant heating or cooling signatures, along with the lack of a hot flux-rope, suggests that CME2 is unlikely to have released heat into the corona. This interpretation is consistent with the FRIS model-derived thermodynamic results, which show that the polytropic index for CME2 remains below or near the adiabatic value during its propagation between 5–15~$R_\odot$.

\subsection{Summary}\label{chap3-sec:summary}
This study highlights the contrasting thermodynamic and dynamic evolution of two CMEs, CME1 (2011 September 24) and CME2 (2018 August 20), that differ significantly in their kinematic profiles at lower coronal heights. Key findings are summarized below:

\begin{enumerate}

\item CME1 is a fast event (1,885~km~s$^{-1}$), exhibiting two-phase expansion: a rapid deceleration phase, followed by nearly constant expansion up to 20~$R_\odot$. In contrast, CME2 is slower (420~km~s$^{-1}$) and displays a gradual acceleration from 5 to 15~$R_\odot$. 

\item Thermodynamic analysis using the FRIS model reveals that CME1 experiences heat release during its initial deceleration, while CME2 shows heat absorption during its early gradual acceleration. Both CMEs later tend toward isothermal behavior, with polytropic indices falling below the adiabatic value. Interestingly, CME1 reaches the isothermal state more rapidly and at a lower height compared to CME2. These results affirm that CMEs do not undergo purely adiabatic expansion but instead experience varying degrees of heating and cooling during their propagation.

\item Force analysis shows that centrifugal and thermal pressure forces primarily drive the radial expansion, while the Lorentz force consistently acts inward, resisting expansion. In CME1's early phase, the thermal pressure force decreases most rapidly, followed by the Lorentz and centrifugal forces. Later, the centrifugal force declines most steeply, making thermal pressure the dominant outward force at larger heights. 

\item Notably, for CME1, the thermal pressure force surpasses the centrifugal force and approaches the Lorentz force by 18~$R_\odot$. This crossover is not observed in CME2 within the studied height range, implying that a similar transition may occur at greater heights for slower CMEs.

\item Multi-wavelength imaging supports these model-derived trends. CME1, associated with an M7.1-class flare and hot flux-rope formation (8–15~MK), shows possible signs of heating at eruption onset. In contrast, CME2 exhibits no obvious hot structure but is linked to a stealth eruption near coronal holes. Although initial heating is not evident, magnetic reconnection in the source region may contribute to energy input and observed acceleration, as reflected in our model results. However, due to observational constraints near the Sun ($<$~2–3~$R_\odot$), early-phase thermodynamics remain unconstrained. 
\end{enumerate}

Our findings suggest that CMEs undergo complex thermodynamic evolution governed by heat exchange processes, rather than ideal adiabatic expansion. Understanding these processes requires the synergy of advanced modeling, multi-wavelength remote sensing, and coordinated in situ observations to resolve the mechanisms driving differential heating in CMEs. By linking CME kinematics to their thermal evolution at coronal heights where direct measurements are sparse, this work underscores the importance of modeling CME dynamics without restrictive assumptions on flux-rope thermal state. While observational and modeling limitations introduce uncertainty, future validation via in situ measurements from PSP, SolO, and other heliospheric missions will be essential.

\section{In Situ Thermal Analysis at 1 au} \label{Ch3-sec:2}

Having examined the near-Sun thermodynamic evolution of the selected CMEs using the FRIS model, we now turn our attention to their corresponding interplanetary counterparts (ICMEs) as observed near 1~au. In this section, we investigate the internal thermal state and turbulence properties of these ICMEs using in situ measurements from the Wind spacecraft. While remote observations offer insights into CME evolution in the low corona, in situ measurements enable direct assessment of plasma properties such as temperature, density, and magnetic field at localized spatial scales. Our objective is to understand how the thermal behavior of CMEs, as inferred near the Sun, manifests in interplanetary space, and whether fast and slow CMEs exhibit distinct thermodynamic signatures at 1~au. In addition, we aim to examine how thermal states and turbulence characteristics vary across different substructures within the ICME plasma, such as the pre-ICME, sheath, magnetic ejecta, and post-ICME regions, and how these are influenced by interactions with the ambient solar wind.

\subsection{Methodology}\label{Ch3-sec2:methods}

To characterize the thermal state of plasma within the two ICMEs, we estimate the polytropic index ($\Gamma$) for both electrons and protons. This parameter serves as a diagnostic of heat exchange processes during ICME expansion, helping to identify whether localized regions are dominated by heating, cooling, or near-isothermal conditions. By comparing the polytropic behavior across different ICME substructures and the surrounding solar wind, as well as between electrons and protons, we aim to understand specific thermodynamic responses under varying plasma conditions.

For this purpose, we use high-resolution in situ measurements from the Wind spacecraft. The 9-second resolution electron number density ($n_e$) and temperature ($T_e$) data from the Wind/SWE are used to derive the electron polytropic index ($\Gamma_e$), while the 92-second resolution proton number density ($n_p$) and temperature ($T_p$) from the same instrument are used to derive the proton polytropic index ($\Gamma_p$). Details of the polytropic index calculation, including the moving window technique, correlation-based filtering, and justification for sub-interval selection, are presented in Section~\ref{sec:chapt2-metho-poly}.

To investigate the turbulent nature of the selected ICMEs, we analyze high-resolution magnetic field data from the Wind/MFI. The 11 Hz (0.092 s) resolution data enable us to resolve fine-scale fluctuations and characterize both inertial and dissipation scale turbulence. Our aim is to extract the spectral slope ($\alpha_B$), magnetic compressibility ($C_B$), and intermittency properties, which together describe the energy cascade and dissipation behavior of the plasma.

The power spectral density (PSD) of the magnetic field is computed using Fast Fourier Transform (FFT), and the spectral slope is estimated in two frequency regimes: the inertial range and the dissipation range. To quantify the compressibility of the turbulence, we calculate $C_B = P_t / P_{tr}$, which serves as a proxy for the Alfvénic versus compressive nature of magnetic fluctuations. Additionally, we use the Partial Variance of Increments (PVI) method to detect localized structures and sharp gradients in the magnetic field, which are indicative of intermittent energy dissipation and potential reconnection sites.

These diagnostics help distinguish between different turbulence regimes and clarify how energy is transferred and dissipated across ICME structures. Details of the turbulence analysis procedure, including smoothing techniques, frequency band selection, and intermittency detection across multiple time lags ($\tau = $ 0.18, 9.2, and 92 s), are described in Section~\ref{sec:chapt2-metho-turbu}.

\subsection{Overview of In Situ Observation at 1AU}

\begin{figure*}[ht]
    \centering
    \begin{adjustbox}{width=1.1\linewidth,center}
        \includegraphics[width=0.52\linewidth]{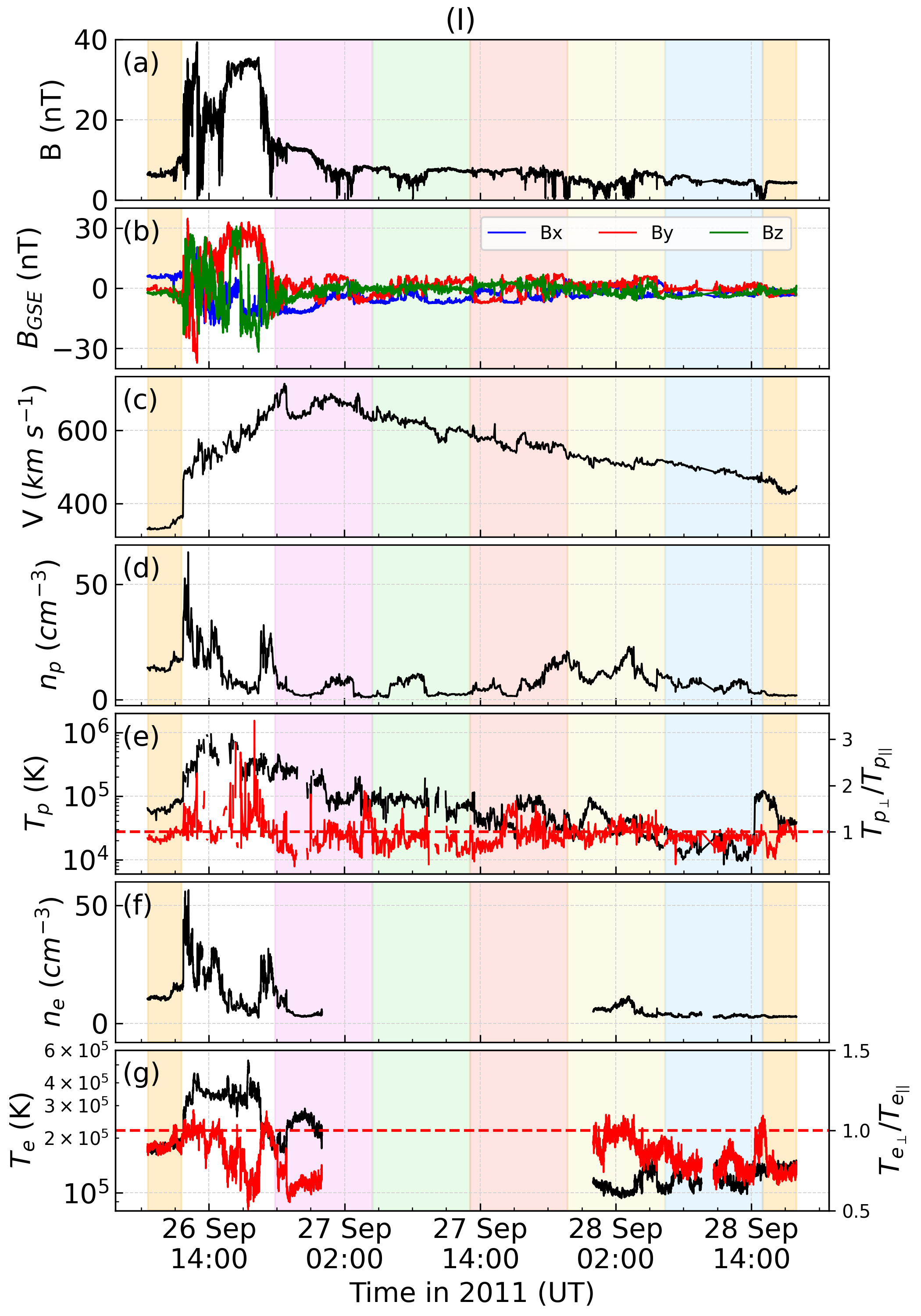}
        \includegraphics[width=0.52\linewidth]{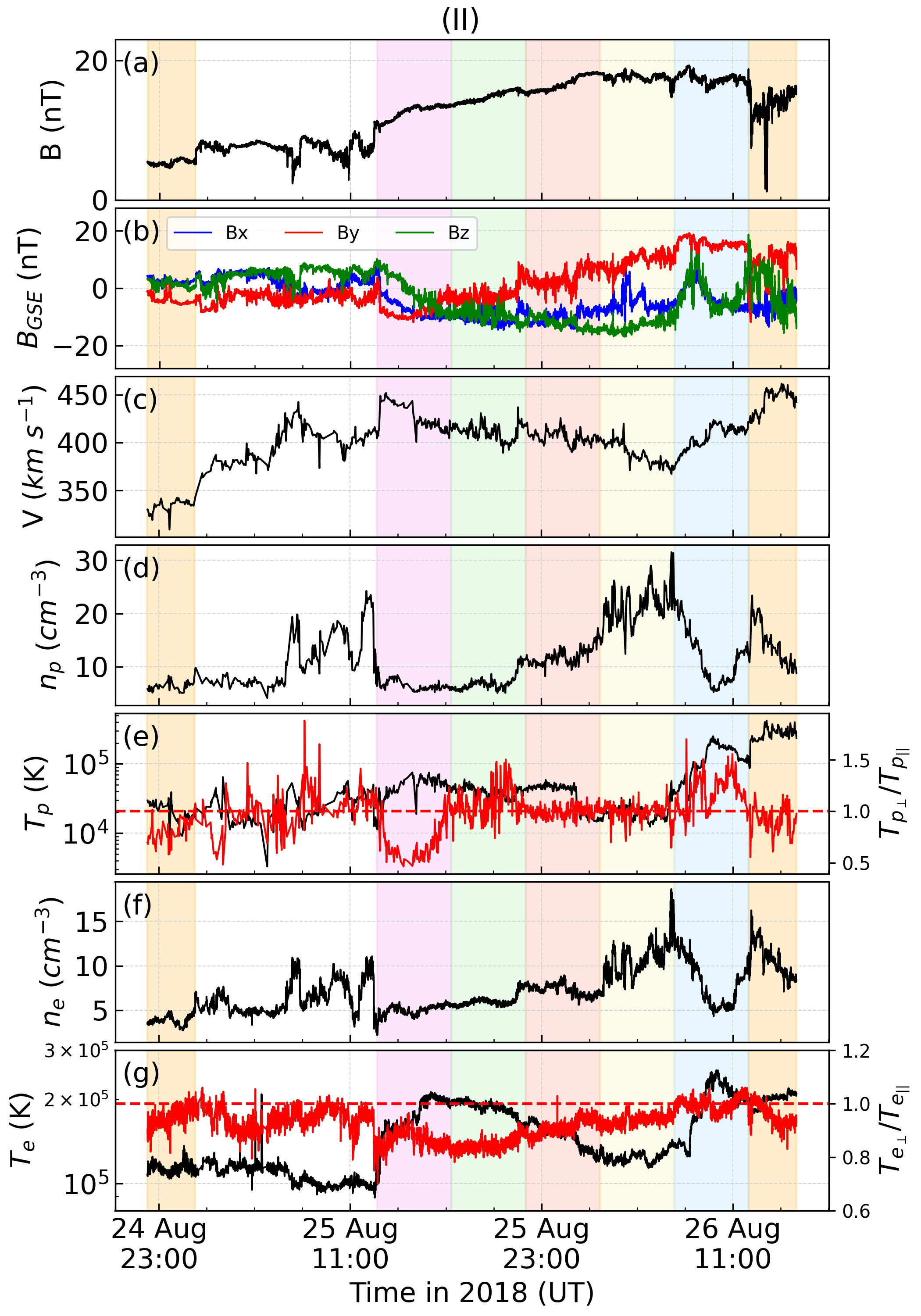}
    \end{adjustbox}

    \caption[Observed in situ parameters for (a) ICME1 and (b) ICME2 ]{Observed in situ parameters for (a) ICME1 and (b) ICME2 using magnetic and plasma data from the Wind spacecraft. The orange-shaded regions at the beginning and end represent the pre and post-ICME solar wind. The white-shaded area marks the sheath region, while the multicolored-shaded region between the sheath and post-ICME solar wind corresponds to the magnetic ejecta, divided into five equal parts.}
    \label{chap3-fig:insitu_wind}
\end{figure*}

Figure~\ref{chap3-fig:insitu_wind} panels~I and~II present the in situ observations of two ICME events, ICME1 on 26 September 2011 and ICME2 on 25 August 2018, recorded by the Wind spacecraft at 1~au. The shock arrival for ICME1 is marked at 11:34~UT on 26 September 2011, coinciding with a sudden increase in total magnetic field strength ($B$), proton bulk velocity ($V$), $n_p$, and $T_p$. The white-shaded region in both events denotes the ICME sheath, characterized by enhanced and fluctuating magnetic field components ($B_x$, $B_y$, $B_z$), elevated plasma velocities, and variable plasma parameters. Following the sheath, the magnetic ejecta (ME) region of the ICME is encountered. For detailed analysis, we divided the ME into five distinct segments, each represented by different colored shaded regions. This segmentation allows us to capture spatial variations within the ME and isolate contributions from different regions. Part~1 may retain some influence from the preceding sheath, while Part~5 might reflect interaction with the trailing solar wind. Part~3, located at the likely center of the ME, is considered to represent the intrinsic properties of the ejecta with minimal influence from the surroundings. Parts~2 and~4 serve as transitional zones between these key segments, bridging the interfaces from Part~1 to Part~3 and from Part~3 to Part~5, respectively. To contextualize the ICME structure with ambient conditions, we also include three hours of upstream (pre-ICME) and downstream (post-ICME) solar wind observations in both cases, indicated by orange-shaded regions. Analyzing the thermal and turbulence properties across these defined segments provides insight into the plasma evolution within the CME and its interaction with the ambient solar wind.

\begin{table*}[ht]
\centering
\caption[Mean values of the observed in situ parameters of various]{Mean values of the observed in situ parameters of various regions associated with ICME1 and ICME2 using magnetic and plasma data from Wind spacecraft.}
\label{chap3-sec2-tab:cme_insitu_table}
\begin{adjustbox}{width=\textwidth}
\begin{tabular}{|c|cc|cc|cc|cc|cc|}
\hline
\multirow{3}{*}{Parameters} & \multicolumn{8}{c|}{Mean values in various regions of ICME1 and ICME2} \\  
\cline{2-9}
& \multicolumn{2}{c|}{Pre-ICME} & \multicolumn{2}{c|}{Sheath} & \multicolumn{2}{c|}{ME} & \multicolumn{2}{c|}{Post-ICME} \\  
\cline{2-9}
& ICME1 & ICME2 & ICME1 & ICME2 & ICME1 & ICME2 & ICME1 & ICME2 \\
\hline
$B$ [nT] & 6.8 & 5.3 & 25.7 & 7.7 & 6.5 & 15.9 & 3.5 & 14.1 \\
\hline
$T_p$ [$10^4$ K] & 6.4 & 2.0 & 44.1 & 2.4 & 7.2 & 5.5 & 7.7 & 26.5 \\
\hline
 $n_p$  [cm$^{-3}$] & 14.5 & 6.3 & 16.9 & 10.3 & 7.3 & 12.0 & 2.3 & 13.7 \\
\hline
$T_e$  [$10^4$ K] & 17.9 & 11.3 & 32.3 & 10.9 & 14.2 & 16.5 & 13.6 & 19.9 \\
\hline
$n_e$ [cm$^{-3}$]  & 12.0 & 3.8 & 17.5 & 6.3 & 4.6 & 7.5 & 2.8 & 10.6 \\
\hline
$v$ [km s$^{-1}$]   & 336 & 334 & 558 & 396 & 578 & 406 & 454 & 443 \\
\hline
$v_{exp}$ [km s$^{-1}$]   & - & - & - & - & 186 & 35 & - & - \\
\hline
\end{tabular}
\end{adjustbox}

\end{table*}

The red curves in panels~(e) and~(g) of Figure~\ref{chap3-fig:insitu_wind} represent the temperature anisotropy ($T_{\text{aniso}} = T_{\perp}/T_{\parallel}$) for protons and electrons, respectively, where $T_{\perp}$ and $T_{\parallel}$ denote temperatures perpendicular and parallel to the magnetic field. In both ICMEs and ambient solar wind, anisotropic heating and cooling mechanisms can preferentially modify one temperature component over the other, leading to $T_{\perp} \neq T_{\parallel}$. Such anisotropies are known to give rise to various plasma instabilities and electromagnetic wave modes \citep{Gary1993, Denton1994}.

For ICME1, the sheath region exhibits $T_{\text{aniso}} > 1$ for protons (Figure~\ref{chap3-fig:insitu_wind}~I(e)), indicating preferential heating in the direction perpendicular to the magnetic field. This condition is conducive to the generation of mirror mode and Alfvén ion cyclotron waves \citep{Liu2006waves, AlaLahti2019, Verscharen2019}. In contrast, electrons in the sheath show $T_{\text{aniso}} < 1$ (Figure~\ref{chap3-fig:insitu_wind}~I(g)), suggesting enhanced parallel temperature, which can trigger firehose instabilities and lead to the generation of low-frequency whistler waves \citep{Hollweg1970, Hellinger2000}. Notably, the electron temperature anisotropy approaches unity ($T_{\text{aniso}} \approx 1$) at the transition boundaries, between pre-ICME and sheath, sheath and magnetic ejecta (ME), and ME and post-ICME regions.

In comparison, ICME2 displays overall lower proton temperature anisotropy (Figure~\ref{chap3-fig:insitu_wind}~II(e)). The sheath region shows a mix of $T_{\perp}$-dominated and $T_{\parallel}$-dominated intervals. Within the ME, the first segment adjacent to the sheath exhibits $T_{\text{aniso}} < 1$, whereas the remaining parts show $T_{\text{aniso}} > 1$. Interestingly, the final two parts of the ME exhibit distinct plasma characteristics: Part~4 contains higher density and lower temperature, while Part~5 shows reduced density and elevated temperature. The denser, cooler material in Part~4 may correspond to embedded filament material, consistent with the filament-associated CME eruption linked to ICME2 \citep{Mishra2015b, Gopalswamy2022}. Additionally, the high density in Part~4 could result from compression by Part~5, which possibly contains fast solar wind material. It is important to note that the post-ICME solar wind is faster than ICME2 and is associated with a co-rotating interaction region (CIR), followed by a high-speed stream (HSS) \citep{Mirko2020}. These interactions likely influence the thermal and dynamic evolution of the ICME plasma.

Table~\ref{chap3-sec2-tab:cme_insitu_table} presents the mean values of key in situ parameters for distinct regions of ICME1 and ICME2, as derived from Wind spacecraft magnetic and plasma data. The sheath region of ICME1 displays significantly enhanced magnetic field strength ($B$), plasma density, and proton and electron temperatures compared to the other regions. Although the sheath of ICME2 also shows elevated magnetic and plasma parameters, the enhancements are notably weaker than those observed in ICME1. This contrast likely stems from the differing dynamics of the magnetic ejecta (ME) in the two ICMEs. Specifically, the ME of ICME1 propagates with a much higher speed relative to the pre-ICME solar wind, causing stronger compression and heating in the sheath. In contrast, ICME2 exhibits a smaller velocity gradient across the shock front, resulting in a more modestly compressed sheath.

Furthermore, the ME of ICME2 is influenced by a trailing high-speed solar wind stream, which limits its deceleration and allows it to maintain a steady propagation speed. Interestingly, the post-ICME region of ICME2 shows higher magnetic field strength and plasma values than that of ICME1, indicating different ambient solar wind conditions surrounding the two events. These environmental differences likely contribute to their divergent propagation characteristics in the interplanetary medium \citep{Manchester2008, Vrsnak2010, Mishra2013}. A notable inverse trend is observed in the properties of the magnetic ejecta: despite being slower, ICME2's ME exhibits stronger magnetic fields and higher plasma densities than the faster ME of ICME1. Additionally, the derived ME expansion speeds highlight a stark contrast; ICME1 expands more vigorously at 186~km~s$^{-1}$, whereas ICME2 shows a modest expansion of 35~km~s$^{-1}$, suggesting a more confined structure. Altogether, these comparisons emphasize the influence of both internal properties and external solar wind conditions on the evolution of ICMEs, as reflected in the distinct features of their pre-ICME, sheath, ME, and post-ICME regions.

\subsection{Derived Thermal States at 1AU}

\begin{figure*}[ht]
    \centering
    \includegraphics[width=\textwidth]{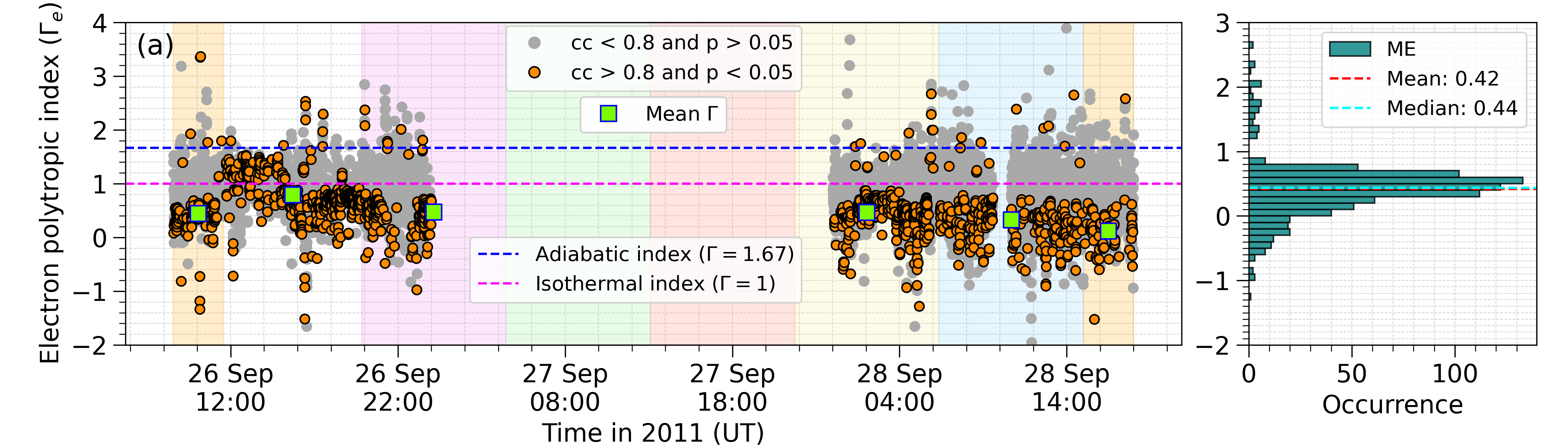}\\
    \includegraphics[width=\textwidth, trim=0 0 0 8, clip]{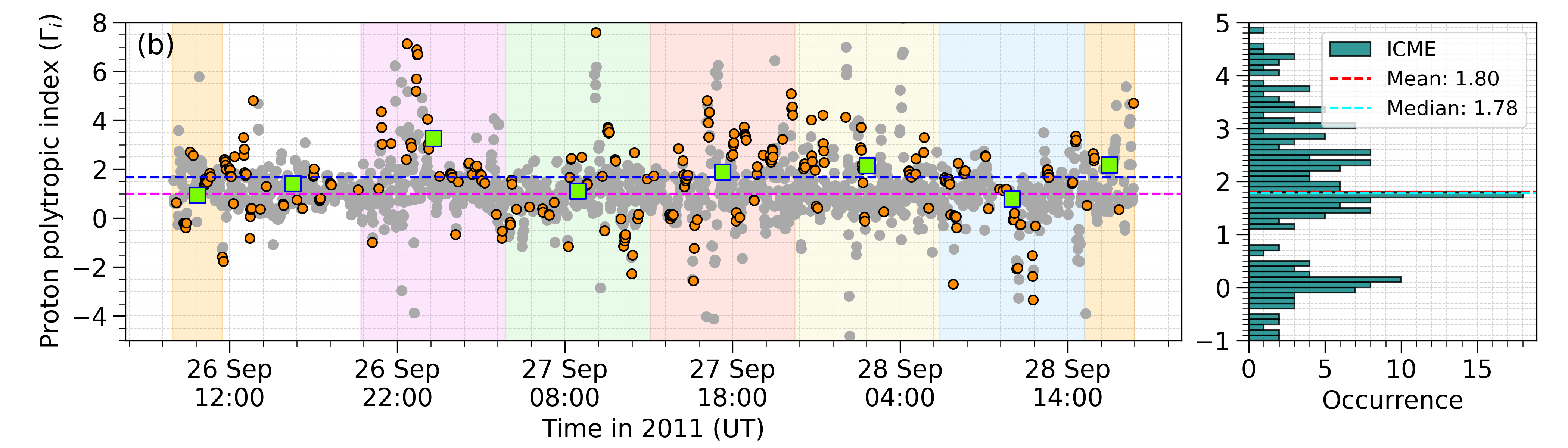}\\
    \includegraphics[width=\textwidth]{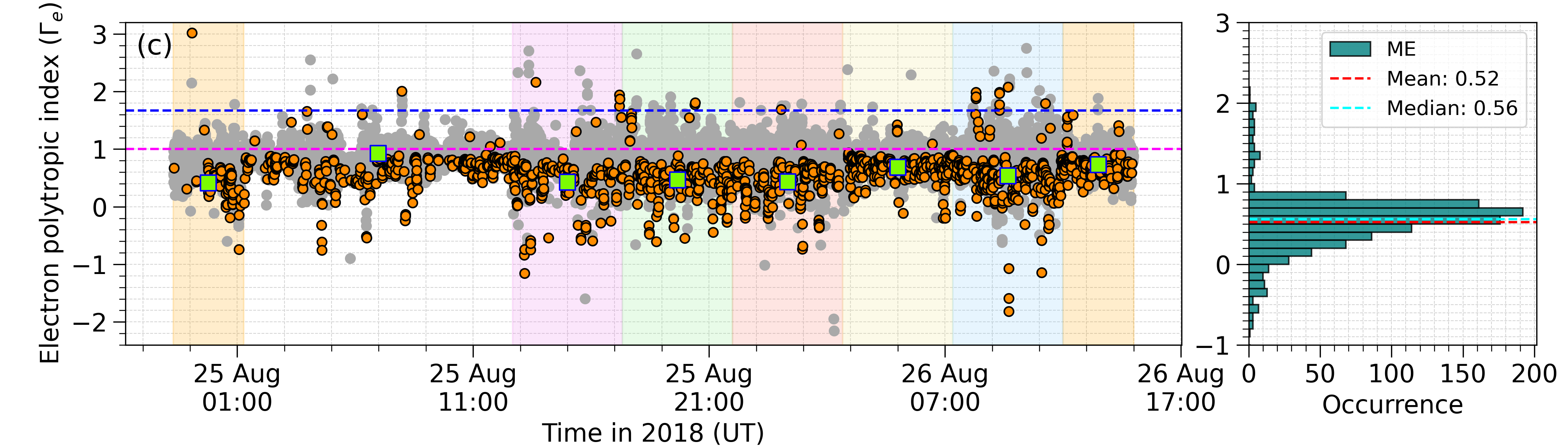}\\
    \includegraphics[width=\textwidth, trim=0 0 0 8, clip]{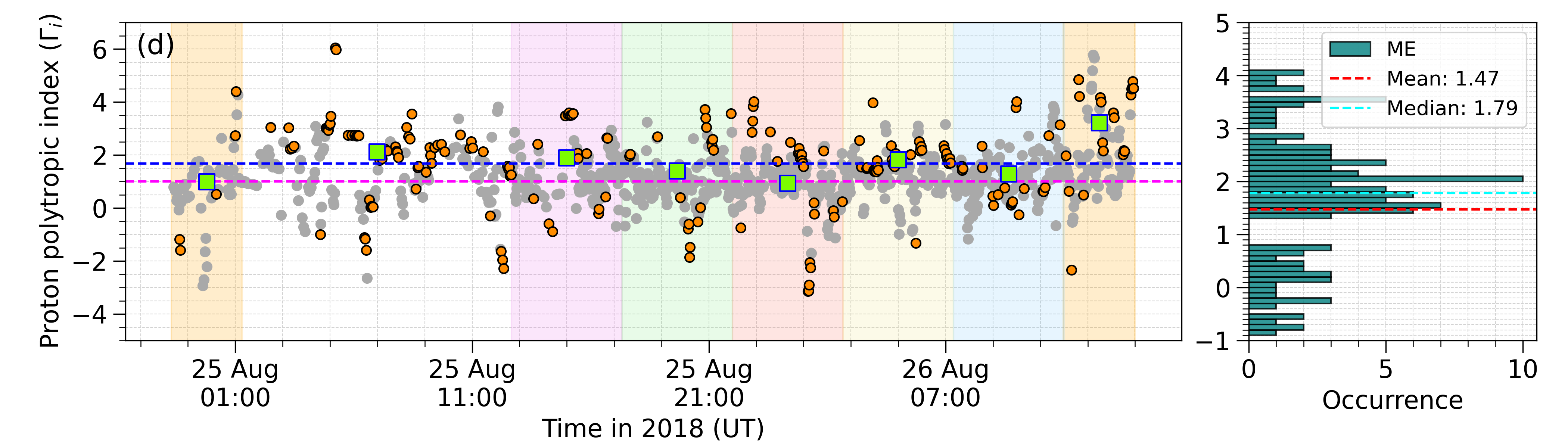}
    \caption[Derived polytropic index ($\Gamma$) values for ICME1]{Derived polytropic index ($\Gamma$) values for ICME1 (top two panels) and ICME2 (bottom two panels). (a) and (c) shows the derived electron polytropic index ($\Gamma_e$) using 9-second resolution electron data, whereas (b) and (d) show the derived proton polytropic index ($\Gamma_p$), using 92-second resolution proton data from Wind/SWE. The orange-shaded regions at the beginning and end represent the pre and post-ICME solar wind. The white-shaded area marks the sheath region, while the multicolored-shaded region between the sheath and post-ICME solar wind corresponds to the magnetic ejecta, divided into five equal parts.}
    \label{chap3-fig:insitu_gamma}
\end{figure*}

Figure~\ref{chap3-fig:insitu_gamma} presents the variations in the polytropic index ($\Gamma$) for both electrons and protons across different regions of the solar wind, pre-ICME, sheath, ME, and post-ICME, for the two events. The background shaded regions represent the structural boundaries of each ICME, as defined earlier. In each panel, the reference lines for the adiabatic ($\Gamma=5/3$) and isothermal ($\Gamma=1$) indices are shown. Both reliable (orange) and unreliable (gray) $\Gamma$ values are plotted, where reliable values satisfy the criteria of correlation coefficient (CC~$>$~0.8) and p-value~$<$~0.5, ensuring statistical confidence in the temperature-density fit. Additionally, mean $\Gamma$ values for each region are marked using green squares, and histograms on the right-hand side summarize the distribution of reliable $\Gamma$ values within the ME.

Panels~(a) and (b) of Figure~\ref{chap3-fig:insitu_gamma} show the electron and proton polytropic indices ($\Gamma_e$ and $\Gamma_p$) for ICME1, respectively. The pre- and post-ICME regions exhibit low mean $\Gamma_e$ values of 0.45 and 0.12, respectively, indicating substantial electron heating in the ambient solar wind surrounding ICME1. Within the sheath and ME, a dominant heating state also persists. Specifically, $\Gamma_e$ gradually decreases across the sheath with an average of 0.79, suggesting moderate electron heating compared to the pre-ICME solar wind. Inside the ME, a further decline in $\Gamma_e$ from the front to the rear indicates progressive electron heating toward the trailing edge. This trend is corroborated by Figure~\ref{chap3-fig:insitu_wind}I(g), which shows a nearly constant electron temperature in Parts 4 and 5 of the ME. The average and median $\Gamma_e$ values within the ME are 0.42 and 0.44, respectively, reinforcing the presence of ongoing heating for electrons.

The observation of $\Gamma_e < 1$ implies a negative correlation between electron temperature and density, a topic that has drawn varying interpretations in past studies. \citet{Osherovich1993} and \citet{Sittler1998} interpret this anti-correlation as indicative of the electron polytropic index in ME plasma. Conversely, \citet{Hammond1996} suggests that such a trend arises from the differing expansion histories of plasma parcels rather than a direct thermodynamic relationship. \citet{Gosling1999} argue that this anti-correlation results from internal structuring within the ME and the plasma’s natural tendency to maintain pressure balance during expansion. In light of these interpretations, we performed a localized correlation analysis between electron density and temperature within

\clearpage

\noindent sub-regions of the ME, rather than over the entire structure, to capture the variability of thermal states more accurately across different spatial intervals.

The thermal behavior of protons shows greater variability than that of electrons. In ICME1, the mean proton polytropic indices ($\Gamma_p$) in the pre- and post-ICME regions are 0.93 and 2.17, respectively. This indicates a heating state for protons in the pre-ICME solar wind and a significant heat-release state post-ICME. Within the sheath region, the mean $\Gamma_p$ is 1.4, pointing to moderate heating that is insufficient to sustain the proton temperature. Immediately following the sheath, Part~1 of the ME exhibits a strong heat-release phase with a mean $\Gamma_p$ of 3.2, consistent with the decline in proton temperature shown in Figure~\ref{chap3-fig:insitu_wind}I(e).

The remaining ME sub-regions, except for Part~5, display a mix of heating and heat-loss states, with $\Gamma_p$ values hovering near the adiabatic index. Part~5 of the ME stands out with a lower mean $\Gamma_p$ of 0.78, indicating a dominant heating state. This is also evident in the nearly constant proton temperature observed in Part~5 (Figure~\ref{chap3-fig:insitu_wind}I(e)). Overall, the mean and median $\Gamma_p$ values across the ME are 1.8 and 1.78, respectively, implying a general trend of heat release, in agreement with the observed decline in proton temperature across ICME1.

Figures~\ref{chap3-fig:insitu_gamma}(c) and (d) show the polytropic indices for electrons ($\Gamma_e$) and protons ($\Gamma_p$), respectively, for ICME2. The mean $\Gamma_e$ values in the pre- and post-ICME2 regions are 0.42 and 0.73, suggesting significant electron heating surrounding the ICME. Within the ME of ICME2, the mean and median $\Gamma_e$ values are 0.52 and 0.56, respectively, confirming sustained electron heating. This thermal state is also reflected in the elevated electron temperatures within the ME compared to the sheath, with only a modest decline observed (Figure~\ref{chap3-fig:insitu_wind}II(h)).

Across ICME2, the proton thermal states appear more mixed. All sub-regions of the ME show $\Gamma_p$ values near the adiabatic index, indicative of both heating and heat-loss processes. The mean and median $\Gamma_p$ within the ME are 1.47 and 1.79, respectively, suggesting minimal heat release and a relatively stable proton temperature profile (Figure~\ref{chap3-fig:insitu_wind}II(e)). Part~5 of the ME has a mean $\Gamma_p$ of 1.28; although this points to mild cooling, an upward trend in temperature is observed, likely due to the region's low proton density, allowing additional heat to be distributed among fewer particles. In the post-ICME solar wind, the mean $\Gamma_p$ reaches 3.2, indicative of rapid heat loss, which could be attributed to compression effects from the trailing corotating interaction region (CIR) and high-speed stream (HSS) \citep{Mirko2020}.

Despite differences in plasma and magnetic parameters between the pre- and post-ICME regions of ICME1 and ICME2 (Table~\ref{chap3-sec2-tab:cme_insitu_table}), their thermal states display similar trends. Both events show electron heating in the surrounding solar wind. While the pre-ICME regions exhibit moderate proton heating, the post-ICME environments reflect strong proton heat release. Within the ME, electron heating is more pronounced in ICME1 than in ICME2, whereas protons experience stronger cooling in ICME1 and a predominantly heating state in ICME2.

\subsection{Connection to The Near-Sun Thermal States}

In the previous section, we investigated the thermal evolution of the CMEs associated with ICME1 and ICME2 using a polytropic framework combined with 3D kinematic data and the analytical FRIS model. In that study, the entire CME was treated as a single thermodynamic structure, with the average temperature defined as \( T = (T_e + T_p)/2 \), and the number density assumed to be equal for electrons and protons (\( n = n_e = n_p \)) under the condition of charge neutrality. To compare our in situ findings at 1~au with those near the Sun, it is necessary to determine an effective polytropic index (\( \Gamma_{\text{eff}} \)) that accounts for the contributions of both electron and proton thermal states within ME.

Based on this framework, the polytropic relation for the ME can be expressed as:
\[
T_e + T_p \propto n^{\Gamma_{\text{eff}} - 1}
\]
Since \( T_e \) and \( T_p \) are governed by their respective polytropic indices \( \Gamma_e \) and \( \Gamma_p \), an effective index can be derived using a temperature-weighted average:
\[
\Gamma_{\text{eff}} \approx \frac{\Gamma_e T_e + \Gamma_p T_p}{T_e + T_p}
\]
This formulation assumes that the thermodynamic influence of each species is proportional to its temperature contribution to the total plasma energy.

Using the mean electron and proton temperatures from the ME (Table~\ref{chap3-sec2-tab:cme_insitu_table}) and the corresponding mean values of \( \Gamma_e \) and \( \Gamma_p \) (Figure~\ref{chap3-fig:insitu_gamma}), we compute:
\begin{itemize}
    \item For ICME1: with \( T_e = 14.2 \times 10^4\,\text{K} \), \( T_p = 7.2 \times 10^4\,\text{K} \), \( \Gamma_e = 0.42 \), and \( \Gamma_p = 1.8 \), we obtain \( \Gamma_{\text{eff}} = 0.88 \).
    \item For ICME2: with \( T_e = 16.5 \times 10^4\,\text{K} \), \( T_p = 5.5 \times 10^4\,\text{K} \), \( \Gamma_e = 0.52 \), and \( \Gamma_p = 1.47 \), we calculate \( \Gamma_{\text{eff}} = 0.76 \).
\end{itemize}

These results indicate that the ME of both ICMEs maintains a near-isothermal state even at 1~au, despite expansion. Notably, a similar thermal state was also reported near the Sun, around \( 20\,R_\odot \) for CME1 and \( 15\,R_\odot \) for CME2, in the previous section (Figures~\ref{fig:Thermo_para_CME1} and \ref{fig:Thermo_para_CME2}). This consistency suggests that CMEs do not undergo purely adiabatic expansion; instead, they experience ongoing heating over large heliocentric distances. The persistence of near-isothermal conditions from the low corona to 1~au highlights the influence of continuous heating processes in governing the thermodynamic evolution of CMEs throughout their propagation.

\subsection{Measured Turbulence Properties at 1AU}

Kinetic processes such as wave-particle interactions and magnetic reconnection are essential in mediating cross-scale energy transfer in magnetized plasmas, linking large-scale dynamics with small-scale dissipation. These mechanisms operate predominantly in collisionless regimes where interactions are governed by collective electromagnetic fields rather than particle collisions. Magnetic flux ropes (MFRs), or magnetic ejecta (ME), are large-scale, organized magnetic structures within the heliosphere that can produce significant geomagnetic disturbances \citep{Gonzalez1999, Tsurutani2011, Echer2013}. However, the turbulence properties within these structures remain poorly understood \citep{Leamon1998, SorrisoValvo2021, Rodriguez2023, Shaikh2024}. In this study, we utilized high-resolution magnetic field measurements from the Wind spacecraft to analyze the turbulence characteristics of the two selected ICMEs that exhibit contrasting kinematic behaviors and interact with distinct solar wind environments.

\begin{figure*}[ht]
    \centering
    \includegraphics[width=\linewidth]{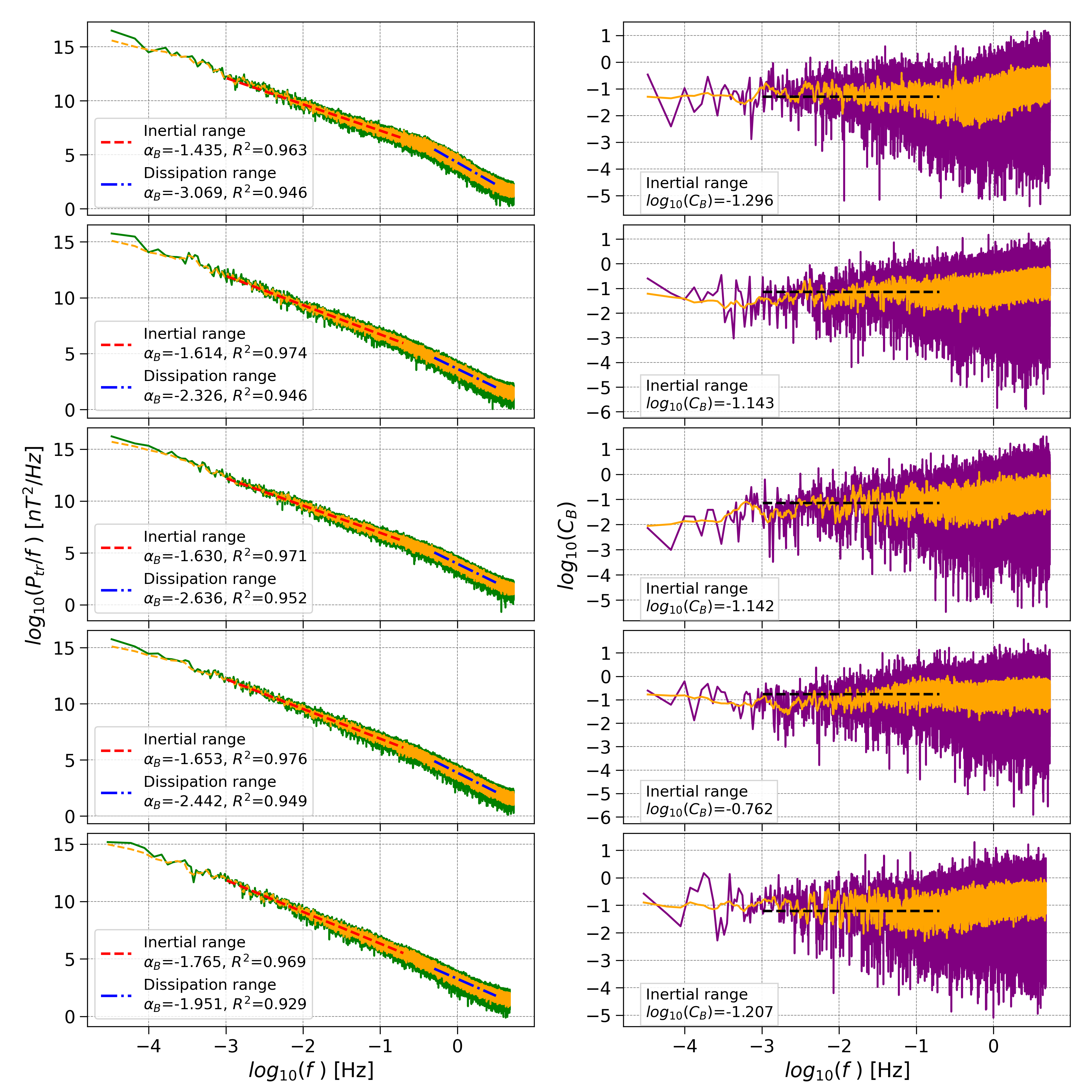}
    \caption[Left panels: variation of trace power spectra ($P_{tr}$) per]{Left panels: variation of trace power spectra ($P_{tr}$) per frequency across five parts in ME of ICME1. The yellow curve represents the 3-point sliding average of $P_{tr}$. We have overplotted the linear fitted dash lines (red and blue) for both inertial ($-3 < \log_{10} f < -0.7 $) and dissipation ($-0.3 < \log_{10} f < 0.5 $) scales along with the corresponding spectral slope ($\alpha_B$) and goodness of fit ($R^2$) are mentioned. Right panels: variation of magnetic compressibility factor ($C_B$) with frequency across five parts in ME of ICME1.  The yellow curve represents 10-point sliding average values of $C_B$. We have overlaid a black dashed line showing the average value of $C_B$ over the inertial scale.}
    \label{chap3-fig:fast_five_me}
\end{figure*}

\begin{figure*}[ht]
    \centering
    \includegraphics[width=0.32\linewidth]{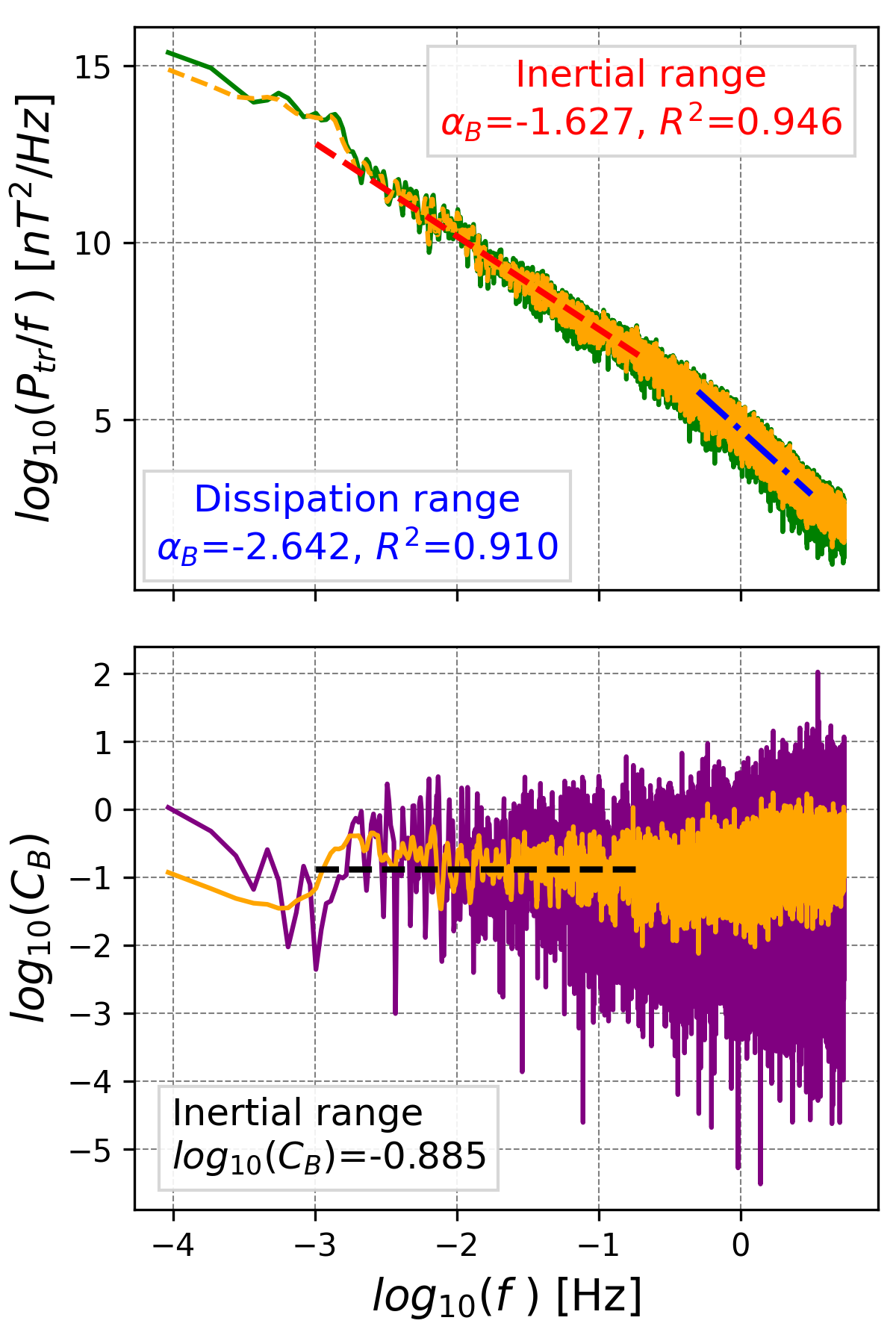}
    \includegraphics[width=0.32\linewidth]{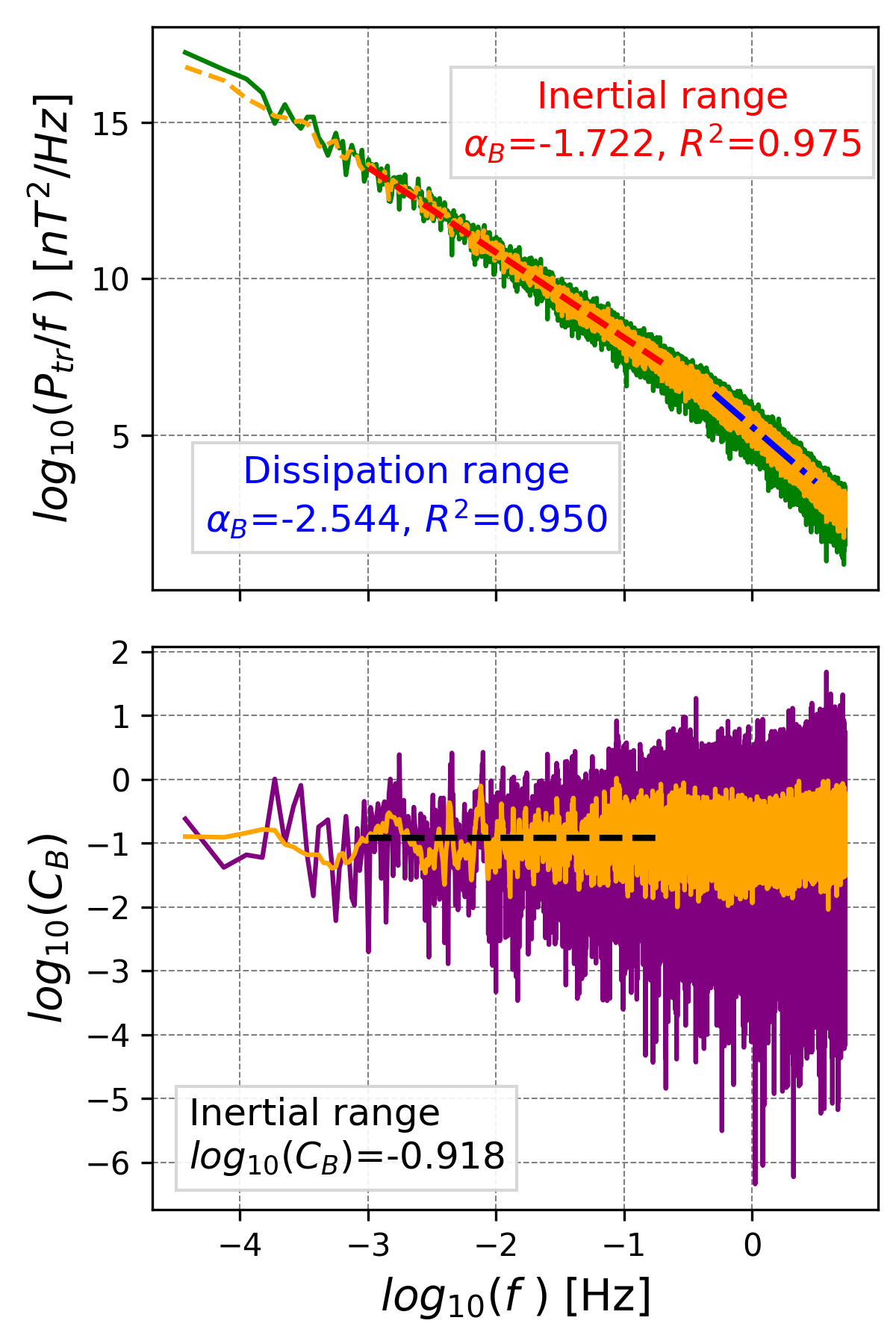}
    \includegraphics[width=0.32\linewidth]{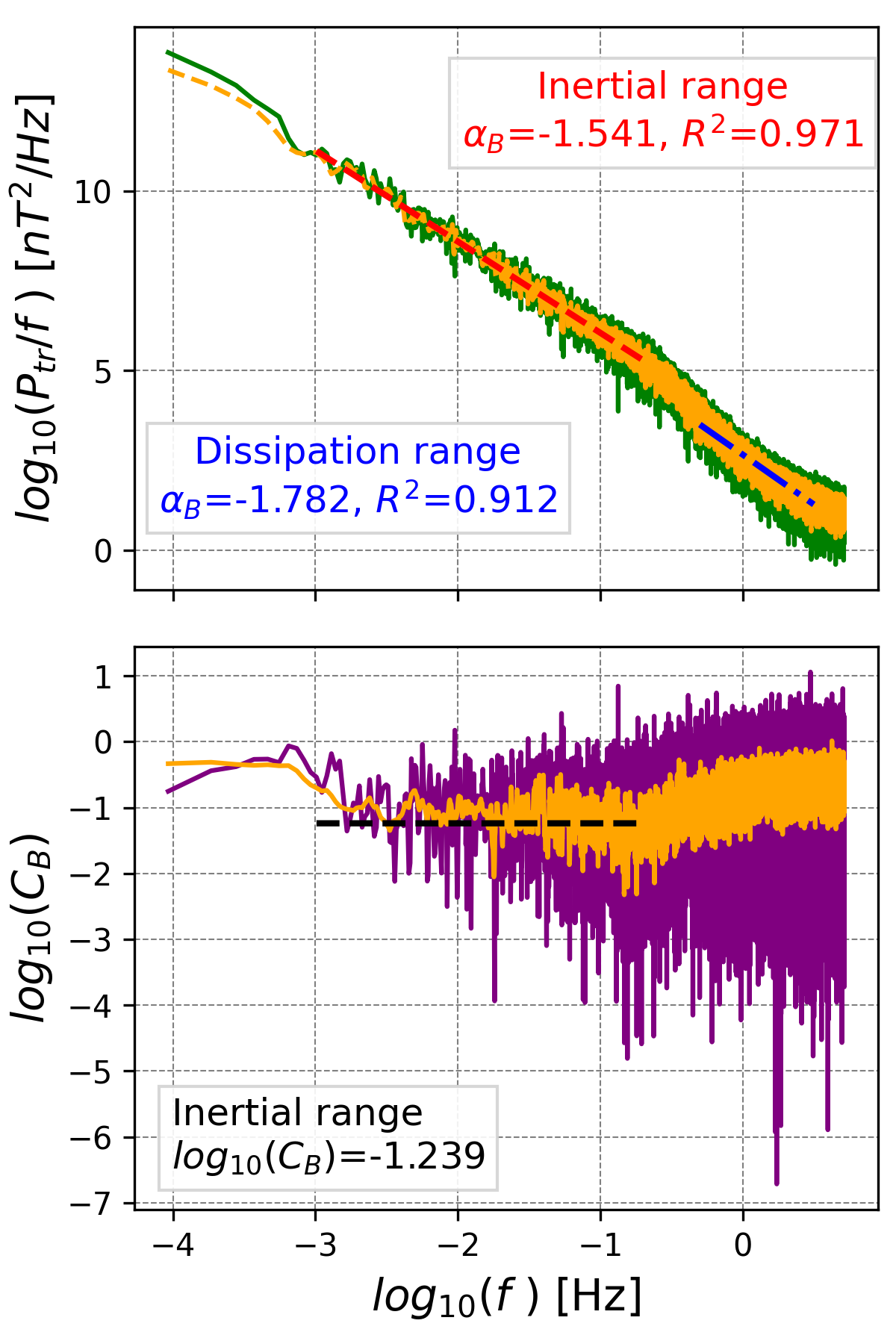}
    \caption[Variation of trace power spectra ($P_{tr}$) per]{Variation of trace power spectra ($P_{tr}$) per frequency (top panels) and magnetic compressibility factor ($C_B$) with frequency (bottom panels) across pre-ICME (left panels), sheath (middle panels) and post-ICME (right panels) regions of ICME1. The top panels include a 3-point sliding average (yellow curve) and linear fits (dashed red and blue lines) for the inertial ($-3 < \log_{10} f < -0.7$) and dissipation ($-0.3 < \log_{10} f < 0.5$) scales, with the spectral slope ($\alpha_B$) and goodness of fit ($R^2$) provided. The bottom panels show 10-point sliding averages of $C_B$ (yellow curve), with a black dashed line indicating the average $C_B$ over the inertial scale.}
    \label{chap3-fig:fast_spec_pre_post}
\end{figure*}

\begin{figure*}[hbt]
    \centering
    \includegraphics[width=\linewidth]{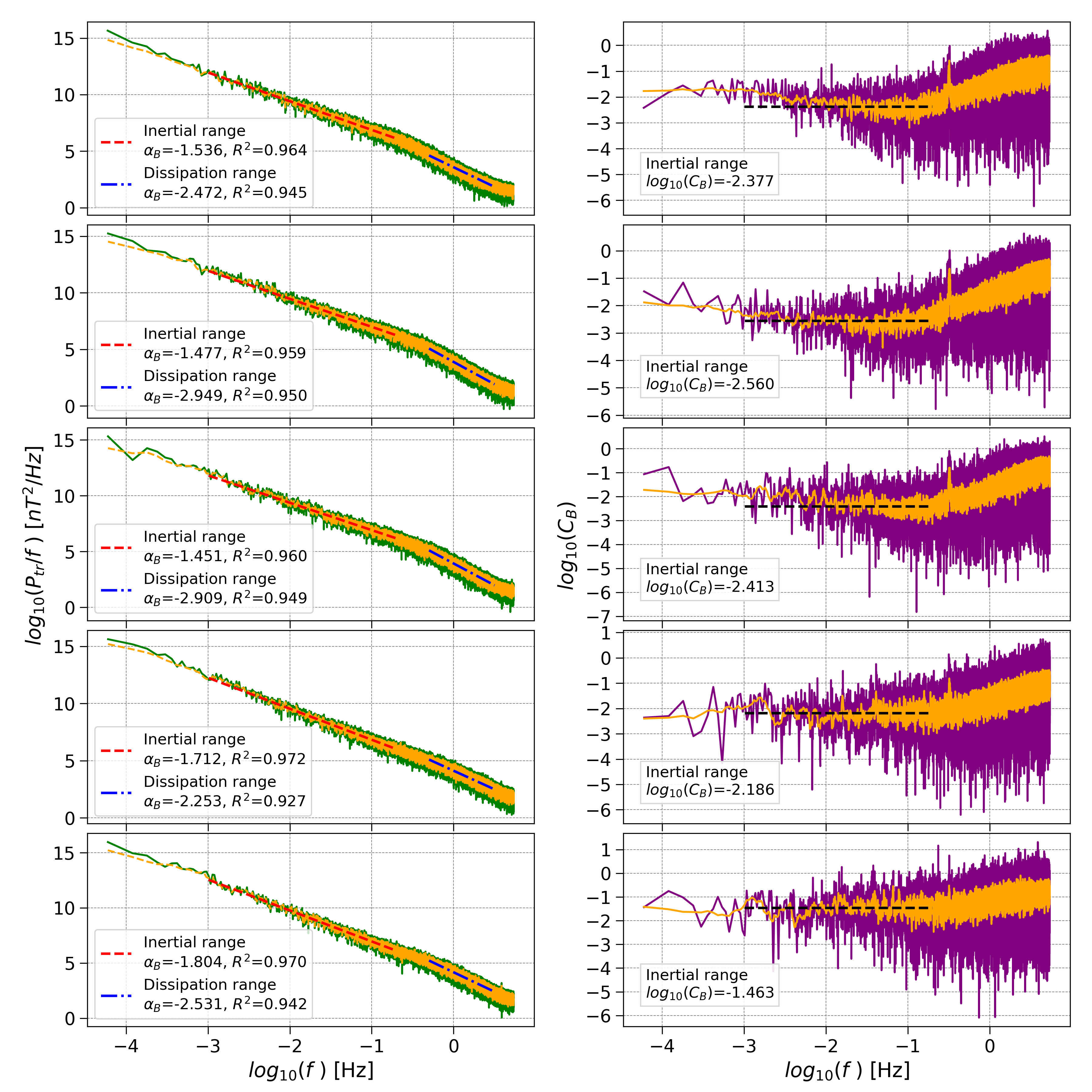}
    \caption[Left panels: variation of trace power spectra ($P_{tr}$) per]{Left panels: variation of trace power spectra ($P_{tr}$) per frequency across five parts in ME of ICME2. The yellow curve represents the 3-point sliding average of $P_{tr}$. We have overplotted the linear fitted dash lines (red and blue) for both inertial ($-3 < \log_{10} f < -0.7 $) and dissipation ($-0.3 < \log_{10} f < 0.5 $) scales along with the corresponding spectral slope ($\alpha_B$) and goodness of fit ($R^2$) are mentioned. Right panels: variation of magnetic compressibility factor ($C_B$) with frequency across five parts in ME of ICME2.  The yellow curve represents 10-point sliding average values of $C_B$. We have overlaid a black dashed line showing the average value of $C_B$ over the inertial scale.}
    \label{chap3-fig:slow_five_me}
\end{figure*}

\begin{figure*}[ht]
    \centering
    \includegraphics[width=0.32\linewidth]{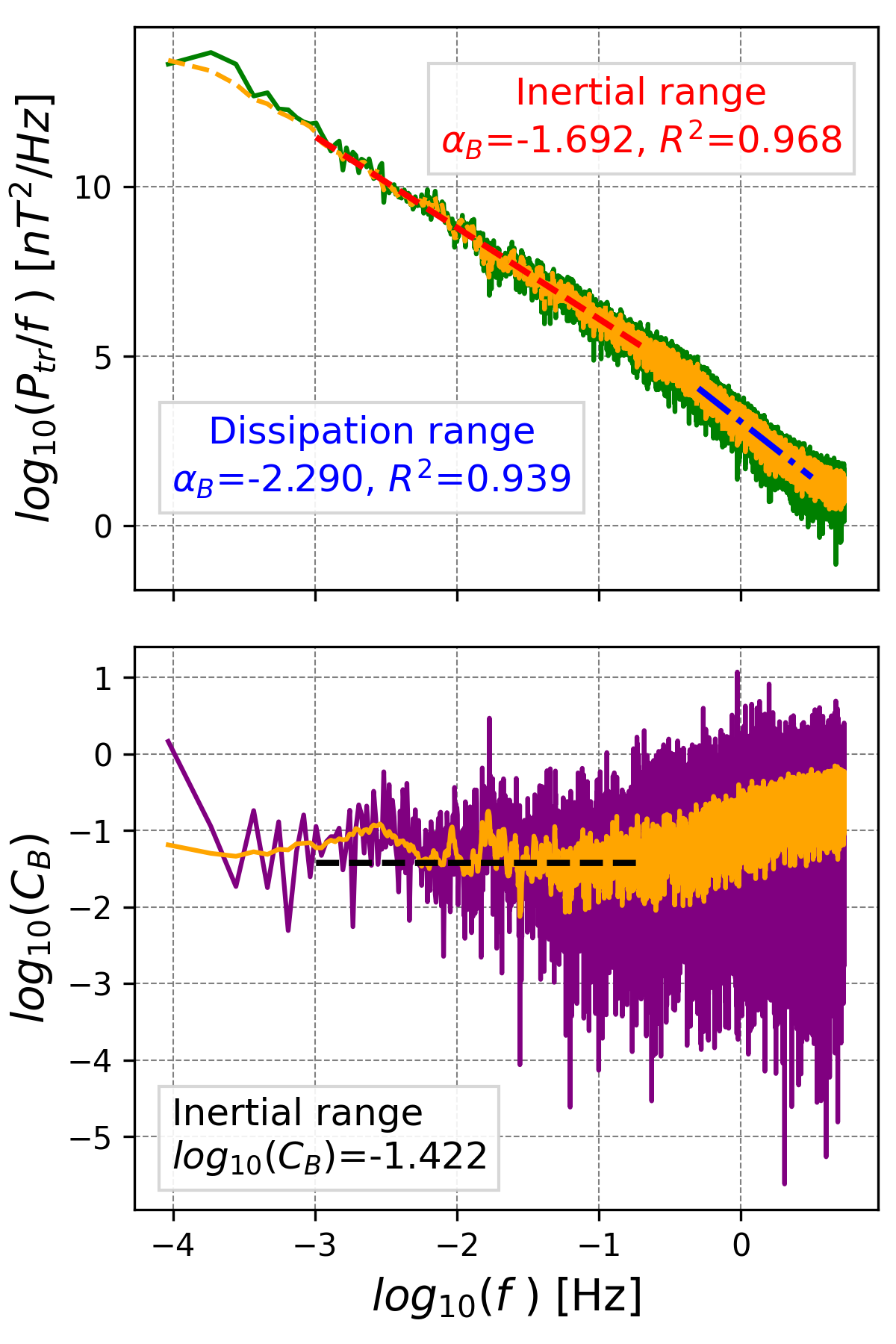}
    \includegraphics[width=0.32\linewidth]{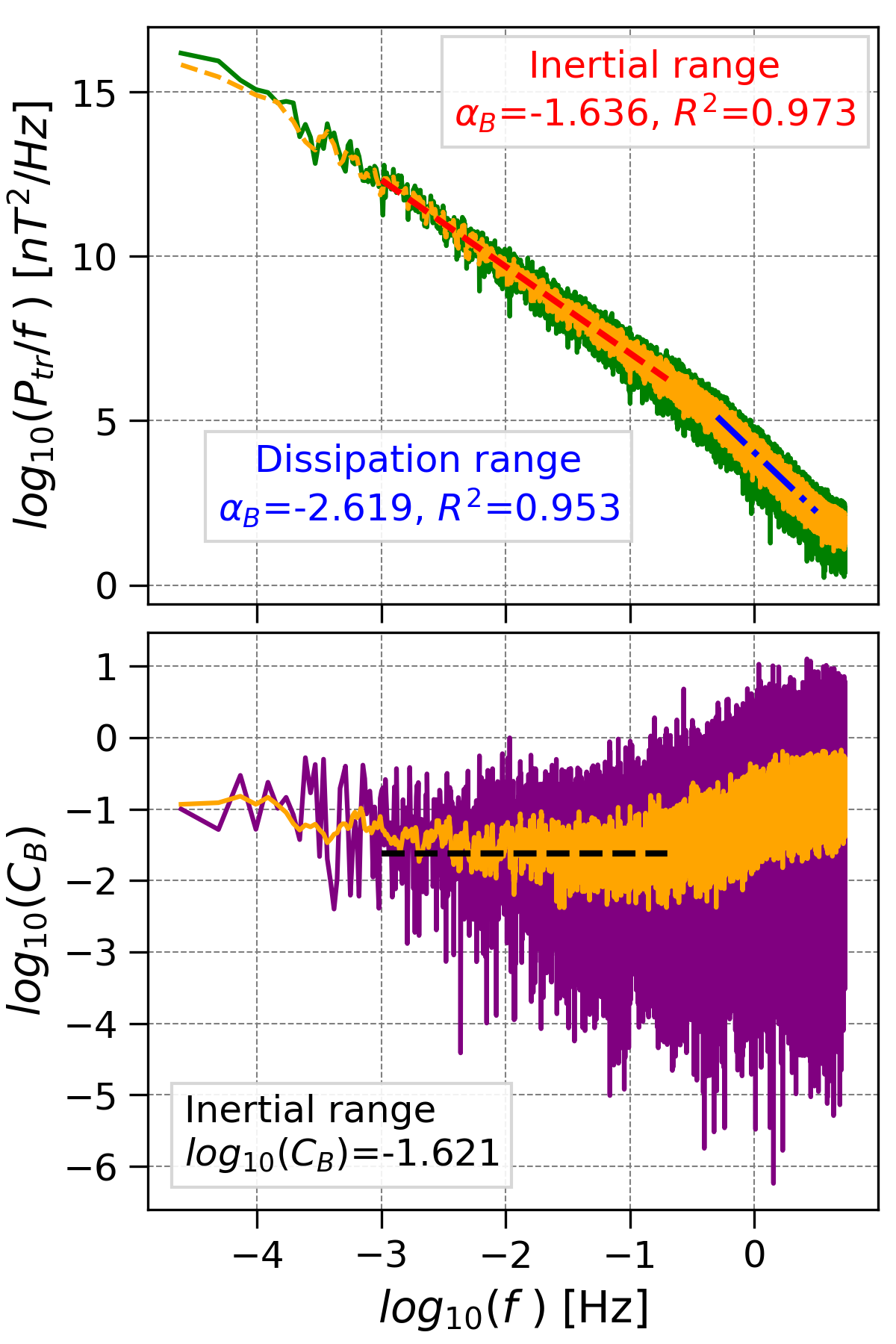}
    \includegraphics[width=0.32\linewidth]{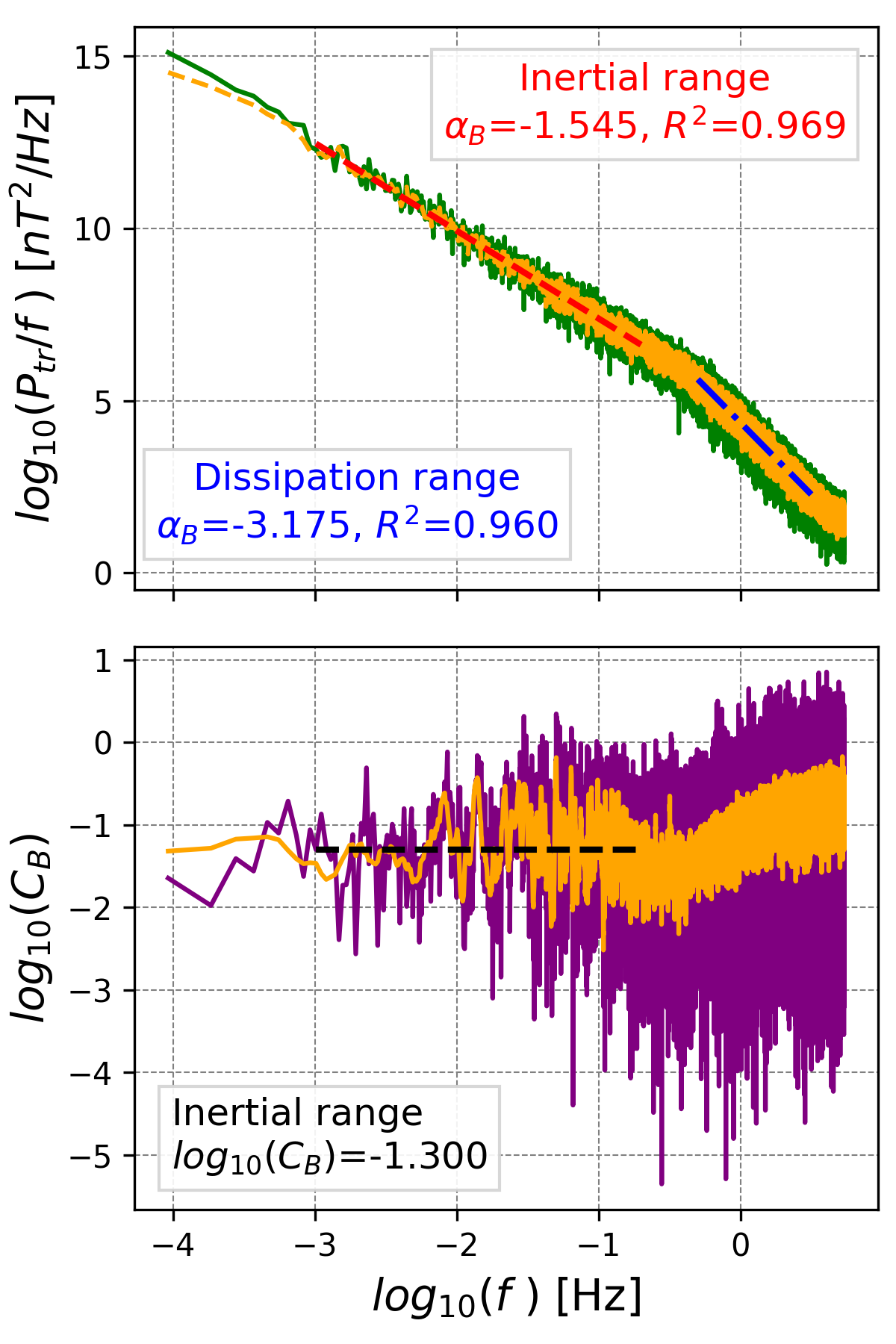}
    \caption[Variation of trace power spectra ($P_{tr}$) per]{Variation of trace power spectra ($P_{tr}$) per frequency (top panels) and magnetic compressibility factor ($C_B$) with frequency (bottom panels) across pre-ICME (left panels), sheath (middle panels) and post-ICME (right panels) regions of ICME2. The top panels include a 3-point sliding average (yellow curve) and linear fits (dashed red and blue lines) for the inertial ($-3 < \log_{10} f < -0.7$) and dissipation ($-0.3 < \log_{10} f < 0.5$) scales, with the spectral slope ($\alpha_B$) and goodness of fit ($R^2$) provided. The bottom panels show 10-point sliding averages of $C_B$ (yellow curve), with a black dashed line indicating the average $C_B$ over the inertial scale. }
    \label{chap3-fig:slow_spec_pre_post}
\end{figure*}

\begin{figure*}[ht]
    \centering
    \includegraphics[width=\textwidth]{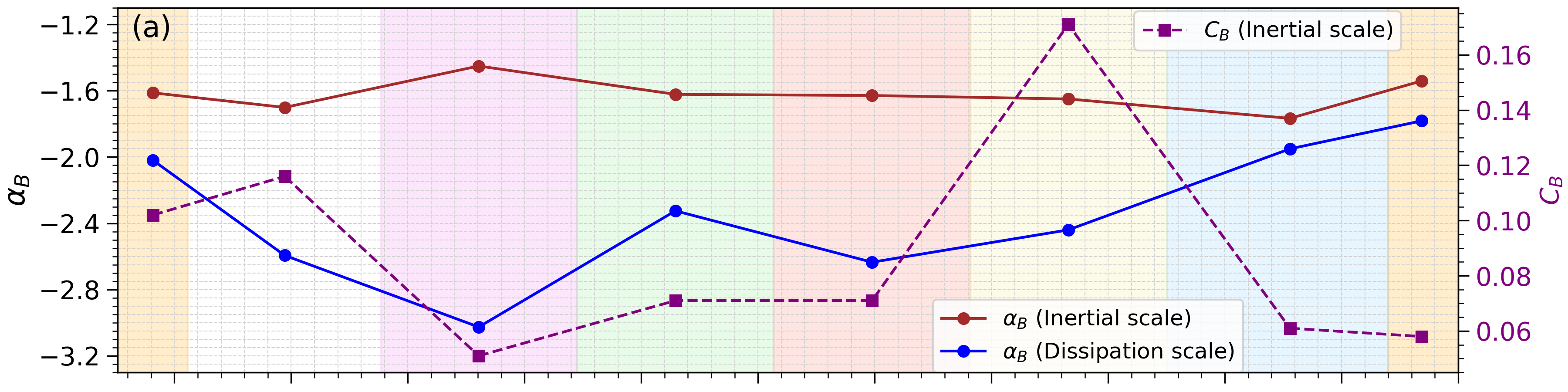}\\ 
    \vspace{-0.1cm}
    \includegraphics[width=\textwidth]{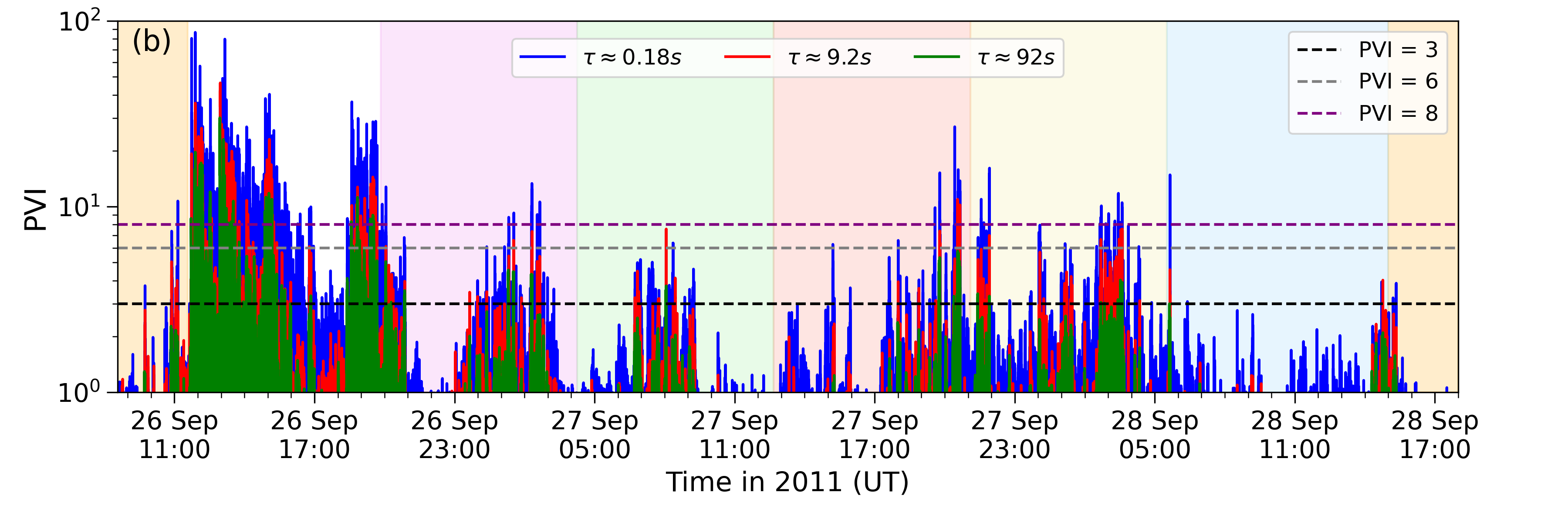}\\
    \includegraphics[width=\textwidth]{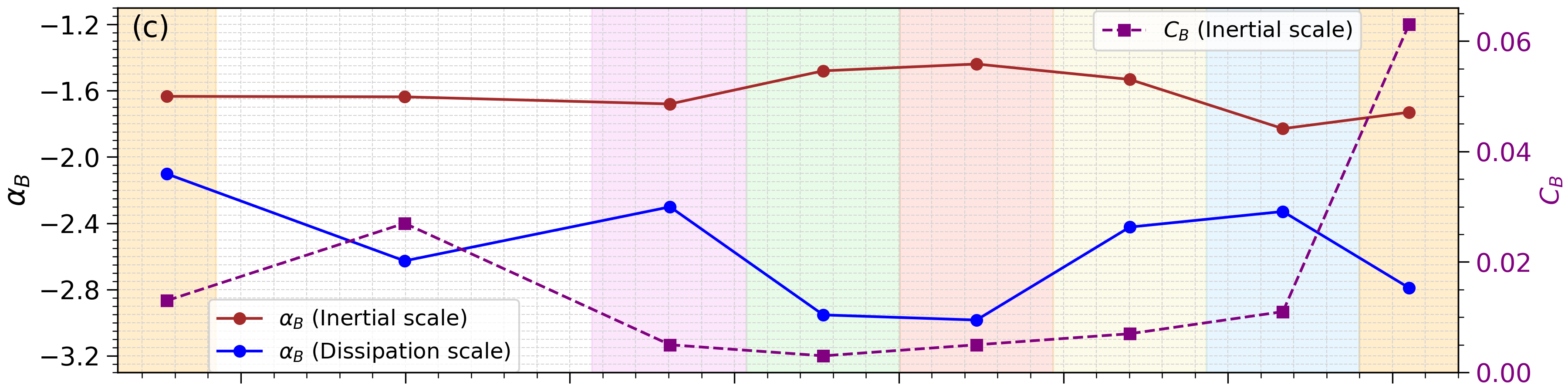}\\
    \vspace{-0.1cm}
    \includegraphics[width=\textwidth]{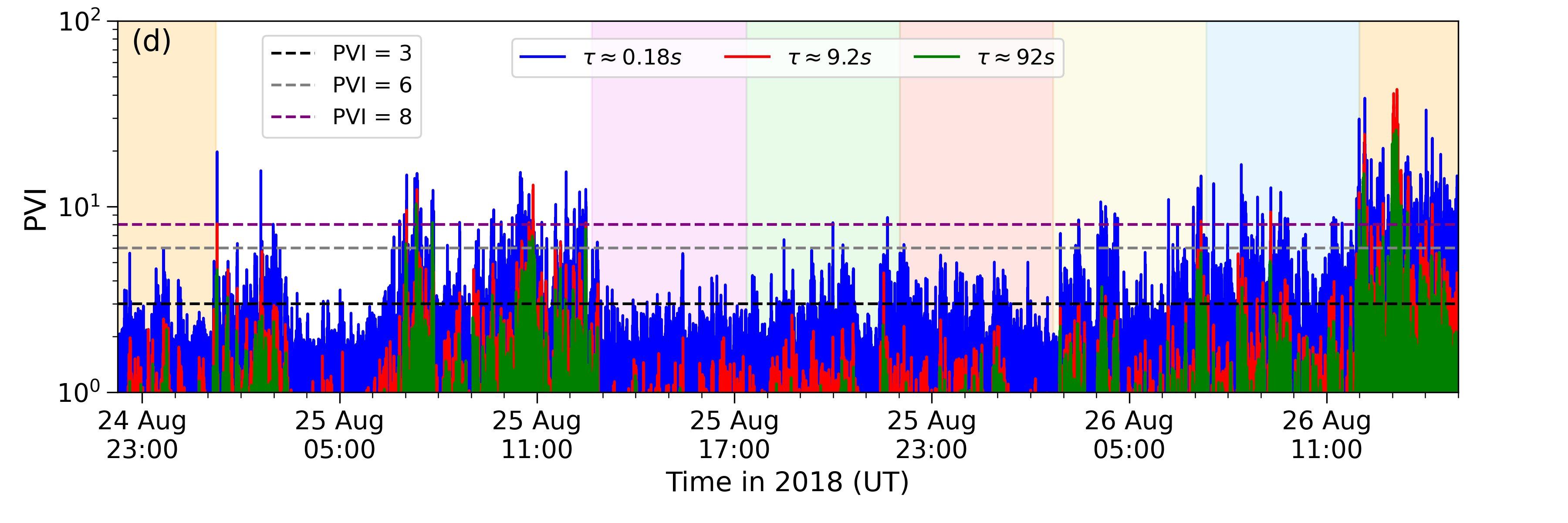}
    \caption[Spectral slope ($\alpha_B$), and magnetic compressibility factor ($C_B$)]{Spectral slope ($\alpha_B$), and magnetic compressibility factor ($C_B$) for ICME1 (a) and ICME2 (c), derived using high-resolution Wind/MFI magnetic data. The derived PVI values for ICME1 (c) and ICME2 (d). The orange-shaded regions at the beginning and end represent the pre and post-ICME solar wind. The white-shaded area marks the sheath region, while the multicolored-shaded region between the sheath and post-ICME solar wind corresponds to the magnetic ejecta, divided into five equal parts. }
    \label{chap3-fig:slope_cb_pvi}
\end{figure*}

To quantify turbulence, we performed linear fits to the three-point sliding average of the trace power spectral density, \( P_{\text{tr}} \), in two frequency regimes: the inertial range (\(-3 < \log_{10} f < -0.7\)) and the dissipation range (\(-0.3 < \log_{10} f < 0.5\)). From these, we extracted the spectral slope (\( \alpha_B \)) and magnetic compressibility (\( C_B \)). The resulting spectral fits for different ICME regions are illustrated in Figures~\ref{chap3-fig:fast_five_me}, \ref{chap3-fig:slow_five_me}, \ref{chap3-fig:fast_spec_pre_post}, and \ref{chap3-fig:slow_spec_pre_post}. Figure~\ref{chap3-fig:slope_cb_pvi} presents the computed values of \( \alpha_B \), \( C_B \), and partial variance of increments (PVI) across various color-shaded regions for both ICME events. The shaded backgrounds in the figure correspond to the regional subdivisions defined in the previous sections. In particular, panels~\ref{chap3-fig:slope_cb_pvi}a and~\ref{chap3-fig:slope_cb_pvi}c display the spectral slopes \( \alpha_B \) in the inertial range (brown dotted lines) and dissipation range (blue dotted lines), along with the mean magnetic compressibility \( C_B \) values calculated within the inertial scale for each region.

In the inertial range, the spectral slope \( \alpha_B \) is approximately $-1.6$ in the pre-ICME solar wind and around $-1.7$ in the sheath region of ICME1 (Figure~\ref{chap3-fig:slope_cb_pvi}a). These values are indicative of fully developed Kolmogorov-type turbulence \citep{Kolmogorov1941} in both regions. A steeper \( \alpha_B \) suggests a more efficient transfer of energy toward smaller scales, signifying enhanced turbulence in the sheath compared to the pre-ICME solar wind. Within the magnetic ejecta (ME), all parts except Part~1 show \( \alpha_B \approx -1.6 \), consistent with typical magnetic flux-rope turbulence spectra at 1~au \citep{Borovsky2019, Good2023}. In contrast, Part~1 of the ME and the post-ICME solar wind exhibit shallower slopes, with \( \alpha_B \approx -1.5 \), suggesting a turbulence regime closer to the Iroshnikov–Kraichnan (IK) model \citep{Iroshnikov1963, Kraichnan1965}, indicative of less-developed or evolving turbulence \citep{Bruno2013}.

At dissipation scales, the turbulent cascade reaches its termination point, and energy is converted into heat. The spectral slope \( \alpha_B \) in this regime becomes progressively steeper from the pre-ICME region through the sheath and into Part~1 of the ME in ICME1 (Figure~\ref{chap3-fig:slope_cb_pvi}a). This trend indicates increasingly efficient dissipation of turbulent energy at smaller scales, with Part~1 of the ME exhibiting the highest dissipation efficiency, followed by the sheath and pre-ICME regions. Beyond Part~1, \( \alpha_B \) gradually becomes shallower in the later parts of the ME and post-ICME solar wind, suggesting a decline in the efficiency of energy dissipation. This observation aligns with the decreasing trends in proton and electron temperatures across these regions, as seen in Figures~\ref{chap3-fig:insitu_wind}I(e) and \ref{chap3-fig:insitu_wind}(g).

The magnetic compressibility, defined as \( C_B = P_t / P_{\text{tr}} \), where \( P_t \) is the power spectral density of the magnetic field magnitude and \( P_{\text{tr}} \) is the trace of the magnetic field vector spectral density \citep{Bavassano1982}, provides insight into the nature of magnetic fluctuations. A low compressibility value (\( C_B \ll 1 \)) implies predominantly directional, non-compressive Alfvénic fluctuations \citep{Bruno1993, Damicis2015}. In ICME1, the pre-ICME region shows a higher \( C_B \) compared to the post-ICME region (Figure~\ref{chap3-fig:slope_cb_pvi}a). Among ME substructures, Part~4 exhibits the highest \( C_B \), indicating a relatively higher presence of compressive fluctuations. However, the ME as a whole shows lower \( C_B \) values than the sheath and pre-ICME regions, highlighting the dominance of Alfvénic turbulence within the magnetic ejecta \citep{Telloni2021}.

\clearpage

In the inertial range, the spectral index \( \alpha_B \) is approximately $-1.6$ for both the pre-ICME solar wind and the sheath of ICME2, indicating the presence of fully developed Kolmogorov-type turbulence (Figure~\ref{chap3-fig:slope_cb_pvi}c). Within the ME, the spectral slope becomes shallower toward the central parts compared to the leading and trailing edges, suggesting less developed turbulence. This implies an inefficient energy cascade from the injection to the dissipation scale in the central region of the ME. Conversely, in the dissipation range, the central part of the ME exhibits a steeper \( \alpha_B \) compared to its edges, indicating enhanced dissipation of turbulent energy exhibits a steeper \( \alpha_B \) compared to its edges, indicating enhanced dissipation of turbulent energy. These findings are consistent with the results of \citet{Riazantseva2019}, who reported that MCs typically exhibit a shallower slope at the inertial scale and a steeper slope at the dissipation scale compared to the broader ME. This distinction aligns well with our observations for ICME2 (MC-like) and ICME1 (ME-dominated).

Furthermore, the magnetic compressibility factor \( C_B \) is lower in the ME compared to the surrounding sheath and post-ICME regions. A higher \( C_B \) in the post-ICME region suggests the presence of compressive structures, such as slow magnetosonic waves, shocks, and density compressions, which can facilitate efficient energy transfer. This interpretation is also supported by the steeper dissipation-scale \( \alpha_B \) observed in the post-ICME region of ICME2. These variations in spectral characteristics between the center and edges of the ME highlight the spatial heterogeneity in turbulence and dissipation, governed by local plasma and magnetic field conditions.

Proton temperature anisotropies can trigger kinetic instabilities such as proton cyclotron or mirror-mode instabilities, which can significantly influence dissipation mechanisms \citep{Gary1993, Denton1994, Verscharen2019}. Higher anisotropies typically enhance dissipation, resulting in steeper spectral slopes in the dissipation range. In ICME1, the sheath exhibits greater proton temperature anisotropy and correspondingly steeper \( \alpha_B \), consistent with stronger dissipation. While the sheath and ME of ICME2 show moderate and comparable proton anisotropies, the ME still exhibits a steeper \( \alpha_B \) in the dissipation scale, likely due to the lower \( C_B \) values.

In Part~1 of the ME in ICME1, both proton and electron anisotropies are lower (Figure~\ref{chap3-fig:insitu_wind}I(e, g)); however, the dissipation remains strong, as reflected in the steep \( \alpha_B \) (Figure~\ref{chap3-fig:slope_cb_pvi}(a)). In ICME2, Part~1 exhibits reduced \( T_\perp \), while Part~5 shows reduced \( T_\parallel \) for protons (Figure~\ref{chap3-fig:insitu_wind}II(e)). Both regions demonstrate shallow dissipation-scale \( \alpha_B \) and low magnetic compressibility (Figure~\ref{chap3-fig:slope_cb_pvi}c). The reduced \( T_\perp \) in Part~1 may indicate diminished perpendicular heating or damping mechanisms like cyclotron resonance or whistler wave interactions \citep{Hellinger2000}, while the lower \( T_\parallel \) in Part~5 could point to reduced parallel heating via Landau damping or more efficient mirror-mode wave activity \citep{AlaLahti2019}. These contrasts suggest region-specific dominance of different dissipation processes.

Overall, the ME of ICME1 exhibits higher proton and electron temperature aniso\-tropies, along with higher \(C_B\) values and shallower dissipation scale \(\alpha_B\), compared to ICME2. This indicates that while ICME1 experiences stronger anisotrop\-ies, its dissipation mechanisms may be less efficient than those operating within ICME2's ME.

\subsection{Measured Intermittencies at 1 au}

In this study, we employed the Partial Variance of Increments (PVI) method to investigate intermittent features in magnetic field fluctuations. PVI is a well-established metric for identifying sharp magnetic field gradients such as discontinuities, current sheets, and potential reconnection sites \citep{Greco2018}. These localized structures are believed to be critical for enhanced energy dissipation, particle heating, and energization in space plasmas \citep{Tessein2013, Chasapis2015}. We analyzed the variation of PVI values in relation to the spectral slope ($\alpha_B$) in both inertial and dissipation scales, as well as with the magnetic compressibility factor ($C_B$). Figures~\ref{chap3-fig:slope_cb_pvi}b and \ref{chap3-fig:slope_cb_pvi}d display the PVI results for various regions of ICME1 and ICME2, respectively.

PVI was calculated using three distinct time lags ($\tau$): 0.18~s, 9.2~s, and 92~s, to capture structures at both inertial and dissipation scales. Horizontal reference lines at PVI thresholds of 3, 6, and 8 are overlaid. Transitions from Gaussian to non-Gaussian behavior in magnetic field increments typically occur around PVI~$\approx$~3 \citep{Osman2011a, Osman2011b}, with higher values (PVI~$>$~6 and 8) corresponding to strong intermittencies, including current sheets and possible reconnection sites \citep{Rohit2020}. All three $\tau$ values identify numerous non-Gaussian events (PVI~$>$~3), though smaller $\tau$ are more sensitive to finer-scale features. In ICME1, the sheath, along with Part~3 and Part~4 of the ME, shows elevated PVI values compared to other regions (Figure~\ref{chap3-fig:slope_cb_pvi}b). Similarly, for ICME2, higher PVI values are observed in the sheath, Parts~4 and 5 of the ME, and the post-ICME solar wind (Figure~\ref{chap3-fig:slope_cb_pvi}d). These findings are consistent with previous case studies \citep{Kilpua2020, Good2020}, which reported enhanced intermittency and compressibility in ICME sheath regions. In contrast, the pre-ICME solar wind exhibits fewer intermittent structures, aligning with the statistical analysis by \citet{Kilpua2021}.

We also find a qualitative correlation between regions of high $C_B$ and high PVI, suggesting a relationship between compressive magnetic field fluctuations and intermittent structures. Since $C_B$ reflects the degree of compressibility in magnetic fluctuations, i.e., how strongly field variations are associated with density perturbations, its positive association with PVI implies that strong gradients and localized structures coincide with compressive behavior. This emphasizes the interconnected role of intermittency and compressibility in mediating turbulence and dissipation in ICMEs. As shown in Figures~\ref{chap3-fig:slope_cb_pvi}(b) and (d), small $\tau$ values are more effective at capturing fine-scale, intermittent events, while longer lags (e.g., $\tau$ = 92~s) yield more smoothed, averaged behavior. \citet{Osman2011a} demonstrated that higher PVI values are associated with elevated proton and electron temperatures and heat fluxes, highlighting the role of intermittency in plasma heating. Their follow-up work \citep{Osman2011b} showed that PVI values are maximized in plasma conditions with extreme parallel beta and proton temperature anisotropy, consistent with our observation of enhanced anisotropy and PVI in the sheaths of both ICMEs.

Comparative analysis of the sheaths reveals that ICME1 exhibits stronger intermittency than ICME2 (Figure~\ref{chap3-fig:slope_cb_pvi}(b, d)), alongside higher mean heating states (Figure~\ref{chap3-fig:insitu_gamma}). Notably, in ICME1’s ME, regions with elevated PVI coincide with stronger heat release (lower $\Gamma_p$), suggesting that intermittency contributes significantly to localized heating. This correlation is less evident in ICME2’s ME, where intermittency and thermal state appear more decoupled. Overall, turbulence and energy dissipation vary significantly across different regions of the ICMEs and their surrounding solar wind. Sheath regions, in particular, exhibit stronger turbulence, compressibility, and intermittency than the ME. The observed relationship between PVI and $C_B$ underscores the coupling between localized intermittent structures and compressive plasma dynamics, emphasizing the importance of such interactions in the overall energy cascade and dissipation processes in ICMEs and interplanetary space.

\subsection{Summary}\label{chap3-sec2:summary}

This study analyzed the thermal and turbulence properties of two ICMEs, ICME1 (fast) and ICME2 (slow), using in situ measurements from the Wind spacecraft at 1 au. Our primary objective was to investigate how thermal states and turbulence properties vary within ICME plasma, and how these are influenced by the surrounding solar wind environment. Key findings are summarized below:

\begin{enumerate}
    \item ICME1 exhibited stronger electron heating, particularly in the ME and sheath, whereas ICME2 showed more pronounced proton heating within the ME. In ICME1, both electron and proton temperatures decreased through the ME, while ICME2 maintained relatively constant temperatures in the leading parts, followed by a decline in the trailing region.

    \item Near-isothermal thermal states were observed both near the Sun (15--20~$R_\odot$) and at 1 au, indicating sustained and efficient heating mechanisms despite the expected cooling from expansion.

    \item At inertial scales, turbulence in ICME1 followed near-Kolmogorov behavior ($\alpha_B \approx -5/3$) across all regions, while ICME2’s ME core exhibited shallower slopes, suggesting underdeveloped turbulence.

    \item At dissipation scales, ICME1 showed a decreasing trend in $\alpha_B$ from the leading to trailing ME, indicating diminishing dissipation. In contrast, ICME2’s ME center showed enhanced dissipation compared to the outer parts.

    \item Enhanced proton temperature anisotropy in the sheath correlated with steeper $\alpha_B$ at dissipation scales, consistent with kinetic instabilities contributing to local energy dissipation.

    \item High PVI values, particularly in sheath regions, aligned with elevated electron and proton temperatures, indicating localized heating and strong intermittency.

    \item A qualitative correlation was observed between high PVI and high magnetic compressibility ($C_B$), suggesting that regions of strong intermittency are also highly compressive and dynamically active.

    \item While thermal states of the pre- and post-ICME solar wind were similar for both events, turbulence characteristics differed. ICME1’s ambient medium showed fewer intermittencies, whereas ICME2’s post-ICME region was more intermittent and dissipative.
\end{enumerate}

These results emphasize the complex interplay between CME internal dynamics and external solar wind conditions. The contrasting behavior of ICME1 and ICME2 illustrates the need to account for both intrinsic properties and environmental solar wind context to better understand and predict ICME evolution.

\section{Chapter Conclusion} \label{Ch3-sec:conclusion}

In this chapter, we present a comprehensive investigation of the thermodynamic and dynamic evolution of two CME–ICME events, combining remote-sensing thermodynamic modeling near the Sun with in situ measurements at 1 au. By analyzing a fast CME (CME1) and a slower CME (CME2), the FRIS-based modeling in the near-Sun domain revealed that CME1 undergoes early heat release during its impulsive deceleration, while CME2 absorbs heat during gradual acceleration. Both CMEs eventually tend toward an isothermal state, with CME1 reaching this condition earlier and at lower heliocentric distances. The dominant forces driving expansion (centrifugal and thermal pressure) were counteracted by the Lorentz force, with their relative contributions varying across radial distances and depending on the CME speed. These insights emphasize the necessity of accounting for continuous heating or cooling processes in CME propagation models, challenging the traditional adiabatic assumption.

At 1 au, in situ analysis further confirmed the persistence of non-adiabatic behavior. ICME1 exhibited stronger electron heating and developed turbulence with Kolmogorov-like spectra, while ICME2 displayed more prominent proton heating within a relatively less turbulent magnetic ejecta. Despite similar near-isothermal conditions derived at coronal and 1 au distances, the nature and distribution of heating and dissipation varied markedly between the events. Intermittency analysis using PVI indicated localized, intense energy dissipation sites, particularly in the sheath and trailing ME regions, where sharp magnetic gradients and enhanced plasma compressibility were prevalent. The connection between temperature anisotropy, spectral slope steepening at dissipation scales, and intermittency reinforces the role of kinetic instabilities in modulating energy transfer. Importantly, our results demonstrate that the ambient solar wind environment, such as high-speed streams or CIRs, modulates the turbulence and thermodynamic properties of ICMEs, especially in the post-ejection phases.

Together, these findings bridge the inner coronal and heliospheric evolution of CMEs, offering insights into how thermal processes evolve with distance. The dual-case approach highlights the diverse pathways CMEs follow depending on their kinematics and interaction with the surrounding plasma, underlining the importance of integrated modeling and multi-point observations in constraining CME thermodynamics. These results have direct implications for understanding the energy budget and space weather impact of CMEs and motivate future investigations using coordinated \textit{in situ} and remote observations from missions.

\newpage

\lhead{\emph{Chapter 4}}  

\quotechapter{4}{`` To understand is to perceive patterns.''\\ \smallskip — Isaiah Berlin}{Thermal Evolution - Statistics}

\lhead{\emph{Chapter 4: Thermal Evolution - Statistics}}

\section{Introduction} \label{Ch4-sec:intro}

Several energetic processes on the Sun induce both short and long-term fluctuations in our geospace conditions \citep{Schwenn2006, Baker2009}. Among these, CMEs stand out as some of the most dynamic and geoeffective solar phenomena, capable of significantly disturbing the heliosphere and initiating intense geomagnetic storms upon interaction with Earth's magnetosphere \citep{Gosling1993, Pulkkinen2007, Webb2012, Temmer2021}. From an observational perspective, CMEs exhibit significant variability in thermal, kinematic, and magnetic characteristics, shaped by their source-region properties and by processes acting during interplanetary propagation \citep{Zurbuchen2006, Gopalswamy2009, Vourlidas2013, Manchester2017}. Near the Sun, polytropic index ($\Gamma$) reflects the balance between thermal input and expansion losses, while at 1~au it encodes the integrated effects of both the eruption and its transit through the heliosphere. Gaining a comprehensive understanding of CMEs necessitates a thorough investigation of their thermal, kinematic, and dynamic evolution from their origin in the low corona to their propagation through the heliosphere up to 1~au and beyond. This chapter offers a multi-faceted investigation of the thermodynamic evolution of fast CMEs near the Sun, statistical plasma and thermal characteristics in the interplanetary medium at 1~au, and their geospace impact as observed through the distribution and recovery behavior of geomagnetic storms.

The chapter is divided into three distinct but interconnected studies. The first section addresses the internal thermal properties and internal force evolution of fast CMEs in the near-Sun region using an analytical modeling approach based on 3D kinematics and flux-rope expansion. The second study focuses on the statistical behavior of ICMEs at 1~au across Solar Cycles 23--25, particularly examining the evolution of their thermal state across ICME structures using the polytropic index framework and its possible implications for geoeffectiveness. In the final study, focusing on intense and stronger-than-intense storms ($\mathrm{Dst}_{\mathrm{min}} \leq -100$~nT) during Solar Cycles 23 and 24, we investigate how storm recovery durations depend on interplanetary drivers (ICMEs, Stream Interaction Regions (SIRs), and their interactions), as well as on storm morphology (single- vs. multiple-peaked). The analysis reveals solar-cycle differences in storm profiles and emphasizes the role of driver structure in shaping recovery characteristics. Together, these studies offer an integrated perspective on CME thermodynamics from initiation to geospace impact, emphasizing the role of thermal evolution in shaping their large-scale dynamics and space weather potential.

\section{Connecting Global Kinematics with Thermodynamics of Fast CMEs} \label{Ch4-part1}

The internal thermodynamics of CMEs play a crucial role in governing their global acceleration, expansion, and propagation characteristics. In our earlier investigation (Section~\ref{Ch3-sec:1}), we applied the FRIS model to a case study involving one fast and one slow CME. The results demonstrated significant differences in their thermodynamic behavior: the fast CME exhibited an initial phase of heat release in the low corona, transitioning to a heat absorption phase during its propagation through the heliosphere. While this finding highlighted the importance of internal heating and cooling processes, the analysis was limited to a single fast event. Therefore, to derive more general and statistically significant conclusions, it is imperative to investigate a broader sample of fast CMEs with diverse propagation speeds and expansion profiles.

 The present study is designed to extend this earlier work by systematically examining the internal thermodynamic evolution of a set of fast CMEs and establishing how their thermal states correlate with global kinematic behavior. We focus specifically on identifying common trends in the evolution of critical internal parameters such as the polytropic index ($\Gamma$), temperature ($T$), and heating rate per unit mass ($dQ/dt$). To achieve this, we employ the FRIS model, using 3D kinematic inputs derived from GCS reconstruction of multi-viewpoint coronagraph data. Additionally, we explore the initial thermal state of the CME flux-rope at the time of its eruption using high-resolution multi-wavelength EUV observations. Investigating this early-stage thermal condition provides essential context for understanding the initial temperature and its influence on subsequent thermodynamic and kinematic evolution. The combination of forward modeling and EUV diagnostics offers a more complete picture of how CME internal heating and expansion dynamics evolve during the early and intermediate phases of CME propagation.

This comprehensive analysis contributes to a deeper understanding of the internal plasma dynamics for fast CMEs and serves to refine assumptions regarding $\Gamma$ often used in both analytical and numerical models. The insights gained here are expected to improve future predictive modeling of CME dynamics and their space weather impacts.

\subsection{Methodology and Event Selection}{\label{chapt4-sec:methods}}

In this study, we employ the analytical FRIS model to explore the internal thermodynamic evolution of CMEs during their outward propagation from the Sun. As discussed in Section~\ref{Ch3-sec:1}, this model was previously applied to analyze the thermal properties and internal dynamics of one fast and one slow CME. A complete derivation of the revised FRIS model and the expressions for all derived physical parameters can be found in Section~\ref{metho:fris_derivation}. Here, we focus on selected output parameters from the model, namely, the polytropic index ($\Gamma$), temperature ($T$), heating rate per unit mass ($dQ/dt$), and internal forces, for a set of fast CMEs.

The model is governed by an equation of motion (Equation~\ref{fitting}) that describes the radial expansion of the CME flux-rope. This equation contains five unknown constants ($c_1$ to $c_5$), which are determined by fitting the equation to observational data. The key input parameters to the FRIS model include the heliocentric distance of the CME flux-rope center ($L$), the flux-rope radius ($R$), and their corresponding time derivatives: propagation speed ($v_c$), propagation acceleration ($a_c$), expansion speed ($v_e$), and expansion acceleration ($a_e$).

Since the FRIS model relies on the 3D kinematics of CMEs as inputs, we first derive these kinematics using the Graduated Cylindrical Shell (GCS) model. The GCS model is a forward-fitting, empirical approach that geometrically represents the flux-rope structure by fitting it to coronagraph images from multiple viewpoints. By incorporating observations from different vantage points, this model mitigates projection effects and provides a more accurate reconstruction of CME morphology and dynamics (Section~\ref{sec:chapt2-metho-gcs}). The height of the CME leading edge ($h$) and the aspect ratio ($\kappa$) obtained from the GCS model are used to estimate the flux-rope radius as $R = \left(\frac{\kappa}{1+\kappa}\right)h$, and the center height as $L = h - R - 1\,R_\odot$.

To derive the time derivatives required for the FRIS model, namely, $v_c$, $a_c$, $v_e$, $a_e$, and $\mathrm{d}a_e/\mathrm{d}t$, we apply a moving-window differentiation technique. Specifically, a linear fit is performed over a three-point moving window to calculate the derivatives (e.g., from $L$ to $v_c$, and then $a_c$), evaluated at the central point of the window. For the endpoints (i.e., the first and last data points), a two-point linear fit is used to compute the derivatives. This approach captures genuine variations in kinematic parameters without reducing the total number of data points in the derivative series.

\begin{table*}
\caption[List of selected fast CMEs from 2010 to 2012]{\label{chapt4-tab:GCS-results} List of selected fast CMEs from 2010 to 2012. Columns 2 to 8 present the GCS model fitted parameters along with manually estimated fitting uncertainties. The second and third columns indicate the time and height range over which the GCS fitting was performed. The final column provides the estimated maximum leading-edge speed ($v$) of each CME within the coronagraphic field of view. }
\centering
\begin{adjustbox}{width=\textwidth}
\begin{tabular} {l|ccccccc|c}
\hline
\textbf{Events} & \textbf{Time (UT)} & \textbf{Height ($R_\odot$)}  & \textbf{Longitude} & \textbf{Latitude} & \textbf{Aspect}  & \textbf{Tilt}   & \textbf{Half}&\textbf{Max Speed} \\

& \textbf{Initial-Final} &   \textbf{Initial-Final} & \textbf{(deg)} & \textbf{(deg)} & \textbf{Ratio} & \textbf{Angle} & \textbf{Angle}& \textbf{(km s$^{-1}$)}\\
 
 & &   & & & &\textbf{(deg)} &\textbf{(deg)}&\\
\hline
CME1 & 09:30-12:39  & 2.5-15.4 & 7$\pm 3$ & -24$\pm 2$ & 0.36$\pm 0.1$ & 13$\pm 6$ & 13$\pm 2$ & 849$\pm 93$\\
\hline
CME2 & 02:05-05:54 & 2.8-15.6 & -5$\pm 3$ & -14$\pm 3$ & 0.30$\pm 0.1$  & 47$\pm 10$ & 14$\pm 3$ & 1131$\pm 146$\\
\hline
CME3 & 04:00-05:54 & 2.6-21.2 & 30$\pm 2$ & 19$\pm 3$ & 0.33$\pm 0.1$  & -87$\pm 8$ & 15$\pm 2$ & 2266$\pm 291$\\
\hline
CME4 & 12:45-14:39 & 2.5-20.1 & -41$\pm 4$ & 13$\pm 3$ & 0.39$\pm 0.1$  & -62$\pm 6$ & 26$\pm 2$ & 1883$\pm 185$\\
\hline
CME5 & 07:15-10:54  & 3.1-20.9  & 39$\pm 3$ & 19$\pm 1$ & 0.45$\pm 0.2$  & -61$\pm 5$ & 42$\pm 4$ & 989$\pm 92$ \\
\hline
CME6 & 01:15-02:54 & 3.9-20.2 & -29$\pm 3$ & -6$\pm 3$ & 0.43$\pm 0.1$  & -56$\pm 6$ & 24$\pm 6$ & 2092$\pm 254$\\
\hline
CME7 & 13:55-17:06 & 2.8-21.4 & -3$\pm 2$ & -27$\pm 2$ & 0.39$\pm 0.1$  & -6$\pm 8$ & 28$\pm 3$ & 1228$\pm 130$ \\
\hline
CME8 & 16:30-18:54 & 2.3-18.2 & -5$\pm 3$ & -4$\pm 3$ & 0.35$\pm 0.1$  & 36$\pm 4$ & 29$\pm 6$ & 1352$\pm 144$\\
\hline
CME9 & 00:05-03:30 & 4.3-25.7 & 24$\pm 4$ & 14$\pm 3$ & 0.52$\pm 0.1$  & -75$\pm 4$ & 43$\pm 4$ & 1341$\pm 122$\\
\hline
\end{tabular}
\end{adjustbox}

\begin{tablenotes}
\small
\it
    \item Note -- CME1: 2010 Apr 03, CME2: 2011 Feb 15, CME3: 2011 Aug 04, CME4: 2011 Sep 24, CME5: 2011 Nov 26, CME6: 2012 Mar 07, CME7: 2012 Jun 14, CME8: 2012 Jul 12 and CME9: 2012 Sep 28
\end{tablenotes}

\end{table*}

We selected nine fast CMEs, each exhibiting a maximum leading-edge speed between 800 and 2300~km~s$^{-1}$, as listed in Table~\ref{chapt4-tab:GCS-results}. The selection was based on minimizing kinematic uncertainties, which is crucial since accurate 3D kinematics are required as input for the FRIS model to derive reliable thermodynamic parameters. Only well-isolated CMEs were chosen, those not interacting with other large-scale structures in the coronagraphic field of view, to avoid ambiguity in tracking and interpretation. Additionally, to ensure precise tracking and 3D reconstruction, we selected events with well-defined, bright leading edges that could be followed to higher coronal heights.

The selected CMEs occurred between 2010 and 2012 and were all Earth-directed. During this period, simultaneous coronagraphic observations were available from three vantage points: STEREO-A, STEREO-B, and SOHO (located near the L1 point). The separation angle between STEREO-A and STEREO-B ranged from approximately $65^{\circ}$ to $125^{\circ}$, providing an optimal configuration for accurate geometric fitting of CMEs using the GCS model. Furthermore, these CMEs have been extensively documented in the literature for their kinematic properties, which allows us to validate our kinematic estimates and confidently focus on deriving their internal thermal characteristics.

\subsection[Measured 3D Kinematics of Selected CMEs]{Measured 3D Kinematics of Selected CMEs}{\label{chapt4-sec:kinem}}

The GCS model employs six free parameters to determine the three-dimensional position and geometric structure of a CME using observations from multiple vantage points in the coronagraphic field of view. In this study, we applied the GCS model to simultaneous coronagraph images from SOHO/LASCO (C2 and C3) and STEREO/COR (COR1 and COR2). To constrain the positional parameters, such as longitude and latitude, at low coronal heights, we utilized information about the source regions of the CMEs. These source regions indicated that the selected CMEs were Earth-directed, consistent with their propagation paths.

The GCS model fitting results for the selected CMEs are presented in Table~\ref{chapt4-tab:GCS-results}. Estimating uncertainties in the GCS model parameters is inherently subjective, as the fitting process involves manual interpretation. According to \citet{Thernisien2009}, the typical uncertainties associated with GCS fitting are approximately: $\pm 4.3^\circ$ in longitude, $\pm 1.8^\circ$ in latitude, $\pm 22^\circ$ in tilt angle, $_{-7^\circ}^{+13^\circ}$ in half angle, $_{-0.04}^{+0.07}$ in aspect ratio, and $\pm 0.48\,R_\odot$ in leading-edge height. More recently, \citet{Verbeke2023} used synthetic CME data to quantify these errors and emphasized the necessity of at least two vantage points to reliably constrain CME parameters.

To estimate uncertainties in the leading-edge height, we considered the fact that CMEs often exhibit a sharp boundary near the Sun but become increasingly diffuse at larger heliocentric distances. Based on repeated fitting attempts, we conservatively estimate a maximum uncertainty of $\pm$10\% in the leading-edge height. This uncertainty was propagated to compute the associated errors in derived kinematic quantities, such as speed and acceleration, which are then used as inputs to the FRIS model. All selected CMEs were determined to be Earth-directed and located near the ecliptic plane. Furthermore, the GCS-derived half-angle and aspect ratio remained approximately constant during their propagation through the coronagraphic field of view. This behavior indicates a self-similar expansion of the CMEs, consistent with one of the primary assumptions of the FRIS model.

\begin{figure*}
         \includegraphics[scale=0.20,trim={5cm 1cm 20cm 0cm}]{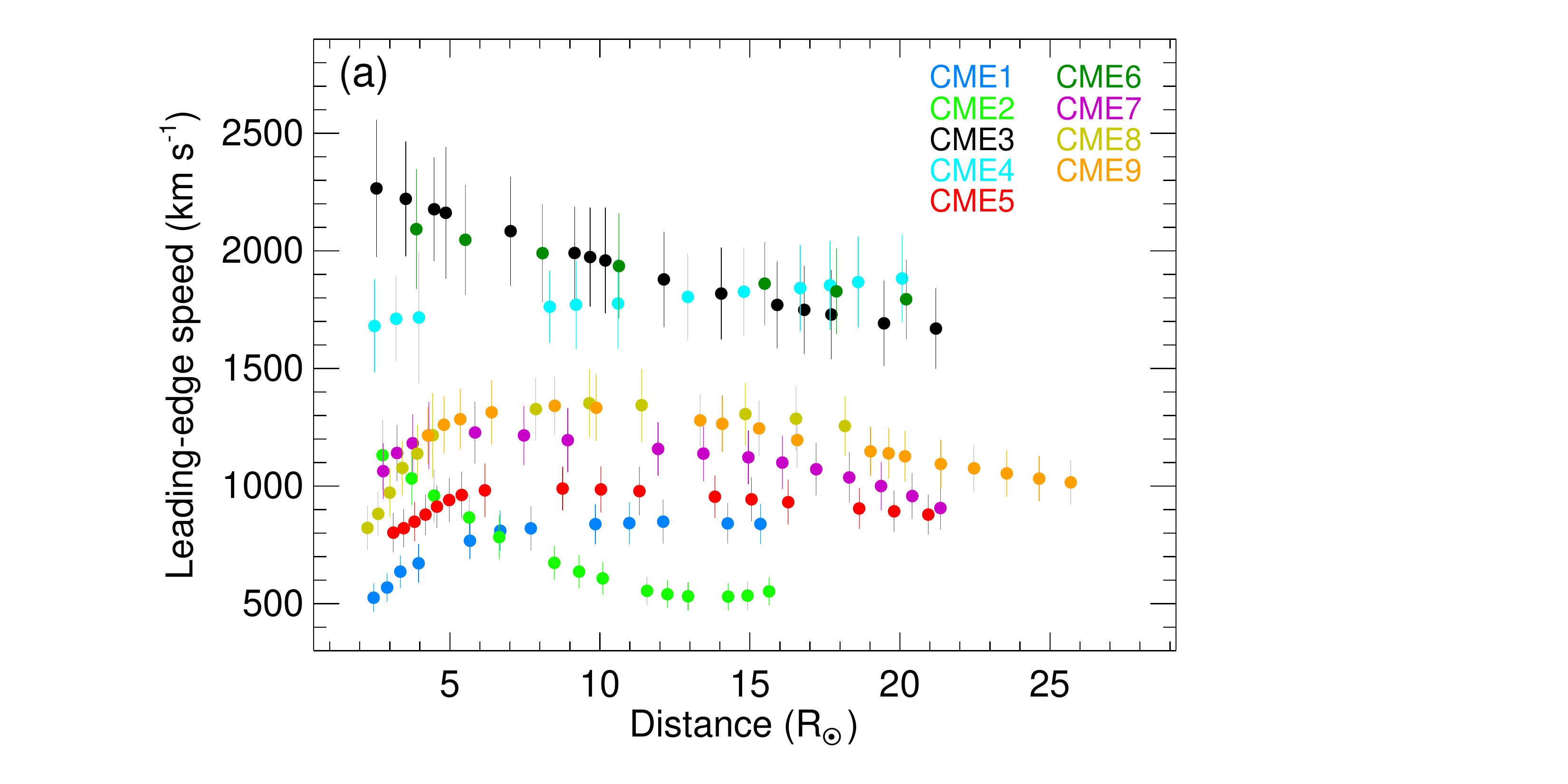}
         \includegraphics[scale=0.20,trim={-1cm 1cm 24cm 0cm}]{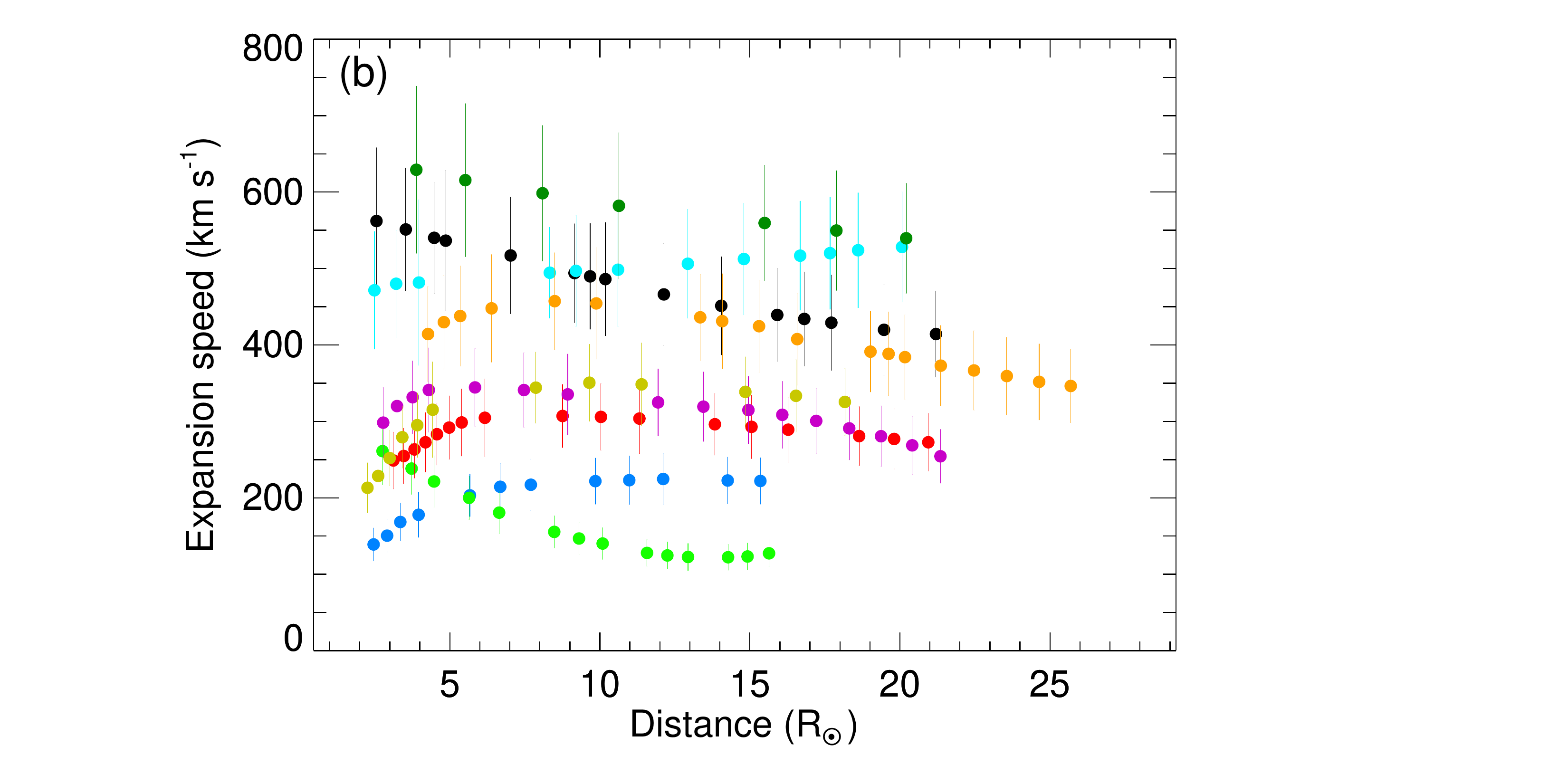}\\ \hspace{-2.5cm}
         \includegraphics[scale=0.20,trim={5cm 1cm 20cm 1cm}]{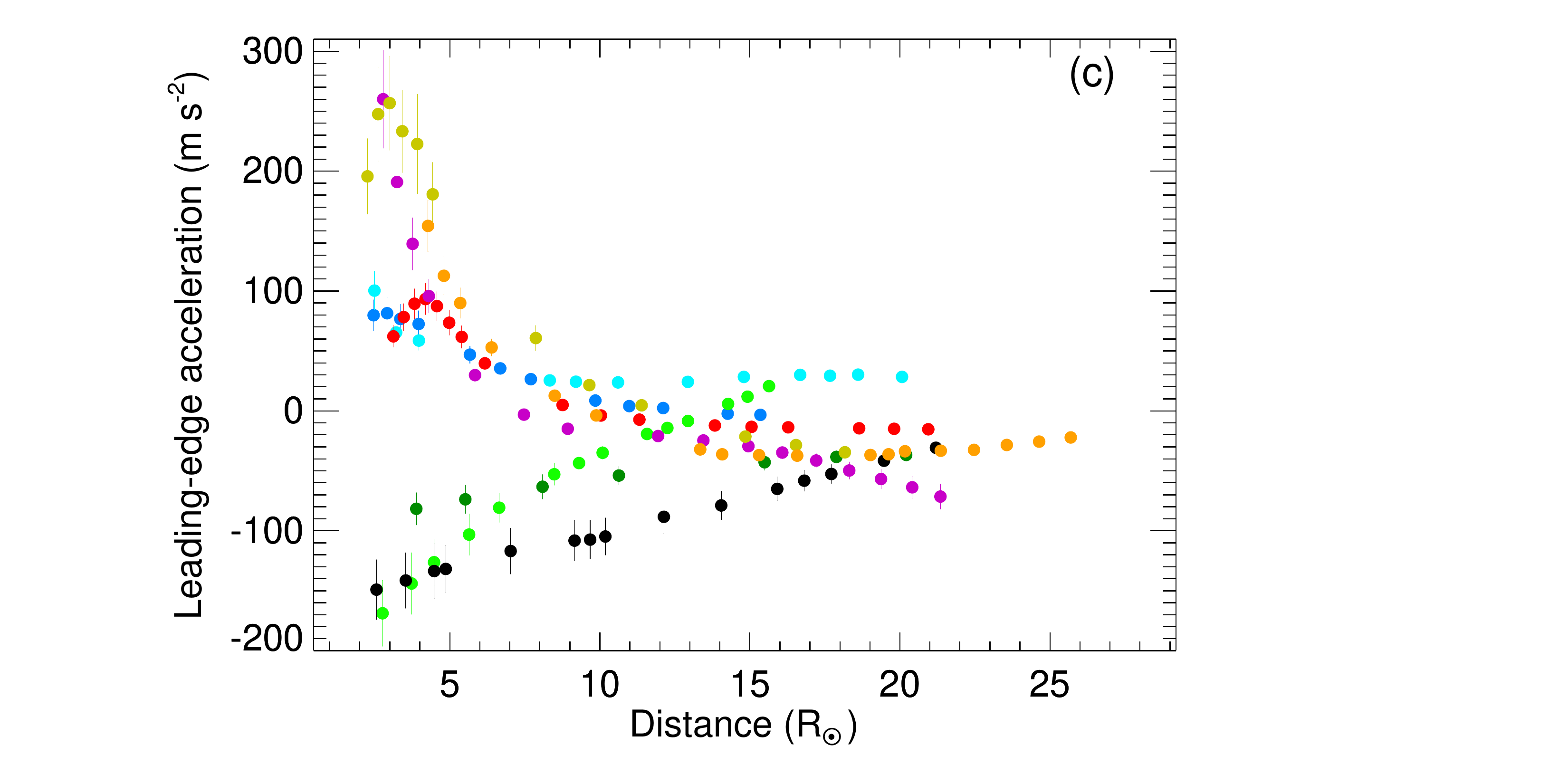}
         \includegraphics[scale=0.20,trim={-1cm 1cm 24cm 1cm}]{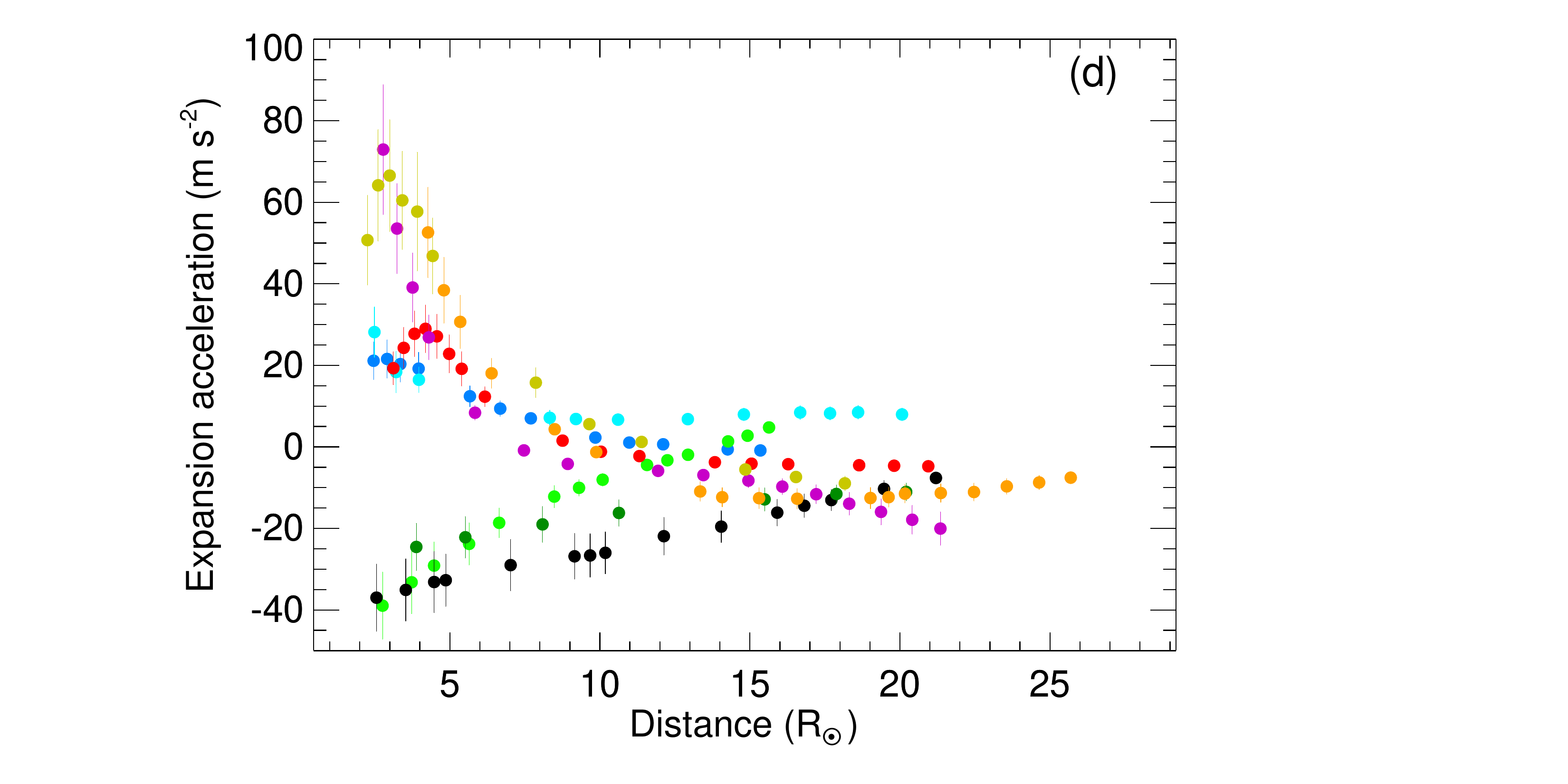}
         \caption[Kinematic evolution of the selected CMEs as a function of]{Kinematic evolution of the selected CMEs as a function of leading-edge height ($h$): (a) leading-edge speed ($v$), (b) expansion speed ($v_e$), (c) leading-edge acceleration ($a$), and (d) expansion acceleration ($a_e$). Error bars at each data point represent uncertainties derived by assuming a 10\% error in the measurement of the flux-rope’s leading-edge height.}
        \label{fig:kinematics}
\end{figure*}

\begin{sidewaysfigure}
\centering
\includegraphics[width=0.30\linewidth,trim={2cm 1.2cm 0cm 1cm},clip]{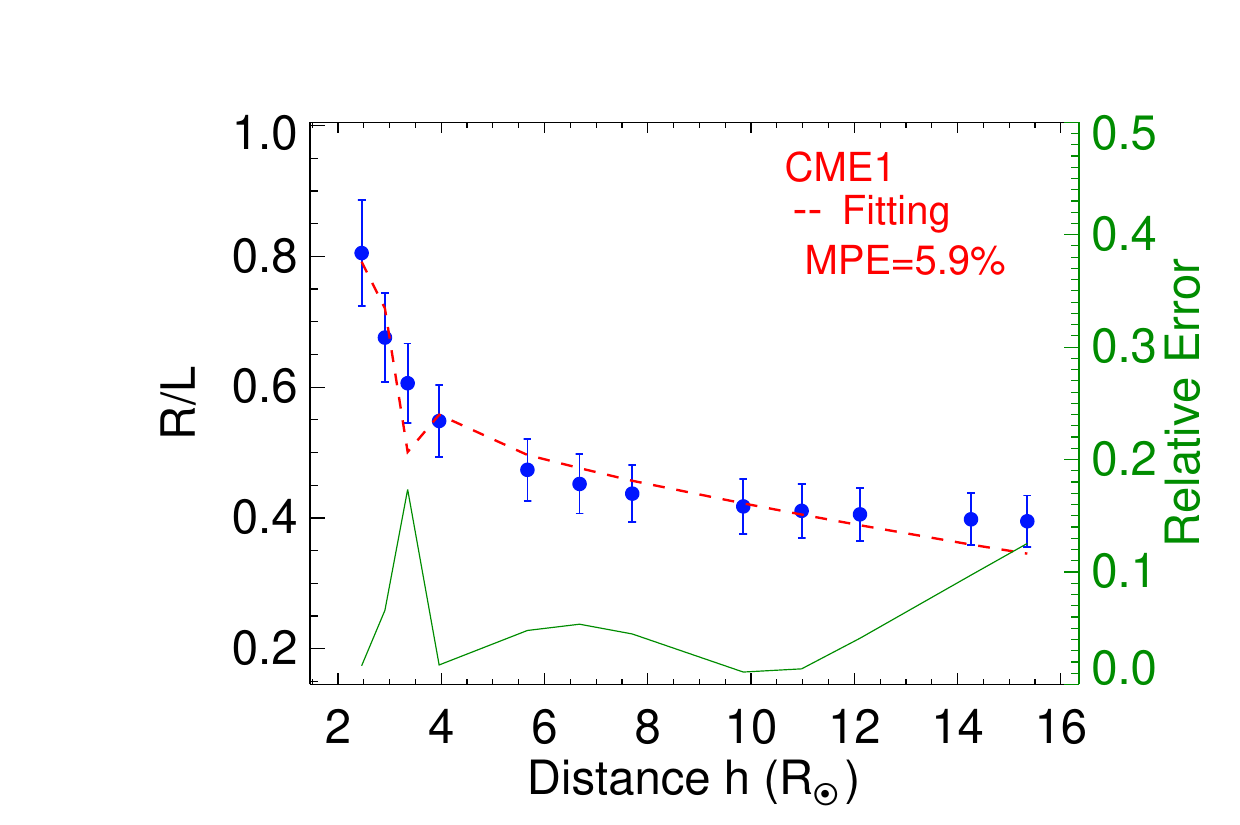}
\includegraphics[width=0.30\linewidth,trim={2cm 1.2cm 0cm 1cm},clip]{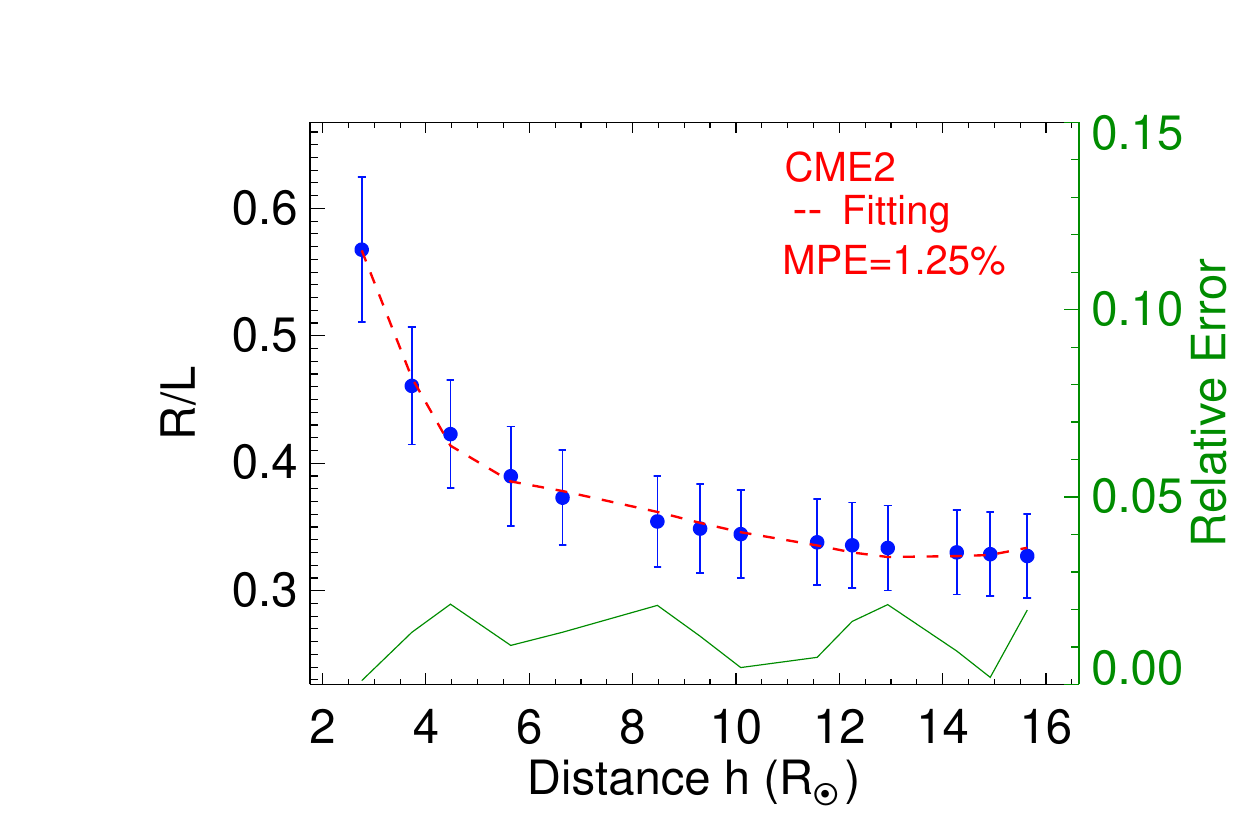}
\includegraphics[width=0.30\linewidth,trim={2cm 1.2cm 0cm 1cm},clip]{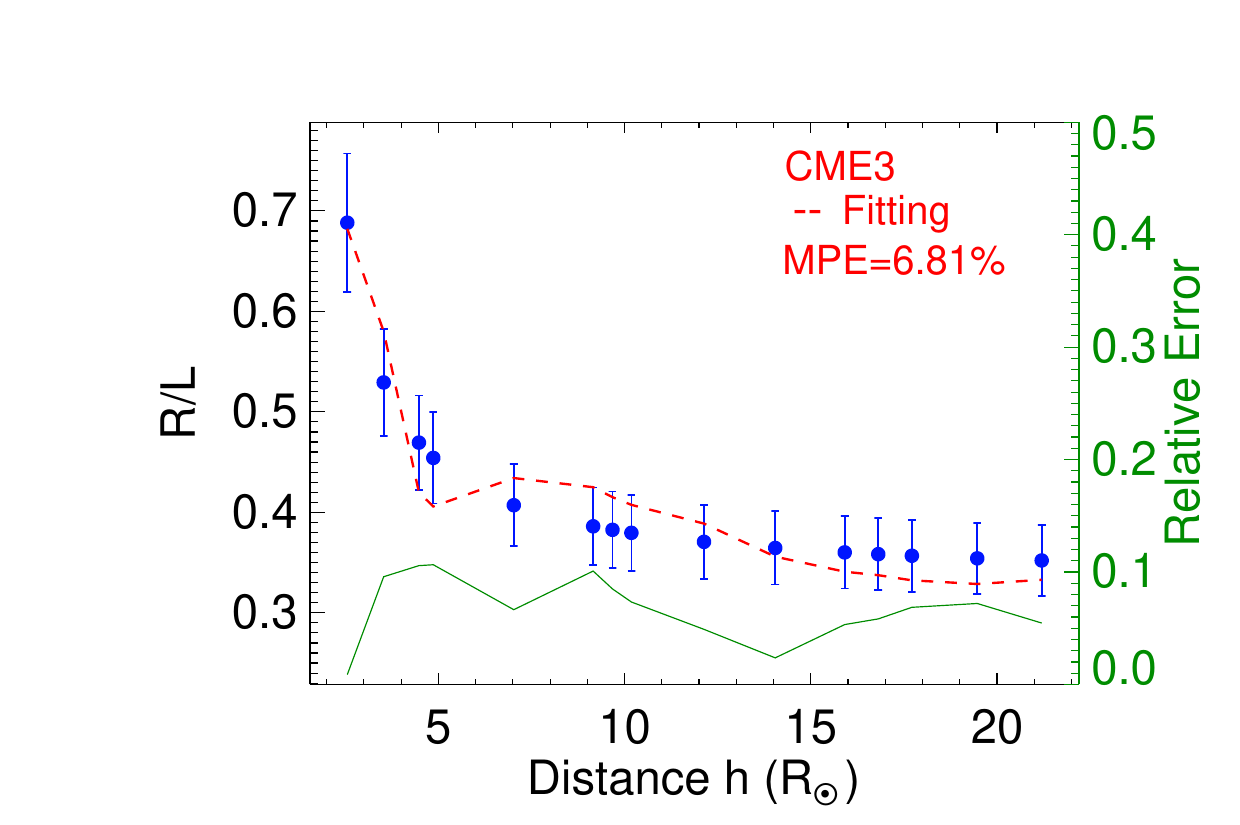}\\
\includegraphics[width=0.30\linewidth,trim={2cm 1.2cm 0cm 1cm},clip]{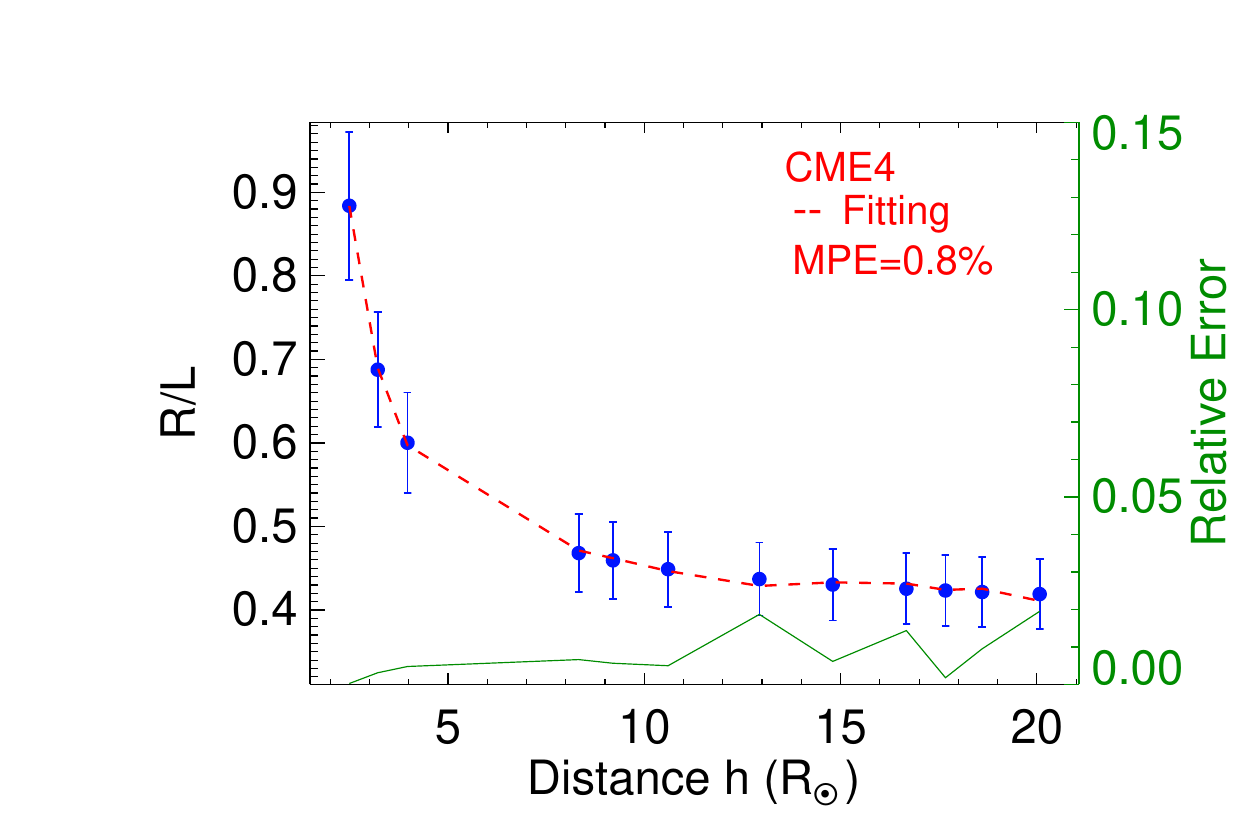}
\includegraphics[width=0.30\linewidth,trim={2cm 1.2cm 0cm 1cm},clip]{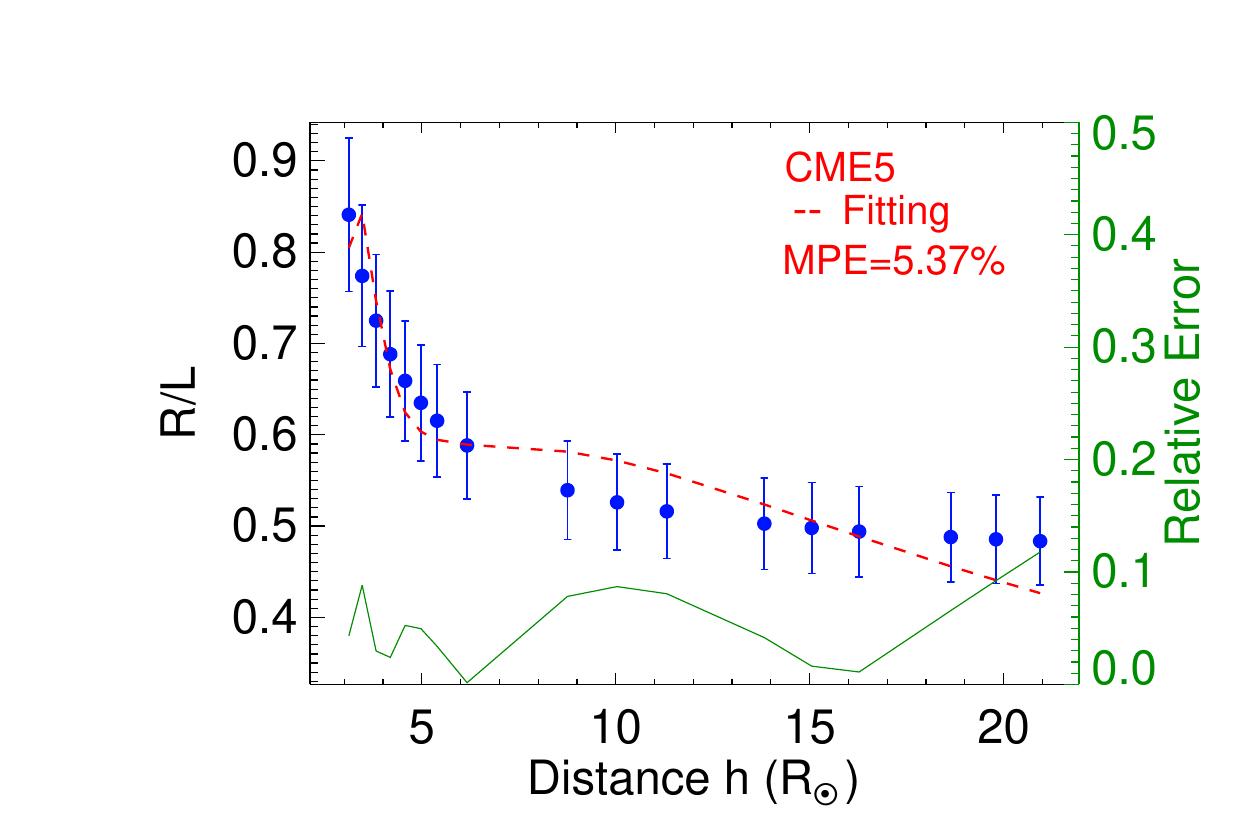}
\includegraphics[width=0.30\linewidth,trim={2cm 1.2cm 0cm 1cm},clip]{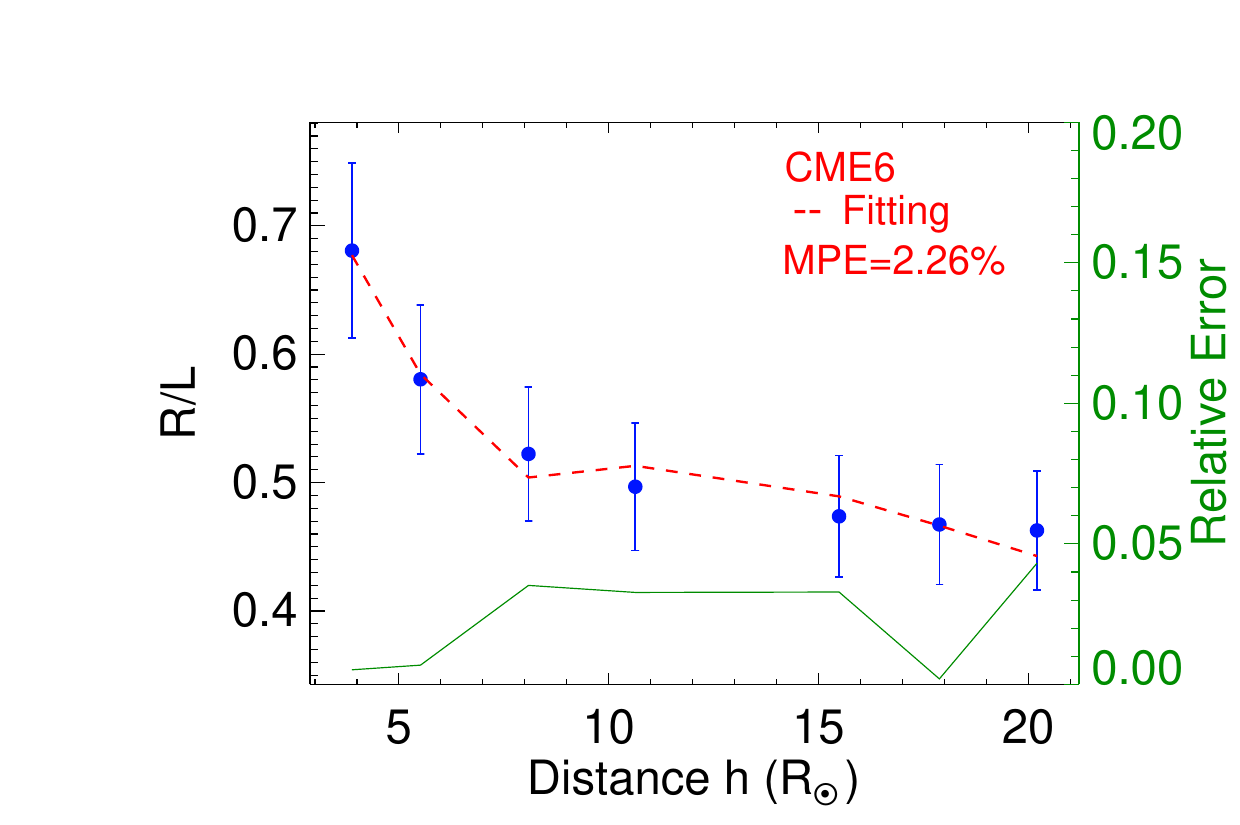}\\
\includegraphics[width=0.30\linewidth,trim={2cm 0cm 0cm 1cm},clip]{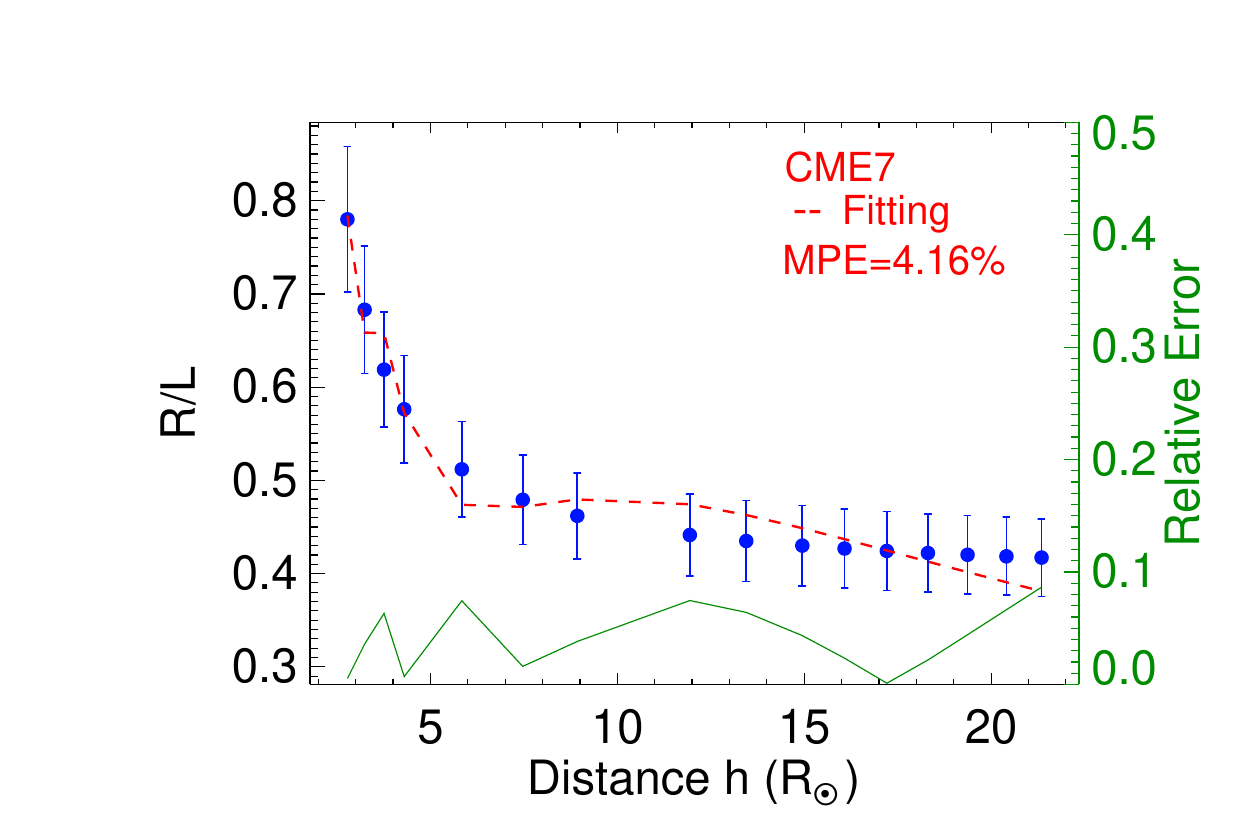}
\includegraphics[width=0.30\linewidth,trim={2cm 0cm 0cm 1cm},clip]{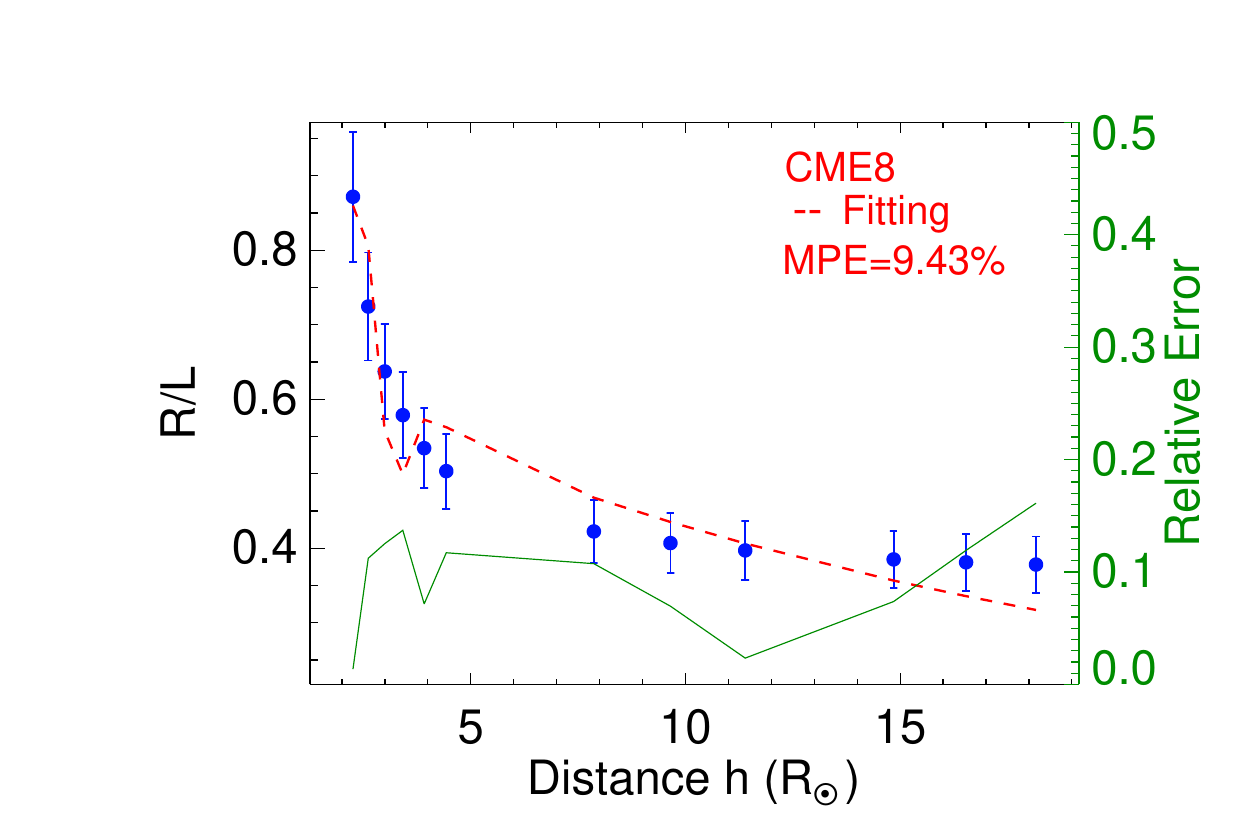}
\includegraphics[width=0.30\linewidth,trim={2cm 0cm 0cm 1cm},clip]{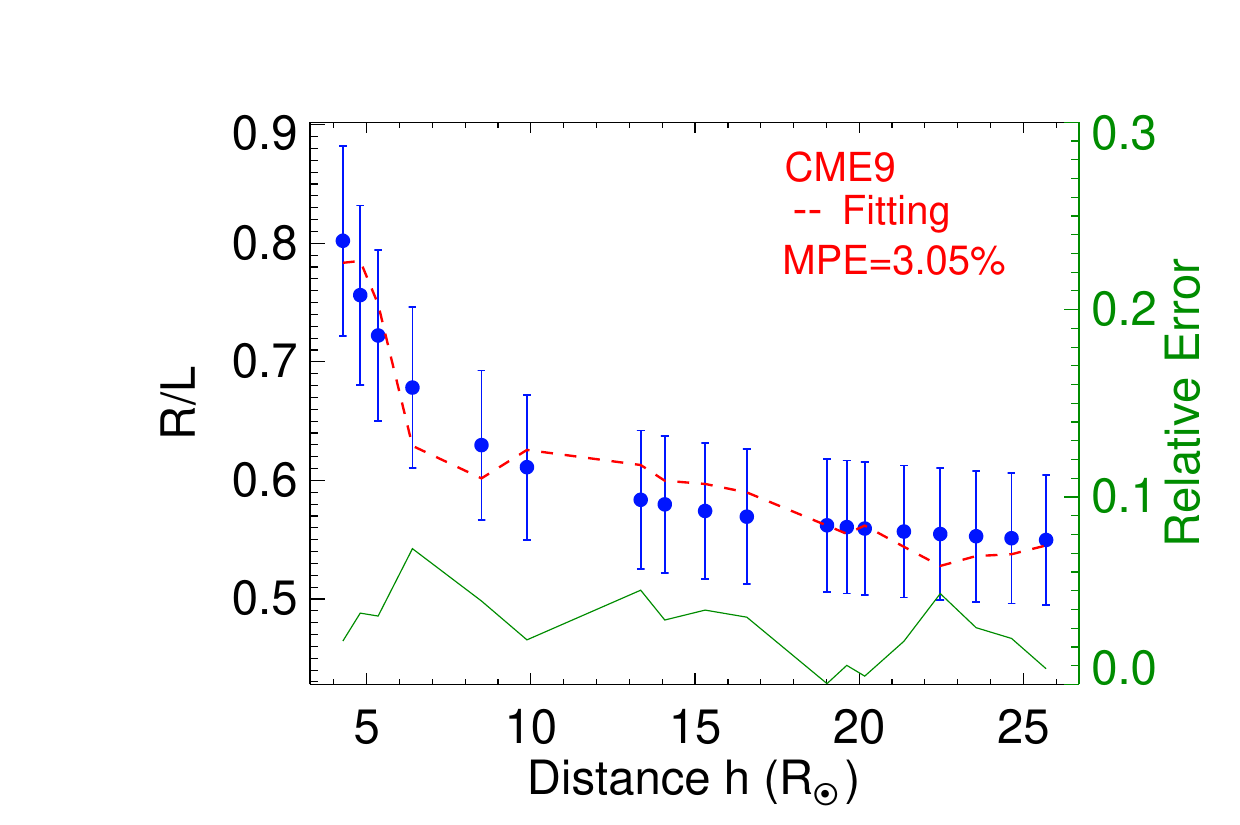}

\caption[Model fitting errors for the selected fast CMEs]{Model fitting errors for the selected fast CMEs. Blue dots represent the left-hand side of Equation~\ref{fitting}, while the red dashed line denotes the corresponding right-hand side. The green line shows the relative fitting error. Blue vertical bars indicate a 10\% perturbation applied to the left-hand side values of Equation~\ref{fitting}. MPE refers to the mean percentage error.}
\label{chapt4-fig:fitting_errors}
\end{sidewaysfigure}

Interestingly, the selected CMEs exhibit notably diverse trends in their speed and acceleration profiles, which may reflect intrinsic differences in their physical properties, such as magnetic field strength, thermal energy content, and temperature. The derived kinematic profiles for these CMEs are shown in Figure~\ref{fig:kinematics}. Uncertainties in the profiles were estimated by assuming a 10\% error in the GCS-derived leading-edge heights. Most of the selected CMEs were tracked up to a leading-edge height of approximately 20~($\pm$2)~$R_\odot$, except for CME1, CME2, and CME9. Specifically, CME1 and CME2 were tracked up to around 16~($\pm$1.6)~$R_\odot$, while CME9 was tracked to higher heights, reaching about 26~($\pm$2.6)~$R_\odot$. Several CMEs, namely CME1, CME5, CME7, CME8, and CME9—show a noticeable increase in both leading-edge speed ($v$) and expansion speed ($v_e$) at lower coronal heights (Figure~\ref{fig:kinematics}a and~\ref{fig:kinematics}b). Beyond this region, the leading-edge speed either remains nearly constant or decreases gradually. Their acceleration profiles follow a similar pattern, with a steep decline at lower heights followed by a more gradual decrease (Figure~\ref{fig:kinematics}c and~\ref{fig:kinematics}d).

CME3 and CME6 exhibit comparable leading-edge speeds but differ significantly in their expansion speeds. Both CMEs show negative propagation and expansion accelerations, consistent with their declining speed profiles. These two events provide an interesting case for comparison, as their main distinguishing factor is their expansion behavior, which may have implications for their internal thermodynamic evolution. CME2 and CME4, in contrast, show markedly different trends. CME4 displays relatively high and steadily increasing values of both leading-edge and expansion speeds, while its acceleration decreases initially before stabilizing. On the other hand, CME2 starts with a lower speed, which decreases early on and then levels off. Its acceleration, both propagation and expansion, begins with negative values, indicating deceleration, and transitions toward positive values at around 15~($\pm$1.5)~$R_\odot$.

In the following sections, we will examine how variations in these global kinematic profiles correspond to changes in the internal thermodynamic properties derived from the FRIS model. CMEs with similar kinematic features will be compared in detail to better understand how their thermal evolution is governed by differences in their propagation and expansion characteristics.

\subsection{Evolution of Thermodynamic Parameters}{\label{chapt4-sec:thermo}}

To apply the FRIS model to the observed CMEs, we performed a non-linear least squares fitting of the model-derived Equation~\ref{fitting}, using the 3D kinematic parameters as inputs. This fitting was carried out using the \texttt{curve\_fit} routine from the \textit{SciPy} library, which provided the best-fit (median) values of the model coefficients ($c_1$–$c_5$) along with their associated standard deviations, as summarized in Table~\ref{tab:fitting_coefficients}. A 10\% perturbation was applied to the left-hand side of Equation~\ref{fitting} during the fitting process to account for observational uncertainties.

\begin{table}
\centering
\resizebox{0.59\linewidth}{!}{
\rotatebox{90}{
\begin{threeparttable}
\caption[The columns show the estimated fitting coefficients]{\label{tab:fitting_coefficients} The columns show the estimated fitting coefficients of Equation \ref{fitting} by considering a perturbation of 10\% in the left-hand side of Equation \ref{fitting} and their standard deviation for each selected CMEs. }
\centering
\begin{tabular} {lccccc}
\hline
\textbf{Events} & \textbf{$c_1$} & \textbf{$c_2$}  & \textbf{$c_3$} & \textbf{$c_4$} & \textbf{$c_5$}\\
\hline
CME1 & 2.91e+09 $\pm$ 1.14e+09 & 1.43e+06 $\pm$ 5.60e+05 & -1.25e+07 $\pm$ 4.90e+06 & 3.60e+03 $\pm$ 5.05e+06 & 8.04e-03 $\pm$ 7.49e+00  \\
\hline
CME2 & 1.05e+06 $\pm$ 6.48e+07  & 4.11e+02 $\pm$ 2.12e+06 & -5.97e+03 $\pm$ 1.60e+07 & 9.15e+05 $\pm$ 1.92e+05 & 8.11e+00 $\pm$ 1.70e+00 \\
\hline
CME3 & 2.45e+06 $\pm$ 1.52e+05 & 1.11e+03 $\pm$ 6.76e+01 & -1.20e+04 $\pm$ 2.28e+02 & 8.69e+05 $\pm$ 2.16e+05 & 1.14e+01 $\pm$ 9.00e-01 \\
\hline
CME4 & 9.21e+05 $\pm$ 2.06e+00 &  3.67e+02 $\pm$ 9.77e-04 & -3.10e+03 $\pm$ 7.34e-03 &  1.64e+06 $\pm$ 7.82e-01 & 1.21e+01 $\pm$ 9.06e-06 \\
\hline
CME5 & 3.18e+09 $\pm$ 4.76e+08 & 1.55e+06 $\pm$ 2.32e+05 & -8.66e+06 $\pm$ 1.30e+06 & 6.94e+03 $\pm$ 2.92e+03  & 1.37e-02 $\pm$ 5.76e-03 \\
\hline
CME6 & 1.15e+10 $\pm$ 1.81e+09 & 6.08e+06 $\pm$ 9.55e+05 & -3.58e+07 $\pm$ 5.62e+06  &  1.27e+03 $\pm$ 3.99e+05  & 5.73e-03 $\pm$ 8.58e-01  \\
\hline
CME7 & 4.18e+10  $\pm$ 1.41e+10 & 2.13e+07 $\pm$ 7.17e+06  & -1.55e+08 $\pm$ 5.21e+07 & 3.01e+02 $\pm$  7.59e+01 & 1.07e-03  $\pm$ 2.70e-04 \\
\hline
CME8 & 1.15e+10 $\pm$ 3.56e+09  & 5.72e+06  $\pm$ 1.76e+06 & -5.28e+07  $\pm$ 1.63e+07 & 5.40e+02  $\pm$ 1.72e+02 & 2.05e-03  $\pm$  6.55e-04 \\
\hline
CME9 & 3.09e+07  $\pm$ 1.32e+09 & 1.70e+04  $\pm$ 5.53e+06  & -6.76e+04  $\pm$ 1.71e+07  & 4.19e+06  $\pm$ 9.28e+05 &  6.54e+00  $\pm$ 1.44e+00 \\
\hline
\end{tabular}
\end{threeparttable}
}}
\end{table}

The model-fitting results for all nine selected CMEs are shown in Figure~\ref{chapt4-fig:fitting_errors}. The derived coefficients were then used to compute various thermodynamic quantities as defined in Table~\ref{chap2_tab:FRIS_parameters}. Among the full set of parameters that the FRIS model can provide, this study focuses on four key properties relevant to the internal thermal state of fast CMEs: the polytropic index ($\Gamma$), heating rate per unit mass ($dQ/dt$), temperature ($T$), and thermal pressure ($p$). The evolution of these parameters for the selected events is illustrated in Figure~\ref{fig:thermo}. Uncertainties in the FRIS model-derived parameters were estimated by incorporating a 10\% error in the input leading-edge height measurements, which was then propagated through the kinematic calculations to derive the final parameter errors.

\subsubsection{Polytropic Index}{\label{chapt4-sec:gamma}}

\begin{figure*}
         \includegraphics[scale=0.20,trim={5cm 1cm 20cm 0cm}]{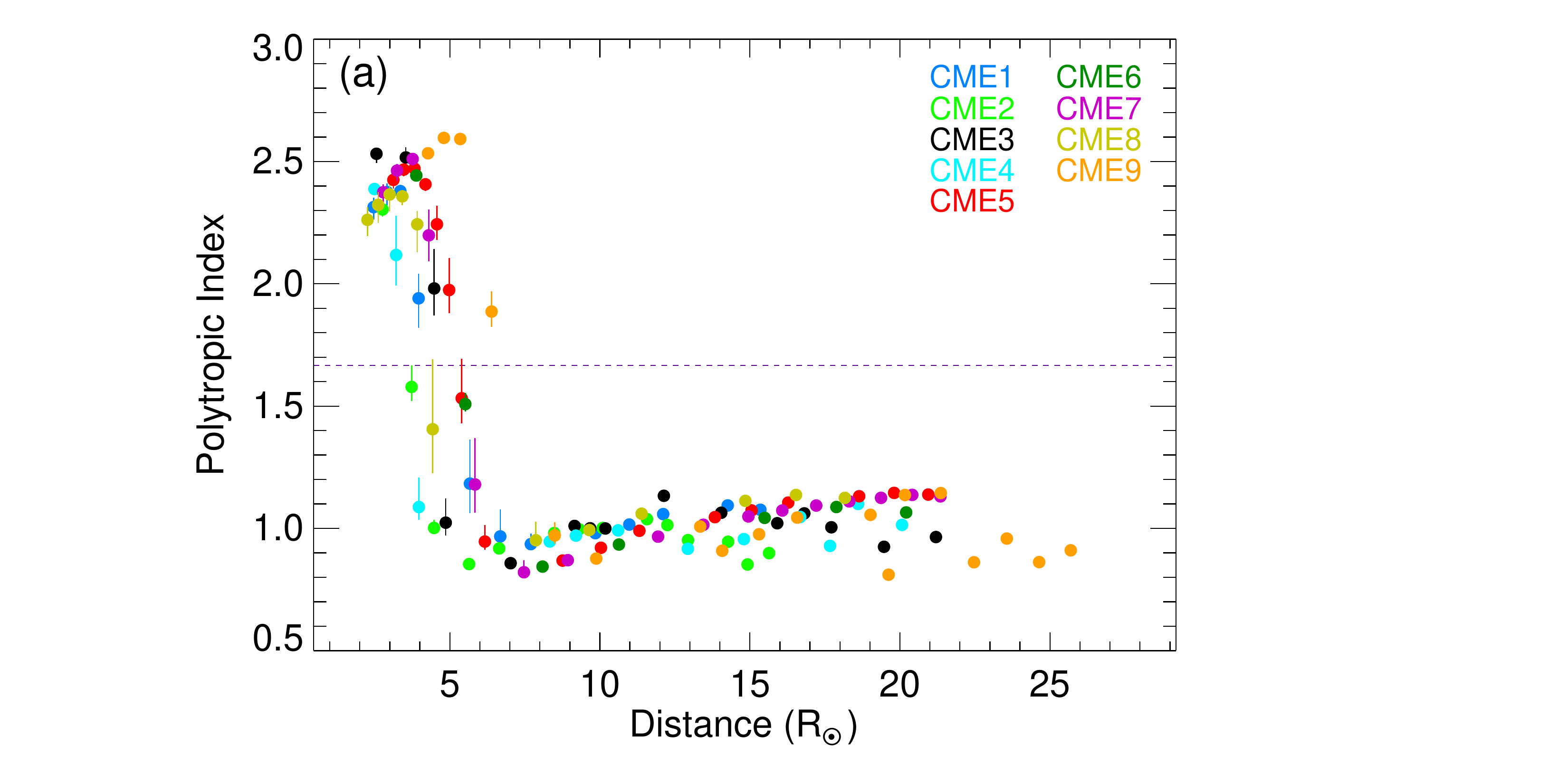}
         \includegraphics[scale=0.20,trim={-1cm 1cm 24cm 0cm}]{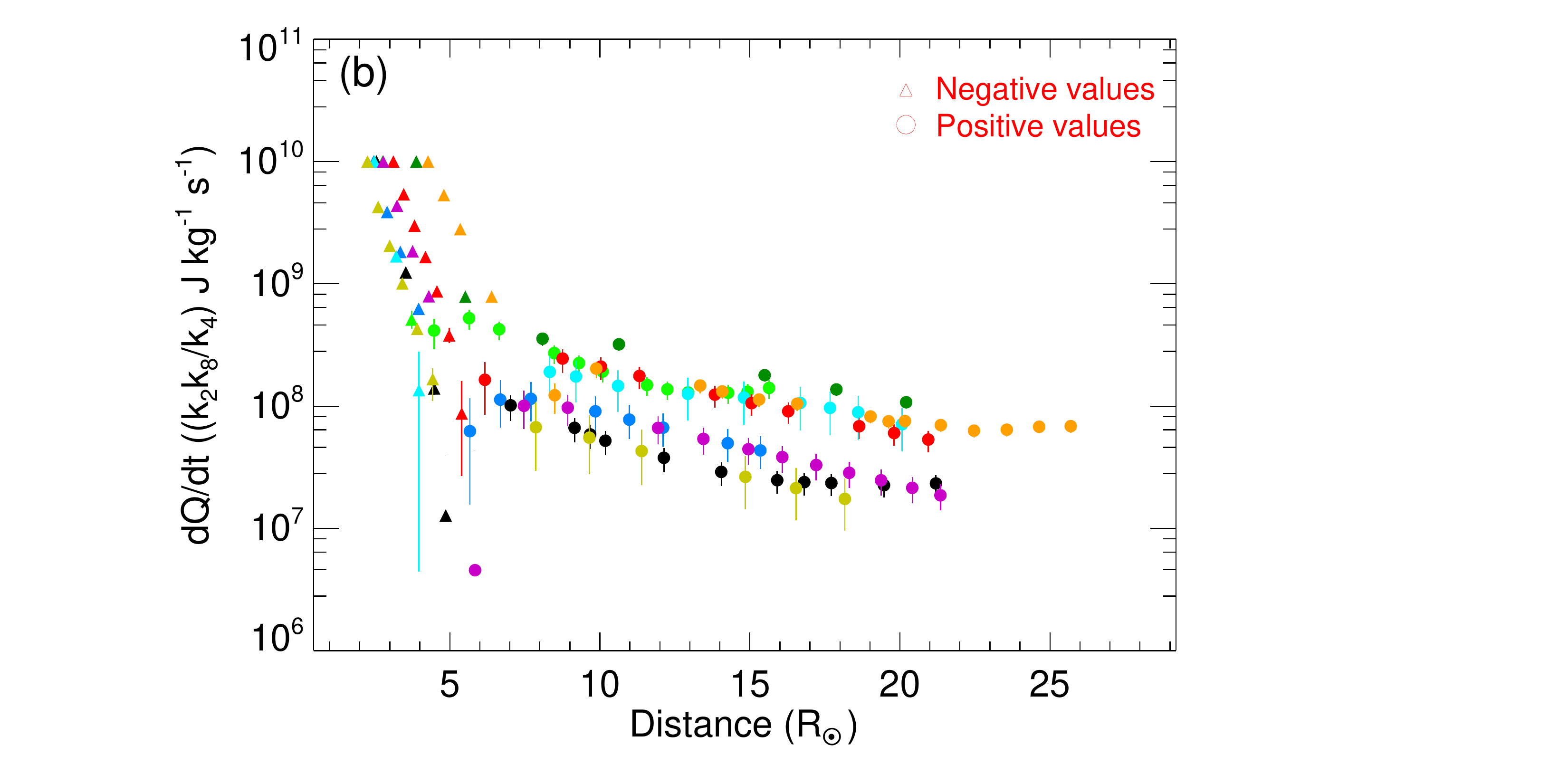}\\ \hspace{-2.5cm}
         \includegraphics[scale=0.20,trim={5cm 1cm 20cm 1cm}]{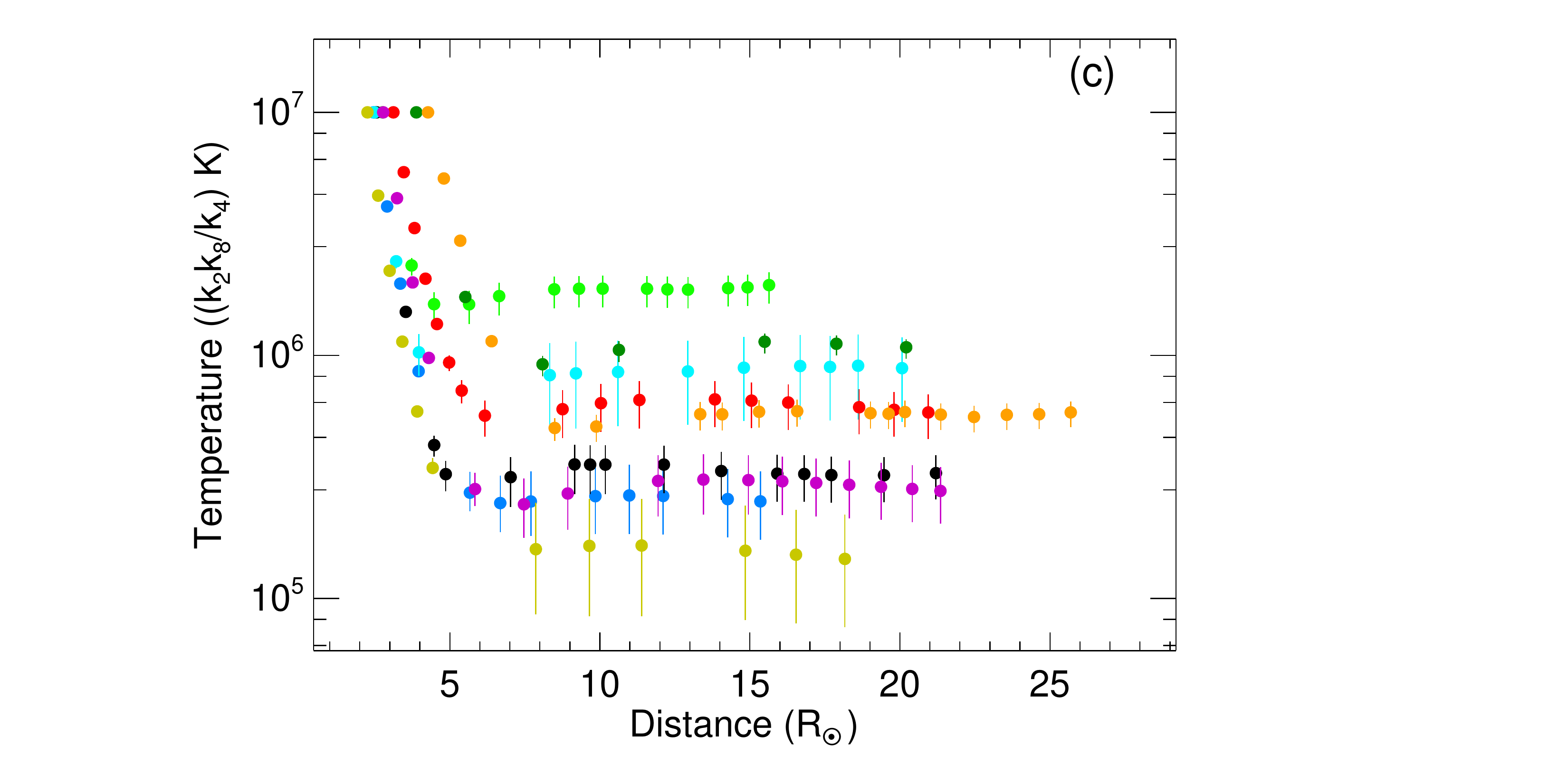}
         \includegraphics[scale=0.20,trim={-1cm 1cm 24cm 1cm}]{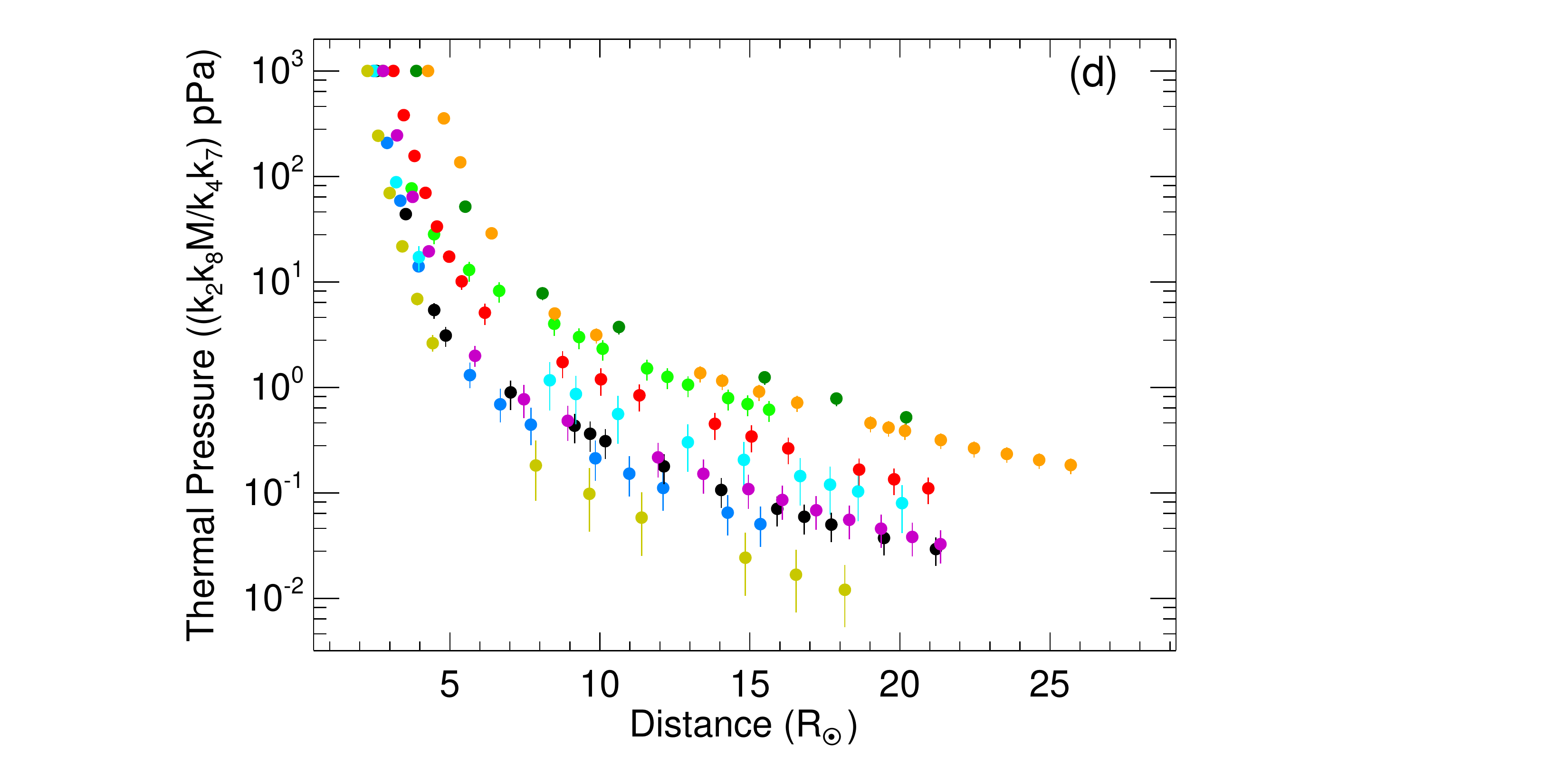}
         \caption[Thermodynamic evolution of the selected CMEs as a function of]{Thermodynamic evolution of the selected CMEs as a function of the heliocentric distance of the leading edge ($h$): (a) polytropic index ($\Gamma$), (b) heating rate per unit mass ($dQ/dt$), (c) temperature ($T$), and (d) thermal pressure ($p$). In panel (a), the dashed horizontal line marks the adiabatic index ($\Gamma = \gamma = 5/3$), corresponding to an ideal adiabatic process with no heat exchange. Vertical error bars represent uncertainties in the FRIS model-derived parameters, estimated by assuming a 10\% error in the input leading-edge height and propagating it through the kinematic calculations.}
        \label{fig:thermo}
\end{figure*}

The variation of the polytropic index ($\Gamma$) for all nine fast CMEs is presented in Figure~\ref{fig:thermo}a. The polytropic index is a key thermodynamic parameter that characterizes the thermal state of a system. For each event, we observe that $\Gamma$ initially starts above the adiabatic value ($\Gamma = 5/3$), indicating a state of heat release near the Sun. Subsequently, $\Gamma$ decreases sharply, crossing the adiabatic threshold at heights ranging from approximately 3.5~($\pm$0.3) to 6.5~($\pm$0.6)~$R_\odot$. Beyond this point, $\Gamma$ typically stabilizes within the range 0.8–1.2, which is close to an isothermal state ($\Gamma = 1$). These results suggest a consistent thermal pattern among the fast CMEs: an initial phase of heat release close to the Sun, followed by a transition into a heat-absorbing or nearly isothermal state during further propagation. Although the FRIS model does not reveal the exact mechanisms of internal heating, we interpret “heat injection” or “heat absorption” as indicative of either external heating sources or internal energy dissipation processes within the CME. Accurate estimation of the heating rate is crucial to identify the physical mechanisms responsible for the observed thermal evolution.

An intriguing feature appears in the $\Gamma$ profiles of CME1, CME5, CME7, CME8, and CME9. These events exhibit an initial rise in $\Gamma$, reaching a local maximum before a sharp decline. This reversal trend is clearly visible in Figure~\ref{fig:thermo}a. An increasing $\Gamma$ above the adiabatic value ($\Gamma > 5/3$) suggests progressively more heat being released at early heights following eruption. The maximum in $\Gamma$, interpreted as the peak of the heat-release phase, occurs between 3~($\pm$0.3) and 5~($\pm$0.5)~$R_\odot$, after which the heat-release rate diminishes as the CME transitions to a heat-absorbing state.

Additionally, for CME5 and CME8, this reversal in $\Gamma$ aligns with a corresponding reversal in their leading-edge or expansion acceleration, occurring at around 4~($\pm$0.4) and 3~($\pm$0.3)~$R_\odot$, respectively (Figure~\ref{fig:kinematics}c and~\ref{fig:kinematics}d). These features persist even when accounting for a 10\% uncertainty in the height estimates. For CME1, CME7, and CME9, although a reversal in $\Gamma$ is also evident, a corresponding reversal in acceleration is not observed. This may be because the acceleration peak occurred at lower heights than the earliest available observations. According to the three-phase kinematic evolution of CMEs described by \citet{Zhang2001}, fast CMEs are expected to exhibit a peak acceleration phase, which may coincide with the peak in $\Gamma$.

As CMEs propagate through the solar wind, the surrounding plasma could act as a potential source of heating. Therefore, comparing the thermodynamic behavior of CMEs (closed magnetic field structures) with that of the solar wind (open field structures) is of interest. Observations of large-scale proton variations in the solar wind indicate a polytropic index typically in the range $1.5 < \Gamma < 5/3$. Interestingly, small-scale fluctuations in the solar wind have been shown to follow a super-adiabatic evolution with $\Gamma \approx 2.7$ in the range 0.17–0.8~au \citep{Nicolaou2020}, consistent with a heat-release regime. Given the large surface area of CMEs, interactions with small-scale solar wind fluctuations could contribute non-negligibly to their internal heating. However, quantifying the heat exchange between CMEs and the ambient solar wind is beyond the scope of the present study.

\subsubsection{Heating Rate}{\label{chapt4-sec:temp}}

While the FRIS model provides the absolute value of $\Gamma$, parameters such as $dQ/dt$, $T$, and $p$ are obtained up to an unknown scaling factor (as noted in the second column of Table~\ref{chap2_tab:FRIS_parameters}). This scaling factor, which varies from event to event, depends on the best-fit coefficients derived from the model for each CME. However, it remains constant over time for a given CME. Due to the indeterminacy of this scaling factor, we cannot assess the absolute magnitudes of $dQ/dt$, $T$, or $p$. To facilitate inter-event comparison, we normalize the values of these thermodynamic quantities relative to their initial observed values. For consistency, the scaling factor for each parameter is chosen such that the first observed data point for all CMEs has the same normalized value. This normalization allows us to meaningfully compare the relative evolution of each thermodynamic property with heliocentric distance across different CMEs.

Figure~\ref{fig:thermo}b displays the variation of the heating rate per unit mass ($dQ/dt$) for the nine selected fast CMEs. To visualize the relative trends, we normalized the data such that the initial value of $dQ/dt$ for each event is set to $10^{10}$~J~kg$^{-1}$~s$^{-1}$. The data points corresponding to negative heating rates (indicating a heat-release state) are marked with stars, while positive heating rates (indicating heat injection or absorption) are shown as circles. All selected CMEs exhibit an initial phase of negative $dQ/dt$, signifying a heat-release state close to the Sun. This phase transitions into a positive heating rate at larger heliocentric distances, implying that the CMEs begin to absorb heat or experience internal heating. The transition from heat release to heat absorption occurs at heights between approximately 3.5~($\pm$0.3) and 6.5~($\pm$0.6)~$R_\odot$, varying across different events. These heating trends are consistent with the earlier-described behavior of the polytropic index $\Gamma$, supporting the interpretation of evolving thermodynamic states during CME propagation.

Furthermore, Figure~\ref{fig:thermo}b shows that both negative and positive heating rates generally decrease with distance from the Sun. However, near the transition region, several CMEs exhibit a brief increase in $dQ/dt$ following the initial heat-release phase, suggesting a short-lived enhancement in heating before entering a gradual decline. A similar feature appears in the evolution of $\Gamma$ (Figure~\ref{fig:thermo}a), where a local dip is observed near the transition to the isothermal index. A detailed discussion on the possible causes of the early-stage heat-release phase will be presented in Section~\ref{chapt4-sec:DEM}.

\subsubsection{Temperature}\label{chapt4-sec:dq/dt}

The temperature values derived from the FRIS model have been scaled such that each CME begins with an initial temperature of $10^7$~K at the first observed data point (Figure~\ref{fig:thermo}c). This normalization allows us to investigate the relative temperature evolution of all selected CMEs during their outward propagation from the Sun. In general, the temperature shows a rapid decline at lower heights, typically between 4~($\pm$0.4) and 8~($\pm$0.8)~$R_\odot$, and tends to stabilize at nearly constant values beyond this range. This trend is consistent with the behavior of previously discussed thermodynamic parameters, such as $\Gamma$ and $dQ/dt$. The rapid drop in temperature at initial heights can be attributed to the dominant heat-release phase ($\Gamma > 5/3$) and the high expansion rates of the CMEs. 

Interestingly, the temperature does not exhibit an immediate increase when the CME transitions into a heat-absorption phase (i.e., $\Gamma < 5/3$ and positive $dQ/dt$). This suggests that the internal heating during this stage may not be sufficient to overcome the adiabatic cooling caused by continued expansion. For instance, in the case of CME5, the $\Gamma$ value crosses below the adiabatic threshold ($\Gamma = 5/3$) near 5.5~($\pm$0.5)~$R_\odot$, indicating the onset of heat absorption. However, the temperature continues to decrease until approximately 6.5~($\pm$0.6)~$R_\odot$, where $\Gamma$ approaches unity. This lag suggests that temperature changes may propagate more gradually due to the finite thermal conduction or magnetoacoustic propagation timescales within the CME structure.

Additionally, we expect that heat injection should increase the internal pressure of the CME, which could enhance its expansion. Conversely, if a CME undergoes limited expansion, possibly due to its intrinsic structural rigidity or a confining external pressure, it may be unable to accommodate the injected heat, leading to enhanced heat release instead. This behavior is evident when comparing CME3 and CME6. Both events have similar propagation speeds (Figure~\ref{fig:kinematics}a), but CME6 exhibits a higher expansion speed than CME3 (Figure~\ref{fig:kinematics}b). As a result, CME3, which expands less, shows a higher rate of heat release (Figure~\ref{fig:thermo}b) and experiences a more pronounced temperature drop (Figure~\ref{fig:thermo}c) compared to CME6.

\begin{figure*}
    \centering
    \includegraphics[scale=0.25,trim={5cm 1cm 10cm 0cm}]{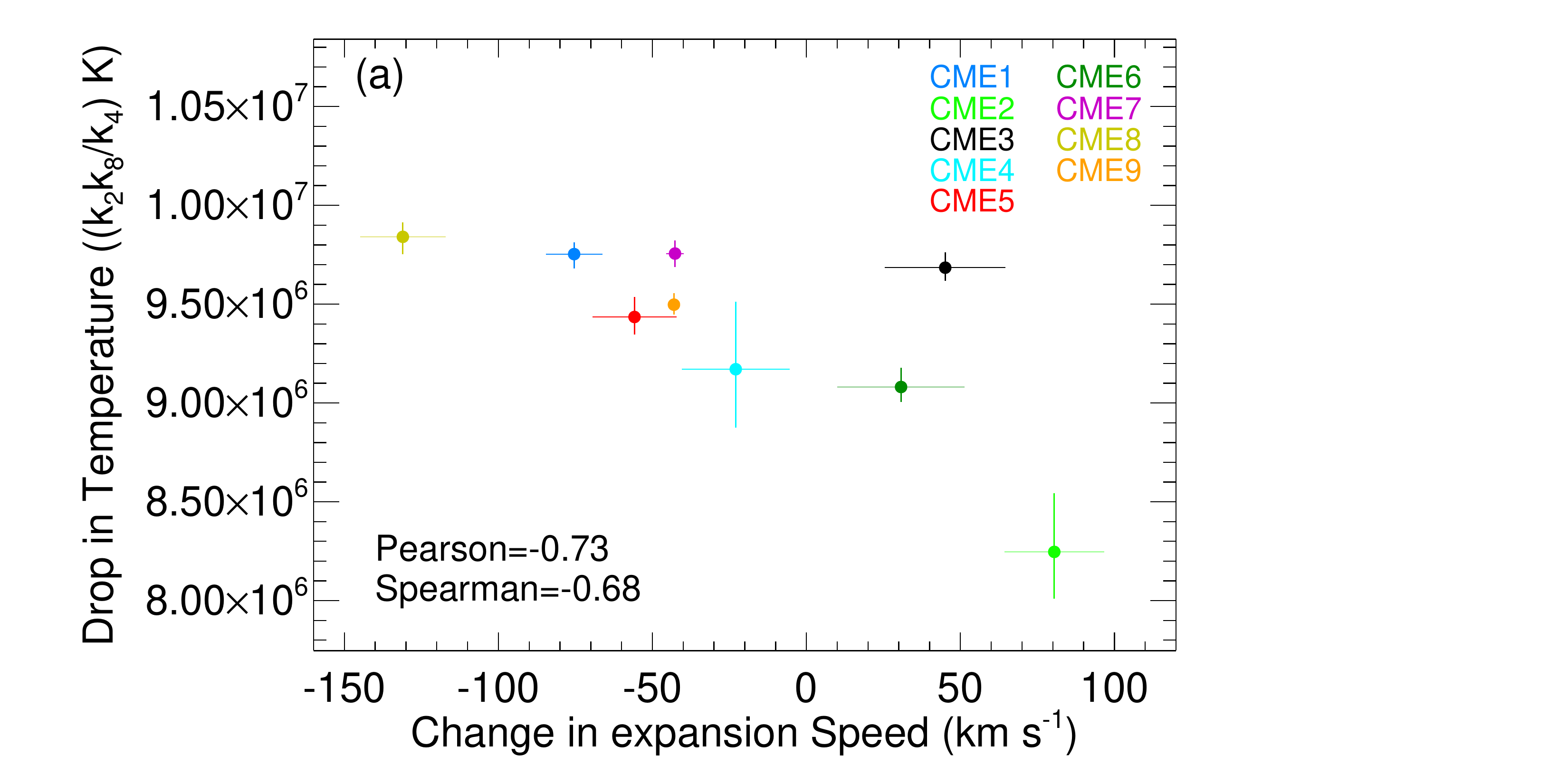}
    
    \includegraphics[scale=0.25,trim={9cm 1cm 15cm 0cm}]{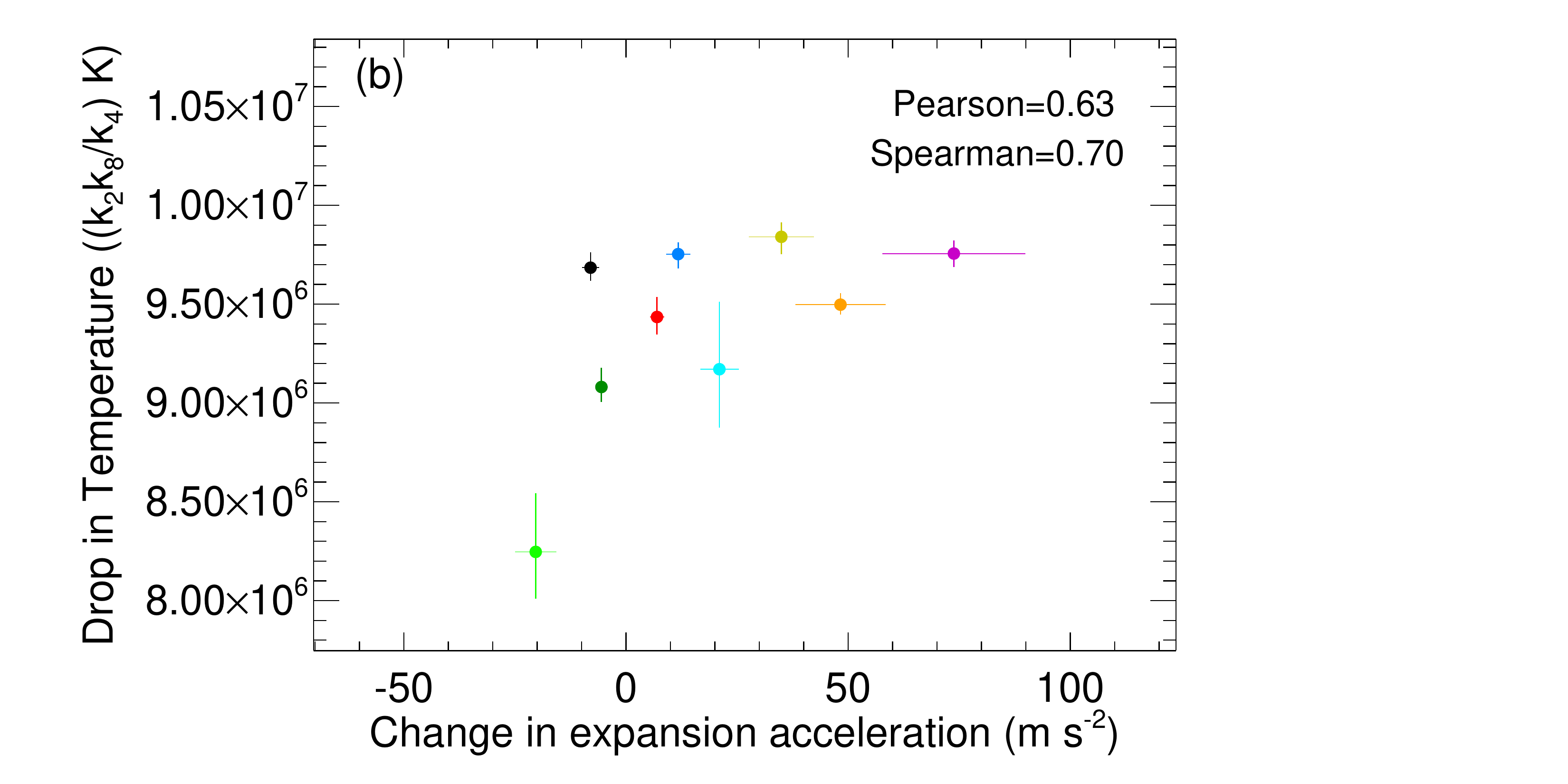}
    \caption[Scatter plots for change in expansion speed and expansion]{(a) Scatter plots for change in expansion speed (a) and expansion acceleration (b) with the initial rapid drop in temperature.}
    \label{fig:corr_tempe_speed_acc}
\end{figure*}

We also observe that CME1 and CME5 exhibit similar magnitudes of leading edge and expansion acceleration within the height range of 5.5~($\pm$0.5) to 6.5~($\pm$0.6)~$R_\odot$, while CME1 has lower values of both leading edge and expansion speeds compared to CME5 (Figure~\ref{fig:kinematics}). As a result, CME1 is expected to release more heat and experience a greater temperature drop. This interpretation is consistent with the observed trends in $dQ/dt$ and $T$ (Figures~\ref{fig:thermo}b and \ref{fig:thermo}c). A similar comparison can be made between CME8 and CME9. Although these two CMEs show comparable acceleration profiles between 5~($\pm$0.5) and 8~($\pm$0.8)~$R_\odot$, CME8 has a lower expansion speed than CME9 (Figure~\ref{fig:kinematics}). Despite being in the heat-absorption phase ($\Gamma<5/3$), both CMEs continue to exhibit a temperature decline in this height range. However, the temperature drop is more pronounced in CME8, suggesting a stronger heat-release response compared to CME9 (Figures~\ref{fig:thermo}b and \ref{fig:thermo}c).

To further understand the relationship between thermal and kinematic behavior, we analyzed the correlation between the initial temperature drop and global expansion parameters. Specifically, we computed the maximum temperature decrease (from initial to final values) before the CME reaches the isothermal regime, and compared this with the corresponding change in expansion speed over the same interval. We also examined the relationship between the temperature drop and the change in expansion acceleration. For both comparisons, we calculated the Pearson and Spearman correlation coefficients to capture linear and non-linear associations, respectively. We find a strong negative Pearson correlation of $-0.73$ between the initial temperature drop and the change in expansion speed (Figure~\ref{fig:corr_tempe_speed_acc}a). This suggests that CMEs exhibiting a greater decrease or weaker increase in expansion speed tend to undergo more substantial early temperature drops. This is consistent with our earlier observation that a higher expansion speed is associated with a more moderate temperature decline. Together, these results imply that CMEs with a high expansion speed but smaller deceleration (or moderate acceleration) tend to maintain their internal temperature more effectively.

Additionally, we observe a moderate positive linear correlation of 0.63 between the temperature drop and the change in expansion acceleration (Figure~\ref{fig:corr_tempe_speed_acc}b). This indicates that CMEs experiencing lesser deceleration or stronger acceleration tend to retain more thermal energy, leading to smaller temperature reductions. It is important to note that the temperature ($T$) derived from the FRIS model represents the mean of proton and electron temperatures, i.e., $T = (T_p + T_e)/2$, and hence provides an average measure of the CME plasma's thermal state. However, beyond the lower corona, proton and electron temperatures may evolve differently. A more accurate description would involve a two-fluid model with separate energy equations for electrons and protons \citep{Jin2013}. Moreover, previous studies have reported temperature anisotropy in both the ICME sheath and ambient solar wind \citep{Maruca2011, Zubair2023}, which may also be present in the CME flux-rope due to turbulence and small-scale instabilities. Turbulent processes could introduce anomalous resistivity, thereby enhancing Joule heating within the CME plasma \citep{Debesh2023}. While these factors are beyond the scope of the present study, our approach, based on a single-fluid analytical model, provides a foundational understanding of CME thermodynamics by capturing the essential features of their large-scale thermal evolution.

\subsubsection{Thermal Pressure}{\label{chapt4-sec:pressure}}

We also analyzed the evolution of thermal pressure for all the selected fast CMEs. The FRIS model-derived thermal pressure values were scaled so that each CME starts with a pressure of $10^3$~pPa, as shown in Figure~\ref{fig:thermo}d. A rapid decrease in thermal pressure is observed during the early phase of CME propagation, up to approximately 5~($\pm$0.5) to 8~($\pm$0.8)~$R_\odot$. This range corresponds to the region where CMEs exhibit strong heat release and a sharp drop in temperature. Beyond this phase, roughly between 6~($\pm$0.6) and 9~($\pm$0.9)~$R_\odot$, the thermal pressure continues to decline but at a much slower rate. This change in behavior aligns with the onset of heat absorption and the subsequent stabilization of temperature. In the following section, we further explore the role of thermal pressure in driving the internal forces and dynamics of CMEs.

\begin{sidewaysfigure}
    \centering
\includegraphics[scale=0.45,trim={1cm 2cm 2cm 1.5cm}]{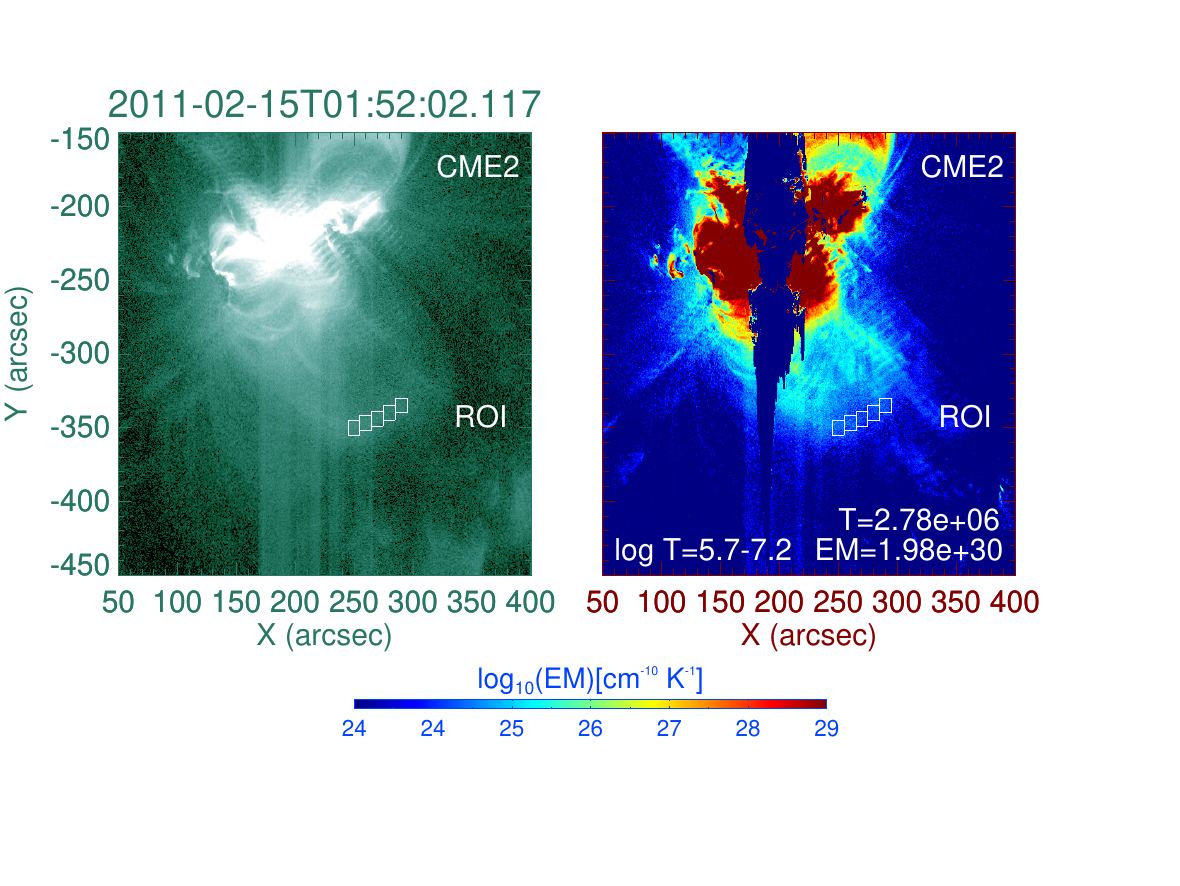}
\includegraphics[scale=0.45,trim={1cm 2cm 2cm 1.5cm}]{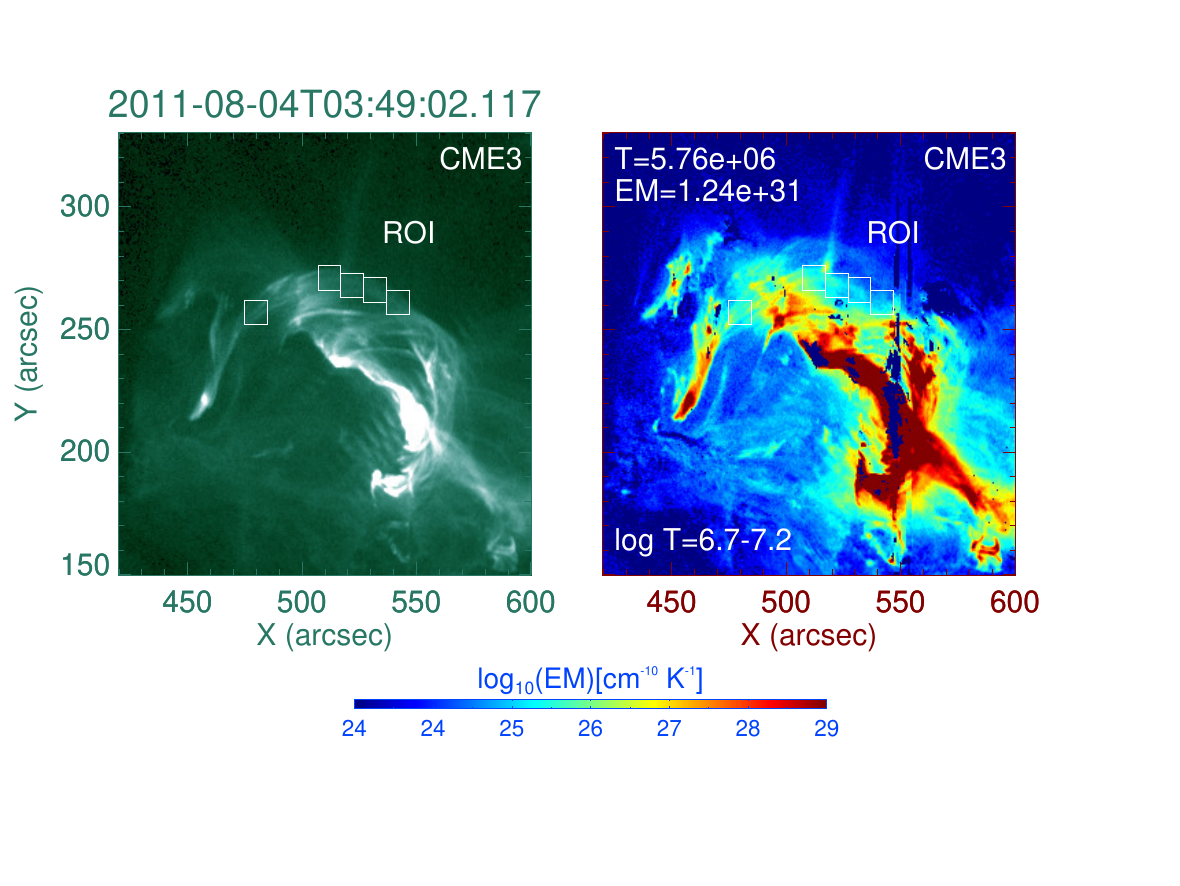}
\includegraphics[scale=0.45,trim={1cm 2cm 2cm 1.5cm}]{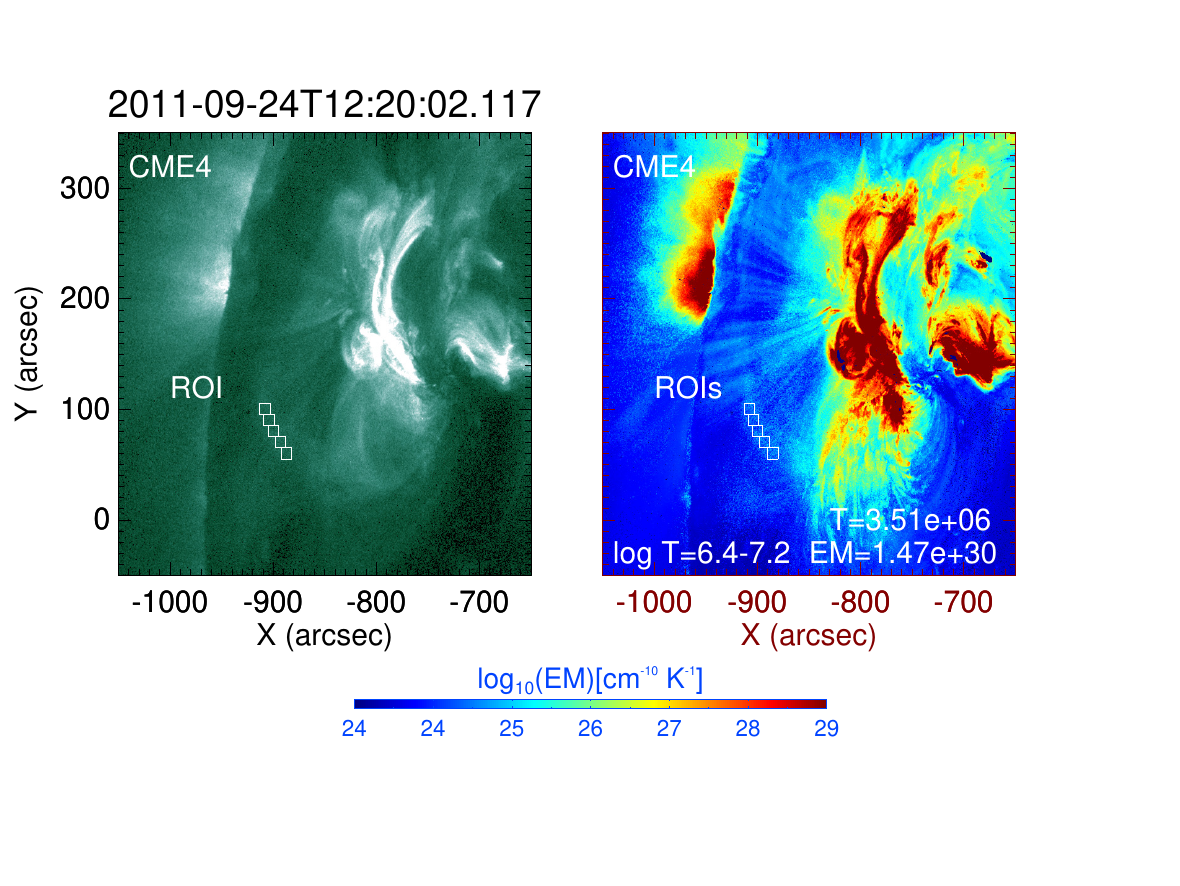}\\
\includegraphics[scale=0.45,trim={1cm 2cm 2cm 1.5cm}]{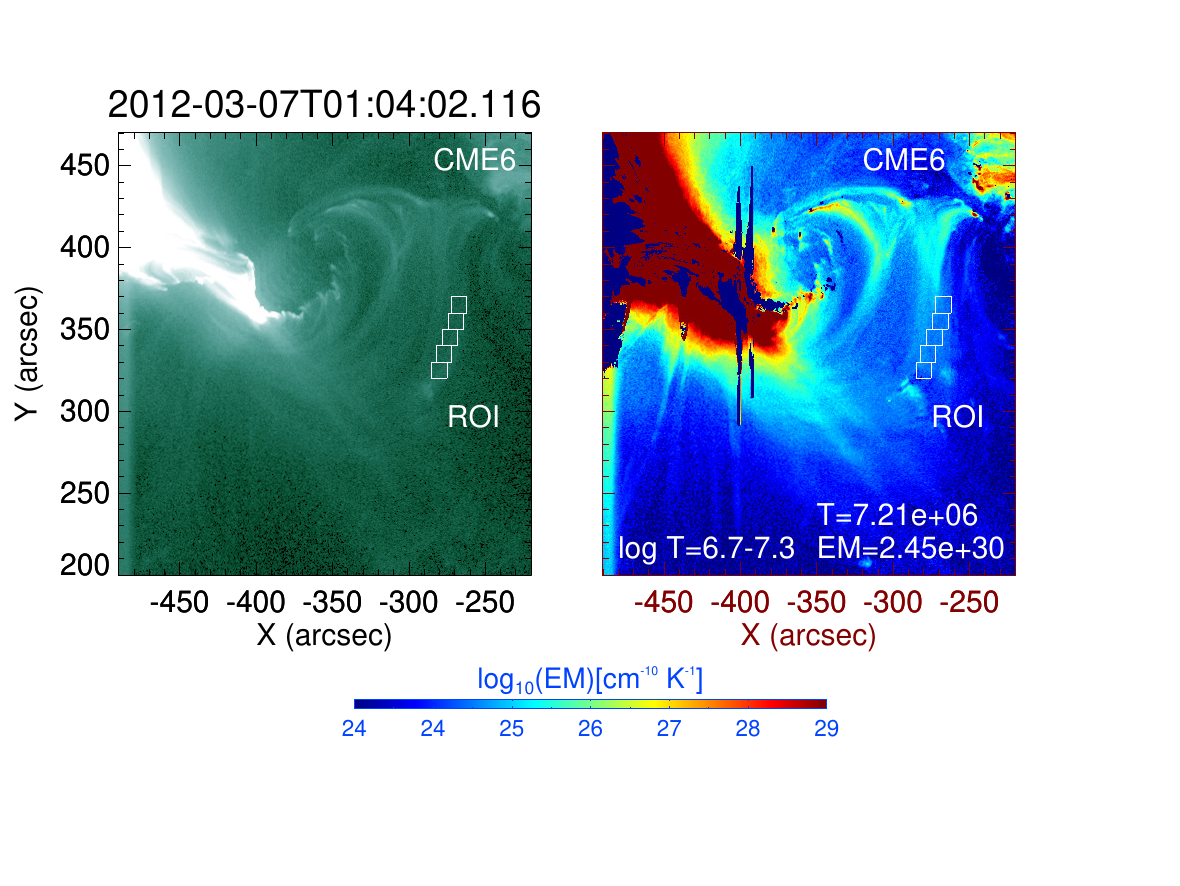}
\includegraphics[scale=0.45,trim={1cm 2cm 2cm 1.5cm}]{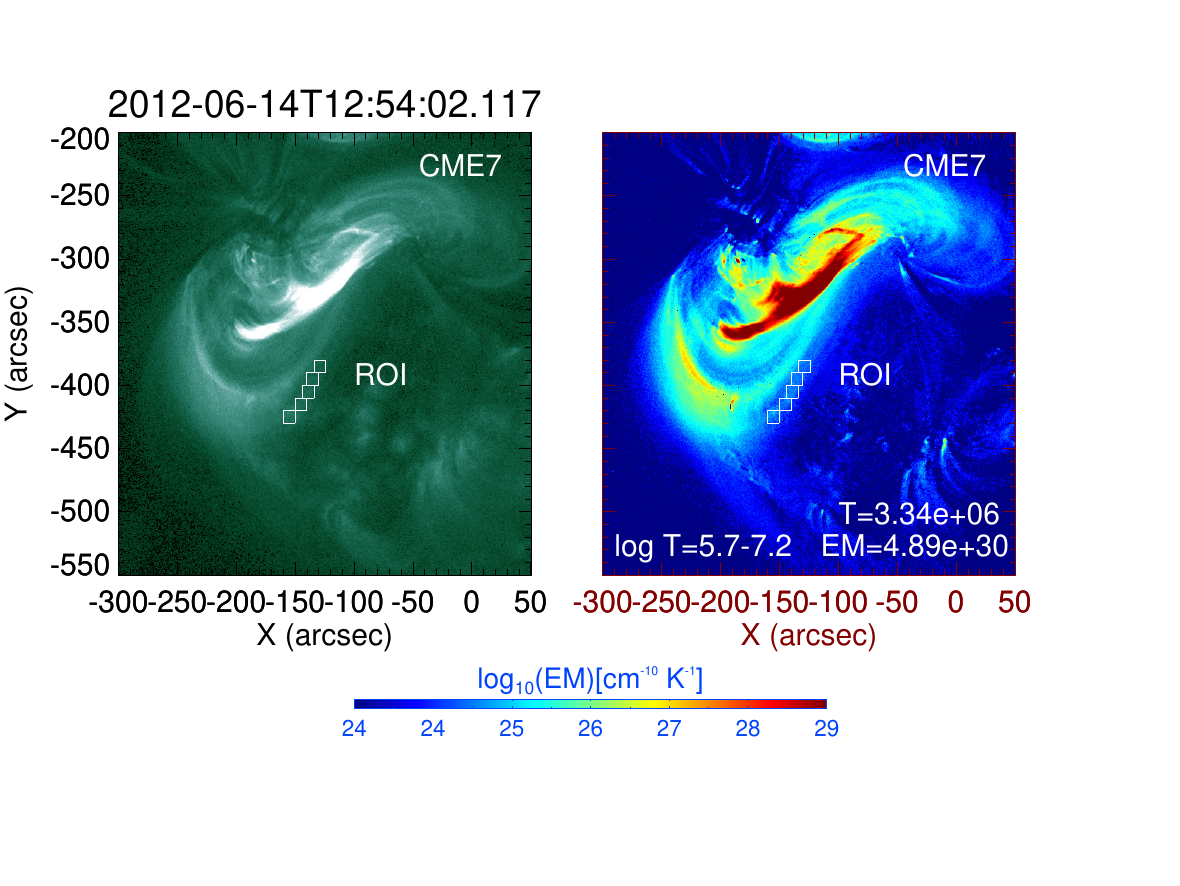}
\includegraphics[scale=0.45,trim={1cm 2cm 2cm 1.5cm}]{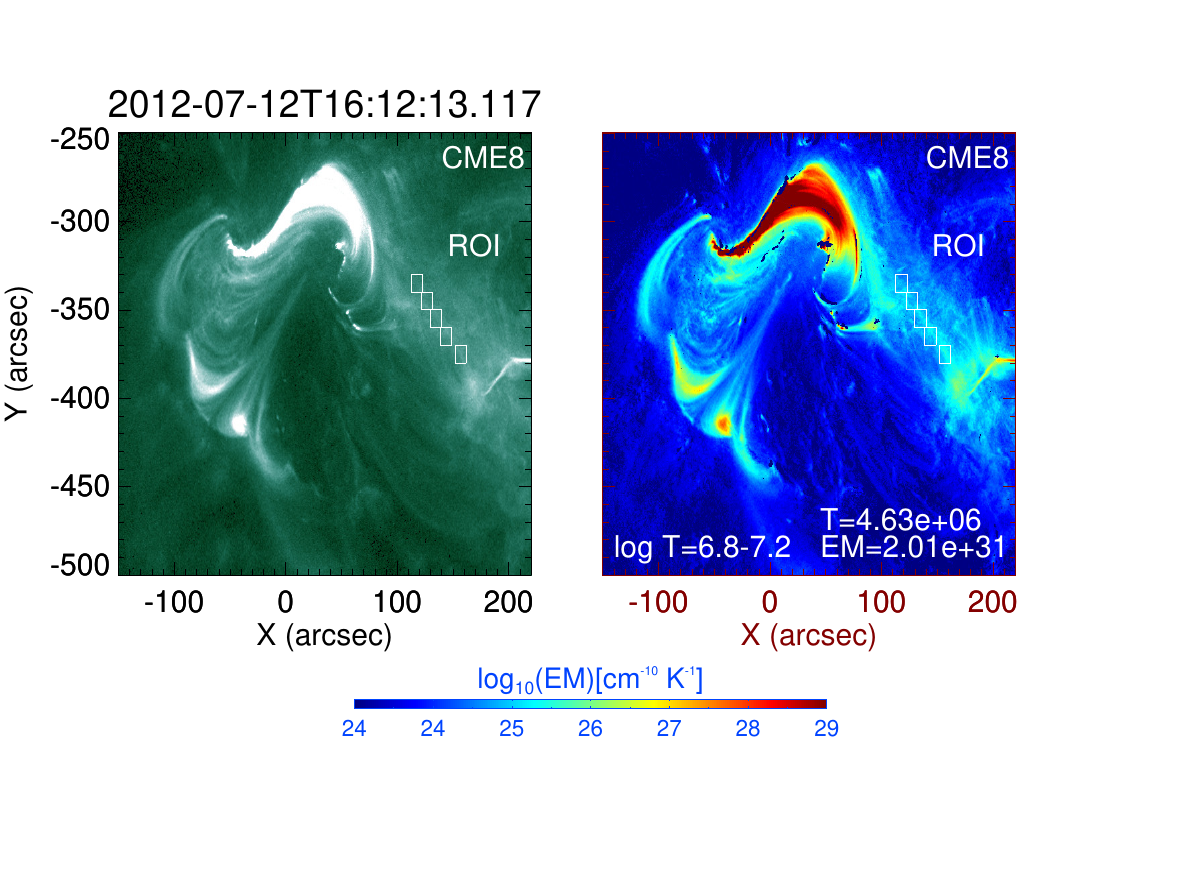}

\caption[SDO/AIA 94~{\AA} images and the corresponding Differential]{SDO/AIA 94~{\AA} images and the corresponding Differential Emission Measure (DEM) maps for the selected CMEs. The white boxes (10 arcsec $\times$ 10 arcsec) mark the regions of interest (ROIs) on the flux-rope where DEM analysis is performed to estimate the average DEM-weighted temperature ($T$) and emission measure (EM). The logarithmic temperature values (log~$T$) indicated on each DEM map denote the selected temperature range optimized for clear visualization of the flux-rope structure.}
\label{fig:aia}
\end{sidewaysfigure}

\subsection{Multi-wavelength Imaging Observations}\label{chapt4-sec:DEM}

All the selected fast-speed CMEs in this study exhibit an early heat-release phase at low heights, as inferred from coronagraphic observations. To better understand this behavior, it is crucial to examine the thermal conditions of the CME flux-rope at its origin, particularly below the coronagraphic field of view. CMEs are fundamentally structured around a magnetic flux-rope (MFR), a coherent bundle of helical magnetic field lines wrapped around a common axis \citep{Chen2017, Wang2017, Wang2018, Green2018}. Ideal MHD models attribute the formation of MFRs in the solar corona to the force-free nature of the magnetic field, where electric currents flow parallel to the field lines. These field-aligned currents generate poloidal magnetic flux, causing the field lines to twist and evolve into flux-rope structures \citep{Kliem2006, Torok2010, Amari2014}. The association of MFRs with CMEs is well-supported by multi-wavelength imaging and magnetic field reconstructions \citep{Chen2017, Wang2017, Gou2019, Duan2019}.

To probe the thermal structure of the flux-rope in the low corona, we employed high-resolution EUV observations from the SDO/AIA instrument across multiple wavelengths. These observations enable us to assess the plasma temperature distribution at the early stages of CME eruption, particularly in the source region, where coronal heating processes are dominant.

We performed DEM analysis to estimate the plasma temperature of the erupting flux-rope (see Section~\ref{chapt2-metho-dem}). The DEM technique characterizes the amount of optically thin plasma along the line of sight as a function of temperature. The identification of CME-associated flux ropes was based on EUV signatures such as flares, post-eruptive arcades, and rising EUV structures \citep{Liu2010a, Cheng2011}. Of the nine selected CMEs, flux ropes could not be reliably identified for two events (CME5 and CME9), and AIA observations were unavailable for CME1. Due to the Earth-directed nature of these CMEs, significant projection effects in AIA images made flux-rope identification challenging in some cases. We utilized the sparse inversion method developed by \citet{Cheung2015}, applying it with a temperature resolution of log $T \leq 0.1$. For each CME, five regions of interest (ROI) of 10 arcsec $\times$ 10 arcsec were selected on the flux-rope, carefully chosen to minimize contamination from surrounding coronal activity. Although these ROIs are smaller than the full flux-rope structure, they are assumed to represent the general thermal state of the eruption.

Figure~\ref{fig:aia} presents the results of the DEM analysis. The derived average electron temperatures within the ROIs range from $2.78 \times 10^6$~K to $7.21 \times 10^6$~K, indicating the presence of a hot flux-rope structure at the eruption site. It is important to note that this temperature is an average over the selected ROIs and may be underestimated due to contributions from cooler background plasma. Nonetheless, the presence of such hot plasma supports the conclusion that the selected fast CMEs are significantly heated at their origin. This initial heating aligns well with the thermodynamic trends observed from the FRIS model: a polytropic index $\Gamma > 5/3$, negative heating rate ($dQ/dt < 0$), and a steep temperature decline at low coronal heights. Finally, it is important to recognize that the DEM analysis reflects the thermal state of electrons only, whereas the FRIS model considers contributions from both electrons and protons. This distinction should be kept in mind when interpreting the heating state derived from these two complementary diagnostics.

\begin{figure}
    \centering
    
    \includegraphics[scale=0.25,trim={5cm 1cm 10cm 0cm}]{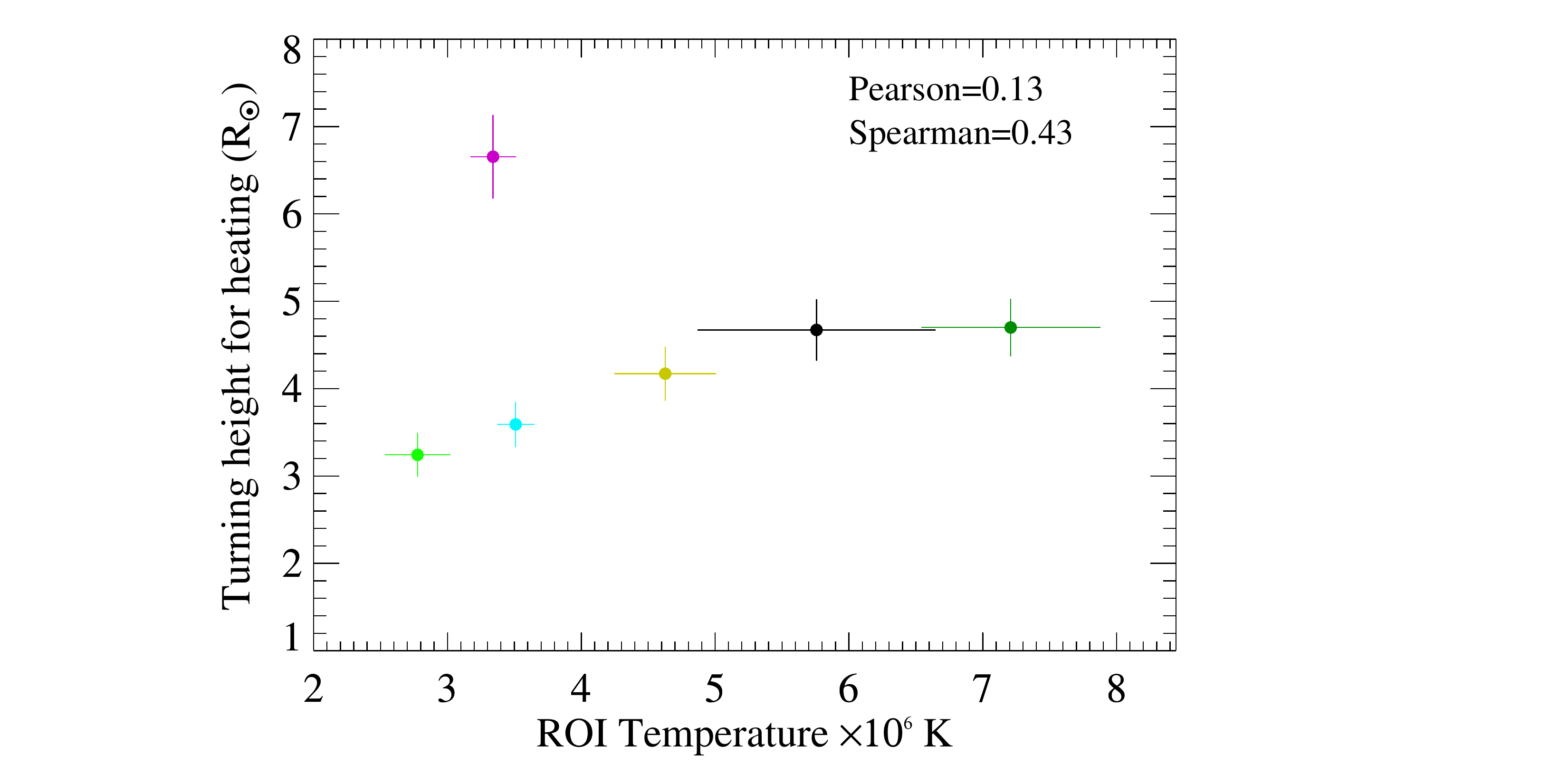}
     
    \caption[Scatter plot illustrating the correlation between the average]{Scatter plot illustrating the correlation between the average DEM-weighted electron temperature of the flux-rope (measured from selected regions of interest) and the turning point in heating, marked by the heliocentric height at which the polytropic index ($\Gamma$) crosses the adiabatic threshold ($\Gamma = 5/3$). Error bars in both directions represent the estimated uncertainties in temperature and turning height.}
    \label{fig:corr_turningH_roiTemp}
\end{figure}

As discussed earlier, each selected fast CME transitions from a heat-release phase to a heat-absorption phase at a specific coronal height, referred to as the turning height for heating, where $\Gamma$ drops below the adiabatic value ($\Gamma = 5/3$). Figure \ref{fig:corr_turningH_roiTemp} presents a scatter plot comparing the average DEM-weighted electron temperature of the flux-rope (measured from regions of interest) with the corresponding turning height. The red error bars represent the maximum and minimum temperatures across the selected ROIs. Our analysis reveals only a weak positive linear correlation coefficient of 0.13, indicating that the inherent thermal state of the flux-rope in the low corona does not strongly influence the height at which the CME transitions into a heat-absorption state. This suggests that the turning height is likely governed by a more complex interplay of factors, potentially a combination of the CME’s initial thermal energy and its evolving kinematic properties. Future investigation into these factors may help better constrain the physical mechanisms that determine the onset of heating in fast CMEs during their early propagation.

\subsection{Evolution of Internal forces}{\label{chapt4-sec:forces}}

\begin{figure*}
         \includegraphics[scale=0.20,trim={5cm 1cm 20cm 0cm}]{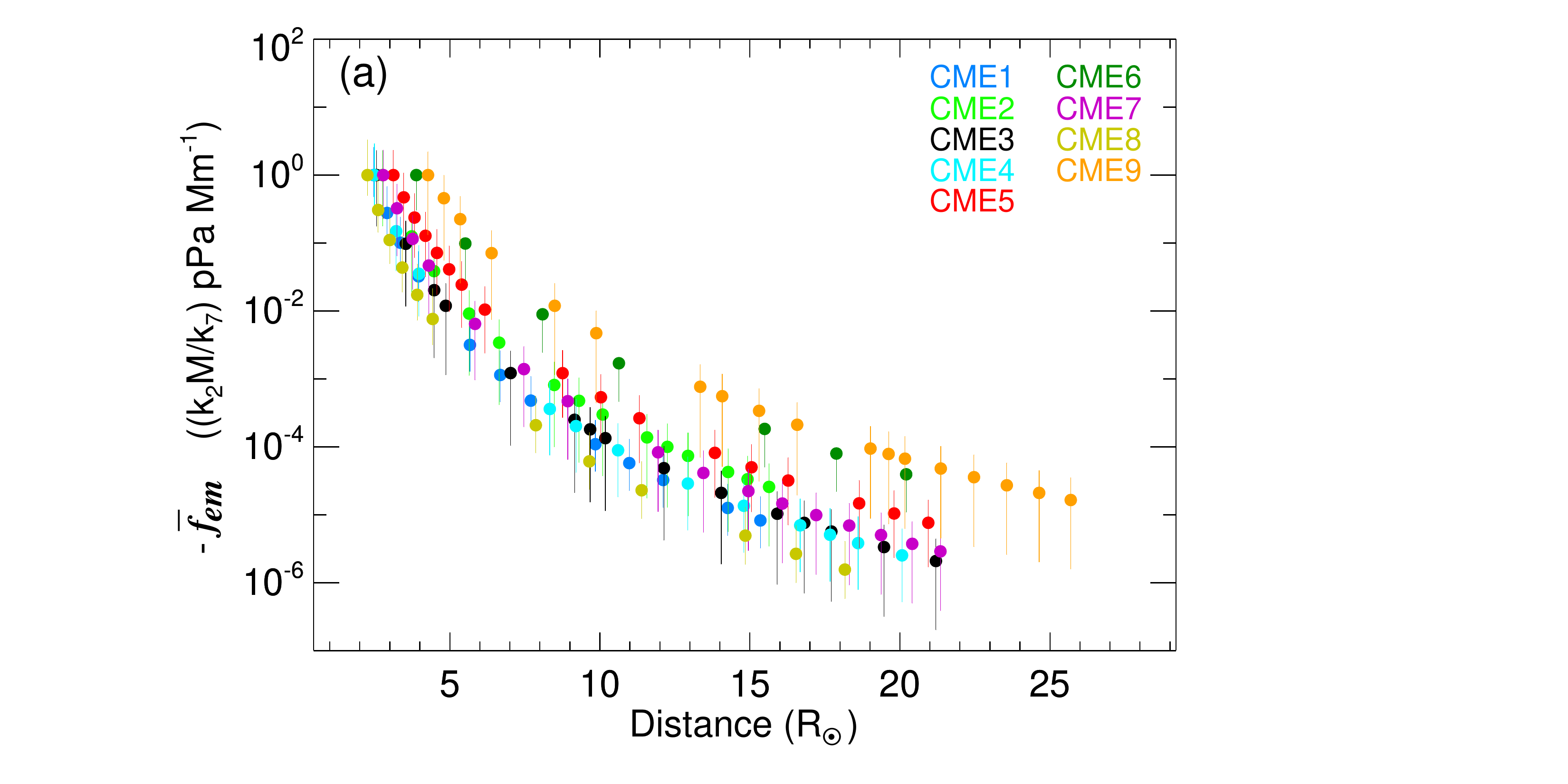} 
         \includegraphics[scale=0.20,trim={-1cm 0cm 24cm 0cm}]{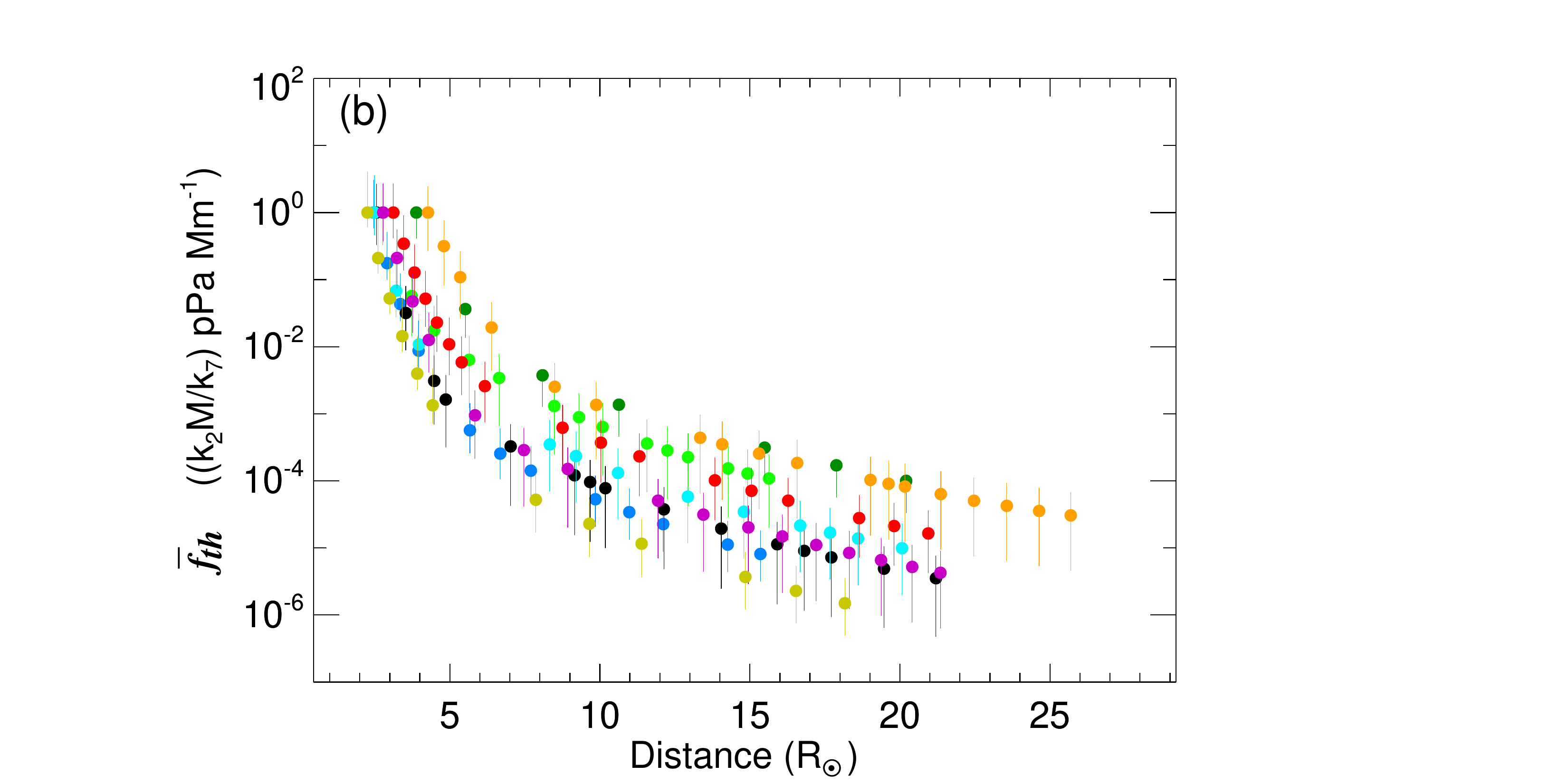}\\ \hspace{-2.5cm}
         \includegraphics[scale=0.20,trim={5cm 1cm 20cm 1cm}]{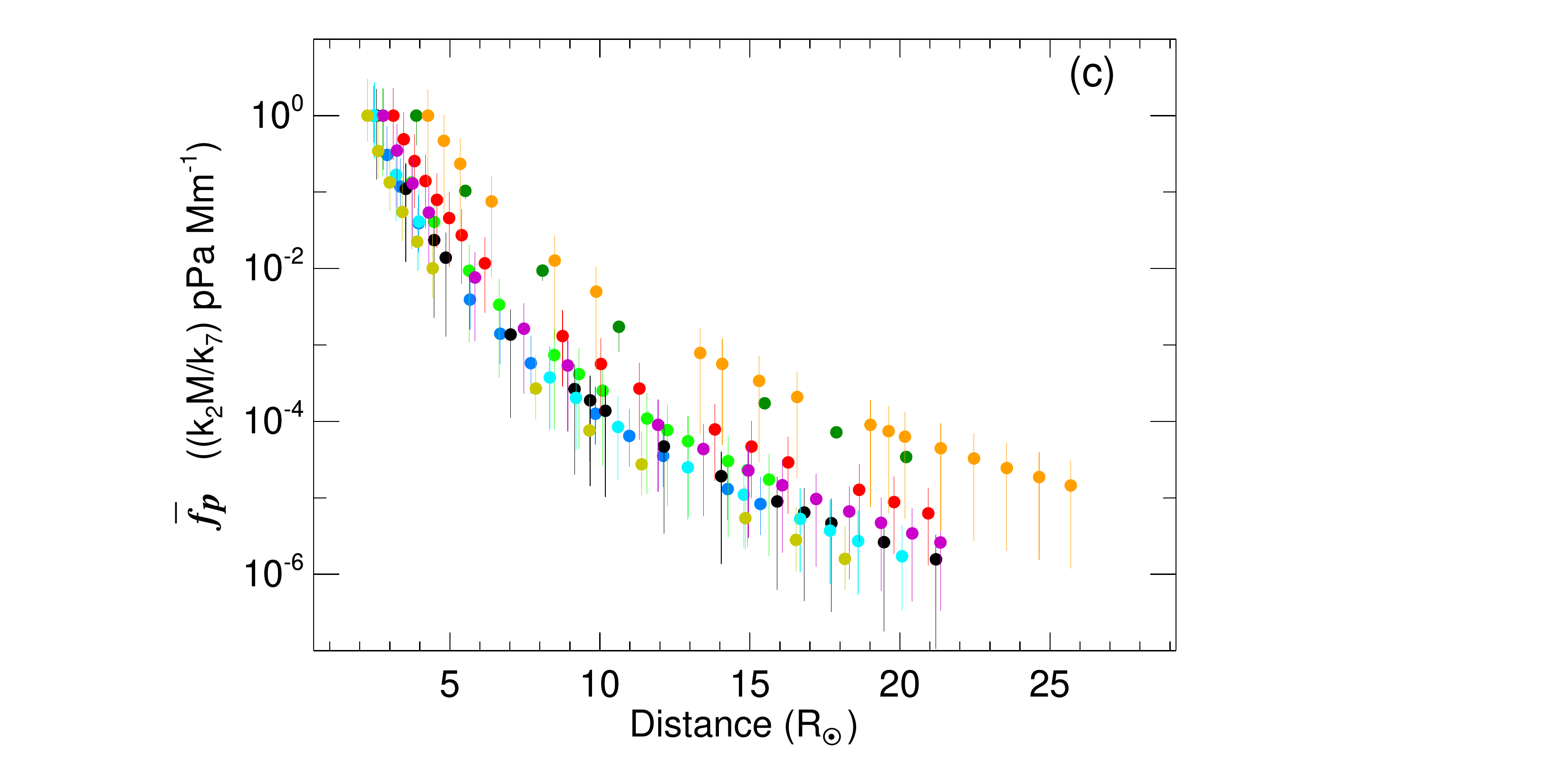} 
         \includegraphics[scale=0.20,trim={-1cm 1cm 24cm 1cm}]{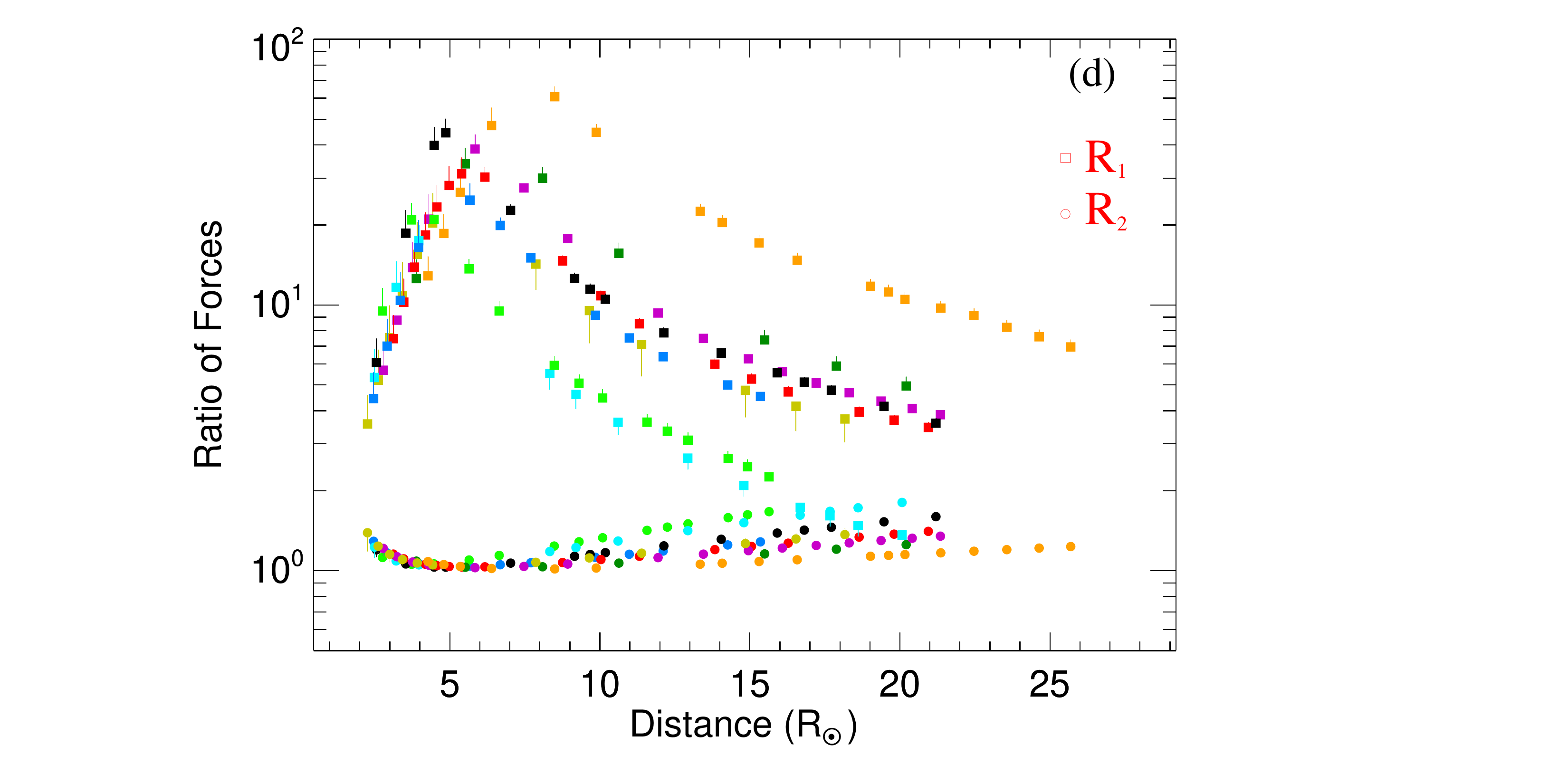}
         \caption[Evolution of internal forces]{Evolution of internal forces: (a) Variation of the Lorentz force ($\bar{f}_{em}$), (b) thermal pressure force ($\bar{f}_{th}$), (c) centrifugal force ($\bar{f}_{p}$), and (d) force ratios $R_1 = \bar{f}_{em} / \bar{f}_{th}$ and $R_2 = \bar{f}_{em} / \bar{f}_{p}$, as functions of the CME’s leading-edge heliocentric distance ($h$). The vertical error bars represent uncertainties derived from a 10\% perturbation in the input leading-edge height and the subsequently propagated errors through the FRIS model.}
        \label{fig:forces}
\end{figure*}

We computed the radial components of the internal forces per unit volume, namely, the Lorentz force (${\bar f}_{em}$), thermal pressure force (${\bar f}_{th}$), and centrifugal force (${\bar f}_{p}$), that govern the expansion dynamics of the CME flux-rope (see Table~\ref{chap2_tab:FRIS_parameters}). Since each force is scaled by the same unknown multiplicative factor for a given CME, we normalized all internal force profiles such that their initial values are set to 1~pPa~Mm$^{-1}$ for each event (Figure~\ref{fig:forces}). This scaling enables both intra- and inter-event comparisons of force evolution.

Our analysis shows that for all selected fast CMEs, the Lorentz force (${\bar f}_{em}$) remains negative, while both thermal pressure (${\bar f}_{th}$) and centrifugal (${\bar f}_{p}$) forces are positive throughout the observational heights. The negative Lorentz force implies an inward-directed force, acting against the radial expansion of the flux-rope, whereas the thermal pressure and centrifugal forces promote expansion. This result is consistent with earlier thermodynamic studies of CMEs \citep{Khuntia2023}.

Although the global magnetic pressure inside CMEs may suggest an outward-directed Lorentz force, local variations can yield a more complex force distribution. In the FRIS model, the sign and magnitude of the Lorentz force are governed by the coefficient $c_2$, which is in turn dependent on the radial profile of the axial magnetic field component, i.e., ${B_z}^2(0) - {B_z}^2(R)$ (see Equations \ref{eqn:fem2} and \ref{eqn:fem_final}). All CMEs analyzed in this study yielded positive values for $c_2$ (Table~\ref{tab:fitting_coefficients}), suggesting a stronger $B_z$ near the flux-rope's axis, resulting in an inward-directed Lorentz force. A steep decline in thermal pressure at lower heights, as discussed in Section~\ref{chapt4-sec:thermo}, also manifests as a sharp drop in the thermal pressure force between 5($\pm$0.5) and 8($\pm$0.8)~$R_\odot$ (Figure~\ref{fig:forces}b).

\begin{figure}
    \centering
    \includegraphics[scale=0.25,trim={10cm 1cm 15cm 0cm}]{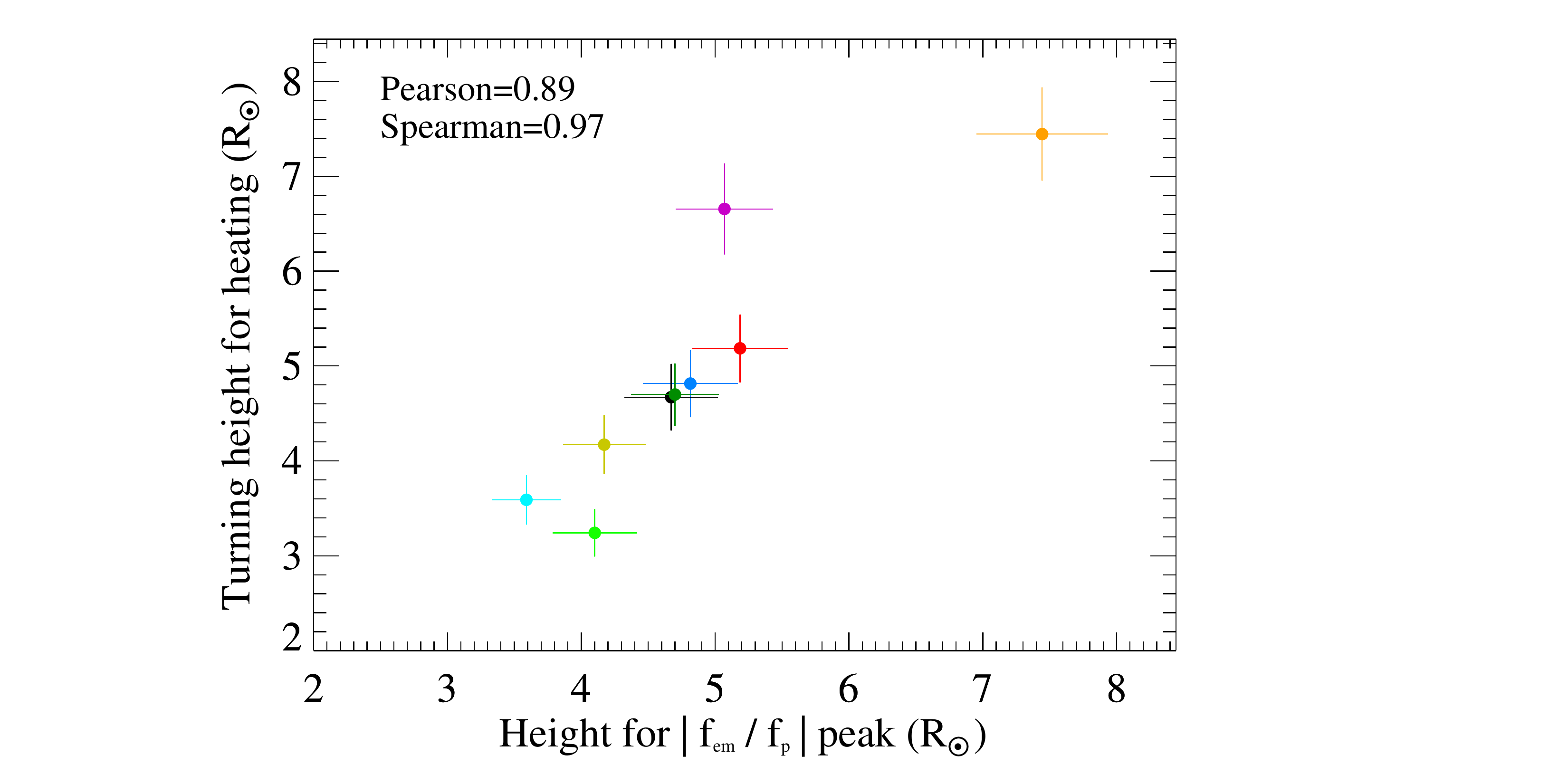}
    \caption[Scatter plot showing the relationship between the height]{Scatter plot showing the relationship between the height at which $R_1$ (the ratio of Lorentz force to thermal pressure force) reaches its peak and the turning height for heating, where $\Gamma$ crosses the adiabatic value. The horizontal and vertical error bars represent the estimated uncertainties in the respective parameters.}
    \label{fig:corr_turningH_fpeak}
\end{figure}

To evaluate the relative decay of internal forces, we computed two force ratios: $R_1 = {\bar f}_{em}/{\bar f}_{th}$ and $R_2 = {\bar f}_{em}/{\bar f}_{p}$ (Figure~\ref{fig:forces}d). For all CMEs, $R_1$ increases initially, peaking at heights between 3($\pm$0.3) and 8($\pm$0.8)~$R_\odot$, followed by a moderate decline. This trend indicates that the thermal pressure force decreases more rapidly than the Lorentz force at early heights, with the trend reversing afterward. Conversely, $R_2$ shows a decreasing trend up to 3($\pm$0.3)–8($\pm$0.8)~$R_\odot$ and then increases, implying that the centrifugal force initially decays more slowly than the Lorentz force but later declines more rapidly.

The force ratios $R_1$ and $R_2$ remain above unity at the lower heliocentric distances, indicating that the Lorentz force dominates over the thermal pressure and centrifugal forces during the early expansion phase of the CME flux-rope. However, since the net radial expansion is governed by the combined effect of all three forces, the evolution of these ratios carries significant implications. The observed decreasing trend in $R_1$ suggests that this ratio will eventually fall below unity at greater heights, beyond which the thermal pressure force overtakes the Lorentz force in driving expansion. Similarly, the increasing trend in $R_2$ indicates that $R_1$ and $R_2$ will intersect at some heliocentric distance, after which the thermal pressure force also dominates over the centrifugal force. Interestingly, for CME2 and CME4, the intersection of $R_1$ and $R_2$ occurs within our observation range, around 17($\pm$1.7)~$R_\odot$, suggesting that thermal pressure plays a significant role in driving CME expansion at greater heights.

Upon comparing the evolution of the polytropic index $\Gamma$ (Figure~\ref{fig:thermo}a) with that of the internal force ratios (Figure~\ref{fig:forces}d), we discovered a strong correspondence between the CME’s thermodynamic transition and the peak value of $R_1$. Specifically, we found that the height at which $R_1$ reaches its maximum closely coincides with the turning height for heating, defined as the point where the polytropic index crosses the adiabatic threshold ($\Gamma = 5/3$). A correlation analysis between these two quantities yields a strong positive linear correlation coefficient of 0.89 (Figure~\ref{fig:corr_turningH_fpeak}). This finding implies that CMEs reaching the peak Lorentz-to-thermal-pressure ratio at greater heights also undergo the thermodynamic transition (from heat release to heat absorption) at correspondingly higher heights.

This result strongly supports the interpretation that all selected fast CMEs are in a heat-release state when the Lorentz force is significantly stronger than the thermal pressure force. Given that the Lorentz force acts inward and inhibits radial expansion, its dominance can suppress expansion-driven cooling or heating effects. In such scenarios, the CME may be unable to expand sufficiently, causing it to release internal heat. During this phase, expansion is instead facilitated primarily by thermal pressure and centrifugal forces. Once the Lorentz force becomes subdominant, beyond the turning height, the CME enters a heat-absorption state and continues expanding freely under the influence of outward forces, consistent with the observed increase in positive heating rate.

\subsection{Summary}{\label{chapt4-part1-sec:summary}}

This study presents a detailed investigation of the radial evolution of thermal properties and internal forces for nine fast CMEs. By combining the FRIS model with 3D kinematic measurements and complementary multi-wavelength EUV observations, we examine the thermodynamic behavior and force balance of these events. The key conclusions of this study are as follows:

\begin{enumerate}
    \item All selected fast CMEs exhibit an initial phase of heat release, transitioning through an adiabatic state at heights between 3($\pm$0.3) and 7($\pm$0.7)~$R_\odot$, and subsequently enter a phase of heat absorption during their outward propagation.

    \item During the later stages of propagation, all CMEs maintain a nearly isothermal state, as indicated by polytropic index values in the range of 0.8–1.2.

    \item Our findings show that CMEs with larger expansion speeds and a greater decrease or smaller increase in expansion speed at lower heights tend to undergo a smaller drop in temperature.

    \item Multi-wavelength imaging and DEM analysis reveal that the CMEs originate from hot flux-rope structures with elevated electron temperatures in the low corona. These observations support the FRIS-derived inference of an initial heat-release state near the Sun.

    \item The expansion of CMEs near the Sun is primarily driven by the combined action of thermal pressure and centrifugal forces. At greater distances, thermal pressure becomes the dominant outward force, while the Lorentz force consistently acts inwards, opposing expansion throughout the observed domain.

    \item Understanding the thermal history of CMEs is essential for identifying the dominant heating and cooling mechanisms that operate at different heliocentric distances. The evolving nature of the polytropic index underscores the importance of constraining $\Gamma$ using both in situ measurements and analytical modeling.

\end{enumerate}

Future work will aim to extend this analysis to greater heliocentric distances by incorporating multi-point in situ observations from missions such as PSP, SolO, and ACE/Wind. These observations will allow for improved constraints on the thermodynamic evolution of CMEs and refinement of model-based diagnostics.

\section{Thermal Properties of Coronal Mass Ejections at 1 au and Their Implications for Geoeffectiveness Over Solar Cycles 23–25}\label{chapt4-part2}

One of the primary objectives in CME research is to forecast their arrival time at Earth and evaluate their potential to trigger geomagnetic storms. Earlier studies have investigated key CME parameters such as magnetic field strength, propagation speed, and radial size/duration, as these are known to play a major role in storm generation \citep{Srivastava2004, Jian2006, Zhang2007, Echer2008}. Given that CMEs consist of magnetized plasma, their global kinematics, including acceleration and expansion, are inherently influenced by their internal thermodynamic state. Although the thermal state of ICMEs at 1~au may not directly determine their geoeffectiveness, it can act as a proxy for other underlying properties that control storm-driving capability. Therefore, a better understanding of additional plasma characteristics is essential for improving both CME arrival-time forecasts and geoeffectiveness assessments. In particular, it is important to investigate whether the thermal properties of CMEs at 1~au show any systematic dependence on the characteristics of the geomagnetic storms they produce. However, our current understanding of CME thermodynamic evolution remains limited, largely due to observational constraints as discussed in \ref{sec:chapt1-gaps}.

To overcome these limitations, we have combined CME 3D kinematics with the FRIS model to derive plasma properties, particularly the thermal state, in the critical range of 2–25~$R_{\odot}$ where direct plasma observations are lacking (Section~\ref{Ch3-sec:1}). These works have shown that CMEs often undergo multiple heat-transfer phases: an initial heat-releasing stage close to the Sun, followed by a near-isothermal heating phase between $\sim$3–9~$R_{\odot}$. This thermal evolution is frequently interpreted using a polytropic approach, where pressure ($p$) and temperature ($T$) relate to density ($\rho$) via $p \propto \rho^{\Gamma}$ and $T \propto \rho^{\Gamma-1}$, with $\Gamma$ being the polytropic index. Such a formulation provides a simplified yet insightful description of CME heating and cooling, avoiding the full complexity of solving the complete energy equation. The choice of $\Gamma$ is critical not only for CME characterization but also for understanding broader coronal and solar wind dynamics, as well as heat transport in the heliosphere \citep{Riley2003, Manchester2004, Desai2020, Cao2013, Liu2022polymhd, Kuzma2023, Cai2025}. In many numerical models, a fixed non-adiabatic $\Gamma$ is assumed, yet simulations demonstrate that varying $\Gamma$ can substantially alter the modeled evolution of both the solar wind and CMEs \citep{Manchester2004, Wu2004, Liou2014, Mayank2022, Liu2022polymhd}. As discussed in Section~\ref{Ch4-part1}, we reported that both fast and slow ICMEs at 1~au exhibit near-isothermal expansion, sustained internal heating, and distinct turbulence characteristics, pointing to the role of Alfvénic fluctuations and intermittent structures in energy dissipation. Evidence for CME heating is also supported by spectroscopic measurements near the Sun \citep{Filippov2002, Lee2017, Reva2023} and ion charge state analyses at 1~au \citep{Rakowski2007, Lepri2012}. Furthermore, We will discuss in later Section~\ref{chap5-sec:nearsun}, where we applied the FRIS model in conjunction with in situ data at 1~au to examine interacting CMEs responsible for a great geomagnetic storm, finding enhanced electron cooling alongside dominant proton heating, indicative of complex thermodynamic interactions during CME–CME encounters. While such studies have examined CME polytropic behavior at specific distances or time intervals, a systematic statistical analysis linking the thermal states of ICMEs and ambient solar wind at 1~au to their geoeffective properties remains scarce. This gap motivates the present investigation, which aims to statistically characterize ICME thermal states at 1~au using in situ measurements.

Previous work has shown that ICME geoeffectiveness is strongly tied to parameters such as southward magnetic field ($B_z$), convective electric field ($E_y = V B_z$), and solar wind dynamic pressure, all of which are key drivers of geomagnetic activity \citep{Echer2008, Richardson2011, Zhao2021}. Additional studies have examined how different solar wind drivers influence storm onset, intensity, growth, and recovery \citep{Echer2008, Richardson2012, Mishra2024}. Ion charge states, although not causative of geomagnetic storm intensity, serve as important markers for distinguishing solar wind and ICME types, which in turn may differ in their likelihood of producing strong storms \citep{Henke1998, Lepri2001, Owens2018}. This suggests that charge state signatures, linked to CME thermal conditions near the Sun, may indirectly relate to geoeffectiveness. However, the role of an ICME’s thermodynamic state near Earth in modulating storm intensity has received comparatively little attention. Because CME thermodynamic evolution influences expansion, density, and pressure gradients, it may also indirectly reflect the CME's efficiency in transferring energy to Earth's magnetosphere. Consequently, examining the thermal properties of both CMEs and the surrounding solar wind at 1~au may offer additional predictive value, complementing established magnetic, velocity, and density parameters, in the context of extreme space weather.

This study aims to statistically investigate the thermal properties of MEs associated with CMEs at 1 au, spanning Solar Cycles 23, 24, and the rising phase of Cycle 25. Using a polytropic framework, we estimate the proton polytropic index ($\Gamma_p$) from in situ measurements and categorize ICMEs based on their heating or cooling state. We then examine how these thermal states relate to the CME’s plasma characteristics, occurrence rate, and geoeffective potential. By linking the CME thermal behavior to storm intensity, we explore whether heating-dominated or cooling-dominated ICMEs are more likely to cause strong geomagnetic storms. This analysis helps deepen our understanding of how thermal evolution modulates CME geoeffectiveness over solar cycles.

\subsection{Data \& Methodology} \label{chapt4-part2-sec:data}

We employed 1-minute resolution data from the OMNI database \citep{King2005}, provided by NASA’s Space Physics Data Facility at the Goddard Space Flight Center (\url{http://omniweb.gsfc.nasa.gov/}). The OMNI dataset combines plasma and magnetic field measurements from multiple spacecraft and ground stations, time-shifted to 1~au for consistent interplanetary analysis. Its continuous coverage from 1963 to the present and uniform processing make it particularly suitable for studying long-term variations in solar wind plasma and interplanetary magnetic field (IMF) properties across multiple solar cycles.

\begin{figure*}[ht]
    \centering
    \includegraphics[width=1\linewidth]{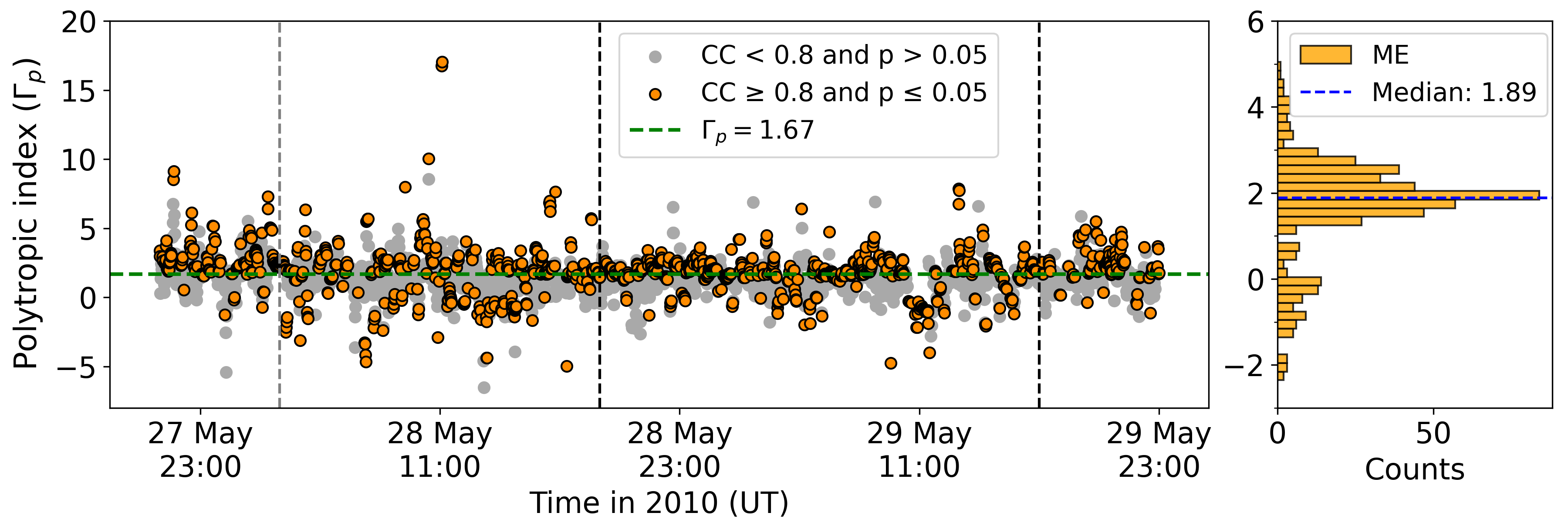}
    \caption[Derived $\Gamma_p$ values across the ICME event on 28 May 2010]{Derived $\Gamma_p$ values across the ICME event on 28 May 2010. The gray dashed vertical line marks the start of the sheath region, while the black dashed vertical lines denote the boundaries of ME. Orange dots represent reliable $\Gamma_p$ values, whereas gray dots indicate less reliable estimates. The green dashed horizontal line marks the adiabatic index ($\Gamma_p = 5/3$) for reference. The right panel shows the histogram of reliable $\Gamma_p$ values within ME. }
    \label{fig:one_gamma}
\end{figure*}

We have used the 1-minute proton number density ($N_p$) and proton temperature ($T_p$) data from the OMNI database to estimate the proton polytropic index ($\Gamma_p$), as described in Section~\ref{sec:chapt2-metho-poly}. Figure~\ref{fig:one_gamma} illustrates a representative case: an ICME observed on 28~May~2010. In the figure, the magnetic ejecta (ME) boundaries are indicated by black dashed lines, the sheath region by a gray dashed line, and 6 hours of pre- and post-ICME ambient solar wind are shown for context. A value of $\Gamma_p = 5/3$ corresponds to adiabatic expansion without heat exchange, $\Gamma_p > 5/3$ indicates a heat-release state, and $\Gamma_p < 5/3$ suggests net heating, implying additional energy input during plasma evolution. For this event, the median $\Gamma_p$ within the ME region was 1.89, signifying a dominant localized heat-release state. This procedure was applied to every ME listed in the RC catalog over the study period (Table~\ref{tab:ICME_list}). Since the $\Gamma_p$ distributions for individual events are generally skewed, the median was chosen as a robust measure of central tendency, being less sensitive to extreme values than the mean.

For event selection, we used the continuously updated RC ICME catalog due to its long temporal coverage and inclusion of recent events, making it well-suited for analyzing ICME properties over Solar Cycles~23,~24, and the rising phase of Cycle~25. This catalog compiles ICMEs arriving at 1~au, identified through visual inspection since 1996, and provides timing for the associated interplanetary disturbance (often marked by a sudden storm commencement) along with start and end times of the ME interval. Each ME is also assigned a classification flag: ‘2’ denotes a well-defined magnetic cloud (MC), ‘1’ indicates partial MC-like features, and ‘0’ signifies an absence of typical MC signatures. In this study, events flagged as ‘1’ or ‘0’ are collectively treated as non-MC ejecta, ensuring a clear separation between classic MCs and other ME structures.

It is important to note that defining ICME boundaries involves some subjectivity, as different studies use varying criteria \citep{Richardson2010}. Determining the end of an ME is particularly challenging, as the gradual transition back to pre-ICME conditions, marked by a slow increase in temperature and decrease in magnetic field strength, often lacks a sharp discontinuity. In the RC catalog, this can lead to timing uncertainties of the order of hours \citep{Richardson2024}. While these uncertainties affect precise duration estimates for individual events, their impact on our results is mitigated by the statistical nature of this study, where the averaging over a large number of events is expected to reduce random boundary errors.

\begin{figure*}[hbt]
    \centering
    \includegraphics[width=1.05\linewidth]{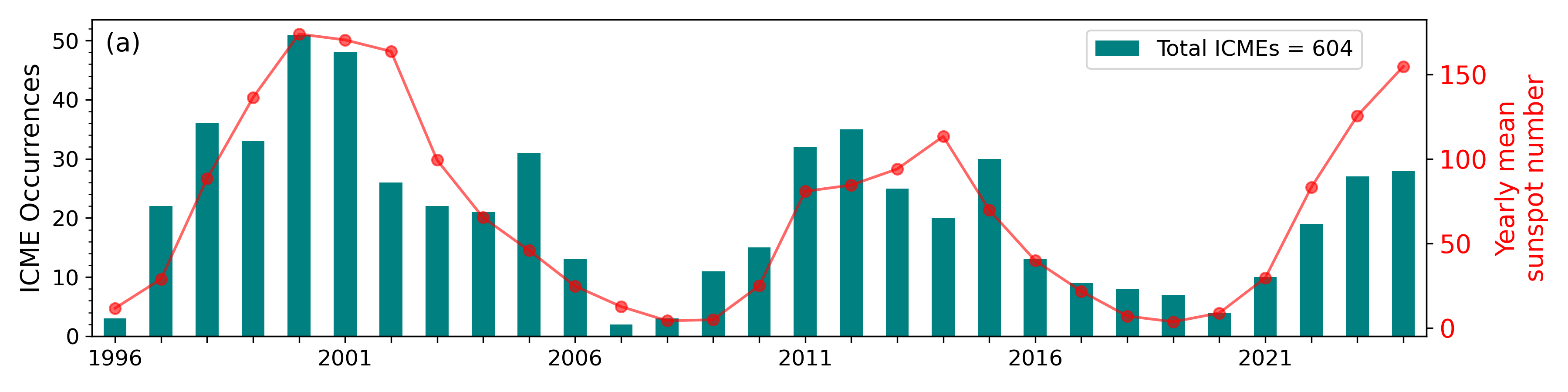}\\
    \vspace{-8pt} 
    \includegraphics[width=1.05\linewidth]{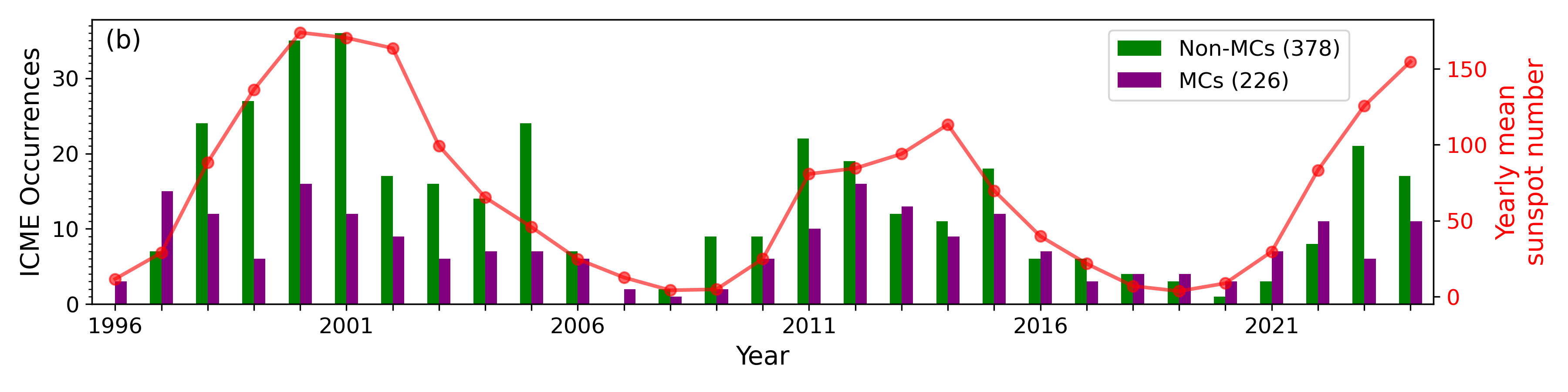}
    \caption[Annual occurrences of (a) ICMEs and (b) MC and non-MC]{Annual occurrences of (a) ICMEs and (b) MC and non-MC events across Solar Cycles 23, 24, and the rising phase of 25. Yearly mean sunspot numbers are overplotted in both panels for comparison with solar activity levels.}
    \label{fig:icme_no}
\end{figure*}

\vspace{-0.5cm}
\subsection{Annual Occurrence of ICMEs and Their Types} \label{chapt4-part2-sec:icme_no}

A brief examination of the annual occurrence of ICMEs and their types offers valuable context for understanding their distribution across solar cycles. Figure~\ref{fig:icme_no}a presents the yearly number of ICMEs from June~1996 to December~2024, alongside the annual mean sunspot number. Over this period, the RC catalog records a total of 604 ICMEs detected at Earth. As expected, ICME occurrence shows a strong positive correlation with solar activity (Pearson coefficient = 0.83). For example, only three ICMEs were reported during the solar minimum of 1996, compared to 51 events in 2000 and 48 in 2001. This overall upward trend follows the Solar Cycle progression, except for a notable dip in 1999. The reduced ICME rate in 1999, also reported by \citet{Cane2000a}, has been linked to an increased prevalence of co-rotating high-speed streams originating from low-latitude coronal holes and a reconfiguration of the near-ecliptic solar wind \citep{Luhmann2002, Richardson2002}. These altered solar wind conditions temporarily reduced the number of CME-related disturbances detected at Earth \citep{Cane2003}. In addition, \citet{Wu2006} suggested that the unusually large heliospheric current sheet (HCS) tilt angle and the occurrence of high-latitude prominence eruptions during 1999 further contributed to the reduction in Earth-directed CMEs. When comparing Solar Cycle~23 (SC23) and Solar Cycle~24 (SC24), the ICME occurrence rate was noticeably lower in SC24. This result aligns with previous studies that characterize SC24 as weaker than SC23 in terms of CME occurrence rate, associated mass loss, expansion speeds, and geo-effective parameters at 1~au \citep{Gopalswamy2015, Mishra2019a, Gopalswamy2020, Mishra2024}. Although the sunspot numbers observed so far in Solar Cycle~25 (SC25) are more comparable to SC23 and clearly exceed those of SC24, the ICME counts during SC25’s rising phase more closely resemble SC24.

As outlined in Section~\ref{chapt4-part2-sec:data}, ICMEs in this study are categorized into magnetic clouds (MCs) and non-magnetic clouds (non-MCs). Across the full study period, 226 MC events and 378 non-MC events were identified. Figure~\ref{fig:icme_no}b shows the annual distribution of these two types. While SC23 was dominated by non-MCs, SC24 exhibited a more balanced distribution between MCs and non-MCs. Notably, the total number of MCs remained relatively consistent between SC23 and SC24 and exhibited a weaker correlation with sunspot number (Pearson coefficient = 0.62) compared to non-MCs (Pearson coefficient = 0.82). This finding supports earlier results indicating that MCs are more prevalent during solar minimum \citep{Cane2003, Jian2006, Richardson2010, Mishra2021a}, likely because interactions between CMEs and other heliospheric structures, such as the HCS, that can modify MC signatures occur less frequently during solar minimum \citep{Wu2015mc}.

\subsection{Yearly Variation of ME's Thermal State at 1 au}\label{chapt4-part2-subsec:yearly_poly}

\begin{figure*}[ht]
    \centering
    \includegraphics[width=1.05\linewidth]{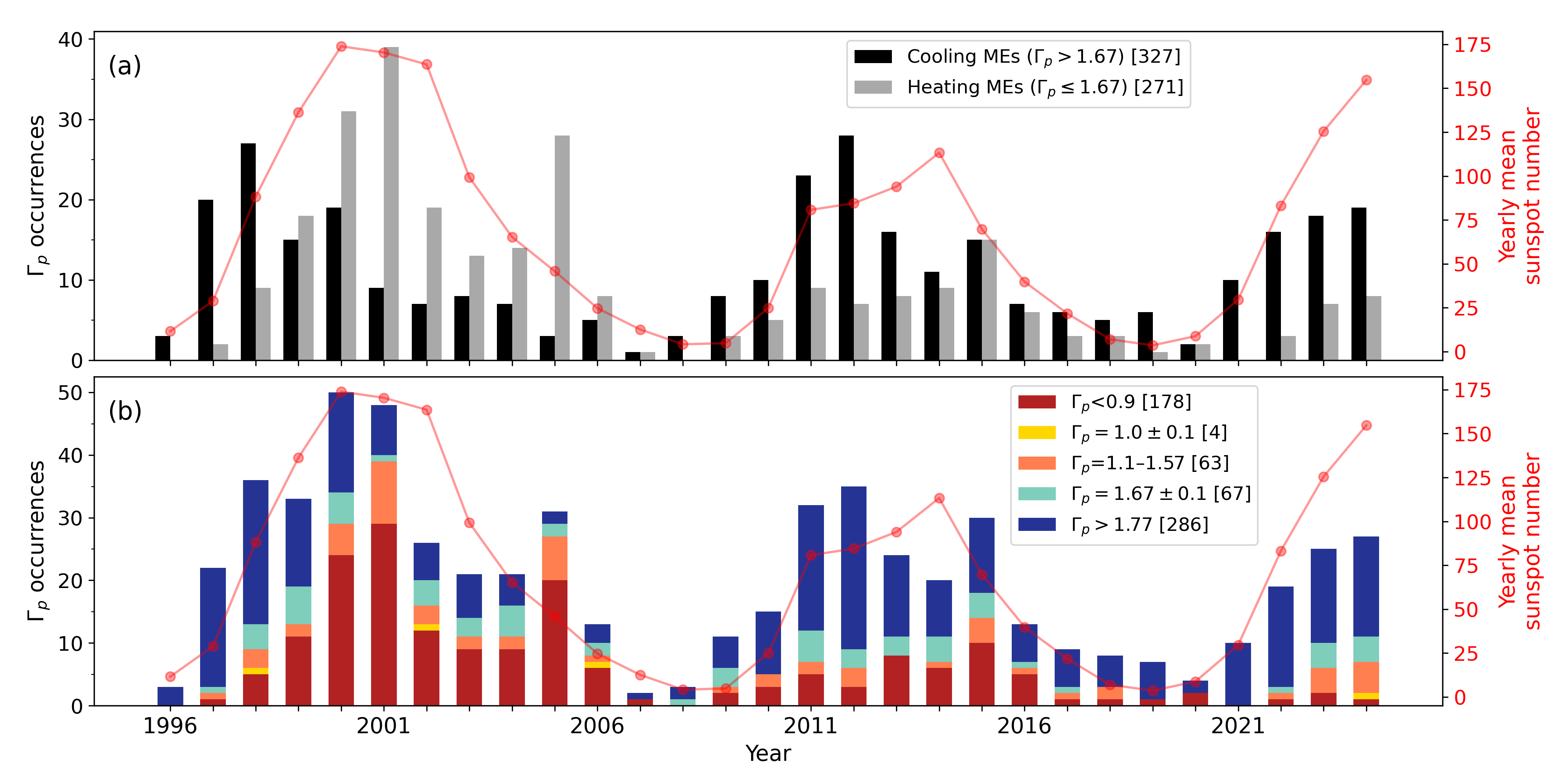}
    \caption[Annual occurrence of heating and cooling ME across SC23,]{(a) Annual occurrence of heating and cooling ME across SC23, 24 and rising phase of SC25. (b) Annual distribution of ME with various categories depending on $\Gamma_p$ values.}
    \label{fig:yearly_gamma}
\end{figure*}

To investigate the annual variation in the dominant thermal states of MEs across Solar Cycles~23 and~24, as well as the rising phase of Cycle~25, we categorized events into two groups: (i) \textit{Heating MEs} (median $\Gamma_p \leq 1.67$) and (ii) \textit{Cooling MEs} (median $\Gamma_p > 1.67$). Figure~\ref{fig:yearly_gamma}a presents the yearly occurrence of these two categories. In SC23, the solar maximum and declining phases were dominated by heating states, whereas the rising phase was characterized by a prevalence of heat-release (cooling) states. An exception occurred in 2005, during the declining phase of SC23, when a notably large fraction of MEs exhibited heating states compared to surrounding years. A similar trend is observed in the rising phases of SC24 and SC25, where cooling states remain dominant. In contrast, during the solar maximum of SC24, no clear preference for either heating or cooling was evident, with both occurring in roughly equal proportions. This contrast between SC23 and SC24 maxima may be linked to intrinsic differences in CME properties at eruption or to variations in the interplanetary medium that affect CME evolution and thermal state at 1~au. Previous studies have reported differences in CME expansion speeds and interplanetary medium characteristics between the maxima of SC23 and SC24 \citep{Gopalswamy2015, Lamy2019, Gopalswamy2020, Mishra2021a}.

A phase-by-phase comparison shows that rising phases tend to be dominated by cooling states, while declining phases exhibit either a more balanced distribution or a slight bias toward heating states. This trend is consistent with earlier statistical studies of ME composition in SC23, which found that elemental charge states and relative abundances increase during the rising phase and remain elevated throughout the maximum and decline, compared to their lower values during solar minima \citep{Gu2020, Song2021}. The ionic charge states within CME plasma serve as a diagnostic of the electron temperature in the source region \citep{Lepri2013}. Since heavy ion charge states and elemental compositions remain largely unchanged during CME propagation \citep{Ko2010, Gruesbeck2011, Lepri2012}, the similarity in trends between composition and thermal state suggests that CMEs originating during maxima and descending phases begin with higher electron (and possibly proton) temperatures. As these CMEs travel to 1~au, such elevated source-region conditions likely contribute to the observed proton heating states, reinforcing the connection between CME origin environments and their interplanetary thermal behavior.

A more refined understanding of ME thermodynamic evolution across Solar Cycles~23, 24, and the rising phase of Cycle~25 emerges from a detailed classification based on the polytropic index $\Gamma_p$ (Figure~\ref{fig:yearly_gamma}b). In this scheme, events are grouped as follows: \textit{Major-heating} ($\Gamma_p < 0.9$), \textit{Isothermal} ($\Gamma_p = 1.0 \pm 0.1$), \textit{Heating} ($1.1 \leq \Gamma_p \leq 1.57$), \textit{Adiabatic} ($\Gamma_p = 1.67 \pm 0.1$), and \textit{Major-cooling} ($\Gamma_p > 1.77$). These categories capture a range of thermal behaviors: Major-heating corresponds to strong energy input, where the ME temperature can increase during expansion; Isothermal indicates that the temperature remains nearly constant; Heating reflects a slower temperature decrease than predicted for adiabatic expansion, implying partial compensation of expansion-driven cooling; Adiabatic represents an idealized expansion with no heat exchange, where cooling results solely from expansion work; and Major-cooling denotes intensified heat-loss processes, with the temperature falling faster than in the adiabatic case.

In SC23, Major-heating events exhibited a clear upward trend during the ascending phase, increasing from a single event in 1997 to 5 events in 1998 and 11 events in 1999 (Table~\ref{tab:thermal_state_perce}). Despite this rise, the ascending phase was still dominated by Major-cooling events, particularly in 1997 and 1998, when they accounted for 86\% and 64\% of all events, respectively. A distinct shift occurred during the solar maximum years of 2000 and 2001, when Major-heating events became dominant, comprising roughly 48\% and 60\% of the total, respectively (Figure~\ref{fig:yearly_gamma}b). This highlights the strong heating associated with CMEs during this high-activity interval. After the peak, the number of Major-heating events declined sharply, dropping to 12 in 2002 and further to 9 in 2003. Notably, 2005 stands out in the declining phase with a resurgence to 20 Major-heating events, representing $\sim$65\% of all events that year, consistent with a pattern of sustained heating behavior well into the descending phase. This trend suggests significant energy input into MEs during their interplanetary propagation, likely linked to elevated electron and proton temperatures at their solar sources during the solar maximum and descending phase of SC23.

In contrast, the ascending phase of SC24 (2008–2012) was characterized by a marked dominance of Major-cooling events, reaching up to 74\% in 2012, indicative of enhanced heat-loss processes or reduced heating efficiency during this period. The solar maximum of SC24 (around 2013–2014) exhibited a more balanced thermal state distribution, with Major-heating and Major-cooling events occurring in comparable proportions. This stands in sharp contrast to SC23’s maximum, where heating was dominant, suggesting that the thermal energy budget and plasma conditions of CMEs during SC24 may have been fundamentally different, possibly reflecting the overall weaker nature of this cycle. In the early rising phase of SC25 (2020–2024), Major-cooling events have remained prevalent, accounting for up to 86\% of events in 2019 and 84\% in 2022. However, 2023 and 2024 show a modest recovery in heating-related states, with 24\%–27\% of events (combining Major-heating, Isothermal, and Heating categories) annually, hinting at a gradual shift in thermal behavior as solar activity intensifies. Across all years, Isothermal MEs are rare, appearing only sporadically and constituting a negligible fraction of the total, underscoring that MEs at 1~au are generally thermodynamically active and deviate markedly from equilibrium expansion.

Overall, these results demonstrate that ME thermal states vary systematically with the Solar Cycle phase. The prevalence of Major-heating states during solar maxima and descending phases points to stronger heating in CME source regions, likely from flare-associated energy release or shock-driven compression, with a portion of this energy retained during heliospheric propagation. Conversely, the dominance of Major-cooling states during rising phases may reflect CMEs originating from cooler coronal environments or possessing insufficient internal energy to counteract adiabatic expansion and radiative losses. This classification thus serves not only to describe ME thermal behavior but also as a diagnostic tool for tracking the large-scale thermodynamic evolution of the solar wind–CME system over successive solar cycles.

\subsection{Solar Cycle Variation of ME's Thermal State at 1 au}\label{chapt4-part2-subsec:sc_poly}

\begin{figure*}[ht]
    \centering
    \includegraphics[width=1.05\linewidth]{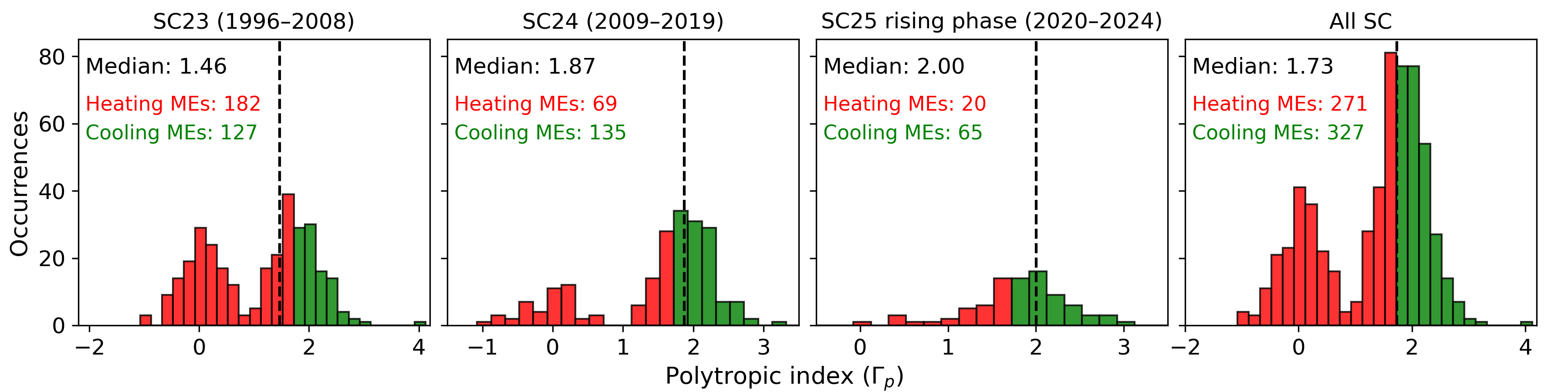}
    \caption{Distribution of $\Gamma_p$ over in SC23, 24 and rising phase of SC25}
    \label{fig:sc_gamma}
\end{figure*}

Figure~\ref{fig:sc_gamma} illustrates the Solar Cycle variation in the distribution of median $\Gamma_p$ values for individual MEs. During SC23, the median $\Gamma_p$ is 1.46, reflecting a clear predominance of Heating MEs, with 182 out of 309 events ($\sim$59\%) exhibiting a heating state at 1~au. In contrast, SC24 shows a higher median $\Gamma_p$ of 1.87, indicating a shift toward heat-release (cooling-dominated) conditions, with only 69 out of 204 events ($\sim$34\%) displaying heating signatures. In the rising phase of SC25, the median $\Gamma_p$ increases further to 2.00, signifying an even stronger dominance of heat-release states, where only 20 out of 85 MEs ($\sim$23\%) exhibit heating behavior.

These findings indicate a pronounced Solar Cycle dependence in the thermal state of ICMEs, with heating being more prevalent during SC23 and progressively diminishing through SC24. The overall median $\Gamma_p$ across all three Solar Cycle phases is 1.73, slightly exceeding the adiabatic index of $5/3$ (1.67). This proximity to the adiabatic threshold suggests that, on average, ME plasma at 1~au is neither subject to intense heating nor rapid cooling, although substantial event-to-event variability is evident. Notably, 271 out of 598 MEs ($\sim$45\%) still display heating signatures, reflecting the inherent complexity and variability in CME thermodynamics under different heliospheric conditions.

Interestingly, \citet{Katsavrias2025} report a comparable average polytropic index of $\Gamma = 1.63$ for MEs, derived from a distinct CME catalog \citep{Teresa2019} covering 1995–2001, in conjunction with Wind spacecraft observations. Their methodology, however, employs a different set of selection filters and focuses on the mean value, whereas our study emphasizes the median of the $\Gamma$ distribution. In our analysis, the $\Gamma_p$ distribution exhibits a clear bimodal character with a noticeable dip near $\Gamma_p = 1$ (Figure~\ref{fig:sc_gamma}), consistent with the results of \citet{Katsavrias2025}. However, isolating a truly isothermal state remains challenging, as filtering criteria can distort the intrinsic distribution, suppressing the tails or enhancing the dip near $\Gamma_p = 1$. Such effects may be further amplified by parametric fitting techniques (e.g., Gaussian or $\kappa$-Gaussian), which impose rigid functional forms that may not capture the full complexity of ICME data.

To address these challenges, we employ a non-parametric, assumption-free approach by using the median, complemented by bootstrap-derived uncertainties, to represent the central tendency of $\Gamma_p$. Unlike parametric estimates, the median is robust to skewness and local depletions, as it depends solely on rank ordering rather than the density of values within narrow intervals. Consequently, unless filtering significantly alters the values near the central rank, which is not the case in our dataset, the median remains a stable and unbiased measure. This approach provides a more reliable characterization of thermodynamic variability in ICMEs, particularly when the underlying distribution deviates from unimodality and may be influenced by filtering artifacts.

Our results indicate that CME thermodynamics are strongly influenced by Solar Cycle activity, with variability likely driven by a combination of factors, including differences in CME initiation mechanisms, coronal background conditions, and CME–solar wind interactions. Determining which of these factors predominantly controls the thermal state of MEs at 1~au, however, remains challenging. Earlier studies have reported that a greater fraction of CMEs in SC24 originated from non-active region (non-AR) sources, such as filament eruptions from quiet Sun regions and higher latitudes, compared to SC23 \citep{Gopalswamy2016, Lamy2019, Mishra2019a}. Moreover, ICME expansion behavior differed markedly between the two cycles, with SC24 experiencing enhanced expansion due to reduced heliospheric background pressure \citep{Gopalswamy2020, Mishra2021a}. This greater expansion is likely a key contributor to the cooling-dominated thermal state observed during SC24. It is also plausible that MEs differed intrinsically at the time of eruption, influenced by their source region characteristics or associated flare properties \citep{Oloketuyi2019, Lin2023, Pandey2023}. Additionally, the lower frequency and intensity of geomagnetic storms (GS) in SC24 compared to SC23 may reflect the generally less energetic nature of CMEs during this period \citep{Mishra2024}. Collectively, these factors provide a coherent explanation for the observed transition from heating-dominated MEs in SC23 to cooling-dominated MEs in SC24.

\subsection{Variations in Plasma Properties of Heating and Cooling Magnetic Ejecta} \label{chapt4-part2-sec:icme_plasma_cool_hot}

\begin{figure*}
    \hspace*{-1.cm}
    \includegraphics[width=1.1\linewidth]{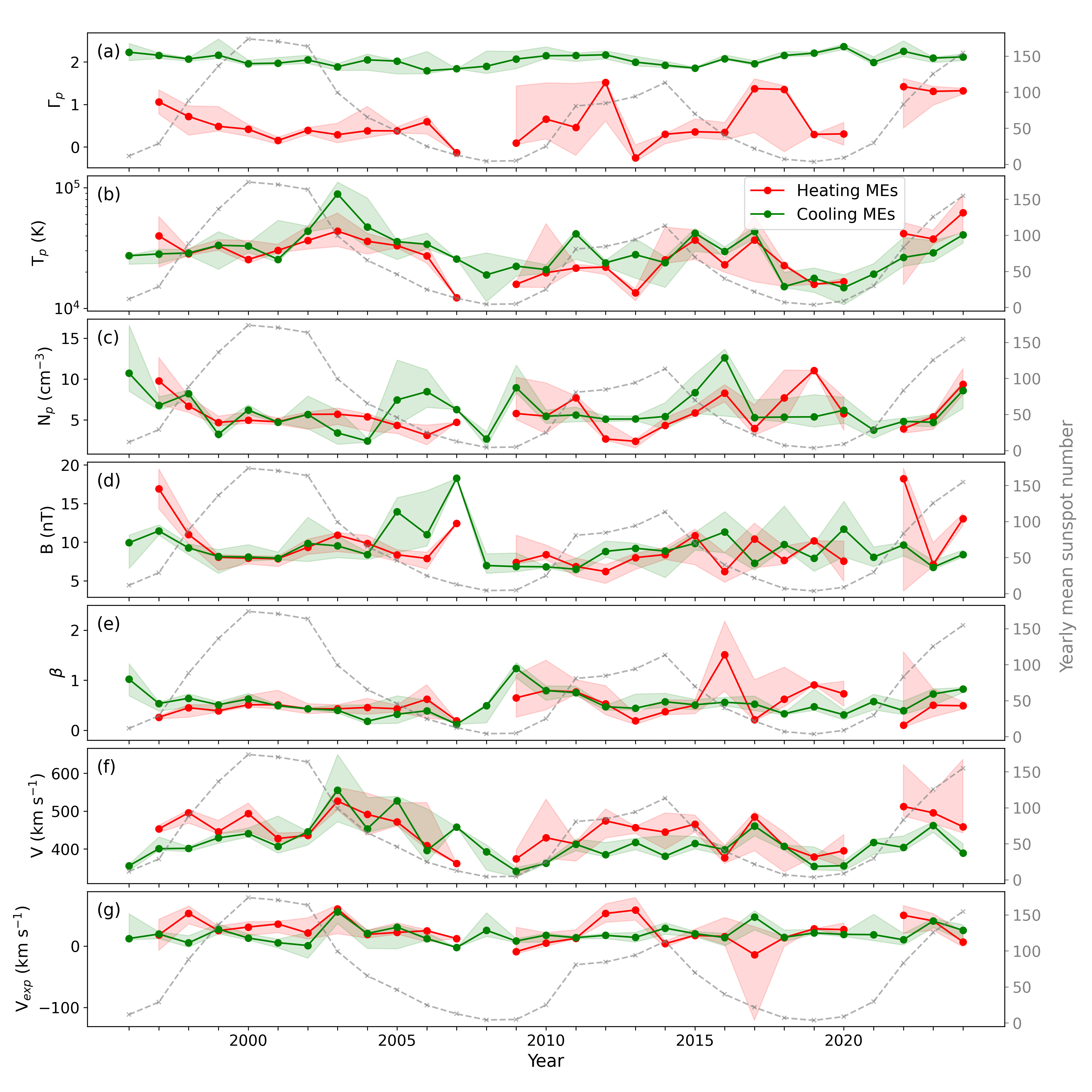}
    \caption[Yearly median values of (a) polytropic index ($\Gamma_p$), (b) proton]{Yearly median values of (a) polytropic index ($\Gamma_p$), (b) proton temperature ($T_p$), (c) proton number density ($N_p$), (d) magnetic field strength ($B$), (e) plasma beta ($\beta$), (f) bulk speed ($V$), and (g) expansion speed ($V_{exp}$) for heating and Cooling MEs across Solar Cycles 23, 24, and the ascending phase of Cycle 25. The background color-shaded regions represent the 68\% (1$\sigma$) confidence intervals obtained via bootstrap resampling. Each panel is overplotted with the yearly mean sunspot numbers to investigate the correlation between solar activity and the thermal, plasma, and magnetic properties of MEs.}
    \label{fig:yearly_heating_cooling_plasma}
\end{figure*}

We categorized the MEs into two groups based on their $\Gamma_p$ values: Heating MEs ($\Gamma_p < 1.67$) and Cooling MEs ($\Gamma_p > 1.67$). For each category, we computed the yearly median of the $\Gamma_p$ distribution. Figure~\ref{fig:yearly_heating_cooling_plasma}a presents the annual median $\Gamma_p$ values for both Heating and Cooling MEs. Cooling MEs consistently maintain a median $\Gamma_p \approx 2$, with minimal year-to-year variation. In contrast, Heating MEs exhibit substantial interannual variability in $\Gamma_p$. However, no strong correlation is found between these median values and solar activity, as represented by annual sunspot numbers.  

During SC23, Heating MEs display a decreasing $\Gamma_p$ trend from the ascending phase toward solar maximum, after which the values remain nearly constant throughout the descending phase. This pattern aligns with our earlier observation (Figure~\ref{fig:yearly_gamma}) of an increased occurrence of Heating MEs during the maximum and declining phases of SC23. In SC24, the trend is more irregular: $\Gamma_p$ for Heating MEs rises during the ascending phase, drops sharply around the solar maximum (2013), and then gradually increases until 2018, near the SC24 minimum. Across different Solar Cycle phases, Heating MEs tend to show stronger heating (lower $\Gamma_p$) during solar maxima and descending phases. In the rising phase of SC25, $\Gamma_p$ remains nearly constant, close to an isothermal value. Notably, during solar minima, especially in 2008 and 2021, only Cooling MEs are observed, with yearly median $\Gamma_p \approx 2$, indicating a pronounced tendency toward heat-release states during these periods.

Figure~\ref{fig:yearly_heating_cooling_plasma}b illustrates the annual median proton temperature ($T_p$) for both Heating and Cooling MEs across Solar Cycles 23, 24, and the rising phase of Cycle 25. For Cooling MEs, $T_p$ remains nearly constant during the ascending and maximum phases of SC23, followed by a gradual rise during the early descending phase, peaking around 2003. Thereafter, $T_p$ steadily decreases toward the 2008 solar minimum. A broadly similar pattern is observed in SC24, though with greater irregularity: $T_p$ increases during the ascending and maximum phases, then declines throughout the descending phase. From the deep solar minimum in 2008, $T_p$ for Cooling MEs rises gradually until 2017, experiences a sharp drop in 2018, and subsequently exhibits a marked increase from 2018 to 2024 during the rising phase of SC25. This long-term variation in median $T_p$ generally follows the Solar Cycle, showing maxima near the middle of the descending phase and minima during solar minima, consistent with previous findings \citep[e.g.,][]{Chi2016, Jian2018} that report elevated solar wind temperatures during active phases and lower values during minima.

Heating MEs exhibit a broadly similar long-term variation in $T_p$ to Cooling MEs, with temperatures generally rising during the ascending, maximum, and early descending phases, followed by a decline. However, some notable differences are evident. For example, in 2013, Figure~\ref{fig:yearly_heating_cooling_plasma}a shows a markedly lower $\Gamma_p$ value for Heating MEs, indicative of stronger heating during interplanetary propagation. Despite this, the in situ measured $T_p$ for the same year is lower than in adjacent years. This underscores an important physical point: the observed $T_p$ in situ reflects a combination of the ME’s initial eruption temperature and the subsequent heating it undergoes en route to 1~au. Thus, even with strong in-transit heating (low $\Gamma_p$), a low $T_p$ can occur if the CME originated with a relatively cool temperature, as appears to have been the case in 2013. A similar pattern is seen during solar minima such as 2007–2009 and 2019–2020, when $\Gamma_p$ values suggest enhanced heating but $T_p$ remains low. This indicates that CMEs during solar minima tend to erupt with cooler intrinsic temperatures, likely due to weaker solar activity and reduced CME energetics \citep{Vourlidas2011err, Gopalswamy2014, Lamy2019, Pant2021}. Therefore, while propagation-driven heating plays a key role in shaping the thermal state of MEs, the initial eruption temperature, strongly influenced by solar activity, remains a decisive factor in determining the final in situ $T_p$. When comparing the annual median $T_p$ between Heating and Cooling MEs, we find that Cooling MEs generally exhibit higher $T_p$ values than Heating MEs during SC23 and SC24, whereas in the ascending phase of SC25, Heating MEs show higher $T_p$, possibly reflecting changes in source region properties or heliospheric conditions.

Figure~\ref{fig:yearly_heating_cooling_plasma}c shows the yearly median proton density ($N_p$) for Heating (red) and Cooling MEs (green). In SC23, Cooling MEs display a generally declining and somewhat irregular trend in $N_p$ through the ascending, maximum, and early descending phases, continuing until 2004. Between 2005 and 2007, $N_p$ rises, before dropping sharply at the deep solar minimum of 2008. In SC24, $N_p$ for Cooling MEs increases from 2008 to 2016, peaking at $\sim$14~cm$^{-3}$, and then remains relatively stable until 2020. A small decrease follows, before another rise during the ascending phase of SC25. In contrast, Heating MEs show a smoother decline in $N_p$ through SC23, reaching a low by 2006. Entering SC24, their $N_p$ values are higher than in SC23’s ascending phase, but reach a minimum around the 2014 solar maximum. This is followed by a steady increase until 2019, with a brief dip in 2017, and then a slight drop after 2019. From 2022 onwards, $N_p$ increases again during SC25’s rising phase. Across all cycles, Heating MEs consistently have lower $N_p$ than Cooling MEs.

The yearly median magnetic field strength ($B$) is shown in Figure~\ref{fig:yearly_heating_cooling_plasma}d. For Cooling MEs, $B$ declines steadily through SC23’s ascending and maximum phases (1997–2001), fluctuating between $\sim$8 and 11~nT, then rises irregularly through the descending phase, peaking near $\sim$18~nT during the 2007 solar minimum. In SC24, $B$ for Cooling MEs increases smoothly from 2008 to a peak of $\sim$11~nT in 2016, followed by a gradual and irregular decline into SC25’s ascending phase. Heating MEs show more distinct, cycle-dependent $B$ variations: a sharp drop in the early ascending phase of SC23 (1997–1999), followed by a near-constant level at solar maximum (1999–2001), a slight increase to 2003, then a decline to 2006. A sharp rise is observed in 2007, coinciding with the solar minimum. Similar to Cooling MEs, Heating MEs show a steady $B$ increase in SC24 from the ascending to early descending phases, peaking around 2015. During SC24’s descending phase, $B$ fluctuates but remains above ascending-phase values. In SC25, Heating MEs display elevated, fluctuating $B$ values, generally higher than in SC24. While no clear solar-cycle dependence emerges, both Heating and Cooling MEs tend to reach their maximum $B$ during descending phases, and their $B$ profiles are broadly similar in magnitude and variability.

Figure~\ref{fig:yearly_heating_cooling_plasma}e reveals a reversal in $\beta$ trends for Heating and Cooling MEs during SC23: $\beta$ decreases for Cooling MEs while increasing for Heating MEs from the ascending to descending phase. Around the 2007–2009 solar minimum, $\beta$ for Cooling MEs rises, whereas in SC24, $\beta$ again decreases from the ascending to descending phase. Unlike SC23, no significant $\beta$ increase is seen at the SC24–SC25 minimum. During SC24’s solar maximum, Cooling MEs have higher $\beta$ than Heating MEs, but this reverses in the descending phase.

Across all solar cycles, $V$ increases during the ascending and maximum phases, peaks in the early descending phase, and then declines toward solar minimum for both Heating and Cooling MEs (Figure~\ref{fig:yearly_heating_cooling_plasma}f). Heating MEs typically have higher speeds than Cooling MEs during the ascending and maximum phases, suggesting that Heating MEs are linked to more energetic eruptions in these periods. Expansion speeds ($V_{exp}$) also show notable behavior (Figure~\ref{fig:yearly_heating_cooling_plasma}g): Heating MEs tend to have higher $V_{exp}$ than Cooling MEs at solar maxima. In 2013, for example, $V_{exp}$ is high despite low $T_p$ and $\Gamma_p$, implying that even with low internal thermal energy, strong magnetic pressure or high source-region eruption energy drove rapid expansion.

\subsection{Exploring Connection between Thermal Properties and Geoeffectiveness of ICMEs} \label{chapt4-part2-sec:dsea}

We applied the Superposed Epoch Analysis (SEA) technique to statistically examine the average temporal evolution of plasma parameters and geomagnetic indices for different categories of ME events. This method is well suited for revealing systematic trends, though it may smooth out individual event-specific features. As detailed in Section~\ref{sec:chapt2-metho-sea}, SEA constructs averaged profiles of physical parameters over the normalized duration of events, thereby enabling a direct comparison across multiple cases of varying lengths. By emphasizing recurring patterns and reducing the influence of event-to-event variability, this approach provides a clearer view of the characteristic behavior of ME structures and their surrounding solar wind environment.

\begin{figure*}
    \centering
    \includegraphics[width=1.05\linewidth]{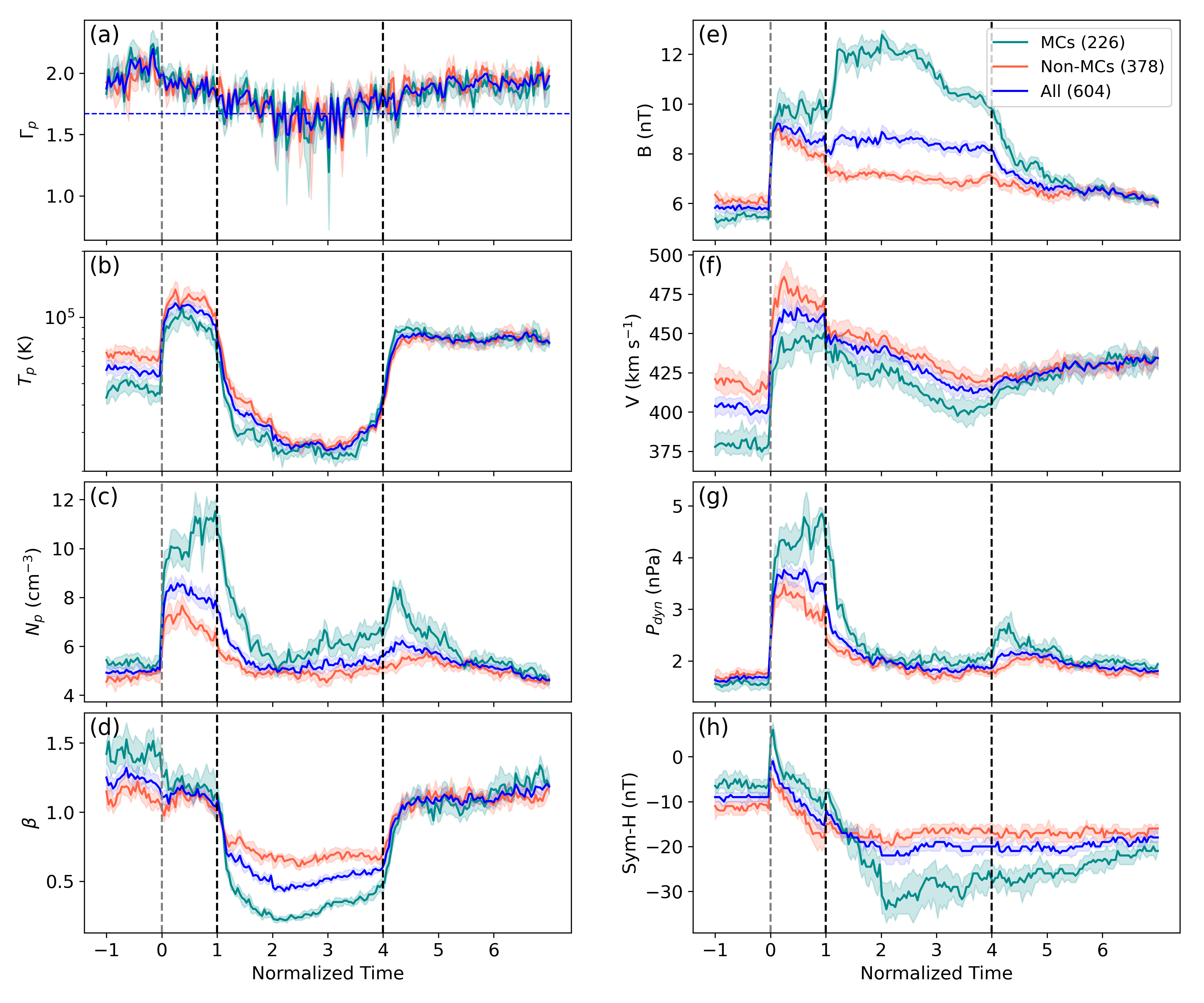}
    \caption[Superposed Epoch Analysis showing the median values ]{Superposed Epoch Analysis showing the median values of parameters across the pre-ICME, sheath, ME, and post-ICME regions: (a) polytropic index ($\Gamma_p$), (b) proton temperature ($T_p$), (c) proton number density ($N_p$), (d) plasma beta ($\beta$), (e) magnetic field strength ($B$), (f) flow speed ($V$), (g) dynamic pressure ($P_{dyn}$), and (h) geomagnetic index (Sym-H). The curves represent: teal for MC, orange for non-MC, and blue for all MEs. Black vertical dashed lines mark the boundaries for the ME region, and gray dashed vertical lines mark the starting sheath region. The background color-shaded regions represent the 68\% (1$\sigma$) confidence intervals obtained via bootstrap resampling.}
    \label{fig:sea_plasma}
\end{figure*}

To rigorously assess differences in the thermodynamic behavior of ICMEs, specifically in the distributions of $\Gamma_p$, across various classes, we employed a suite of complementary statistical techniques. The \textit{Kolmogorov--Smirnov (KS) test} was used to evaluate differences in the overall shape and cumulative distribution of $\Gamma_p$. In this context, a test statistic $D > 0.1$ combined with a $p$-value $< 0.05$ is considered strong evidence for a significant difference in distributional form. The \textit{Mann--Whitney U test}, a non-parametric method, was applied to compare central tendencies (e.g., medians) without assuming Gaussian behavior; here, a $p$-value $< 0.05$ similarly indicates statistical significance. To quantify both the magnitude and the direction of these differences, we calculated \textit{Cliff’s delta}, which measures the probability that a randomly chosen $\Gamma_p$ value from one group exceeds that from another. Values close to zero imply substantial overlap between distributions, whereas values approaching $\pm 1$ indicate near-complete separation, with one group consistently exhibiting higher or lower values.

In addition, we determined the \textit{percentage of time bins} in the superposed epoch analysis (SEA) for which the median $\Gamma_p$ was below the heating threshold ($\Gamma_p < 1.67$), providing a measure of the temporal persistence and frequency of enhanced heating states. Finally, we applied a \textit{permutation test} to the median $\Gamma_p$ values, wherein group labels were randomly shuffled to generate a null distribution. If the observed median difference falls within the extreme tail of this null distribution (e.g., $p < 0.01$), it provides strong evidence that the difference is unlikely to arise from random variation. Collectively, these statistical approaches enable a robust evaluation of whether two populations differ, the strength and direction of those differences, and the temporal characteristics of any heating-related deviations.

\subsubsection{Evolution in Plasma Properties across MEs (MCs and Non-MCs)}\label{chapt4-part2-subsec:dsea_plasma}

To examine the characteristic evolution of plasma and magnetic field parameters during ICMEs, we performed a Superposed Epoch Analysis (SEA) on 604 events from the RC catalog. As previously described, these events were classified into magnetic clouds (MCs, flag 2) and non-magnetic clouds (non-MCs, flags 0 and 1), comprising 226 MCs and 378 non-MCs. Figure~\ref{fig:sea_plasma} presents the median time profiles of $\Gamma_p$, $T_p$, $N_p$, $\beta$, $B$, $V$, $P_{\mathrm{dyn}}$, and Sym-H, with shaded regions indicating the 68\% (1$\sigma$) confidence intervals for the median values. These confidence intervals were derived using a non-parametric bootstrap resampling approach (1000 iterations), providing a robust estimate of statistical variability that accounts for sample dispersion within each epoch bin. The corresponding median values for each region, pre-ICME, sheath, ME, and post-ICME, are summarized in Table~\ref{tab:median} for the different event categories.

Across all ICMEs, $\Gamma_p$ displays a distinct four-phase variation: it is highest in the pre-ICME region (2.02), decreases in the sheath (1.83), reaches a minimum within the ME (1.73), and partially recovers in the post-ICME region (1.91) (Table~\ref{tab:median}). This sequence reflects a transition from an enhanced cooling state upstream, to near-adiabatic expansion within the ME, followed by a return to enhanced cooling conditions downstream. A similar pattern is evident for MCs and non-MCs considered separately, although MCs exhibit slightly higher pre-ICME $\Gamma_p$ (2.09) and slightly lower post-ICME values (1.88) compared to non-MCs (1.95 and 1.93, respectively).

Statistical comparisons between MCs and non-MCs reveal a subtle but statistically significant difference in $\Gamma_p$ distributions. The Kolmogorov–Smirnov (KS) test yields $D = 0.030$ with $p \approx 1.38 \times 10^{-47}$, indicating a difference in distribution shapes. However, the effect size is minimal: Cliff’s delta of $-0.009$ implies that a randomly chosen non-MC has a marginally higher $\Gamma_p$ than an MC in only 50.4\% of cases. The Mann–Whitney U test also returns a highly significant $p$-value ($p \approx 2.26 \times 10^{-4}$) but confirms the small magnitude of the difference. A permutation test on median values supports these results, showing a median difference of $\Delta \Gamma_p = -0.0519$ with $p < 10^{-5}$. Kernel density estimates (Figure~\ref{fig:kde}a) further illustrate that both distributions are bimodal with nearly identical peaks near $\Gamma_p \approx 2$ and closely matched tails. Taken together, these findings indicate that while large sample sizes enable detection of a statistical difference, the thermodynamic properties inferred from $\Gamma_p$ are broadly similar for MCs and non-MCs. This suggests that factors beyond magnetic cloud topology are likely more influential in determining the compressive and heating characteristics of ICMEs.

Furthermore, $T_p$ exhibits its highest values in the sheath and lowest values within the ME (Figure~\ref{fig:sea_plasma}b). For all ICMEs, the sheath median temperature reaches $11.5 \times 10^4$~K, whereas the ME drops to $3.0 \times 10^4$~K. This pattern reflects the strong heating from compression and turbulence in the sheath, followed by substantial cooling during ME expansion. Comparing event types, MCs consistently display lower $T_p$ than non-MCs in the pre-ICME ($4.5 \times 10^4$~K vs.\ $6.3 \times 10^4$~K), sheath ($10.1 \times 10^4$~K vs.\ $12.8 \times 10^4$~K), and ME ($2.8 \times 10^4$~K vs.\ $3.1 \times 10^4$~K) regions, indicating that MCs are generally cooler. In the post-ICME region, however, temperatures converge ($7.5 \times 10^4$~K vs.\ $7.7 \times 10^4$~K), suggesting similar downstream recovery conditions (Table~\ref{tab:median}). 

The proton density $N_p$ shows the expected enhancement in the sheath and reduction in the ME (Figure~\ref{fig:sea_plasma}c), consistent with shock compression followed by expansion. For all ICMEs, the sheath median $N_p$ is 8.6~cm$^{-3}$, decreasing to 5.3~cm$^{-3}$ in the ME. MCs exhibit even higher densities in both regions (11.4 and 6.3~cm$^{-3}$, respectively) compared to non-MCs (7.2 and 5.0~cm$^{-3}$), implying that MCs are more strongly compressed and contain denser plasma.

The plasma beta ($\beta$), representing the ratio of thermal to magnetic pressure, varies inversely with $T_p$ (Figure~\ref{fig:sea_plasma}d). For all ICMEs, $\beta$ peaks at 1.1 in the sheath before falling to 0.5 within the ME, signifying magnetic pressure dominance. MCs have an even lower ME $\beta$ (0.3) compared to non-MCs (0.7), consistent with their stronger magnetic field strengths (Figure~\ref{fig:sea_plasma}e). Across all ICMEs, the pre-ICME median $B$ is 8.6~nT, similar to 8.5~nT in the ME. For MCs, however, the ME $B$ reaches 11.3~nT, in contrast to only 7.1~nT for non-MCs, underscoring the stronger magnetic cores in MCs.

Flow speed $V$ also shows type-dependent differences (Figure~\ref{fig:sea_plasma}f). Non-MCs tend to travel faster, with median ME speeds of 438~km~s$^{-1}$ compared to 421~km~s$^{-1}$ for MCs, and sheath speeds of 479~km~s$^{-1}$ versus 455~km~s$^{-1}$, respectively. Despite these lower speeds, MCs exhibit higher dynamic pressures $P_{\mathrm{dyn}}$ in both the sheath (4.9~nPa vs.\ 3.4~nPa) and ME (2.2~nPa vs.\ 1.9~nPa) (Figure~\ref{fig:sea_plasma}g), indicating greater momentum density.

The consistently higher $P_{\mathrm{dyn}}$ observed in the sheath regions of MCs points to stronger momentum fluxes associated with these structures. This elevated pressure at shock arrival frequently produces a sharp positive spike in the Sym-H index, a phenomenon known as Sudden Storm Commencement (SSC) \citep{Burlaga1969,Veenadhari2012}. Our results show that this positive excursion in Sym-H is generally more pronounced for MCs than for non-MCs (Figure~\ref{fig:sea_plasma}h). Following SSC, the Sym-H index typically declines as the ME passes, particularly when strong southward $B_z$ components enhance magnetic reconnection, thereby intensifying geomagnetic storm activity \citep{Dungey1961,Tsurutani1997,Kamide1997}. While the Sym-H decline rate in the sheath is comparable for both MCs and non-MCs, a key distinction arises within the ME: MCs display a steeper drop and reach lower minimum Sym-H values. This sequence, initial high $P_{\mathrm{dyn}}$ in the sheath triggering SSC, followed by a marked Sym-H decrease during the ME phase, is consistent with the established progression of geomagnetic storm development. Overall, MCs are more likely to generate stronger SSCs and drive more intense geomagnetic storms compared to non-MCs \citep{Wu2007,Echer2008}. Furthermore, once the minimum Sym-H value is reached during the storm’s main phase, the subsequent recovery to pre-ICME levels is typically slower for MC-driven storms, indicating a more prolonged disturbance to Earth’s magnetosphere.

Both MCs and non-MCs exhibit similar expansion speeds, with $V_{\mathrm{exp}} \approx 20$~km~s$^{-1}$, implying that the overall radial growth rates of the ME structures are nearly identical. To investigate the ME internal structure in more detail, we used the distortion parameter (DiP) defined by \citet{Nieves2018}. Specifically, DiP-B quantifies the asymmetry in the magnetic field strength profile, while DiP-N provides an analogous measure for the proton number density profile. A DiP of 0.5 represents a perfectly symmetric profile centered at the ME midpoint. Values $< 0.5$ indicate stronger compression toward the ME front, whereas values $> 0.5$ suggest compression toward the rear. For the entire ICME set, both DiP-B and DiP-N have median values of 0.48, indicating slight asymmetry with compression concentrated at the leading edge. Separating events by type, MCs and non-MCs share the same DiP-B median (0.48), suggesting comparable magnetic field asymmetry. However, their density structures differ slightly: MCs have a lower DiP-N (0.46) than non-MCs (0.49), implying that MCs possess a more front-loaded density distribution, whereas non-MCs exhibit a more symmetric density profile.

\begin{figure*}
    \centering
    \hspace{-1cm}  
    \includegraphics[width=1.06\linewidth]{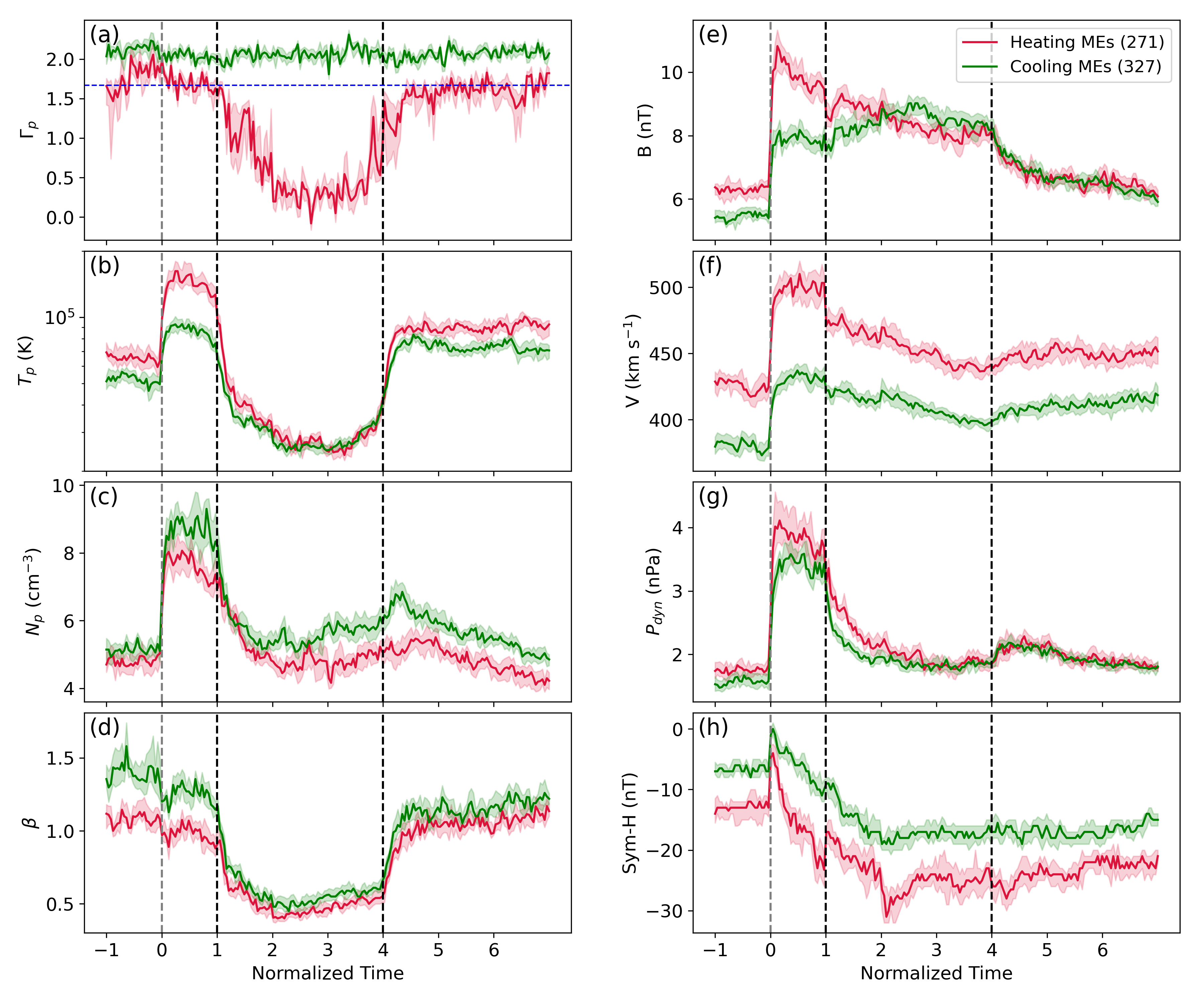}
    \caption[Superposed Epoch Analysis showing the median values]{Superposed Epoch Analysis showing the median values of parameters across the pre-ICME, sheath, ME, and post-ICME regions: (a) polytropic index ($\Gamma_p$), (b) proton temperature ($T_p$), (c) proton number density ($N_p$), (d) plasma beta ($\beta$), (e) magnetic field strength ($B$), (f) flow speed ($V$), (g) dynamic pressure ($P_{dyn}$), and (h) geomagnetic index (Sym-H). The curves represent: red for Heating ME and green for Cooling ME. Black vertical dashed lines mark the boundaries for the ME region, and gray dashed vertical lines mark the starting sheath region. The background color-shaded regions represent the 68\% (1$\sigma$) confidence intervals obtained via bootstrap resampling.}
    \label{fig:sea_heating_cooling}
\end{figure*}

\subsubsection{Comparative Plasma Evolution in Heating and Cooling MEs}\label{chapt4-part2-subsec:dsea_hot_cool}

We conducted a comparative SEA between Heating and Cooling MEs, classified according to their $\Gamma_p$ values: Heating MEs with $\Gamma_p < 1.67$ and Cooling MEs with $\Gamma_p > 1.67$ (Figure~\ref{fig:sea_heating_cooling}). This categorization highlights notable physical contrasts in plasma and magnetic field behavior across the four key regions, pre-ICME, sheath, ME, and post-ICME, revealing the thermodynamic and dynamical processes that distinguish the two populations.

For Heating MEs, $\Gamma_p$ decreases progressively from a median of 1.84 in the pre-ICME region to 1.64 in the sheath (Table~\ref{tab:median}). The drop becomes particularly steep within the ME, where $\Gamma_p$ falls to a median of 0.42, indicating strong heating and pronounced deviation from adiabatic expansion (Figure~\ref{fig:sea_heating_cooling}a). In the post-ICME region, $\Gamma_p$ partially recovers to 1.59, suggesting a relaxation phase following the central heating episode inside the ME. In contrast, Cooling MEs maintain consistently elevated $\Gamma_p$ values across all regions, starting at 2.16 in the pre-ICME solar wind, dipping slightly to 2.07 within the ME, and returning to 2.14 post-ICME, indicative of sustained cooling throughout their passage.

Statistical testing confirms that the $\Gamma_p$ distributions of Heating and Cooling MEs differ markedly. The KS test yields a large statistic ($D = 0.325$) with a p-value effectively zero, demonstrating a significant difference in distribution shapes. The Mann–Whitney U test similarly returns $p < 10^{-10}$, indicating a shift in central tendency independent of Gaussian assumptions. Cliff’s delta of $-0.361$ corresponds to a medium–large effect size, implying that a randomly selected Cooling ME has a higher $\Gamma_p$ than a Heating ME in roughly 68\% of cases. This pattern is evident in the KDE plot (Figure~\ref{fig:kde}b), where Cooling MEs show a sharp peak near $\Gamma_p \approx 2$, whereas Heating MEs display a broader, bimodal distribution with enhanced occurrence at very low $\Gamma_p$ values (near 0), characteristic of strongly compressive or sub-adiabatic conditions. A permutation test on median $\Gamma_p$ values further reinforces these findings: the median difference between Heating and Cooling MEs is substantial ($\Delta \Gamma_p = -1.82$) with a p-value effectively zero, ruling out random chance. Overall, the results indicate that Heating MEs are defined by substantially lower polytropic indices, reflecting enhanced internal energy deposition, while Cooling MEs exhibit near-adiabatic or thermally relaxing evolution.

The $T_p$ profiles further emphasize the contrast between Heating and Cooling MEs. In Heating MEs, $T_p$ rises sharply from $6.2 \times 10^4$~K in the pre-ICME solar wind to $15.3 \times 10^4$~K in the sheath, indicative of shock- or compression-driven heating at the leading edge (Table~\ref{tab:median}). Inside the ME, the temperature drops to $3.1 \times 10^4$~K, typical for ME interiors—but the exceptionally low $\Gamma_p$ suggests that this reduction is not due to adiabatic cooling alone, but may involve non-adiabatic energy exchange processes. The post-ICME region shows a rebound to $8.6 \times 10^4$~K. In contrast, Cooling MEs exhibit a smoother $T_p$ variation: starting at $4.9 \times 10^4$~K pre-ICME, increasing to $9.3 \times 10^4$~K in the sheath, then decreasing to $2.9 \times 10^4$~K within the ME, followed by a rise to $6.8 \times 10^4$~K post-ICME. This gradual cooling pattern aligns with their consistently high $\Gamma_p$ values ($>1.67$) across all regions.

The $N_p$ evolution reinforces these distinctions. Heating MEs show a moderate increase from 5.0~cm$^{-3}$ pre-ICME to 7.7~cm$^{-3}$ in the sheath due to compression, followed by a drop to 5.0~cm$^{-3}$ in the ME and a nearly unchanged post-ICME value of 4.9~cm$^{-3}$. Cooling MEs begin with a slightly higher pre-ICME density (5.0~cm$^{-3}$), experience a stronger sheath enhancement (9.4~cm$^{-3}$), and retain moderately elevated densities in the ME (5.6~cm$^{-3}$) and post-ICME (5.7~cm$^{-3}$), consistent with reduced expansion and stronger compression overall (Figure~\ref{fig:sea_heating_cooling}).

Magnetic field strength ($B$) is enhanced in the sheath for both categories, 10.2~nT for Heating and 8.3~nT for Cooling MEs, likely due to compression, but decreases within the ME. Notably, Cooling MEs retain a slightly stronger median $B$ (8.5~nT) inside the ME compared to Heating MEs (8.3~nT), consistent with less expansion and greater magnetic field preservation. Flow speed ($V$) also differentiates the two groups: Heating MEs start at 423~km~s$^{-1}$, peak at 505~km~s$^{-1}$ in the sheath, and decrease to 454~km~s$^{-1}$ in the ME and 448~km~s$^{-1}$ post-ICME, reflecting expansion-driven deceleration. Cooling MEs start slower at 377~km~s$^{-1}$, rise to 437~km~s$^{-1}$ in the sheath, and slow further to 410~km~s$^{-1}$ in the ME and post-ICME regions. This indicates that Heating MEs undergo stronger expansion, whereas Cooling MEs remain more compact and slower-moving.

The $\beta$ profiles reveal similar patterns inside the ME and post-ICME regions but diverge upstream and in the sheath. In Heating MEs, $\beta$ is $\sim$1.05 in the pre-ICME and post-ICME solar wind, decreases to 0.94 in the sheath, and drops further to 0.48 inside the ME, reflecting the dominance of magnetic pressure in the ejecta before returning to ambient conditions afterward. Cooling MEs display a sharper drop in $\beta$ inside the ME (0.5) from higher values in the pre-ICME (1.3) and post-ICME (1.1) regions. The key difference lies in the upstream and sheath conditions: Cooling MEs are preceded by plasma that is less magnetically dominated, consistent with the weaker magnetic fields observed in those regions (Figure~\ref{fig:sea_heating_cooling}).

The $P_{dyn}$ profiles closely follow the velocity evolution, peaking in the sheath for both categories, 4.0~nPa for Heating MEs and 3.6~nPa for Cooling MEs, before declining through the ME and post-ICME regions. Interestingly, despite differences in $V$ and $N_p$, the post-ICME dynamic pressure remains identical for both categories at 2.0~nPa. While Heating MEs exhibit stronger $P_{dyn}$ in the sheath, Cooling MEs tend to produce a more pronounced Sym-H spike at the time of SSC, implying that factors beyond dynamic pressure, such as the magnitude and orientation of the upstream magnetic field, also influence SSC strength. Nevertheless, a closer inspection reveals that the jump in Sym-H from the pre-ICME baseline is larger for Heating MEs, consistent with enhanced magnetospheric compression resulting from higher dynamic pressures at shock arrival (Figure~\ref{fig:sea_heating_cooling}).

Following SSC, the Sym-H index declines as the sheath and ME pass (Figure~\ref{fig:sea_heating_cooling}h). This decline is noticeably steeper in the sheath of Heating MEs compared to Cooling MEs, suggesting a more rapid onset of magnetic reconnection and geomagnetic disturbance during the early storm phase. Moreover, Heating MEs reach lower Sym-H minima within the ME, indicating a greater tendency to drive intense geomagnetic storms. This is consistent with the elevated magnetic field strengths observed in the sheath and leading edge of Heating MEs. During the recovery phase, both categories display a similar return towards baseline values, implying that the post-ICME solar wind environment is largely independent of the preceding ME’s heating or cooling nature. Notably, pre-ICME regions associated with Heating MEs exhibit more negative Sym-H values than those preceding Cooling MEs, suggesting a more disturbed magnetospheric state prior to ME arrival, potentially due to earlier solar wind structures. Overall, Heating MEs are associated with stronger dynamic pressure, deeper Sym-H depressions, and more pronounced pre- and post-event activity than Cooling MEs.

The expansion speed $V_{exp}$ is slightly higher for Heating MEs (24~km~s$^{-1}$) than for Cooling MEs (19~km~s$^{-1}$). Although modest, this difference indicates that Heating MEs undergo stronger radial expansion, which may suppress cooling and promote energy retention within the ME, a trend also identified for CMEs near the Sun by \citet{Khuntia2024}. The distortion parameters DiP-B and DiP-N for both categories are close to 0.5, indicating largely symmetric internal structures (Table~\ref{tab:median}). However, Heating MEs exhibit marginally lower values (0.47 for DiP-B, 0.46 for DiP-N), pointing to slight compression toward the front of the ME. Cooling MEs, with slightly higher values (both 0.49), display near-central symmetry with a minor tendency for rear-side compression. These subtle differences suggest that Heating MEs may be more dynamically evolving or shaped by upstream solar wind compression, whereas Cooling MEs, being more magnetically organized, maintain a more balanced radial profile.

\begin{figure*}
    \centering
    \hspace{-1cm}  
    \includegraphics[width=1.05\linewidth]{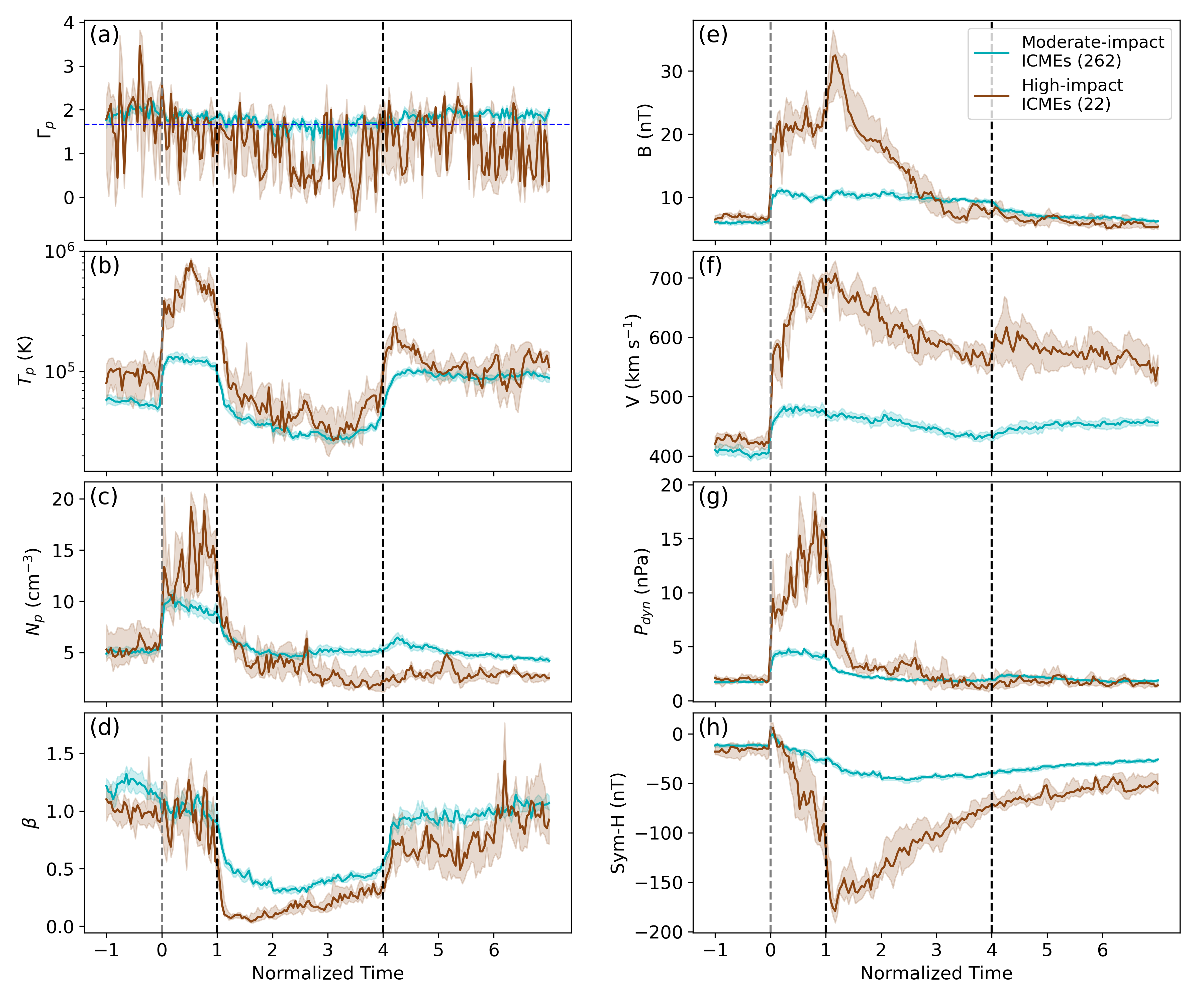}
    \caption[Superposed Epoch Analysis showing the median values]{Superposed Epoch Analysis showing the median values of parameters across the pre-ICME, sheath, ME, and post-ICME regions: (a) polytropic index ($\Gamma_p$), (b) proton temperature ($T_p$), (c) proton number density ($N_p$), (d) plasma beta ($\beta$), (e) magnetic field strength ($B$), (f) flow speed ($V$), (g) dynamic pressure ($P_{dyn}$), and (h) geomagnetic index (Sym-H). The curves represent: teal for Moderate-impact ICMEs and brown for High-impact ICMEs. Black vertical dashed lines mark the boundaries for the ME region, and gray dashed vertical lines mark the starting sheath region.The background color-shaded regions represent the 68\% (1$\sigma$) confidence intervals obtained via bootstrap resampling.}
    \label{fig:sea_storm}
\end{figure*}

\subsubsection{Variation in Plasma Evolution Across Different Storms Classes}\label{chapt4-part2-subsec:dsea_storm}

To assess the geoeffectiveness of ICMEs, we classified events into two categories based on their geomagnetic impacts:  
(i) \textit{High-impact ICMEs}, which drive storms with Sym-H~$<$~–200~nT, and  
(ii) \textit{Moderate-impact ICMEs}, which drive storms with –200~nT~$<$~Sym-H~$<$~–50~nT.  
Events with Sym-H~$>$~–50~nT were excluded to avoid weak cases that could dilute the median profiles and mask clear trends in plasma and magnetic field parameters. This classification is designed to highlight how variations in solar wind and ICME properties, particularly thermal and magnetic characteristics, relate to storm intensity.

Our analysis reveals a distinct difference in the thermodynamic behavior, as measured by $\Gamma_p$, between High-impact and Moderate-impact ICMEs (Figure~\ref{fig:sea_storm}a). In High-impact cases, $\Gamma_p$ drops sharply from 1.97 in the pre-ICME region to 1.56 in the sheath, and further to 0.59 within the ME, before partially recovering to 1.21 in the post-ICME region (Table~\ref{tab:median}). This pronounced decline within the ME indicates strong heating processes and significant deviations from adiabatic behavior. Moreover, High-impact ICMEs display substantial variability in $\Gamma_p$, suggesting the presence of localized heating. In comparison, Moderate-impact ICMEs also show a decrease in $\Gamma_p$ from 1.98 (pre-ICME) to 1.79 (sheath), but their values remain consistently above the adiabatic index and higher than those of High-impact ICMEs. Inside the ME, $\Gamma_p$~$=$~1.69 for Moderate-impact cases, consistent with near-adiabatic expansion, while the post-ICME value of 1.88 reflects enhanced cooling relative to adiabatic conditions. This contrast suggests that High-impact ICMEs undergo more pronounced thermodynamic changes, likely driven by stronger compression and greater magnetic energy dissipation, compared to their Moderate-impact counterparts.

\begin{figure*}
    \centering
    \includegraphics[width=0.47\linewidth]{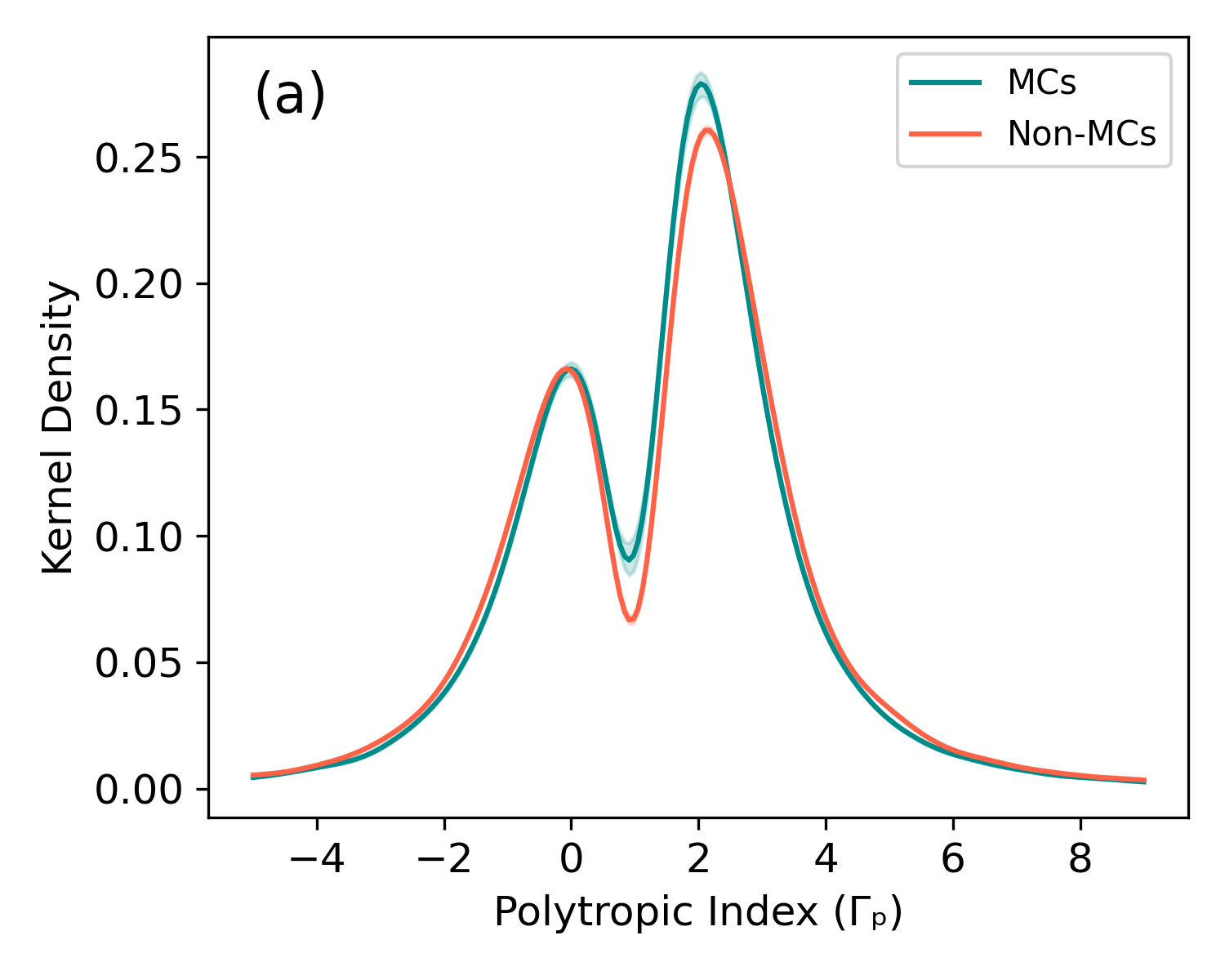}
     \includegraphics[width=0.47\linewidth]{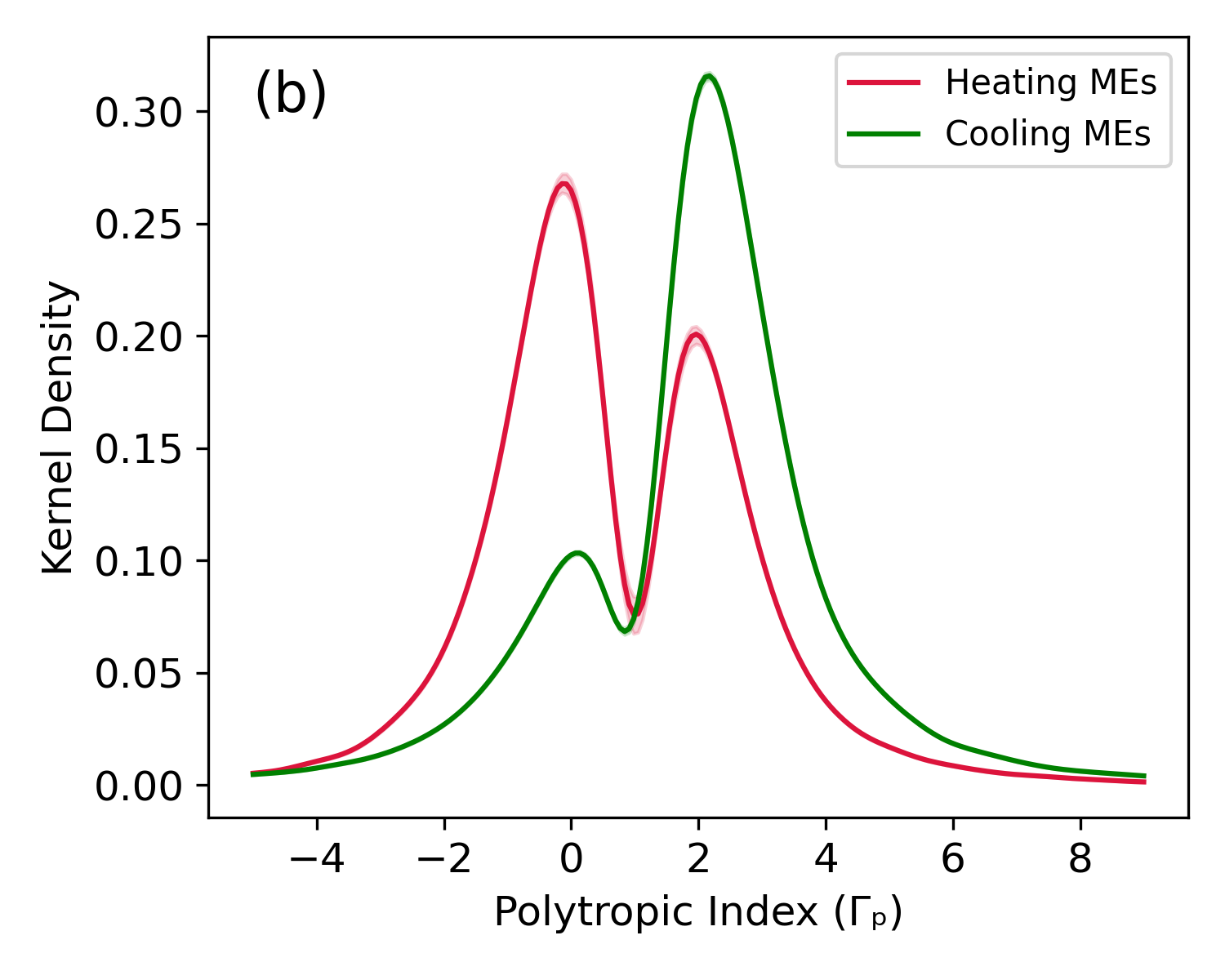}
      \includegraphics[width=0.47\linewidth]{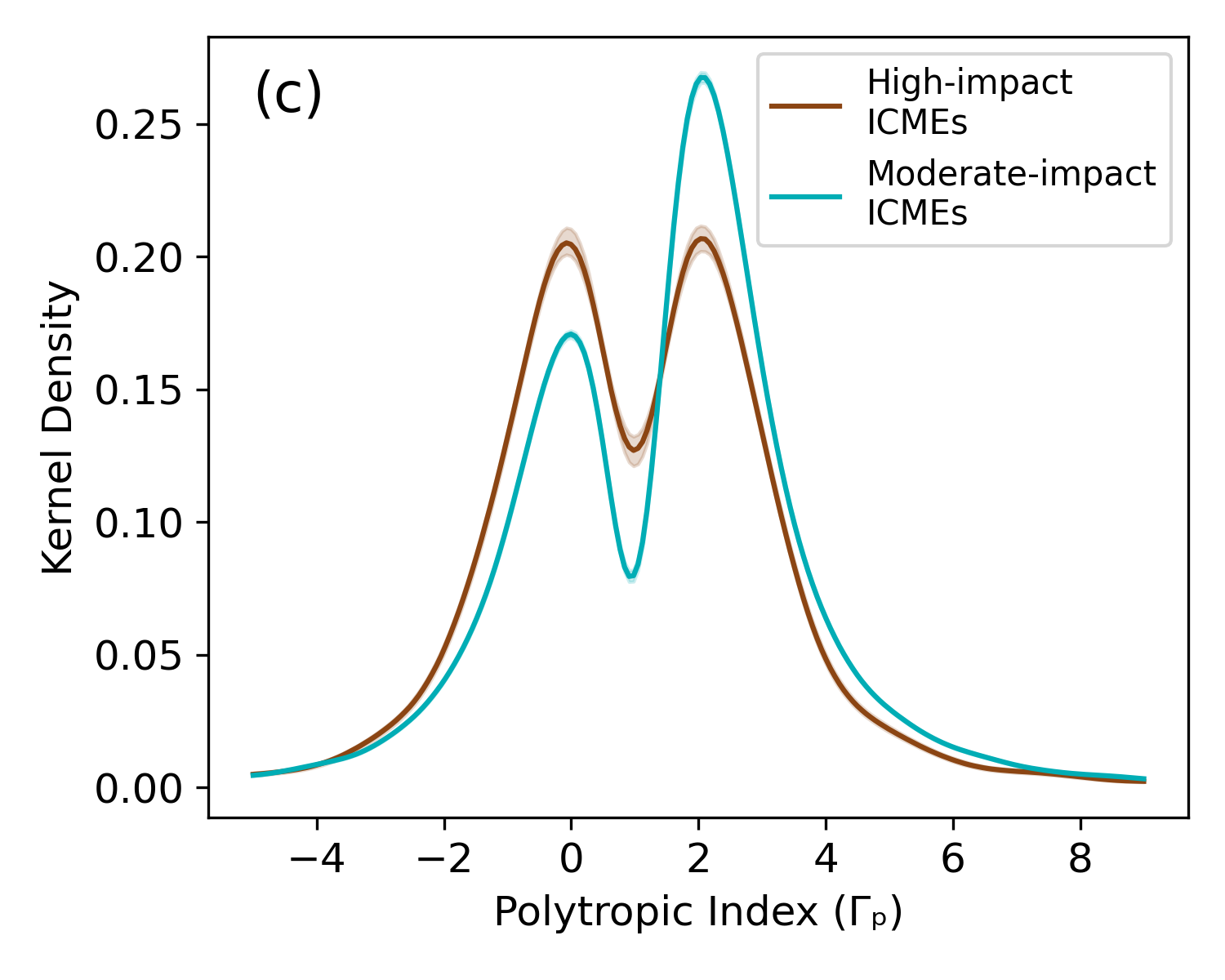}
    \caption[Comparison of $\Gamma_p$ distributions using KDE plots]{Comparison of $\Gamma_p$ distributions using KDE plots with 68\% bootstrap confidence bands for ME part of (a) MCs vs. Non-MCs, (b) Heating vs. Cooling MEs, and (c) High-impact vs. Moderate-impact ICMEs. These plots highlight differences in thermodynamic properties across ICME subsets.}
    \label{fig:kde}
\end{figure*}

A direct comparison of the ME regions in High-impact and Moderate-impact ICMEs reveals statistically significant differences in their thermodynamic behavior, as characterized by $\Gamma_p$, though the magnitude of these differences is modest. The KS test returns a statistic of $D = 0.106$ with a p-value of $9.52 \times 10^{-98}$, indicating a clear difference in the shapes of the two $\Gamma_p$ distributions. The Mann–Whitney U test corroborates this result, yielding a p-value of $1.09 \times 10^{-89}$ and confirming a shift in central tendency. The effect size, however, is small: Cliff’s Delta of $-0.116$ suggests that a randomly chosen Moderate-impact ICME has a higher $\Gamma_p$ than a High-impact ICME in roughly 56\% of cases. This is consistent with the observed median difference of $\Delta \Gamma_p = -0.56$ between the two populations, which is statistically significant according to the permutation test ($p < 10^{-5}$). The KDE plot (Figure~\ref{fig:kde}c) visually captures this subtle but persistent separation, showing that Moderate-impact events cluster more densely in the higher-$\Gamma_p$ range. SEA results reinforce this pattern: during the main storm phase, the fraction of time bins with $\Gamma_p < 1.67$ is markedly greater for High-impact events (66\%) than for Moderate-impact events (19\%), indicating more frequent heating in the stronger storms. Conversely, Moderate-impact ICMEs retain $\Gamma_p > 1.67$ in 81\% of time bins, suggesting a tendency toward adiabatic evolution. Collectively, these findings indicate that High-impact ICMEs are statistically more likely to exhibit lower polytropic indices, pointing to stronger internal heating or energy dissipation processes that may contribute to their enhanced geoeffectiveness.

\begin{table*}[ht]
\caption[Median value of plasma and magnetic field parameters]{Median value of plasma and magnetic field parameters of various regions associated with different categories of ME.}
\label{tab:median}
\centering
\hspace{-1cm}
\begin{adjustbox}{width=1.05\textwidth}
\begin{tabular}{llcccccccccc}
\hline
Category & Regions & $\Gamma_p$ & $T_p$ & $N_p$ & $V$ & $\beta$ & $B$ & P$_{dyn}$ & DiP-B & DiP-N & $V_{exp}$ \\
 &  &  & ($10^4$ K) & (cm$^{-3}$) & (km s$^{-1}$) &  & (nT) & (nPa) & & & (km s$^{-1}$) \\
\hline
\multirow{4}{3cm}{All ICME}
 & pre & 2.02 & 5.3 & 5.0 & 400 & 1.2 & 8.6 & 1.7 & - & - & - \\
 & sheath & 1.83 & 11.5 & 8.6 & 469 & 1.1 & 9.3 & 3.8 & - & - & -  \\
 & me & 1.73 & 3.0 & 5.3 & 433 & 0.5 & 8.5 & 2.0 & 0.48 & 0.48 & 21\\
 & post & 1.91 & 7.6 & 5.4 & 428 & 1.1 & 6.6 & 2.0 & - & - & - \\
 \hline
\multirow{4}{3cm}{MC}
 & pre & 2.09 & 4.5 & 6.6 & 375 & 1.4 & 5.6 & 1.6 & -  & -  &  -  \\
 & sheath & 1.83 & 10.1 & 11.4 & 455 & 1.1 & 10.4 & 4.9 & -  & -  & -   \\
 & me & 1.75 & 2.8 & 6.3 & 421 & 0.3 & 11.3 & 2.2 & 0.48 & 0.46 &  21\\
 & post & 1.88 & 7.5 & 5.7 & 432 & 1.1 & 6.8 & 2.1 & -  &  - &  - \\
 \hline
\multirow{4}{3cm}{Non-MC} 
 & pre & 1.95 & 6.3 & 4.8 & 417 & 1.1 & 6.1 & 1.7 & -  & -  & -   \\
 & sheath & 1.83 & 12.8 & 7.2 & 479 & 1.1 & 8.7 & 3.4 &  - & -  & -  \\
 & me & 1.72 & 3.1 & 5.0 & 438 & 0.7 & 7.1 & 1.9 & 0.48 & 0.49 &  20\\
 & post & 1.93 & 7.7 & 5.2 & 428 & 1.1 & 6.6 & 1.9 & -  & -  & -  \\
 \hline
\multirow{4}{2.5cm}{Heating MEs}  
& pre   & 1.84 & 6.2 & 5.0 & 423 & 1.0 & 6.6 & 1.8 & - & - & - \\
& sheath & 1.64 & 15.3 & 7.7 & 505 & 0.9 & 10.2 & 4.0 & - & - & - \\
& me     & 0.42 & 3.1 & 5.0 & 454 & 0.5 & 8.3 & 2.2 & 0.47 & 0.46 & 24 \\
& post   & 1.59 & 8.6 & 4.9 & 448 & 1.0 & 6.7 & 2.0 & - & - & - \\
\hline
\multirow{4}{2.5cm}{Cooling MEs}  
& pre   & 2.16 & 4.9 & 5.0 & 377 & 1.3 & 5.5 & 1.6 & - & - & - \\
& sheath & 1.99 & 9.3 & 9.4 & 437 & 1.2 & 8.3 & 3.7 & - & - & - \\
& me     & 2.07 & 2.9 & 5.6 & 412 & 0.5 & 8.5 & 2.0 & 0.49 & 0.49 & 19 \\
& post   & 2.14 & 6.8 & 5.7 & 410 & 1.1 & 6.6 & 2.0 & - & - & - \\
\hline
\multirow{4}{2.5cm}{High-impact ICMEs}
 & pre & 1.97 & 9.1 & 7.3 & 422 & 1.1 & 6.8 & 1.8 & - & - & - \\
 & sheath & 1.56 & 58.4 & 17.2 & 676 & 1.0 & 22.8 & 14.5 & - & - & - \\
 & me & 0.59 & 4.5 & 3.2 & 615 & 0.1 & 13.4 & 3.0 & 0.30 & 0.38 & 70 \\
 & post & 1.21 & 10.8 & 3.1 & 573 & 0.8 & 6.2 & 2.0 & - & - & - \\
 \hline
\multirow{4}{3cm}{Moderate-impact ICMEs}
 & pre & 1.98 & 4.9 & 5.4 & 409 & 1.1 & 6.2 & 1.8 & - & - & - \\
 & sheath & 1.79 & 13.1 & 9.7 & 479 & 1.0 & 10.7 & 4.7 & - & - & -\\
 & me & 1.69 & 3.3 & 5.2 & 455 & 0.4 & 10.0 & 2.1 & 0.46 & 0.45 & 22 \\
 & post & 1.88 & 9.1 & 5.0 & 456 & 1.0 & 7.0 & 2.0 & - & - & - \\
 \hline
\end{tabular}
\end{adjustbox}
\end{table*}

The $T_p$ profiles further emphasize this contrast. High-impact ICMEs display a pronounced temperature peak in the sheath region, with a median of $58.4 \times 10^4$~K, far exceeding the $13.1 \times 10^4$~K observed for Moderate-impact events (Table~\ref{tab:median}). This elevated sheath temperature in strong events reflects intense shock-driven compression ahead of the ME. Upstream conditions are also hotter in High-impact cases, with pre-ICME temperatures of $9.1 \times 10^4$~K compared to $5.0 \times 10^4$~K for Moderate-impact events. Within the ME, temperatures drop sharply in both categories, to $4.5 \times 10^4$~K (High-impact) and $3.3 \times 10^4$~K (Moderate-impact), consistent with expansion-related cooling in the magnetically dominated ejecta. Post-ICME temperatures remain elevated, at $10.8 \times 10^4$~K for High-impact and $9.1 \times 10^4$~K for Moderate-impact ICMEs.

Density evolution ($N_p$) follows the expected pattern of enhancement in the sheath and reduction inside the ME, but with notable differences in magnitude. In High-impact events, the sheath density surges to 17.1~cm$^{-3}$ from 7.3~cm$^{-3}$ in the pre-ICME region, then drops steeply to 3.1~cm$^{-3}$ inside the ME, remaining at this low level post-ICME. In contrast, Moderate-impact ICMEs exhibit a more modest sheath density of 9.7~cm$^{-3}$ and sustain higher densities within the ME (5.2~cm$^{-3}$) and post-ICME region (5.0~cm$^{-3}$). These differences indicate that High-impact ICMEs undergo stronger sheath compression but also more significant rarefaction within the ME, consistent with enhanced expansion effects.

Notably, $\beta$ distinctly highlights the magnetic dominance within the ME regions of the two storm-driving categories (Figure~\ref{fig:sea_storm}d). In High-impact ICMEs, $\beta$ drops sharply from 1.1 in the pre-ICME solar wind and 1.0 in the sheath to an exceptionally low value of 0.1 inside the ME, before rising to 0.8 in the post-ICME region. Moderate-impact ICMEs follow a similar general pattern but with much smaller variations, maintaining $\beta$ values of 1.1 (pre-ICME), 1.0 (sheath), 0.4 (ME), and 1.0 (post-ICME). The extremely low $\beta$ inside the MEs of High-impact ICMEs reflects a stronger magnetic pressure and higher magnetic flux content, both of which directly enhance their storm-driving potential. Magnetic field strength ($B$) is also significantly higher in High-impact events, peaking at 22.8~nT in the sheath and 13.4~nT in the ME, compared to 10.7~nT and 10.1~nT, respectively, for Moderate-impact ICMEs (Figure~\ref{fig:sea_storm}e). Within the ME of High-impact events, $B$ declines rapidly from the leading to the trailing edge, a feature also evident in the DiP-B parameter, which has a low value of 0.30, indicative of strong magnetic compression at the ME front (Table~\ref{tab:median}). The corresponding DiP-N value of 0.38 shows that density compression is present but less pronounced than magnetic compression. Such front-loaded asymmetry in High-impact ICMEs is likely the result of stronger dynamic interactions with the ambient solar wind and enhanced expansion, producing sharper, more compressed fronts and more depleted trailing regions. In contrast, Moderate-impact ICMEs have DiP-B = 0.46 and DiP-N = 0.45, pointing to more symmetric magnetic and density profiles within the ME. These elevated magnetic fields in both the sheath and ME are key contributors to the development of intense geomagnetic storms, especially when accompanied by sustained southward magnetic field components.

As expected, $V$ is substantially higher in High-impact ICMEs, reflecting their faster propagation speeds (Figure~\ref{fig:sea_storm}f). For these events, $V$ rises from 422~km~s$^{-1}$ in the pre-ICME region to 676~km~s$^{-1}$ in the sheath, peaks at 615~km~s$^{-1}$ inside the ME, and remains elevated at 573~km~s$^{-1}$ post-ICME. Moderate-impact ICMEs, by comparison, show a more subdued profile: 409~km~s$^{-1}$ (pre), 479~km~s$^{-1}$ (sheath), 455~km~s$^{-1}$ (ME), and 456~km~s$^{-1}$ (post). The larger velocity enhancements in High-impact ICMEs indicate greater kinetic energy and geoeffectiveness, consistent with their association with faster CMEs. Notably, $V_{\mathrm{exp}}$ is significantly greater in High-impact events, with a median of 70~km~s$^{-1}$ versus 22~km~s$^{-1}$ for Moderate-impact events, suggesting higher internal pressure and a more dynamically evolving structure. Dynamic pressure ($P_{dyn}$) also exhibits striking differences: in High-impact ICMEs, it surges from 1.8~nPa in the pre-ICME region to 14.5~nPa in the sheath, before falling to 3.0~nPa in the ME and 2.0~nPa post-ICME. Moderate-impact ICMEs show a smaller sheath enhancement (4.7~nPa from 1.8~nPa pre-ICME), with ME and post-ICME values of 2.1~nPa and 2.0~nPa, respectively. These results indicate that High-impact ICMEs drive a much more intense initial compression of the magnetosphere.

The geomagnetic response, represented by the Sym-H index, also differs markedly between the two categories, particularly during and after storm onset (Figure~\ref{fig:sea_storm}h). In the pre-ICME region, both High-impact and Moderate-impact events exhibit similar Sym-H levels, indicating comparable quiet geomagnetic conditions prior to arrival. Upon impact, both produce a sudden positive spike in Sym-H (SSC), despite the much stronger $P_{dyn}$ associated with High-impact shocks, suggesting that SSC magnitude depends on additional coupling conditions beyond $P_{dyn}$ alone. Following the SSC, the sheath phase shows a steep drop in Sym-H for High-impact ICMEs, signifying more intense magnetic reconnection and energy transfer into the magnetosphere. The lowest Sym-H values, marking the main storm phase, occur near the ME leading edge in High-impact events. The recovery phase typically begins shortly thereafter, extending through the remainder of the ME and into the post-ICME solar wind.

\begin{table*}
\centering
\resizebox{0.78\linewidth}{!}{
\rotatebox{90}{
\begin{threeparttable}
\caption[List of ICMEs studied from June 1996 to December 2024,]{List of ICMEs studied from June 1996 to December 2024, covering solar cycles 23 to 25. The table includes the corresponding derived median value of the polytropic index ($\Gamma_p$), expansion speed ($V_{exp}$), and distortion parameters for magnetic field (DiP-B) and proton number density (DiP-N).}
\label{tab:ICME_list}
\begin{tabular}{rccccccccc}
\hline
\textbf{No. }& \textbf{Disturbance time} & \textbf{ME start time} & \textbf{ME end time} & \textbf{ME type} & \textbf{Year} & \textbf{Median $\Gamma_p$} & \textbf{$V_{exp}$} &\textbf{ DiP-B} & \textbf{DiP-N} \\
 & &  &  &  &  &  & ($\mathrm{km\ s^{-1}}$) &  &  \\
\hline
1 & 1996-07-01 13:20 & 1996-07-01 18:00 & 1996-07-02 11:00 & 2 & 1996 & 2.439 & 12 & 0.522 & 0.525 \\
2 & 1996-08-07 06:00 & 1996-08-07 12:00 & 1996-08-08 10:00 & 2 & 1996 & 2.230 & 10 & 0.578 & 0.439 \\
3 & 1996-12-23 16:00 & 1996-12-23 17:00 & 1996-12-25 11:00 & 2 & 1996 & 2.032 & 53 & 0.589 & 0.515 \\
4 & 1997-01-10 01:04 & 1997-01-10 04:00 & 1997-01-11 02:00 & 2 & 1997 & 0.776 & 44 & 0.501 & 0.777 \\
5 & 1997-02-09 13:21 & 1997-02-10 02:00 & 1997-02-10 19:00 & 2 & 1997 & 2.117 & 32 & 0.520 & 0.195 \\
6 & 1997-04-10 17:45 & 1997-04-11 06:00 & 1997-04-11 19:00 & 2 & 1997 & 1.342 & -7 & 0.467 & 0.521 \\
7 & 1997-04-21 06:00 & 1997-04-21 10:00 & 1997-04-23 04:00 & 2 & 1997 & 2.208 & 31 & 0.512 & 0.444 \\
8 & 1997-05-15 01:59 & 1997-05-15 09:00 & 1997-05-16 00:00 & 2 & 1997 & 1.967 & -12 & 0.443 & 0.400 \\
9 & 1997-05-26 09:57 & 1997-05-26 16:00 & 1997-05-27 10:00 & 2 & 1997 & 1.936 & 11 & 0.495 & 0.443 \\
10 & 1997-06-08 16:36 & 1997-06-08 18:00 & 1997-06-10 00:00 & 2 & 1997 & 2.038 & 19 & 0.538 & 0.402 \\
\hline
\end{tabular}

\begin{tablenotes}
\small
\item Note: This table is available in its entirety in the \href{https://drive.google.com/file/d/1YwEyPO1zbwrVcNjuJPZi2j5F_gbAuMmc/view?usp=sharing}{G-drive link}. A portion is shown here for illustrative purposes, indicating the form and content. The date time format used in columns 2–4 is \textit{YYYY-MM-DD HH:MM}.
\end{tablenotes}

\end{threeparttable}}}

\end{table*}

\subsection{Summary} \label{chapt4-part2-sec:summary}

This study provides a comprehensive thermal characterization of MEs and their associated plasma properties across Solar Cycles 23, 24, and the rising phase of 25, using $\Gamma_p$ as a key diagnostic parameter. By classifying MEs into Heating and Cooling types based on $\Gamma_p$, and applying a refined categorization scheme, we examined trends within and around ICMEs, revealing critical insights into their Solar Cycle dependence, thermodynamic evolution, plasma behavior, and geoeffectiveness. The main findings are as follows:

\begin{enumerate}
    
    \item Heating MEs are most prevalent during solar maxima and descending phases (particularly in SC23), reflecting strong thermal energy input during eruption. In contrast, the rising phases of SC23, SC24, and SC25 are dominated by Cooling MEs, suggesting either reduced heating efficiency or distinct source-region conditions during these times.
    
    \item Median $\Gamma_p$ increases progressively from 1.46 in SC23 to 1.87 in SC24, marking a shift from heating-dominated to cooling-dominated thermal states. The proportion of Heating MEs drops from $\sim$59\% in SC23 to 34\% in SC24, pointing to differences in CME properties near the Sun and/or their evolution en route to 1~au.
    
    \item The refined $\Gamma_p$-based classification (Major-heating, Isothermal, Adiabatic, Major-cooling) shows that MEs rarely evolve adiabatically or isothermally, instead behaving as thermodynamically active systems. Most MEs exhibit extreme $\Gamma_p$ values ($<0.9$ or $>1.77$), indicating intense energy exchange, either substantial energy gain or pronounced thermal loss.
    
    \item Heating MEs display strong Solar Cycle modulation in $\Gamma_p$, $T_p$, and $V_{exp}$, indicating active heating processes tied to solar activity and eruption energetics. Large year-to-year variability in $\Gamma_p$, with minima (strong heating) near solar maxima and minima (especially in SC23 and SC24), reflects periods of enhanced heating. Even in years with low in situ $T_p$ (e.g., 2013), high $V_{exp}$ and low $\Gamma_p$ suggest significant in-transit heating, implying that both initial eruption conditions and interplanetary evolution control ME thermal states.
    
    \item Cooling MEs remain thermodynamically stable across cycles, with nearly constant $\Gamma_p \approx 2$, implying enhanced cooling beyond adiabatic expansion. Despite this, their $T_p$ is often comparable to that of Heating MEs, suggesting higher initial thermal content or better retention of internal energy during transit to 1~au.
    
    \item Geoeffective ICMEs are predominantly MCs, characterized by strong magnetic fields, low plasma beta, and distinct thermal properties such as low $T_p$ and $\Gamma_p < 1.67$. Their coherent magnetic structure and magnetic pressure dominance contribute to deeper Sym-H depressions compared to non-MC ICMEs.
    
    \item Heating and Cooling MEs follow fundamentally different thermodynamic pathways. Heating MEs exhibit strongly non-adiabatic behavior with very low $\Gamma_p$, elevated dynamic and magnetic pressures, significant internal energy deposition, and enhanced geoeffectiveness. They also show front-loaded compression, internal density asymmetry (low DiP), and faster expansion speeds. Cooling MEs, by contrast, exceed adiabatic expectations (high $\Gamma_p$), maintain magnetic coherence, and generally produce milder geomagnetic responses.
    
    \item High-impact ICMEs are typically Heating MEs (predominantly MCs) with median $\Gamma_p \approx 0.59$, stronger magnetic fields, lower plasma beta, higher bulk and expansion speeds, and greater dynamic compression in the sheath. The thermal state of an ME, combined with its magnetic and plasma properties, is a useful predictor of its potential to drive strong geomagnetic storms. In many cases, storm intensity results not only from ME properties but also from significant contributions of the preceding sheath.
    
\end{enumerate}

Overall, these results emphasize the importance of incorporating the thermal state of ICMEs, alongside other plasma and magnetic parameters, into studies of their interplanetary evolution, classification, and space weather impact potential at 1~au. The statistical behavior of $\Gamma_p$ offers a novel perspective for interpreting CME energetics and may aid in improving space weather forecasting capabilities.

\begin{table*}
\centering
\caption[Annual distribution of magnetic ejecta with various thermal]{Annual distribution of magnetic ejecta with various thermal states at 1 au.}
\label{tab:thermal_state_perce}
\begin{adjustbox}{width=0.7\textwidth}
\begin{tabular}{cccccc}
\hline
\textbf{Year} & 
\textbf{Major-heating} & 
\textbf{Isothermal} & 
\textbf{Heating} & 
\textbf{Adiabatic} & 
\textbf{Major-Cooling} \\
& \textbf{Count (\%)} & \textbf{Count (\%)} & \textbf{Count (\%)} & \textbf{Count (\%)} & \textbf{Count (\%)} \\
\hline
1996 & 0 (0) & 0 (0) & 0 (0) & 0 (0) & 3 (100) \\
1997 & 1 (5) & 0 (0) & 1 (5) & 1 (5) & 19 (86) \\
1998 & 5 (14) & 1 (3) & 3 (8) & 4 (11) & 23 (64) \\
1999 & 11 (33) & 0 (0) & 2 (6) & 6 (18) & 14 (42) \\
2000 & 24 (48) & 0 (0) & 5 (10) & 5 (10) & 16 (32) \\
2001 & 29 (60) & 0 (0) & 10 (21) & 1 (2) & 8 (17) \\
2002 & 12 (46) & 1 (4) & 3 (12) & 4 (15) & 6 (23) \\
2003 & 9 (43) & 0 (0) & 2 (9) & 3 (14) & 7 (33) \\
2004 & 9 (43) & 0 (0) & 2 (10) & 5 (24) & 5 (24) \\
2005 & 20 (65) & 0 (0) & 7 (23) & 2 (6) & 2 (6) \\
2006 & 6 (46) & 1 (8) & 1 (8) & 2 (15) & 3 (23) \\
2007 & 1 (50) & 0 (0) & 0 (0) & 0 (0) & 1 (50) \\
2008 & 0 (0) & 0 (0) & 0 (0) & 1 (33) & 2 (67) \\
2009 & 2 (18) & 0 (0) & 1 (9) & 3 (27) & 5 (45) \\
2010 & 3 (20) & 0 (0) & 2 (13) & 0 (0) & 10 (67) \\
2011 & 5 (16) & 0 (0) & 2 (6) & 5 (16) & 20 (63) \\
2012 & 3 (9) & 0 (0) & 3 (9) & 3 (9) & 26 (74) \\
2013 & 8 (33) & 0 (0) & 0 (0) & 3 (13) & 13 (54) \\
2014 & 6 (30) & 0 (0) & 1 (5) & 4 (20) & 9 (45) \\
2015 & 10 (33) & 0 (0) & 4 (13) & 4 (13) & 12 (40) \\
2016 & 5 (38) & 0 (0) & 1 (8) & 1 (8) & 6 (46) \\
2017 & 1 (11) & 0 (0) & 1 (11) & 1 (11) & 6 (67) \\
2018 & 1 (13) & 0 (0) & 2 (25) & 0 (0) & 5 (63) \\
2019 & 1 (14) & 0 (0) & 0 (0) & 0 (0) & 6 (86) \\
2020 & 2 (50) & 0 (0) & 0 (0) & 0 (0) & 2 (50) \\
2021 & 0 (0) & 0 (0) & 0 (0) & 0 (0) & 10 (100) \\
2022 & 1 (5) & 0 (0) & 1 (5) & 1 (5) & 16 (84) \\
2023 & 2 (8) & 0 (0) & 4 (16) & 4 (16) & 15 (60) \\
2024 & 1 (4) & 1 (4) & 5 (19) & 4 (15) & 16 (59) \\
\hline
\end{tabular}
\end{adjustbox}
\end{table*}


\section{Recovery Phase of Intense and Stronger-than-intense Geomagnetic Storms During Solar Cycles 23 and 24}

Geomagnetic storms are one of the most prominent manifestations of solar-terrestrial coupling, characterized by large-scale disturbances in Earth's magnetic field. These storms are primarily driven by the arrival of interplanetary structures such as CMEs and Stream Interaction Regions (SIRs) \citep{Gosling1993a, Wimmer1997}. SIRs arise from the interaction between fast solar wind streams emanating from coronal holes and slower-moving ambient solar wind. When such structures persist over several solar rotations, they are often referred to as Co-rotating Interaction Regions (CIRs) \citep{Kilpua2017, Richardson2018}. In addition to isolated drivers, complex interactions between CMEs or between CMEs and SIRs can also result in geomagnetic storms \citep{Zhang2007}. When these interplanetary structures, ICMEs or SIRs, encounter Earth's magnetosphere, they can transfer energy into it, particularly when their magnetic field possesses a strong southward component. This southward field couples effectively with Earth's northward magnetic field on the dayside magnetopause, initiating magnetic reconnection and energy injection into the magnetosphere \citep{Dungey1961, Daglis1999, Tsurutani2006}. The resulting magnetospheric response involves the generation of a westward ring current, which weakens Earth's magnetic field and is recorded as a decrease in the Dst index \citep{Daglis1999, Kozyra2003}. The final stage of the storm, the recovery phase, corresponds to the decay of the ring current and the gradual return of Earth's magnetic field to its quiet state \citep{Akasofu1963, Akasofu2021}. The temporal evolution of the Dst index is governed by the relative efficiency of the ring current injection and its decay processes.

Several mechanisms contribute to the decay of the ring current, including charge exchange with neutral atoms, Coulomb collisions, drift and bounce loss, and wave-particle interactions such as scattering by electromagnetic ion cyclotron (EMIC) and whistler-mode waves \citep{Campbell1996, Daglis1999}. Additionally, other current systems, such as ionospheric, field-aligned, and magnetotail currents, can influence the net ring current behavior. The ring current composition is also critical; for example, O$^+$ ions have a shorter charge exchange lifetime than H$^+$ ions, affecting the decay rate. SIRs, which are typically embedded with large-amplitude Alfvén waves, are known to produce storms with longer recovery durations \citep{Richardson2018, Telloni2021}, and their intensity is often underestimated by empirical models \citep{Liemohn2008}.

It has been well-established that fast CMEs are the dominant drivers of geomagnetic storms near solar maximum, while SIRs become more prominent during solar minimum \citep{Gonzalez1999}. Therefore, understanding the recovery phase characteristics of geomagnetic storms across different phases of solar cycles and for different solar wind drivers is essential. While individual ICMEs and CIRs typically generate single-step (classical) storms, characterized by a main phase decline followed by a smooth recovery, interacting structures (e.g., CME-CME or CME-SIR) often lead to complex storms with multi-step main phases and extended recoveries \citep{Burlaga1987, Gonzalez1999, Zhang2007}. Multiple studies have examined the role of CME-CME and CME-SIR interactions in shaping storm dynamics and intensity \citep{Gopalswamy2001, Harrison2012, Mishra2015, Mishra2017, Mishra2021, Palmerio2022}. Furthermore, it has been observed that nearly all geomagnetic storms, regardless of their drivers, exhibit a two-phase recovery: an initial rapid phase followed by a slower decay \citep{Gonzalez1994, Tsurutani2006}. Given that interacting solar wind structures often result in atypical storm profiles, comparing their recovery characteristics with those of classical storms remains a key area of investigation \citep{Farrugia2006, Lugaz2017}.

Most earlier works have focused on the onset and main phase of geomagnetic storms, analyzing their intensity and association with solar wind parameters \citep{Gonzalez1994, Tsurutani2009, Srivastava2004}. While the storm peak (Dst minimum) is important for assessing the severity of geomagnetic disturbances, the duration of the magnetospheric disturbance, captured by the recovery phase, is equally vital for space weather assessment. However, only limited studies have explored the recovery dynamics in detail, and those that have mostly targeted specific storm cases or limited subsets of intense events \citep{Aguado2010, Yermolaev2012, Cid2013, Telloni2021}.

In this study, we focus on the recovery phase of geomagnetic storms that occurred during Solar Cycles 23 and 24. These two cycles differ in several aspects, including sunspot numbers and the frequency of ICMEs, SIRs, and average radial sizes of ICMEs and MCs at 1 au \citep{Chi2018, Richardson2018, Mishra2019, Gopalswamy2020, Mishra2021a}. Our objective is to analyze and compare the recovery characteristics of both one-step and multi-step geomagnetic storms driven by isolated and interacting interplanetary structures. Specifically, we estimate the recovery durations and examine their dependence on the main phase parameters, including the rate of Dst decrease, main phase duration, and storm intensity. The availability of continuous in situ measurements of the near-Earth solar wind, interplanetary magnetic field, and Dst index throughout Solar Cycles 23 and 24 provides a valuable opportunity for such a comprehensive analysis. This study aims to fill the gap in current understanding by offering a broader statistical perspective on storm recovery behavior across multiple solar cycles and under varying solar wind conditions.

\subsection{Data \& Methodology} 

We utilize the 1-hour resolution Dst index data from the OMNI database, hosted by the National Space Science Data Center of NASA \citep{King2005}, with the original source being the World Data Center for Geomagnetism, Kyoto (\url{https://wdc.kugi.kyoto-u.ac.jp/wdc/Sec3.html}). The Dst index is employed to analyze the occurrence frequency of various classes of geomagnetic storms during solar cycles~23 and~24. Storm classification follows the criteria outlined by \citet{Gonzalez2011}, based on the peak Dst minimum value reached during the main phase:  
\begin{itemize}
    \item Weak geomagnetic storms: $-50\;\text{nT} < \text{Dst} \leq 0\;\text{nT}$  
    \item Moderate geomagnetic storms: $-100\;\text{nT} < \text{Dst} \leq -50\;\text{nT}$  
    \item Intense geomagnetic storms: $-200\;\text{nT} < \text{Dst} \leq -100\;\text{nT}$  
    \item Severe geomagnetic storms: $-350\;\text{nT} < \text{Dst} \leq -200\;\text{nT}$  
    \item Great geomagnetic storms: $\text{Dst} \leq -350\;\text{nT}$  
\end{itemize}

\begin{figure*}
        \centering
        \includegraphics[width=0.7\textwidth]{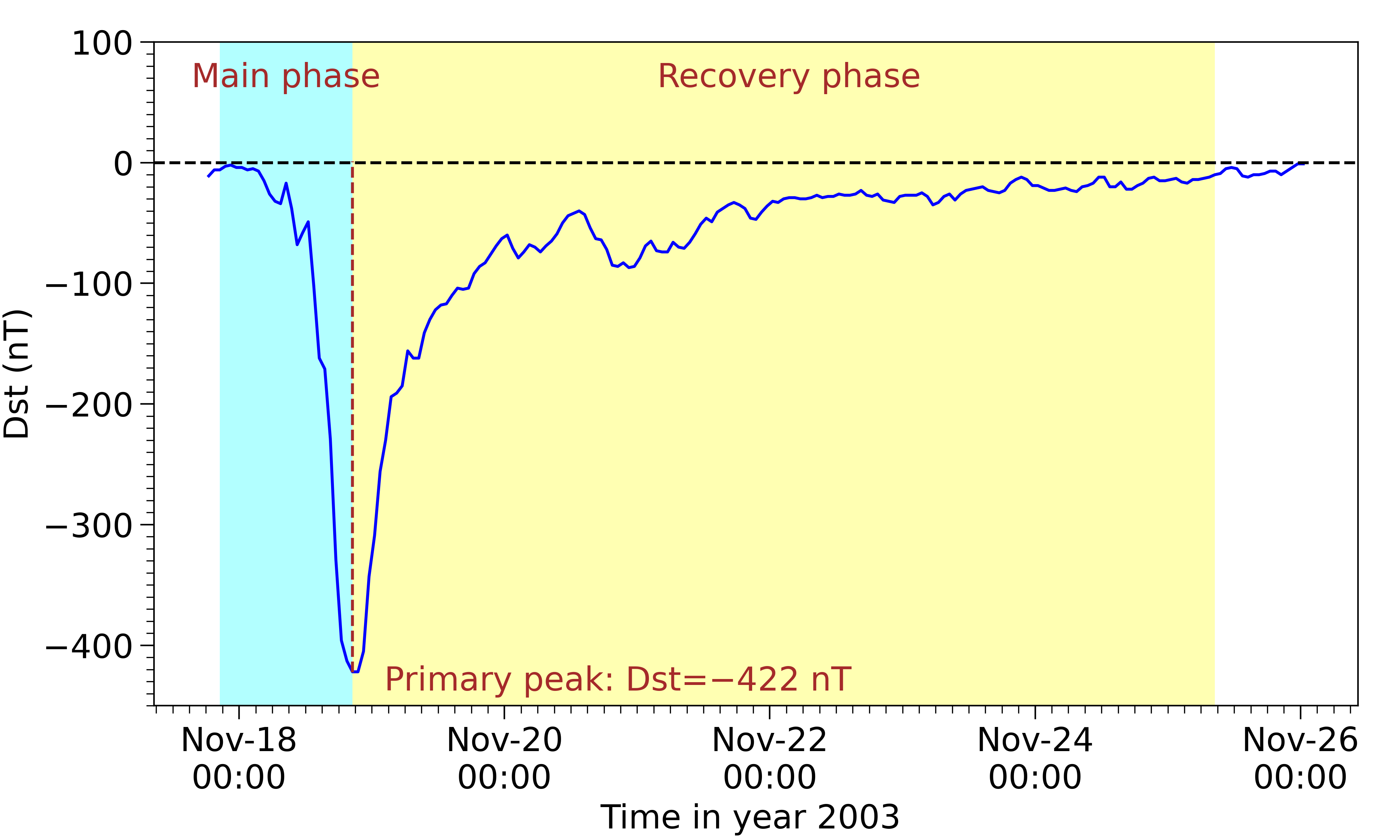} 
        \includegraphics[width=0.7\textwidth]{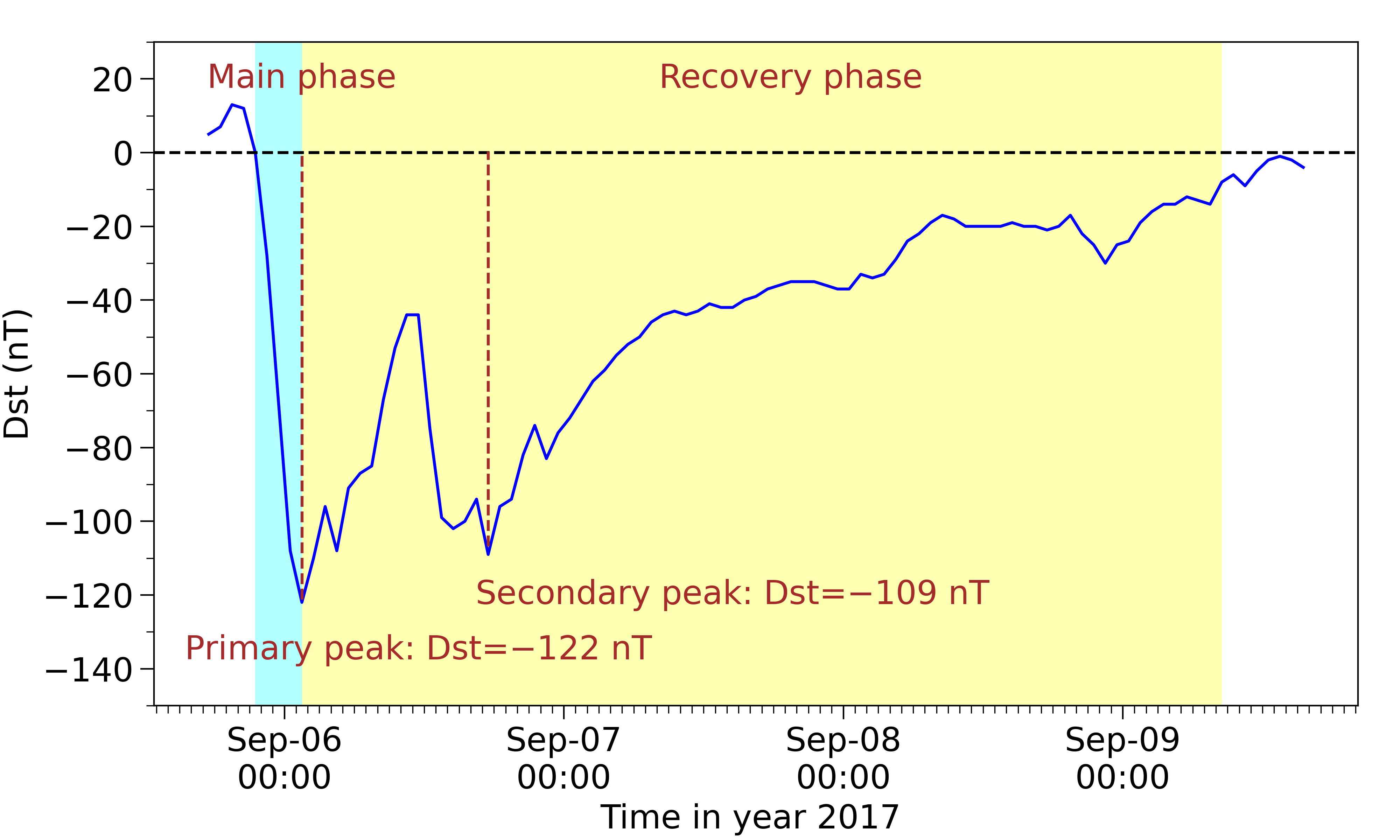} 
       \caption[Example of a single-peaked geomagnetic storm (top)]{Example of a single-peaked geomagnetic storm (top) and a multiple-peaked storm (bottom). The top panel shows the intense geomagnetic storm of November 2003, with a smooth main and recovery phase and a minimum Dst value of $-422\;\text{nT}$. The bottom panel illustrates the multiple-peaked storm of September 2017, where the primary Dst minimum reached $-122\;\text{nT}$. The cyan and yellow shaded regions indicate the main and recovery phases, respectively, for both events.}
        \label{sm_storms2324}
   
\end{figure*}

Analyzing the recovery phase of geomagnetic storms in relation to storm characteristics is essential for understanding the decay of the magnetospheric ring current. However, identifying the precise onset and conclusion of a storm’s recovery interval is inherently challenging due to the frequent presence of multiple Dst minima during both the main and recovery phases. This complexity is illustrated in the top panel of Figure~\ref{sm_storms2324}, which presents examples of both single- and multiple-peaked geomagnetic storms. The top-left panel depicts an isolated storm characterized by a single prominent Dst minimum. In contrast, the top-right panel displays a multiple-peaked storm, where both primary and secondary Dst minima are evident. To address these complexities, we employ a robust automated algorithm capable of reliably distinguishing between isolated and multi-peaked storm profiles.

\begin{figure*}
        \centering
        \includegraphics[width=0.45\textwidth, trim={0.3cm 1.7cm 0.3cm 0.2cm},clip]{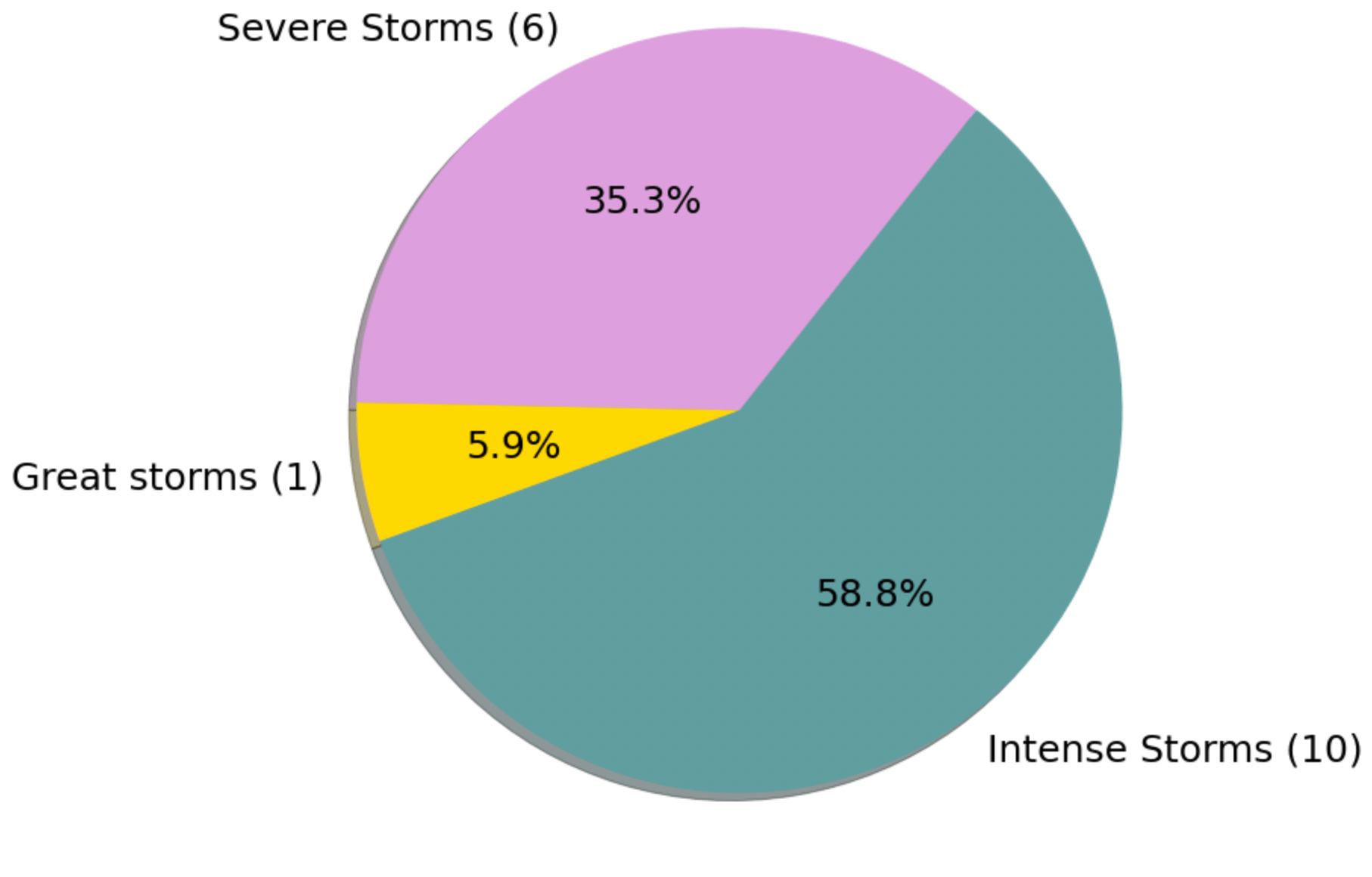}
        \includegraphics[width=0.50\textwidth, trim={0.2cm 0cm 0.3cm 0.2cm},clip]{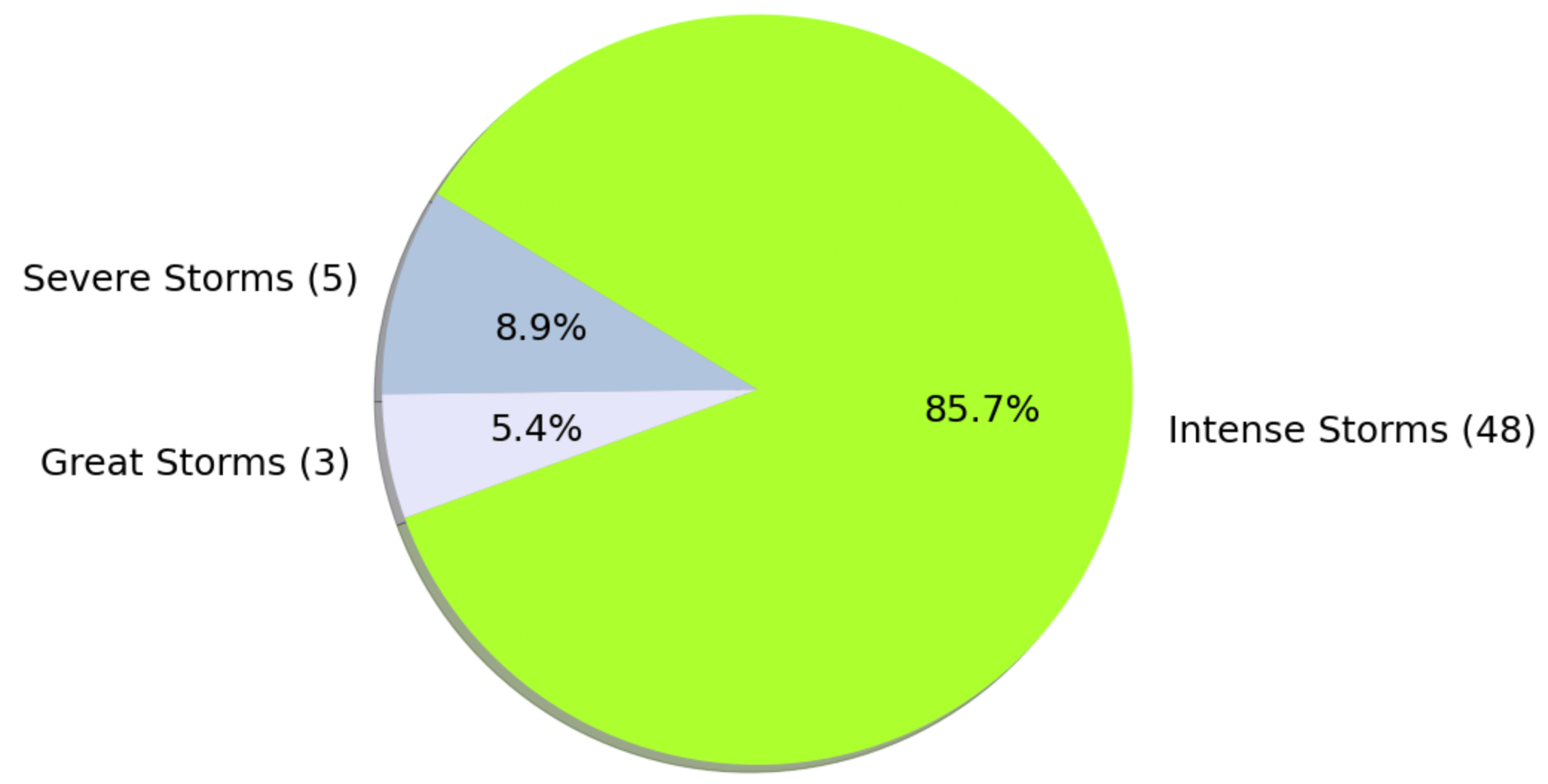}  \\
        \vspace{0.5cm}
        \includegraphics[width=0.49\textwidth]{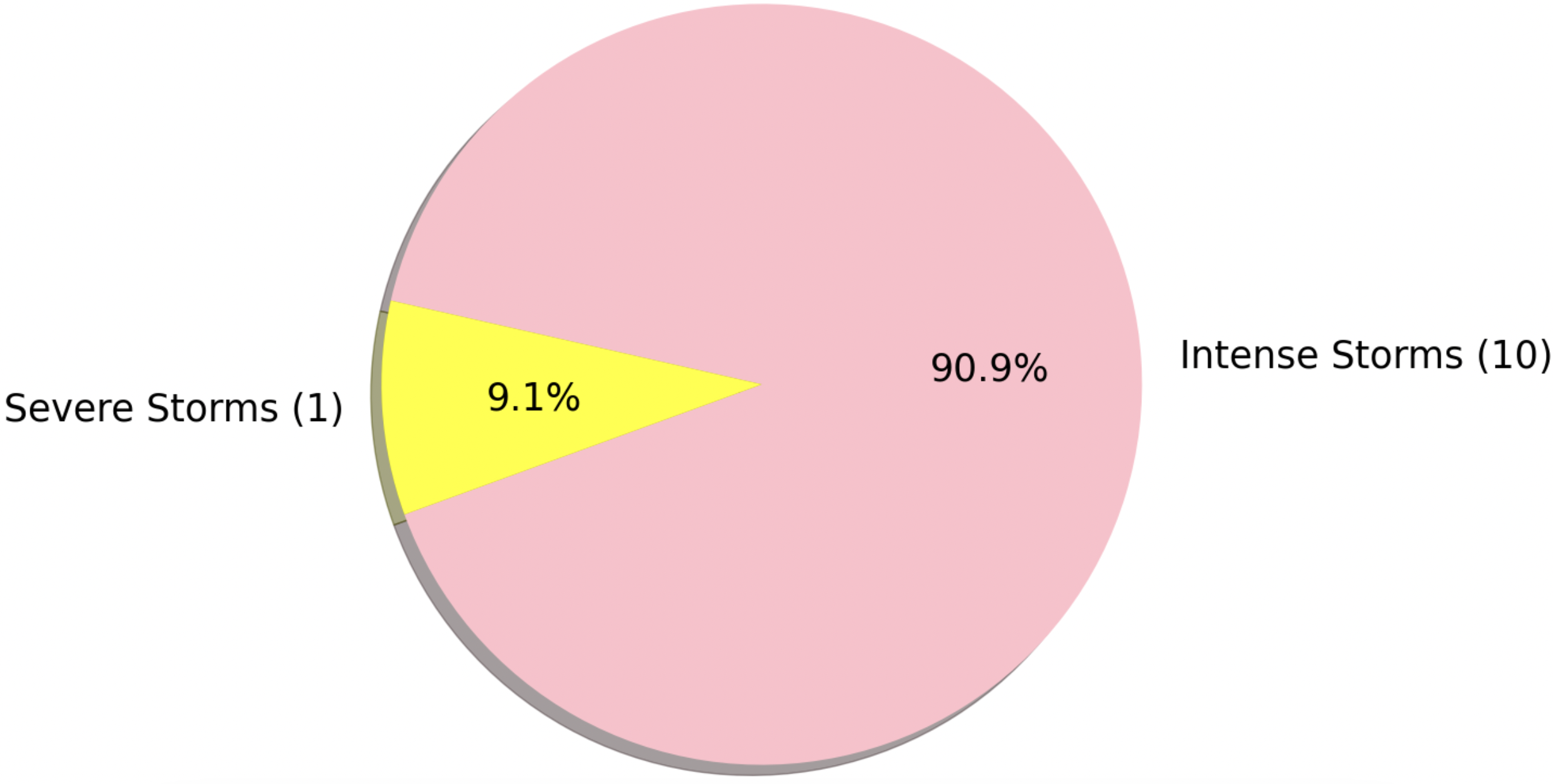}
        \includegraphics[width=0.49\textwidth]{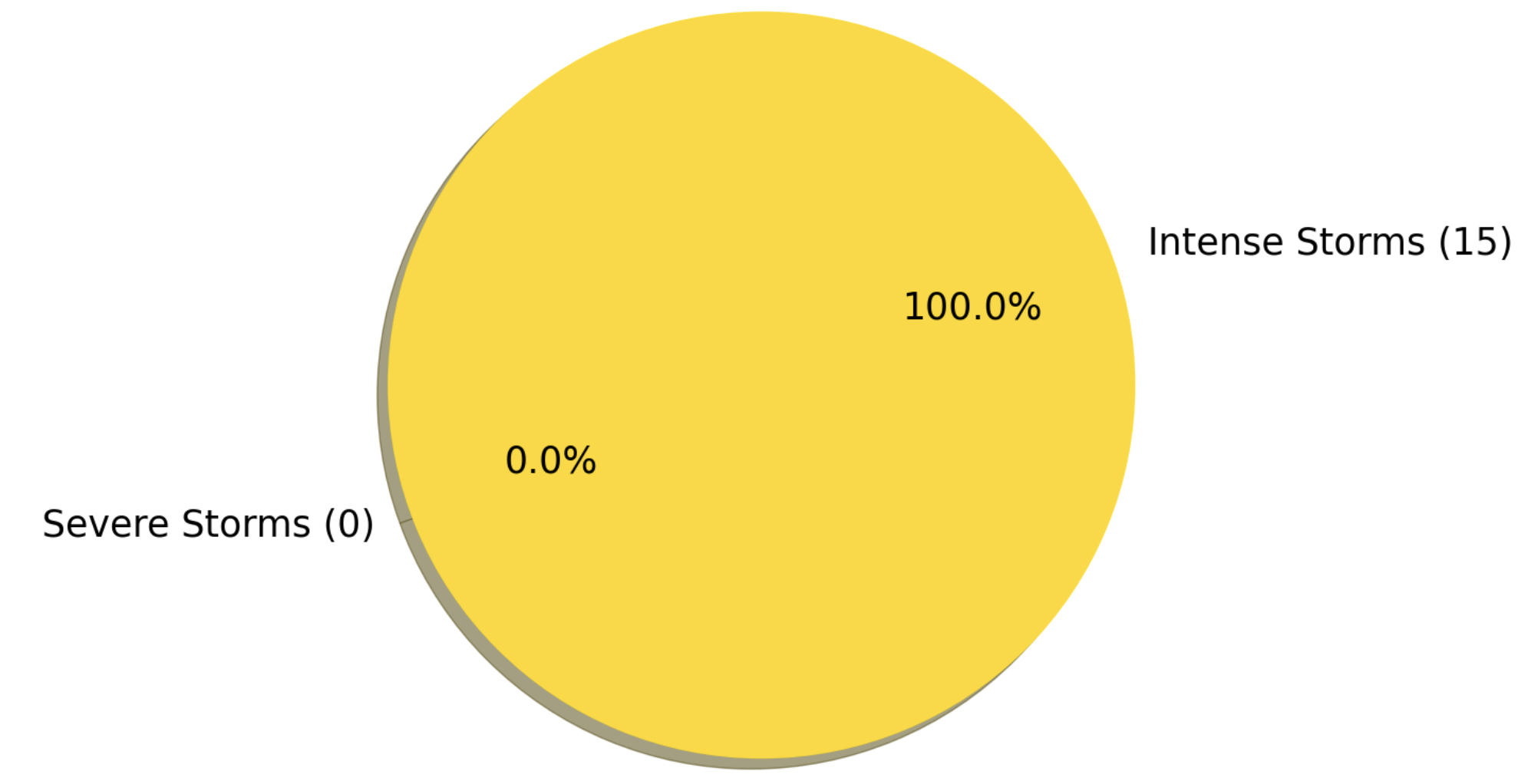}
       \caption[Top panel: Pie charts showing the distribution of]{Top panel: Pie charts showing the distribution of single-peaked and multiple-peaked storms during Solar Cycle 23 (left and right panels, respectively). Bottom panel: Same as the top panel, but for Solar Cycle 24.}
        \label{sm_storms2324_2}
   
\end{figure*}

The classification of geomagnetic storms into single-peaked and multiple-peaked categories relies on the threshold used to detect secondary peaks during the main or recovery phase. \citet{Kamide1998} proposed that, for identifying multiple peaks, the intervening recovery should remain below 90\% of the initial peak and the peaks must be separated by at least three hours. In our study, we first determine the Dst minimum associated with each storm and segment the storm interval into its main and recovery phases. Using an automated algorithm, we identify the minimum Dst values corresponding to intense and stronger-than-intense storms (Dst$_{\text{min}} \leq -100$~nT). For storms exhibiting multiple peaks, we use the most negative Dst value as the primary criterion for intensity classification.

Next, we extract sub-intervals from both the main and recovery phases, focusing on Dst values between 40\% and 95\% of the storm's primary peak magnitude. Within these intervals, we search for additional peaks that show prominent and sustained fluctuations. A peak is considered significant if it involves a reversal in the Dst trend of $\leq -5$~nT within one hour, followed by a consistent increase or decrease lasting at least five hours. Although our method may miss some closely spaced peaks, we believe this limitation is less critical for the intense storms that are the primary focus of this study. The underlying rationale for identifying multiple-peaked storms is based on the assumption that such trend reversals in the Dst profile are unlikely to occur without substantial changes in geoeffective solar wind parameters, often associated with different sub-structures of an ICME/SIR or multiple interacting drivers \citep{Farrugia2006}. Applying these criteria, we successfully classified the storms in Solar Cycles 23 and 24 as either single- or multiple-peaked.

\subsection{Distribution of Single-peaked and Multiple-peaked Storms}

We further analyze the distribution of intense and stronger-than-intense geomagnetic storms by categorizing them based on the intensity corresponding to their primary Dst peak values. For Solar Cycle 23, the distribution of single- and multiple-peaked storms is shown in the top panel of Figure~\ref{sm_storms2324_2}, with the left and right panels representing single-peaked and multiple-peaked storms, respectively. The corresponding distribution for Solar Cycle 24 is presented in the bottom panel of the same figure. In Solar Cycle 23, the majority of single-peaked storms (10 events) fall under the intense category, while a smaller number (6 events) are classified as severe, and only one storm qualifies as a great storm. In contrast, Solar Cycle 24 exhibits a total of 11 intense single-peaked storms, only one severe storm, and no great storms, reflecting the comparatively weaker solar activity during this cycle. For multiple-peaked storms, Solar Cycle 23 shows a substantially higher number, with 48 intense storms, 5 severe storms, and 3 great storms. Meanwhile, Solar Cycle 24 registers only 12 intense multiple-peaked storms, with no occurrences in the severe or great storm categories. This trend reinforces the connection between stronger solar cycles and the occurrence of more powerful geomagnetic storms, particularly those exhibiting multiple peaks, likely due to successive or interacting large-scale solar wind structures depositing substantial energy into the Earth's magnetosphere \citep{Liu2014}.

During Solar Cycle~23, a total of 73 storms with $Dst \leq -100\;\text{nT}$ were recorded, compared to only 26 such storms in Solar Cycle~24, highlighting the markedly reduced geomagnetic activity in the latter. Among all intense and stronger-than-intense storms, only about 23\% were single-peaked in cycle~23, whereas the proportion increased to 42\% in cycle~24. For reference, \citet{Richardson2008a} estimated that approximately 59\% of storms could be classified as single-step events, while \citet{Kamide1998} reported a lower estimate of 29\%. Our results indicate that the stronger Solar Cycle~23 exhibited a substantially higher fraction of multiple-peaked storms compared to the weaker cycle~24. It is important to note that the classification of storms by step count depends on the adopted criteria for identifying multiple peaks, including the magnitude and duration of Dst fluctuations. Future investigations combining detailed Dst profiles with geoeffective solar wind parameters could provide deeper insights into the origins of such fluctuations or dips.

Each storm, whether classified as single-peaked or multiple-peaked based on Dst profiles, is further categorized into four distinct groups according to its identified interplanetary drivers: (i) single ICME (denoted as I-ICME), (ii) interacting ICMEs (denoted as M-ICME), (iii) stream interaction regions (SIR), and (iv) interacting structures involving both ICME and SIR (denoted as ICME-SIR). Driver identification for each storm is based on three authoritative catalogs: the Richardson and Cane ICME catalog \citep{Cane2003, Richardson2010}, the ICME catalog by \citet{Shen2017}, and the SIR catalog by \citet{Chi2018}. It is worth noting that certain storms identified as single-peaked may still be associated with multiple interplanetary structures and vice versa, which introduces a degree of uncertainty in the classification. In Solar Cycle 23, we found that out of 73 storms with a $Dst\leq-100$ nT,  37 storms were produced by I-ICMEs, 23 by M-ICMEs, 9 by SIR, and 4 by the interacting ICME-SIR drivers. In Solar Cycle 24, out of 26 storms with a $Dst\leq-100$ nT, there are 15 storms due to I-ICME, 4 storms by M-ICME, 2 by SIR, and 5 storms resulting from the interaction between ICME and SIR.

During Solar Cycle 23, we identify 17 single-peaked geomagnetic storms with Dst peak values $\leq -100$~nT. Among these, 12 were driven by I-ICMEs, 3 by M-ICMEs, 1 by a SIR, and 1 by an ICME-SIR interaction. In contrast, there were 56 multiple-peaked storms during the same cycle, with 25 linked to I-ICMEs, 20 to M-ICMEs, 8 to SIRs, and 3 to ICME-SIR interactions. In Solar Cycle 24, we find 11 single-peaked intense storms. Of these, 6 were driven by I-ICMEs, 1 by M-ICME, 1 by SIR, and 3 by ICME-SIR interactions. The same cycle also witnessed 15 multiple-peaked storms, of which 9 were associated with I-ICMEs, 3 with M-ICMEs, 1 with a SIR, and 2 with ICME-SIR interactions. Interestingly, while only about 23\% of the storms in Cycle 23 were single-peaked, this fraction increases to roughly 42\% in Cycle 24.

Notably, approximately 23\% and 36\% of the single-peaked storms in Cycles 23 and 24, respectively, were driven by multiple interacting interplanetary structures. On the other hand, 59\% and 67\% of the multiple-peaked storms in Cycles 23 and 24, respectively, were associated with isolated ICMEs or SIRs. This indicates that even in the weaker Cycle 24, a significant portion of the complex, multiple-peaked storms originated from single interplanetary structures. These cases suggest that multiple dips in the Dst profile may result from intermittent intervals of southward magnetic fields within a single ICME or SIR, separated by less geoeffective conditions. For instance, both the sheath and ejecta regions of an ICME can contribute independently to geomagnetic disturbances \citep{Gonzalez1999, Huttunen2002}. Moreover, we observe that some single-peaked storm, particularly in Cycle 24, can be linked to complex interactions involving multiple ICMEs or ICME-SIR combinations. This underlines that the number of peaks in the Dst profile is not always a direct indicator of the number of solar wind structures involved \citep{Richardson2008a}.

\subsection{Main Phase and Recovery Phase Parameters of Storms}

We investigate both the main (buildup) and recovery phases of geomagnetic storms that occurred during Solar Cycles 23 and 24. For each storm, the end of the recovery phase is defined as the point where the Dst index rises to values greater than or equal to $-10$~nT following the storm's minimum Dst peak. The recovery phase duration is computed as the time difference between the Dst minimum and this recovery endpoint.

To determine the duration of the main phase, we identify two key points using an automated algorithm. The onset of the main phase is defined as the first transition point where the Dst index shifts from positive to negative values prior to reaching the minimum, while the endpoint corresponds to the Dst minimum itself. In many cases, the entire main phase lies within negative Dst values. In such instances, while the algorithm accurately identifies the Dst minimum, we manually review the Dst profiles and examine the associated solar wind drivers to more reliably determine the onset of the main phase. The top panel of Figure~\ref{sm_storms2324} illustrates the main and recovery phase durations for a representative single-peaked storm, highlighted in cyan and yellow, respectively. The bottom panel presents a similar depiction for a typical multiple-peaked storm.

To further analyze the recovery phase, we divide it into two segments: a fast decay phase and a slow decay phase \citep{Aguado2010, Cid2013, Yermolaev2016}. The fast decay phase is defined as the time taken for the Dst index to recover by 75\% from its minimum value. The remaining portion of the recovery interval is considered the slow decay phase. The duration of the slow decay phase is obtained by subtracting the fast decay duration from the total recovery time.

To understand the relationship between various storm characteristics, we perform correlation analyses involving parameters such as the main phase duration, main phase buildup rate, total recovery time, recovery rate, and durations of the fast and slow decay phases. For instance, the main phase buildup rate (in nT/hr) is calculated by dividing the minimum Dst value by the main phase duration. We utilize both Pearson and Spearman correlation coefficients to quantify these relationships. Pearson’s $r$ measures linear correlation between two variables, while Spearman’s $r$ assesses monotonic relationships based on the rank order of the data, making it suitable for capturing non-linear dependencies as well.

\subsection{Recovery Phase of Single-peaked and Multiple-peaked Storms of Solar Cycle 23}

\begin{figure*}

        \centering
        \includegraphics[width=0.47\textwidth]{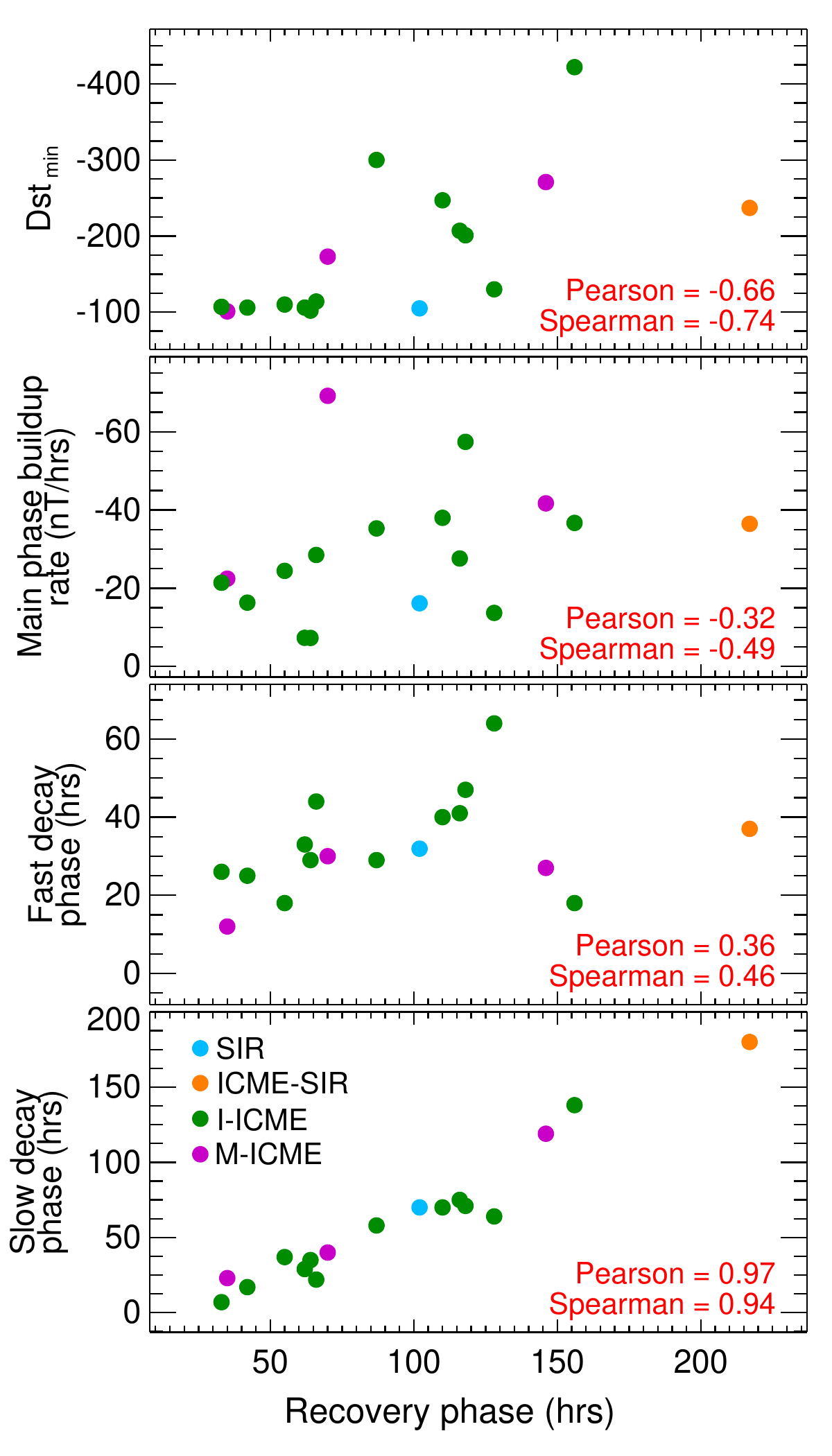}
        \includegraphics[width=0.505\textwidth]{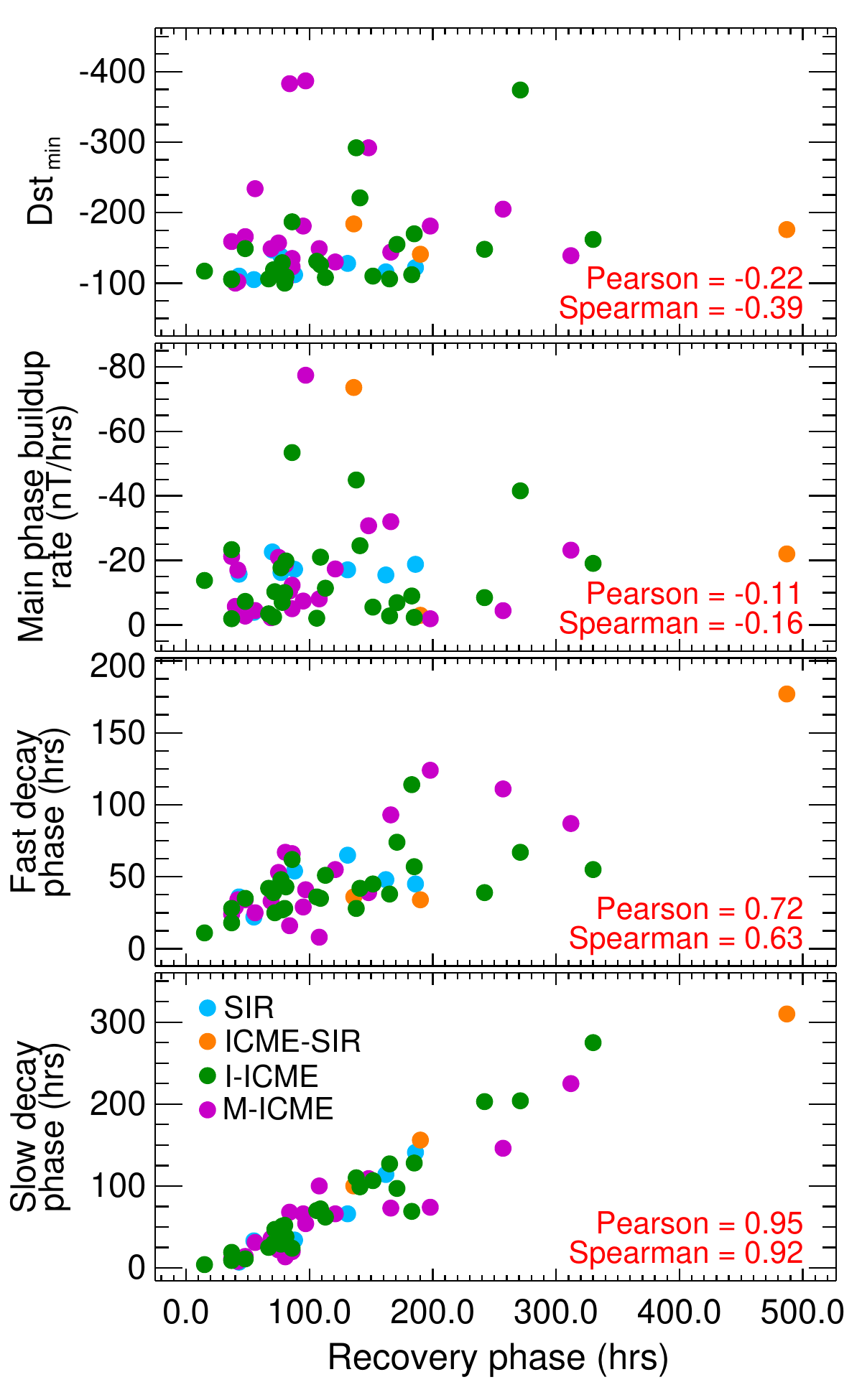}
        
        \caption[Left panel: Correlation plots illustrating the relationship]{Left panel: Correlation plots illustrating the relationship between the recovery phase duration of single-peaked storms during Solar Cycle 23 and various storm characteristics. From top to bottom, the panels show correlations with (i) peak Dst index, (ii) main phase buildup rate, (iii) duration of the fast decay phase, and (iv) duration of the slow decay phase. Right panel: Same as the left panel, but for multiple-peaked storms.}
        \label{sc23_recov}
  
\end{figure*}

From our analysis of the single-peaked storms during Solar Cycle~23, we obtained the following Pearson’s (Spearman’s) correlation coefficients between the duration of the recovery phase and other storm characteristics: $-0.66$ ($-0.74$) with the peak Dst index, $0.11$ ($0.23$) with the main phase duration, $-0.32$ ($-0.49$) with the main phase buildup rate, $0.47$ ($0.43$) with the recovery rate, $0.36$ ($0.46$) with the fast decay phase, and $0.97$ ($0.94$) with the slow decay phase. We consider a correlation coefficient between $0.4$–$0.6$ as indicative of moderate dependency, and between $0.6$–$0.8$ as indicative of strong dependency. These results show that the duration of the recovery phase has a strong dependency on the Dst peak and the slow decay phase, and a moderate dependency on the main phase buildup rate and fast decay phase. In essence, stronger storms with prolonged slow decay phases tend to result in extended recovery durations.

The end of the main phase is associated with reversing the polarity of IMF Bz from the southward to northward condition. This signals the onset of the recovery phase when the ring current ceases to receive the supply of energetic ions through the nightside earthward convection. On many occasions, when the IMF Bz suddenly turns northward after being stable and southward for some time, it generates an over-shielding electric field in the entire inner magnetosphere \citep{Chakrabarty2006}. In addition, substorms can also generate an overshielding-like scenario in the inner magnetosphere \citep{Hashimoto2017}. Overshielding events, in general, mark the beginning of the recovery phase of the storm. During substorms, the tail current collapses \citep{Mcpherron1973,Kepko2015}, leading to the reduction in the negative horizontal component measured at the ground. \citet{Huang2004} showed that the Dst/Sym-H index shows similar variations after each substorm onset and proposed that dipolarizations of the nightside magnetic field related to substorm onsets cause these changes. Given the above scenario, it is possible that the fast decay of the recovery phase is associated with the onset of substorms, and hence, relatively poor correlations are found compared to the slow decay (longer duration), particularly in Cycle 23.  The roles of CME and CIR-triggered substorms in the early recovery phase of geomagnetic storms need to be investigated in greater detail to understand these aspects.

For multiple-peaked storms of Solar Cycle~23, the corresponding Pearson’s (Spearman’s) correlations are: $-0.22$ ($-0.39$) with the Dst peak, $-0.01$ ($-0.01$) with the main phase duration, $-0.11$ ($-0.16$) with the main phase buildup rate, $0.58$ ($0.84$) with the recovery rate, $0.72$ ($0.63$) with the fast decay phase, and $0.95$ ($0.92$) with the slow decay phase. These results indicate that the recovery phase duration for multiple-peaked storms is strongly governed by the slow and fast decay phases, whereas the peak Dst and main phase buildup rate do not significantly influence the overall recovery duration. We note that for all cases where the correlation coefficient exceeds $0.5$, the corresponding $p$-values are much less than $0.05$, indicating the statistical significance of those relationships. Conversely, correlations with weaker coefficients do not pass the $p < 0.05$ threshold, further supporting the reliability of our findings.

Figure~\ref{sc23_recov} illustrates the correlation between recovery phase duration and various storm characteristics for single-peaked (left) and multiple-peaked (right) storms during Solar Cycle~23. The associated interplanetary drivers for each storm are also indicated in the figure. For both storm categories, recovery duration strongly depends on the slow decay phase and, to a moderate extent, on the fast decay phase. Notably, for multiple-peaked storms, both the fast and slow decay phases significantly determine the total recovery duration. In the case of single-peaked storms, the recovery phase also shows a moderate correlation with the main phase buildup rate and a strong correlation with the Dst peak, trends that are absent for multiple-peaked storms. This distinction may stem from the higher geoeffectiveness of many multiple-peaked storms, often associated with multiple ICMEs, suggesting that these storms experience an early rapid recovery followed by a prolonged slow recovery. 

Previous studies have shown that the fast decay phase may be driven by contributions from tail current and O$^{+}$ charge-exchange losses, whereas the slow decay phase is primarily governed by ring current and H$^{+}$ charge-exchange losses \citep{Feldstein2000, Choraghe2021}. The study of \citet{Choraghe2021} suggests that the magnitude of the recovery rate during the slow decay phase is proportional to the Dst peak; however, in our study, it is only true for single-peak storms of cycle 23. Our study suggests that the loss mechanism for magnetospheric ring currents responsible for multiple-peak storms could be different than for single-peak storms, and therefore, the origin of multiple peaks needs to be investigated further. We also note that ICMEs, especially interacting ICMEs, are more geoeffective than SIRs, but the SIRs-led storms have a larger recovery time \citep{Chi2018}. It is evident that a stronger geomagnetic storm does not necessarily have a longer main or recovery phase duration, especially for multiple-peak storms, as different current systems are involved at different steps, and their decay rate can vary from storm to storm.

\subsection{Recovery Phase of Single-peaked and Multiple-peaked Storms of Solar Cycle 24}

In our analysis of single-peaked geomagnetic storms during Solar Cycle~24, we computed the Pearson’s (Spearman’s) correlation coefficients between the duration of the recovery phase and various storm parameters. The results are as follows: $-0.49$ ($-0.32$) with the Dst peak, $0.15$ ($0.07$) with the main phase duration, $-0.47$ ($-0.52$) with the main phase buildup rate, $0.70$ ($0.85$) with the recovery rate, $0.78$ ($0.78$) with the fast decay phase, and $0.88$ ($0.77$) with the slow decay phase. These correlations indicate that the recovery phase duration for single-peaked storms during Cycle~24 is strongly governed by both the fast and slow decay phases, while exhibiting moderate dependencies on the main phase buildup rate and the Dst peak.

\begin{figure*}
  
        \centering
        \includegraphics[width=0.49\textwidth]{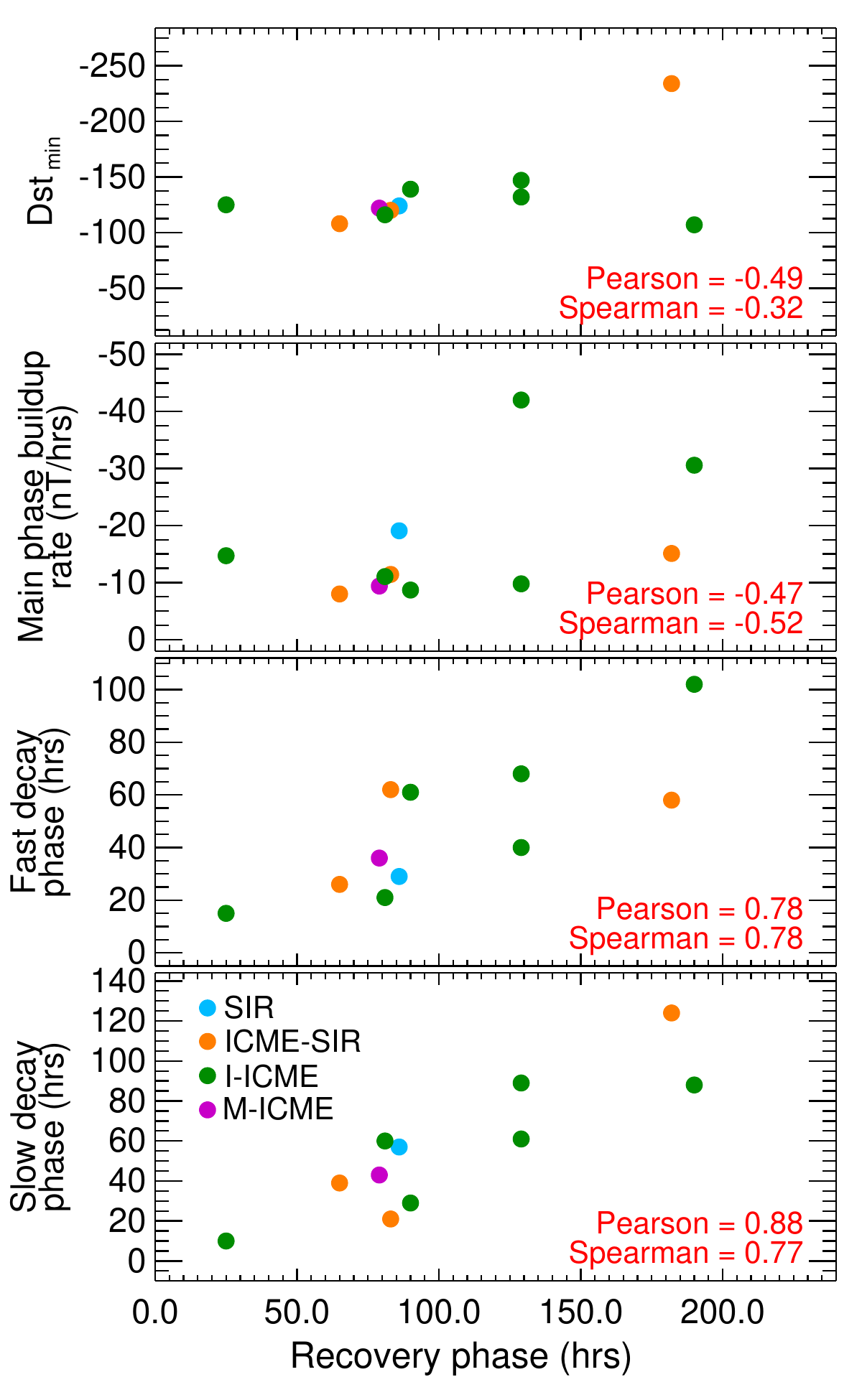}
        \includegraphics[width=0.49\textwidth]{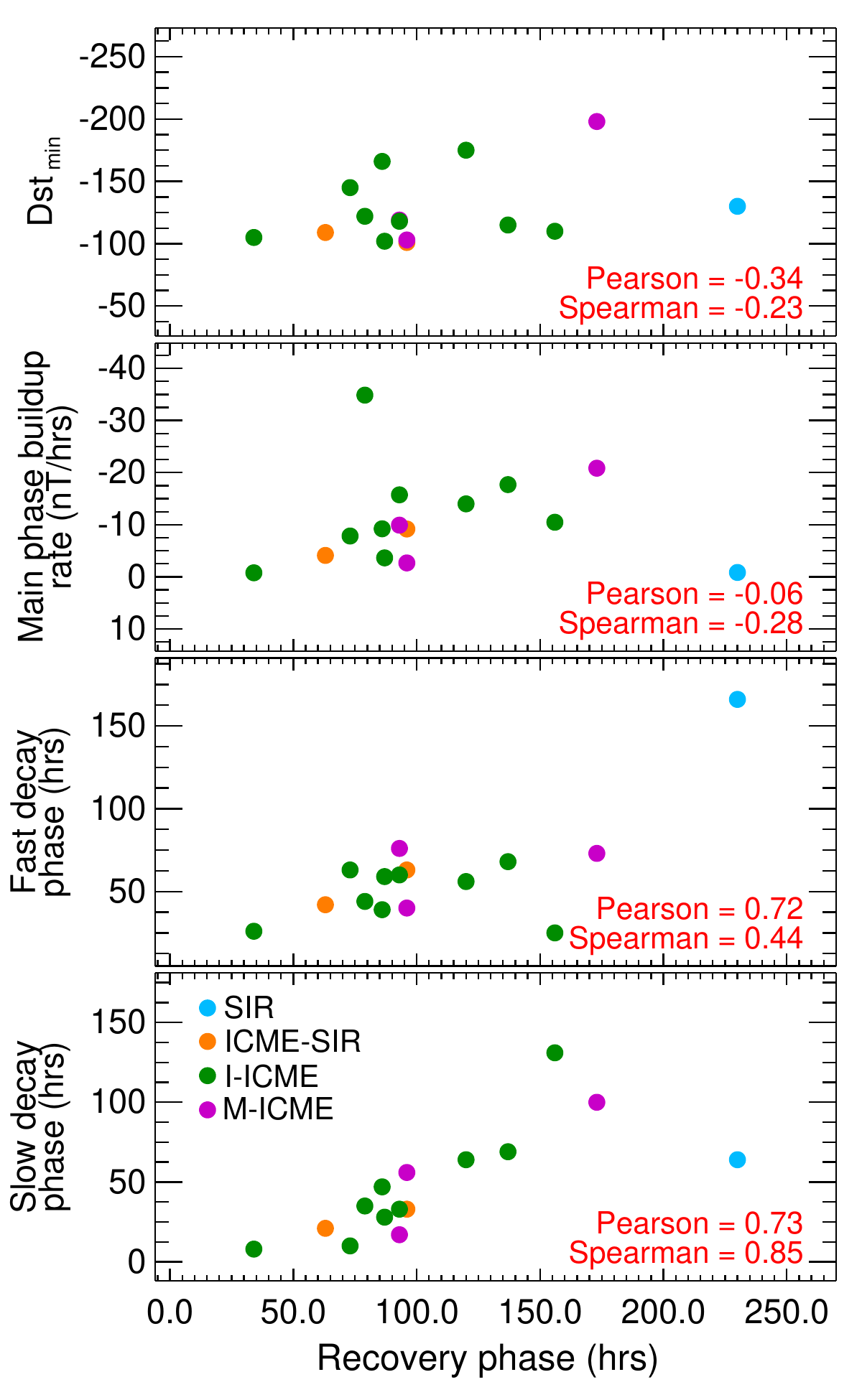}
        
        \caption[Left panel: Correlation between the recovery phase]{Left panel: Correlation between the recovery phase duration of single-peaked storms during Solar Cycle~24 and various storm parameters. From top to bottom, the plots show the relationship with the peak Dst index, main phase buildup rate, fast decay phase duration, and slow decay phase duration. Right panel: Same as the left, but for multiple-peaked storms.}

        \label{sc24_recov}
  
\end{figure*}

For multiple-peaked storms in Cycle~24, the correlations with recovery phase duration are: $-0.34$ ($-0.23$) with the Dst peak, $0.21$ ($-0.21$) with the main phase duration, $-0.05$ ($-0.28$) with the main phase buildup rate, $0.76$ ($0.87$) with the recovery rate, $0.72$ ($0.44$) with the fast decay phase, and $0.73$ ($0.85$) with the slow decay phase. These findings underscore that the recovery phase duration in multiple-peaked storms strongly correlates with both the fast and slow decay phases, while showing negligible or no meaningful dependency on the Dst peak and main phase buildup rate. Across both storm categories in Cycle~24, the characteristics of the fast and slow decay phases emerge as the dominant factors influencing the duration of the recovery phase. Importantly, all correlation coefficients exceeding 0.5 have $p$-values significantly lower than $0.05$, affirming the statistical robustness of these correlations. In contrast, weaker correlations do not meet this threshold, further validating the credibility of our results.

Figure~\ref{sc24_recov} illustrates the relationships between the recovery phase duration and various storm attributes for single-peaked (left) and multiple-peaked (right) storms during Solar Cycle~24. The corresponding interplanetary transients responsible for each storm are also annotated. Notably, the strong correlation of recovery phase duration with both decay phases (fast and slow) is consistent across storm categories. In the case of multiple-peaked storms, the association between recovery phase and fast decay duration appears to be non-parametric, as indicated by a higher Spearman correlation compared to the Pearson correlation.

It is worth emphasizing that while the fast decay phase showed only a moderate correlation with the recovery duration in single-peaked storms of Cycle~23, it exhibits a strong dependency in all other scenarios. Similarly, the recovery phase duration in Cycle~24 shows a moderate dependency on the Dst peak for single-peaked storms, whereas Cycle~23 single-peaked storms demonstrated a stronger correlation. Although this difference may appear minor, it underscores the impact of multiple peaks on the overall storm recovery process. Finally, we acknowledge the inherent limitations in comparing both solar cycles, as the smaller sample size in Cycle~24 may affect the strength and reliability of correlation coefficients relative to the more statistically robust dataset of Cycle~23.

Further, we adopt a binary classification by combining the interplanetary drivers of each storm: (i) storms driven by ICMEs (including both isolated and interacting ICMEs) and (ii) storms driven by SIRs (either standalone or in interaction with ICMEs). We examined their characteristic features across solar cycles 23 and 24. During Solar Cycle~23, SIR-driven storms exhibited an average main phase duration of approximately 12 hours, whereas ICME-driven storms showed a longer average main phase duration of about 19 hours. The recovery phase also differed between the two drivers: SIR-driven storms had a mean recovery duration of roughly 145 hours, compared to 135 hours for ICME-driven storms. The average minimum Dst index ($\text{Dst}_{\text{min}}$) for ICME-driven events reached -170~nT, while for SIR-driven storms it was approximately -140~nT.

In Solar Cycle~24, the trend persists, with SIR-driven storms generally exhibiting longer main phase durations than ICME-driven ones. The recovery phase duration for SIR-driven storms averaged around 120 hours, slightly longer than the ~100 hours observed for ICME-driven storms. The geoeffectiveness also appeared reduced, with $\text{Dst}_{\text{min}}$ values averaging -130~nT for ICME-driven storms and -115~nT for those driven by SIRs. The extended recovery times associated with SIR-driven storms in both cycles can be attributed to their association with large-amplitude Alfvén waves and fast solar wind streams, which introduce fluctuating magnetic fields \citep{Echer2013}. Our findings are consistent with earlier results that emphasize the higher geoeffectiveness of ICMEs compared to SIRs \citep{Mouikis2019}. Notably, the majority of intense storms ($\text{Dst}_{\text{min}} \leq -100$~nT) in both cycles were triggered by ICMEs. However, we also observed a few purely SIR-driven storms that reached $\text{Dst}_{\text{min}}$ values approaching -150~nT, indicating that SIRs can occasionally drive extreme space weather events. Overall, the comparative analysis reinforces the conclusion that Solar Cycle~24 was significantly weaker than Cycle~23 in terms of both storm intensity and geoeffectiveness.

Earlier studies have focused on modeling the complete recovery phase of storms by a mathematical function, and those studies are limited to only a few great or severe storms \citep{Cid2013,Choraghe2021}. In our study, we estimate the main phase and the recovery phase duration of all the storms with $\text{Dst}_{\text{min}}$ $\leq-100\;\text{nT}$ for both solar cycles 23 and 24. Our study also divides the storm's recovery phase into two intervals (early and late recovery) and examines its dependence on its main phase characteristics. This approach is also taken in \cite{Yermolaev2016}, but our finding differs from that reported. \cite{Yermolaev2016} indicated a correlation between storm recovery time and main phase buildup rate in events induced by CIR and sheath compression regions, but not in those induced ICMEs. Our study has classified the storms as single or multiple-peaked storms to understand the recovery phase dependent on different drivers.

\subsection{Summary}\label{chapt4-part3-sec:summary}

This study investigates the recovery phase of geomagnetic storms during solar cycles 23 and 24, emphasizing the influence of storm drivers (ICMEs, SIRs, and their interactions) and the differences between single-peaked and multiple-peaked storms. Using 1-hour resolution Dst index data from the OMNI database and interplanetary driver catalogs, we classified intense and stronger-than-intense storms ($\text{Dst}_{\text{min}}\leq -100$~nT) and analyzed their main and recovery phase characteristics.

\begin{enumerate}
    \item Our results reveal that Solar Cycle 23 experienced significantly more stronger-than-intense storms (73 events) compared to cycle 24 (26 events), confirming the latter's reduced geomagnetic activity. In cycle 23, multiple-peaked storms dominated (77\%), whereas in cycle 24, the proportion of single-peaked storms was higher (42\%). This difference suggests that stronger solar cycles favor multiple-peak events, possibly due to more frequent complex or interacting interplanetary structures.

    \item The recovery phase duration shows a strong dependence on the slow decay phase and, in most cases, also on the fast decay phase, regardless of cycle or storm type. Single-peaked storms in cycle 23 exhibit strong correlations between recovery duration and $\text{Dst}_{\text{min}}$, whereas this relation weakens in cycle 24. For multiple-peaked storms, $\text{Dst}_{\text{min}}$ and main phase buildup rate have little influence on recovery time, suggesting that internal storm structure and decay processes are more important. Recovery is generally longer for SIR-driven storms than ICME-driven storms in both cycles, likely due to the presence of large-amplitude Alfvén waves and fluctuating magnetic fields within SIRs. Nevertheless, ICMEs remain more geoeffective overall, producing stronger storms on average.
   
    \item The analysis also shows that multiple-peak storms can be driven by isolated structures (ICMEs or SIRs), indicating that the number of Dst peaks does not necessarily reflect the number of interplanetary drivers. This behavior may result from different substructures within a single driver contributing separate southward IMF intervals.
\end{enumerate}

Our investigation into the recovery phase of storms induced by various drivers is confined to intense and stronger-than-intense storms ($\text{Dst}_{\text{min}}\leq-100\;\text{nT}$). Consequently, the findings we have drawn may not be directly applicable to moderate storms. It would be interesting to explore the recovery phase characteristics of moderate storms in future research to determine if they exhibit similar patterns. We acknowledge that categorizing storms into single and multiple-steps, using a specific criterion, oversimplifies their complexity. In our analysis, we have identified the driving factors for each storm from distinct catalogs \citep{Richardson2010, Shen2017, Chi2018}, without distinguishing between the sheath, leading, or trailing portions of the ejecta as potential drivers for the storms. Our study does not consider whether interacting structures arrive at Earth as individual entities or complex ejecta \citep{Burlaga2002, farrugia2004, Lugaz2014, Mishra2014}. Such differentiation may offer insights into the multiple peaks observed in storm profiles and their subsequent recovery phases. Our future research plans include delving deeper into this direction and conducting a detailed analysis of in situ solar wind observations near Earth. Furthermore, a comprehensive and in-depth investigation into modeling the development and decay of the Dst index for storms driven by diverse factors is warranted.

\section{Chapter Conclusion}\label{chap4-sec:conclusion}

This chapter presented an integrated investigation of CME thermodynamics and their geospace consequences, spanning from the low corona to 1~au and through the recovery phase of geomagnetic storms. The three studies, near-Sun thermodynamic modeling, statistical in situ characterization, and geomagnetic storm recovery analysis, together provide new insights into the CME evolution and their role in space weather.

From the near-Sun analysis using the FRIS model, we find that fast CMEs typically undergo an initial heat-release phase, pass through an adiabatic regime at low coronal heights, and subsequently experience heat absorption during interplanetary expansion. The transition to a near-isothermal state at larger heliocentric distances highlights the interplay between expansion-driven cooling and ongoing heating. The force balance analysis reveals that thermal and centrifugal forces dominate CME expansion close to the Sun, while the Lorentz force persistently acts inward, resisting expansion. These results emphasize the need to consider evolving force contributions in CME propagation models.

The statistical characterization of ICMEs across three solar cycles demonstrates that the thermal state, quantified by $\Gamma_p$, is a robust discriminator of CME evolution and geoeffectiveness. Heating MEs, prevalent during solar maxima and descending phases, exhibit strong non-adiabatic behavior, higher expansion speeds, enhanced magnetic and dynamic pressures, and are more frequently associated with high-impact geomagnetic storms. Cooling MEs, dominant in rising phases, evolve in a more adiabatic or over-adiabatic manner, maintain magnetic coherence, and generally produce milder geomagnetic responses. The $\Gamma_p$ distribution shows clear solar-cycle modulation, suggesting that large-scale solar magnetic field configurations and coronal conditions influence CME thermal evolution. Importantly, high-impact ICMEs are predominantly Heating MEs with low $\Gamma_p$, stronger magnetic fields, and faster speeds, indicating that the thermal state is a valuable predictor of storm potential.

The geomagnetic storm recovery analysis reveals that recovery durations are influenced by both storm morphology and driver type. Cycle 23 produced more multiple-peaked storms, reflecting a higher occurrence of complex drivers and interacting structures, whereas Cycle 24 saw a greater proportion of single-peaked events. SIR-driven storms exhibit longer recoveries than ICME-driven storms, likely due to sustained Alfvénic fluctuations and intermittent energy input. Multiple-peaked storms can arise even from single isolated structures, indicating that internal driver substructure, not just the number of distinct drivers, plays a critical role in shaping geomagnetic activity.

Future work should pursue a more seamless connection between near-Sun CME diagnostics and in situ thermal classifications, ideally using coordinated multi-point observations from missions such as \textit{Parker Solar Probe}, \textit{Solar Orbiter}, and L1 monitors. Extending the polytropic index framework to include electron and heavy-ion populations may further refine our understanding of CME energetics. Additionally, modeling efforts that couple CME internal thermodynamics with magnetospheric response models will be essential for improving forecasts of storm intensity and recovery dynamics.

\newpage

\lhead{\emph{Chapter 5}}  

\quotechapter{5}{`` Complexity is the result of simple interactions occurring over time.''\\ \smallskip —  Mitchell Waldrop}{Thermo-Kinematic Evolution of Interacting CMEs}

\lhead{\emph{Chapter 5: Thermo-Kinematic Evolution of Interacting CMEs}}

\section{Introduction} \label{Ch5-sec:intro}

Over the years, forecasting the arrival time and geoeffectiveness of CMEs has become one of the central goals in heliophysics and space weather prediction. The dynamic interaction between a CME's magnetic field and the geomagnetic field can lead to significant disturbances in Earth's space environment, resulting in geomagnetic storms. These storms can disrupt satellite operations, communication systems, GPS signals, and power grids, posing major risks to modern technological infrastructure \citep{Gonzalez1994, Pulkkinen2007}. However, the properties of CMEs are not static. As they propagate through the heliosphere, they undergo complex evolution in their kinematic, thermodynamic, and magnetic characteristics due to interactions with the ambient solar wind or with other CMEs \citep{Liu2014, Lugaz2017, Desai2020, Mishra2021, Koehn2022, Temmer2023}. This evolution influences not only their arrival time at Earth but also their potential to cause geomagnetic storms. CME-CME interactions, in particular, have garnered significant attention in recent years. When multiple CMEs erupt in quick succession and interact in interplanetary space, they can merge to form complex ejecta or amplify the internal magnetic field, potentially enhancing their geoeffectiveness \citep{Liu2014, Lugaz2017, Desai2020, Mishra2021, Koehn2022, Temmer2023}. Despite several well-documented case studies, a comprehensive understanding of how such interactions influence the thermodynamic state and structural evolution of CMEs near Earth remains elusive.

The geoeffectiveness of a CME is often linked to the orientation and strength of its magnetic field. A strong, sustained southward component of the magnetic field in the CME can reconnect efficiently with Earth's northward magnetic field, facilitating the transfer of energy into the magnetosphere and triggering geomagnetic storms \citep{Dungey1961, Tsurutani1988}. The severity of these storms is commonly quantified using the disturbance storm time (Dst) index, with values below −350 nT classified as “great” storms \citep{Gonzalez2011, Mishra2024}. Since the beginning of the space age, such great storms have been exceptionally rare, with only a handful recorded \citep{Meng2019}. Their infrequency poses a challenge for building robust statistical models of their drivers, making case-by-case analyses vital.

In May 2024, a rare and great geomagnetic storm was recorded, reaching a minimum Dst value of $−406$ nT, the most extreme event of its kind in over two decades. This event triggered a cascade of space weather effects, including enhanced ionospheric disturbances, increased auroral activity at low latitudes, satellite drag, and communication disruptions \citep{Gonzalez-Esparza2024, Hajra2024, Hayakawa2024, Lazzus2024, Spogli2024, Parker2024, Jain2025}. \citet{Liu2024} described the event as a ``perfect storm," emphasizing the differences in magnetic field configurations and geoeffectiveness observed from two vantage points, despite only a mesoscale spatial separation.

This chapter presents a comprehensive analysis of the CMEs responsible for the May 2024 geomagnetic storm, utilizing both remote-sensing and in situ observations. In the first part of the study, we investigate the near-Sun evolution of successive CMEs, deriving their 3D kinematics and examining potential interaction heights. Thermodynamic properties such as polytropic index ($\Gamma$) and temperature ($T$) evolution are also assessed using the FRIS model and remote imaging data. In the second part, we examine in situ measurements near Earth to disentangle the complex structures associated with the storm and evaluate the plasma and thermal state across different regions of the ICME. Our goal is to understand how CME-CME interactions can modify the internal properties of the ejecta. This event serves as a valuable case study to explore the role of interplanetary evolution in governing the thermal state and geoeffectiveness of complex solar eruptions.

\section{Near-Sun Remote Measurements}{\label{chap5-sec:nearsun}}

In May 2024, the Sun exhibited intense activity driven by the emergence of a complex active region with a complex $\beta\gamma\delta$ magnetic configuration. This region was formed through the merging of NOAA AR 13664 and 13668 and remained flare-productive even after completing a full solar rotation. Around the onset of the geomagnetic storm on 10 May 2024, in situ measurements at 1~au recorded high-speed solar wind flows in the range of $\approx$700--1000~km~s$^{-1}$. Motivated by this, we investigated halo and partial-halo CME events launched from the Sun during 8--9 May 2024 as potential drivers of the storm.

We utilized coronagraph observations from STEREO‐A/COR2 and SOHO/LASCO to identify Earth-directed CMEs that might have erupted 2--3 days prior to the geomagnetic storm. During this period, STEREO‐A was positioned approximately $12^{\circ}$ west of the Sun-Earth line, providing a viewing geometry comparable to that of SOHO for tracking these CMEs. To determine their solar sources, we also examined EUV images from SDO/AIA, enabling us to correlate each CME with its corresponding on-disc activity.

Through this analysis, we identified six candidate CMEs, summarized in Table~\ref{chap5-tab:GCS}, with first appearances in LASCO/C2 as noted in the second column. The source regions and associated flare or filament activities are listed in the third and fourth columns. Of these, CME3 was linked to a filament eruption from AR 13667, located at N25E14, while the remaining events were associated with flare activity originating from AR 13664.

\begin{figure*}[[tbp]
   \centering
   \includegraphics[scale=0.20]{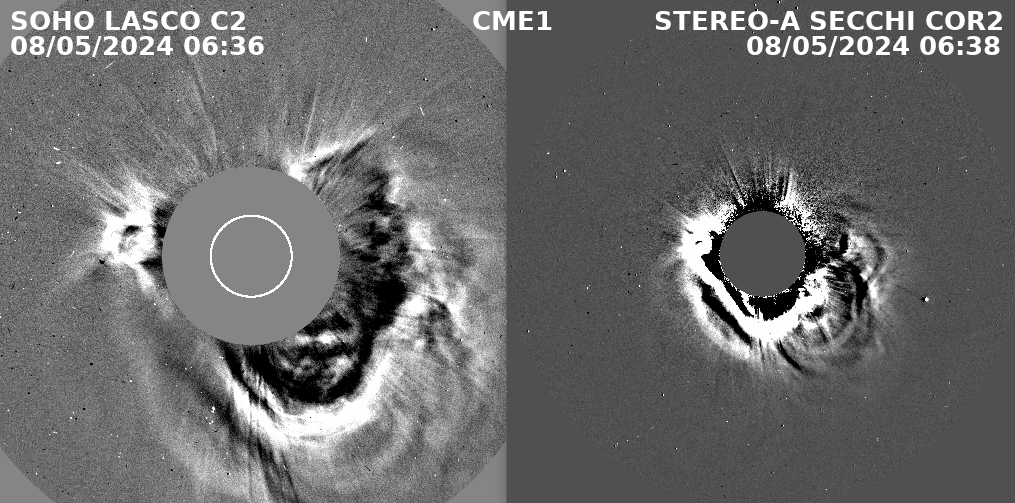}
   \includegraphics[scale=0.20]{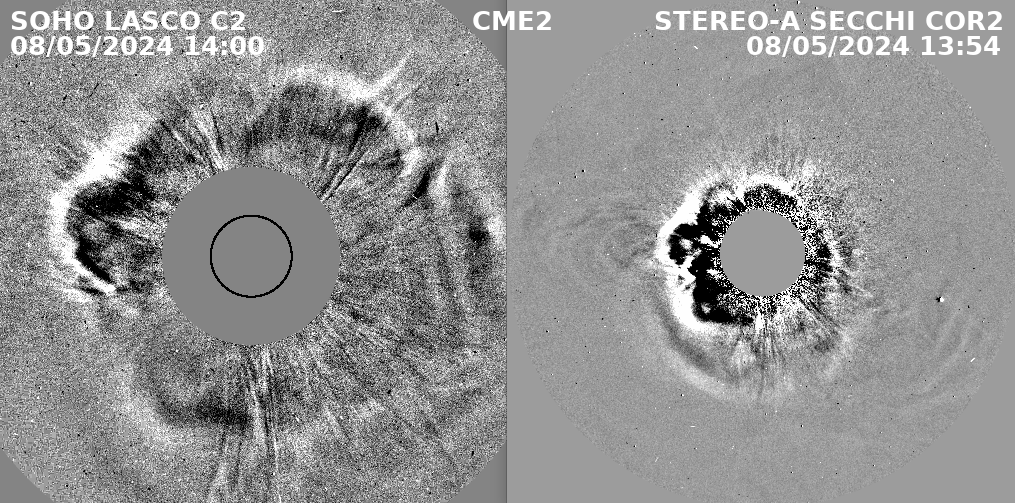}\\ \vspace{-1mm}
   \includegraphics[scale=0.20,trim={0cm 0cm 0cm 2cm}, clip]{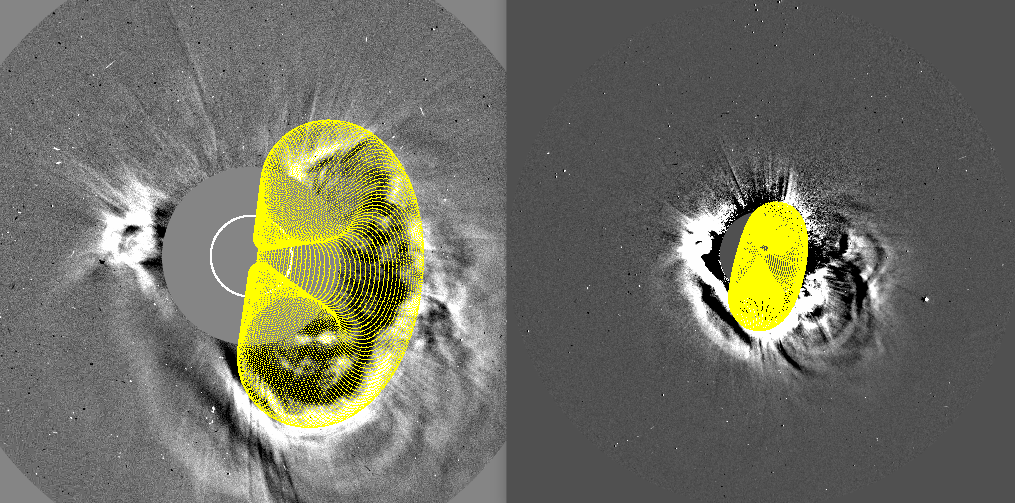}
   \includegraphics[scale=0.20,trim={0cm 0cm 0cm 2cm}, clip]{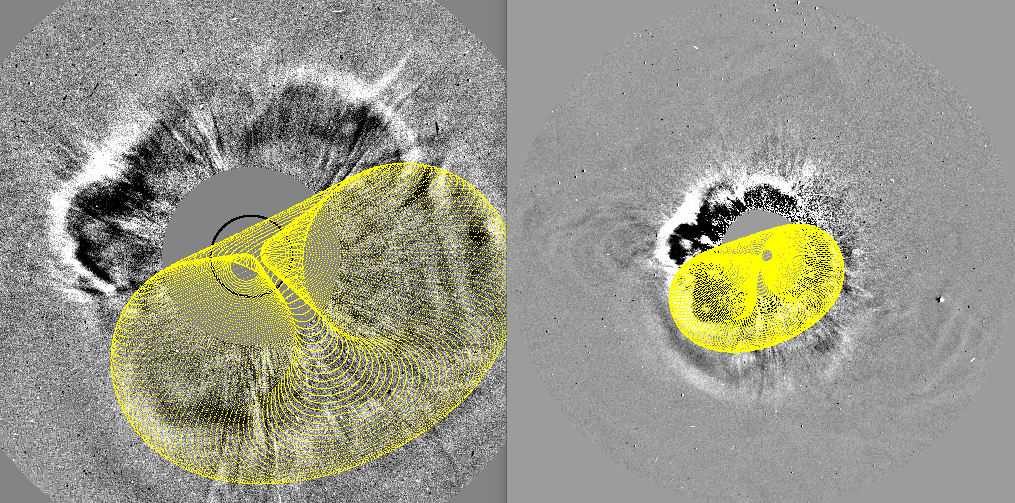}\\
   \includegraphics[scale=0.20]{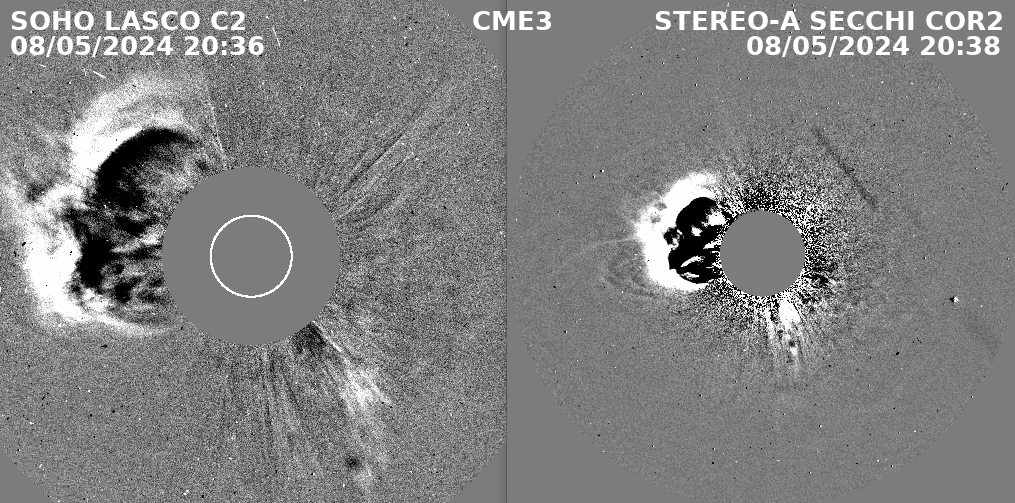}
   \includegraphics[scale=0.20]{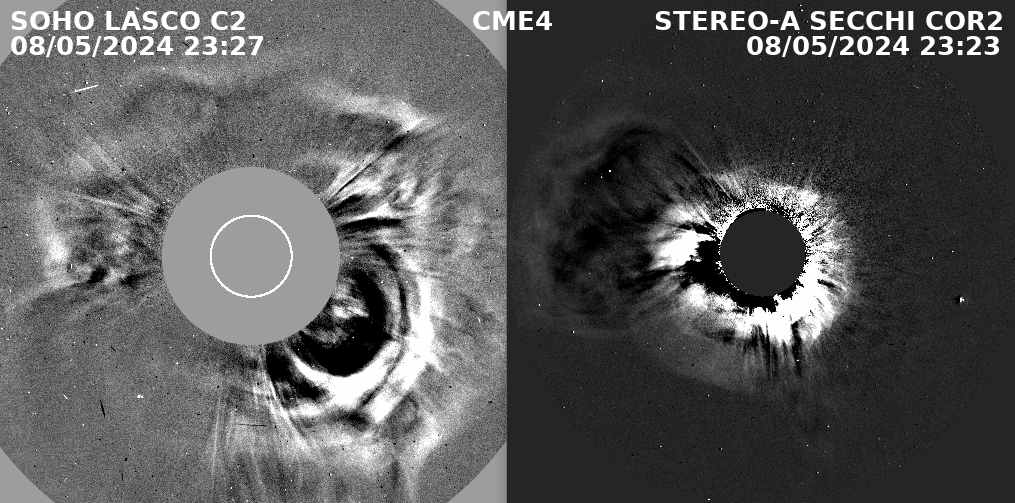}\\
   \vspace{-1mm}
   \includegraphics[scale=0.20,trim={0cm 2cm 0cm 0cm}, clip]{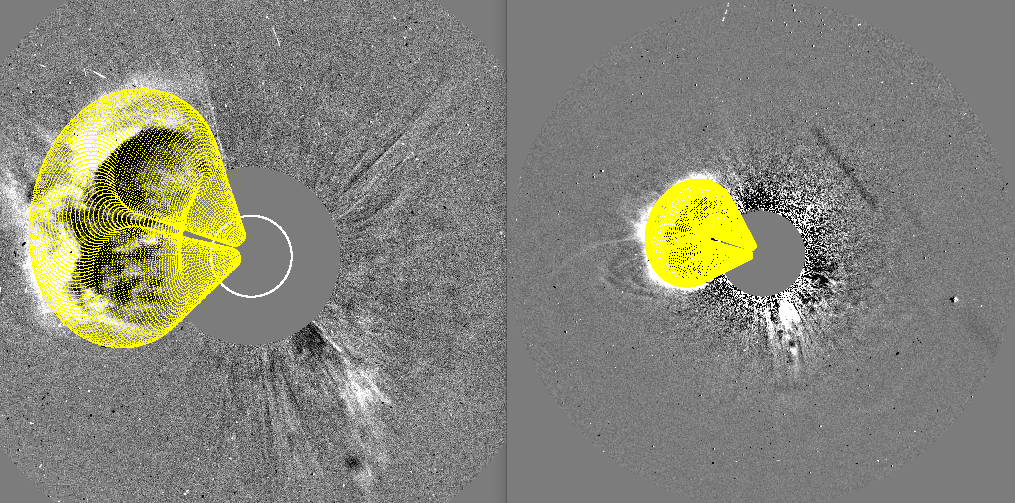}
   \includegraphics[scale=0.20,trim={0cm 0cm 0cm 2cm}, clip]{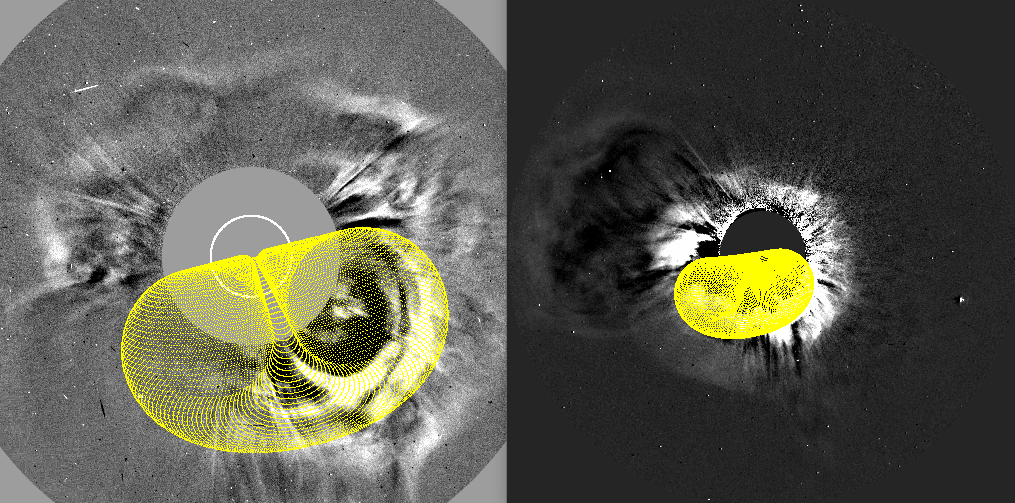}\\
   \includegraphics[scale=0.20,trim={0cm 1cm 0cm 0cm}, clip]{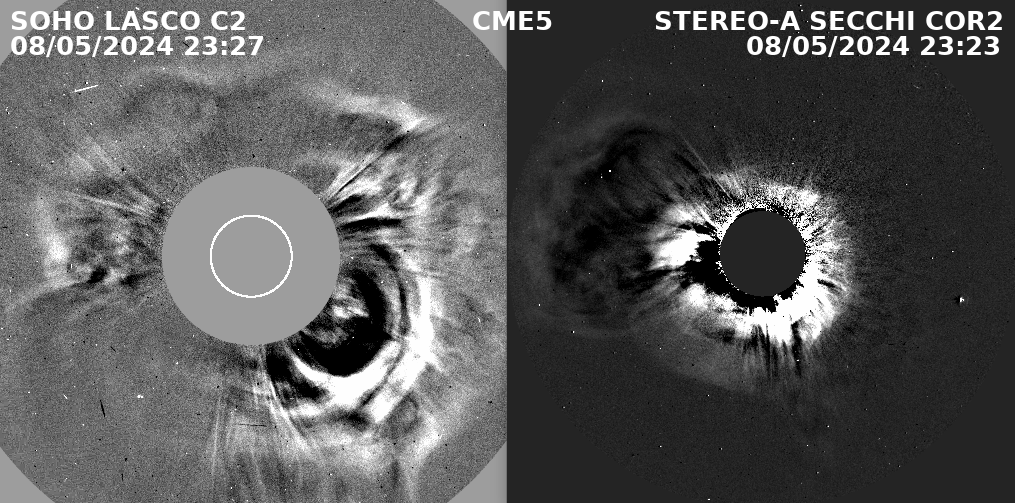}
   \includegraphics[scale=0.20,trim={0cm 1cm 0cm 0cm}, clip]{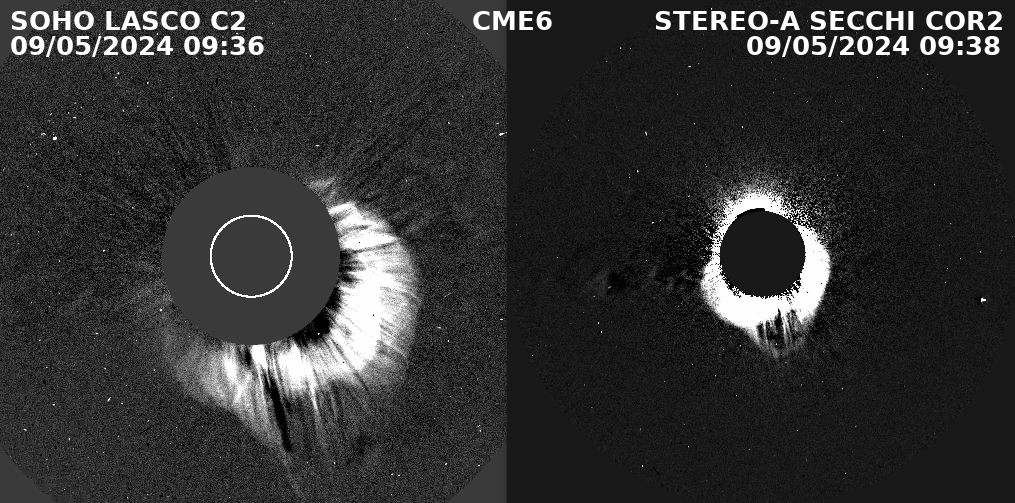}\\ \vspace{-1mm}
   \includegraphics[scale=0.20,trim={0cm 1cm 0cm 2cm}, clip]{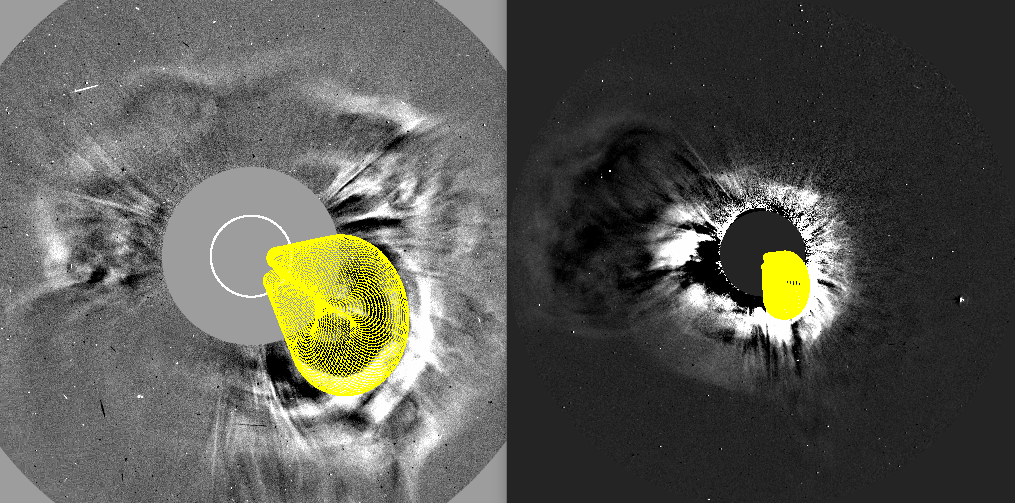}
   \includegraphics[scale=0.20,trim={0cm 1cm 0cm 2cm}, clip]{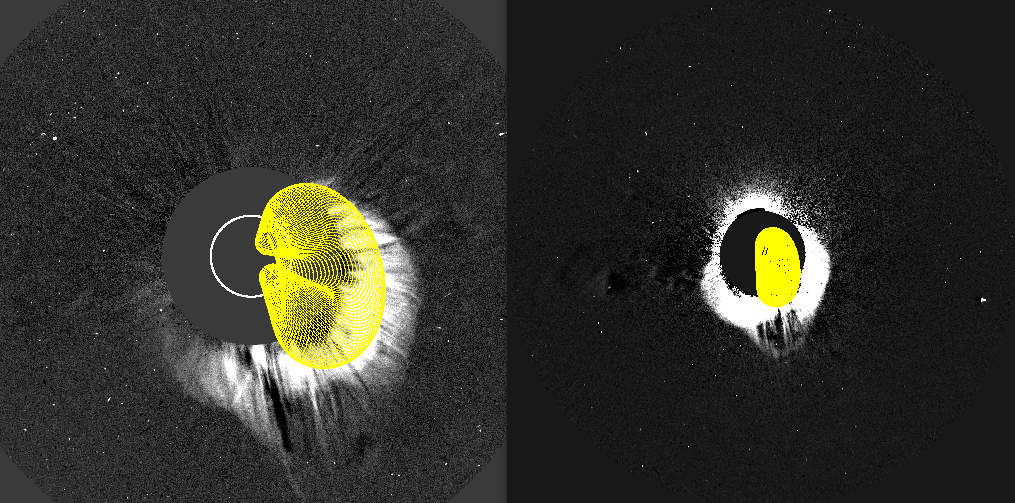}\\
   
   \caption[The GCS model fitted Earth-impacting CMEs, responsible]{The GCS model fitted Earth-impacting CMEs, responsible for the great event on 10 May 2024. The top panel of each corresponding CME shows the running-difference imaging observations of CMEs by SOHO/LASCO-C2 and STEREO-A/SECCHI-COR2, whereas in the bottom panel, the fitted GCS model is overlaid as the yellow wire frame.}
   \label{chap5-fig:cmes_gcs}
\end{figure*}

\subsection{3D kinematics using White-light Observations }{\label{chap5-subsec:kinematics}}

We employed the GCS model \citep{Thernisien2006, Thernisien2011} to estimate the three-dimensional leading edge (LE) height ($h$) and propagation direction of the selected CMEs (see Section~\ref{sec:chapt2-metho-gcs}). The GCS model offers a geometric approximation of a CME as a magnetic flux-rope structure in three-dimensional space. By fitting the outer envelope of a CME using simultaneous observations from multiple vantage points, the model enables a more accurate reconstruction of CME kinematics and morphology. In this study, we used contemporaneous coronagraph images from STEREO-A/COR2 and SOHO/LASCO-C2 \& C3 to constrain the GCS parameters. The resulting model fits for each CME are shown in Figure~\ref{chap5-fig:cmes_gcs}.

The derived LE heights and associated geometrical and positional parameters from the GCS fitting are listed in Table~\ref{chap5-tab:GCS}. Among the analyzed events, CME3 was found to erupt from the northeastern quadrant of the solar disk, with a model-estimated source location at N04E24. In contrast, all other CMEs originated from the southwestern quadrant. Within the observed field of view, we did not observe significant variation in the tilt angle, aspect ratio, or half angle for any of the events. Uncertainties in the fitted parameters, as reported in Table~\ref{chap5-tab:GCS}, were estimated by repeating the fitting procedure multiple times. However, since the GCS fitting is performed manually and depends on the operator's experience and interpretation, quantifying exact uncertainties remains challenging. \citet{Thernisien2009} estimated the mean uncertainties in the GCS model to be approximately $\pm4.3^\circ$ (longitude), $\pm1.8^\circ$ (latitude), $\pm22^\circ$ (tilt angle), $_{-7^\circ}^{+13^\circ}$ (half angle $\alpha$), $_{-0.04}^{+0.07}$ (aspect ratio $k$), and $\pm0.48\,R_\odot$ (LE height).

\begin{table}
\centering
\resizebox{0.63\linewidth}{!}{
\rotatebox{90}{
\begin{threeparttable}
\caption[The selected list of responsible CMEs causing the great]{\label{chap5-tab:GCS} The selected list of responsible CMEs causing the great geomagnetic storm on 10 May 2024.}
\centering
\begin{tabular} {llcc|ccccccc|c}
\hline
Events & Date, Time & Flare/ & Source  & Time (UT) & Height ($R_\odot$)  & Longitude & Latitude & Aspect  & Tilt  & Half &Max Speed\\

& (UT) & filament & region &Initial-Final & Initial-Final & (deg) & (deg) & Ratio & Angle & Angle & (km s$^{-1}$)\\
 
 & & & & & & & & (deg) & (deg) &\\
\hline
\hline
CME1 & 8, 5:36 & X1.0 & S18W10& 05:48-10:18  & 5.8-26.9 & 16$\pm 3$ & -8$\pm 2$ & 0.27$\pm 0.1$ & 84$\pm 6$ & 24$\pm 3$ & 967\\
\hline
CME2 & 8, 12:24 & M8.7& S19W11  & 12:36-15:54 & 5.4-22.4 & 13$\pm 2$ & -16$\pm 4$ & 0.34$\pm 0.1$  & 27$\pm 5$ & 23$\pm 6$ & 1142\\
\hline
CME3 & 8, 19:12 & Filament & N25E14 & 19:36-23:42 & 5.2-24.7 & -27$\pm 5$ & 4$\pm 6$ & 0.24$\pm 0.1$ & 79$\pm 2$ & 15 $\pm 5$ & 991 \\
\hline
CME4 & 8, 22:36 & X1.02 & S20W14 & 23:06-25:54 & 8.9-28.3 & 6$\pm 7$ & -18$\pm 3$ & 0.26$\pm 0.12$ & 15$\pm 10$ & 18 $\pm 7$& 1406 \\
\hline
CME5 & 8, 22:36 & M9.87 & S19W28  & 23:06-26:42  &  4.6 -22.5  & 38$\pm 6$ & -15$\pm 3$ & 0.15$\pm 0.4$ & -83$\pm 8$ & 16$\pm 8$ & 1103 \\
\hline
CME6 & 9, 9:24 & X2.2 & S20W22 & 09:24-12:06 & 3.9-25.4 & 27$\pm 3$ & -14$\pm 2$ & 0.28$\pm 0.1$ & -77$\pm 4$ & 23$\pm 3$ & 1746\\
\hline

\hline


\end{tabular}
\begin{tablenotes}
\small
\item Note: The second column lists the date and time of each CME’s first appearance in the SOHO/LASCO-C2 field of view. The fourth and fifth columns indicate the time and height ranges used for the GCS fitting. Columns 5--12 present the GCS-derived geometrical and positional parameters. The final column provides the maximum leading edge (LE) speed ($v$) estimated from the height-time measurements within the coronagraphic field of view.
\end{tablenotes}
\end{threeparttable}}}
\end{table}

\begin{figure*}[tbp]
   \centering
   \includegraphics[width=0.48\textwidth,trim={6cm 1cm 13cm 0cm}]{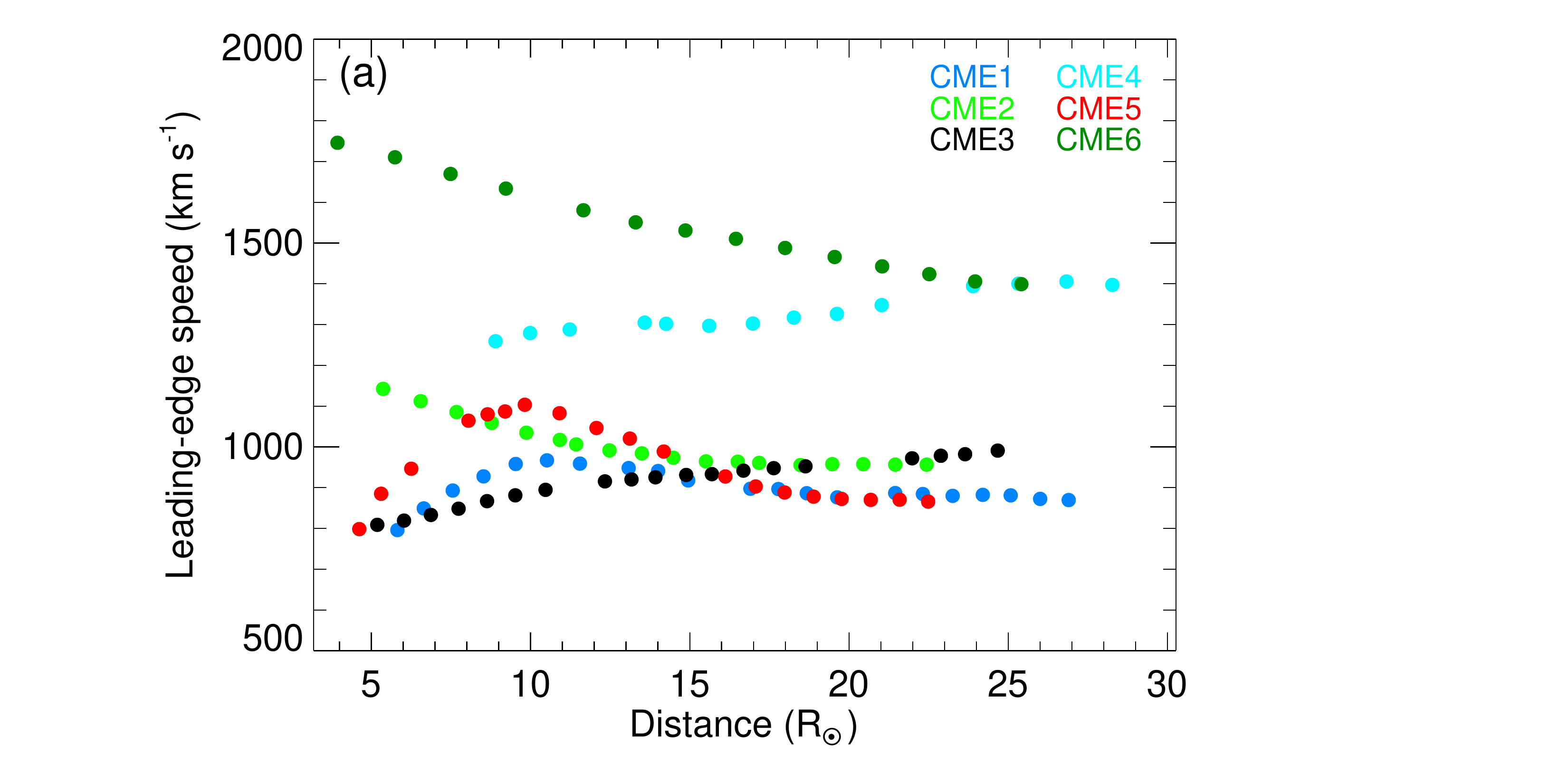}
   \includegraphics[width=0.48\textwidth,trim={5cm 1cm 14cm 0cm}]{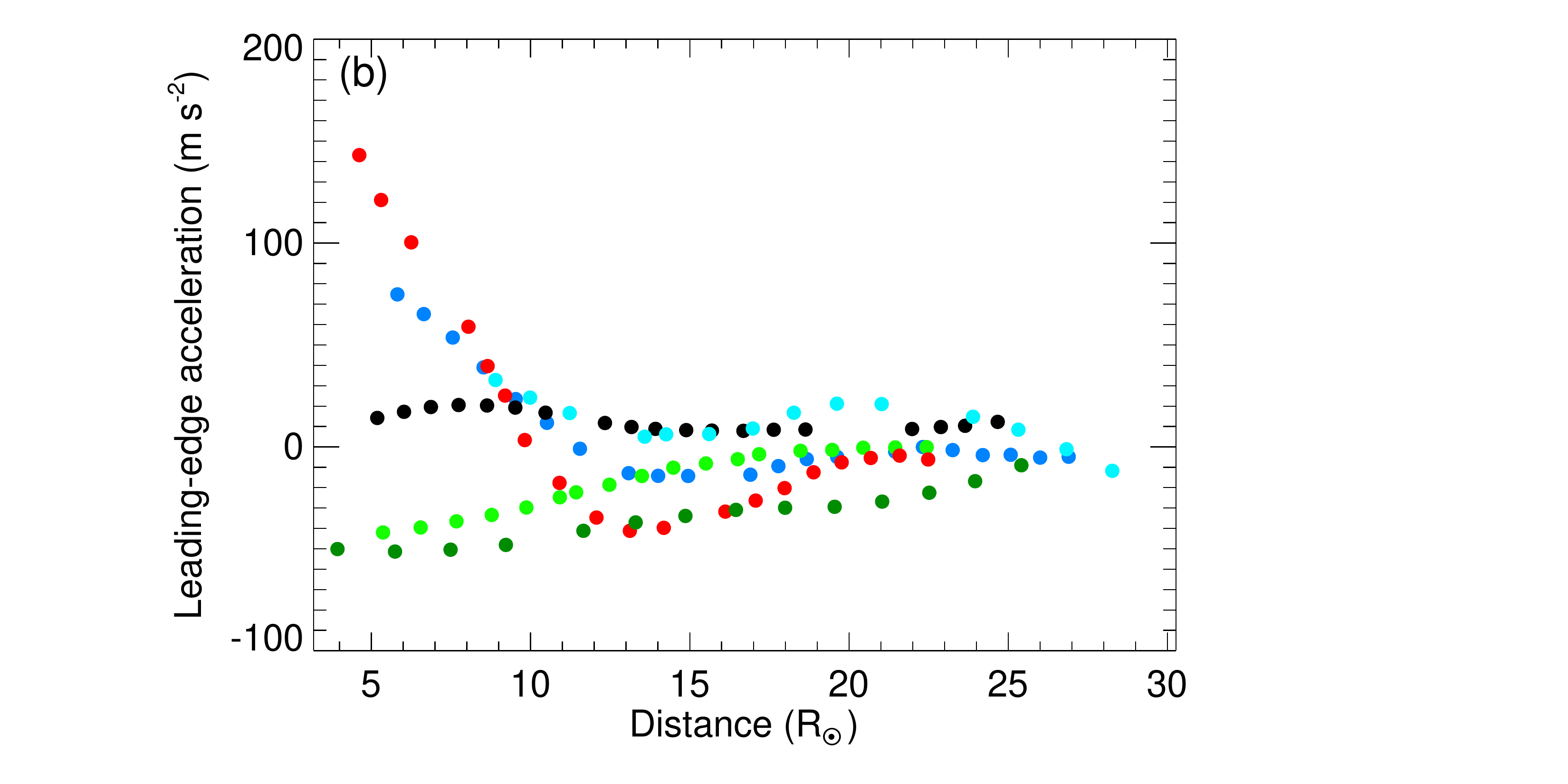}\\
    \includegraphics[width=0.95\textwidth,trim={1cm 0cm 3cm 0cm}]{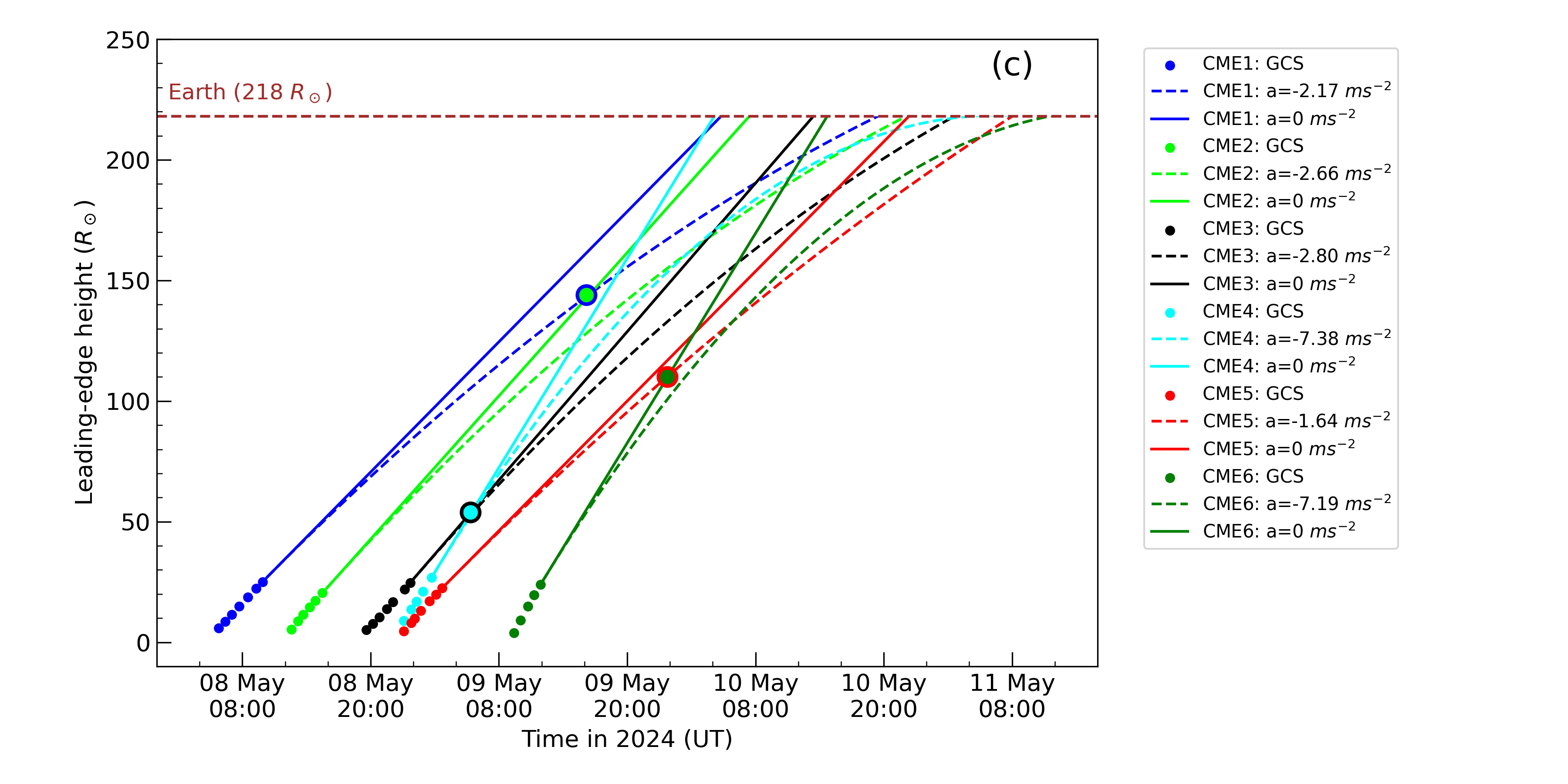}
      \caption[Propagation characteristics of the selected CMEs based]{Propagation characteristics of the selected CMEs based on GCS-derived 3D LE heights. Panel (a) shows the LE speed, and panel (b) presents the corresponding acceleration profiles. Panel (c) illustrates the height-time evolution of the CMEs using two extrapolation approaches: constant speed (solid lines) and constant average deceleration (dashed lines). Colored filled circles mark the estimated heights where CME-CME interactions may have begun. The initial data points represent the GCS-fitted heights; for visual clarity, every alternate point has been omitted in the plots.}
         \label{chap5-fig:kine}
   \end{figure*}

The leading edge (LE) speed ($v$) and acceleration ($a$) of the selected CMEs (Figure~\ref{chap5-fig:kine}) were derived by performing successive time derivatives of the measured three-dimensional LE height ($h$). We employed a moving three-point window technique, where a linear fit was applied within each window to estimate the derivatives at the central point. For the boundary points (i.e., the first and last time steps), a two-point window was used to perform the same linear fitting. This approach, described in detail by \citet{Agarwal2024}, effectively captures realistic variations in speed and acceleration while preserving the temporal resolution of the dataset.

The maximum speed attained by each CME within the field of view is presented in the final column of Table~\ref{chap5-tab:GCS}, as calculated from the GCS-derived height profiles. To estimate uncertainties in the derived kinematics, we accounted for the sharpness of the CME front near the Sun and its increasingly diffuse structure at larger heliocentric distances. Based on repeated fitting trials, we estimated a maximum uncertainty of $\pm10\%$ in the LE height, which was subsequently propagated to compute the uncertainties in speed and acceleration. Interestingly, as shown in Figure~\ref{chap5-fig:kine}a, CME6 exhibited a notably higher speed than the other events. This elevated velocity may indicate the possibility of CME-CME interaction occurring at greater heights in the corona or interplanetary space.

Furthermore, we extrapolated the GCS-derived 3D leading edge (LE) heights of each CME out to a distance of 218~$R_\odot$ using a constant acceleration assumption. Two different approaches were adopted for this extrapolation: (i) assuming zero acceleration, i.e., the CMEs propagate with constant speed beyond the final GCS-tracked height, and (ii) applying the classical equation of motion with constant acceleration, $s = ut + \frac{1}{2} a t^2$, where $u$ is the LE speed at the last GCS-tracked height, $s$ is the height difference between that point and 218~$R_\odot$, and $t$ is the time interval between the last tracked height and the in situ arrival time of the CME LE. The arrival time of each CME LE at near-Earth in situ spacecraft was determined using various plasma and magnetic field signatures, as discussed in Section~\ref{chap5-sec:near-earth}. In the case of non-zero acceleration, the extrapolation was performed such that the modeled arrival time matched the observed in situ LE arrival time. By contrast, extrapolation with zero acceleration does not yield accurate arrival times at 1~au, especially if CME-CME interactions occur en route. However, the comparison of both approaches provides valuable insight: while the zero-acceleration case is more applicable before interaction begins, it also establishes an upper bound for the CME's transit speed and helps constrain the possible height range over which CME-CME interactions may have occurred.

 Typically, fast CMEs attain their maximum speed within about 10~$R_\odot$ and complete most of their acceleration by 20~$R_\odot$ \citep{Zhang2004, Vrsnak2007, Temmer2010}. This is evident in Figure~\ref{chap5-fig:kine}b, where the LE acceleration profiles of all CMEs show a clear decline toward lower values at greater heights. Beyond this region, CMEs primarily experience aerodynamic drag from the ambient solar wind, which tends to decelerate fast CMEs and accelerate slower ones \citep{Gopalswamy2000, Vrsnak2007}. Therefore, the zero-acceleration extrapolation can be considered a reasonable upper bound for CME kinematic evolution. Without employing complex modeling to account for the varying ambient solar wind, we estimated the kinematic evolution of the selected CMEs beyond the coronagraphic field of view by extrapolating their GCS-derived kinematics. This approach enables us to infer the likely height ranges where CME-CME interactions may have occurred. However, it is important to acknowledge that the post-interaction dynamics of CMEs cannot be accurately captured by such extrapolations, as different CMEs may experience diverse preconditioning of the interplanetary medium and exhibit varying acceleration profiles during both pre- and post-interaction phases \citep{Shen2012, Mishra2017}.

Despite these limitations, a simple extrapolation provides useful constraints on possible interaction regions and aids in interpreting the in situ observations at 1~au during the geomagnetic storm. Figure~\ref{chap5-fig:kine}c shows the extrapolated height-time profiles of the CMEs up to 218~$R_\odot$. At a heliocentric distance of 22~$R_\odot$, CME1 and CME2 display similar leading-edge (LE) speeds (Figure~\ref{chap5-fig:kine}a). However, since CME1 is leading, it is likely subjected to stronger drag forces from the ambient solar wind, resulting in greater deceleration compared to CME2. Their source longitudes are estimated to be $16^\circ \pm 3^\circ$ for CME1 and $13^\circ \pm 2^\circ$ for CME2, implying a potential scenario where the LE of CME2 may catch up with the trailing structure of CME1. Using the derived accelerations of $a = -1.78$~m~s$^{-2}$ (CME1) and $a = -0.71$~m~s$^{-2}$ (CME2), their interaction is estimated to occur at a height of approximately 144~$R_\odot$.

An earlier interaction is inferred between CME3 and CME4 at an LE height of $\sim$54~$R_\odot$, under the assumption of zero acceleration. Based on their source longitudes, this interaction is more likely to occur at the flanks rather than a central encounter. This is also supported by observations from the LASCO C3 coronagraph. Furthermore, the extrapolated kinematics suggest that CME5 and CME6 interacted around 110~$R_\odot$, primarily due to the higher speed of CME6 and the proximity of their source longitudes. If CME6 retained its elevated speed post-interaction, as indicated by the fast trailing magnetic ejecta observed in in situ data, this supports the scenario that all six CMEs may have interacted en route to Earth. Such interactions would have contributed to the formation of a large-scale, complex magnetic structure at 1~au, which will be discussed in detail in subsequent sections.

\begin{figure*}[ht]
   \centering
   \includegraphics[width=0.45\textwidth,trim={3cm 1cm 1cm 0cm}]{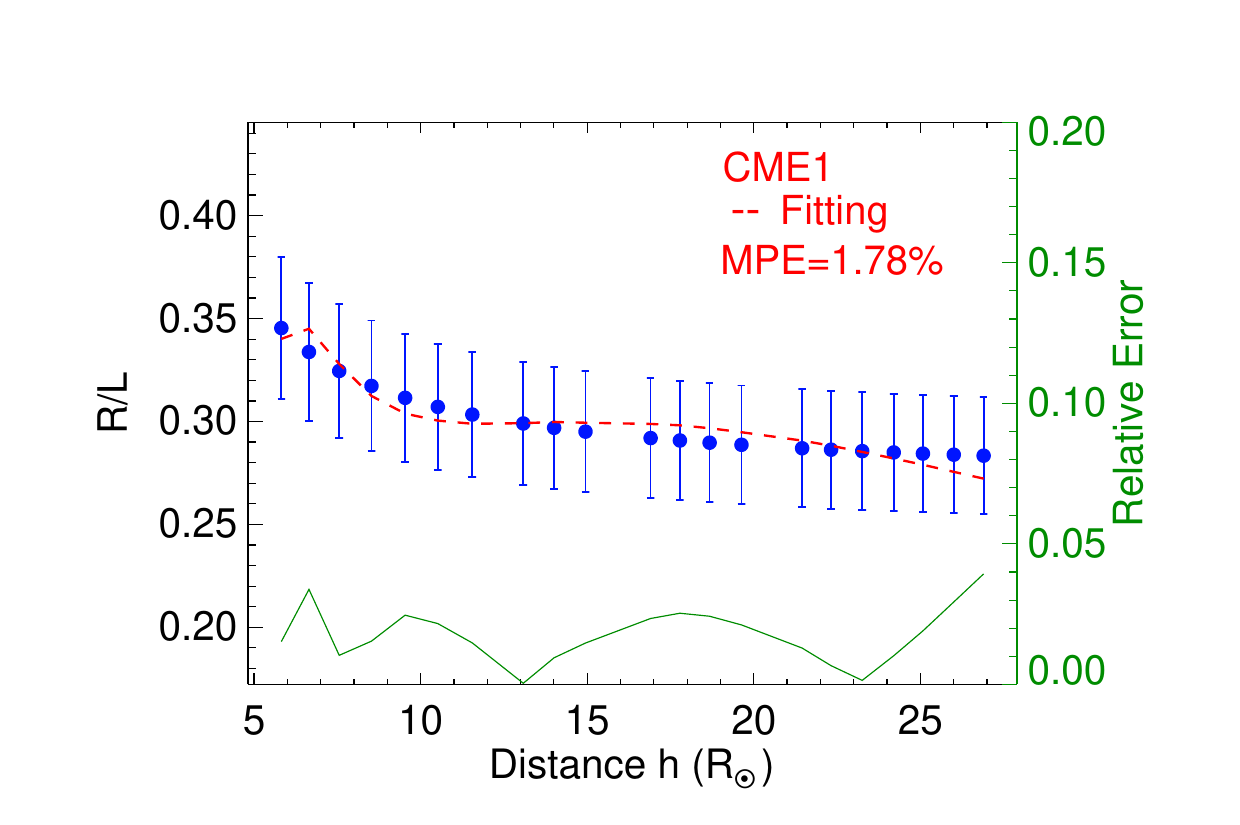}
   \includegraphics[width=0.45\textwidth,trim={1cm 1cm 3cm 0cm}]{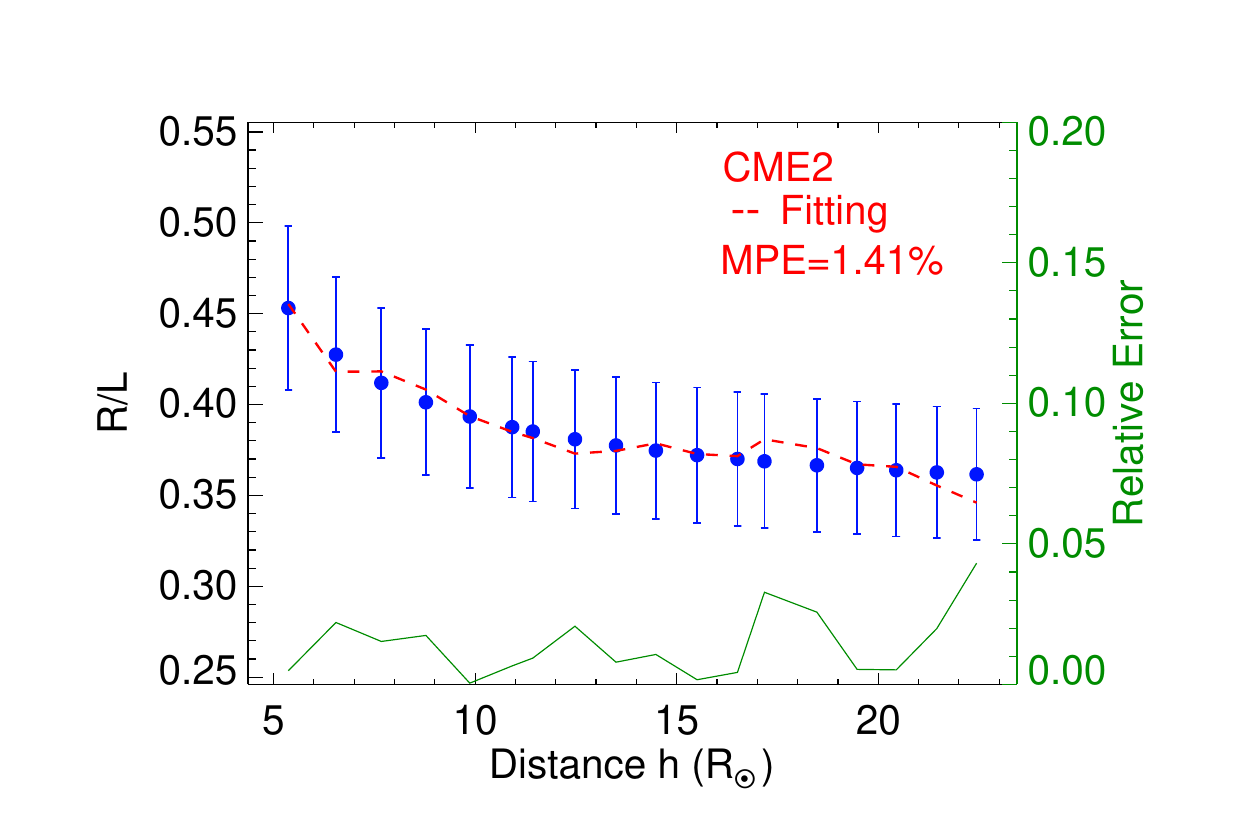}\\
    \includegraphics[width=0.45\textwidth,trim={3cm 1cm 1cm 0cm}]{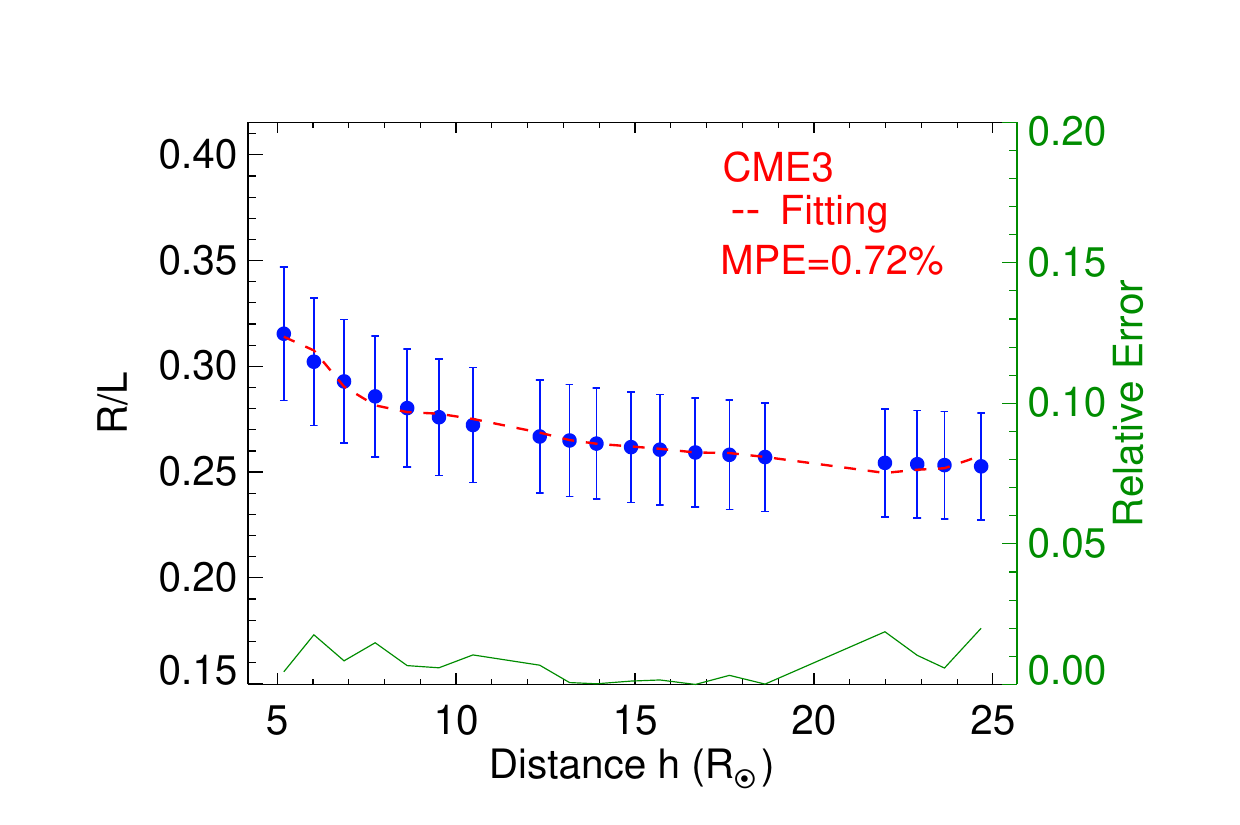}
     \includegraphics[width=0.45\textwidth,trim={1cm 1cm 3cm 0cm}]{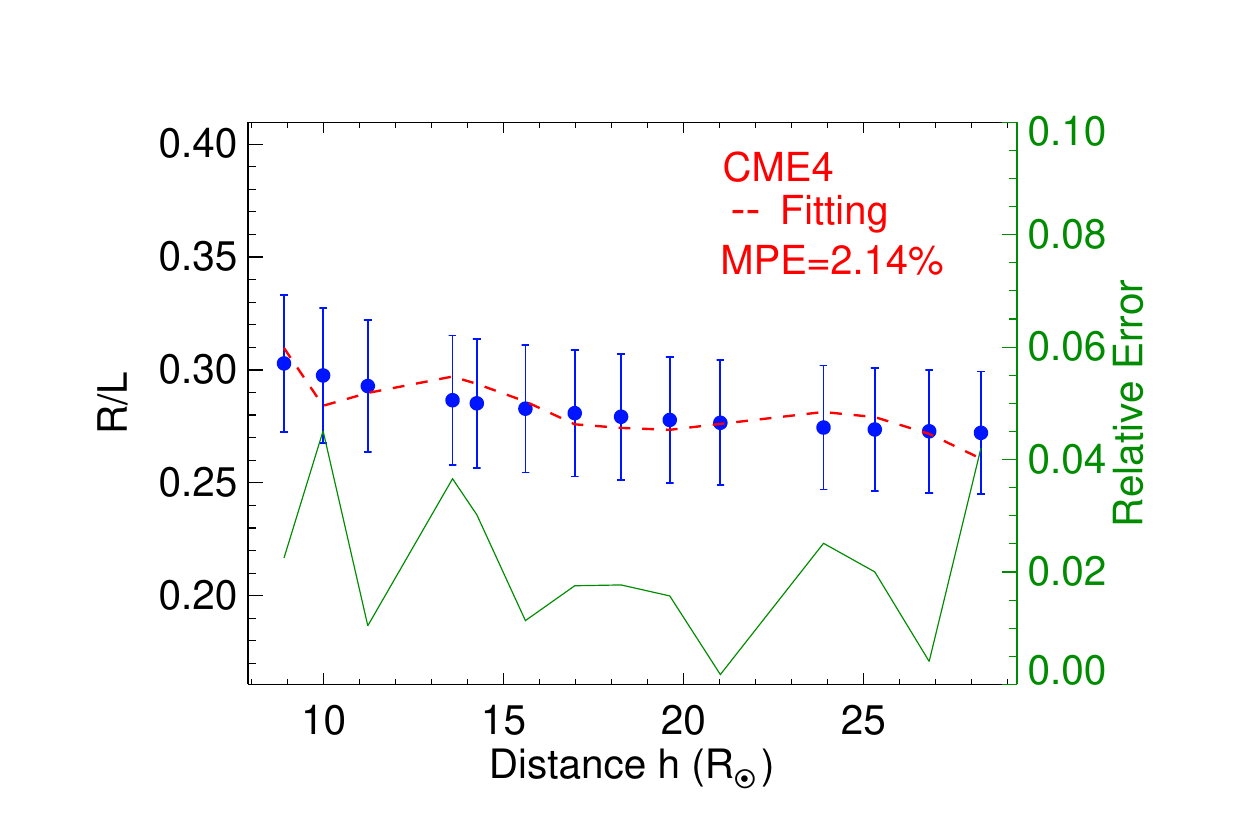}\\
     \includegraphics[width=0.45\textwidth,trim={3cm 1cm 1cm 0cm}]{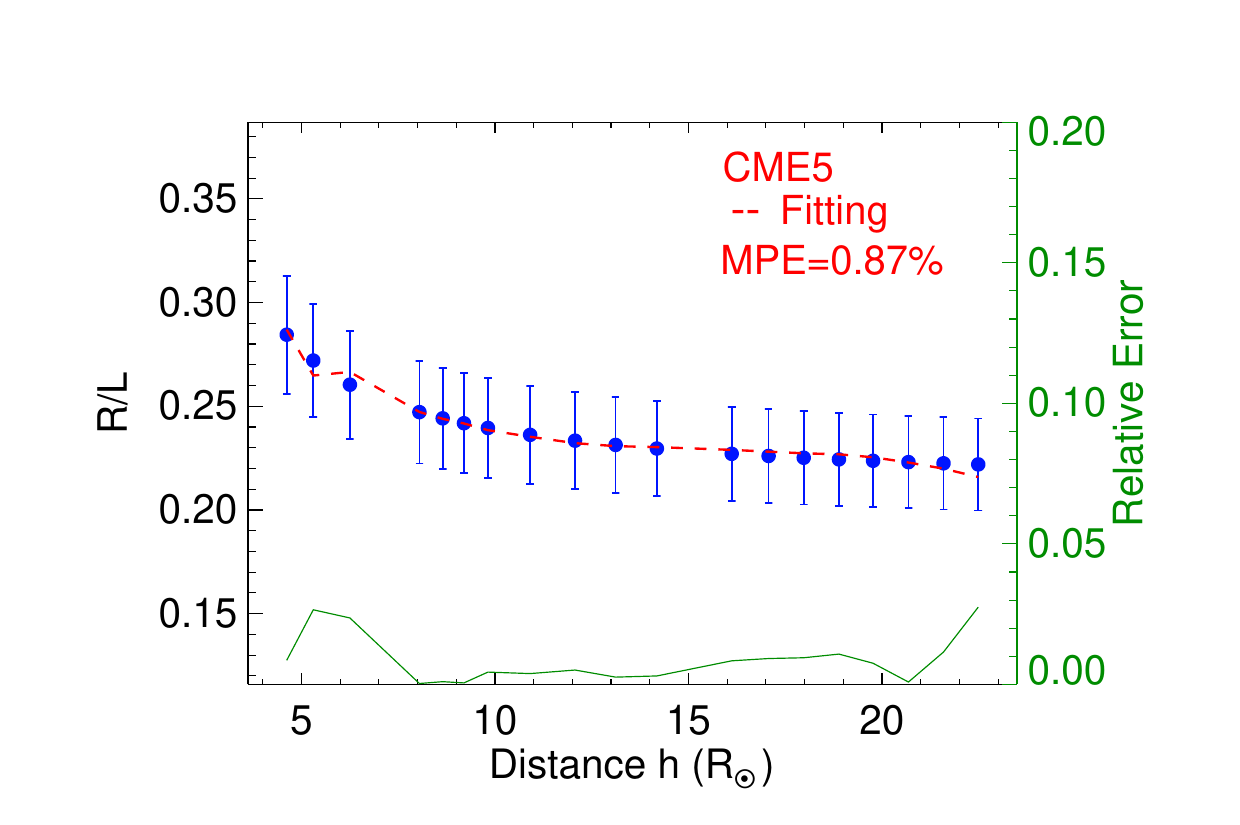}
      \includegraphics[width=0.45\textwidth,trim={1cm 1cm 3cm 0cm}]{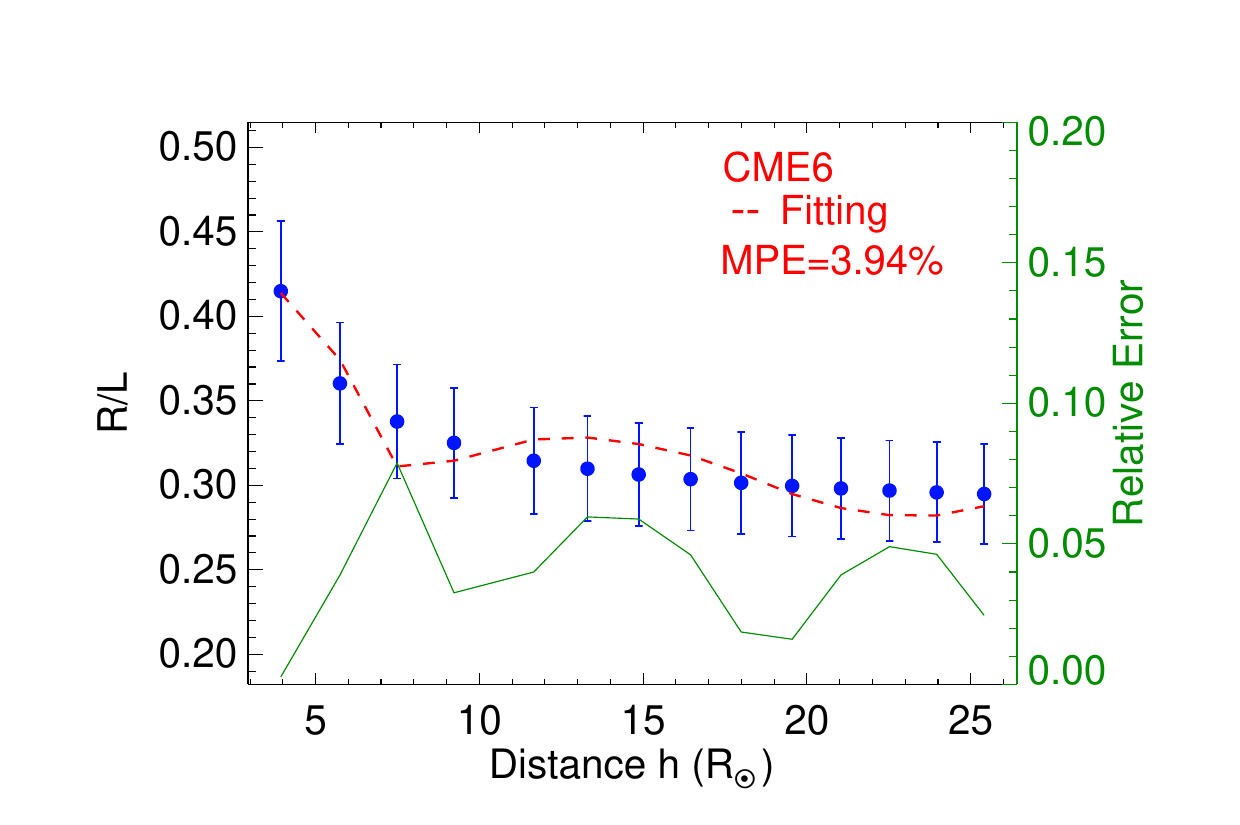}
      \caption[Model fitting errors for the selected fast CMEs]{Model fitting errors for the selected fast CMEs. Blue dots represent the left-hand side of Equation~\ref{fitting}, the red dashed line corresponds to the right-hand side of Equation~\ref{fitting}, and the green line shows the relative fitting errors. MPE denotes the mean percentage error.}
      \label{chap5-fig:FRIS_error}
\end{figure*}

It is important to note that our approach assumes a constant acceleration during the propagation of CMEs beyond the coronagraphic field of view to estimate a broad range of possible distances for CME-CME interactions. However, the estimated interaction heights between different CME pairs may vary if additional physical processes are considered, such as drag forces between individual CMEs and the ambient solar wind, momentum exchange during CME-CME interactions, and magnetic interactions between closely spaced CMEs. A more comprehensive analysis incorporating Heliospheric Imager (HI) observations, along with a detailed assessment of momentum transfer and the nature of CME interactions or collisions \citep{Mishra2015, Mishra2017}, as well as their geoeffectiveness \citep{Lugaz2014, Scolini2020}, could offer deeper insights into the evolution of such complex ejecta. Moreover, any modification in CME kinematics due to these effects will likely leave a signature in their thermal properties, which will be addressed in the following section.

\subsection{Thermodynamics using the measured 3D kinematics}{\label{chap5-subsec:thermal}}

The thermodynamic evolution of successively launched CMEs from the same active region remains poorly understood. In particular, there are limited studies that quantify the thermodynamic properties of individual substructures within complex ejecta formed due to possible CME-CME interactions before their arrival at 1~au. To address this, we employed the analytical FRIS model (see Section~\ref{metho:fris_derivation}) to estimate the internal thermal properties of the selected CMEs. This model utilizes global kinematic parameters, such as the height and radius of the CME flux-rope, to derive various internal quantities, which are summarized in Table~\ref{chap2_tab:FRIS_parameters}.

As described in Section~\ref{metho:fris_derivation}, we solve Equation~\ref{fitting}, where $c_1$–$c_5$ represent unknown constant coefficients. These coefficients are determined by fitting the equation to the observed 3D kinematic profiles of each CME. The results of this fitting procedure for all six CMEs are shown in Figure~\ref{chap5-fig:FRIS_error}.

Among the various internal properties derivable through the FRIS model, this study primarily focuses on the evolution of two key thermodynamic parameters: the polytropic index ($\Gamma$) and the temperature ($T$) of the CMEs. The model-derived expression for $\Gamma$ and $T$ are,
\begin{equation}\label{chap5-eqn:gamma_eqn}
    \Gamma =  \gamma +   \frac{ln{\frac{\lambda (t)}  {\lambda (t+\Delta t)}}}  {ln[(\frac{L(t+\Delta t)}  {L(t)})[\frac{R(t+\Delta t)}  {R(t)}]^2]} 
\end{equation}
\begin{equation}\label{chap5-eqn:temp_eqn}
     {T}=  \frac{ k_2 k_8 }{k_4 } \biggl[ \frac{\pi \sigma}   {(\gamma -1)} {\lambda (LR^2)^{1{-\gamma}}  }\biggr].
\end{equation}
Excluding $L$, $R$, and $\gamma$, all other quantities in the above equations are unknown. However, by determining the fitting coefficients in Equation~\ref{fitting}, several of these unknowns can be evaluated. In particular, the model-derived temperature ($T$), apart from its dependence on $\gamma$, is scaled by a multiplicative factor of $\left( \frac{k_2 k_8}{k_4} \right)$, the absolute value of which cannot be determined directly from the model.

This scaling factor varies for each CME, as it depends on the specific set of fitted coefficients, but remains constant over time for any individual CME. As a result, the model does not yield absolute temperature values, limiting our ability to interpret the actual thermal state. To facilitate a comparative analysis of temperature trends across the different CMEs, we normalize each temperature profile to an initial value of $10^7$~K at the first observational data point. The normalization factor for each CME is chosen such that all CMEs have the same starting temperature, enabling a meaningful comparison of their relative thermal evolution with increasing heliocentric distance.

\begin{figure*}[tbp]
   \centering
   \includegraphics[width=0.48\textwidth,trim={8cm 1cm 13cm 0cm}]{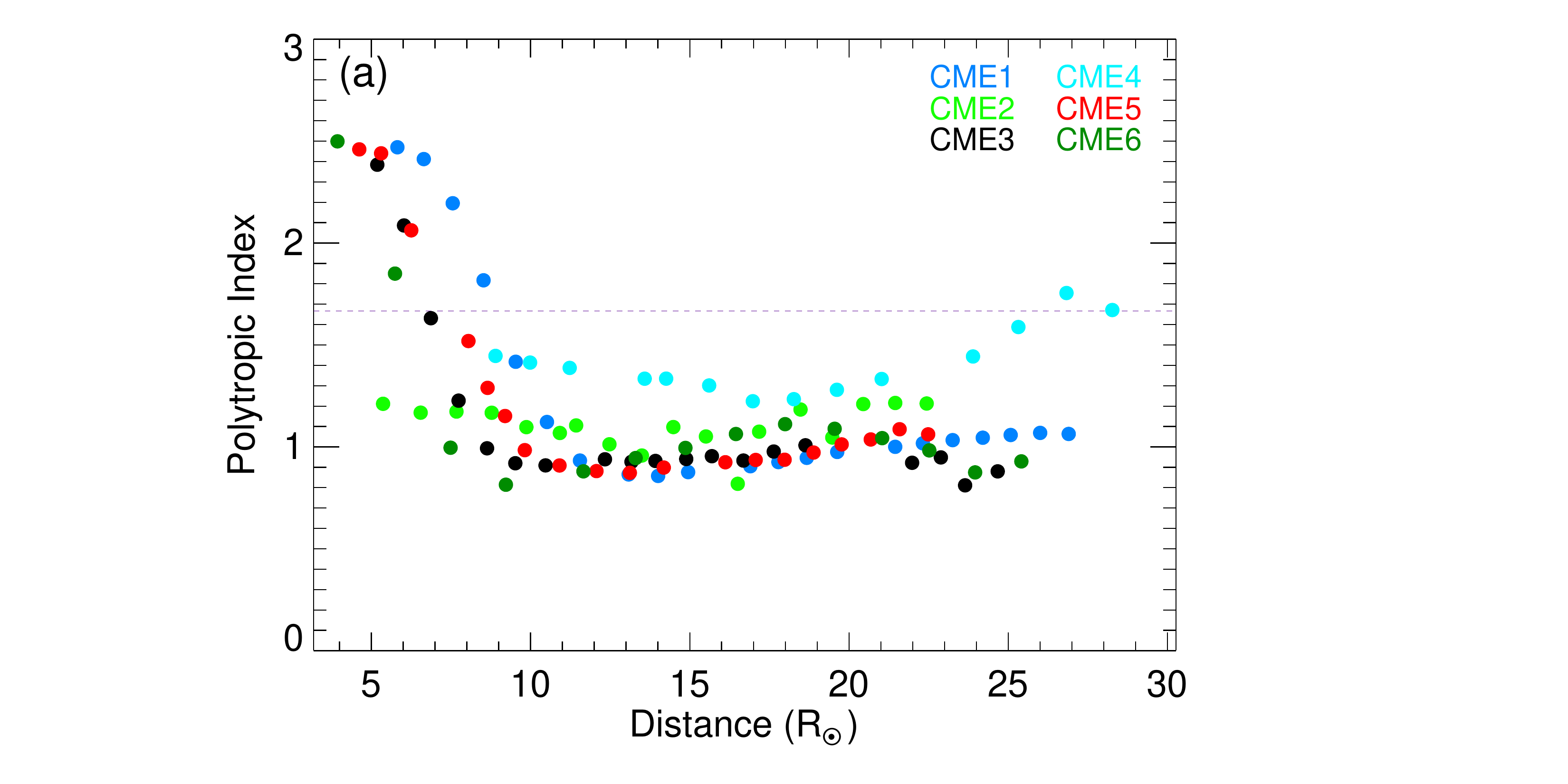}
   \includegraphics[width=0.48\textwidth,trim={6cm 1cm 15cm 0cm}]{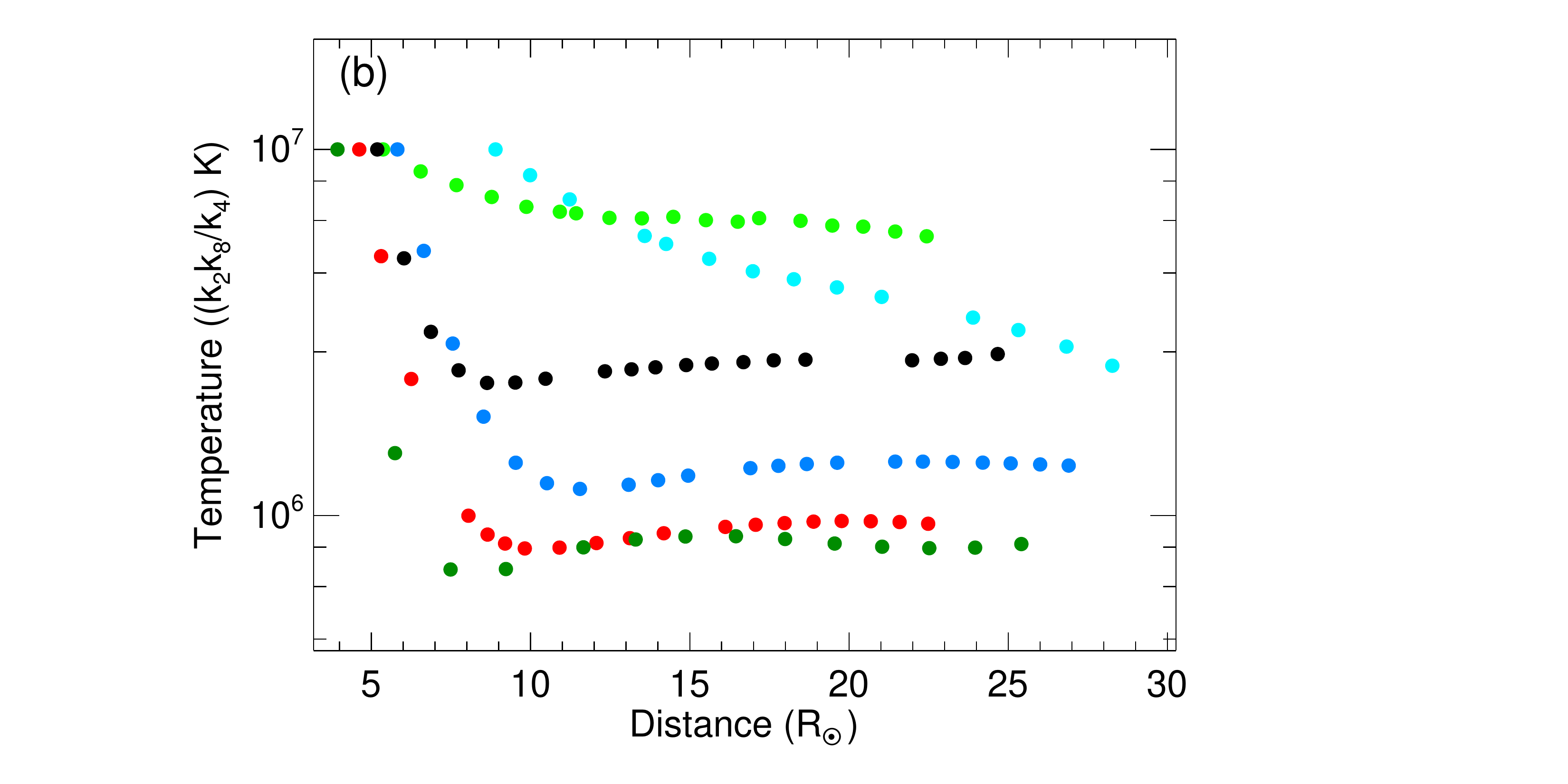}
    
  \caption[FRIS model-derived (a) polytropic index ($\Gamma$) as]{FRIS model-derived (a) polytropic index ($\Gamma$) as a function of CME leading edge (LE) height, with the black dashed line indicating the adiabatic index ($\gamma = 1.67$), and (b) temperature ($T$) profile versus LE height for the six selected CMEs.}
   \label{chap5-fig:gamma}
   
\end{figure*}

The FRIS model-derived polytropic index ($\Gamma$) and temperature ($T$) for the selected CMEs are presented in Figures~\ref{chap5-fig:gamma}a and \ref{chap5-fig:gamma}b, respectively. $\Gamma$ provides insight into the thermal state of the CME plasma without requiring a detailed solution of the energy conservation equations. A $\Gamma$ value lower than the adiabatic index ($\gamma = 1.67$) indicates a heating state, whereas a value higher than $\gamma$ suggests a heat-releasing state. For the majority of the CMEs analyzed (except CME2 and CME4), the initial $\Gamma$ values exceed the adiabatic index, indicating a heat-release process in the early phase of their evolution. In contrast, CME2 and CME4 exhibit $\Gamma$ values below $\gamma$ even at lower heights, implying ongoing heating during their early propagation. At larger heliocentric distances, all CMEs, except CME4, tend toward a $\Gamma$ value close to 1, suggesting an approximately isothermal evolution. CME4, however, shows a distinct behavior, maintaining a $\Gamma$ value near the adiabatic index at higher altitudes. This deviation may be attributed to a notable decrease in the CME’s propagation or expansion speed (see Figure~\ref{chap5-fig:kine}b). When the system is already undergoing heating, a decline in propagation or expansion acceleration can shift the thermal response toward a heat-release regime.

As discussed earlier, the FRIS model-derived temperature values have been normalized such that each CME begins with a temperature of $10^7$~K at the start of the observational interval (Figure~\ref{chap5-fig:gamma}b). This normalization allows for a comparative analysis of the relative temperature evolution among the selected CMEs. For most CMEs, except CME2 and CME6, the temperature decreases rapidly at lower heights and then stabilizes. CME2 exhibits a heating state ($\Gamma < 1$) throughout its observed evolution, leading to a relatively slower decrease in temperature. In contrast, CME4 approaches an adiabatic state at greater heights, which is reflected in its steadily decreasing temperature profile. This analysis suggests that, although successive CMEs may alter the local solar wind environment, the thermodynamic evolution, specifically the variation of the polytropic index and temperature, remains broadly consistent across the studied CMEs. This indicates that CMEs may retain distinct thermal characteristics determined during their initiation, regardless of moderate interaction with preceding ejecta or ambient conditions. Interactions among multiple CMEs launched in quick succession are inherently complex, influenced by local magnetic configurations, relative orientation, and propagation direction.

For instance, CME3, CME4, and CME5 were all launched within a $\sim$4-hour interval. Given their longitudinal separation and the faster propagation speed of CME4, a flank interaction between CME3 and CME4 is likely. Our extrapolated kinematic profiles indicate that their leading edges (LEs) begin to overlap around 50~$R_\odot$. However, using the aspect ratio $k = 0.24$ (Table~\ref{chap5-tab:GCS}) and a leading edge height of $\sim$30~$R_\odot$ (Figure~\ref{chap5-fig:kine}c) for CME3, we calculate the trailing edge height as $h - 2\left(\frac{k}{1 + k} \times h\right) \approx 18.4~R_\odot$. This suggests that CME4 begins interacting with the trailing part of CME3 around 20~$R_\odot$, which corresponds to a noticeable drop in LE acceleration (Figure~\ref{chap5-fig:kine}b).

Despite CME3 and CME4 being launched in different directions, their combined angular widths, calculated using $2\alpha + 2 \sin^{-1}k$ \citep{Thernisien2011}, are 58$^\circ$ and 66$^\circ$, respectively. The wider angular extents and the higher speed of CME4 support the possibility of interaction. This is further corroborated by a distinct change in the polytropic index of CME4 beyond $\sim$22~$R_\odot$ (Figure~\ref{chap5-fig:gamma}a), indicating a thermal response possibly triggered by the interaction with CME3’s trailing edge. On the other hand, CME4 and CME5 appear less likely to interact, as CME5 propagates more slowly in the initial phase. The thermal parameters derived for CME5 show no significant variation within the observed height range. However, interaction between CME4 and CME5 could still occur at larger distances if CME4 decelerates following its interaction with CME3 or if CME5 accelerates, potentially due to interaction with the faster CME6. In situ measurements near 1~au will therefore be essential to further investigate the thermal evolution of CME5 and the other CMEs at greater heliocentric distances.

In the following sections, we will analyze the in situ observations of the resulting complex ICME structures and derive their thermal states to gain deeper insight into the nature and consequences of CME-CME interactions.

 \begin{figure*}
   \centering
   \hspace{-0.8cm}    
   \includegraphics[width=1.03\textwidth,trim={4cm 0cm 3cm 1cm}]{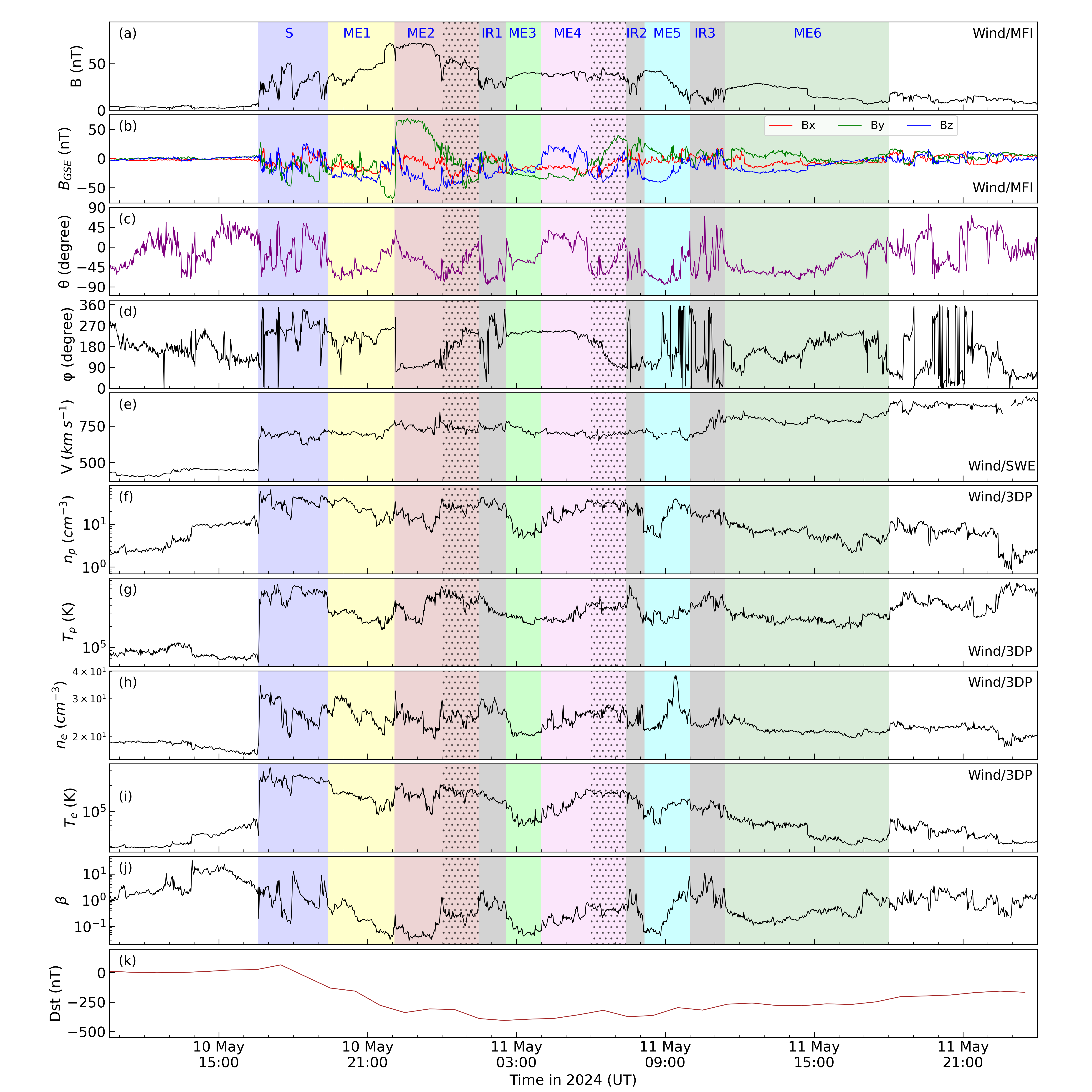}
    \captionsetup{width=\textwidth}
      \caption[In situ observations from the Wind spacecraft of CME]{In situ observations from the Wind spacecraft of CME structures responsible for the intense geomagnetic storm on 10–11 May 2024. Panels (a) and (b) display the total magnetic field strength and its components in GSE coordinates, respectively. Panels (c) and (d) show the inclination angle ($\theta$) with respect to the ecliptic plane and the azimuthal angle ($\phi$, with $0^\circ$ directed toward the Sun) of the magnetic field. Panels (e)–(j) present the bulk solar wind speed, proton number density, proton temperature, electron number density, electron temperature, and plasma beta, respectively. Panel (k) shows the Dst index during the same period. Colored vertical bars across all panels indicate key structures: sheath (S), magnetic ejecta (ME), and interaction regions (IR). The dotted intervals within ME2 and ME4 highlight regions exhibiting signatures of double flux-rope structures, which may have originated during eruption or formed later due to CME-CME interactions.}
         \label{chap5-fig:insitu}
   \end{figure*}

\section{Near-Earth In Situ Measurements}{\label{chap5-sec:near-earth}}

We identify the large-scale solar wind structures observed in situ at 1~au that are associated with the candidate CMEs driving the intense geomagnetic storm of 10–11 May 2024. Figure~\ref{chap5-fig:insitu} presents the near-Earth solar wind conditions measured by the Wind spacecraft, located at the first Lagrange point (L1), along with the corresponding ground-based Dst index during this period. The magnetic field data (1-minute resolution), bulk solar wind speed (92-second resolution), and plasma measurements (92-second resolution) are obtained from the MFI, SWE, and 3DP instruments onboard Wind, respectively. The Dst index (1-hour resolution) is sourced from the World Data Center for Geomagnetism, Kyoto \citep{Nose2015}.

To assist in distinguishing the substructures within the complex magnetic ejecta, we compute the inclination angle $\theta$, defined with respect to the ecliptic plane, using the total magnetic field magnitude ($B$) and its $z$-component ($B_z$), according to:
\[
\theta = \sin^{-1} \left( \frac{B_z}{B} \right).
\]
The azimuthal angle $\phi$, which describes rotation within the ecliptic plane (ranging from $0^\circ$ to $360^\circ$), is derived using the $x$ and $y$ components of the magnetic field ($B_x$, $B_y$) as:

- For \( B_x > 0 \) and \( B_y > 0 \), $\phi = \tan^{-1} \left(\frac{B_y}{B_x}\right).$

- For \( B_x < 0 \) and \( B_y > 0 \), $ \phi = \tan^{-1} \left(\frac{B_y}{B_x}\right) + 180^\circ.$

- For \( B_x < 0 \) and \( B_y < 0 \), $\phi = \tan^{-1} \left(\frac{B_y}{B_x}\right) + 180^\circ.$

- For \( B_x > 0 \) and \( B_y < 0 \), $\phi = \tan^{-1} \left(\frac{B_y}{B_x}\right) + 360^\circ.$

\noindent
The calculated values of $\theta$ and $\phi$ are shown in Figures~\ref{chap5-fig:insitu}b and \ref{chap5-fig:insitu}c, respectively. The shaded regions in the figure, marked with different colors, denote distinct solar wind structures, including the sheath (S), interaction region (IR), and magnetic ejecta (ME), as indicated along the top axis of Figure~\ref{chap5-fig:insitu}. Note that these shaded regions are not applied to the bottom panel of Figure~\ref{chap5-fig:insitu}, which displays the Dst index, since the Dst represents the geomagnetic response measured at Earth and is not temporally aligned with the in situ measurements taken by the Wind spacecraft at L1.

A sudden enhancement in the magnetic field strength ($B$) and solar wind speed ($V$) is observed at approximately 16:35~UT on 10 May 2024, indicating the arrival of a shock front, likely associated with CME1. The impact of this shock on the Earth's magnetosphere leads to a compression of the magnetopause, causing a sudden increase in the dayside geomagnetic field, an event known as a sudden storm commencement (SSC). This SSC is evident in the Dst index profile (Figure~\ref{chap5-fig:insitu}k), where the Dst rises sharply to $\sim$61~nT. The SSC phase persisted for nearly two hours before the Dst index began to decline, indicating the commencement of the main phase of the geomagnetic storm.

Following the shock, a turbulent sheath region (denoted as region S and shaded in purple in Figure~\ref{chap5-fig:insitu}) is observed, characterized by enhanced and fluctuating magnetic field strength ($B$), rapid variations in magnetic field components ($B_{\rm GSE}$), elevated proton density ($n_p$), and increased proton temperature ($T_p$). The first magnetic ejecta (ME1), likely associated with CME1, arrived at 19:25~UT on 10 May 2024. This region (shaded yellow) exhibits strong magnetic fields, smooth rotation in magnetic field vectors ($B_{\rm GSE}$), decreased $n_p$, lower temperature, and plasma beta ($\beta$) below unity, typical features of a magnetic cloud.

Immediately following ME1, a distinct shift in the magnetic field orientation ($\theta$ and $\phi$) is observed, along with continued smooth rotation and low $\beta$, suggesting the presence of a second magnetic ejecta, labeled ME2. Notably, both CME1 and CME2 have comparable leading-edge speeds and accelerations at 20~$R_\odot$ (Figure~\ref{chap5-fig:kine}), but CME1, by clearing the ambient solar wind ahead of it, may reduce the drag experienced by CME2. This setup enhances the potential for CME1–CME2 interaction.

Interaction regions (IRs) between successive MEs are identified by analyzing the in situ plasma and magnetic field signatures following the methodology outlined in \citet{Mishra2015}. The first such region, IR1 (gray shaded in Figure~\ref{chap5-fig:insitu}), lies between ME2 and ME3. It is characterized by a drop in magnetic field strength, abrupt changes in magnetic field orientation, elevated proton density, and increased plasma $\beta$. As discussed in Section~\ref{chap5-sec:nearsun}, CME3 has a relatively narrow half-angle ($15^\circ$) and a source region located $\sim$27$^\circ$ east of the Sun-Earth line, implying a likely flank encounter with Earth. The corresponding structure in the in situ data is identified as ME3.

We distinguish ME4 from ME3 based on several parameters: a reduction in proton and electron densities, variation in magnetic field direction, and a distinct $\theta$ angle profile. An additional interaction region, IR2, is found between ME4 and ME5. It shows reduced magnetic field strength, abrupt magnetic vector changes, increased proton temperature, and higher plasma $\beta$. Unlike IR1, which is marked by a significant increase in $n_p$ but no clear temperature enhancement, IR2 shows a rise in $T_p$ without a corresponding density increase.

CME5, according to GCS modeling, originated from a source longitude $\sim$38$^\circ$ west of the Sun-Earth line, suggesting a flank encounter at Earth. However, the in situ observations for ME5 reveal a smooth magnetic field rotation, low $\beta$, and reduced $T_p$, indicating that CME5 may have been deflected toward the Sun-Earth line, possibly due to an interaction with the faster-following CME6. Such interactions are known to cause slower CMEs to realign along the trajectory of faster ones. A third interaction region, IR3, is identified between ME5 and ME6, exhibiting abrupt variations in magnetic field direction and enhancements in plasma density, temperature, and $\beta$. These sequential interactions and deflections likely contributed to the formation of a large, complex ejecta structure observed at 1~au, driving the intense geomagnetic storm of May 2024.

The gray-dotted regions observed in the trailing parts of ME2 and ME4 (Figure~\ref{chap5-fig:insitu}) can be attributed to interaction-induced structural changes. \citet{Lugaz2013} showed that a relative inclination of approximately $90^{\circ}$ between two ejecta increases the likelihood of their merging into a single, more complex structure. Based on GCS model estimates, the tilt angles of CME2 and CME3 are $27^{\circ}$ and $79^{\circ}$, respectively. While this difference is not exactly $90^{\circ}$, interaction with CME4 may have modified the tilt of CME3, facilitating its interaction with CME2. Similarly, the tilt angles for CME4 and CME5 were found to be $15^{\circ}$ and $-83^{\circ}$, respectively, which also favors significant interaction.

Prior to each interaction region (IR1 and IR2), the preceding magnetic ejecta exhibit clear deformation in magnetic field orientation, along with enhanced plasma parameters, including proton and electron densities ($n_p$, $n_e$), temperatures ($T_p$, $T_e$), and plasma beta ($\beta$), as highlighted in the gray-dotted regions of Figure~\ref{chap5-fig:insitu}. Due to the early interaction between CME3 and CME4, the shock associated with CME4 may have traversed through CME3. Likewise, during the subsequent interaction between CME3 and CME2, the same shock could have further propagated through CME2. According to \citet{Nlugaz2005, Lugaz2009}, when a trailing shock overtakes a leading magnetic ejecta (ME), the dense sheath behind the shock remains between the two MEs, even as the shock continues propagating through the leading structure. This scenario explains the presence of interaction regions between ME2 and ME3, and between ME4 and ME5, as observed in this study. Detailed numerical simulations \citep[e.g.,][]{Nlugaz2005, Xiong2007} could provide deeper insights into the interaction dynamics, though such an investigation is beyond the current scope.

Since CMEs are magnetized plasma structures traversing the ambient solar wind, understanding their interaction or collision remains a complex problem. Several studies have examined CME-CME collisions and their effects on propagation speed, expansion, and structural evolution \citep{Temmer2014, Mishra2014, Mishra2016, FShen2016}. In our study, all CMEs except CME3 exhibited visible shock fronts in coronagraphic observations. Based on this, we expect that faster CMEs with shocks may accelerate preceding CMEs \citep{Schmidt2004, Nlugaz2005}. This is consistent with our in situ observations, where the combined ejecta show a gradual increase in speed.

Interestingly, distinct shock signatures for each fast CME are not observed in the in situ data, despite the identification of individual magnetic ejecta within the complex structure at 1~au. This could be due to early interactions occurring well before 1~au, as well as the propagation of trailing shocks through preceding ejecta, which may weaken or erase typical shock features \citep{Nlugaz2005, Lugaz2009}. Overall, our analysis of the interacting CMEs reveals that the composite structure exhibits a sustained increase in bulk speed, accompanied by a gradual decline in both plasma density and temperature. These trends suggest that the trailing magnetic ejecta continue to compress and accelerate the leading components of the structure as they propagate through interplanetary space.z

\subsection{Thermodynamics using In Situ Measurements}{\label{chap5-subsec:thermal_insitu}}

Several studies have attempted to investigate the thermodynamic behavior of CMEs using in situ observations. However, such analyses remain limited for complex ejecta resulting from the interaction of multiple CMEs. Assuming that the CME plasma evolves through a polytropic process, $\Gamma$ serves as a key parameter to characterize its thermal state. The methodology for determining $\Gamma$ from in situ measurements is detailed in Section~\ref{sec:chapt2-metho-poly}.

In this study, we use high-resolution (92-second) plasma measurements from the Wind spacecraft to analyze the thermodynamics of the identified complex ejecta, along with the surrounding ambient solar wind measured during 6-hour intervals before and after the ejecta passage. These pre- and post-ejecta intervals provide a comparative baseline to evaluate how the thermodynamic properties of the complex structure differ from the ambient medium. Plasma parameters (including ion and electron densities and temperatures) are taken from the Wind/3DP instrument, with data accessed via the Coordinated Data Analysis Web (CDAWeb).

To ensure the reliability of the derived $\Gamma$ values, we apply a statistical filter by selecting only those subintervals with a Pearson correlation coefficient (CC) $\geq 0.8$ and $p$-value $\leq 0.05$ between $\log(n)$ and $\log(T)$. These stringent criteria ensure that each $\Gamma$ value represents a consistent thermal relationship over the respective subinterval. Unlike some prior studies, we do not apply the Bernoulli integral filter, which is commonly used to ensure that a solar wind parcel follows a single streamline. Instead, we argue that the strong correlation criteria used here sufficiently ensure that the derived $\Gamma$ values reliably capture the underlying thermodynamic behavior, even if the Bernoulli integral varies within the interval.

\begin{figure*}[ht]
   \centering
   \includegraphics[width=\textwidth,trim={2cm 0cm 2cm 0cm}]{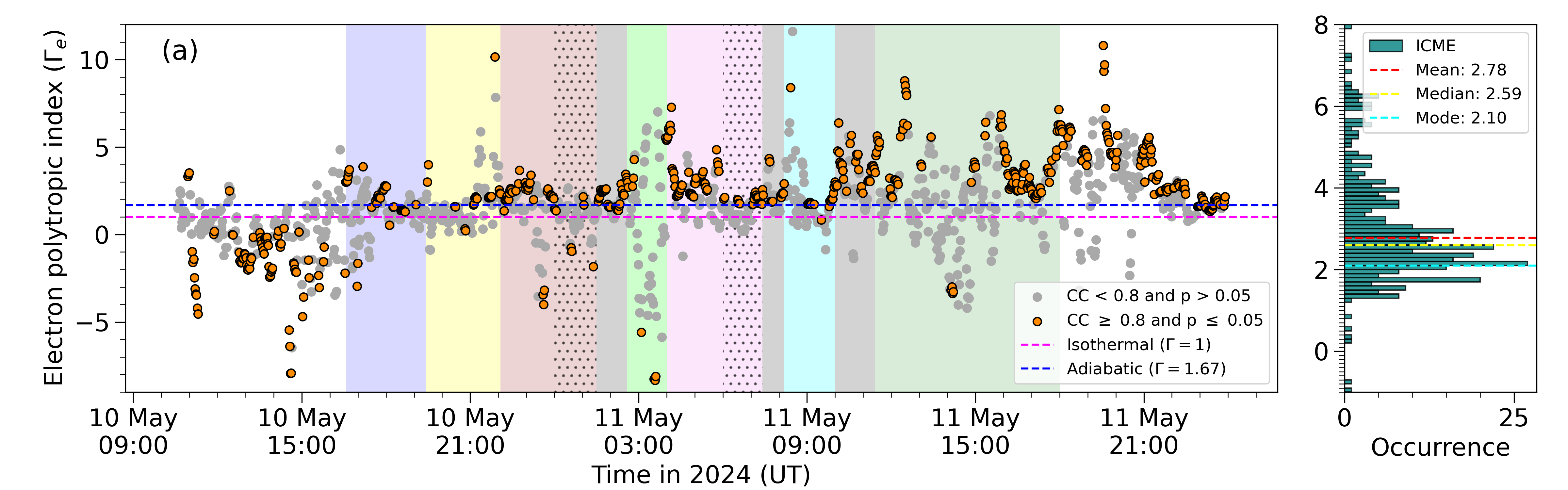}\\
   \includegraphics[width=\textwidth,trim={2cm 0cm 2cm 0cm}]{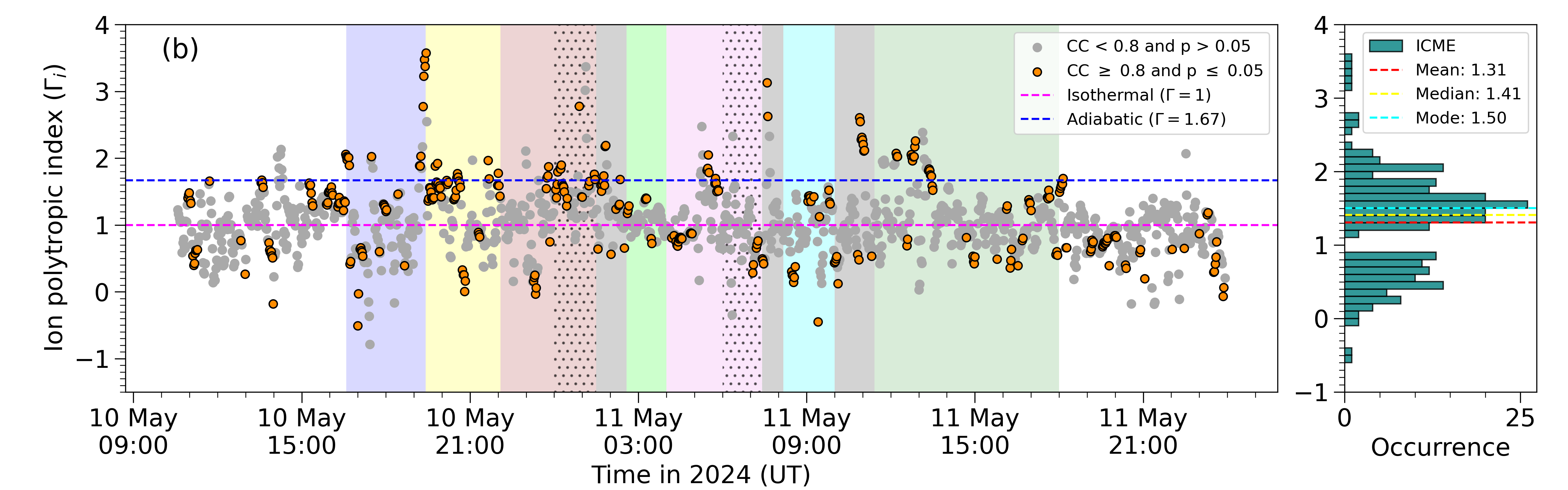}
    \caption[(a) Electron and (b) proton polytropic index for]{(a) Electron and (b) proton polytropic index for the ICME structures measured using Wind data.}
    \label{chap5-fig:insitu_thermo}
\end{figure*}

Figures~\ref{chap5-fig:insitu_thermo}a and~\ref{chap5-fig:insitu_thermo}b illustrate the variation of $\Gamma$ within the complex ejecta, as well as in the pre- and post-ejecta intervals of the ambient solar wind, for electrons and ions, respectively. The background shaded regions correspond to the various substructures of the complex ejecta, as identified in the previous section. In both panels, horizontal dashed lines mark the reference values for the adiabatic index ($\Gamma = 1.67$) and the isothermal condition ($\Gamma = 1$). Each plot includes both reliable (orange) and unreliable (gray) $\Gamma$ values obtained from the log–log fit between density and temperature. Reliable values indicate statistically significant fits that are suitable for interpreting the thermal state. On the right-hand side of each panel, histograms show the distribution of reliable $\Gamma$ values within the entire duration of the complex ejecta.

A clear contrast is observed between the electron polytropic index ($\Gamma_e$) values inside the complex ejecta and those in the surrounding ambient solar wind. The pre-ejecta interval is dominated by $\Gamma_e$ values below 1.67, indicating ongoing heating. In contrast, most of the $\Gamma_e$ values within the complex ejecta and the post-ejecta interval exceed 1.67, suggesting a transition to a heat-releasing state. The mean and median $\Gamma_e$ values inside the complex ejecta were found to be 2.78 and 2.59, respectively. The bottom panel displays the ion polytropic index ($\Gamma_i$) across the same intervals. Similar to electrons, $\Gamma_i$ shows predominantly heating behavior (i.e., $\Gamma_i < 1.67$) in the pre- and post-ejecta ambient solar wind. However, within the complex ejecta, the distribution of $\Gamma_i$ exhibits a bimodal character, indicating a mixture of heating and heat-releasing processes. The mean and median $\Gamma_i$ values within the complex ejecta were calculated to be 1.31 and 1.41, respectively. These results suggest that the ion thermal state within the ejecta is dominated by heating, but also contains significant regions undergoing heat release.

An examination of the thermal states associated with individual substructures within the complex ejecta reveals a distinct trend: each successive magnetic ejecta (ME) exhibits, on average, a higher electron polytropic index ($\Gamma_e$) compared to its predecessor (Figure~\ref{chap5-fig:insitu_thermo}a). This pattern suggests that CMEs contributing later to the complex ejecta, i.e., those that interacted more recently and thus spent less time en route before being sampled in situ at 1 au, tend to exhibit stronger heat-releasing characteristics. A similar trend is observed in the interaction regions (IR1, IR2, and IR3). Notably, IR2, which formed later than IR1, displays more prominent heat-release signatures, as indicated by its elevated $\Gamma_e$ values. In contrast, the ion polytropic index ($\Gamma_i$) does not exhibit a consistent trend across successive CMEs. However, the overall $\Gamma_i$ profile within the complex ejecta is dominated by values indicative of heating, interspersed with short-duration intervals suggestive of localized heat release.

Previous statistical studies of single CMEs have reported $\Gamma_i$ values typically ranging between 1.1 and 1.3 over heliocentric distances from 0.3 to 20 au \citep{Liu2005, Liu2006}, pointing toward sustained heating during CME propagation. On a broader scale, $\Gamma_i$ values in the ambient solar wind tend to lie between 1.5 and 1.67, whereas smaller-scale analyses often yield values as high as 2.7 \citep{Nicolaou2020}. In our case, the large-scale structure under investigation, a complex ejecta formed from multiple interacting CMEs, exhibits distinct thermodynamic behavior shaped by these interactions. The observed differences between $\Gamma_e$ and $\Gamma_i$ are expected. Electrons, being significantly less massive than ions, respond more rapidly and energetically to physical processes that drive thermal changes. As a result, electron-based polytropic indices often reflect sharper thermal perturbations, particularly in regions of recent CME-CME interaction.

As discussed in Section~\ref{chap5-subsec:kinematics}, the estimated CME-CME interaction heights are well within the inner heliosphere, well before reaching 1~au. Specifically, interactions are expected around 144~$R_\odot$ for CME1 and CME2, 54~$R_\odot$ for CME3 and CME4, and 110~$R_\odot$ for CME5 and CME6. It is important to emphasize that these estimates are based on a simplified extrapolation approach, given the absence of direct kinematic measurements beyond the coronagraphic field-of-view and particularly during the post-interaction phases. Consequently, establishing a direct one-to-one correspondence between CME sequences and speeds inferred from remote observations and those identified in in situ data remains challenging.

From in situ observations, we do not identify distinct IRs between ME1 and ME2 or between ME3 and ME4 (Figure~\ref{chap5-fig:insitu}). This is consistent with previous reports noting the occasional absence of IRs in observations of interacting CMEs at 1~au \citep{Mishra2015, Mishra2015a}. Such absences may arise due to specific characteristics of the following CME that render it inefficient in compressing the preceding CME or in forming a dense sheath region. Additionally, the presence or absence of IRs could be influenced by the nature of the collision, whether it is head-on, flank, or glancing, which remains difficult to constrain without detailed knowledge of 3D kinematics before and after interaction.

Despite these limitations, our remote sensing analysis indicates a high likelihood of CME-CME interaction, which is further supported by in situ measurements showing enhanced plasma temperatures and reduced spatial extent of MEs within the complex ejecta. This aligns with previous studies suggesting that CME-CME interactions contribute to heating and compression of the ejecta \citep{Liu2012, Temmer2014, Mishra2014}. The derived $\Gamma_e$ predominantly shows values indicating a heat-release state, which is expected if the electrons were heated during interaction and later transitioned into a cooling phase. In contrast, the heavier ions, which are less responsive to rapid thermal adjustments, exhibit a mixed thermal state with signatures of both ongoing heating and heat release. These results suggest that $\Gamma_e$ may serve as a more sensitive proxy for CME-CME interaction compared to $\Gamma_i$.

We also highlight that the regions marked as gray-dotted intervals in ME2 and ME4 (Section~\ref{chap5-sec:near-earth}) exhibit characteristics resembling double flux-rope (FR) structures. The $\phi$ angle profiles reveal opposing orientations in the two segments of ME2 and ME4. ME2 displays an eastward-directed first FR and a westward-directed second FR, whereas ME4 exhibits the opposite. Correspondingly, $\theta$ values show a north-to-south rotation in the first FR and a south-to-north rotation in the second, implying NES and SWN configurations for ME2 and NWS and SEN configurations for ME4 \citep{Bothmer1998}. Moreover, both flux ropes in ME2 exhibit right-handed helicity, while those in ME4 are left-handed, suggesting consistent handedness within each ME. Although the magnetic field rotation in the leading part of ME4's first FR is not very prominent, a rotation is still evident.

The plasma parameters, proton density ($n_p$), proton temperature ($T_p$), electron density ($n_e$), electron temperature ($T_e$), and plasma beta ($\beta$), show clear differences between the two FRs in both ME2 and ME4. Notably, the second FR in each ME exhibits a higher plasma density than the first. These signatures are consistent with those expected from CME-CME interactions or from trailing portions of one CME merging with the leading structures of another, thereby mimicking a double FR configuration. Multiple studies have reported the presence of such structures in both remote and in situ observations \citep{Ogilvie1997, Hu2003, Marubashi2007, Farrugia2011, Nieves2020, Hu2021, Wood2021}. \citet{Osherovich2013} presented one of the first observational examples of a double helix/double flux-rope structure in both eruptive and in situ contexts. Similar evidence has been obtained through Grad-Shafranov (GS) reconstruction methods applied to ICMEs \citep{Hu2003, Hu2021}.

We specifically examine the thermal states of the substructures resembling double flux ropes. In ME2, the mean $\Gamma_e$ values are 1.9 and 0.4 for the first and second FRs, respectively. In ME4, the corresponding values are 3.0 and 2.0. The significantly lower $\Gamma_e$ in the second FRs, coincident with elevated densities, suggests a weaker correlation between temperature and density in those intervals. In fact, the first and second FRs of ME2 exhibit an inverse correlation between these quantities. Taken together with the magnetic and plasma data, the thermal profiles lend further support to the interpretation of these intervals as double flux-rope structures. Our findings are consistent with those of \citet{Osherovich2013}, who showed that double FRs can exhibit distinct $\Gamma_e$ values within a single ICME.

We note, however, that these observed structures could either represent genuine double flux ropes formed near the Sun or be the product of interaction-driven evolution during interplanetary propagation. Discerning the true origin of such structures, whether formed intrinsically or generated through CME-CME interaction, requires dedicated multi-perspective analysis, incorporating remote observations, in situ data, and numerical modeling. Such investigation lies beyond the scope of this study but remains a promising direction for future research.

\subsection{Comparison of Thermal States}{\label{chap5-subsec:thermal_fris_insitu}}

The FRIS model-derived $\Gamma$, described in Section~\ref{chap5-subsec:thermal}, was calculated under a polytropic approximation by considering the entire CME with an average temperature, $T = (T_e + T_p)/2$, and a number density, $n = n_e = n_p$. To enable a meaningful comparison between the model-derived $\Gamma$ near the Sun and the thermal state of the corresponding ME at 1~au, we define an effective polytropic index, $\Gamma_{\mathrm{eff}}$, that accounts for both electron and proton contributions.

Since $T_e$ is associated with $\Gamma_e$ and $T_p$ with $\Gamma_p$, the effective polytropic index can be approximated by a temperature-weighted average as follows:

\begin{equation}
\Gamma_{\mathrm{eff}} \approx \frac{\Gamma_e T_e + \Gamma_p T_p}{T_e + T_p}
\end{equation}

Here, the weights are proportional to the thermal energies of the respective populations. In our analysis, we assume that the mean values of $\Gamma_e$, $\Gamma_p$, $T_e$, and $T_p$, estimated over selected subintervals within each ME, represent the average thermal state of that ejecta at 1~au. The computed mean values are summarized in Table~\ref{chap5-tab:eff_gamma}. Using these mean values, the derived $\Gamma_{\mathrm{eff}}$ values for ME1 through ME6 are found to be: ME1 = 1.77, ME2 = 1.47, ME3 = 0.80, ME4 = 1.63, ME5 = 1.26, and ME6 = 1.75. Based on these results, ME1 and ME6 exhibit a heat-release state ($\Gamma_{\mathrm{eff}} > \gamma$), ME2 and ME4 are consistent with a near-adiabatic heating state ($\Gamma_{\mathrm{eff}} \approx \gamma$), while ME3 and ME5 reflect a near-isothermal heating state ($\Gamma_{\mathrm{eff}} \approx 1$).

\begin{table*}[ht]
\centering
\caption[Mean values of plasma parameters for different magnetic]{Mean values of plasma parameters for different magnetic ejecta and the calculated effective polytropic index ($\Gamma_{eff}$) from in situ measurements at 1 au.}
\label{chap5-tab:eff_gamma}
\begin{tabular}{|c|c|c|c|c|c|}
\hline
\multirow{2}{*}{Magnetic Ejecta No.} & \multicolumn{4}{c|}{Mean values for each magnetic ejecta} & \multirow{2}{*}{$\Gamma_{eff}$} \\  
\cline{2-5}
& $T_e$ ($10^5$ K) \rule{0pt}{2.2ex} & $T_p$ ($10^5$ K) & $\Gamma_e$ & $\Gamma_p$ & \\
\hline
ME1 & 1.46  & 2.74  & 2.42 & 1.43  & 1.77  \\
ME2 & 1.62  & 4.35  & 1.70  & 1.38  & 1.47  \\
ME3 & 0.83 & 2.5  & -0.31  & 1.17 & 0.8  \\
ME4 & 1.41 & 3.24 & 2.96  & 1.05 & 1.63  \\
ME5 & 1.1 & 1.81  & 2.33  & 0.85  & 1.26  \\
ME6 & 0.61 & 2.56 & 3.74 & 1.27 & 1.75 \\
\hline
\end{tabular}
\end{table*}

A comparison between the effective polytropic index ($\Gamma_{\mathrm{eff}}$) estimated at 1~au and the model-derived $\Gamma$ values near the Sun reveals significant differences for ME1 and ME6, negligible differences for ME3, ME4, and ME5, and a moderate discrepancy for ME2. Such a direct comparison, relying on only two spatial points, one close to the Sun and the other near 1~au, may yield meaningful insights only for a single interacting CME pair. However, in the present study, the ejecta under consideration results from multiple interacting CMEs, which likely experience several episodes of heating and heat-release throughout their propagation. Consequently, establishing a one-to-one correspondence between the in situ-derived and model-derived $\Gamma$ values becomes challenging and may not accurately capture the thermal evolution. Our findings underscore the complexity of thermodynamic processes in interacting CMEs. A comprehensive understanding requires tracking the continuous thermal evolution of CMEs across their pre-interaction, interaction, and post-interaction phases. We suggest that future investigations should incorporate HI observations in conjunction with in situ measurements spanning a wide range of heliocentric distances. Such combined analyses would offer a more complete picture of the thermodynamic evolution and interaction-driven thermal dynamics of CMEs.

\section{Summary}{\label{chap5-sec:summary}}
In this study, we investigated a sequence of six CMEs launched from the Sun on 8–9 May 2024, which collectively contributed to the onset of a major geomagnetic storm on 10 May 2024. Utilizing observations from SOHO/LASCO-C2, STEREO-A/COR2, and SDO/AIA, we tracked the evolution of these CMEs and applied the GCS model to derive their 3D kinematics. This enabled us to assess their interactions en route to Earth, where they arrived as a complex ejecta. In addition to the kinematic analysis, we examined the thermodynamic behavior of these CMEs using both remote and in situ observations. The key findings of our study are summarized below:

\begin{enumerate}

    \item Based on the derived 3D kinematics, we identified likely interaction points among CME pairs, CME1 and CME2 at $\sim$144~R$_\odot$, CME3 and CME4 at $\sim$54~R$_\odot$, and CME5 and CME6 at $\sim$110~R$_\odot$, indicating the formation of a complex ejecta well before reaching 1~au.

    \item In situ observations revealed multiple MEs associated with the remotely tracked CMEs. Signatures of CME-CME interactions were evident, including enhanced heating and compression, highlighting the impact of such interactions on the global structure and geoeffectiveness of CMEs.

    \item We identified IRs between successive MEs in in situ data, confirming ongoing interactions during their transit. Additionally, ME2 and ME4 exhibited properties consistent with double flux-rope structures, seen in magnetic field rotation, polytropic index variations, and density-temperature relationships, suggesting that CME-CME interactions can result in such composite configurations.

    \item The thermal profiles of CMEs during their early evolution varied significantly. While most CMEs approached an isothermal state at larger heights, CME4 exhibited a transition toward adiabatic behavior. FRIS model results indicated that the thermal evolution, whether heating or heat-release, is closely linked to CME expansion dynamics, with slower expansion (as in CME4) leading to heat release. Notably, despite traveling through different pre-conditioned solar wind environments, the CMEs showed no major differences in their late-phase thermodynamic properties in the corona.

    \item At 1~au, in situ electron polytropic indices ($\Gamma_e$) within the complex ejecta predominantly exceeded 1.67, indicating a heat-release state. Moreover, while the post-ejecta solar wind exhibited similar heat-release characteristics, the pre-ejecta region showed signs of heating, suggesting that distinct thermodynamic processes operate inside and outside of interacting CME structures.

    \item The ion polytropic index ($\Gamma_i$) within the complex ejecta showed a bimodal distribution, with dominant heating signatures interspersed with intervals of heat-release. This contrasts with electron behavior, which may reflect the relatively slower thermal response of ions and the influence of the time elapsed since interaction events.

\end{enumerate}

Overall, the combined kinematic and thermodynamic analysis provides important insights into the evolution of interacting CMEs. Although the focus of this work was not extensively on the geomagnetic consequences, the study demonstrates that interactions among multiple CMEs can significantly influence their plasma properties and, in turn, their space weather impact. Future investigations using heliospheric imaging, in conjunction with in situ measurements, are essential to trace the continuous thermal evolution of CMEs through their pre-, during-, and post-interaction phases. Moreover, detailed analysis of substructures, such as interaction regions and double flux ropes, within complex ejecta is needed to better understand their role in driving rare and intense geomagnetic storms.

\section{Chapter Conclusion}\label{chap5-sec:conclusion}

This chapter presented a comprehensive analysis of a series of six interacting CMEs launched on 8–9 May 2024, which collectively formed a complex ejecta observed at 1~au and triggered a great geomagnetic storm on 10 May 2024. By combining multi-viewpoint remote sensing observations with in situ solar wind and geomagnetic measurements, we tracked the evolution, interactions, and thermodynamic signatures of the CMEs near the Sun and at 1~au. The application of the GCS model enabled us to determine the 3D kinematics and estimate interaction heights for different CME pairs, indicating that the complex ejecta formed well before their arrival near Earth. In situ measurements confirmed the presence of multiple magnetic ejecta and interaction regions, as well as structures exhibiting characteristics akin to double flux ropes, likely shaped by CME-CME interactions. These observations underscore the dynamic and complex nature of CME evolution in interplanetary space, especially under successive eruptions and collisions.

The thermal characterization of these interacting CMEs, using polytropic index estimations from in situ data, revealed distinct differences between electrons and ions. Electrons predominantly exhibited a heat-release state, especially within the complex ejecta, while ions showed a bimodal distribution, suggesting ongoing heating and partial equilibration. This divergence highlights different thermal responses of charged species in CME plasmas and their sensitivity to interaction history. Comparisons with model-derived thermodynamic parameters near the Sun revealed limitations in capturing the full complexity of CME evolution using two-point measurements alone. Our results advocate for the need to resolve the continuous thermal history of CMEs, especially during their pre- and post-interaction phases. This study demonstrates that CME-CME interactions significantly alter the magnetic and thermal structure of ejecta and may enhance their geoeffectiveness, emphasizing the importance of integrating remote and in situ diagnostics for advancing space weather forecasting capabilities.
\newpage

\lhead{\emph{Chapter 6}} 
\quotechapter{6}{`` Science is not finished until it is communicated.''\\ \smallskip — Mark Walport}{Conclusions and Future Work}

\lhead{\emph{Chapter 6: Conclusions and Future Work}}

\section{Conclusions} \label{Ch6-sec:conclusion}

This thesis addresses the long-standing challenge of understanding the thermodynamic evolution of CMEs from their initiation in the low corona to their interplanetary manifestations and geospace impacts. While decades of research have advanced our understanding of CME kinematics, propagation, and geoeffectiveness, the internal thermal processes governing their expansion, the balance of different forces, and heliospheric interactions remain comparatively less understood. As outlined in Section~\ref{sec:chapt1-gaps}, earlier studies often assumed fixed polytropic indices or simplified energy treatments, treating thermodynamics as secondary to kinematics. Observational limitations, remote sensing constrained to the low corona, and in situ measurements restricted to single radial points, further hindered a continuous description of CME thermal evolution. The work presented here sought to close these gaps by combining analytical modeling, multi-instrument remote reconstructions, and statistical analyses to establish a framework that consistently links CME kinematics, thermodynamics, and geoeffectiveness.

A central contribution of this work is the development and refinement of the FRIS model. This analytical framework connects remotely observed CME kinematics with their internal thermal parameters, including the polytropic index ($\Gamma$), pressure, temperature, and internal force balance. Through detailed case studies of fast and slow CMEs, constrained by 3D kinematics, we demonstrated that the internal energy budgets of CMEs are highly dynamic. Fast CMEs exhibited significant heat release during early deceleration, while slower CMEs displayed net heat absorption during gradual acceleration. Both tended toward nearly isothermal states at larger heliocentric distances, though on different scales. These results show that CME thermodynamics are sensitive to initial conditions and early kinematic profiles, thereby reconciling two perspectives in the literature: adiabatic expansion versus persistent heating. Our results
show that both thermal behaviors appear, but at different phases or distances from the Sun, and they are interconnected with the changing internal force balance and global kinematics rather than by a single, fixed polytropic regime. Our results indicate that CME evolution is governed by a distance-dependent force balance, where thermal pressure and centrifugal forces drive expansion and Lorentz tension provides inward restraint. The interplay among these forces evolves with distance and is coupled to global kinematics.

By connecting remote sensing with in situ sampling, we further quantified non-adiabatic heating in both electrons and protons out to 1~au. Enhanced electron heating relative to protons, together with turbulence and intermittency diagnostics, revealed localized dissipation sites, particularly in sheaths and trailing ejecta. These results show that variations in $\Gamma$ are not mere fitting parameters but encode underlying dissipation processes such as wave–particle interactions, reconnection-driven heating, and anisotropic conduction. This addresses the gap in the physical interpretation of the polytropic index and highlights its diagnostic value in capturing energy transport along the CME trajectory.

The broader generality of these findings was confirmed through a set of nine fast CME events. Their radial thermodynamic evolution followed a consistent sequence: initial heat release in the low corona, a nearly adiabatic regime at intermediate heights (3--7~$R_\odot$), and renewed heating at larger distances. Importantly, $\Gamma$ converged toward values close to isothermal ($\sim 0.8$--1.2), indicating that expansion cooling is largely balanced by sustained heating. The magnitude of temperature decrease was shown to depend on expansion speed and deceleration, establishing a quantitative thermo–kinematic coupling. These results directly address two research gaps: the need for a radially varying treatment of CME thermodynamics, and the lack of an observational link between remote-sensed 3D kinematics and model-derived internal states.

Extending beyond case studies, statistical analyses of ICMEs across Solar Cycles 23–25 revealed systematic trends in $\Gamma_p$, temperature, and expansion properties. We found that the distribution of $\Gamma_p$ differs significantly between ICMEs and the background solar wind and varies with Solar Cycle phase and storm class. By classifying ejecta into the categories of Heating and Cooling types, we showed that CME thermal states are modulated by Solar Cycle activity: Heating ejecta dominate around solar maxima, while Cooling ejecta prevail during rising phases. Importantly, geoeffective ICMEs, especially magnetic clouds, were associated with very low $\Gamma_p$, enhanced magnetic fields, strong compression, and high expansion speeds. This demonstrates that the thermal state of CMEs is an important determinant of geoeffectiveness, complementing the traditionally emphasized magnetic and plasma parameters. In this way, $\Gamma_p$ along with other plasma and magnetic parameters emerges as a novel diagnostic for CME classification and space-weather forecasting.

The work was further extended from CME-driven plasma properties to the magnetospheric response, focusing on the recovery phase of intense and stronger-than-intense storms during Solar Cycles 23 and 24. Three coherent patterns emerged: (i) reduced storm occurrence and weaker recovery in Cycle 24 reflect diminished heliospheric input; (ii) recovery duration depends more on internal decay processes than on main-phase intensity, indicating that recovery is not a passive response to interplanetary drivers but reflects magnetospheric–ionospheric coupling efficiency; and (iii) while ICMEs are the dominant storm drivers, SIRs produce longer recovery phases due to embedded fluctuations. These results address the gap in driver-specific storm recovery and show that storm evolution cannot be generalized by Dst minimum alone. Together with the CME-focused analyses, this completes the chain from CME thermodynamics to solar wind variability to geomagnetic response.

Finally, the thermodynamic consequences of CME–CME interactions were explored using the great geomagnetic storm that began on 10 May 2024. Interacting CMEs were shown to merge into complex ejecta, reorganizing magnetic topology, compressing plasma, and redistributing energy between species. Their thermodynamic states diverged early but converged toward near-isothermal profiles at larger distances. In situ observations showed double flux-rope structures and species-dependent heating signatures (electrons predominantly in heat-release
states; ions exhibiting bimodal heating), demonstrating that interactions amplify thermodynamic complexity and play a critical role in extreme space-weather events. This addresses the gap in CME interactions, showing that their thermodynamic consequences are as important as their kinematic signatures.

\vspace{1cm}
\textbf{Probable implications:}\\Collectively, the results of this thesis establish a framework linking CME kinematics, thermodynamics, and their space-weather consequences. The implications are twofold. From a scientific perspective, the findings show that CME thermal evolution cannot be captured by fixed assumptions, but instead requires radially varying, species-specific, and observation-constrained treatments. Future models should therefore incorporate dynamic polytropic behavior, internal force balance, and non-adiabatic heating processes to reproduce CME evolution more realistically. From an operational perspective, diagnostics such as the polytropic index and internal-force-modulated expansion, when analyzed alongside plasma and magnetic parameters, emerge as valuable predictors of CME impact. Incorporating these diagnostics into forecasting can refine CME classification, improve impact prediction, and provide more realistic expectations for storm development and recovery. In this way, the thesis advances both the physical understanding and the predictive capability of space-weather science.

\textbf{Limitations:}\\ The conclusions of this thesis should be interpreted in light of several methodological and observational limitations. Thermodynamic inference in the low corona is constrained by EUV/UV diagnostics that are largely confined below $\sim$2~R$_\odot$ and subject to line-of-sight ambiguities, while the linkage to interplanetary evolution relies on single-point in situ sampling (primarily Wind/ACE) at certain distances from the Sun; consequently, remote–in situ associations and continuous radial thermodynamic profiles retain uncertainties.

The FRIS framework, although revised and observationally constrained, assumes simplified flux-rope geometry/force decomposition; explicit energy transport terms (e.g., anisotropic conduction, radiative losses, parameterized turbulent/wave heating). Similarly, physical effects like erosion by reconnection with the ambient wind, deflection/rotation, and cross-sectional distortion are not modeled directly and are instead folded into an effective $\Gamma$, limiting physical specificity. Furthermore, FRIS model–derived results depend on kinematic inputs, and thus the estimated thermodynamic and force parameters inherently carry the limitations and uncertainties associated with GCS model outputs, including projection effects, cadence gaps, and subjective manual fitting.

The nine-event near-Sun sample emphasizes fast CMEs, which may bias inferences about typical radial thermal evolution, and the statistical study at 1~au, though spanning Solar Cycles 23--25, relies on catalog classifications that do not uniformly separate sheath, shock, leading/trailing ME contributions, nor do they fully resolve complex ejecta. Similarly, the storm-recovery analysis is restricted to intense and stronger-than-intense events (hourly Dst), so results may not fully generalize to moderate storms or to higher-resolution morphology captured by the SYM-H index. 

Finally, the thermodynamic diagnostics proposed here (e.g., $\Gamma$) should not be interpreted in isolation as direct indicators of geoeffectiveness; rather, they need to be considered in conjunction with other plasma and magnetic field parameters to meaningfully assess the role of CME thermal state in driving geomagnetic responses. Nevertheless, the thesis findings provide a solid foundation for future investigations to achieve a more comprehensive modeling of CME evolution and to develop improved space weather forecasting techniques.

\section{Future Work} \label{Ch6-sec:fw}

The investigations presented in this thesis open up several avenues for future research, both in terms of scientific exploration and practical applications in space-weather prediction.

\begin{itemize}
    \item \textbf{Improved thermodynamic modeling:} Analytical models such as FRIS can be extended to non-symmetric formulations and multi-fluid frameworks that separately track proton, electron, and heavy-ion temperatures. Such an approach would enable direct comparison with observed species-dependent heating signatures.

    \item \textbf{Novel diagnostics:} Future EUV/UV spectrographs and radio propagation techniques (e.g., Faraday rotation, radio occultation) have the potential to provide direct measurements of CME thermal structure in transit, thereby constraining theoretical models and validating inferred $\Gamma$ values.

    \item \textbf{Continuous thermodynamic tracking:} The present work highlights the variability of $\Gamma$ across heliocentric distances and ICME populations. Future studies should systematically investigate $\Gamma$ evolution using coordinated multi-point observations from missions such as Aditya-L1, PSP, SolO, and Polarimeter to Unify the Corona and Heliosphere (PUNCH). These assets will enable radially resolved thermodynamic profiles of CMEs, improving constraints on heating mechanisms (e.g., turbulence-driven dissipation, reconnection, wave–particle interactions) and their dependence on CME speed, size, and solar-cycle phase. Coordinated use of these multi-scale observations can reconstruct a CME’s complete heating and cooling history, directly addressing the remote–in situ disconnect.

    \item \textbf{Expanded statistical analyses:} Extending thermodynamic surveys to future solar cycles will test the universality of heating and cooling patterns. Sub-1~au datasets from PSP and SolO will be particularly valuable in revealing how $\Gamma$ evolves with radial distance for different CME types.

    \item \textbf{Storm recovery analysis:} A valuable next step is to analyze in depth the statistical results of this thesis with ring-current and radiation belt models, explicitly linking ICME thermal parameters to magnetospheric energy budgets. This would clarify whether CME thermodynamics, beyond magnetic structure alone, significantly modulate recovery timescales and profiles.

    \item \textbf{CME–CME interaction physics:} Further work is needed to quantify how CME mergers redistribute energy between thermal, magnetic, and kinetic components. This requires heliospheric imaging and multi-point in situ sampling to identify localized heating sites within complex ejecta.

    \item \textbf{Operational forecasting integration:} While this thesis demonstrates the diagnostic value of $\Gamma$, such thermodynamic parameters have not yet been fully embedded into operational space-weather models. A concrete future direction is to assimilate these diagnostics into global heliospheric MHD models (e.g., ENLIL, EUHFORIA, SWASTi), testing their predictive capability alongside conventional plasma and magnetic-field metrics. This would clarify whether thermodynamic information improves CME arrival-time forecasts, impact strength estimation, and recovery-phase characterization.
\end{itemize}

Advancing along these directions will require a concerted synthesis of high-cadence imaging and in situ observations, physics-based modeling, and targeted case studies. By incorporating thermodynamic parameters more systematically into CME research, the community can progress towards a more comprehensive understanding of the evolution of CMEs and their space weather impacts.

\newpage

\addtocontents{toc}{\vspace{2em}}

\backmatter
\newpage

\lhead{\emph{Bibliography}}

\setstretch{0}

\end{document}